\documentclass[11pt]{article}
\usepackage{booktabs}
\usepackage{capt-of}
\usepackage{fullpage}
\usepackage{titlesec}

\usepackage{amsmath,amsfonts,amsthm,mathrsfs,xspace,graphicx}
\usepackage[backref,colorlinks,citecolor=blue,bookmarks=true]{hyperref}
\usepackage{mathpazo}
\usepackage{endnotes}
\usepackage{color}
\usepackage{float}
\usepackage{xcolor}
\usepackage{mdframed}
\usepackage{bbm}
\usepackage{suffix} \usepackage{times}
\usepackage{tabularx}
\usepackage{makecell}
\usepackage{amssymb,latexsym}
\usepackage[capitalize]{cleveref}
\usepackage{enumitem}
\usepackage{tikz}
\usepackage{tikz-cd}
\usepackage{multirow}
\usepackage[section]{placeins}
\usepackage[affil-it]{authblk}

\mdfdefinestyle{figstyle}{ linecolor=black!7, backgroundcolor=black!7, innertopmargin=10pt, innerleftmargin=25pt, innerrightmargin=25pt, innerbottommargin=10pt }

\theoremstyle{plain}
\newtheorem{theorem}{Theorem}[section]
\newtheorem{proposition}[theorem]{Proposition}

\newtheorem{lemma}[theorem]{Lemma}
\newtheorem{claim}[theorem]{Claim}
\newtheorem{fact}[theorem]{Fact}
\newtheorem{corollary}[theorem]{Corollary}

\newtheorem{remark}[theorem]{Remark}

\theoremstyle{definition}
\newtheorem{definition}[theorem]{Definition}
\newtheorem{example}[theorem]{Example}

\newcommand{\code}{\mathcal{C}}

\newcommand{\ket}[1]{|#1\rangle}
\newcommand{\bra}[1]{\langle#1|}
\newcommand{\ketbra}[2]{\ket{#1}\!\bra{#2}}
\newcommand{\ip}[2]{\langle #1  | #2 \rangle}
\newcommand{\proj}[1]{\ket{#1}\!\bra{#1}}
\newcommand{\Tr}{\mbox{\rm Tr}}
\newcommand{\Id}{\ensuremath{I}}
\DeclareMathOperator*{\Expectation}{\mathbb{E}}
\newcommand{\Es}[1]{\Expectation_{#1}}

\newcommand{\ol}[1]{\overline{#1}}

\newcommand{\C}{\ensuremath{\mathbb{C}}}
\newcommand{\N}{\ensuremath{\mathbb{N}}}

\newcommand{\complex}{\ensuremath{\mathbb{C}}}

\newcommand{\qp}{\tau}

\newcommand{\F}{\ensuremath{\mathbb{F}}}

\newcommand{\ot}{\otimes}
\newcommand{\Fp}{\F_p}
\newcommand{\Fq}{\F_q}

\newcommand{\ld}{\textsc{ld}}
\newcommand{\downsize}{\kappa}

\newcommand{\downsizem}{\chi}

\newcommand{\R}{\ensuremath{\mathbb{R}}}

\newcommand{\mA}{\ensuremath{\mathcal{A}}}
\newcommand{\mB}{\ensuremath{\mathcal{B}}}
\newcommand{\mC}{\ensuremath{\mathcal{C}}}

\newcommand{\mG}{\ensuremath{\mathcal{G}}}
\newcommand{\mH}{\ensuremath{\mathcal{H}}}

\newcommand{\cM}{\ensuremath{\mathcal{M}}}

\newcommand{\mR}{\ensuremath{\mathcal{R}}}

\newcommand{\mX}{\ensuremath{\mathcal{X}}}
\newcommand{\mY}{\ensuremath{\mathcal{Y}}}

\newcommand{\ind}{\ensuremath{\mathrm{ind}}}

\DeclareMathOperator{\poly}{poly}

\newcommand{\val}{\ensuremath{\mathrm{val}}}
\newcommand{\valco}{\ensuremath{\mathrm{val}^{\mathrm{co}}}}
\newcommand{\ia}{\Id_\alice}
\newcommand{\ib}{\Id_\bob}

\newcommand{\TIMEOUT}{\textsc{TIMEOUT}}
\newcommand{\desc}[1]{\overline{\cal{#1}}}
\newcommand{\supp}{\textsc{Supp}}

\newcommand{\eps}{\varepsilon}

\newcommand{\GX}{\textsc{Gap-Maxcut}}
\newcommand{\GNI}{\textsc{Graph Nonisomorphism}}

\DeclareMathOperator{\polylog}{polylog}

\newcommand{\abs}[1]{\left\vert {#1} \right\vert}
\newcommand{\norm}[1]{\left\| {#1} \right\|}

\DeclareMathOperator{\tr}{tr}

\newcommand{\E}{\mathop{\mathbb{E}}\displaylimits} 

\newcommand{\Matrix}{\mathrm{M}}
\newcommand{\End}{\mathrm{End}}

\newcommand{\game}{\mathfrak{G}}
\newcommand{\sampler}{\mathcal{S}}
\newcommand{\decider}{\mathcal{D}}
\newcommand{\verifier}{\mathcal{V}}
\newcommand{\strategy}{\mathscr{S}}

\newcommand{\type}{\mathcal{T}}

\newcommand{\gamestyle}[1]{\ensuremath{\textsc{#1}}\xspace}
\newcommand{\qld}{\gamestyle{QLD}}
\newcommand{\ms}{\gamestyle{MS}}
\newcommand{\pauli}{\gamestyle{Pauli}}
\newcommand{\ora}{\gamestyle{Orac}}
\newcommand{\pcp}{\gamestyle{PCP}}
\newcommand{\ar}{\gamestyle{AR}}
\newcommand{\intro}{\gamestyle{Intro}}

\newcommand{\labelstyle}[1]{\ensuremath{\textsc{#1}}\xspace}
\newcommand{\EPR}{\labelstyle{EPR}}
\newcommand{\aux}{\labelstyle{aux}}
\newcommand{\ancilla}{\labelstyle{anc}}
\newcommand{\msc}{\labelstyle{MC}}
\newcommand{\msv}{\labelstyle{MV}}
\newcommand{\vertex}[1]{\labelstyle{V#1}}
\newcommand{\edge}[1]{\labelstyle{N#1}}

\newcommand{\xpt}{\labelstyle{X}}
\newcommand{\zpt}{\labelstyle{Z}}
\newcommand{\rxpt}{\labelstyle{R}_\xpt}
\newcommand{\rzpt}{\labelstyle{R}_\zpt}
\newcommand{\dir}[1]{\labelstyle{V#1}}
\newcommand{\coord}{\labelstyle{I}}

\newcommand{\lnf}{\labelstyle{Ln}}

\newcommand{\tvarstyle}[1]{\mathsf{#1}}
\newcommand{\tvar}{\ensuremath{\tvarstyle{t}}}
\newcommand{\lvar}{\ensuremath{\tvarstyle{u}}}
\newcommand{\rvar}{\ensuremath{\tvarstyle{v}}}

\newcommand{\trole}{\ensuremath{v}} 

\newcommand{\alice}{\labelstyle{A}}
\newcommand{\bob}{\labelstyle{B}}

\newcommand{\oracle}{\labelstyle{Oracle}}
\newcommand{\ab}{\{\alice, \bob\}}

\newcommand{\typestyle}[1]{\ensuremath{\textsc{#1}}\xspace}

\renewcommand{\line}{\mathbf{\ell}}

\newcommand{\Point}{\typestyle{Point}}

\newcommand{\Line}{\typestyle{Line}}
\newcommand{\ALine}{\typestyle{ALine}}
\newcommand{\DLine}{\typestyle{DLine}}
\newcommand{\Pair}{\typestyle{Pair}}
\newcommand{\Constraint}{\typestyle{Constraint}}
\newcommand{\Variable}{\typestyle{Variable}}
\newcommand{\Pauli}{\typestyle{Pauli}}
\newcommand{\Sample}{\typestyle{Sample}}
\newcommand{\Read}{\typestyle{Read}}

\newcommand{\Hide}[1]{\typestyle{Hide}_{#1}}

\newcommand{\Oracle}{\typestyle{Oracle}}
\newcommand{\Introspect}{\typestyle{Intro}}
\newcommand{\Intro}{\typestyle{Intro}}

\newcommand{\Anchor}{\typestyle{Anchor}}
\renewcommand{\Game}{\typestyle{Game}}
\newcommand{\AB}{\{\alice, \bob\}}

\newcommand{\abc}[1][\delta]{\otimes I_\bob \simeq_{#1} I_\alice \otimes}

\newcommand{\ldc}{k} 

\newcommand{\class}[1]{\ensuremath{\mathsf{#1}}\xspace}
\newcommand{\NP}{\class{NP}} %
\newcommand{\IP}{\class{IP}} %
\newcommand{\NEXP}{\class{NEXP}} %
\newcommand{\QMIP}{\class{QMIP}} %
\WithSuffix\newcommand\QMIP*{\ensuremath{\class{QMIP}^*}} %
\newcommand{\PSPACE}{\class{PSPACE}} %
\newcommand{\MIP}{\class{MIP}} %
\newcommand{\MIPco}{\class{MIP}^{\mathrm{co}}} %
\newcommand{\RE}{\class{RE}} %
\newcommand{\coRE}{\class{coRE}}
\newcommand{\NEEXP}{\class{NEEXP}} %
\newcommand{\NEEEXP}{\class{NEEEXP}}
\WithSuffix\newcommand\MIP*{\ensuremath{\class{MIP}^*}} %

\newcommand{\Ent}{\mathscr{E}}
\newcommand{\compr}{\textsc{Compr}}
\newcommand{\halt}{\textsc{Halt}}
\newcommand{\machine}{\cal{M}}
\providecommand{\cal}{}
\renewcommand{\cal}[1]{\mathcal{#1}}

\newcommand{\qldparams}{\mathsf{qldparams}}
\mathchardef\mhyphen="2D

\newcommand{\introparams}{\mathsf{introparams}}
\newcommand{\ldparams}{\mathsf{ldparams}}

\newcommand{\pcpparams}{\mathsf{pcpparams}}

\newcommand{\BoundedHalting}{\mathrm{BH}}

\newcommand{\TIME}{\mathsf{TIME}}

\newcommand{\MS}{\mathrm{MS}}

\newcommand{\anch}{\gamestyle{Anch}}

\newcommand{\tmstyle}[1]{\ensuremath{\mathsf{#1}}}
\newcommand{\Compress}{\tmstyle{Compress}}

\newcommand{\ComputeSampler}{\tmstyle{ComputeSampler}}

\newcommand{\ComputeIntroSampler}{\tmstyle{ComputeIntroSampler}}

\newcommand{\ComputeIntroDecider}{\tmstyle{ComputeIntroDecider}}
\newcommand{\ComputeIntroVerifier}{\tmstyle{ComputeIntroVerifier}}
\newcommand{\ComputeOracleVerifier}{\tmstyle{ComputeOracleVerifier}}
\newcommand{\ComputeAnsVerifier}{\tmstyle{ComputeAnsRedVerifier}}
\newcommand{\ComputeParrepVerifier}{\tmstyle{ComputeRepeatedVerifier}}

\newcommand{\detype}{\tmstyle{Detype}}
\newcommand{\PACK}{\tmstyle{PACK}}
\newcommand{\PK}{\tmstyle{P}}

\newenvironment{gamespec}{
  \begin{mdframed}[style=figstyle]}{
  \end{mdframed}}

\newcommand{\polymeas}[3]{\mathrm{PolyMeas}(#1,#2,#3)}
\newcommand{\simulpolymeas}[4]{\mathrm{PolyMeas}(#1,#2,#3, #4)}

\newcommand{\eval}{\mathrm{eval}}

\newcommand{\succinctdecider}{\ensuremath{\mathsf{SuccinctDecider}}}
\newcommand{\circuit}{\mathcal{C}}
\newcommand{\formula}{\mathcal{F}}

\newcommand{\pcpeval}{\Xi}
\newcommand{\pcpverifier}{\mathcal{M}_\ar}
\newcommand{\qlen}{Q}
\DeclareMathOperator{\ev}{eval}

\newcommand{\coded}{\mathrm{Dec}}

\newcommand{\soundness}{\mathrm{sound}}

\newcommand{\rep}{\gamestyle{Rep}}
\newcommand{\sep}{\gamestyle{Sep}}

\newcommand{\binary}[1]{\mathrm{binary}_{#1}}

\newcommand{\notename}[2]{}

\newcommand{\pnote}[1]{}
\newcommand{\tnote}[1]{}
\newcommand{\jnote}[1]{}
\newcommand{\anote}[1]{}
\newcommand{\znote}[1]{}
\newcommand{\hnote}[1]{}
\newcommand{\ftnote}[1]{}
\newcommand{\tdnote}[1]{}

\title{$\MIP^* = \RE$}
\author[1]{Zhengfeng Ji\footnote{Email: jizhengfeng@tsinghua.edu.cn}}
\author[2]{Anand Natarajan\footnote{Email: anandn@mit.edu}}
\author[3]{Thomas Vidick\footnote{Email: thomas.vidick@epfl.ch}}
\author[4]{John Wright\footnote{Email: jswright@berkeley.edu}}
\author[5]{Henry Yuen\footnote{Email: hyuen@cs.columbia.edu}}
\affil[1]{Department of Computer Science and Technology, Tsinghua University}
\affil[2]{Department of Electrical Engineering and Computer Science, Massachusetts Institute of Technology}
\affil[3]{Department of Information and Communication Systems, \'Ecole Polytechnique Fédérale de Lausanne}
\affil[4]{Department of Electrical Engineering and Computer Sciences, University of California, Berkeley}
\affil[5]{Department of Computer Science, Columbia University}

\date{}

\begin{document}

\maketitle

\begin{abstract}
We show that the class $\MIP^*$ of languages that can be decided by a
classical verifier interacting with multiple all-powerful
quantum provers sharing entanglement is equal to the class $\RE$ of recursively
enumerable languages. Our proof builds upon the quantum low-degree test of (Natarajan and Vidick, FOCS 2018) and the classical low-individual degree test of (Ji, et al., 2020) by integrating recent developments from (Natarajan and Wright, FOCS 2019) and combining them with the recursive compression framework of (Fitzsimons et al., STOC 2019).

An immediate byproduct of our result is that there is an efficient reduction from the Halting Problem to the problem of deciding whether a two-player nonlocal game has entangled value $1$ or at most $\frac{1}{2}$.
Using a known connection, undecidability of the entangled value implies a negative answer to Tsirelson's problem: we show, by providing an explicit example, that the closure $C_{qa}$ of the set of quantum tensor product correlations is strictly included in the set $C_{qc}$ of quantum commuting correlations. Following work of (Fritz, Rev. Math. Phys. 2012) and (Junge et al., J. Math. Phys. 2011) our results provide a refutation of Connes' embedding problem from the theory of von Neumann algebras.

\end{abstract}

\newpage

\tableofcontents

\section{Introduction}
\label{sec:intro}

For integer $n,k\geq 2$ define the quantum (spatial) correlation set $C_{qs}(n,k)$ as the subset of $\R^{n^2k^2}$ that contains all tuples $(p_{abxy})$ representing families of bipartite distributions that can be locally generated in nonrelativistic quantum mechanics. Formally, $(p_{abxy}) \in C_{qs}(n,k)$ if and only if there exist separable Hilbert spaces $\mH_\alice$ and $\mH_\bob$, for every $x\in\{1,\ldots,n\}$ (resp. $y\in\{1,\ldots,n\}$), a collection of orthogonal projections $\{A^x_a\}_{a\in\{1,\ldots,k\}}$ on $\mH_\alice$ (resp. $\{B^y_b\}_{b\in\{1,\ldots,k\}}$ on $\mH_\bob$) that sum to identity, and a state (unit vector) $\psi \in \mH_\alice \otimes \mH_\bob$ such that 
\begin{equation}\label{eq:intro-qs}
  \forall x,y\in \{1, 2, \ldots, n\}\;,\quad\forall a,b\in\{1, 2, \ldots, k\}\;,
  \qquad p_{abxy} = \psi^* \big( A^x_a \otimes B^y_b\big) \psi\;.
\end{equation}
Note that due to the normalization conditions on $\psi$ and on $\{A^x_a\}$ and
$\{B^y_b\}$, for each $x,y$, $(p_{abxy})$ is a probability distribution on
$\{1, 2, \ldots, k\}^2$.
By taking direct sums it is easy to see that the set $C_{qs}(n,k)$ is convex. Let $C_{qa}(n,k)$ denote its closure (it is known that $C_{qs}(n,k) \neq C_{qa}(n,k)$, see~\cite{slofstra2019set}).

Our main result is that the family of sets $\{C_{qa}(n,k)\}_{n,k \in \N}$ is extraordinarily complex, in the following computational sense. For any $0<\eps<1$ define the $\eps$-\emph{weak membership problem} for $C_{qa}$ as the problem of deciding, given $n,k \in \N$ and a point $p=(p_{abxy}) \in \R^{n^2 k^2}$, whether $p$ lies in $C_{qa}(n,k)$ or is $\eps$-far from it in $\ell_1$ distance, promised that one is the case. Then we show that for any given $0<\eps<1$ the  $\eps$-{weak membership problem} for $C_{qa}$ cannot be solved by a Turing machine that halts with the correct answer on every input.

We show this by directly reducing the Halting problem to the weak membership problem for $C_{qa}$: we show that for all $0 < \eps < 1$ and any Turing machine $\cM$ one can efficiently compute integers $n,k\in \N$ and a linear functional $\ell_\cM$ on $\R^{n^2k^2}$ such that, whenever $\cM$ halts it holds that 
\begin{equation}\label{eq:intro-win-1}
\sup_{p\in C_{qa}(n,k)} \big|\ell_\cM(p)\big| \,=\,1\;,
\end{equation}
 whereas if $\cM$ does not halt then 
\begin{equation}\label{eq:intro-win-2}
\sup_{p\in C_{qa}(n,k)} \big|\ell_\cM(p)\big| \,\leq\, 1 - \eps\;.
\end{equation}
By standard results in convex optimization, this implies the aforementioned claim on the undecidability of the $\eps$-weak membership problem for $C_{qa}$ (for any $0<\eps<1$). 

Our result has interesting consequences for long-standing conjectures in quantum information theory and the theory of von Neumann algebras. Through a connection that follows from the work of Navascues, Pironio, and Acin \cite{navascues2008convergent} the undecidability result implies a negative answer to Tsirelson's problem \cite{Tsi06}. Let $C_{qc}(n,\allowbreak k)$ denote the set of quantum commuting correlations, which is the set of tuples $(p_{abxy})$ arising from operators $\{A^x_a\}$ and $\{B^y_b\}$ acting on a single Hilbert space $\mH$ and a state $\psi \in \mH$ such that
\begin{equation}\label{eq:intro-qc}
\begin{aligned}
&\forall x,y\in \{1,\ldots,n\}\;,\;\forall a,b\in\{1,\ldots,k\}\;,\\
 &p_{abxy} = \psi^* \big( A^x_a B^y_b\big) \psi \quad\text{and}\quad  \big[A^x_a,\,B^y_b\big]=0\;.
\end{aligned}
\end{equation}
Then \emph{Tsirelson's problem} asks if, for all $n,k$, the sets $C_{qa}(n,k)$ and $C_{qc}(n,k)$ are equal. 
Using results from~\cite{navascues2008convergent} we give integer $n,k$ and an explicit linear function $\ell$ on $\R^{n^2k^2}$ such that 
\[ \sup_{p\in C_{qc}(n,k)} \big|\ell(p)\big| \,=\,1\;,\qquad\text{but}\qquad \sup_{p\in C_{qa}(n,k)} \big|\ell(p)\big| \,\leq\,\frac{1}{2}\;,\]
which implies that $C_{qa}(n,k)\neq C_{qc}(n,k)$. 
By an implication of Fritz~\cite{fritz2012tsirelson} and Junge et al.~\cite{junge2011connes} we further obtain that Connes' Embedding Problem~\cite{connes1976classification} is false; in other words, there exist type II$_1$ von Neumann factors that do not embed in an ultrapower of the hyperfinite II$_1$ factor. We explain these connections in more detail in Section~\ref{sec:consequences} below.

Our approach to constructing such linear functionals on correlation sets goes through the theory of 
interactive proofs from complexity theory. 
To explain this connection we first review the concept of interactive proofs. The reader familiar with interactive proofs may skip the next section to arrive directly at a formal statement of our main complexity-theoretic result in Section~\ref{sec:result}. 

\subsection{Interactive proof systems}
An \emph{interactive proof system} is an abstraction that generalizes the familiar notion of \emph{proof}. Intuitively, given a formal statement $z$ (for example, ``this graph admits a proper $3$-coloring''), a proof $\pi$ for $z$ is information that enables one to check the validity of $z$ more efficiently than without access to the proof (in this example, $\pi$ could be an explicit assignment of colors to each vertex of the graph). 

Complexity theory formalizes the notion of proof in a way that emphasizes the role played by the verification procedure. To explain this, first recall that in complexity theory a \emph{language} $L$ is a subset of $\{0,1\}^*$, the set of all bit strings of any length, that intuitively represents all problem instances to which the answer should be ``yes.'' For example, the language $L=\textsc{3-Coloring}$ contains all strings $z$ such that $z$ is the description (according to some pre-specified encoding scheme) of a $3$-colorable graph $G$. We say that a language $L$ admits efficiently verifiable proofs if there exists an algorithm $V$ (formally, a two-input Turing machine that runs in time polynomial in the length of the first input) that satisfies the following two properties: (i) for any $z\in L$ there is a string $\pi$ such that $V(z,\pi)$ returns $1$ (we say that $V$ ``accepts''), and (ii) for any $z\notin L$ there is no string $\pi$ such that $V(z,\pi)$ accepts. Property (i) is generally referred to as the \emph{completeness} property, and (ii) is the \emph{soundness}. The set of all languages $L$ with both these completeness and soundness properties is denoted by the complexity class $\NP$.

Research in complexity and cryptography in the 1980s and 1990s led to a significant generalization of the notion of ``efficiently verifiable proof.'' The first modification is to allow \emph{randomized} verification procedures by relaxing (i) and (ii) to \emph{high probability} statements: every $z\in L$ should have a proof $\pi$ that is accepted \emph{with probability at least $c$} (the completeness parameter), and for no $z\notin L$ should there be a proof $\pi$ that is accepted \emph{with probability larger than $s$} (the soundness parameter).
A common setting is to take $c=\frac{2}{3}$ and $s=\frac{1}{3}$; standard amplification techniques reveal that the exact values do not significantly affect the class of languages that admit such proofs, provided that they are chosen within reasonable bounds. 

The second modification is to allow \emph{interactive} verification. 
Informally, this means that instead of receiving a proof string $\pi$ in its entirety and making a decision based on it, the verification algorithm (called the ``verifier'') instead communicates with another algorithm called a ``prover,'' and based on the communication decides whether $z\in L$. There are no restrictions on the computational power of the prover, whereas the verifier is required to run in polynomial time.\footnote{The reader may find the following mental model useful: in an interactive proof, an all-powerful prover is trying to convince a skeptical, but computationally limited, verifier that a string $z$ (known to both) lies in the set $L$, even when it may be that in fact $z \notin L$. By interactively interrogating the prover, the verifier can reject false claims, i.e.,\ determine with high statistical confidence whether $z \in L$ or not. Importantly, the verifier is allowed to probabilistically and adaptively choose its messages to the prover. }


To understand how randomization and interaction can help for proof checking, consider the following example of an interactive proof for the language $\GNI$, which contains all pairs of graphs $(G_0,\allowbreak G_1)$ such that $G_0$ and $G_1$ are \emph{not} isomorphic.\footnote{Here and in the rest of the section, we implicitly assume that graphs and tuples of graphs have a canonical encoding as binary strings.}
It is not known if \[\textstyle \GNI\in \NP,\] because it is not clear how to give an efficiently verifiable proof string that two graphs $G_0$ and $G_1$ are not isomorphic. (A proof of isomorphism is, of course, trivial: given a bijection from the vertices of $G_0$ to those of $G_1$ it is straightforward to verify that the bijection induces an isomorphism.) However, consider the following \emph{randomized}, \emph{interactive} verification procedure. Suppose the input to the verifier and prover is a pair of $n$-vertex graphs $(G_0,\allowbreak G_1)$ (if the graphs do not have the same number of vertices, they are trivially nonisomorphic and the verification procedure can automatically accept). The verifier first selects a uniformly random $b\in \{0,\allowbreak 1\}$ and a uniformly random permutation $\sigma$ of $\{1,\allowbreak \ldots,\allowbreak n\}$ and sends the graph $H = \sigma( G_b)$ to the prover. The prover is then supposed to respond with a bit $b'\in \{0,\allowbreak 1\}$; if $b'=b$  the verifier accepts and if $b'\neq b$ it rejects. 

Clearly, if $G_0$ and $G_1$ are not isomorphic then there exists a prover {strategy} to compute $b$ from $H$ with probability $1$: using its unlimited computational power, the prover can determine whether $H$ is isomorphic to $G_0$ or to $G_1$. However, if $G_0$ and $G_1$ \emph{are} isomorphic then the distribution of $H$ is uniform over the isomorphism class of $G_0$, which is the same as the isomorphism class of $G_1$, and the prover (despite having unlimited computational power) cannot distinguish between whether the verifier generated $H$ using $G_0$ or $G_1$. Thus the probability that \emph{any} prover can correctly guess $b'=b$ is exactly $\frac{1}{2}$. As a result, we have shown that the graph nonisomorphism problem has an interactive proof system with completeness $c=1$ and soundness $s=\frac{1}{2}$. Note how little ``information'' is communicated by the prover: a single bit!
The extreme succinctness of the ``proof'' comes from the fact that whether $G_0$ is isomorphic to $G_1$ determines whether a prover can reliably compute, given the data available to it (which is $G_0$,$G_1$, and $H$), the correct bit $b$.

We denote by $\IP$ the class of languages that admit randomized interactive proof systems such as the one just described. 
The class $\IP$ is easily seen to contain $\NP$, but it is thought to be a much larger class: one of the famous results of complexity theory is that $\IP$ is exactly the same as $\mathsf{PSPACE}$~\cite{lund1990algebraic,shamir1990ip}, the class of languages decidable by Turing machines using polynomial space.\footnote{The reason $\mathsf{PSPACE}$ is considered a ``difficult'' class of problems is because many computational problems believed to require super-polynomial or exponential time (such as $\textsc{3-Coloring}$ or deciding whether a quantified Boolean formula is true) can be solved using a polynomial amount of space.} Thus a polynomial-time verifier, when augmented with the ability to interrogate an all-powerful prover and use randomization, can solve computational problems that are (believed to be) vastly more difficult than those that can be checked using static, deterministic proofs (i.e., $\mathsf{NP}$ problems).

\paragraph{Multiprover interactive proofs.}
We now discuss a generalization of interactive proofs called \emph{multiprover interactive proofs}. Here, a polynomial-time verifier can interact with \emph{two} (or more) provers to decide whether a given instance $z$ is in a language $L$ or not. In this setting, after the verifier and all the provers receive the common input $z$, the provers are not allowed to communicate with each other, and the verifier ``cross-interrogates'' the provers in order to decide if $z \in L$. The provers may coordinate a joint strategy ahead of time, but once the protocol begins the provers can only interact with the verifier. As we will see, the extra condition that the provers cannot communicate with each other is a powerful constraint that can be leveraged by the verifier.



Consider the computational problem of deciding membership in a \emph{promise} language called $\GX$. 
A promise language $L$ is specified by two disjoint subsets $L_{yes},L_{no}\subseteq\{0,1\}^*$, and the task is to decide whether a given instance $z$ is in $L_{yes}$ or $L_{no}$, promised that $z \in L_{yes} \cup L_{no}$. In a proof system for a promise language, the completeness case consists of accepting with probability at least $c$ if $z \in L_{yes}$, and the soundness case consists of accepting with probability at most $s$ if $z \in L_{no}$. If $z \notin L_{yes} \cup L_{no}$, then there are no constraints on the behavior of the verifier.



The promise language $\GX$ is defined as follows:\newline $\GX_{yes}$ (resp. $\GX_{no}$) is the set of all graphs $G$ 
such that there is a \emph{cut} (i.e., a bipartition of the vertices) such that at least $90\%$ of edges cross the cut (resp.\ for every cut, at most $60\%$ of edges cross the cut).\footnote{The specific numbers $90\%$ and $60\%$ are not too important; the only thing that really matters is that the first one is strictly less than $100\%$ and the second strictly larger than $50\%$, as otherwise the problem becomes much easier.} For simplicity, we also assume that all graphs in $\GX_{yes}\cup \GX_{no}$ are regular, i.e.,\ the degree is a constant across all vertices in the graph. 

The $\GX$ problem clearly lies in $\NP$, since given a candidate cut it is easy to count the number of edges that cross it and verify that it is at least $90\%$ of the total number of edges. Observe that the length of the proof and the time required to verify it are linear in the size of the graph (the number of vertices and edges). \emph{Finding} the proof is of course much harder, but we are only concerned with the complexity of the verification procedure.

Now consider the following simple \emph{two-prover interactive proof system} for $\GX$. Given a graph $G$, the verification procedure first samples a uniformly random edge $e=\{u,v\}$ in $G$. It then sends a uniformly random $x\in\{u,v\}$ to the first prover, and a uniformly random $y\in \{u,v\}$ to the second prover. Each prover sees its respective question only and is expected to respond with a single bit, $a,b\in\{0,1\}$, respectively. The verification procedure accepts if and only if $a=b$ if $x=y$, and $a\neq b$ if $x\neq y$.

We claim that the verification procedure described in the preceding paragraph is a \emph{multiprover interactive proof system} for the language $\GX$, with completeness $c=0.95$ and soundness $s=0.9$, in the following sense. First, whenever $G\in \GX_{yes}$ then there is a successful strategy for the provers: specifically, the provers can fix an optimal bipartition and consistently answer ``$0$'' when asked about a vertex from one side of the partition, and ``$1$'' when asked about a vertex from the other side; assuming there exists a cut that is crossed by at least $90\%$ of the edges, this strategy succeeds with probability at least $\frac{1}{2} + \frac{1}{2}0.9$, where the first factor $\frac{1}{2}$ arises from the case when both provers are sent the same vertex, in which case they always succeed. 

Conversely, suppose given a strategy for the provers that is accepted with probability $p = \frac{1}{2} + \frac{1}{2}(1-\delta)$ when the verification procedure is executed on a (regular) $n$-vertex graph $G$. 
We then claim that $G$ has a cut crossed by at least a $1-2\delta$ fraction of all edges. To show this, we leverage the noncommunication assumption on the provers. Since either prover's question is always a single vertex, their strategy can be represented by a function from the vertices of $G$ to answers in $\{0,1\}$. Any such function specifies a bipartition of $G$. While the provers'  bipartitions need not be identical, the fact that they succeed with high probability, for the case when they are sent the same vertex, implies that they must be consistent with high probability. Finally, the fact that they also succeed with high probability when sent opposite endpoints of a randomly chosen edge implies that either prover's bipartition must be cut by a large number of edges. Taking the contrapositive establishes the soundness property.

We denote by $\MIP$ the class of languages that have multiprover interactive proof systems such as the one described in the preceding paragraph. 
Note that, in comparison to the $\NP$ verification procedure for $\GX$ considered earlier, the interactive, two-prover verification is much more efficient in terms of the effort required for the verifier.
 Assuming the graph is provided in a convenient format,\footnote{For example, the graph can be specified via a circuit that takes as input an edge index --- using some arbitrary ordering --- and returns labels for the two endpoints of the edge.} it is possible to sample a random edge and verify the provers' answers in time and space that scales \emph{logarithmically} with the size of the graph. 
 This exponential improvement in the efficiency of the verification procedure serves as the starting point for another celebrated result from complexity theory: $\mathsf{MIP}$ is exactly the same as the class $\mathsf{NEXP}$~\cite{babai1991non}, which are problems that admit \emph{exponential-time checkable proofs}.\footnote{An example of such a problem is the language $\textsc{Succinct-3-Coloring}$, which contains descriptions of polynomial-size circuits $C$ that specify a $3$-colorable graph $G_C$ on \emph{exponentially many} vertices.} The class $\NEXP$ contains $\PSPACE$, but is believed to be much larger; this suggests that the ability to interrogate more than one prover enables a polynomial-time verifier to verify much more complex statements.
 

\paragraph{Nonlocal games.}
In this paper we will only be concerned with multiprover interactive proof systems that consist of a single round of communication with two provers: the verifier first sends its questions to each of the provers, the provers respond with their answers, and the verifier decides whether to accept or reject. The class of problems that admit such interactive proofs is denoted $\MIP(2,1)$, and it is known that $\MIP=\MIP(2,1)$~\cite{feige1992two}. Such proof systems have a convenient reformulation using the language of \emph{nonlocal games}, that we now explain. 

In a nonlocal game, we say that a verifier interacts with multiple noncommunicating \emph{players} (instead of provers --- there is no formal difference between the two terms). An $n$-question, $k$-answer nonlocal game $\game$ is specified by two procedures: a \emph{question sampling} procedure that samples a pair of questions $(x,y)\in \{1,\ldots,n\}^2$ for the players according to a distribution $\mu$ (known to the verifier and the players), and a \emph{decision} procedure that takes as input the players' questions and their respective answers $a,b\in\{1,\ldots,k\}$ and evaluates a predicate $D(x,y,a,b)\in \{0,1\}$ to determine the verifier's acceptance or rejection. In classical complexity theory, the main quantity associated with a nonlocal game $\game$ is its \emph{classical value}, which is the maximum success probability that two cooperating but noncommunicating players have in the game. Formally, the classical value is defined as
\begin{equation}\label{eq:class-val}
 \val(\game) = \sup_{p\in C_c(n,k)} \sum_{x,y} \mu(x,y) \sum_{a,b} D(x,y,a,b) p_{abxy}\;,
\end{equation}
where the set $C_c(n,k)$ is the set of \emph{classical correlations}, which are tuples $(p_{abxy})$ such that there exists a set $\Lambda$ with probability measure $\nu$ and for every $\lambda \in \Lambda$ functions $A^\lambda,B^\lambda: \{1,2,\ldots,n\} \to \{1,2,\ldots,k\}$ such that
\[
\begin{aligned}
&\forall x,y \in \{1,2,\ldots,n\}, \quad \forall a,b \in \{1,2,\ldots,k\}, \\
 &p_{abxy} = \Pr_{\lambda \sim \nu} (A^\lambda(x) = a \wedge B^\lambda(y) = b).
\end{aligned}
\]
%
 This definition captures the intuitive notion that a classical strategy for the players is specified by (i) a distribution $\nu$ on $\Lambda$ that represents some probabilistic information shared by the players that is independent of the verifier's questions, and (ii) two functions $A^\lambda,B^\lambda$ that represent each players' ``local strategy'' for answering given their shared randomness $\lambda$ and question $x$ or $y$. 
\footnote{For the functional analyst we briefly note that if we define a tensor 
\[L  = \sum_{x,y,a,b} \mu(x,y)D(x,y,a,b) e_{xa}\otimes e_{yb} \in \R^{nk}\otimes \R^{nk}\]
then $\val(\game) = \|L\|_{\ell_1^n(\ell_\infty)^k \otimes_\epsilon \ell_1^n(\ell_\infty)^k}$, with $\otimes_\epsilon$ denoting the injective tensor norm of Banach spaces. (For more connections between interactive proofs, nonlocal games and tensor norms we refer to the survey~\cite{palazuelos2016survey}.)} Note that due to the shared randomness $\lambda$, the set $C_c(n,k)$ is a (closed) convex subset of $[0,1]^{n^2k^2}$. 

To make the connection with interactive proof systems, observe that the assertion that $L\in \MIP(2,1)$ precisely amounts to the specification of an efficient mapping from problem instances $z$ to games $\game_z$ such that whenever $z\in L$ then  $\val(\game_z) \geq \frac{2}{3}$, whereas if $z \notin L$ then $\val(\game_z) \leq \frac{1}{3}$.\footnote{Here by ``efficient'' we mean that there should be a polynomial-time Turing machine that on input $z$ returns (i) a polynomial-size randomized circuit that samples from $\mu$, and (ii) a polynomial-size circuit that evaluates the predicate $D$; see item 1 in Definition~\ref{def:mipstar}.} Thus the complexity of the optimization problem~\eqref{eq:class-val} captures the complexity of the decision problem $L$. The aforementioned characterization of $\MIP$ as the class $\NEXP$ by~\cite{babai1991non} shows that in general this optimization problem is very difficult: it is as hard as deciding any language in $\NEXP$. 


\subsection{Statement of result}
\label{sec:result}

We now introduce the main complexity class that is the focus of this paper: $\MIP^*$, the  ``entangled-prover'' analogue of the class $\MIP$ considered earlier. Informally the class $\MIP^*$, first introduced in~\cite{cleve2004consequences}, contains all languages that can be decided by a classical poly\-nomial-time verifier interacting with multiple \emph{quantum} provers sharing \emph{entanglement}. We focus on the class $\MIP^*(2,\allowbreak 1)$, which corresponds to the setting of one-round protocols with two  provers. Equivalently, a language $L$ is in $\MIP^*(2,\allowbreak 1)$ if and only if there is an efficient mapping from instances $z\in\{0,\allowbreak 1\}^*$ to nonlocal games $\game_z$ such that if $z \in L$, then $\val^*(\game_z) \geq 2/3$ and otherwise $\val^*(\game_z) \leq 1/3$. Here, for an $n$-question, $k$-answer game $\game$, we let $\val^*(\game)$ denote its \emph{entangled value}, which is defined as
\begin{equation}\label{eq:ent-val}
 \val^*(\game) = \sup_{p\in C_{q}(n,k)} \sum_{x,y} \mu(x,y) \sum_{a,b} D(x,y,a,b) p_{abxy}\;,
\end{equation}
where the set $C_{q}(n,k)$ is the set of all finite-dimensional quantum correlations, i.e.,\ correlations of the form~\eqref{eq:intro-qs} where $\mH_\alice$ and $\mH_\bob$ are restricted to be finite-dimensional. Although the sets $C_q(n,k)$ and $C_{qs}(n,k)$ in general are distinct~\cite{coladangelo2018unconditional}, \tnote{edit for ref:}it is well known (see, e.g.,~\cite[Theorem 3]{scholz2008tsirelson}) that these two sets have the same closure $C_{qa}(n,k)$, and, therefore, the supremum in~\eqref{eq:ent-val} can equivalently be taken over $C_q(n,k)$, $C_{qs}(n,k)$ or $C_{qa}(n,k)$. We use $C_q$ for convenience in the analysis.

Since $C_c(n,k) \subseteq C_{q}(n,k)$ for all $n,k$, we have that $\val(\game) \leq \val^*(\game)$; in other words, quantum spatial strategies can perform at least as well as classical strategies in a nonlocal game. 


The consideration of quantum strategies and the set $C_{qs}(n,k)$ for the definition of $\MIP^*$ is motivated by a long line of works in the foundations of quantum mechanics around the topic of \emph{Bell inequalities}, that are linear functionals which separate the sets $C_{c}(n,k)$ and $C_{qs}(n,k)$. The simplest such functional is the \emph{CHSH inequality}~\cite{clauser1969proposed}, that shows $C_c(2,2) \subsetneq C_{qs}(2,2)$. The CHSH inequality can be reformulated as a game $\game$ such that $\val^*(\game) > \val(\game)$. This game is very simple: it is defined by setting $\mu(x,y)=\frac{1}{4}$ for all $x,y\in\{0,1\}$ and $D(x,y,a,b)=1$ if and only if the parity $a\oplus b = x\wedge y$. It can be shown that $\val(\game) = \frac{3}{4}$ and $\val^*(\game) = \frac{1}{2} + \frac{1}{2\sqrt{2}} > \frac{3}{4}$. The study of Bell inequalities is a large area of research not only in foundations, where they are a tool to study the nonlocal properties of entanglement, but also in quantum cryptography, where they form the basis for cryptographic protocols for e.g., quantum key distribution~\cite{ekert1991quantum}. 

The introduction of entanglement in the setting of interactive proofs has interesting consequences for complexity theory; indeed it is not \textup{a priori} clear how the class $\MIP^*$ compares to $\MIP$. Take a language $L\in \MIP(2,1)$, and let $z$ be an instance. Then the associated game $\game_z$ is such that $\val(\game_z)\geq\frac{2}{3}$ if $z\in L$, and $\val(\game_z)\leq\frac{1}{3}$ otherwise. The fact that in general $\val^*(\game_z) \geq \val(\game_z)$ (and that as demonstrated by the CHSH game inequality can be strict) cuts both ways. On the one hand, the soundness property can be affected, so that instances $z\notin L$ could have $\val^*(\game_z) = 1$, meaning that we would not be able to establish that $L \in \MIP^*$.
On the other hand, a language $L \in \MIP^*(2,1)$ may not necessarily be in $\MIP$, because for $z\in L$ the fact that $\val^*(\game_z)\geq\frac{2}{3}$ does not automatically imply $\val(\game_z)> \frac{1}{3}$ (in other words, the game $\game_z$ may require the players to use a quantum strategy in order to win with probability greater than $1/3$). Just as the complexity of the class $\MIP$ is characterized by the complexity of approximating the classical value of nonlocal games (the optimization problem in~\eqref{eq:class-val}), the complexity of $\MIP^*$ is intimately related to the complexity of approximating the entangled value of games (the optimization problem in~\eqref{eq:ent-val}).

In~\cite{ito2012multi} the first nontrivial lower bound on $\MIP^*$ was shown, establishing that $\MIP = \NEXP \subseteq \MIP^*$. (Earlier results~\cite{kempe2011entangled,ito2009oracularization} gave more limited hardness results, for approximating the entangled value up to inverse polynomial precision.) This was proved by arguing that for the specific games constructed by~\cite{babai1991non} that show $\NEXP \subseteq \MIP$, the classical and entangled values are approximately the same. In other words, the classical soundness and completeness properties of the proof system of~\cite{babai1991non} are maintained in the presence of shared entanglement between the provers. 
Following~\cite{ito2012multi} a sequence of works~\cite{vidick2016three,ji2016classical,natarajan2018two,ji2017compression,natarajan2018low,fitzsimons2018quantum} established progressively stronger lower bounds on the complexity of approximating the entangled value of nonlocal games, culminating in~\cite{NW19} which showed that approximating the entangled value is at least as hard as $\NEEXP$, the collection of languages decidable in nondeterministic \emph{doubly exponential} time. This proves that $\NEEXP \subseteq \MIP^*$, and since it is known that $\NEXP\subsetneq \NEEXP$ it follows that $\MIP \neq \MIP^*$.


In contrast to these increasingly strong lower bounds the only upper bound known on $\MIP^*$ is the trivial inclusion $\MIP^* \subseteq \RE$, the class of recursively enumerable languages, i.e., languages $L$ such that there exists a Turing machine $\cM$ such that $x\in L$ if and only if $\cM$ halts and accepts on input $x$. This inclusion follows since the supremum in~\eqref{eq:ent-val} can be approximated from below by performing an exhaustive search in increasing dimension and with increasing accuracy. We note that, in addition to containing all decidable languages, this class also contains undecidable problems such as the Halting problem, which is to decide whether a given Turing machine eventually halts.

Our main result is a proof of the reverse inclusion: $\RE \subseteq \MIP^*$. Combined with the preceding observation it follows that 
\[ \MIP^* \,=\, \RE \;,\]
which is a full characterization of the power of entangled-prover interactive proofs. In particular, for any $0 < \eps < 1$, it is an undecidable problem to determine whether a given nonlocal game has entangled value $1$ or at most $1 - \eps$ (promised that one is the case).

\paragraph{Proof summary.}
The proof of the inclusion $\RE \subseteq \MIP^*$ is obtained by designing an entangled-prover interactive proof for the Halting problem, which is complete for the class $\RE$. Specifically, we design an efficient transformation that maps any Turing machine $\cal{M}$ to a 
nonlocal game $\game_{\cal{M}}$ such that, if $\cal{M}$ halts (when run on an empty input tape) then there is a quantum strategy for the provers that succeeds with probability $1$ in $\game_{\cal{M}}$ (i.e.,\ $\val^*(\game_{\cal{M}})=1$), whereas if $\cal{M}$ does not halt then no quantum strategy can succeed with probability larger than $\frac{1}{2}$ in the game (i.e.,\ $\val^*(\game_{\cal{M}})\leq \frac{1}{2}$). Furthermore, the game $\game_\cal{M}$ has the property that whenever $\val^*(\game_\cal{M})=1$ then this fact is witnessed by a \emph{synchronous} strategy, i.e.,\ a strategy where the players always give the same answer when simultaneously asked the same question. Synchronous strategies, or correlations, were first introduced in~\cite{paulsen2016estimating} and have played an important role in approaches to CEP based on quantum information and the study of nonlocal games; see, e.g.,~\cite{kim2018synchronous} and references therein. 
In the paper we use a terminology of \emph{projective, consistent and commuting} (PCC) strategies (see Definition~\ref{def:spcc} in Section~\ref{sec:strat}), a notion which implies the notion of being synchronous. 


A very rough sketch of this construction is as follows (we give a detailed overview in Section~\ref{sec:overview}). Given an infinite family of games $\{ \game_n \}_{n \in \N}$, we say that the family is \emph{uniformly generated} if there is a polynomial-time Turing machine that on input $n$ returns a description of the game $\game_n$. Given a game $\game$ and $p\in[0,1]$ let $\Ent(\game,p)$ denote the minimum local dimension of an entangled state shared by the players in order for them to succeed in $\game$ with probability at least $p$. 

We proceed in two steps. First, we design a \emph{compression} procedure for a specific class of nonlocal games that we call \emph{normal form}. Given as input a uniformly generated family $\{ \game_{n} \}_{n \in \N}$ of normal form games, the compression procedure returns another uniformly generated family $\{\game'_{n}\}_{n \in \N}$ of normal form games with the following properties: (i) for all $n$, if $\val^*(\game_{2^n})=1$ then $\val^*(\game'_n)=1$, and (ii) for all $n$, if $\val^*(\game_{2^n})\leq \frac{1}{2}$ then $\val^*(\game'_n)\leq \frac{1}{2}$ and, moreover,
\[ \Ent(\game'_{n},\frac{1}{2})\,\geq\, \max \Big\{ \Ent\Big(\game_{2^n},\frac{1}{2}\Big), \,2^{2^{\Omega(n)}} \Big\}\;.\] 

The construction of this compression procedure is our main contribution. Informally, it combines the recursive compression technique developed in~\cite{ji2017compression,fitzsimons2018quantum} with the so-called ``introspection'' technique of~\cite{NW19} that was used to prove $\NEEXP \subseteq \MIP^*$. The introspection technique itself relies heavily on the quantum low individual degree test of~\cite{natarajan2018low,ML20}\tnote{added ref. to ML doc here} to robustly self-test certain distributions that arise from constructions of \emph{classical} probabilistically checkable proofs. The quantum low-degree test and the introspection technique allow us to avoid the shrinking gap limitation of the results from~\cite{fitzsimons2018quantum}.


In the second step, we use the compression procedure in an iterated fashion to construct an interactive proof system for the Halting problem. Fix a Turing machine $\cM$ and consider the following family  of nonlocal games $\{\game_{\cM,n}^{(0)}\}_{n\in\N}$: for all $n \in \N$, if $\cM$ halts in at most $n$ steps (when run on an empty input tape), then $\val^*(\game_{\cM,n}^{(0)})=1$, and otherwise  $\val^*(\game_{\cM,n}^{(0)}) \leq \frac{1}{2}$.\footnote{There is nothing special about the choice of $\frac{1}{2}$; this can be set to any constant that is less than $1$.} 

Constructing such a family of games is trivial; furthermore, they can be made in the  ``normal form'' required by the compression procedure. However, consider applying the compression procedure to obtain a family of normal form games $\{ \game_{\cM,n}^{(1)} \}_{n\in\N}$. Then for all $n \in \N$, it holds that if $\cM$ halts in at most $2^n$ steps then $\val^*(\game_{\cM,n}^{(1)})=1$, and otherwise $\val^*(\game_{\cM,n}^{(1)})\leq \frac{1}{2}$, and, furthermore, any strategy that achieves a value of at least $\frac{1}{2}$ requires an entangled state of dimension at least $2^{2^{\Omega(n)}}$.


Intuitively, one would expect that iterating this procedure and ``taking the limit'' gives a family of games $\{ \game_{\cM,n}^{(\infty)} \}_{n\in\N}$ such that if $\cM$ halts then $\val^*(\game_{\cM,n}^{(\infty)})=1$ for all $n \in \N$, whereas if $\cM$ does not halt then no finite-dimensional strategy can succeed with probability larger than $\frac{1}{2}$ in $\game_{\cM,n}^{(\infty)}$, for all $n \in \N$; in particular, $\val^*(\game_{\cM,n}^{(\infty)})\leq \frac{1}{2}$. 
Formally, we do not take such a limit but instead define directly the family of games $\{ \game_{\cM,n}^{(\infty)} \}_{n\in\N}$ as a \emph{fixed point} of the Turing machine that implements the compression procedure. The game $\game_\cM$ can then be taken as $\game_{\cM,1}^{(\infty)}$. We describe this in more detail in Section~\ref{sec:overview}.

\subsection{Consequences}
\label{sec:consequences}

	Our result is motivated by a connection with Tsi\-relson's problem from quantum information theory, itself related to Connes' Embedding Problem in the theory of von Neumann algebras~\cite{connes1976classification}. 
In a celebrated sequence of papers, Tsirelson~\cite{tsirelson1993some} initiated the systematic study of quantum correlation sets. Recall the definition of the set of quantum spatial correlations 
\begin{equation}
\begin{aligned}
\label{eq:spatial}
	&C_{qs}(n,k) = \big\{ (p_{abxy} ) \mid p_{abxy} = \bra{\psi} A^x_a \otimes B^y_b \ket{\psi}, \\
 &\ket{\psi} \in  \mH_\alice \otimes \mH_\bob,\,\;\forall xy,\, \{A^x_a\}_a,\{B^y_b\}_b\text{ POVM}\big\}\;,
\end{aligned}
\end{equation}
where here $\ket{\psi}$ ranges over all unit norm vectors $\ket{\psi} \in \mH_\alice \otimes \mH_\bob$ with $\mH_\alice$ and $\mH_\bob$ arbitrary (separable) Hilbert spaces, and a POVM is defined as a collection of positive semidefinite operators that sum to identity. (From now on we use the Dirac ket notation $\ket{\psi}$ for states.) Recall the closure $C_{qa}(n,k)$ of $C_{qs}(n,k)$.

Tsirelson observed that there is a natural alternative definition to the quantum
spatial correlation set, called the \emph{quantum commuting correlation set} and
defined as
\begin{equation}
  \label{eq:commuting}
  C_{qc}(n,k) = \bigl\{ (p_{abxy}) \mid p_{abxy} = \bra{\psi} A^x_a\,  B^y_b
  \ket{\psi} \bigr\}\;,
\end{equation}
where $\ket{\psi}\in \mH$ is a quantum state, $\{A^x_a\}$ and $\{B^y_b\}$ are
POVMs for all $x, y$, and $[A^x_a, B^y_b] = 0$ for all $a, b, x, y$.
Note the key difference with spatial correlations is that
in~\eqref{eq:commuting} all operators act on the same (separable) Hilbert space.
The requirement that operators associated with different inputs (questions)
$x,y$ commute is arguably a minimal requirement within the context of quantum
mechanics for there to not exist any causal connection between outputs (answers)
$a,b$ obtained in response to the respective input.

The set $C_{qc}(n,k)$ is closed and convex, and it is easy to see that $C_{qa}(n,k)\subseteq C_{qc}(n,k)$ for all $n,k\geq 1$. When Tsirelson initially introduced these sets he claimed that equality holds. However, it was later pointed out that this is not obviously true. The question of equality between $C_{qc}$ and $C_{qa}$ (for all $n,k$) is now known as \emph{Tsirelson's problem}~\cite{Tsi06}. Let $	C_q(n,k) $ denote the same as $C_{qs}(n,k)$ except that both $\mH_\alice$ and $\mH_{\bob}$ in~\eqref{eq:spatial} are restricted to finite-dimensional spaces. Then more generally one can consider the following chain of inclusions
\begin{equation}
\label{eq:corr-sets-chain}
	C_q(n,k) \subseteq C_{qs}(n,k) \subseteq C_{qa}(n,k) \subseteq C_{qc}(n,k)\;,
\end{equation}
for all $n,\allowbreak k \in \N$, and ask which (if any) of these inclusions are strict. We let $C_q,\allowbreak C_{qs},\allowbreak C_{qa},\allowbreak C_{qc}$ denote the disjoint union of $C_q(n,\allowbreak k),\allowbreak C_{qs}(n,\allowbreak k),\allowbreak C_{qa}(n,\allowbreak k),\allowbreak C_{qc}(n,\allowbreak k)$, respectively, over all integers $n,\allowbreak k \in \N$. 
In a breakthrough work, Slofstra established the first separation between these four correlation sets by proving that $C_{qs} \neq C_{qc}$~\cite{slofstra2019tsirelson}; he later proved the stronger statement that $C_{qs} \neq C_{qa}$~\cite{slofstra2019set}. As a consequence of the technique used to demonstrate the separation Slofstra also obtains the complexity-theoretic statement that the problem of determining whether an element $p$ lies in $C_{qc}$, even promised that if it does, then it also lies in $C_{qa}$, is undecidable. Interestingly, this is shown by reduction from the \emph{complement} of the halting problem; for our result we reduce from the halting problem (see Section~\ref{sec:open} for further discussion of this point). 
Since his work, simpler proofs of Slofstra's results have been found~\cite{dykema2019non,musat2018non,coladangelo2019two}. In~\cite{coladangelo2018unconditional}, Coladangelo and Stark showed that $C_{q} \neq C_{qs}$ by exhibiting a $5$-input, $3$-output correlation that can be attained using infinite-dimensional spatial strategies (i.e.,\ infinite-dimensional Hilbert spaces, a state and POVMs satisfying~\eqref{eq:spatial}) but cannot be attained via finite-dimensional strategies.

As already noted in~\cite{fritz2014can} (and further elaborated on by~\cite{fitzsimons2018quantum}), the undecidability of $\MIP^*(2,1)$ 
implies the separation $C_{qa} \neq C_{qc}$.\footnote{Technically~\cite{fritz2014can} make the observation for the commuting-prover analogue $\MIPco(2,1)$, discussed further in Section~\ref{sec:open}, but the reasoning is the same.} This follows from the observation that if $C_{qa} = C_{qc}$, then there exists an algorithm that can correctly determine if a nonlocal game $\game$ satisfies $\val^*(\game)=1$ or $\val^*(\game)\leq  \frac{1}{2}$ and always halts: this algorithm interleaves a hierarchy of semidefinite programs providing outer approximations to the set $C_{qc}$~\cite{navascues2008convergent,doherty2008quantum} with a simple exhaustive search procedure providing inner approximations to $C_{q}$. Our result that $\RE \subseteq \MIP^*(2,1)$ implies that no such algorithm exists, thus resolving Tsirelson's problem in the negative.\footnote{For the formal argument, we refer to e.g.,~\cite[Section 7]{goldbring2024universal}.}

We, furthermore, exhibit an explicit nonlocal game $\game$ such that $\val^*(\game) < \valco(\game) = 1$, where $\valco(\game)$ is defined as $\val^*(\game)$ except that the supremum is taken over the set $C_{qc}(n,k)$ in~\eqref{eq:commuting}. This in turn yields an explicit correlation that is in the set $C_{qc}$ but not in $C_{qa}$. 
This game closely resembles the game $\game_\machine$ described in the sketch of the proof that $\RE \subseteq \MIP^*$, where $\machine$ is the Turing machine that runs the hierarchy of semidefinite programs on the game $\game_\machine$ and halts if it certifies that $\valco(\game_\machine) < 1$. It is in principle possible to determine an upper bound on the parameters $n,k$ for our separating correlation from the proof. While we do not provide such a bound, there is no step in the proof that requires it to be astronomical; e.g.,\ we believe (without proof) that $10^{20}$ is a clear upper bound.

\paragraph{Connes' Embedding Problem.} 
Connes' Embedding Problem (CEP)~\cite{connes1976classification} is a conjecture in the theory of von Neumann algebras. Briefly, CEP posits that every \tnote{added:}separable type II$_1$ von Neumann factor embeds into an ultrapower of the hyperfinite II$_1$ factor. This question rose to prominence most notably through the work of Kirchberg~\cite{kirchberg1993non} who showed its equivalence with a range of notable conjectures in operator algebras, including the so-called Kirchberg QWEP conjecture. 
We refer to~\cite{ozawa2013connes} for a precise formulation of the problem and connections to other conjectures in operator algebras, such as  Kirchberg's QWEP conjecture. In independent work Fritz~\cite{fritz2012tsirelson} and Junge et al.~\cite{junge2011connes} showed that a positive answer to CEP would imply a positive resolution of Tsirelson's problem, i.e.,\ that $C_{qa}(n,k)=C_{qc}(n,k)$ for all $n,k$. (This was later promoted to an equivalence by Ozawa~\cite{ozawa2013connes}.) Since our result disproves this equality for some $n,k$ it also implies that CEP does not hold. In work that appeared subsequently to the first announcement of our results, Goldbring and Hart~\cite{goldbring2016computability,goldbring2024universal} show using arguments from continuous logic that the uncomputability of approximating $\val^*(\game)$ refutes the CEP. Interestingly, their argument uses general elementary considerations from logic and completely bypasses Tsirelson's problem and its equivalence with Kirchberg's QWEP conjecture. We note that our result is not sufficiently constructive to immediately lead to an explicit description of a factor that does not embed into an ultrapower of the hyperfinite II$_1$ factor, as this would a priori require a description of a commuting strategy that demonstrates $\valco(\game) = 1$ for a $\game$ such that  $\val^*(\game) < \valco(\game)$, and we do not give such a description. 

\paragraph{Entanglement tests.} As a step towards showing our result for any integer $n\geq 1$ we construct a game $\game_n$, with question and answer length polynomial in the size of the smallest Turing machine $\cal{M}_n$ that halts (on the empty tape) in exactly $n$ steps (i.e.,\ the \emph{Kolmogorov complexity} of $n$), such that $\val^*(\game_n)=1$ yet any quantum strategy that succeeds in $\game_n$ with probability larger than $\frac{1}{2}$ must use an entangled state whose Schmidt rank is at least $2^{\Omega(n)}$. This is by far the most efficient entanglement test that we are aware of.

\paragraph{Prover and round reduction for $\MIP^*$ protocols.} Let $\MIP^*(k,r)$ denote the collection of languages decidable by $\MIP^*$ protocols with $k\geq 2$ provers and $r$ rounds. Prior to our work it was known how to perform \emph{round reduction} for $\MIP^*$ protocols, at the cost of adding provers; it was shown by~\cite{ji2017compression,fitzsimons2018quantum} that $\MIP^*(k,r) \subseteq \MIP^*(k + 15,1)$ for all $k, r$. However, it was an open question whether the complexity of the class $\MIP^*$ increases if we add more provers.
Our main complexity-theoretic result implies that $\MIP^* = \MIP^*(2,1)$. This follows from the following chain of inclusions: for all polynomially-bounded functions $k,r$,
 \[
 	\MIP^*(2,1) \subseteq \MIP^*(k,r) \subseteq \RE \subseteq \MIP^*(2,1)\;.
\]
The first inclusion follows since the verifier in an $\MIP^*$ protocol can always ignore extra provers and rounds; the second inclusion follows from a simple exhaustive-search procedure that enumerates over strategies for a given $\MIP^*(k,r)$ protocol; the third result is proven in this paper.\footnote{In fact, we note that the second term $\MIP^*(k,r)$ can be replaced by $\QMIP^*(k,r)$, which is the analogous class with a quantum verifier and quantum messages, since the first inclusion is trivial and the second remains true. As a result, we obtain that $\QMIP^* = \MIP^*(2,1)$ as well.}

However, this method of reducing provers and rounds in a given $\MIP^*$ protocol is indirect; it involves first converting a given $\MIP^*$ protocol into a Turing machine that accepts if and only if the $\MIP^*$ protocol has value larger than $\frac{1}{2}$, and then constructing an $\MIP^*(2,1)$ protocol to decide whether the Turing machine halts. In particular, this transformation does not generally preserve the complexity of the provers and verifier in the original protocols. We leave it as an open question to find a more direct method for reducing the number of provers in an $\MIP^*$ protocol.

%
	%
	%

\subsection{Open questions}
\label{sec:open}

We mention several questions left open by our work.

\paragraph{Explicit constructions of counter-examples to Connes' Embedding Problem.} We provide an explicit counter-example to Tsirelson's problem in the form of a game whose entangled value differs from its commuting-operator value. Through the aforementioned connection with Connes' embedding Problem \cite{fritz2012tsirelson,junge2011connes,ozawa2013connes}, the counter-example may lead to the construction of interesting objects in other areas of mathematics. A first question is whether it can lead to an explicit description of a type II$_1$ factor that does not satisfy the Connes embedding property. Such a construction could be obtained along the lines of~\cite{kim2018synchronous}, using the fact that our game $\game$ such that $\val^*(\game) < \valco(\game) = 1$ has the property of being \emph{synchronous}, i.e.,\ perfect strategies in the game are required to return the same answer when both parties are provided the same question. 

Going further, one may ask if the example can eventually lead to a construction of a group that is not hyperlinear. A group is hyperlinear if and only if its group von Neumann algebra embeds into the hyperfinite $II_1$ factor $\mathcal{R}^\omega$. Thus, if the nonembeddable $II_1$ factor whose existence follows from our work arose from a group, then this group would be nonhyperlinear. Currently it is not known if all groups are hyperlinear.\footnote{After the completion of this works, it was determined by automated methods that there exists a non-sofic group~\cite{OAI}. Being sofic is a related but potentially stronger condition than being hyperlinear. Very recent subsequent work led to further concrete counter-examples~\cite{alekseev2026spectral,gao2026explicit}; these works are very recent and heavily AI-assisted and we are unable to survey them in any detail.} We refer to the survey~\cite{capraro2015introduction} for more on this connection.  

\paragraph{The complexity of variants of $\MIP^*$.} Our result characterizes the complexity class $\MIP^*$ as the set of recursively enumerable languages. One can also consider the complexity class $\MIPco$, which stands for \emph{multiprover interactive proofs in the commuting-operator model}. For the sake of the discussion we consider only two-prover one-round protocols; a language $L$ is in $\MIPco(2,1)$ if there exists an efficient reduction that maps $z \in \{0,1\}^*$ to a nonlocal game $\game_z$ such that if $z \in L$ then $\valco(\game_z) \geq \frac{2}{3}$, and otherwise $\valco(\game_x) \leq \frac{1}{3}$.

The semidefinite programming hierarchy of~\cite{navascues2008convergent,doherty2008quantum} can be used to show that $\MIPco(2,1)$ is contained in the \emph{complement} of $\RE$, denoted as $\coRE$: to certify that $z \notin L$ it suffices to run the hierarchy until it obtains a certificate that $\valco(\game_z) < \frac{2}{3}$. Since it is known that $\RE \neq \coRE$,\footnote{$\RE \neq \coRE$ follows from the fact that $\RE \cap \coRE$ is the set of decidable languages and $\RE$ contains undecidable languages.} this implies that $\MIP^*(2,1) \neq \MIPco(2,1)$.

It is thus plausible that $\MIPco = \coRE$,\footnote{We note that the ``co'' modifier on both sides of the equation $\MIPco = \coRE$ refer to different things!} which would provide a very pleasing ``dual'' complexity characterization to $\MIP^* = \RE$. One possible route to proving this would be to adapt our gap-preserving compression framework to the commuting-operator setting by showing that each of the steps (question reduction, answer reduction, and parallel repetition) remain sound against commuting-operator strategies. Using the connection established in~\cite{fritz2014can}, this would imply that the operator norm over the maximal $C^*$ algebra $C^*(F_2 \times F_2)$, where $F_2$ is the free group on two elements, is uncomputable. This question was subsequently resolved by Lin~\cite{lin2025mipco}.

Another interesting open question concerns the \emph{zero gap} variants of $\MIP^*$ and $\MIPco$, which we denote by $\MIP^*_0$ and $\MIPco_0$, respectively. These classes capture the complexity of deciding whether a nonlocal game $\game$ has entangled value (or commuting-operator value, respectively) \emph{exactly} equal to $1$. In~\cite{slofstra2019set}, Slofstra shows that there is an efficient reduction from Turing machines $\cal{M}$ to nonlocal games $\game_\machine$ such that $\machine$ does not halt if and only if $\val^*(\game_\machine) = \valco(\game_\machine) = 1$. This implies that $\coRE = \MIPco_0(2,1)$ and, furthermore, $\coRE \subseteq \MIP^*_0$. However, since $\RE \subseteq \MIP^*(2,1) \subseteq \MIP^*_0(2,1)$, this implies that $\MIP^*_0(2,1)$ is \emph{strictly} bigger than both $\RE$ and $\coRE$. In fact, it was shown by Mousavi, et al.~\cite{mousavi2020complexity} that the class $\MIP^*_0(3,1)$ (the complexity class corresponding to determining whether $\val^*(\game) = 1$ for \emph{three-player} nonlocal games) is equal to $\Pi_2^0$, which is the set of all languages $L$ such that $x \in L$ if and only if $\forall\, y, \, \exists z \, R(x,y,z) = 1$ for a computable function $R$ that depends on $L$.\footnote{The class $\Pi_2^0$ is also characterized as being part of the second level of the \emph{arithmetical hierarchy} from computability theory, where $\RE = \Sigma_1^0$ and $\coRE = \Pi_1^0$ form the first level.} Thus it is plausible that the complexity landscape of nonlocal games looks like the following: $\MIP^* = \RE$, $\MIP^*_0 = \Pi_2^0$, and $\MIPco = \MIPco_0 = \coRE$. Such statements about the complexity of $\MIP^*$ versus $\MIPco$, in both the gapped and zero-gap cases, may reveal additional insights into the difference between the tensor product and commuting-operator models of correlations.

In subsequent work, Bowen et al.\ used a similar approach to the one described in this work, but with a different notion of value and a restricted class of games, to resolve the Aldous-Lyons conjecture~\cite{bowen2024aldous,bowen2024aldousb}.

\paragraph{Statement on AI use.} AI was used in the last stages of revision of this paper; for targeted scientific edits that were reviewed by the authors and copy-editing.

\paragraph{Lean Formalization.} The authors used general LLM methods to formalize the main statement of the paper (namely, Corollary~\ref{cor:mip-re}) in the proof verification tool Lean. The complete formalization was made available in the Palomar registry~\cite{palomar-2026-10-09-000012-v1}. There are minor departures between the formalized proof and the one presented here; in particular the formalization makes use of the general parallel repetition theorem and its formalization by OpenAI~\cite{OAI2}. We thank Sirui Lu, Ruixuan Deng and Zhengfeng Ji for allowing us to use their earlier formalization of the quantum soundness of the low-indidividual degree test~\cite{ML20}, which is also avilable on Palomar~\cite{palomar-2026-08-18-000001-v1}.

\paragraph{Acknowledgments.} We thank Matthew Coudron, William Slofstra and Jalex Stark for enlightening discussions regarding possible consequences of our work. We thank William Slofstra and Jalex Stark for suggestions regarding explicit separations between $C_{qa}$ and $C_{qc}$. We thank Peter Burton, William Slofstra and Jalex Stark for comments on a previous version. We thank Lewis Bowen and Mikael de la Salle for pointing out typos and minor errors in a previous version. We are grateful to the Annals reviewers for innumerable corrections and suggestions that greatly improved our manuscript. We acknowledge the use of AI methods, in the form of Large Language Models, to help us revise the manuscript. 

Zhengfeng Ji is supported by National Key Research and Development Program of China (Grant No.\ 2023YFA1009403) and National Natural Science Foundation of China (Grant No.\ 12347104).
Part of this work was done while Zhengfeng Ji was at the University of
Technology Sydney, where he was supported by an ARC Discovery Project
(DP200100950).
Anand Natarajan is supported by a Sloan Research Fellowship and NSF
CAREER Grant CCF-2339948. 
Most of Anand Natarajan's work on this project was conducted while
he was a member at the California Institute of Technology, where he
was supported by IQIM, an NSF Physics Frontiers Center (NSF Grant PHY-1733907). Thomas Vidick is supported by the Swiss State Secretariat for Education, Research and Innovation (SERI). 
Parts of this work was completed while Thomas Vidick was at the California Institute of Technology, where he was supported by NSF CAREER Grant CCF-1553477, AFOSR YIP award number FA9550-16-1-0495, a CIFAR Azrieli Global Scholar award, MURI Grant FA9550-18-1-0161 and the IQIM, an NSF Physics Frontiers Center (NSF Grant PHY-1125565) with support of the Gordon and Betty Moore Foundation (GBMF-12500028). 
Henry Yuen is supported by a Sloan Research Fellowship and NSF CAREER Grant CCF-2144219.
Most of Henry Yuen's work on this project was conducted while he was a member of the Departments of Computer Science and Mathematics at the University of Toronto, where he was supported by NSERC Discovery Grant 2019-06636.
John Wright is supported by a Sloan Research Fellowship and NSF CAREER Grant CCF-233971. 
Most of John Wright's work on this project was conducted while he was at the Massachusetts Institute of Technology, the California Institute of Technology, and the University of Texas at Austin,
where he was supported by IQIM, an NSF Physics Frontiers Center (NSF Grant PHY-1733907), and by ARO contract W911NF-17-1-0433.


\section{Proof overview}
\label{sec:overview}

In this section we give an overview of the proof of the inclusion $\RE \subseteq \MIP^*$. Since all interactive proof systems considered in the paper involve a single-round interaction between a classical verifier and two quantum provers sharing entanglement we generally use the language of nonlocal games to describe such proof systems, and often refer to the provers as ``players.'' In a nonlocal game $\game$ (or simply ``game'' for short), the verifier can be described as the combination of two procedures: a \emph{question sampling} procedure that samples a pair of questions $(x,y)$ for the players according to a distribution $\mu$ (known to the verifier and the players), and a \emph{decision} procedure (also known to all parties) that takes as input the players' questions and their respective answers $a,b$ and evaluates a predicate $D(x,y,a,b)\in \{0,1\}$ to determine the verifier's acceptance or rejection. Given a description of a nonlocal game $\game$, recall that $\val^*(\game)$ denotes the \emph{entangled value} of the game, which is defined as the supremum~\eqref{eq:ent-val} of the players' success probability in the game over all finite-dimensional tensor product strategies. (We refer to Section~\ref{sec:games} for definitions regarding nonlocal games.) 

Our results establish the existence of transformations on \emph{families} of nonlocal games $\{\game_n\}_{n\in \N}$ having certain properties. In order to keep track of efficiency (and ultimately, computability) properties it is important to have a way to specify such families in a {uniform} manner. Towards this we introduce the following formalism. A \emph{uniformly generated family of games} is specified through a pair of Turing machines $\verifier = (\sampler,\decider)$ that satisfy certain conditions, in which case the pair is called a \emph{normal form verifier}. The Turing machine $\sampler$ (called a \emph{sampler}) takes as input an index $n \in \N$ and returns the description of a procedure that can be used to sample questions $(x,y)$ in the game (this procedure itself obeys a certain format associated with ``conditionally linear'' distributions, defined below). The Turing machine $\decider$ (called a \emph{decider}) takes as input an index $n$, questions $(x,y)$, and answers $(a,b)$, and returns a single-bit decision. For the sake of this proof overview we assume that the sampling and decision procedures run in time polynomial in the index $n$; we refer to the running time of these procedures as the \emph{complexity} of the verifier. 
	Given a normal form verifier $\verifier= (\sampler,\decider)$ we associate to it an infinite family of nonlocal games $\{\verifier_n \}$ indexed by positive integers in the natural way.

The main technical result of this paper is a \emph{gap-preserving compression} transformation on normal form verifiers. The following theorem presents an informal summary of the properties of this transformation. Recall that for a  game $\game$ and probability $0\leq p \leq 1$, $\Ent(\game,p)$ denotes the minimum local dimension of an entangled state shared by the players in order for them to succeed in $\game$ with probability at least $p$.

\begin{theorem}[Gap-preserving compression of normal form verifiers, informal]
\label{thm:compression-informal}
There exists a polynomial-time Turing machine $\Compress$ that, when given as input the description of a normal form verifier $\verifier = (\sampler,\decider)$, outputs the description of another normal form verifier $\verifier' = (\sampler',\decider')$ that satisfies the following properties: for all $n \in \N$, letting $N = 2^n$,
\begin{enumerate}
\item (\textbf{Completeness}) If $\val^*(\verifier_N) = 1$ then $\val^*(\verifier'_n) = 1$.
\item (\textbf{Soundness}) If $\val^*(\verifier_N) \leq \frac{1}{2}$ then $\val^*(\verifier'_n) \leq \frac{1}{2}$.
\item (\textbf{Entanglement lower bound}) $\Ent(\verifier_n',\allowbreak \frac{1}{2}) \geq \max \{ \Ent(\verifier_N,\allowbreak \frac{1}{2}),\allowbreak  2^{2^{\Omega(n)}} \}$. 
\end{enumerate}
\end{theorem}
\noindent The formal version of this theorem is stated in Section~\ref{sec:compression} as Theorem~\ref{thm:compression}. The terminology \emph{compression} is motivated by the fact, implicit in the informal statement of the theorem, that the time complexity of the verifier's sampling and decision procedures in the game $\verifier_n'$, which is polynomial in $n$, is exponentially smaller than the time complexity of the verifier in the game $\verifier_N$, which is polynomial in $N$ and thus exponential in $n$. 

Before giving an overview of the proof of Theorem~\ref{thm:compression-informal} we sketch how the existence of a Turing machine $\Compress$ with the properties stated in the theorem implies the inclusion $\RE\subseteq \MIP^*$. Recall that the complexity class $\RE$ consists of all languages $L$ such that there is a Turing machine $\cal{M}$ that accepts instances $x$ in $L$, and does not accept instances $x$ that are not in $L$ (but is not required to terminate on such instances). 
To show $\RE \subseteq \MIP^*$ we give an $\MIP^*$ protocol for the Halting Problem, which is a complete problem for $\RE$. The Halting Problem is the language that consists of all Turing machine descriptions $\cal{M}$ such that $\cal{M}$ halts when run on an empty input tape. (For the purposes of this overview, we blur the distinction between a Turing machine and its description as a string of bits.) We give a procedure that given a Turing machine $\cal{M}$ as input returns the description of a normal form verifier $\verifier^{\cal{M}} = (\sampler^\machine,\decider^\machine)$ with the following properties. First, if $\cal{M}$ does eventually halt on an empty input tape, then it holds that for all $n \in \N$, $\val^*(\verifier^{\cal{M}}_n) = 1$. Second, if $\cal{M}$ does not halt then for all $n \in \N$, $\val^*(\verifier^{\cal{M}}_n) \leq \frac{1}{2}$. 

We describe the procedure that achieves this. Informally, the procedure returns the specification of a verifier $\verifier^{\cal{M}} = (\sampler^\machine,\decider^\machine)$ such that $\decider^\machine$ proceeds as follows: on input $(n,x,y,a,b)$ it first executes the Turing machine $\cal{M}$ for $n$ steps. If $\cal{M}$ halts, then $\decider^\machine$ accepts. Otherwise, $\decider^\machine$ computes the description of the compressed verifier $\verifier' = (\sampler',\decider')$ that is the output of $\Compress$ on input $\verifier^\machine$, then executes the decision procedure $\decider'(n,x,y,a,b)$ and accepts if and only if $\decider'$ accepts.\footnote{The fact that the decider $\decider^\machine$ can invoke the $\Compress$ procedure on itself follows from a well-known result in computability theory known as \emph{Kleene's recursion theorem} (also called \emph{Roger's fixed point theorem})~\cite{Kleene1954,Rogers1987}.} 

To show that this procedure achieves the claimed transformation, consider two cases. First, observe that if $\cal{M}$ eventually halts in some number of time steps $T$, then by definition $\val^*(\verifier^\machine_n) = 1$ for all $n\geq T$. Using Theorem~\ref{thm:compression-informal} along with an inductive argument it then follows that $\val^*(\verifier^\machine_n) = 1$ for all $n \geq 1$. Second, if $\cal{M}$ never halts, then observe that for any $n\geq 1$ Theorem~\ref{thm:compression-informal} implies two separate lower bounds on the amount of entanglement required to win the game $\verifier^\machine_n$ with probability at least $\frac{1}{2}$: the dimension is (a) at least $2^{2^{\Omega(n)}}$, and (b) at least the dimension needed to win the game $\verifier^\machine_{2^n}$ with probability at least $\frac{1}{2}$. Applying an inductive argument it follows that an \emph{infinite} amount of entanglement is needed to win the game $\verifier_n$ with any probability greater than $\frac{1}{2}$. Thus, a sequence of finite-dimension strategies for $\verifier_n$ cannot lead to a limiting value larger than $\frac{1}{2}$, and $\val^*(\verifier^\machine_n) \leq \frac{1}{2}$. 

We continue with an overview of the ideas behind the proof of Theorem~\ref{thm:compression-informal}.

\newcommand{\CompressNW}{\mathsf{Compress}^{\mathsf{NW}}}
 
\paragraph{Compression by introspection.}
To start, it is useful to review the protocol introduced in~\cite{NW19} to show
the inclusion $\NEEXP \subseteq \MIP^*$.
Fix an $\NEEXP$-complete language $L$.
The $\MIP^*$ protocol for $\NEXP$ from~\cite{natarajan2018two}, when scaled up
to decide languages from $\NEEXP$, yields a family of nonlocal games $\{ \game_z
\}$ that are indexed by instances $z \in \{0,1\}^*$.
The family of games decides $L$ in the sense that for all $z$, the game
$\game_z$ has entangled value $1$ if $z \in L$, and has entangled value at most
$\frac{1}{2}$ if $z \notin L$.
Furthermore, if $n = |z|$ is the length of $z$, the verifier of the game
$\game_z$ has complexity $\poly(N)=\exp(|z|)$ (recall that we use this
terminology to refer to an upper bound on the running time of the verifier's
sampling and decision procedure).
Thus, this family of games does not by itself yield an $\MIP^*$ protocol for
$L$.
To overcome this the main contribution in~\cite{NW19} is the design of an
efficient compression procedure $\CompressNW$ that applies specifically to the
family of games $\{\game_z \}$.
When given as input the description of $\game_z$, $\CompressNW$ returns the
description of a game $\game_z'$ such that if $\val^*(\game_z) = 1$, then
$\val^*(\game_z') = 1$, and if $\val^*(\game_z) \leq \frac{1}{2}$, then
$\val^*(\game_z') \leq \frac{1}{2}$.
Furthermore, the complexity of the verifier for $\game_z'$ is $\poly(n)$.
Thus the family of games $\{ \game_z' \}$ decides $L$ and this shows that
$\NEEXP \subseteq \MIP^*$, which is the best lower bound known on $\MIP^*$
prior to our work.

Presented in this way, it is natural to suggest iterating the procedure $\CompressNW$ to achieve e.g.,\ the inclusion $\NEEEXP \subseteq \MIP^*$. To explain the difficulty in doing so, we give a little more detail on the compression procedure. It consists of two main steps: starting from $\game_z$, perform (1) \emph{question reduction}, and (2) \emph{answer reduction}. The goal of (1) is to reduce the length of the questions generated by the verifier in $\game_z$ from $\poly(N)$ to $\poly(n)$. The goal of (2) is to achieve the same with respect to the length of answers expected from the players. Furthermore, the complexity of the verifier of the resulting game $\game_z'$ should be reduced from $\poly(N)$ to  $\poly(n)$.
	
	Part (1) is achieved through a technique referred to as ``introspection'' where, rather than sampling questions $(x,y)$ of length $\poly(N)$ as in the game $\game_z$, the verifier instead executes a carefully crafted nonlocal game with the players that (a) requires questions of length $\poly(n)$, (b) checks that the players share $\poly(N)$ EPR pairs, and (c) checks that the players measure the EPR pairs in such a way as to \emph{sample for themselves} a question pair $(x,y)$ such that one player gets $x$ and the other player gets $y$. In other words, the players are essentially forced to introspectively ask themselves the questions of $\game_z$. 
	
	After question reduction, the players still respond with $\poly(N)$-length answers, which the verifier has to check satisfies the decision predicate of the original game $\game_z$. The goal of Part (2) is to enable the decision procedure to implement the verification procedure while not requiring the entire full-length answers from the players. In the answer reduction scheme of~\cite{NW19} this is achieved by having the verifier run a \emph{probabilistically checkable proof} (PCP) with the players so that they succinctly \emph{prove} that first, they have introspected questions $(x,y)$ from the correct distribution, and second, that they are able to generate $\poly(N)$-length answers $(a,b)$ that would satisfy the decision predicate of the original game $\game_z$ when executed on $(x,y)$ and $(a,b)$. Since the questions and answers in the PCP are of length $\poly(n)$, this achieves the desired answer length reduction.
	
	Iterating this scheme presents a number of immediate difficulties that have to do with the fact that the sampling and decision procedures of the verifier in $\game'_z$ do not have such a nice form as those in $\game_z$. First of all, the compression procedure of~\cite{NW19} can only ``introspect'' a specific question distribution of a nonlocal game from~\cite{natarajan2018two}; we call this distribution a ``low-degree test distribution.''\footnote{The nonlocal game of~\cite{natarajan2018two} is part of a more general family of games called ``low-degree tests,'' which have been studied extensively for their applications in complexity theory~\cite{babai1991non}, property testing~\cite{rubinfeld1996robust}, and PCPs~\cite{arora1998probabilistic}. Generally, the question distribution of a low-degree test is a random question pair $(x,y)$ where $y$ is a randomly chosen affine subspace in $\F^m$ and $x$ is a uniformly random point on $y$, where $\F$ is a finite field and $m \geq 2$ is an integer. The specific low-degree test game in~\cite{natarajan2018two} (which is based on the low-degree test of~\cite{raz1997sub}) uses two-dimensional subspaces; thus the question distribution of~\cite{raz1997sub,natarajan2018two} is referred to as the ``plane-point distribution.''} However, the resulting question distribution of the question-reduced verifier, which is used to check the introspection, has a much more complex structure. A similar issue arises with the modifications required to perform answer reduction. In the PCP employed to achieve this the question distribution appears to be much more complex than the low-degree test distribution (this is in large part due to the need for a specially tailored PCP procedure that encodes separately different chunks of the witness, corresponding to answers from different players). As a result it is entirely unclear at first whether the question distribution used by the verifier in $\game'_z$ can be ``introspected'' for a second time. 
	
	To overcome these difficulties we identify a natural class of question distributions, called \emph{conditionally linear distributions}, that generalize the low-degree test distribution. We show that conditionally linear distributions can be ``introspected'' using conditionally linear distributions only, enabling recursive introspection. (In particular, they are a rich enough class to capture the types of question distributions produced by the compression scheme of~\cite{NW19}.) We define normal form verifiers by restricting their sampling procedure to generate conditionally linear question distributions, and this allows us to obtain the compression procedure on normal form verifiers described in Theorem~\ref{thm:compression-informal}. 
	
	Conceptually,  the identification of a natural class of distributions that is ``closed under introspection'' is a key step that enables the introspection technique to be applied recursively. (As we will see later, other closure properties of conditionally linear distributions, such as taking direct products, play an important role as well.) Since conditionally linear distributions are central to our construction we describe them next.

\paragraph{Conditionally linear distributions.} 
Fix a vector space $V$ that is identified with $\F^m$, for a finite field $\F$ and integer $m$. Informally (see Definition~\ref{def:cl-func} for a precise definition), a function $L$ on $V$ is \emph{conditionally linear} (CL for short) if it can be evaluated by a procedure that takes the following form: (i) read a substring $z^{(1)}$ of $z$; (ii) evaluate a linear function $L_1$ on $z^{(1)}$; (iii) repeat steps (i) and (ii) with the remaining coordinates $z\backslash z^{(1)}$, such that the next steps are allowed to depend in an arbitrary way on $L_1(z^{(1)})$ but not directly on $z^{(1)}$ itself. What distinguishes a function of this form from an arbitrary function is that we restrict the number of iterations of (i)---(ii) to a constant number (at most $9$, in our case). (One may also think of CL functions as ``adaptively linear'' functions, where the number of ``levels'' of adaptivity is the number of iterations of (i)---(ii).) A distribution $\mu$ over pairs $(x,y) \in V\times V$ is called {conditionally linear}  if it is the image under a pair of {conditionally linear functions} $L^\alice,L^\bob: V\to V$ of the uniform distribution on $V$, i.e.,\ $(x,y)\sim (L^\alice(z),L^\bob(z))$ for uniformly random $z\in V$. 

An important class of CL distributions are low-degree test distributions, which are distributions over question pairs $(x,y)$ where $y$ is a randomly chosen affine subspace of $\F^m$ and $x$ is a uniformly random point on $y$. We explain this for the case where the random subspace $y$ is one-dimensional (i.e.,\ a line). Let $V = V_\xpt \oplus V_{\dir{}}$ where $V_\xpt = V_{\dir{}} = \F^m$. Let $L^\alice$ be the projection onto $V_\xpt$ (i.e., it maps $(x,v) \to (x,0)$ where $x \in V_\xpt$ and $v \in V_{\dir{}}$). Define $L^\bob: V \to V$ as the map $(x,v) \mapsto (L^\lnf_v(x),v)$ where $L^\lnf_v: V_\xpt \to V_\xpt$ is a linear map that, for every $v \in V_{\dir{}}$, projects onto a complementary subspace to the one-dimensional subspace of $V_\xpt$ spanned by $v$ (one can think of this as an ``orthogonal subspace'' to the span of $\{ v \}$). $L^\bob$ is conditionally linear because it can be seen as first reading the substring $v \in V_{\dir{}}$ (which can be interpreted as specifying the \emph{direction} of a line), and then applying a linear map $L^\lnf_v$ to $x \in V_\xpt$ (which can be interpreted as specifying a canonical point on the line $\ell = \{ x + tv : t \in \F \}$). It is not hard to see (and shown formally in Section~\ref{sec:ld-game}) that the distribution of $(L^\alice(z),L^\bob(z))$ for $z$ uniform in $V$, is identical (up to relabeling) to the low-degree test distribution $(x,\ell)$ where $\ell$ is a uniformly random affine line in $\F^m$, and $x$ is a uniformly random point on $\ell$.



Our main result about CL distributions, presented in Section~\ref{sec:introspection}, is that any CL distribution $\mu$, associated with a pair of CL functions $(L^\alice,L^\bob)$ over a linear space $V=F^m$, can be ``introspected'' using a CL distribution that is ``exponentially smaller'' than the initial distribution. Slightly more formally, to any CL distribution $\mu$ we associate a two-player game $\game_\mu$ (called the ``introspection game'') in which questions from the verifier are sampled from a CL distribution $\mu'$ over $\F^{m'}$ for some $m' = \poly\log(m)$ and such that in any successful strategy for the game $\game_\mu$, when the players are queried on a special question labeled $\Introspect$, they must respond with a pair $(x,y)$ that is approximately distributed according to $\mu$.
(The game allows us to do more: it allows us to conclude \emph{how} the players obtained $(x,y)$ --- by measuring shared EPR pairs in a specific basis --- and this will be important when using the game as part of a larger protocol that involves other checks.) 
Crucially for us, the distribution $\mu'$ only depends on a size parameter associated with $(L^\alice,L^\bob)$ (essentially, the integer $m$ together with the number of ``levels'' of adaptivity of $L^\alice$ and $L^\bob$), but not on any other structural property of $(L^\alice,L^\bob)$. Only the decision predicate for the introspection game $\game_\mu$ depends on the entire description of $(L^\alice,L^\bob)$.

We say a few words about the design of $\mu'$ and the associated introspection game, which borrow heavily from~\cite{NW19}. Building on the ``quantum low-degree test'' introduced in~\cite{natarajan2018low} it is already known how a verifier can force a pair of players to measure $m$ EPR pairs in either the computational or Hadamard basis and report the (necessarily identical) outcome $z$ obtained, all the while using questions of length polylogarithmic in $m$ only. The added difficulty is to ensure that a player obtains, and returns, precisely the information about $z$ that is contained in $L^\alice(z)$ (resp. $L^\bob(z)$), and not more. A simple example is the line-point distribution described earlier: there, the idea to ensure that e.g.,\ the ``point'' player only obtains the first component, $x$ of $(x,v) \in V_\xpt \oplus V_{\dir{}}$, the verifier demands that the ``point'' player measures their qubits associated with the space $V_{\dir{}}$ in the Hadamard, instead of computational, basis; due to the uncertainty principle this has the effect of ``erasing'' the outcome in the computational basis.  The case of the ``line'' player is a little more complex: the goal is to ensure that, conditioned on the specification of the line $\ell$ received by the ``line'' player, the point $x$ received by the ``point'' player is uniformly random within $\ell$. This was shown to be possible in~\cite{NW19}.

We can now describe how samplers of normal form verifiers are defined: these are
Turing machines $\sampler$ that specify an infinite family of CL distributions
$\{ \mu_n \}$ by computing, for each index $n$, 
the CL functions $(L^{\alice,\, n},L^{\bob,\, n})$ associated with $\mu_n$. 
(See Definition~\ref{def:sampler} for a formal definition of samplers.)
Thus, the question distributions of a normal form verifier $\verifier =
(\sampler,\decider)$ are the CL distributions corresponding to $\sampler$.

\paragraph{Question reduction.} Just like the compression procedure of~\cite{NW19}, the compression procedure $\Compress$ of Theorem~\ref{thm:compression-informal} begins with performing question reduction on the input game. Given a normal form verifier $\verifier = (\sampler,\decider)$, the procedure $\Compress$ first computes a normal form verifier $\verifier^\intro = (\sampler^\intro,\decider^\intro)$ where for all $n \in \N$, the game $\verifier^\intro_n$ consists of playing the original game $\verifier_N$ where $N = 2^n$, except that instead of sampling the questions according to the CL distribution $\mu_N$ specified by the sampler $\sampler_N$, the verifier executes the introspection game $\game_{\mu_N}$  described in the previous subsection. Thus, in the game $\verifier^\intro_n$, when both players receive the question labeled $\Intro$ they are expected to sample $(x,y)$, respectively, according to $\mu_N$, and respond with the sampled question together with answers $a,b$, respectively. The decider $\decider^\intro$ on index $n$ evaluates $\decider(N,x,y,a,b)$ and accepts if and only if $\decider$ accepts. As a result the time complexity of decider $\decider^\intro$ on index $n$ remains that of $\decider$, i.e.,\ $\poly(N)$. However, the length of questions asked in $\verifier^\intro_n$ and the complexity of the sampler $\sampler^\intro$ are exponentially reduced, to $\poly(n)$.

For convenience we refer to the questions asked by the verifier in the ``question-reduced'' game $\verifier^\intro_n$ as ``small questions,'' and the questions that are introspected by the players in $\verifier^\intro_n$ (equivalently, the questions asked in the original game $\verifier_N$) as ``big questions.''

\paragraph{Answer reduction.} 
Having reduced the complexity of the question sampling, the next step in the compression procedure $\Compress$ is to reduce the complexity of decider $\decider^\intro$ from $\poly(N)$ to $\poly(n)$ (which necessarily implies reducing the answer length to $\poly(n)$). To achieve this the compression procedure computes a normal form verifier $\verifier^\ar = (\sampler^\ar,\decider^\ar)$ from $\verifier^\intro$ such that both the sampler and decider complexity in $\verifier^\ar$ are $\poly(n)$ (here, $\ar$ stands for ``answer reduction'').

Similarly to the answer reduction performed in~\cite{NW19}, at a high level this is achieved by composing the game $\verifier^\intro_n$ with a probabilistically checkable proof (PCP). In our context a PCP is a proof encoding that allows a verifier to check whether, given Turing machine $\cal{A}$ and time bound $T$ provided as input, there exists some input $x$ that $\cal{A}$ accepts in time $T$. The PCP proof can be computed from $\cal{A}$, $T$, and the accepting input (if it exists) and has length polynomial in $T$ and the description length $|\cal{A}|$ of $\cal{A}$. Crucially, the verifier can check a purported proof while only reading a constant number of symbols of it, each of length $\polylog(T,|\cal{A}|)$, and executing a verification procedure that runs in time $\polylog(T,|\cal{A}|)$. 

We use PCPs for answer reduction as follows. The verifier in the game $\verifier^\ar_n$ samples questions as $\verifier^\intro_n$ would and sends them to the players. Instead of receiving the introspected questions and answers $(x,y,a,b)$ for the original game $\verifier_N$ and running the decision procedure $\decider(N,x,y,a,b)$, the verifier instead asks the players to compute a PCP $\Pi$ for the statement that the original decider $\decider$ accepts the input $(N,x,y,a,b)$ in time $T=\poly(N)$. The verifier then samples additional questions for the players that ask them to return specific entries of the proof $\Pi$. Finally, upon receipt of the players' answers, the verifier executes the PCP verification procedure. Because of the efficiency of the PCP, both the sampling of the additional questions and the decision procedure can be executed in time $\poly(n)$.\footnote{This idea is inspired by the technique of composition in the PCP literature, in which the complexity of a verification procedure can be reduced by composing a proof system (often a PCP itself) with another PCP.}

This very rough sketch presents some immediate difficulties. A first difficulty is that in general no player by themselves has access to the entire input $(N,x,y,a,b)$ to $\decider$, so no player can compute the entire proof $\Pi$. We discuss this issue in the next paragraph. A second difficulty is that a black-box application of an existing PCP, as done in~\cite{NW19}, results in a question distribution for $\verifier_n^\ar$ (i.e.,\ the sampling of the proof locations to be queried) that is rather complex --- and, in particular, it may no longer fall within the framework of CL distributions for which we can do introspection. 
To avoid this, we design a bespoke PCP based on the classical MIP for NEXP (in particular, we borrow and adapt techniques from~\cite{ben2005simple,ben2006robust}). Two essential properties for us are that (i) the PCP proof is a collection of several low-degree polynomials,
    two of which are low-degree encodings of each player's big answer in the game $\verifier^\intro_n$, and (ii) verifying the proof only requires (a) running low-degree tests, (b) querying all polynomials at a uniformly random point, and (c) performing simple consistency checks. Property (i) allows us to eliminate the extra layer of encoding in~\cite{NW19}, who had to consider a PCP of proximity for a circuit applied to the
low-degree encodings of the players' big answers. Property (ii) allows us to ensure that the question distribution employed by $\verifier_n^\ar$ remains conditionally linear. 

\paragraph{Oracularization.}
The preceding paragraph raises a nontrivial difficulty. In order for the players to compute a proof for the claim that $\decider(N,x,y,a,b)=1$ they need to know the entire input $(x,y,a,b)$. However, in general a player only has access to their own question and answer: one player only knows $(x,a)$ and the other player knows $(y,b)$. The standard way of circumventing this difficulty is to consider an ``oracularized'' version of the game, where one player gets both questions $(x,y)$ and is able to determine both answers $(a,b)$, while the other player only gets one of the questions at random, and is only asked for one of the answers, that is then checked for consistency with the first player's answer. 

While this technique works well for games with classical players, when the players are allowed to use quantum strategies using entanglement oracularization does not, in general, preserve the completeness property of the game. To ensure that completeness is preserved we need an additional property of a completeness-achieving strategy for the original game: that there exists a \emph{commuting and consistent strategy} on all pairs of questions $(x,y)$ that are asked in the game with positive probability. Here commuting means that the measurement $\{A^x_a\}_a$ performed by the player receiving $x$ commutes with the measurement $\{B^y_b\}_b$ performed by the player receiving $y$.\footnote{We stress that the commuting property only applies to question pairs that occur with positive probability, and does not mean that \emph{all} pairs of measurement operators are required to commute; indeed this would imply that the strategy is effectively classical.} Consistent means that if both players perform measurements associated with the same question they obtain the same answer. If both properties hold then in the oracularized game when one player receives a pair $(x,y)$ and the other player receives the question $x$ (say), the first player can simultaneously measure both $\{A^x_a\}_a$ and $\{B^y_b\}_b$ on their own space to obtain a pair of answers $(a,b)$, and the second player can measure $\{A^x_a\}_a$ to obtain a consistent answer $a$.

For answer reduction to be possible it is thus applied to the {oracularized} version of the introspection game $\verifier^\intro_n$. This in turn requires us to ensure that the introspection game $\verifier^\intro_n$ has a commuting and consistent strategy achieving value $1$ whenever it is the case that $\val^*(\verifier_n^\intro)=1$. For this property to hold we verify that it holds for the initial game that is used to seed the compression procedure (this is true because we can start with an $\MIP^*$ protocol for $\NEXP$ for which there exists a perfect classical strategy) and we also ensure that each of the transformations of the compression protocol (question reduction, answer reduction, and parallel repetition described next) maintains it.

\paragraph{Parallel repetition.}
The combined steps of question reduction (via introspection) and answer reduction (via PCP composition) result in a game $\verifier_n^\ar$ such that the complexity of the verifier is $\poly(n)$. Furthermore, if the original game $\verifier_N$ has value $1$, then $\verifier_n^\ar$ also has value $1$. Unfortunately the sequence of transformations incurs a loss in the soundness parameters: if $\val^*(\verifier_N) \leq \frac{1}{2}$, then we can only establish that $\val^*(\verifier^\ar_n) \leq 1 - C$ for some positive constant $C < \frac{1}{2}$ (we call $C$ the \emph{soundness gap}). Such a loss would prevent us from recursively applying the compression procedure $\Compress$ an arbitrary number of times, which is needed to obtain the desired complexity results for $\MIP^*$. 

To overcome this we need a final transformation to restore the soundness gap of the games after answer reduction to a constant larger than $\frac{1}{2}$. To achieve this we use the technique of {parallel repetition}. The parallel repetition of a game $\game$ is another nonlocal game $\game^k$, for some number of repetitions $k$, which consists of playing $k$ independent and simultaneous instances of $\game$ and accepting if and only if all $k$ instances accept. Intuitively, parallel repetition is meant to decrease the value of a game $\game$ exponentially fast in $k$, provided $\val^*(\game) < 1$ to begin with. However, it is an open question of whether this is generally true for the entangled value $\val^*$. 

Nevertheless, some variants of parallel repetition are known to achieve exponential amplification. We use  a variant  called ``anchored parallel repetition'' and introduced in~\cite{bavarian2017hardness}. This allows us to devise a transformation that efficiently amplifies the soundness gap to a constant. The resulting game $\verifier^\rep_n$ has the property that if $\val^*(\verifier^\ar_n) = 1$, then $\val^*(\verifier^\rep_n) = 1$ (and, moreover, this is achieved using a commuting and consistent strategy), whereas if $\val^*(\verifier^\ar_n) \leq 1 - C$ for some universal constant $C > 0$ then $\val^*(\verifier^\rep_n) \leq \frac{1}{2}$. Furthermore, we have the additional property, essential for us, that good strategies in the game $\verifier^\rep_n$ require as much entanglement as good strategies in the game $\verifier^\ar_n$ (which in turn require as much entanglement as good strategies in the game $\verifier_N$). The complexity of the verifier in $\verifier^\rep_n$ remains $\poly(n)$.

The anchored parallel repetition procedure, when applied to a normal form verifier, also yields a normal form verifier: this is because the direct product of CL distributions is still conditionally linear.

\paragraph{Putting it all together.} This completes the overview of the transformations performed by the compression procedure $\Compress$ of Theorem~\ref{thm:compression-informal}. To summarize, given an input normal form verifier $\verifier$, question reduction is applied to obtain $\verifier^\intro$, answer reduction is applied to the oracularized version of $\verifier^\intro$ to obtain $\verifier^\ar$, and  anchored parallel repetition is applied to obtain $\verifier^\rep$, which is returned by the compression procedure. Each of these transformations preserves completeness (including the commuting and consistent properties of a value-1 strategy) as well as the entanglement requirements of each game; moreover, the overall transformation preserves soundness. 


\section{Preliminaries}
\label{sec:prelim}

\paragraph{Notation.}
We use $\Sigma$ to denote a finite alphabet.
$\N$ is the set of positive integers.
For $w\in\{0,1\}$, $\overline{w}$ denotes $1-w$.
For $w\in\{\alice,\bob\}$, $\overline{w}=\bob$ if $w=\alice$ and
$\overline{w}=\alice$ otherwise.
(For notational convenience we often implicitly make the identifications
$1\leftrightarrow \alice$ and $2\leftrightarrow \bob$.)
We use $\F$ to denote a finite field.
We write $M_n(\F)$ to denote the set of $n\times n$ matrices over $\F$.
We write $I$ to denote the identity operator on a vector space.
We write $\Tr(\cdot)$ for the matrix trace.
We write $\mH$ to denote a separable Hilbert space.
For a linear operator $T$, $\|T\|$ denotes the operator norm.

\paragraph{Asymptotics.}
All logarithms are base $2$.
We use the notation $O(\cdot)$, $\poly(\cdot)$, and $\polylog(\cdot)$ in the
following way.
For $f,g:\N\to\R_+$ (resp.\ $f,g:[0,1]\to\R_+$) we write $f(n)=O(g(n))$ (resp.\ $f(\delta)=O(g(\delta))$, omitting the integer $n$ or real $\delta$ when they are
clear from context, to mean that there exists a constant $C>0$ such that for all
$n\in\N$, $f(n)\leq C g(n)$ (resp. for all $\delta\in[0,1], f(\delta)\leq C g(\delta)$).

When we write $f(a_1,\ldots,a_k) = \poly(a_1,\ldots,a_k)$, this indicates that
there exists a universal constant $C > 0$ (which may vary each time the notation
is used in the paper) such that $f(a_1,\ldots,a_k) \leq C( a_1 \cdots a_k)^C$
for all positive $a_1,\ldots,a_k$.
Similarly, when we write $f(a_1,\ldots,a_k) = \polylog(a_1,\ldots,a_k)$, there
exists a universal constant $C$ such that $f(a_1,\ldots,a_k) \leq C\prod_{i =
  1}^k \log^C (1+ a_i)$ for all positive $a_1,\ldots,a_k$.
	
Finally, we write $\log(a_1,\ldots,a_k)$ as short hand for $\prod_{i = 1}^k \log
(1 + a_i)$.\footnote{The additional $1$ in the argument of the $\log(\cdot)$ is
  to ensure that this quantity is strictly positive.}

\subsection{Turing machines}
\label{sec:tms}

Turing machines are a model of computation introduced
in~\cite{turing1937computable}, and play a central role in our modeling of verifiers for nonlocal
games. Here we give an overview of certain aspects of Turing machines that are relevant for
this paper. For an in-depth treatment of Turing machines, we refer the reader to the textbooks of 
Papadimitriou~\cite{Pap94} or Sipser~\cite{sipser2012introduction}.

The tapes of a Turing machine are infinite one-dimensional arrays of cells that
are indexed by natural numbers.
A Turing machine $\cal{M}$ has one input tape, one work tape,
and one output tape. 
Each cell of a tape has symbols taken either from the set $\{0,1\}$ or the blank
symbol $\sqcup$.
At the start of the execution of a Turing machine, the work and output tapes are
initialized to have all blank symbols and the input tape contains the input. 
A Turing machine \emph{halts} when it enters a designated halt state.
The \emph{output} of a Turing machine, when it halts, is the binary string that
occupies the longest initial stretch of the output tape that does not have a
blank symbol.
If the first symbol on the output tape is a blank symbol, then by convention we say
that the Turing machine's output is $0$.

\tnote{Edited this.
  Before we talked about $k$-input TM's, but I didn't find such a concept in the
  literature, only $k$-tape, which is quite different.
  So I added text to explain what we mean:}We often want to consider that a
Turing machine $\cal{M}$ has $k$ inputs, for some $k\geq 1$.
For this we introduce the following convenient terminology.
A \emph{$k$-input} Turing machine $\cal{M}$ is the same as a standard $3$-tape
Turing machine, except that we will write the contents of its input tape as a
$k$-tuple $x \in (\{0,1\}^*)^k$.
What this effectively means is that the input tape of the Turing machine is
initialized using some unambiguous binary encoding $\mathrm{enc}_k(x)$ of
$k$-tuples $x \in (\{0,1\}^*)^k$.
We systematically use the following canonical choice of encoding: a tuple
$(x_1,\ldots,x_k)$ is encoded into a concatenation of two-bit encoding
of each $x_i$ by expanding every bit of $x_i$ via the map $0 \mapsto 00$,
$1 \mapsto 01$ and indicating the end of the string $x_i$ by $10$.\znote{Updated the encoding so that it is consistent with Cook-Levin in answer reduction. We use blank to separate the components in the tuple.}
For example, the pair $(00,1)$ is encoded as $0000100110$.
Note that, using our encoding, the length of the input $(x_1,\ldots,x_k)$ is
$2(|x_1|+\cdots+|x_k|+k)$.

Every $k$-input Turing machine $\cal{M}$ computes a (partial) function $f: (\{0,\allowbreak 
1\}^*)^k \to \{0,\allowbreak 1\}^*$ where the function is only defined on subset $S
\subseteq (\{0,\allowbreak 1\}^*)^k$ of inputs $x$ on which $\machine$ halts.
We use $\cal{M}(x_1,\allowbreak  x_2,\allowbreak  \ldots,\allowbreak  x_k)$ to denote the output of a $k$-input
Turing machine $\cal{M}$ on input tuple $(x_1,\allowbreak \ldots,\allowbreak x_k)$. 
If $\cal{M}$ does not halt on an input $x$, then we define $\cal{M}(x)$ to be
$\bot$.

The \emph{time complexity} of a Turing machine $\cal{M}$ on input $x = (x_1,
x_2, \ldots, x_k)$, denoted by $\TIME_{\cal{M},\, x}$, is the number of time
steps that $\cal{M}$ takes on input $x$ before it enters its designated halt
state; if $\cal{M}$ never halts on input $x$, then we define $\TIME_{\cal{M},x}
= \infty$. \hnote{Added this May 20, 2026.} Many Turing machines used in this paper have a distinguished first input,
denoted by $n$.  If $\cal M$ is such a Turing machine, with remaining input
components denoted by $z_1,\ldots,z_k$, we write
\[
  \TIME_{\cal M}(n)
  =
  \sup_{z_1,\ldots,z_k\in\{0,1\}^*}
  \TIME_{\cal M,(n,z_1,\ldots,z_k)} .
\]
The supremum ranges over \emph{all} length strings $(z_1,\ldots,z_k)$, so, in particular,
$\TIME_{\cal M}(n)$ may be infinite if there are some inputs $(n,z_1,\ldots,z_k)$ on which $\cal M$ does not halt.

\tnote{Added a little more specific explanations here, as requested by reviewer:}Every Turing machine $\cal{M}$ has a canonical encoding as a bit string $\desc{\cal{M}} \in \{0, 1\}^*$, called
the \emph{canonical description of $\cal{M}$}. The canonical encoding describes the 
finite number of states and transition rules of $\cal{M}$. The details of this 
encoding are not important for this paper; for the sake of concreteness we use the encoding described in~\cite[Section 9.1.2]{10.5555/1196416}. Using this encoding, each state is represented in unary and the bit length of the encoding is polynomial (in fact, quadratic) in the total number of 
states of $\cal{M}$. The length of the description
$\desc{\cal{M}}$ is denoted by $|\desc{\cal{M}}|$. 
For every string $\alpha \in \{0, 1\}^*$, the
Turing machine described by $\alpha$ is denoted $[\alpha]_k$.

Throughout this paper we frequently give high-level descriptions of Turing machines in English language, 
rather than explicitly describing them in terms of states and transition functions. In doing so we implicitly assume that all such descriptions can be formally converted into a description in terms of states and transition functions such that there is no hidden blow-up in the complexity of the representation. 

We elaborate on several properties of Turing machines that are implicitly used throughout the paper.

\paragraph{Input representations.} 
  Although the inputs and outputs of a Turing machine are strictly speaking
  binary strings, we oftentimes slightly abuse notation and specify Turing
  machines that treat their inputs and outputs as objects with more structure,
  such as finite field elements, integers, symbols from a larger alphabet, and
  so on.
  In this case we implicitly assume that the Turing machine specification uses a
  consistent convention to represent these structured objects as binary strings.
  Conventions for objects such as integers are straightforward.
  For representations of finite field elements, we refer the reader to
  Section~\ref{sec:ff-representations}.

\paragraph{Basic computations.} \hnote{Added: May 20, 2026.} 
When describing Turing machines, we freely use standard algorithms for basic
finite computations.  For instance, two binary strings of length at most $L$ can
be copied, compared, concatenated, or parsed according to a fixed tuple format
in time $\poly(L)$.  Integers with binary representations of length at most $L$
can be added, multiplied, compared, and exponentiated to powers of at most
$L$ in time $\poly(L)$. Finite-field elements are represented as specified in
Section~\ref{sec:ff-representations}; the field operations used in this paper
are performed in time polynomial in the representation length.

\paragraph{Subroutine simulations.}
\hnote{NEW: }
We often describe a Turing machine $\cal M$ as calling other (standard three-tape) Turing machines $\cal R_1,\cal R_2,\ldots,\cal R_r$ as subroutines; each subroutine can be called multiple times and in different orders. We specify what this means in greater detail. We first describe $\cal M$ as a six-tape Turing machine, with one input tape, one work tape, one output tape, and three additional tapes for simulating the subroutines. The parent machine $\cal M$ makes a call to one of the subroutines $\cal R_j$ by writing the input to $\cal R_j$ on a separate ``subroutine input'' tape, and then simulating the work and output tapes of $\cal R_j$ on two additional tapes. Before each subroutine is called, the parent machine erases the contents of the extra work tapes. We assume only one subroutine can be called at a time.

Suppose that the description length of the transition rules, the states, etc. of the ``outer computation'' of $\cal M$ is $\sigma$. Then the complete description length of $\cal M$, as a six-tape Turing machine, is $\sigma +O(1) + \sum_{i=1}^r |\desc{\cal R}_i|$. This is because we simply concatenate the transition rules of the subroutines $\cal R_i$ to the transition rules of the outer computation, and add a constant number of additional states and transition rules to implement the subroutine calls.

Now we bound the time complexity of $\cal M$ as a six-tape Turing machine. Fix an input $x$ to $\cal M$. Suppose that the ``outer computation,'' where we treat the calls to the subroutines as taking a single time step, takes $T_0$ steps. Now suppose that the calls to all the subroutines take $T_1,\ldots,T_m$ steps, respectively (note that $m$ may be different from $r$, if subroutines are called multiple times). Then the total time taken by $\cal M$ on input $x$ is $T_0 + O(\sum_{j=1}^m T_j)$, where the $O(\cdot)$ comes from the overhead of clearing the subroutine tapes. 

Finally, we can simulate a six-tape Turing machine by a three-tape Turing machine~\cite[Theorem 2.1]{Pap94}. This simulation increases the description length and time complexity of $\cal M$ by at most a polynomial factor; thus the description length of $\cal M$ as a three-tape Turing machine is $\poly(\sigma + \sum_{i=1}^r |\desc{\cal{R}}_i|)$, and the time complexity of $\cal M$ on input $x$ is $\poly(T_0 + \sum_{j=1}^m T_j)$. The polynomials $\poly(\cdot)$ are universal and do not depend on the specific Turing machines $\cal M,\cal R_1,\cal R_2,\ldots$.

\paragraph{Turing machines with timeout counters.}

\newcommand{\AddPolynomialTimeoutCounter}{\mathsf{AddPolyTimeout}}
\newcommand{\AddExponentialTimeoutCounter}{\mathsf{AddExpTimeout}}
\newcommand{\IdentifyTimeoutCounter}{\mathsf{IdentifyTimeout}}

We introduce the concept of a Turing machine with a timeout counter. This is a Turing machine which by design halts on all inputs within a specified timeout bound; in this paper this bound will either be some polynomial or exponential function. Furthermore, the timeout bound can be efficiently ``read off'' from the description of the Turing machine. A number of Turing machines (in particular, deciders) in the compression procedure will be of this form. 

We define two polynomial-time Turing machines
$\AddPolynomialTimeoutCounter$ and $\AddExponentialTimeoutCounter$. On input
$(\desc{\cal M},c_0,c_1,c_2)$, where $\cal M$ is a Turing machine with distinguished
first input $n$ and $c_0,c_1,c_2\in\N$ are represented in binary, each transformation outputs the
description of a three-tape Turing machine $\cal N$.

The machine $\cal N$ output by $\AddPolynomialTimeoutCounter$ simulates $\cal M$ for at most
$c_0 (c_1 n)^{c_2}$ steps. The machine output by
$\AddExponentialTimeoutCounter$ simulates $\cal M$ for at most $c_0 (c_1 2^n)^{c_2}$ steps. Let $B_{\cal N}(n)$ denote the timeout bound of $\cal N$, which is either $c_0 (c_1 n)^{c_2}$ or $c_0 (c_1 2^n)^{c_2}$. 

We first describe $\cal N$ as a high-level multitape Turing
machine with an additional work tape for the timeout counter. Given input with
first component $n$, the machine computes the corresponding timeout bound $B_{\cal N}(n)$ and writes it on the counter tape in unary. It then runs the underlying computation $\cal M$ step by
step and decrements the counter after each of $\cal M$'s transitions.\footnote{We note that we are not treating $\cal M$ as a subroutine here, so there is no overhead from preparing the input to $\cal M$; it directly reads the input tape.} If $\cal M$
halts before the counter reaches zero, then $\cal N$ halts with the same
output. Otherwise, $\cal N$ enters a distinguished halting state $\TIMEOUT$.
The standard multitape simulation (as given by, e.g.,~\cite[Theorem 2.1]{Pap94}) converts this high-level implementation into
the three-tape machine $\cal N$ that is output by the transformation.

\paragraph{Running time.}
Computing the timeout bound $B_{\cal N}(n)$ (which can be done using standard integer arithmetic), simulating the underlying computation and decrementing the counter together take time $\poly(B_{\cal N}(n))$. 
Since the standard
three-tape simulation introduces only polynomial overhead, the resulting
machines satisfy $\TIME_{\cal N}(n) = \poly(B_{\cal N}(n))$. 


\paragraph{Description length.}
To construct the description of $\cal N$, the transformations copy the
description of $\cal M$ and add transition rules for computing the
timeout bound, maintaining the counter, and simulating $\cal M$. The transition rules for computing the timeout bound and maintaining the counter are independent of $\cal M$ and depend only on the hardwired parameters $c_0,c_1,c_2$, and, furthermore, the number of these transition rules are polynomial in $\log c_0, \log c_1,\log c_2$.  Converting
the resulting high-level multitape description into the three-tape model
increases its length by at most a polynomial factor. Consequently, $|\desc{\cal N}|=\poly(|\desc{\cal M}|,\log c_0, \log c_1,\log c_2)$.

\paragraph{Standard timeout-counter form.}
The implementations of $\AddPolynomialTimeoutCounter$ and
$\AddExponentialTimeoutCounter$ are such that the timeout parameters $c_0,c_1,c_2$ and whether it is polynomial or exponential are easily identifiable from the description of the output machine. More precisely, there is a polynomial-time Turing
machine $\IdentifyTimeoutCounter$ with the following behavior. When given the
description of a machine output by $\AddPolynomialTimeoutCounter$ with
parameters $c_0, c_1,c_2$, it outputs $(\mathsf{poly},c_0,c_1,c_2)$. When given the description of a
machine output by $\AddExponentialTimeoutCounter$ with parameters $c_0, c_1,c_2$, it
outputs $(\mathsf{exp},c_0,c_1,c_2)$. On all other inputs, it outputs $\bot$. 
This way, we define a Turing machine $\cal N$ to be in \emph{standard timeout-counter form} if $\IdentifyTimeoutCounter(\desc{\cal N}) \neq \bot$. 

Throughout this paper, when constructing Turing machines with timeout counters we will not explicitly mention the use of $\AddPolynomialTimeoutCounter$ or $\AddExponentialTimeoutCounter$. Instead, for brevity, we will directly specify the underlying computation and the timeout bound $B_{\cal N}(n)$ together. 

We will sometimes write the timeout bound $B_{\cal N}(n)$ in asymptotic notation. It will be clear whether it is polynomial or exponential. For example, sometimes we will write bounds that depend on other parameters such as $B_{\cal N}(n) = \poly((\lambda n)^\mu, \sigma)$. What we mean by this is there exists parameters $c_0 = \poly(\sigma),c_1 = \poly(\lambda),c_2 = O(\mu)$ such that $B_{\cal N}(n) \leq c_0 (c_1 n)^{c_2}$ for all $n$. Here, although $c_0,c_1,c_2$ depend polynomially on $\lambda,\sigma,\mu$, they do not depend on $n$. Another example is if we write $B_{\cal N}(n) = \poly(2^{\lambda n}, \ell)$; then there exists $c_0 = \poly(\ell),c_1 = 1,c_2 = O(\lambda)$ such that $B_{\cal N}(n) \leq c_0 (c_1 2^n)^{c_2}$ for all $n$.

\paragraph{Universal Turing machine.}
In the proof of the compression theorem (see Section~\ref{sec:compression}), a Turing machine receives the description of another Turing machine as
part of its input and simulates it.  For this purpose we use a standard
universal Turing machine.

\begin{theorem}[Efficient universal Turing machine]
  \label{thm:universal-tm}
  There exists a three-tape Turing machine $\cal{U}$ such that for every $x \in \{0,1\}^*$
  and $\alpha \in \{0,1\}^*$, if the input tape of $\cal{U}$ is initialized with
  $(\alpha,x)$ and $[\alpha]$ halts within $T$ steps on input $x$, then
  $\cal{U}$ halts in at most $C(|\alpha|+T)^c$ steps and the contents of its
  output tape is the output of $[\alpha]$ on input $x$. Here, $C,c\geq 1$ are universal constants.
\end{theorem}

\begin{proof}
  We refer the reader to the proof in~\cite[Theorem 1.9]{arora2009computational}.
\end{proof}

\begin{remark}
  \label{rmk:universal-tm}
  We note that the time complexity of the universal Turing machine $\cal{U}$ on
  input $(\alpha,x)$ only depends on $x$ insofar as the time complexity of
  $\alpha$ depends on $x$.  That is, if the Turing machine described by
  $\alpha$ halts in $T$ steps without reading the entire input $x$, then the
  universal Turing machine $\cal{U}$ will halt in time polynomial in $T$ and the
  length of $\alpha$ without reading the entire string $x$.
\end{remark}

\paragraph{Hardwiring parameters.}
Throughout this paper, certain integer parameters play a special role in our
complexity accounting.  For example, the parameter $\lambda$ will often control
running-time bounds for samplers and deciders, and the parameter $\ell$ will
often control the number of levels in a conditionally linear sampler.  We will
describe Turing machines whose behavior depends on such parameters.  We will
also describe Turing machines that take these parameters as input and output
descriptions of other Turing machines.

This can be a source of confusion, so we spell out the convention with a simple
example.  An illustrative situation is the following.

\begin{quote}
\emph{Let $\cal A$ be a Turing machine that, on input $\lambda$ written in
binary, outputs the description of a Turing machine $\cal M_\lambda$.  The
machine $\cal M_\lambda$, on input $(n,x)$, parses $x$ as an integer written in
binary, computes the binary representation of the integer $y=\lambda n x$ and accepts if $y$ is even.}
\end{quote}

Let us unpack this example.  The machine $\cal A$ is a fixed Turing machine; its
description does not depend on $\lambda$.  On input $\lambda$, it writes down
the transition rules of a new Turing machine $\cal M_\lambda$.  The description
of $\cal M_\lambda$ contains a copy of the binary representation of $\lambda$,
together with fixed instructions for multiplication and the parity check.  Thus
the description length of $\cal M_\lambda$ is $O(\log \lambda)+O(1)$, and
$\cal A$ runs in time polynomial in $\log\lambda$.

After $\cal M_\lambda$ has been constructed, $\lambda$ is no longer an input to
$\cal M_\lambda$; it is part of the finite description of the machine.  Thus,
when we bound the running time of $\cal M_\lambda$ on input $(n,x)$, the
dependence on $\lambda$ comes from the size of the hardwired binary string and
from the arithmetic performed using it.  In this example, $\cal M_\lambda$ runs
in time $\poly(\log \lambda,\log n,|x|)$ on input $(n,x)$.

In general, when we construct a Turing machine (such as $\cal M_\lambda$) that has hardwired parameters such as $\lambda$ or $\ell$, we mean that its description contains that data, and the description can itself be computed by a fixed ``constructor'' Turing machine (e.g., the machine $\cal A$ above) in time polynomial in the length of the parameters.

\subsection{Linear spaces}
\label{sec:linear-spaces}

Linear spaces considered in the paper generally take the form $V = \F^n$ for a
finite field $\F$ and integer $n \geq 1$.
In particular, when we write ``let $V$ be a linear space,'' unless explicitly
stated otherwise we always mean a space of the form $\F^n$.
Let $\hat{E} = \{ \hat{e}_1, \hat{e}_2, \ldots, \hat{e}_n\}$ denote the standard basis of $V$, where for
$i \in\{ 1, 2, \ldots, n\}$,
\begin{equation*}
  \hat{e}_i = (0, \ldots, 0, 1, 0, \ldots, 0)
\end{equation*}
has a $1$ in the $i$-th coordinate and $0$'s elsewhere.
We write $\End(V)$ to denote the set of linear transformations from $V$ to
itself.
\begin{definition}[Register subspace]
  A \emph{register subspace} $S$ of $V$ is a subspace that is the span of a
  subset of the standard basis of $V$.\footnote{The use of the term ``register''
    is meant to create an analogy for how the space of multiple qubits is often
    partitioned into ``registers'' containing a few qubits each.}
  We often represent such a subspace as an indicator vector $u\in \{0,1\}^s$,
  where $s=\dim(V)$, such that if $\{\hat{e}_1,\ldots,\hat{e}_s\}$ is the standard basis of
  $V$ then $S = \text{span}\{\hat{e}_i|\,u_i=1\}$.
\end{definition}

\begin{definition}
  Let $\hat{E} = \{ \hat{e}_i\}$ be the standard basis of $V = \F^n$.
  For two vectors $u = \sum_{i=1}^n u_i \hat{e}_i$, $v = \sum_{i=1}^n v_i \hat{e}_i$ in $V$,
  define the \emph{dot product}
  \begin{equation*}
    u \cdot v = \sum_{i=1}^n u_i v_i \in \F\;.
  \end{equation*}
  Let $S$ be a subspace of $V$. The \emph{subspace orthogonal to $S$ in $V$} is
  \begin{equation*}
    S^\perp \,=\, \big\{ u \in V : u \cdot v = 0
    \text{ for all }  v \in S \big\}\;.
  \end{equation*}
  Although the notation $S^\perp$ does not explicitly refer to $V$, the ambient
  space will always be clear from context.
\end{definition}

We note that over finite fields, the notion of orthogonality does not possess
all of the same intuitive properties of orthogonality over fields such as $\R$
or $\C$; for example, a nonzero subspace $S$ may be orthogonal to itself (e.g.,\
$\mathrm{span} \{ (1,1) \}$ over $\F_2$).
However, the following remains true over all fields.

\begin{lemma}
  \label{lem:perp_perp}
  Suppose $S$ is a subspace of $V$. Then $\dim(S) + \dim(S^\perp) =
  \dim(V)$, and 
  \begin{equation*}
    (S^\perp)^\perp = S\;.
  \end{equation*}
\end{lemma}

\begin{proof}
  The statement about the dimensions follows from the fact that vectors in $S^\perp$ are
  the solution to a feasible linear system of equations with $\dim(S)$ linearly
  independent rows; this implies that the solution space has dimension exactly
  $\dim(V) - \dim(S)$.
  Next, we argue that $S \subseteq (S^\perp)^\perp$.
  Let $u \in S$.
  Since all vectors $v \in S^\perp$ are orthogonal to every vector in $S$,
  in particular, $u$, this implies that $u \in (S^\perp)^\perp$.
  By dimension counting, it follows that $(S^\perp)^\perp = S$.
\end{proof}

\begin{definition}\label{def:complementary}
  Given a linear space $V$, two subspaces $S$ and $T$ of $V$ are said to form a
  pair of \emph{complementary subspaces} of $V$ if
\begin{equation*}
  S \cap T = \{ 0 \}, \quad S + T = V\;.
\end{equation*}
In this case, we write $V = S \oplus T$.
Any $x\in V$ can be written as $x = x^S + x^T$ for $x^S \in S$ and $x^T \in T$
in a unique way.
We refer to $x^S$ (resp.~$x^T$) as the \emph{projection of $x$ onto $S$ parallel
  to $T$} (resp.~\emph{onto $T$ parallel to $S$}).
We call the unique linear map $L: V \to V$ that maps $x \mapsto x^S$ the
\emph{projector onto $S$ parallel to $T$}.
\end{definition}

A given subspace may have many different complementary subspaces: consider the
example of $S = \mathrm{span}\{(1,1) \}$ in $\F_2^2$.
Different complementary subspaces include $T = \mathrm{span}\{(1,0)\}$ and
$T' = \mathrm{span}\{(0,1)\}$.
It is convenient to define the notion of a \emph{canonical complement} of a
subspace $S$, given a basis for $S$.

\begin{definition}
  \label{def:canonical-complement}
  Let $E$ be the standard basis of linear space $V = \F^n$.
  Let $F = \{v_1, v_2, \ldots, v_m\} \subset V$ be a set of $m$ linearly
  independent vectors in $V$.
  The \emph{canonical complement $F^\perp$ of $F$} is the set of $n-m$ independent
  vectors defined as follows.
  Write $v_i = \sum_{j=1}^n a_{i, j}\, \hat{e}_j$.
	Using a canonical algorithm for Gaussian elimination that works over arbitrary
  fields, transform the $m\times n$ matrix $(a_{i, j})$ to reduced row echelon
  form $(b_{i, j})$.
  Let $J$ be the set of $m$ column indices of the leading $1$ entry in each row
  of $(b_{i, j})$.
  The canonical complement is defined as $F^\perp = \{ \hat{e}_j : j \not\in J \}$.
\end{definition}

\begin{remark}\label{rk:cancomp}
We emphasize that the canonical complement $F^\perp$ is a \emph{set} (rather than
a subspace), and it is defined with respect to a \emph{set} of linearly
independent vectors $F$. In particular, is easy to see that $F^\perp$ depends on $F$ only and not on the order of the vectors that specify $F$.\footnote{An equivalent definition proposed by a referee is that $F^\perp$ is the set of $\hat{e}_j$ where $j$ is such that $u_j=u_{j+1}$, where $u_j$ is the dimension of $\text{span}(F) \cap \text{span}(\hat{e}_j,\dots,\hat{e}_n)$.}
\end{remark}

\begin{remark}
  Let $\hat{E}$ be the standard basis of $V$.
  Suppose subspace $S$ is a register subspace of $V$ spanned by
  $\hat{E}_0 \subseteq \hat{E}$.
  Then it is not hard to verify that the canonical complement of $\hat{E}_{0}$ is
  $\hat{E}\setminus \hat{E}_0$ and the span of the canonical complement coincides with
  $S^\perp$.
\end{remark}

\begin{lemma}
  \label{lem:canonical-complement}
  Let $S$ be the span of linearly independent vectors $F = \{v_1,\ldots,v_m \}
  \subseteq V$ and let $F^\perp$ be the canonical complement of $F$ as defined
  in Definition~\ref{def:canonical-complement}.
  Let $T = \mathrm{span}(F^\perp)$.
  Then
  \begin{equation*}
   \qquad S + T = V\;,  S \cap T = \{ 0 \} \;.
  \end{equation*}
\end{lemma}

\begin{proof}
To show the first item, we must show that every vector in $V$ can be
written as a sum of an element of $S$ and an element of $T$. To do
this, let $A = (a_{i, j})$ be the $m\times n$ matrix over $\F$ whose
rows are the vectors $v_i$, as in Definition~\ref{def:canonical-complement}.
  Write $A = U B$ where $U$ is invertible and $B$ is in reduced row echelon
  form.
  Let $J$ be the set of column indices of the leading 1 entries in
  $B$, as in Definition~\ref{def:canonical-complement}.  
  Now, the rows of $B$ are linearly independent and span the subpsace $S$, and the restriction of
  these rows to the columns in $J$ are still linearly independent and
  span $\F^m$. Thus, for any vector $u \in V$, there is some linear
  combination $v$ of rows of $B$ such that $u$ and $v$ agree
  restricted to the columns in $J$. In other words, there exists $v
  \in S$ such that $u_j = v_j$ for all $j \in J$. Hence, $u = v + w$
  for $w \in T$, where $T$ is the canonical complement.
  This shows $S + T = V$.
  Counting dimensions shows that necessarily $S\cap T =
  \{0\}$.
\end{proof}

\begin{definition}
  \label{def:cl-canonical}
  Let $F \subseteq V$ be a set of linearly independent vectors.
  Let $F^\perp$ be the canonical complement of $F$.
  Define the \emph{canonical linear map $L \in \End(V)$ with kernel basis $F$}
  as the projector onto $T$ parallel to $S$, where $S = \mathrm{span}(F)$ and $T
  = \mathrm{span}(F^\perp)$.
  We refer to this map as the
  \emph{canonical linear map with kernel $S$,} which does not depend on $F$ by Remark~\ref{rk:cancomp}.
\end{definition}

\begin{definition}
  \label{def:Lperp}
  Let $L \in \End(V)$ be a linear map, and let $F$ be a basis for
  $\ker(L)^\perp$.
  Define $L^\perp : V \rightarrow V$ as the canonical linear map with kernel
  basis $F$.
\end{definition}

\begin{lemma}
  \label{lem:L_perp_perp}
  Let $L \in \End(V)$ be a linear map and $F$ a basis for $\ker(L)^\perp$.
  Let $L^\perp \in \End(V)$ be the linear map defined in
  Definition~\ref{def:Lperp}.
  Then $\ker(L^\perp) = \ker(L)^\perp$.
\end{lemma}

\begin{proof}
  First, to set notation, let $S = \mathrm{span}(F) = \ker(L)^\perp$ and $T =
  \mathrm{span}(F^\perp)$. By Lemma~\ref{lem:canonical-complement},
  any vector $v \in V$ can be uniquely decomposed as $v = v^S +
  v^T$. 

  Now we will show the lemma.  First we show that if $\ker(L^\perp) \subseteq \ker(L)^\perp$. Let
  $v \in \ker(L^\perp)$, and write $v = v^S + v^T$.  By the definition
  of $L^\perp$, it follows that $v^T = 0$, and hence $v \in S = \ker(L)^\perp$.

  It remains to show the other direction, that is $\ker(L)^\perp \subseteq
  \ker(L^\perp)$. Let $v \in \ker(L)^\perp = S$. Then $v = v^S + v^T$
  where $v^T = 0$, and hence by the definition of $L^\perp$, $L^\perp
  v = 0$. Hence $v \in \ker(L^\perp)$.

\end{proof}

\subsection{Finite fields}
\label{sec:finite-fields}

Let $p$ be a prime and $q = p^k$ be a prime power.
We denote the finite fields of $p$ and $q$ elements by $\Fp$ and $\Fq$,
respectively.
The prime $p$ is the characteristic of field $\Fq$, and field $\Fp$ is the
\emph{prime subfield} of $\Fq$.
We sometimes omit the subscript and simply use $\F$ to denote the finite field
when the size of the field is implicit from context.
For general background on finite fields, and explicit algorithms for elementary
arithmetic operations, we refer to~\cite{mullen2013handbook}.

\subsubsection{Subfields and bases}
\label{sec:subfields}
Let $q$ be a prime power and $k$ an integer. 
The field $\Fq$ is a subfield of $\F_{q^k}$ and $\F_{q^k}$ is a linear space of
dimension $k$ over $\Fq$.
Let $\{ e_i \}_{i=1}^k$ be a basis of $\F_{q^k}$ as a linear space over $\Fq$.
Introduce a bijection $\downsize_q : \F_{q^k} \rightarrow \Fq^k$ between
$\F_{q^k}$ and $\Fq^k$ defined with respect to the basis $\{ e_i \}_{i=1}^k$ by
\begin{equation*}
  \downsize_q : a \mapsto (a_i)_{i=1}^k
\end{equation*}
where $a = \sum_{i=1}^k a_i e_i$.
This map satisfies several nice properties.
First, the map is an isomorphism of $\Fq$-vector spaces: it is $\Fq$-linear and addition in $\F_{q^k}$
naturally corresponds to vector addition in $\Fq^k$.
Namely, for all $a,b \in \F_{q^k}$,
\begin{equation*}
	\downsize_q (a + b) = \downsize_q(a) + \downsize_q(b)\;.
\end{equation*}
Second, multiplication by a field element in $\F_{q^k}$ corresponds to a linear
map on $\F_{q}^k$.
For all $a \in \F_{q^k}$, there exists a matrix $K_a \in \Matrix_k (\Fq)$ such
that for all $b \in \F_{q^k}$,
\begin{equation*}
	\downsize_q (a b) = K_a \, \downsize_q(b)\;.
\end{equation*}
The matrix $K_a$ is called the \emph{multiplication table of $a$ with respect to
  basis $\{e_i\}_{i=1}^k$}.

We extend the map $\downsize_q$ to vectors, matrices and sets over $\F_{q^k}$.
For $v = (v_1, v_2, \ldots, v_n) \in \F_{q^k}^n$, define
\begin{equation*}
  \downsize_q (v) = \bigl( \downsize_q (v_i) \bigr)_{i=1}^n \in \Fq^{kn}\;.
\end{equation*}
Similarly, for matrix $M = (M_{i, j}) \in \Matrix_{m, n}(\F_{q^k})$, define
\begin{equation*}
  \downsizem_q (M) = \bigl( K_{M_{i, j}} \bigr) \in \Matrix_{mk, nk} (\Fq)\;,
\end{equation*}
the block matrix whose $(i,j)$-th block is the multiplication table $K_{M_{i,
    j}}$ of $M_{i, j}$ with respect to basis $\{e_i\}_{i=1}^k$.
For a set $S$ of vectors in $\F_{q^k}^n$, define
\begin{equation*}
	\downsize_q(S) = \{ \downsize_q(v) : v \in S \}\;.
\end{equation*}
We omit the subscript and write $\downsize$ and $\downsizem$ for $\downsize_q$
and $\downsizem_q$, respectively, when $q$ equals to $p$, the characteristic of
the field.

The \emph{trace of\/ $\F_{q^k}$ over\/ $\Fq$} is defined as
\begin{equation}\label{eq:def-trace}
  \tr_{q^k \to q} :\; a \,\mapsto \, \Tr(K_a)
\end{equation}
for $a \in \F_{q^k}$, where $\Tr(K_a)$ is the trace of the multiplication table
of $a$ with respect to the basis $\{e_i\}$.
By definition, the trace is an $\Fq$-linear map from $\F_{q^k}$ to
$\Fq$. The trace is in fact independent of the choice of basis, which
can be seen from the following equivalent definition~\cite[Definition 2.1.80]{mullen2013handbook}:
\[
	\tr_{q^k \to q}(a) = \sum_{j = 0}^{k-1} a^{q^j}\;.
\]

A \emph{dual basis} $\{e_1',\allowbreak  e_2',\allowbreak  \ldots,\allowbreak  e_k' \}$ of $\{e_1,\allowbreak  e_2,\allowbreak  \ldots,\allowbreak 
e_k\}$ is a basis such that \[\textstyle \tr_{q^k \to q}(e^{}_i e'_j) = \delta_{i,j}\] for
all $i,\allowbreak  j \in \{1,\allowbreak  2,\allowbreak  \ldots,\allowbreak  k\}$.
A \emph{self-dual basis} is one that is equal to its dual.
If for some $\alpha \in \F_{q^k}$ the set $\{ \alpha^{q^j} \}_{j=0}^{k-1}$ forms a
basis of $\F_{q^k}$ over $\Fq$, the basis is called a \emph{normal
  basis}. Self-dual and normal bases do not exist for all fields but are guaranteed to exist under
certain conditions; in particular, if $q = 2$ and $k$ is odd, as will
be shown in Lemma~\ref{lem:efficient_basis}.

We record some convenient facts about the maps $\downsize(\cdot)$ and
$\downsizem(\cdot)$ for self-dual bases.

\begin{lemma}
  \label{lem:downsize_field}
  Let $q$ be a prime power, $k$ an integer and $\{e_i\}$ a self-dual basis for
  $\F_{q^k}$ over $\F_q$.
  The map $\downsize_q(\cdot)$ corresponding to $\{e_i\}$ satisfies the
  following properties:
  \begin{enumerate}
  \item For all $x \in \F_{q^k}$, $\downsize_q(x) = \bigl( \tr_{q^k \to
      q}(xe_1),\ldots,\tr_{q^k \to q}(xe_k) \bigr)$.
  \item For all $x,y \in \F_{q^k}$, $\tr_{q^k \to q}(xy) = \downsize_q(x) \cdot
    \downsize_q(y)$.
  \item For all $M \in \Matrix_{m, n}(\F_{q^k})$ and $v \in
    \F_{q^k}^n$, we have $\downsizem_q(M) \downsize_q(v) = \downsize_q(Mv)$.
  \end{enumerate}
\end{lemma}
 
\begin{proof}
  We show each property by direct calculation.
  \begin{enumerate}
    \item Let $\downsize_q(x) = (x_1, \dots, x_k)$. Then by the definition of
    $\downsize_q$, $x = \sum_i x_i e_i$. Multiplying both sides by $e_j$
    and taking the trace yields $\tr_{q^k \to q}(x e_j) = \sum_i x_i
    \tr_{q^k \to q}(e_i e_j) = x_j$, where the last equality is by
    self-duality of the basis.
    \item Write $\downsize_q(x) = (x_1, \dots, x_k)$ and $\downsize_q(y) =
      (y_1, \dots, y_k)$. Then
      \begin{align*}
        \tr_{q^k \to q} (xy) &= \tr_{q^k \to q}(\sum_{ij} x_i y_j e_i
                               e_j) \\
                             &= \sum_{ij} x_i y_j \tr_{q^k \to q}(  e_i
                               e_j) \\
                             &= \downsize_q(x) \cdot \downsize_q(y),
      \end{align*}
      where the last equality uses the self-duality of the basis.
    \item
      We compute the $i$th $\F_q^k$-block of $\downsize_q(Mv)$:
      \begin{align*}
        \downsize_q(Mv)_i &= \sum_{ij} \downsize_q(M_{ij} v_j) \\
                          &= \sum_{ij} K_{M_{ij}} \downsize_q(v_j) \\
                          &= \downsizem_q(M)_{ij} \downsize_q(v_j) ,
      \end{align*}
      where we have applied the definition of the multiplication table
      $K_{M_{ij}}$ and the map $\downsizem_q(\cdot)$.
  \end{enumerate}
\end{proof}

For $z\in \F^n$ and $V,W$ a pair of complementary subspaces, recall from
Definition~\ref{def:complementary} the notation $z^V$ for the projection of $z$
onto $V$ and parallel to $W$.

\begin{lemma}
  \label{lem:downsize_subspace}
  Let $\downsize_q(\cdot)$ denote the map corresponding to a
  basis 
  $\{e_i\}$ for $\F_{q^k}$ over $\F_q$.
  Let $V$ be a subspace of $\F_{q^k}^n$ with linearly independent basis
  $\{b_1,\ldots,b_t\} \subseteq \F_{q^k}^n$.
  Then the following hold:
  \begin{enumerate}
 	\item $\downsize_q(V)$ is a subspace of $\,\F_{q}^{nk}$.
	\item $\{ \downsize_q(e_i b_j) \}_{i,j}$ is a linearly independent basis of
    $\downsize_q(V)$ over $\F_q$.
	\item Let $V, W$ be complementary subspaces of $\F_{q^k}^n$.
    Then $V' = \downsize_q(V)$ and $W' = \downsize_q(W)$ are complementary
    subspaces of $\F_q^{kn}$, and, furthermore, for all vectors $z \in
    \F_{q^k}^n$, we have $\downsize_q(z^V) = \downsize_q(z)^{V'}$ and
    $\downsize_q(z^W) = \downsize_q(z)^{W'}$.
  \end{enumerate}
\end{lemma}

\begin{proof}
  For the first item, we first verify that $\downsize_q(V)$ is a subspace.
  Since $V$ is a subspace, it contains $0 \in \F_{q^k}^n$, and, therefore, $\downsize_q(0)
  = 0$ is also in $\downsize_q(V)$.
  Let $u', v' \in \downsize_q(V)$.
  Using that $\downsize_q$ is a bijection there exist $u,v \in \F_{q^k}^n$ such that
  $u' = \downsize_q(u)$ and $v' = \downsize_q(v)$. Therefore,
  \[
    u' + v' = \downsize_q(u) + \downsize_q(v) =
    \downsize_q(u + v) \in \downsize_q(V)\;,
  \]
  where the inclusion follows because $V$ is a subspace and thus contains $u +
  v$.
  Finally, for all $x' \in \F_{q}$, for all $v \in V$, we have that $x'
  \downsize_q(v) = \downsize_q(x' v)\in \downsize_q(V)$ where we used that $V$
  is closed under scalar multiplication by $\F_{q^k}$ and thus by $\F_q$ (since
  $\F_q$ is a subfield of $\F_{q^k}$).
  Thus $\downsize_q(V)$ is closed under scalar multiplication by $\F_{q}$.

  For the second item, note that an element $v \in V$ can be expressed uniquely
  as $v = \sum_{i = 1}^t v_i b_i$ for $v_i \in \F_{q^k}$.
  The element $v_i$ can further be written as $\sum_j v_{i, j} e_j$ where $v_{i,
    j} \in \F_q$.
  Thus $v$ is a linear combination of the vectors $\{ e_j b_i \}$, and, therefore,
  $\downsize_q(v)$ is a linear combination of the vectors $\{ \downsize_q (e_j
  b_i) \}$.
  To establish that the vectors $\{ \downsize_q(e_j b_i) \}$ are linearly
  independent, suppose towards contradiction that they are not.
  Then there would exist $\alpha_{i, j} \in \F_q$ such that at least one
  $\alpha_{i, j}$ is nonzero and
  \begin{align*}
    0 & = \sum_{i,j} \alpha_{i, j} \downsize_q(e_j b_i) \\
      & = \downsize_q \Bigl( \sum_i \bigl (\sum_j \alpha_{i, j}
        e_j \bigr) b_i \Bigr) \\
      & = \downsize_q \left ( \sum_i \beta_i b_i \right ),
  \end{align*}
  where we define $\beta_i = \sum_j \alpha_{i, j} e_j$.
  Since at least one $\alpha_{i, j} \neq 0$ and the $\{e_j\}$ are linearly
  independent over $\F_q$, there exists $i$ such that $\beta_i \neq 0$, which
  means that there is a nontrivial linear combination of the basis elements
  $b_i$ that equals $0$ under $\downsize_q(\cdot)$.
  Since $\downsize_q(\cdot)$ is injective, we get a contradiction with linear
  independence of the $\{b_i\}$.

  For the third item, we observe that $\downsize_q(V)$ and $\downsize_q(W)$ must
  be complementary because $\downsize_q(\cdot)$ is a linear map as well as a
  bijection.
  Let $\{v_1,\allowbreak \ldots,\allowbreak v_m\}$ and $\{v_{m+1},\allowbreak \ldots,\allowbreak v_n\}$ denote bases for $V$ and
  $W$, respectively.
  Thus the set $\{v_1,\allowbreak \ldots,\allowbreak v_n\}$ forms a basis for $\F_{q^k}^n$, and from the
  previous item, the set $\{ \downsize_q(e_j v_i) \}_{i,\, j}$ is a basis for
  $\F_q^{kn}$.
  Furthermore, the sets $\{ \downsize_q(e_j v_i) \}_{j,\,i = 1,\ldots m}$ and
  $\{ \downsize_q(e_j v_i) \}_{j,\, i = m+1,\ldots n}$ are bases for
  $\downsize_q(V)$ and $\downsize_q(W)$, respectively.

  There is a unique choice of coefficients $\alpha_{i, j} \in \F_q$ such that
  $\downsize_q(z) = \sum_{i, j} \alpha_{i, j} \downsize_q(e_j v_i)$.
  But then
  \begin{align*}
    \downsize_q(z) & = \downsize_q \Bigl( \sum_i \bigl ( \sum_j
                     \alpha_{i, j} e_j \bigr) v_i \Bigr) \\
                   & = \downsize_q \Bigl( \sum_i \alpha_i v_i \Bigr)\;,
  \end{align*}
  where we define $\alpha_i = \sum_j \alpha_{i, j} e_j$.
  Since $\downsize_q(\cdot)$ is a bijection, this implies that $z = \sum_i
  \alpha_i v_i$, and, therefore, $z^V = \sum_{i=1}^m \alpha_i v_i$ (and similarly
  $z^W = \sum_{i=m+1}^n \alpha_i v_i$).
  This implies that
  \begin{equation*}
	  \downsize_q(z)^{V'} = \sum_{i=1}^m \sum_{j} \alpha_{i, j}
    \downsize_q(e_j v_i) = \downsize_q (z^V)\;,
  \end{equation*}
  and similarly $\downsize_q(z)^{W'} = \downsize_q(z^W)$.
  This completes the proof of the lemma.
\end{proof}

\subsubsection{Bit string representations}
\label{sec:ff-representations}

As mentioned at the end of Section~\ref{sec:tms}, we sometimes treat the
inputs and outputs of Turing machines as representing elements of a finite
field, or a vector space over a finite field.
We discuss some important details about bit representations of finite field
elements and arithmetic over finite fields.

In the paper we only consider fields $\F_{2^k}$ where $k$ is odd. 

\begin{definition}\label{def:admissible-size}
  A field size $q$ is called an \emph{admissible} field size if $q=2^k$ for odd
  $k$.
\end{definition}

Elements of $\F_2$ are naturally represented using bits. 
To represent elements of $\F_{2^k}$ as binary strings we require the
specification of a basis of $\F_{2^k}$ over $\F_2$.
Given a basis $\{e_i\}_{i=1}^k$ of $\F_{2^k}$, every element $a \in \F_{2^k}$
has a unique expansion $a = \sum_{i=1}^k a_i e_i$ and can be represented as the
$k$-bit string corresponding to $\downsize(a) \in \F_2^k$.
Note that we omitted the subscript $2$ of $\downsize$ as it maps to the linear
space over the prime subfield $\F_2$.
Thus the \emph{binary representation} of $a \in \F_{2^k}$ is defined as the
natural binary representation of $\downsize(a) \in \F_2^k$ (which in turn is the
\emph{$\F_2$-representation of $a$}).
Throughout the paper we freely associate between the binary representation of a
field element $a \in \F_{2^k}$ and its $\F_2$-representation,
although---technically speaking---these are distinct objects.

Given the representations $\downsize(a), \downsize(b)$ of $a, b \in \F_{2^k}$, to
compute the binary representation of $a+b$ it suffices to compute the addition
bit-wise, modulo $2$.
Computing the multiplication of elements $a, b$ requires the specification of
the multiplication tables $\{K_{e_i} \in \Matrix_k (\F_2)\}_{i=1}^k $ for the
basis $\{e_i\}$.
Given representations $\downsize(a) = (a_i)_{i=1}^k$, $\downsize(b) =
(b_i)_{i=1}^k$ for $a, b \in \F_{2^k}$, respectively, the representation
$\downsize(ab)$ of the product $ab$ is computed as
\begin{equation}
  \label{eq:eq-mult}
  \downsize(ab) = \sum_{i=1}^k \, a_i \, \downsize \bigl( e_i b \bigr) =
  \sum_{i=1}^k \, a_i \bigl( K_{e_i}\, \downsize (b) \bigr)\;.
\end{equation}
Thus, using our representation for field elements, efficiently performing finite
field arithmetic in $\F_{2^k}$ reduces to having access to the multiplication
table of some basis of $\F_{2^k}$ over $\F_2$.

The following fact provides an efficient deterministic algorithm for computing a
self-dual normal basis for $\F_{2^k}$ over $\F_2$ and the corresponding
multiplication tables for any odd $k$.

\begin{lemma}
  \label{lem:efficient_basis}
  There exists a deterministic algorithm that given an odd integer $k > 0$,
  outputs a self-dual normal basis of $\F_{2^k}$ over $\F_2$ and the
  multiplication tables of the basis in $\poly(k)$ time.
\end{lemma}

\begin{proof}
	The algorithm of Shoup~\cite[Theorem 3.2]{shoup1990new} shows that for prime
  $p$, an irreducible polynomial in $\Fp[X]$ of degree $k$ can be computed in
  time $\poly(p, k)$.
  Then, the algorithm of Lenstra~\cite[Theorem 1.1]{lenstra1991finding} shows
  that given such an irreducible polynomial, the multiplication table of a
  normal basis of $\F_{p^k}$ over $\Fp$ can be computed in $\poly(k,\log p)$
  time.
  Finally, the algorithm of Wang~\cite{wang1989algorithm} shows that for odd $k$
  and a multiplication table $K$ of a normal basis of $\F_{2^k}$ over $\F_2$, a
  multiplication table $K'$ for a self-dual normal basis of $\F_{2^k}$ over
  $\F_2$ can be computed in $\poly(k)$ time.
  Putting these three algorithms together yields the claimed statement.
\end{proof}

For $q=2^k$ for $k$ odd (i.e.,\ $q$ is an admissible field size) we use the shorthand $\tr$ for $\tr_{q\to 2}$.

\begin{lemma}
  \label{lem:one}
  Let $k$ be an odd integer and $\{e_i\}_{i=1}^k$ be a self-dual normal basis of
  $\F_{2^k}$ over $\F_2$.
  Then $\tr(e_i) = 1$ for all $i$, and, furthermore, the representation
  $\downsize(1)$ of the unit $1 \in \F_{2^k}$ is the all ones vector in
  $\F_2^k$.
\end{lemma}

\begin{proof}
  Since $\{e_i \}$ is a normal basis, $e_i = \alpha^{2^i}$ for some element
  $\alpha \in \F_{2^k}$.
  Furthermore, for every element $b \in \F_{2^k}$, we have that $\tr(b^2) =
  \tr(b)$.
  This is because
  \[
    \tr(b^2) = \sum_{i=0}^{k-1} b^{2^{i+1}} =
    \sum_{i = 0}^{k-1} b^{2^i} = \tr(b)\;,
  \]
  where we use that $b^{2^k} = b$ for all $b \in \F_{2^k}$.
  Since $e_{i+1} = e_i^2$, we get that $\tr(e_i) = \tr(e_j)$ for all $i,j$.
  It cannot be the case that $\tr(e_i) = 0$ for all $i$.
  Suppose that this were the case.
  This would imply that $\tr(b) = 0$ for all $b \in \F_{2^k}$.
  But then for all $j \in \{1, \ldots, k\}$ and for some $b \neq 0$, we would
  also have by item 1 of Lemma~\ref{lem:downsize_field} that $b_j = \tr(b e_j) = 0$ where $b = \sum_j b_j e_j$ with $b_j \in
  \F_2$.
  This implies that $b$ is the all zero element of $\F_{2^k}$, which is a
  contradiction.
  Thus $\tr(e_i) = 1$ for all $i=1, 2, \ldots, k$.

	The ``furthermore'' part follows item 1 of
        Lemma~\ref{lem:downsize_field}:
        \begin{equation*}
          \downsize(1) = (\tr(1 \cdot e_1), \dots, \tr(1 \cdot e_k))
          = (1, \dots, 1).
        \end{equation*}
\end{proof}

\begin{lemma}
  \label{lem:efficient_arithmetic}
  For any odd integer $k$, let $\{e_i\}_{i=1}^k$ denote the self-dual normal basis
  of $\F_{2^k}$ over $\F_2$ that is returned by the algorithm specified in
  Lemma~\ref{lem:efficient_basis} on input $k$.
  Then the following can be computed in time $\poly(k)$ on input $k$:
  \begin{enumerate}
  \item The representation $\downsize(a+b)$ of the sum $a+b$ given the
    representations $\downsize(a)$ and $\downsize(b)$ of $a, b \in \F_{2^k}$.
	\item The representation $\downsize(ab)$ of the product $ab$ given the
    representations $\downsize(a)$ and $\downsize(b)$ of $a, b \in \F_{2^k}$.
  \item The multiplication table $K_a \in \Matrix_k(\F_2)$ given the
    representation $\downsize(a)$ of  $a\in \F_{2^k}$.
  \item The representation $\downsize(a^{-1})$ of the multiplicative inverse of
    $a \in \F_{2^k}$, given the representation $\downsize(a)$.
	\item The trace $\tr(a)$ given the multiplication table $K_a$ of $a \in \F_{2^k}$.
  \end{enumerate}
  Furthermore, for all integers $n$, the representations of projections
  $\downsize(x^S)$ and $\downsize(x^T)$ of $x \in \F_{2^k}^n$ for complementary
  subspaces $S,T$ of $\F_{2^k}^n$ can be computed in $\poly(k,n)$ time, given
  the representations $\downsize(x)$, $\{\downsize(v_1), \downsize(v_2), \ldots,
  \downsize(v_m)\}$ and $\{\downsize(w_1), \downsize(w_2), \ldots,
  \downsize(w_{n - m})\}$ where $\{v_i\}$ and $\{w_j\}$ are bases for $S$ and
  $T$, respectively.
\end{lemma}

\begin{proof}
  Given an odd integer $k$ as input, by Lemma~\ref{lem:efficient_basis} it is
  possible to compute the self-dual normal basis $\{e_i\}_{i=1}^k$ together with the
  multiplication tables $K_{e_i}$ for $i = 1, 2, \ldots, k$.
  Addition is performed component-wise, and multiplication is done
  using Eq.~\eqref{eq:eq-mult}.
  For the multiplication table $K_a$ it suffices to compute the $k$ products
  $\downsize(a e_i)$ for $i\in\{1,\ldots,k\}$.
  To compute inverses, observe that $\downsize(1) = \downsize(aa^{-1}) = K_a
  \downsize(a^{-1})$.
  The matrices $K_a$ are invertible over $\F_2$, so, therefore, $\downsize(a^{-1}) =
  K_a^{-1} \downsize(1)$; moreover, $\kappa(1)$ is the all-ones vector
  by Lemma~\ref{lem:one} and can thus be efficiently computed.
  Inverting the matrix can be done in $\poly(k)$ time via Gaussian elimination.
  The trace of an element $a \in \F_{2^k}$ is by definition the trace of the
  multiplication table $K_a$.

  For the ``Furthermore'' part, we observe that since $\{v_1,\allowbreak  v_2,\allowbreak  \ldots,\allowbreak  v_m
  \} \cup \{w_1,\allowbreak  w_2,\allowbreak  \ldots,\allowbreak  w_{n-m} \}$ forms a basis for $\F_{2^k}^n$, there
  is a unique way to write $x$ as a $\F_{2^k}$ linear combination of $\{ v_i\}$
  and $\{w_j\}$.
  Via Gaussian elimination over $\F_{2^k}$, the $\F_2$-representation of the
  coefficients of this linear combination can be computed in $\poly(n,\allowbreak k)$ time.
  Here we use that addition, multiplication and division over $\F_{2^k}$ can be
  performed in time $\poly(k)$ using the previous items of the Lemma.
\end{proof}

\begin{remark}
  \label{rmk:tm_fields}
  Throughout this paper, whenever we refer to Turing machines that perform
  computations with elements of a field $\F_q$ for an admissible field size $q =
  2^k$, we mean that that the Turing machines are representing elements of
  $\F_q$ as vectors in $\{0,1\}^{k}$ using the basis specified by the
  algorithm of Lemma~\ref{lem:efficient_basis} and performing arithmetic as
  described in Lemma~\ref{lem:efficient_arithmetic}.
\end{remark}

\subsection{Polynomials and the low-degree code}
\label{sec:ld-encoding}

In this section, we introduce some basic definitions about polynomials over finite fields. 
An $m$-variate polynomial $f$ over $\F_q$ is a function of the form
\[
	f(x_1,\ldots,x_m) = \sum_{\alpha \in \{0,1,\ldots,q-1\}^m} c_\alpha
  x_1^{\alpha_1} \cdots x_m^{\alpha_m}
\]
where $\{c_\alpha\}$ is a collection of coefficients in $\F_q$.
We say that $f$ has \emph{individual degree (at most) $d$} if $c_\alpha\neq 0$ only if
$\alpha_i \leq d $ for all $1\leq i\leq m$, and that it has \emph{total degree
  (at most) $d$} if $c_\alpha\neq 0$ only if $\sum_i \alpha_i \leq d$.
(Affine) \emph{multilinear} polynomials are polynomials with individual degree
$1$.

Low-degree polynomials play an important role in this paper.
For one, the classical and quantum low-degree tests of Section~\ref{sec:ldt} are
nonlocal games that efficiently certify that the players' answers are consistent
with low-degree polynomials.
Low-degree polynomials are also crucial in the probabilistically checkable
proofs (PCP) construction that are used in ``answer reduction'' transformation
(Section~\ref{sec:ans}).

We recall the Schwartz-Zippel lemma: 

\begin{lemma}[Schwartz-Zippel lemma~\cite{Sch80,Zip79}]
  \label{lem:schwartz-zippel}
  Let $f, g: \F_q^m \to \F_q$ be two unequal polynomials with total degree at most $d$. Then
  \begin{equation*}
    \Pr_{x \sim \F_q^m}[f(x) = g(x)] \leq d/q\;.
  \end{equation*}
\end{lemma}

\paragraph{The low-degree code.}
The Schwartz-Zippel lemma implies that the set of low-degree polynomials form an
error-correcting code with good distance, which we call the \emph{low-degree
  code}.\footnote{It is also known as the \emph{generalized Reed-Muller code} in
  the coding theory literature.}

Fix an integer $m \in \N$ and let $M = 2^m$.
For every $y \in \{0,1\}^m$ define the following $m$-variate multilinear
polynomial over $\F_q$:
\[
	\ind_{m,y}(x) = \prod_{i: y_i = 1} x_i \cdot \prod_{i: y_i = 0} (1 - x_i) \;.
\]
Here, we identify $\{0,1\}$ as a subset of $\F_q$.
Notice that, when restricted to the subcube $\{0,1\}^m$, $\ind_{m, y}$ is zero
everywhere except for when $x = y$.
For $a\in \F_q^M$, label the coordinates of $a$ as $a_y$ for $y\in \{0,1\}^m$
(identifying the latter set with $\{1,\ldots,M\}$ using, say, the lexicographic
ordering on strings).
For any such $a$ define the multilinear polynomial $g_a: \F_q^m \to \F_q$,
called the \emph{low-degree encoding of $a$}, as follows:
\begin{equation}
\label{eq:ld-encoding}
	 g_a(x) = \sum_{y \in \{0,1\}^m} a_y \cdot \ind_{m,y}(x)\;.
\end{equation}
Note that for any $y \in \{0,1\}^m$, $g_a(y) = a_y$.
Furthermore, the map $a\mapsto g_a$ is linear: for every $x \in \F_q^m$,
\begin{equation}
\label{eq:low-degree-encoding-definition}
	g_a(x) = a \cdot \ind_m(x)\;,
\end{equation}
where $\ind_m(x)$ is the vector $(\ind_{m,y}(x))_{y \in \{0,1\}^m}\in\F_q^M$.

\begin{lemma}
\label{lem:ld-encoding-complexity}
Let $q$ be an admissible field size, and let $M = 2^m$ for some integer $m$.
The complexity of computing the evaluation of the low-degree encoding $g_a(x)$,
given $a \in \F_q^M$ and $x \in \F_q^m$ represented in binary, is $\poly(M,\log
q)$.
\end{lemma}
\begin{proof}
	Computing $g_a(x)$ requires computing the sum of products $a_y \cdot
  \ind_{m,y}(x)$ over all $y \in \{0,1\}^m$.
  Via Lemma~\ref{lem:efficient_arithmetic}, evaluating $\ind_{m,y}(x)$ takes time
  $\poly(m,\log q)$ because it requires performing $m$ multiplications of $\F_q$
  elements, and, therefore, the product $a_y \cdot \ind_{m,y}(x)$ requires
  $\poly(m,\log q)$ time.
  Computing the sum over all $y \in \{0,1\}^m$ requires $M \cdot \poly(m,\log q)
  = \poly(M,\log q)$ time, as claimed.
\end{proof}

Finally, for any $H \subseteq \F_q$ we define the \emph{decoding} map
$\coded_H(\cdot)$ which takes as input a polynomial $g : \F_q^m \to \F_q$ and
returns the following vector $a \in \F_q^M$: for all $y \in \{0,1\}^m$, if $g(y)
\in H$, then set $a_y = g(y)$, and otherwise set $a_y = 0$.
Note that for all $a \in H^M$ we have $\coded_H(g_a) = a$ where $g_a$ is the
low-degree encoding of $a$.
For notational brevity we often write $\coded(\cdot)$ to denote the boolean
decoding map $\coded_{\{0,1\}}(\cdot)$. A convenient property of this
decoding map is that $\coded(g) \in \{0, 1\}^{M}$ for all polynomials
$g$.

\subsection{Linear spaces and registers}
\label{sec:lin-reg}

For a set $V$, we write $\C^V$ for the complex vector space of dimension $|V|$.
The space $\C^V$ is endowed with a canonical orthonormal basis
$\{\ket{x}\}_{x\in V}$.
By ``a quantum state on $V$'' we mean a unit vector
\begin{equation*}
  \ket{\psi}_V \in \complex^V\;.
\end{equation*}
If $V = \bigoplus_{i=1}^k V_k$ is the direct sum of subspaces $V_k$ over $\F$,
then $\complex^V$ can be identified with $\bigotimes_{i=1}^k
\complex^{V_i}$, by the identification of $\ket{x_1 \oplus \dots \oplus
  x_k}$ with $\ket{x_1} \otimes \dots \otimes \ket{x_k}$.
As a special case, if $\{e_i\}$ is a basis of $V$ the decomposition $V =
\bigoplus_{i=1}^k ( \F e_i)$ yields the tensor product decomposition $\C^V =
\bigotimes_{i=1}^k \C^{\abs{\F}}$.
We sometimes refer to the spaces $\C^{\abs{\F}}$ as the ``qudits'' of $\C^V$ (or
of a state on it).

\begin{definition}
  \label{def:EPR}
  For a linear space $V$ over a finite field $\F$, define the \EPR state on $\C^V
  \otimes \C^V$ by
  \begin{equation*}
    \ket{\EPR}_V = \frac{1}{\sqrt{\abs{V}}} \sum_{x\in V}\ket {x}\otimes \ket{x}\;.
  \end{equation*}
  We also write $\ket{\EPR}_{\F_q}$ as $\ket{\EPR_q}$ and $\ket{\EPR_2}$ as
  $\ket{\EPR}$.
\end{definition}

\subsection{Measurements and observables}

Quantum measurements are modeled as positive operator-valued
measures (POVMs).
A POVM consists of a set of positive semidefinite operators $\{ M_a \}_{a\in S}$
indexed by outcomes $a\in S$ that satisfy the condition $\sum_a M_a = I$. If the latter condition is relaxed to $\sum_a M_a \leq I$ then we refer to $\{M_a\}$ as a \emph{sub-measurement}.
We sometimes use the same letter $M$ to refer to the collection of operators
defining the POVM.
The probability that the measurement returns outcome $a$ on state $\ket{\psi}$
is given by
\begin{equation*}
  \Pr(a) = \bra{\psi} M_a \ket{\psi}\;.
\end{equation*}
A POVM $M=\{M_a\}$ is said to be \emph{projective} if each operator $M_a$ is a
projector ($M_a^2 = M_a$). This automatically implies that operators
$M_a, M_b$ for distinct outcomes $a \neq b$ are orthogonal ($M_a M_b =
M_b M_a = 0$). An \emph{observable} is a unitary matrix.
A \emph{binary observable} is an observable $O$ such that $O^2 = \Id$, i.e.,\ $O$
has eigenvalues in $\{-1,1\}$.

We follow the convention that subscripts of the measurement index
the outcome and superscripts of the measurement are used to index different
measurements.
For example, we use $\{M^x_{a,\, b}\}$ to represent a measurement indexed by $x$
whose outcome consists of two parts $a$ and $b$.
In this case, by slightly abusing the notation, we use $\{M^x_a\}$ and $\{M^x_b\}$
to denote
\begin{equation*}
  M^x_a = \sum_b M^x_{a,\, b}\;, \quad M^x_b = \sum_a M^x_{a,\, b}\;.
\end{equation*}
For any $x$, $\{M^x_a\}$ and $\{M^x_b\}$ are POVMs sometimes referred to as the
``mar\-ginals'' of $\{M^x_{a,\, b}\}$.

\begin{definition}
  \label{def:bracket}
  Let $\{M^x_a\}_{a\in A}$ be a family of POVMs indexed by $x\in \cal{X}$.
  Let $f:A\to B$ be an arbitrary function.
  We write $\bigl\{ M^x_{[f(\cdot)=b]} \bigr\}$ for the POVM derived from
  $\{M^x_a\}$ by applying the function $f$ before returning the outcome.
  More precisely,
\begin{equation*}
  M^x_{[f(\cdot) = b]} = \sum_{a : f(a)=b} M^x_a\;.
\end{equation*}
If $b$ is not in the image of $f$, then we define $M^x_{[f(\cdot) = b]}$ to be
$0$. In cases where the outcome $a$ has a natural interpretation as a tuple $(a_1,\ldots,a_k)$ we sometimes slightly abuse notation and write $M^x_{[a_i=b]}$ for some $i\in \{1,\ldots,k\}$ to denote 
\[ M^x_{[a_i=b]} \,=\,\sum_{(a_1,\ldots,a_k):\,a_i=b} M^x_{(a_1,\ldots,a_k)}\;.\]
\end{definition}

We will make frequent use of Naimark dilation. In our setting, it can be formulated as follows. For a proof, see~\cite[Theorem 5.1]{ML20}

\begin{theorem}[Naimark dilation]\label{thm:naimark}
Let $\ket{\psi}$ be a state in $\mH_{\mathrm{A}} \ot \mH_{\mathrm{B}}$.
Let $A = \{A^x_a\}$ be a sub-measurement acting on $\mH_{\mathrm{A}}$
and $B=\{B^y_b\}$ be a sub-measurement acting on $\mH_{\mathrm{B}}$.
Then there exists
\begin{enumerate}
\item Hilbert spaces $\mH_{\mathrm{A}_{\mathsf{aux}}}$ and $\mH_{\mathrm{B}_{\mathsf{aux}}}$,
\item a state $\ket{\mathsf{aux}} \in \mH_{\mathrm{A}_{\mathsf{aux}}} \ot \mH_{\mathrm{B}_{\mathsf{aux}}}$,
\item and two projective sub-measurements  $\widehat{A} = \{\widehat{A}^x_a\}$ and $\widehat{B}=\{\widehat{B}^y_b\}$
acting on $\mH_{\mathrm{A}} \ot \mH_{\mathrm{A}_{\mathsf{aux}}}$
and $\mH_{\mathrm{B}} \ot \mH_{\mathrm{B}_{\mathsf{aux}}}$,
respectively,
\end{enumerate}
such that the following is true.
If we write $\ket{\widehat{\psi}} = \ket{\psi} \otimes \ket{\mathsf{aux}}$,
then for all $x,y, a, b$,
 \begin{equation*}
 \bra{\psi} A^x_a \ot B^y_b \ket{\psi}
 = \bra{\widehat{\psi}} \widehat{A}^x_a \ot \widehat{B}^y_b \ket{\widehat{\psi}}.
 \end{equation*}
 In addition, $\ket{\mathsf{aux}}$ is a \emph{product state},
 meaning that we can write it as
 \begin{equation*}
 \ket{\mathsf{aux}}= \ket{\mathsf{aux}_{\mathrm{A}}} \ot \ket{\mathsf{aux}_{\mathrm{B}}},
 \end{equation*}
for $\ket{\mathsf{aux}_{\mathrm{A}}}$ in $\mH_{\mathrm{A}_{\mathsf{aux}}}$
and $\ket{\mathsf{aux}_{\mathrm{B}}}$ in $\mH_{\mathrm{B}_{\mathsf{aux}}}$.
\end{theorem}

\subsection{Generalized Pauli observables}
\label{sec:generalized-pauli}

For prime number $p$, the generalized Pauli operators over $\Fp$ are a
collection of observables indexed by a basis setting $X$ or $Z$ and an element
$a$ or $b$ of $\Fp$, with eigenvalues that are $p$-th roots of unity.
They are given by
\begin{equation}
  \label{eq:pauli-fp}
  \sigma^X(a) = \sum_{j \in \Fp} \ket{j + a} \bra{j}\qquad \text{and} \qquad
  \sigma^Z(b) = \sum_{j \in \Fp} \omega^{bj} \ket{j} \bra{j}\;,
\end{equation}
where $\omega = e^{\frac{2\pi i}{p}}$, and addition and multiplication are over
$\Fp$.
These observables obey the ``twisted commutation'' relations
\begin{equation}
  \label{eq:twisted-fp}
  \forall a,b\in\Fp\;,\qquad \sigma^X(a)\, \sigma^Z(b) =
  \omega^{-ab} \,\sigma^Z(b)\,\sigma^X(a)\;.
\end{equation}
Similarly, over a field $\Fq$ with $q$ a power of $p$, we can consider a set of generalized Pauli
operators, indexed by a basis setting $X$ or $Z$ and an element of $\Fq$ and
with eigenvalues that are $p$-th roots of unity.
For $a, b\in\Fq$ they are given by
\begin{equation*}
  \qp^X(a) = \sum_{j \in \Fq} \ket{j+a}\bra{j} \qquad \text{and} \qquad
  \qp^Z(b)= \sum_{j \in \Fq} \omega^{\tr(b j)} \ket{j}\bra{j}\;,
\end{equation*}
where addition and multiplication are over $\Fq$.
For all $W \in \{X,Z\}$, $a, a' \in \Fq$, and $b \in \Fp$, powers of these
observables obey the relations
\begin{equation*}\label{eq:pauli-product-power}
  \bigl(\qp^W(a)\qp^W(a')\bigr) = \qp^W(a+a') \qquad \text{and}
  \qquad \bigl( \qp^W(a) \bigr)^b = \qp^W(a b) \;.
\end{equation*}
In particular, since $pa=0$ for any $a\in \Fq$ we get that that $(\qp^W(a))^p =
I$ for any $a\in\Fq$.
The observables obey analogous ``twisted commutation'' relations
to~\eqref{eq:twisted-fp},
\begin{equation}
  \label{eq:twisted-fq}
  \forall a, b \in \Fq\;, \qquad \qp^X(a)\, \qp^Z(b) =
  \omega^{-\tr(a b)} \, \qp^Z(b)\, \qp^X(a)\;.
\end{equation}
It is clear from the definition that all of the $\qp^X$ operators commute with
each other, and similarly all the $\qp^Z$ operators commute with each other.
Thus, it is meaningful to speak of a common eigenbasis for all $\qp^X$
operators, and a common eigenbasis for all $\qp^Z$ operators.
The common eigenbasis for the $\qp^Z$ operators is the computational
basis $\{\ket{j}\}_{j \in \F_q}$.
To map this basis to the common eigenbasis of the $\qp^X$ operators, one can
apply the Fourier transform
\begin{equation}
  \label{eq:fourier-f}
  F \,=\, \frac{1}{\sqrt{q}} \sum_{a, b\in \Fq} \omega^{-\tr(ab)}
  \ket{a}\bra{b}\;.
\end{equation}
Explicitly, the eigenbases consist of the vectors $\ket{e_W}$ labeled by an
element $e \in \Fq$ and $W \in \{X, Z\}$, given by
\begin{equation*}
  \ket{e_X} = \frac{1}{\sqrt{q}} \sum_j \omega^{- \tr(e j)} \ket{j}\;, \qquad
  \ket{e_Z}   = \ket{e}\;.
\end{equation*}
We denote the POVM whose elements are projectors onto basis vectors of the
eigenbasis associated with the observables $\qp^W$ by $\{\qp^W_e\}_e$.
Then for all $W \in \{X,Z\}$ and $a\in \Fq$, the observables $\qp^W(a)$ can be
written as
\begin{equation}\label{eq:pauli-obs-proj-single}
  \qp^W(a) = \sum_{b \in \Fq} \omega^{\tr(a b)} \qp^W_b\;.
\end{equation}

To invert this relation, we first quote the following useful Fourier fact.
\begin{lemma}[Fact 3.2 of \cite{NW19}]
  \label{lem:cancellation}
  Let $V$ be a subspace of $\F^k$. For all $v \not\in V^\perp$, 
  \begin{equation*}
    \E_{u \sim V} \omega^{\tr(u \cdot v)} = 0\;,
  \end{equation*}
  where the expectation is over a uniformly random vector $u$ from $V$.
\end{lemma}

We may now invert \eqref{eq:pauli-obs-proj-single}:
\begin{equation}\label{eq:pauli-inversion-0-single}
  \E_{a\in \F_q} \omega^{-\tr(ab)} \qp^W(a) = \sum_{b'\in \F_q} \E_{a\in \F_q}
  \omega^{\tr(a(b'-b))} \qp^W_{b'} = \qp^W_b\;,
\end{equation}
where the second step follows from Lemma~\ref{lem:cancellation}.

For systems with many qudits, we will consider tensor products of the operators
$\qp^W$.
Slightly abusing notation, for $W \in \{X, Z\}$ and $a \in \Fq^n$ we denote by
$\qp^W(a)$ the tensor product $\qp^W(a_1) \ot \dots \ot \qp^W(a_n)$.
These obey the twisted commutation relations
\begin{equation*}
  \forall a,b\in \Fq^n,\qquad \qp^X(a)\, \qp^Z(b) \,=\, \omega^{-\tr (a \cdot
    b)}\, \qp^Z(b)\, \qp^X(a)\;,
\end{equation*}
where $a \cdot b = \sum_{i=1}^{n} a_i b_i \in \Fq$.
For $W\in\{X,Z\}$ and $e \in \Fq^n$ define the eigenstates
\[ \ket{e_W} = \ket{(e_1)_W} \ot \dots \ot \ket{(e_n)_W}\;, \]
and associated rank-$1$ projectors $\qp^W_e$. Analogous versions of \eqref{eq:pauli-obs-proj-single}
and \eqref{eq:pauli-inversion-0-single} relate the multi-qubit observables
and projectors:
\begin{align}
  \qp^W(a) &= \sum_{b \in \Fq^n} \omega^{\tr(a \cdot b)}
  \qp^W_b\;, \label{eq:pauli-obs-proj} \\
    \E_{a\in \F_q^n} \omega^{-\tr(a \cdot b)} \qp^W(a) &= \sum_{b'\in \F_q^n} \E_{a\in \F_q^n}
  \omega^{\tr(a\cdot (b'-b))} \qp^W_{b'} = \qp^W_b\;. \label{eq:pauli-inversion-0}
\end{align}

Since we only consider finite fields $\F_q$ such that $q = 2^k$ the maximally
entangled state $\ket{\EPR_q}$ and the corresponding qudit Pauli
observables/projectors are isomorphic to a tensor product of maximally entangled
states $\ket{\EPR_2}$ and \emph{qubit} Pauli observables/projectors,
respectively; this is shown in the next lemma.
This is used to argue that the Pauli basis test (described in
Section~\ref{sec:pauli-verifier}) gives a self-test for Pauli observables and
maximally entangled states over qubits.

\begin{lemma}
  \label{lem:pauli-binary}
  For all admissible field sizes $q = 2^k$ and integers $L$, there exists
  an isomorphism $\phi: (\C^q)^{\otimes L} \to (\C^2)^{\otimes L k}$
  such that
\begin{equation}
    \label{eq:epr-q-to-epr-2}
    \phi \otimes \phi \, \ket{\EPR_q}^{\otimes L} =
    \ket{\EPR_2}^{\otimes L k}\;.\text{\footnotemark}
  \end{equation}
\footnotetext{ Here we have applied the natural identifications between
  $(\C^q)^{\ot L} \ot (\C^q)^{\ot L}$ and $(\C^q \ot \C^q)^{\ot L}$,
  and between $(\C^2)^{\ot L k} \ot (C^2)^{\ot Lk}$ and $(\C^2 \ot
  \C^2)^{\ot Lk}$.}
  Moreover, for all $W \in \{X,Z\}$ and for all $u \in \F_q^L$
	\begin{equation}
	\label{eq:pauli-q-to-pauli-2}
  	\tau^W_u = \phi^\dagger\, \Bigl (\bigotimes_{i=1}^L
    \bigotimes_{j=1}^{k} \sigma^{W}_{u_{ij}} \Bigr) \, \phi\;.
	\end{equation}
	Here, the $(u_{ij})_{i,j}$ denotes a vector of $\F_2$ values such that $u_i =
  \sum_j u_{ij} e_j$ for all $i \in \{1,\ldots,L\}$ with
  $\{e_1,\ldots,e_{k}\}$ being the self-dual normal basis of $\F_q$ over $\F_2$
  specified by Lemma~\ref{lem:efficient_basis}.
  For $i\in\{1,\ldots,L\}$ and $j\in\{1,\ldots,q\}$ the $(i,j)$-th factor
  $\sigma^{W}_{u_{ij}}$ denotes the projector $\frac{1}{2} \left ( \Id +
    (-1)^{u_{ij}} \sigma^W(1) \right)$ acting on the $s$-th qubit of
  $\ket{\EPR_2}^{\otimes L k}$, where $s = (i-1) k + j$.
\end{lemma}

\begin{proof}
  Define the isometry $\theta : \C^q \to (\C^2)^{\otimes k}$ as $\theta\,
  \ket{a} = \ket{a_1} \otimes \ket{ a_2} \cdots \ket{a_k}$ where $\kappa(a) = (a_1, a_2, \ldots, a_k)
  \in \F_2^k$ is the bijection introduced in Section~\ref{sec:finite-fields}
  corresponding to the basis $\{e_1,\ldots,e_k\}$. Observe that
  $\theta \otimes \theta \ket{\EPR_{2^k}} = \ket{\EPR_2}^{\ot k}$.

  Let $a \in \F_q$, and let $\kappa(a) = (a_1,\ldots,a_k) \in \F_2^k$.
  Then from~\eqref{eq:pauli-inversion-0}, we get
  \begin{equation}
    \label{eq:qudit-to-qubit-pauli-1}
    \begin{split}
      \tau^W_a
      & = \E_{b \in \F_q} (-1)^{\tr(a b)} \, \tau^W(b) \\
      & = \E_{b \in \F_q} (-1)^{\tr(\sum_j a_j e_j b)} \, \tau^W(b) \\
      & = \E_{b \in \F_q} (-1)^{\sum_j a_j b_j} \,
      \tau^W \left ( \sum_j b_j e_j \right) \;,
    \end{split}
  \end{equation}
  where $b = \sum_j b_j e_j$; since the basis $\{e_j \}$ is self-dual, we have
  that $b_j = \tr(b e_j)$. From~\eqref{eq:qudit-to-qubit-pauli-1} and
  the product relation in~\eqref{eq:pauli-product-power} we get that
  \begin{equation}
		\label{eq:qudit-to-qubit-pauli}
    \begin{split}
      \tau^W_a & = \E_{b_1,\, \ldots,\, b_k \in \F_2}\,
      \prod_{j=1}^k (-1)^{a_j b_j} \, \tau^W(b_j e_j) \\
      & = \prod_{j=1}^k \E_{b_j \in \F_2} (-1)^{a_j b_j} \, \tau^W(b_j e_j)\;,
    \end{split}
  \end{equation}
  where we have the used the fact that the expectation is over iid
  $b_1, \dots, b_k$ to take it inside the product in the second line.
  Next we claim that for all $c \in \F_2$, we have $\tau^W(c e_j) =
  \theta^\dagger \sigma^{W,j} (c) \theta$ where $\sigma^{W,j}(c) = \Id$ if $c =
  0$, and otherwise is the Pauli $W$ observable acting on the $j$-th qubit of
  $(\C^2)^{\otimes k}$. This can be verified by comparing the actions of both
  operators on the basis states of $\C^q$.
  
  Thus we obtain that the right-hand side of~\eqref{eq:qudit-to-qubit-pauli} is
  equal to
  \begin{align}
		\prod_{j=1}^k \E_{b_j \in \F_2} (-1)^{a_j b_j} \,
    \Big [ \theta^\dagger \sigma^{W,j}(b_j) \theta \Big ]
    & = \theta^\dagger \, \bigotimes_{j=1}^k
      \Big ( \E_{b_j \in \F_2} (-1)^{a_j b_j} \, \sigma^{W}(b_j) \Big) \theta \\
		& = \theta^\dagger \Big (\bigotimes_{j=1}^k \sigma^W_{b_j} \Big) \theta.
		\label{eq:qudit-to-qubit-pauli-2}
  \end{align}
  Define $\phi = \theta^{\otimes L}$. It is evident
  that~\eqref{eq:epr-q-to-epr-2} holds for this choice of $\phi$, and
  it remains to show~\eqref{eq:pauli-q-to-pauli-2}. The projector $\tau^W_u$ can be
  decomposed as the tensor product $\bigotimes_{i=1}^L \tau^{W}_{u_i}$
  where $\tau^{W}_{u_i}$ acts on the $i$-th factor of $(\C^q)^{\otimes L}$.
  Express each $u_i$ as $\sum_j u_{ij} e_j$ where $u_{ij} \in \F_2$. Then from
  Equation~\eqref{eq:qudit-to-qubit-pauli-2} we get that
  \begin{equation}
  	\tau^W_u = \bigotimes_{i=1}^L \tau^W_{u_i}
    = \phi^\dagger \left( \bigotimes_{i=1}^L
      \bigotimes_{j=1}^k \sigma^W_{u_{ij}} \right) \phi,
  \end{equation}
  which establishes Equation~\eqref{eq:pauli-q-to-pauli-2}. 
\end{proof}


\section{Conditionally linear functions, distributions, and samplers}
\label{sec:linear}

\subsection{Conditionally linear functions and distributions}
\label{sec:cl}

We first introduce \emph{conditionally linear functions}, which are used to
specify the question distribution for games considered in the paper in a way
that the question distribution can be ``introspected,'' as described in
Section~\ref{sec:introspection}.
Intuitively, a conditionally linear function takes as input an element
$x\in V = \F^n$ for some $n\geq 0$, and applies linear maps $L_j$ sequentially
on $x^{V_j}$ where $V_1,V_2,\ldots $ are a sequence of complementary register
subspaces such that both the linear maps $L_j$ and the subspace $V_j$ for
$j \ge 2$ may depend on the values taken by previous linear maps $L_1(x^{V_1})$,
$L_2(x^{V_2})$, etc.

In the remainder of the section we use $V$ to denote the linear space $\F^n$ for
some integer $n \geq 0$.
For ease of notation we extensively use the subscript range notation.
For example, if $V_1, V_2, \ldots, V_\ell$ are fixed subspaces of $V$ and $k \in
\{1, 2, \ldots, \ell\}$ we write
\begin{equation*}
  V_{<k} = \bigoplus_{j\,:\,1 \le j<k} V_j\;,\quad
  V_{> k} = \bigoplus_{j\,:\, \ell\ge j > k} V_j\;,
\end{equation*}
and it is understood that $V_{\le k}$ and $V_{\ge k}$ are identical to
$V_{<k+1}$ and $V_{>k-1}$, respectively.
Moreover, if $V'$ is a register subspace of $V$, $F: V' \to V'$ is a function,
and $x \in V$, we write $x^F$ to denote $F(x^{V'})$.
For example, in the following definition $x^{L_1}$ is used as shorthand notation
for $L_1(x^{V_1})$.

\begin{definition}
  \label{def:cl-func}
	Let $V$ be $\F^n$ for some $n\geq 0$.
  For all integers $\ell\geq 0$ the collection of \emph{$\ell$-level
    conditionally linear functions} (implicitly, \emph{on $V$}) is defined
  inductively as follows.
  \begin{enumerate}
  \item There is a single $0$-level conditionally linear function, which is the
    $0$ function on $V$.

  \item Let $\ell \geq 1$ and suppose the collection of $(\ell-1)$-level
    conditionally linear functions has been defined.
    The collection of $\ell$-level conditionally linear functions on $V$ consists
    of all functions $L$ on $V$ that can be expressed in the following form.
    There exist complementary register subspaces $V_1$ and $V_{>1}$ of $V$, a
    linear function $L_1$ on $V_1$, and for all $v \in L_1(V_1)$, an
    $(\ell-1)$-level conditionally linear function $L_{>1,\, v}$ on $V_{>1}$,
    such that for all $x\in V$,
    \begin{equation*}
      L(x) = x^{L_1} + L_{>1,\, x^{L_1}} (x^{V_{>1}})\;.
    \end{equation*}
  \end{enumerate}
\end{definition}

The concept of conditionally linear functions is simple even though the
notations seem complicated.
It is best understood using an operational definition illustrated in
Fig.~\ref{fig:cl-functions}.
We give an example of a $2$-level CL function in Example~\ref{ex:cl-function}.

\begin{remark}
  \label{rk:level-1-is-linear}
  It follows from the above definition that all $1$-level CL functions are linear.
\end{remark}

\begin{example}\label{ex:cl-function}
Consider $V = \F_{2}^{3}$. Then the following map
\begin{equation*}
  L:(x_{0}, x_{1}, x_{2})
  \mapsto (0, x_{0}x_{2} + x_{1}x_{2} + x_{0},  x_{2})
\end{equation*}
on $V$ is a $2$-level CL function by choosing $V_{1} = \mathrm{span}\{(0,0,1)\}$, $L_{1}$
the identity function on $V_{1}$, $V_{2} = V_{1}^{\perp}$, and
\begin{equation*}
  L_{2,\, x_{2}}: (x_{0}, x_{1}, 0)
  \mapsto (0, x_{0}x_{2} + x_{1}x_{2} + x_{0}, 0),
\end{equation*}
which is linear for all $x_{2} \in \F_{2}$.
Note that intuitively the function $L$ selects either $x_{0}$ or $x_{1}$ in the
second coordinate conditioning on the value of $x_{2}$.
\end{example}

\begin{figure}[!htbp]
  \centering
  \begin{tikzpicture}


    \node[draw, minimum width=2cm, anchor=east] at (0,0) (xv1) {$x^{V_{1}}$};
    \node[draw, minimum width=6cm, anchor=west] at (-.4pt,0) {$x^{V_{>1}}$};
    \node[draw, minimum width=2cm, anchor=east] at (0,-2) (xl1) {$x^{L_{1}}$};
    \draw[-stealth] (xv1.south) -- ++(0,-1) node[above right] {$L_{1}$} -- (xl1.north);

    \node[draw, minimum width=2cm, anchor=west] at (0,-3) (xv2) {$x^{V_{2}}$};
    \node[draw, minimum width=4cm, anchor=west] at (2cm,-3) {$x^{V_{>2}}$};

    \node[draw, minimum width=2cm, anchor=west] at (0,-5) (xl2) {$x^{L_{2}}$};

    \draw[-stealth] (xv2.south) -- ++(0,-1) node[above right]
      {$L_{2,\,x^{{L_{1}}}}$} -- (xl2.north);
    \draw[dashed, -stealth, to path={|- (\tikztotarget)}] (xl1) edge (xv2.west);

    \draw[dashed, -stealth, to path={|- (\tikztotarget)}] (xl2) edge (3cm,-6);
    \draw[dashed, -stealth, to path={|- (\tikztotarget)}] (xl1) edge (3cm,-6);
    \node[right] at (3cm,-6) {$\;\cdots\cdots$};

  \end{tikzpicture}
  \caption{An illustration of an $\ell$-level CL function $L$.
    First a register subspace $V_{1}$ and a linear map $L_{1}$ are chosen and
    applied to obtain $x^{L_{1}} = L_{1}(x^{V_{1}}) \in V_{1}$.
    Then depending the value of $x^{L_{1}}$, a register subspace $V_{2}$ and a
    linear map $L_{2,\,x^{L_{1}}}$ are chosen and applied to obtain
    $x^{L_{2}} = L_{2,\,x^{{L_{1}}}}(x^{V_{2}}) \in V_{2}$ and so on.
    Finally, $L(x)$ is defiend to be
    $\sum_{k=1}^{\ell} x^{L_{k}}$.}\label{fig:cl-functions}
\end{figure}

\begin{remark}
  \label{rk:higher-level}
  Note that for any integer $\ell\geq 1$ the collection of $\ell$-level CL
  functions trivially contains the collection of $(\ell-1)$-level CL functions:
  for this it suffices to note that the $0$ function, which is a $0$-level CL
  function, is also a $1$-level CL function by setting $V_1=V$, $V_{>1}= \{0\}$,
  $L_1(x)=0$ for all $x\in V$, and $L_{>1,\, x^{L_1}}$ is the $0$ map for all
  $x\in V$.
\end{remark}

\begin{definition}
  \label{def:cl-dist}
  Let $L, R: V \to V$ be conditionally linear functions.
  The \emph{conditionally linear distribution $\mu_{L,R}$ corresponding to $(L,
    R)$} is defined as the distribution over pairs $(L(x), R(x)) \in V \times V$
  for $x$ drawn uniformly at random from $V$.
\end{definition}

Throughout the paper we abbreviate ``conditionally linear functions'' and
``conditionally linear distributions'' as \emph{CL functions} and \emph{CL
  distributions}, respectively.

The following lemma elucidates structural properties of $\ell$-level CL
functions.
Recall that using our shorthand notation, $x^{L_{<k}}$ and $x^{L_k}$ in the
lemma denote $L_{<k}(x)$ and $L_{k,u}(x^{V_{k,\, u}})$ where $u = L_{<k}(x)$.

\begin{lemma}
  \label{lem:cl-kth}
	Let $\ell \geq 1$ and $V = \F^n$ for some integer $n \geq 0$.
  A function $L: V \to V$ is an $\ell$-level CL function if and only if the
  following collection of functions and subspaces exists:
  \begin{enumerate}
  \item \label{enu:cl-k-marginal} For each $k \in \{1, 2, \ldots, \ell\}$, a
    function $L_{\le k} : V \to V$ called the \emph{$k$-th marginal of $L$};
  \item \label{enu:cl-k-space} For each $k \in \{1, 2, \ldots, \ell\}$ and $u
    \in L_{<k} (V)$, a register subspace $V_{k,\, u}$ of $V$ called the
    \emph{$k$-th factor space with prefix $u$};
  \item \label{enu:cl-k-map} For each $k \in \{1, 2, \ldots, \ell\}$ and $u \in
    L_{<k} (V)$, a linear map $L_{k,\, u} : V_{k,\, u} \to V_{k,\, u}$ called
    the \emph{$k$-th linear map of $L$ with prefix $u$};
  \end{enumerate}
  such that the following conditions hold for all $k \in \{1, 2, \ldots,
  \ell\}$.
  \begin{enumerate}
  \item \label{enu:cl-k-cl} $L_{\leq k}$ is a $k$-level CL function on $V$;
	\item \label{enu:cl-space-sum} $V = \bigoplus_{i = 1}^\ell V_{i,\,
      x^{L_{<i}}}$ for all $x \in V$;
  \item \label{enu:cl-map-sum} $L_{\leq k}(x) = \sum_{i = 1}^{k} x^{L_i}$ for
    all $x \in V$, where $L_i$ is shorthand notation for $L_{i,\,
      x^{L_{<i}}}$;
  \item \label{enu:cl-last-ell} $L = L_{\leq \ell}$.
  \end{enumerate}
  As in Item~\ref{enu:cl-map-sum}, we sometimes use $V_k$ and $L_k$ to denote
  $V_{k,\, u}$ and $L_{k,\, u}$, respectively, leaving the prefix $u$ implicit.
\end{lemma}

\begin{proof}
	We first prove the ``if'' direction: if there exist spaces and functions
  satisfying the conditions in the lemma, the fact that $L$ is an $\ell$-level
  CL function follows from Items~\ref{enu:cl-k-cl} and~\ref{enu:cl-last-ell} of the lemma
  statement.

  We now prove the ``only if'' direction.
  Given a CL function $L$ on $V$, we construct the $k$-th family of subspaces
  and functions for all $k \in \{1, \ldots, \ell\}$ by induction on the level
  $\ell$.
  First consider the base case $\ell=1$.
  Since $L_{<1} = 0$, we omit the mentioning of the prefix $u \in L_{<1}(V)$.
  Define $L_{\leq 1} = L$, the factor space $V_1 = V$, and the linear map $L_1 =
  L$.
  It is straightforward to verify that the conditions in the lemma hold for
  these choices of linear maps and spaces.
  
  Now, assume that the lemma holds for CL functions of level at most $\ell - 1$,
  and we prove the lemma for $\ell$-level CL functions.
  By definition, an $\ell$-level CL function $L$ can be written as
  \begin{equation*}
  	L(x) = x^{L_1} + L_{>1,\, x^{L_1}}(x^{V_{>1}})
  \end{equation*}
  for some linear map $L_1 : V_1 \to V_1$ and a family of $(\ell-1)$-level CL
  functions
  \begin{equation*}
    \bigl \{ L_{>1,\, v} : V_{>1} \to V_{>1} \bigr \}_{\, v \in L_1(V_1)}
  \end{equation*}
  where $V_1$ and $V_{>1}$ are complementary register subspaces of $V$.
  Next, using the inductive hypothesis on the $(\ell-1)$-level CL function
  $L_{>1,\, v}$ we get that for all $v \in L_1(V_1)$ and all $k \in \{1, 2,
  \ldots, \ell-1\}$ there exist $k$-th marginal functions $L'_{v,\, \le k}:
  V_{>1} \to V_{>1}$, $k$-th factor spaces $V'_{v,\, k,\, u}$, and $k$-th linear
  maps $L'_{v,\, k,\, u}$ of $L_{>1,\, v}$ with prefix $u \in L'_{v,\,
    <k}(V_{>1})$ such that the conditions of the lemma for $L_{>1,\, v}$ hold.
  
  Define the marginal functions $L_{\le k} : V \to V$, factor spaces
  $V_{k,\, u}$ and linear maps $L_{k,\, u}$ for $L$ as follows.
  \begin{enumerate}
  \item Define $L_{\le 1} = L_1$ and the first factor space to be $V_1$;
  \item For all $k \in \{2, 3, \ldots, \ell\}$, define
    \begin{equation}
      \label{eq:cl-k-def}
      L_{\le k}: x \mapsto x^{L_1} + L'_{x^{L_1},\, <k} (x^{V_{>1}})
      \quad\text{for } x\in V;
    \end{equation}
	\item For all $k \in \{2, 3, \ldots, \ell\}$ and $u \in L_{< k}(V)$, define
    $V_{k,\, u} = V'_{v,\, k-1,\, w}$ and $L_{k,\, u} = L'_{v,\, k-1,\, w}$
    where $v = u^{V_1} \in L_{1}(V_{1})$ and $w = u^{V_{>1}} \in L'_{v,< k-1}(V_{> 1})$.
  \end{enumerate}
  We verify that the conditions of the lemma are satisfied.
  Since $L'_{v,\, < k}$ is by assumption a $(k-1)$-level CL function on $V_{>
    1}$, we get that $L_{\leq k}$ is a $k$-level CL function on $V$ from
  \eqref{eq:cl-k-def}, establishing Item~\ref{enu:cl-k-cl}.
  By the induction hypothesis, we have for all $v\in L_1(V_1)$ and $y \in
  V_{>1}$,
  \begin{equation}
    \label{eq:cl-k-1}
    V_{>1} = \bigoplus_{i=1}^{\ell-1}
    V'_{\smash{v,\, i, \, y^{L'_{v,\, <i}}}},
  \end{equation}
  which implies that for all $x \in V$ and $v = x^{L_1}$,
  \begin{equation*}
    V = V_1 \oplus \Big ( \bigoplus_{i=1}^{\ell-1}
    V'_{\smash{v,\, i, \, x^{L'_{v,\, <i}}}} \Big)
    = V_1 \oplus \Big (\bigoplus_{i=2}^{\ell} V_{i,\, x^{L_{<i}}} \Big)
    = \bigoplus_{i=1}^\ell V_{i,\, x^{L_{<i}}}\;.
  \end{equation*}
  The first equality follows from \eqref{eq:cl-k-1} while the second and third
  equalities follow from the definition of $V_{k,\, u}$.
  This establishes Item~\ref{enu:cl-space-sum}.

  Next, we have that for all $x \in V$, $v = x^{L_1}$, and $k \in
  \{1,2,\ldots,\ell\}$,
  \begin{align}
    \label{eq:cl-last-ell}
    L_{\leq k}(x)
    & = v + L'_{v,\, <k} (x^{V_{>1}}) \\
    & = v + \sum_{i=1}^{k-1} x^{L'_{v,\, i}}
    = \sum_{i = 1}^{k} x^{L_{i}}, \label{eq:cl-last-ell2}
  \end{align}
  where $L'_{v,\, i}$ is the $i$-th linear map of $L_{>1,\, v}$ with prefix
  $L'_{v,\, <i}(x)$ and $L_i$ is the $i$-th linear map of $L$ with prefix
  $L_{<i}(x)$.
  The second equality follows from the inductive hypothesis applied to $L'_{v,\,
    <k}$ and the third equality follows from the definition of $L_{i}$.
  Line~\eqref{eq:cl-last-ell2} implies Item~\ref{enu:cl-map-sum} of the lemma.
  
  Finally, Item~\ref{enu:cl-last-ell} follows from~\eqref{eq:cl-last-ell} where we
  set $k = \ell$ and observe that $L'_{x^{L_1},\, \le \ell-1}$ is equal to
  $L_{>1,\, x^{L_1}}$ by the inductive hypothesis.
  This shows that $L_{\le k}$, $V_{k,\, u}$, and $L_{k,\, u}$ satisfy the
  conditions of the lemma and completes the induction.
\end{proof}

We note that the marginal functions, factor spaces, and linear maps of a given
CL function $L$ may not be unique; for example, consider the identity function
on a linear space $V = \F^n$.
This is clearly a $1$-level CL function, but it can also be viewed as a
$k$-level CL function for $k \in \{2, \ldots, n\}$ with an arbitrary partition
of $V$ into factor spaces.

\begin{lemma}
  \label{lem:cl-concat}
  Let $\ell, k \ge 0$ be integers and $U = \F^n$, $V = \F^m$ be linear spaces.
  Suppose $L$ is a $k$-level CL function on $U$ and $R_u$ is an $\ell$-level CL
  function on $V$ for each $u \in L(U)$.
  Then the \emph{concatenation} $T$ of $L$ and $\{R_u\}_u$ defined as
  \begin{equation*}
    T (x) = L(x^U) + R_{L(x^U)} (x^V)
  \end{equation*}
  is a $(k+\ell)$-level conditionally linear function on $U\oplus V$.
\end{lemma}

\begin{proof}
  We prove the claim by induction on $k$.
  The case $k = 0$ follows from the Definition~\ref{def:cl-func}.
  Assume that the lemma holds for $L$ being at most $(k-1)$-level.
  By Definition~\ref{def:cl-func}, there are complementary register subspaces
  $U_1$ and $U_{>1}$ of $U$, a linear function $L_1$ on $U_1$, and a family of
  $(k-1)$-level CL functions $L_{>1,\, v}$ for $v \in L_1(U_1)$ such that
  \begin{equation*}
    L(x^U) = x^{L_1} + L_{>1,\, x^{L_1}}(x^{U_{>1}}).
  \end{equation*}
  For all $x^{L_1}$, define the function $T_{>1,\, x^{L_1}}$ on $U_{>1} \oplus V$ as
  \begin{equation*}
    T_{>1,\, x^{L_1}}(x^{U_{>1} \oplus V}) = L_{>1,\, x^{L_1}}(x^{U_{>1}}) +
    R_{x^{L_1} + x^{L_{>1}}}(x^V),
  \end{equation*}
  the concatenation of $L_{>1,\, x^{L_1}}$ and $\{ R_{L(x^U)} \}_{x^{L_{>1}}}$
  where $L_{>1}$ is the shorthand notation of $L_{>1,\, x^{L_1}}$.
  By the induction hypothesis, $T_{>1,\, x^{L_1}}$ is $(k+\ell-1)$-level
  conditionally linear.
  The lemma follows from Definition~\ref{def:cl-func}.
\end{proof}
\begin{lemma}[Direct sums of CL functions]
  \label{lem:cl-func-prod}
  Let $V^{(1)}, V^{(2)}, \ldots, V^{(m)}$ be register subspaces of $V$ such that
  $V = \bigoplus_{j=1}^m V^{(j)}$.
  Suppose that, for each $j\in\{1, 2, \ldots, m\}$, $L^{(j)}$ is an
  $\ell_j$-level conditionally linear function on $V^{(j)}$.
  Then the direct sum $L = \bigoplus_{j=1}^m L^{(j)}$ is an $\ell$-level CL
  function over $V$ for $\ell = \max_j \{\ell_j\}$, where $L$ is defined by
  \begin{equation*}
    L(x) = \sum_{j=1}^m  L^{(j)}(x^{(j)})
  \end{equation*}
  for all $x = \sum_{j=1}^m x^{(j)} \in \bigoplus_{j=1}^m V^{(j)}$.
\end{lemma}

\begin{proof}
  It is easy to see that an $\ell$-level CL function is also $k$-level
  conditionally linear for all $k \ge \ell$.
  Hence, it suffices to prove the claim where $\ell_j = \ell$ for $j=1, 2,
  \ldots, m$.

  We prove the theorem by an induction on $\ell$.
  For $\ell = 1$, the functions $L^{(j)}$ are linear and the claim follows by
  the fact that the direct sum of linear maps is linear.

  Assume now the theorem holds for conditionally linear functions of level at
  most $\ell-1$ and $L^{(j)}$ are $\ell$-level conditionally linear functions
  for $j=1, 2, ,\ldots, m$.
  By definition, $L^{(j)}$ is the concatenation of conditionally linear
  functions $L^{(j)}_{1}$ on $V^{(j)}_1$ and $\{ L^{(j)}_{>1,\, v_j} \}_{v_j}$
  on $V^{(j)}_{>1}$ of levels $1$, and $\ell-1$, respectively.
  Furthermore,
  \begin{equation*}
    L(x) = \sum_{j=1}^m L^{(j)} (x^{(j)}) = \sum_{j=1}^m \Bigl( v_j +
    L^{(j)}_{>1,\, v_j}
    \Bigl((x^{(j)})^{V^{(j)}_{>1}} \Bigr) \Bigr),
  \end{equation*}
  where $v_j = L^{(j)}_{1}\Bigl( (x^{(j)})^{V^{(j)}_1}
    \Bigr)$.
  By the induction hypothesis,
  \begin{equation*}
    L_{1} (x^{V_1}) = \sum_{j=1}^m L^{(j)}_{1}\Bigl( (x^{(j)})^{V^{(j)}_1}
    \Bigr),\quad L_{>1,\, v} (x^{V_{>1}}) = \sum_{j=1}^m L^{(j)}_{>1,\, v_j}
    \Bigl((x^{(j)})^{V^{(j)}_{>1}} \Bigr)
  \end{equation*}
  are $1$-level and $(\ell-1)$-level conditionally linear, respectively, for $v =
  \sum_j v_j$, $V_1 = \bigoplus_{j=1}^m V^{(j)}_1$, and $V_{>1} =
  \bigoplus_{j=1}^m V^{(j)}_{>1}$.
  This proves that $L$ is $\ell$-level conditionally linear.
\end{proof}

\begin{lemma}
  \label{lem:cl-dist-prod}
  For each $i \in \{1,\ldots,m\}$ let $L^{(i)}, R^{(i)} : V^{(i)} \to V^{(i)}$
  be $\ell_i$-level conditionally linear functions and let $L, R: V \to V$ be the
  direct sums $L = \bigoplus_i L_i$ and $R = \bigoplus_i R_i$, respectively, as
  defined in Lemma~\ref{lem:cl-func-prod}.
  Then the conditionally linear distribution $\mu_{L, R}$ is the product
  distribution $\prod_{i=1}^m \mu_{L^{(i)},\, R^{(i)}}$ over $V \times V$.
\end{lemma}

\begin{proof}
	The distribution $\mu_{L,R}$ is the distribution over pairs $(L(x),R(x))$
  where $x$ is sampled uniformly from $V \times V$.
  By Lemma~\ref{lem:cl-func-prod}, this is equivalent to the distribution over
  pairs $\bigl( (L_i(x^{V_i}))_{i=1}^m, (R_i(x^{V_i}))_{i=1}^m \bigr)$ where $x$
  is chosen uniformly at random from $V$.
  This distribution is exactly the product of the distributions $
  \mu_{L^{(i)},\, R^{(i)}}$ for $i = 1, 2, \ldots, m$.
\end{proof}

CL functions used in the paper are frequently defined over a ``large'' field
$\F_{2^t}$ (e.g., the CL functions used in the low degree tests of
Section~\ref{sec:ldt}).
However, the introspection protocol in Section~\ref{sec:introspection} handles
CL functions defined over $\F_2$.
The following definition and lemma show that CL functions over prime power
fields can be viewed as CL functions over the prime field via a ``downsizing''
operation.

\begin{definition}[Downsizing CL functions]
  \label{def:cl-downsize}
  Let $V = \F_{q}^n$ be a linear space for a prime power $q=p^t$ for odd $t$.
  Let $L: V \to V$ be a function.
  Let $\downsize(\cdot)$ denote the downsize map from
  Section~\ref{sec:finite-fields} corresponding to the basis
  $\{e_1,\ldots,e_t\}$ of $\F_q$ over $\F_p$ specified by
  Lemma~\ref{lem:efficient_basis}.
  In particular, $\downsize$ is linear over $\F_p$, and by
  Lemma~\ref{lem:downsize_subspace}, the set $\downsize(V)$ is the linear space
  $\F_p^{nt}$.
  Define the \emph{downsized function} $L^\downsize : \downsize(V) \to
  \downsize(V)$ by $L^\downsize = \downsize \circ L \circ \downsize^{-1}$.
\end{definition}

\begin{lemma}
  \label{lem:cl-downsize}
	Let $V = \F_q^n$ for a prime power $q = p^t$.
  Let $L: V \to V$ be an $\ell$-level CL function over $V$ for some integer
  $\ell \ge 0$.
  Let $L_{\leq j}$, $V_{j,\, u}$, and $L_{j,\, u}$ denote the $j$-th marginal
  functions, factor spaces, and linear maps corresponding to $L$ as guaranteed
  by Lemma~\ref{lem:cl-kth}.
  Then $L^\downsize: \downsize(V) \to \downsize(V)$ is an $\ell$-level CL
  function on $V^\downsize = \downsize(V) = \F_p^{nt}$ with marginal functions
  $L^\downsize_{\leq j}$, factor spaces $V^\downsize_{j,\, v}$, and linear maps
  $L^\downsize_{j,\, v}$ that satisfy the following for all
  $j\in\{1,\ldots,\ell\}$.
  \begin{enumerate}
  \item The $j$-th marginal function $L^\downsize_{\leq j}$ of $L^\downsize$ is
    equal to $\downsize \circ L_{\leq j} \circ \downsize^{-1}$.
	\item For all $u \in L_{< j} (V)$, the $j$-th factor space $V_{j,\,
      \downsize(u)}^\downsize$ and the $j$-th linear map $L_{j,\,
      \downsize(u)}^\downsize$ of $L^\downsize$ are equal to $\downsize(V_{j,\,
      u})$ and $\downsize \circ L_{j,\, \downsize(u)} \circ \downsize^{-1}$, respectively.
  \end{enumerate}
\end{lemma}

\begin{proof}
  We prove the lemma by induction on $\ell$.
  Let $L:V \to V$ be an $\ell$-level CL function.
  For the base case $\ell = 1$, observe that since $\downsize$ is a linear
  bijection between $\F_q$ and $\F_p^t$ as linear spaces over $\F_p$, the
  function $L^\downsize$ is linear, and thus a $1$-level CL function over
  $\downsize(V) = \F_p^{nt}$.
  Furthermore, the first marginal function $L^\downsize_{\leq 1} = L^\downsize =
  \downsize \circ L_{\leq 1} \circ \downsize^{-1}$; the factor space
  $V^\downsize_1 = \downsize(V_1) = \downsize(V)$, and $L^\downsize_1 = L =
  \downsize \circ L_1 \circ \downsize^{-1}$.

  Assume that the statement of the lemma holds for some $\ell - 1 \geq 1$.
  Let $L_{\leq j}$, $V_{j,\, u}$, and $L_{j,\, u}$ denote the marginal
  functions, factor spaces, and linear maps corresponding to $L$ as guaranteed
  by Lemma~\ref{lem:cl-kth}.
  Recursively define the following functions and spaces, for $j \in
  \{1,\ldots,\ell\}$.
  \begin{enumerate}
	\item $L^\downsize_{\leq j} = \downsize \circ L_{\leq j} \circ
    \downsize^{-1}$.
	\item For all $u \in L_{<j} (V)$, set $V_{j,\, \downsize(u)}^\downsize =
    \downsize(V_{j,\, u})$ and set $L_{j,\, \downsize(u)}^\downsize = \downsize
    \circ L_{j,\,\downsize( u)} \circ \downsize^{-1}$.
  \end{enumerate}
  We argue that $\{L^\downsize_{\leq j}\}$, $\{V^\downsize_{j,\, v}\}$, and
  $\{L_{j,\, v}^\downsize \}$ satisfy the conditions of Lemma~\ref{lem:cl-kth}
  for the function $L^\downsize$, which implies that $L^\downsize$ is an
  $\ell$-level CL function over $\downsize(V)$.

  We first establish Item~\ref{enu:cl-last-ell} of Lemma~\ref{lem:cl-kth}.
  Since $L_{\leq \ell} = L$, this implies
  \begin{equation*}
	  L^\downsize = \downsize \circ L \circ \downsize^{-1}
    = \downsize \circ L_{\leq \ell} \circ \downsize^{-1}
    = L^{\downsize}_{\leq \ell},
  \end{equation*}
  as desired.
  Next, for all $j \in \{1, 2, \ldots, \ell\}$, for all $y \in \downsize(V)$
  with $y = \downsize(x)$ for some $x \in V$, letting $u_j = L_{<j}(x)$, we have
  \begin{equation*}
	  \downsize(V) = \downsize \biggl ( \bigoplus_{j=1}^\ell V_{j,\, \smash{x^{L_{<j}}}} \biggr)
    = \bigoplus_{j=1}^\ell \downsize \bigl( V_{j,\, \smash{x^{L_{<j}}}} \bigr)
    = \bigoplus_{j=1}^\ell V^{\,\downsize}_{j,\, \smash{\downsize(x^{L_{<j}})}}\,.
  \end{equation*}
  The first equality follows from Item~\ref{enu:cl-space-sum} of Lemma~\ref{lem:cl-kth},
  the second equality follows from Lemma~\ref{lem:downsize_subspace}, and the third
  equality follows by definition.
  Since $\downsize(x^{L_{<j}}) = L^\downsize_{<j} \circ \downsize (x) =
  L^\downsize_{<j} (y)$, this establishes Item~\ref{enu:cl-space-sum} of
  Lemma~\ref{lem:cl-kth}.

  Next, we have for all $j \in \{1, 2, \ldots, \ell\}$ and all $y \in
  \downsize(V)$ with $y = \downsize(x)$ for some $x \in V$,
  \begin{equation*}
    \begin{split}
      L_{\leq j}^\downsize(y)
      & = \bigl( \downsize \circ L_{\leq j} \circ \downsize^{-1} \bigr) (y)
      = \bigl( \downsize \circ L_{\le j} \bigr) (x)
      = \sum_{i=1}^{j} \downsize \bigl( x^{L_{i,\, x^{L_{<i}}}} \bigr)\\
      & = \sum_{i=1}^{j} L^\downsize_{i,\, v_i}
      \bigl( \downsize(x^{V_{i,\, x^{L_{<i}}}}) \bigr)
      = \sum_{i=1}^{j} L^\downsize_{i,\, v_i} (y^{V^\downsize_{i,\, v_i}})
    \end{split}
  \end{equation*}
  where $v_i = L^\downsize_{< i}(y) = \downsize(x^{L_{<i}})$.
  The first equality follows from definition of $L^\downsize_{\leq j}$, the
  second equality follows from $y = \downsize(x)$, the third equality follows
  from Item~\ref{enu:cl-map-sum} of Lemma~\ref{lem:cl-kth} applied to $L_{\leq j}$, the
  fourth equality follows from the definition of the linear map
  $L^\downsize_{i,\, v}$, and the fifth equality follows from
  Lemma~\ref{lem:downsize_subspace}.
  This establishes Item~\ref{enu:cl-map-sum} of Lemma~\ref{lem:cl-kth} for
  $L^\downsize_{\leq j}$.

  Finally, since $L_{\leq j}$ is a $j$-level CL function over $V$, using the
  inductive hypothesis we have that $L^\downsize_{\leq j}$ is a $j$-level CL
  function over $\downsize(V)$ when $j \in \{1,2,\ldots,\ell-1\}$.
  It remains to establish that $L_{\leq \ell}^\downsize$ is an $\ell$-level CL
  function.
  Since $L$ is an $\ell$-level CL function, there exists register subspaces
  $V_1, V_{> 1}$ such that $V = V_1 \oplus V_{> 1}$, a linear map $L_1 : V_1 \to
  V_1$ and a collection of $(\ell-1)$-level CL functions $\{ L_{>1,\, v} : V_{>
    1} \to V_{> 1} \}_{v \in L_1(V_1)}$ such that $L(x) = x^{L_1} + L_{>1,\,
    x^{L_1}}(x^{V_{>1}})$ for all $x \in V$.
  Observe that $L_1^\downsize$ is a $1$-level CL function on $V_1^\downsize =
  \downsize(V_1)$, and for $v' = \downsize(v) \in L_1^\downsize(V_1^\downsize)$,
  the inductive hypothesis implies the function $L_{>1,\, v'}^\downsize$ is an
  $(\ell-1)$-level CL function on $V_{>1}^\downsize = \downsize(V_{>1})$.
  Furthermore, since $L^\downsize_{\leq \ell} = L^\downsize$, we have that for
  all $y \in \downsize(V)$ with $y = \downsize(x)$ for some $x \in V$,
  \begin{gather*}
  	L^\downsize_{\leq \ell}(y) = L^\downsize(y) = L_1^\downsize(y) +
    L^\downsize_{>1,\, L_1^\downsize(y)}(y^{V_{>1}^\downsize})
  \end{gather*} 
  which implies that $L^\downsize_{\leq \ell}$ is an $\ell$-level CL function
  over $\downsize(V) = \downsize(V_1) \oplus \downsize(V_{>1})$.
  This establishes Item~\ref{enu:cl-k-cl} of Lemma~\ref{lem:cl-kth}, and completes
  the induction.
\end{proof}

\begin{lemma}\label{lem:downsize-cl-dist}
  Let $V = \F_q^n$ for some integer $n$ and prime power $q=p^t$ for odd $t$.
  Let $L,R : V\to V$ be CL functions.
  Let $L^\downsize, R^\downsize : \downsize(V)\to \downsize(V)$ be the
  associated downsized CL functions, as defined in
  Definition~\ref{def:cl-downsize}.
  Then the distribution $\mu_{L^\downsize,R^\downsize}$ over $\downsize(V)\times
  \downsize(V)$ defined in Definition~\ref{def:cl-dist} is identical to the
  distribution of $(x,y) \in \downsize(V)\times \downsize(V)$ obtained by first
  sampling $(x',y')$ according to $\mu_{L,R}$ and then returning
  $(\downsize(x'),\downsize(y'))$.
\end{lemma}

\begin{proof}
  The fact that $L^\downsize, R^\downsize$ are well-defined CL functions follows
  from Lemma~\ref{lem:cl-downsize}.
  The lemma is immediate from the definition of $\mu_{L^\downsize,R^\downsize}$
  and the fact that $\downsize$ is a bijection.
\end{proof}

\subsection{Conditionally linear samplers}
\label{sec:cls}

Samplers are Turing machines that perform computations corresponding to CL
functions defined in Section~\ref{sec:cl}.
The inputs and outputs of the sampler are binary strings that are interpreted as
representing data of different types (integers, bits, vectors in $\F_q^s$,
etc.).
See Section~\ref{sec:ff-representations} and, in particular,
Remark~\ref{rmk:tm_fields} for an in-depth discussion of representing structured
objects on a Turing machine.

\begin{definition}
	A function $q: \N \to \N$ is an \emph{admissible field size function} if for
  all $n \in \N$, $q(n)$ is an admissible field size as defined in
  Definition~\ref{def:admissible-size}.
\end{definition}

\begin{definition}[Conditionally linear samplers]
  \label{def:sampler}
	Let $q:\N \to \N$ be an admissible field size function, and let $s: \N \to \N$
  be a function.
  A $6$-input Turing machine $\sampler$ is a \emph{$\ell$-level conditionally
    linear sampler with field size $q(n)$ and dimension $s(n)$} if for all $n
  \in \N$, letting $q = q(n)$ and $s = s(n)$, there exist $\ell$-level CL
  functions $L^{\alice,\, n}, L^{\bob,\, n} : \F_q^s \to \F_q^s$ with marginal functions
  $\{ L^{w,\, n}_{\leq j}\}$ and factor spaces $\{V^{w,\, n}_{j,\, u}\}$ for $w \in \AB$
  satisfying the conditions of Lemma~\ref{lem:cl-kth}, such that for all $w \in
  \AB$, $j \in \{1,\ldots,\ell\}$, $z,u,y \in \F_q^s$ \hnote{Mar 12, 2026: added quantification over u,y}:
  \begin{itemize}
  \item On input $(n, \gamestyle{dimension})$, the sampler $\sampler$ outputs the
    dimension $s(n)$.
  \item On input $(n,\gamestyle{field})$, the sampler $\sampler$ outputs the field size $q(n)$ in binary. 
  \item On input $(n, w, \gamestyle{marginal}, j, z)$, the sampler $\sampler$
    outputs the binary representation of $L^{w,\, n}_{\leq j}(z)$,\footnote{Here and henceforth, the binary representation of a vector is understood as a list of coefficients in the standard basis. In case an input is not formatted as described (e.g., $z$ has more than $s$ coordinates, etc.), then the sampler halts with a fixed default output; no correctness condition is imposed.}
	 \item On input $(n, w, \gamestyle{linear}, j, u, y)$, the sampler $\sampler$
    outputs the binary representation of $L^{w,\, n}_{j,\, u}(y)$, where $u$ is
    interpreted as an element of $L^{w,\, n}_{<j}(V)$,
	\item On input $(n, w, \gamestyle{factor}, j, u)$, the sampler $\sampler$
    outputs the $j$-th factor space $V^{w,\, n}_{j,\, u}$ of $L^{w,\, n}$ with
    prefix $u \in L^{w,\, n}_{<j}(V)$, represented as a vector in
    $\{0,1\}^s$ indicating which elementary basis vectors of $\F_q^s$ span the factor space.
  \end{itemize}
  Furthermore, the sampler is required to halt on every input of the form $(n,\ldots)$.
  More precisely, on every input of the form $(n,\ldots)$, the sampler computes
  $s(n)$ and $q(n)$, and then performs any necessary format and
  length checks prescribed by the query type before parsing any auxiliary input
  as an element of $\F_q^{s(n)}$, an index, or a player label.  If any check
  fails, the sampler halts and outputs a fixed default string.

  We call $\F_{q(n)}^{s(n)}$ the \emph{ambient space of $\sampler$} on index $n$.
  We call the CL functions $L^{w,\, n}$ for $w \in \ab$ the \emph{CL functions
    of $\sampler$ on index $n$}. A conditionally linear sampler, or \emph{sampler} for short, is an $\ell$-level conditionally linear sampler for some $\ell\geq 1$.
\end{definition}

\begin{remark}
  \label{rmk:sampler-inputs}
  Conditionally linear samplers are defined as $6$-input Turing machines, but
  depending on the input, not all elements of the input tuple are read.
  For example, if the second input is $\gamestyle{dimension}$,
  then the remaining elements of the input tuple are ignored.
  For notational convenience we write samplers with different numbers of
  arguments, depending on the type of argument it gets.
  The number of arguments is always at most $6$, however.
\end{remark}

\begin{remark}
  \label{rmk:sampler-totality}
  The preceding definition makes every sampler a Turing machine that halts on all inputs
  of the form $(n,\ldots)$.  In particular, the quantity
  $\TIME_\sampler(n)$ is finite for every sampler $\sampler$ and every
  $n\in\N$, even though the supremum in its definition ranges over all auxiliary
  strings, including malformed or excessively long ones.
\end{remark}


The following definition shows how samplers naturally correspond to
conditionally linear distributions.
\begin{definition}[Distribution of a sampler]
  \label{def:sampler-sample}
  Let $\sampler$ be a sampler with field size $q(n)$, dimension $s(n)$.
  For each $n \in \N$, let $L^{\alice,\, n}, L^{\bob,\, n}$ denote the CL
  functions of $\sampler$ on index $n$.
  Let $\mu_{\sampler,\, n}$ denote the CL distribution $\mu_{L^{\alice, n},\,
    L^{\bob, n}}$ corresponding to $(L^{\alice,\, n}, L^{\bob,\, n})$, as
  defined in Definition~\ref{def:cl-dist}.
  We call $\mu_{\sampler,\, n}$ the \emph{distribution of sampler $\sampler$ on
    index $n$}.
\end{definition}

The following provides a definition of a ``downsized'' sampler that can be
obtained from any sampler $\sampler$ over an admissible field $\F_q$. \hnote{added: March 12, 2026}
Throughout this paper we will construct families of nonlocal games (such as the
introspection games from Section~\ref{sec:introspection}) where the question
distributions are ``natively'' defined over fields $\F_q$ of growing size. However, it will be 
convenient (especially in the compression theorem) to only refer to samplers over $\F_2$;
in fact we define normal form verifiers (see Section~\ref{sec:normal-form}) in terms of samplers
over $\F_2$.

\begin{definition}[Downsized sampler]
  \label{def:downsize_sampler}
	Let $q: \N \to \N$ be an admissible field size function.
  Let $\sampler$ be an $\ell$-level sampler with field size $q(n)$ and dimension
  $s(n)$.
  Define $\downsize(\sampler)$ as the following Turing machine.
  For all $n \in \N$, $w \in \{\alice,\bob\}$, $j \in \{1,\ldots,\ell\}$, and $z, u, y
  \in \F_2^{s \log q}$ where $q = q(n)$ and $s = s(n)$:
	\begin{itemize}
  \item On input $(n, \gamestyle{dimension})$, the sampler returns the output of
    $\sampler(n, \gamestyle{dimension})$ multiplied by $\log q$ where $q$ is the output of $\sampler(n, \gamestyle{field})$.
	
  \item On input $(n,\gamestyle{field})$, the sampler outputs $2$. 
  
	\item On input $(n, w, \gamestyle{marginal}, j, z)$, the sampler
    $\downsize(\sampler)$ returns the output of $\sampler(n, w,
    \gamestyle{marginal}, j, z)$.
		
  \item On input $(n, w, \gamestyle{linear}, j, u, y)$, the sampler
    $\downsize(\sampler)$ 
    returns
    the output of $\sampler(n, w, \gamestyle{linear}, j, u, y)$.
		
	\item On input $(n, w, \gamestyle{factor}, j, u)$, the sampler
    $\downsize(\sampler)$ computes 
    the
    indicator vector
    \[
      C = \sampler(n, w, \gamestyle{factor}, j, u) \in \{0,1\}^s\;,
    \]
    and returns the expanded indicator vector $(D_1, D_2, \ldots, D_s) \in (\{0,
    1\}^{\log q})^s$ where $D_i$ is the all ones vector in $\{0, 1\}^{\log q}$
    if $C_i = 1$ and $D_i$ is the all zeroes vector otherwise.
  \end{itemize}
\hnote{added June 3, 2026} Before executing any nondimension query above, $\downsize(\sampler)$ implements
the format and length-check convention from Definition~\ref{def:sampler}.  It
first calls $\sampler(n,\gamestyle{dimension})$ and $\sampler(n,\gamestyle{field})$ to compute $s(n)$ and $q(n)$, uses
$s(n)\log q(n)$ as its own dimension, and then performs the checks prescribed by
the requested query type.  If any check fails, it halts with a fixed default
output.
\end{definition}

\begin{remark}
\hnote{added: March 12, 2026} The inputs $z, u, y$ to the downsized sampler
$\downsize(\sampler)$ are interpreted as binary strings in $\F_2^{s \log q}$. 
However, when they are passed to the ``upsized'' sampler $\sampler$ over $\F_q$,
we will interpret $z,u,y$ as binary representations of vectors in $\F_q^s$. 
See Section~\ref{sec:ff-representations} for discussion about binary representations
of finite field elements.
\end{remark}

The next lemma establishes that $\downsize(\sampler)$ is a well-defined CL
sampler, in the sense that it can be derived from a family of CL functions as in
Definition~\ref{def:sampler}.

\begin{lemma}
  \label{lem:downsize_sampler}
  Let $\ell \geq 1$ be such that $\sampler$ is an $\ell$-level CL sampler, and
  let $q(n)$ and $s(n)$ be as in Definition~\ref{def:downsize_sampler}.
  Then the Turing machine $\downsize(\sampler)$ is an $\ell$-level CL sampler
with field size $2$, dimension $s'(n)=s(n)\log q(n)$, and time complexity
\[
  \TIME_{\downsize(\sampler)}(n) = \poly(\TIME_\sampler(n)).
\]
Moreover, the description of $\downsize(\sampler)$ can be computed from the
description of $\sampler$ in time $\poly(|\sampler|)$.
  Furthermore, for every integer $n \in \N$ the CL functions of
  $\downsize(\sampler)$ on index $n$ are $(L^{\alice,\, n})^\downsize$ and
  $(L^{\bob,\, n})^\downsize$, where $L^{\alice,\, n}, L^{\bob,\, n}$ are the CL
  functions of $\sampler$ on index $n$.
\end{lemma}

\begin{proof} 
  First we observe that the dimension $s(n)$ and field size $q(n)$ of $\sampler$ can be computed by querying $\sampler$ on input $(n,\gamestyle{dimension})$ and $(n,\gamestyle{field})$. 

  To show that $\downsize(\sampler)$ is an $\ell$-level CL sampler we first show
  the ``Furthermore'' part, i.e.,\ verify that for any integer $n\geq 1$ the CL
  functions $(L^{\alice,\, n})^\downsize$ and $(L^{\bob,\, n})^\downsize$ are its
  associated CL functions on index $n$, as defined in
  Definition~\ref{def:sampler}.

  Observe that for $z\in V$, the binary representation of $z$ as an element of
  $\{0,1\}^{s\log q}$ passed as input to $\sampler$ is, by definition (see
  Section~\ref{sec:ff-representations}), identical to the binary representation
  of $\downsize(z)$.
  Using the definition $(L^{w,\, n})^\downsize = \downsize \circ L^{w,\, n}
  \circ \downsize^{-1}$ for $w \in \ab$ this justifies that
  $\downsize(\sampler)$ returns the correct output when executed on inputs of
  the form $(n,\gamestyle{dimension})$, $(n,w,\gamestyle{marginal},j,z)$ and
  $(n, w, \gamestyle{linear}, j, u', y)$.

  Next, if $T$ is a register subspace of $\F_q^s$ with indicator vector $C \in
  \{0,1\}^s$, then $\downsize(T)$ is a register subspace of $\F_2^{s\log q}$
  with indicator vector $D$ defined from $C$ as in
  Definition~\ref{def:downsize_sampler}.
  Thus the output of $\downsize(\sampler)$ on input $(n, w, \gamestyle{factor},
  j, u')$ is equal to the indicator vector of $\downsize(V^{w,\, n}_{j,\, u})$,
  which is the $j$-th factor space of $L^{w,\, n}$ with prefix $u' =
  \downsize(u)$.

  \hnote{march 25, 2026, edited:} The time complexity of $\downsize(\sampler)$ follows from the subroutine
simulation convention of Section~\ref{sec:tms}.  The initial call to
$\sampler(n,\allowbreak \gamestyle{dimension})$ contributes \[\textstyle \poly(\TIME_\sampler(n))\] to
the running time of the downsized sampler.  Each subsequent call to the upsized sampler contributes
$\poly(\TIME_\sampler(n))$ as well.  Before parsing any input string as a vector, $\downsize(\sampler)$ performs the
bounded checks required by Definition~\ref{def:sampler}, so malformed or
excessively long inputs do not affect the claimed bound. After this check, all
parsing, conversion between the binary representations over $\F_q^s$ and
$\F_2^{s\log q}$, and padding of factor-space indicator vectors takes time
$\poly(s(n),\allowbreak \log q(n))$. Since $\max \{ s(n),\allowbreak  \log q(n) \} \leq \TIME_\sampler(n)$ as mentioned above, this is also bounded by $\poly(\TIME_\sampler(n))$.
\end{proof}


\section{Nonlocal games and $\MIP^*$}
\label{sec:games}

We introduce definitions associated with nonlocal games and strategies that will
be used throughout.

\subsection{Games and strategies}
\label{sec:strat}

\begin{definition}[Two-player one-round games]
  \label{def:game}
  A \emph{two-player one-round game} $\game$ is specified by a tuple
  $(\cal{X}, \cal{Y}, \cal{A}, \cal{B}, \mu, D)$ where
  \begin{enumerate}
  \item $\cal{X}$ and $\cal{Y}$ are finite sets (called the \emph{question
      alphabets}),
  \item $\cal{A}$ and $\cal{B}$ are finite sets (called the \emph{answer
      alphabets}),
  \item $\mu$ is a probability distribution over $\cal{X} \times \cal{Y}$
    (called the \emph{question distribution}), and
  \item $D: \cal{X} \times \cal{Y} \times \cal{A} \times \cal{B} \to \{0,1\}$ is
    a function (called the \emph{decision predicate}).
  \end{enumerate}
\end{definition}

\begin{definition}[Tensor product strategies]
  \label{def:tensor-product-strategy}
  A \emph{tensor product strategy} $\strategy$ for a game $\game = (\cal{X},
  \cal{Y}, \cal{A}, \cal{B}, \mu, D)$ is a tuple $(\ket{\psi}, A, B)$ where
  \begin{itemize}
	\item $\ket{\psi}$ is a pure quantum state, i.e.,\ a unit vector in $\mH_A \otimes \mH_B$ for finite-dimensional complex Hilbert spaces $\mH_A, \mH_B$,
	\item $A$ is a set $\{A^x\}$ such that for every $x \in \cal{X}$, $A^x =
    \{A^x_a \}_{a \in \cal{A}}$ is a POVM over $\mH_A$, and
	\item $B$ is a set $\{B^y\}$ such that for every $y \in \cal{Y}$, $B^y =
    \{B^y_b \}_{b \in \cal{B}}$ is a POVM over $\mH_B$.
\end{itemize}
\end{definition}

\begin{definition}[Tensor product value]
  \label{def:tensor-product-value}
	The \emph{tensor product value} of a tensor product strategy $\strategy =
  (\ket{\psi},\allowbreak  A,\allowbreak  B)$  with respect to a game $\game=(\cal{X},\allowbreak  \cal{Y},\allowbreak  \cal{A},\allowbreak 
  \cal{B},\allowbreak  \mu,\allowbreak  D)$ is defined as
  \begin{equation*}
		\val^*(\game, \strategy) = \sum_{x,\, y,\, a,\, b} \, \mu(x,y)\, D(x,y,a,b)\,
    \bra{\psi} A^x_a \otimes B^y_b\, \ket{\psi}\;.
  \end{equation*}
	For $v\in[0,1]$ we say that the strategy $\strategy$ \emph{passes (or wins)
    $\game$ with probability $v$ if} $\val^*(\game, \strategy) \geq v$.
  The \emph{tensor product value} of $\game$ is defined as
  \begin{equation*}
		\val^*(\game) = \sup_\strategy \val^*(\game, \strategy)\;,
  \end{equation*}
	where the supremum is taken over all tensor product strategies $\strategy$ for
  $\game$.
\end{definition}

\begin{remark}
  Unless specified otherwise, all strategies considered in this paper are tensor
  product strategies, and we simply call them \emph{strategies}.
  Similarly, we refer to $\val^*(\game)$ as the \emph{value} of the game
  $\game$.
\end{remark}

\begin{definition}[Projective strategies]
  \label{def:projective-strategy}
  We say that a strategy $\strategy = (\ket{\psi},\allowbreak  A,\allowbreak  B)$ is \emph{projective} if
  all the measurements $\{A^x_a\}_a$ and $\{B^y_b\}_b$ are projective.
\end{definition}

\begin{definition}\label{rem:symmetric-games}
	A game $\game = (\cal{X}, \cal{Y}, \cal{A}, \cal{B}, \mu, D)$ is
  \emph{symmetric} if the question and answer alphabets are the same for both
  players (i.e.,\
  $\cal{X} = \cal{Y}$ and $\cal{A} = \cal{B}$), the distribution $\mu$ is
  symmetric (i.e.,\
  $\mu(x,y) = \mu(y,x)$), and the decision predicate $D$ treats both players
  symmetrically (i.e.,\ for all $x,y,a,b$, $D(x,y,a,b) = D(y,x,b,a)$).
 
We call a strategy $\strategy = (\ket{\psi}, A, B)$
  \emph{symmetric} if $\ket{\psi}$ is a (pure) state in $\mH \otimes \mH$, for some
  Hilbert space $\mH$, that is invariant under permutation of the two factors,
  and the measurement operators of both players are identical.
\end{definition}

We often specify symmetric games $\game$ and symmetric strategies $\strategy$ using
  a compact notation: we write $\game = (\cal{X}, \cal{A}, \mu, D)$ and
  $\strategy = (\ket{\psi}, M)$ where $M$ denotes the set of measurement
  operators for both players.

\begin{lemma}\label{lem:symmetric-strat}
  \leavevmode\par Let $\game = (\cal{X},\allowbreak  \cal{A},\allowbreak   \mu,\allowbreak  D)$ be a symmetric game
  such that $\val^*(\game) = 1-\eps$ for some $\eps\geq 0$.
  Then there exists a symmetric and projective strategy
  $\strategy=(\ket{\psi},\allowbreak M)$ such that $\val^*(\game,\allowbreak \strategy)\geq 1-\eps$.
	 Moreover, for any strategy $\strategy'$ for $\game$ with state
  $\ket{\psi'}$, the construction below produces a symmetric projective
  strategy with value at least $\val^*(\game,\allowbreak \strategy')$ whose state has
  Schmidt rank at most twice that of $\ket{\psi'}$.
\end{lemma}

\begin{proof}
  By definition for any $\eps'>\eps$ there exists a strategy $\strategy'= (\ket{\psi'},A,B)$ such
  that $\val^*(\game,\strategy')\geq 1-\eps'$.
Using Naimark's theorem (Theorem~\ref{thm:naimark}) we can assume
without loss of generality that $\ket{\psi'}\in\C_{\alice'}^d\otimes \C_{\bob'}^d$ for some integer $d$
  and that for every $x$, $A^x$ and $B^x$ are projective measurements.
  Let
  \[
    \ket{\psi} = \frac{1}{\sqrt{2}} \big( \ket{0}_\alice \ket{1}_\bob
    \ket{\psi'}_{\alice'\bob'} + \ket{1}_\alice \ket{0}_\bob
    \ket{\psi'_\tau}_{\alice'\bob'}\big) \in (\C_{\alice}^2 \otimes
    \C_{\alice'}^d) \otimes (\C_\bob^2 \otimes \C_{\bob'}^d)\;,
  \]
  where $\ket{\psi'_\tau}$ is obtained from $\ket{\psi}$ by permuting the two
  players' registers.
  Observe that $\ket{\psi}$ is invariant under permutation of $\alice\alice'$
  and $\bob\bob'$.

  For any question $x\in \cal{X}$ define the measurement $M^x = \{M^x_a\}_{a \in \cal{A}}$
  acting on the Hilbert space $\C^2 \otimes \C^d$ as follows:
  \[
  	M^x_a = \ketbra{0}{0} \otimes A^x_a + \ketbra{1}{1} \otimes B^x_a \;,
  \]
	i.e.,\ for $\ket{\varphi} \in \C^d$, $M^x_a(\ket{0}\ket{\varphi})=\ket{0}A^x_a\ket{\varphi}$ and $M^x_a(\ket{1}\ket{\varphi})=\ket{1}B^x_a\ket{\varphi}$.
  When Alice receives question $x$, she measures $M^x$ on registers $\alice \alice'$, and when Bob receives 
  question $y$, he measures $M^y$ on registers $\bob \bob'$.
  Using that by assumption the decision predicate $D$ for $\game$ is symmetric,
  it is not hard to verify that $\val^*(\game,\strategy) \geq 1-\eps'$.
\end{proof}

\begin{definition}
  \label{def:comm-strategy}
  Let $\game = (\cal{X}, \cal{Y}, \cal{A}, \cal{B}, \mu, D)$ be a game, and let
  $\strategy = (\ket{\psi}, A, B)$ be a strategy for $\game$ such that  $\mH_A = \mH_B$.
  Let $S \subseteq \cal{X} \times \cal{Y}$ denote the support of the question
  distribution $\mu$, i.e.,\ the set of $(x,y)$ such that $\mu(x,y)>0$.
  We say that $\strategy$ is a \emph{commuting strategy for $\game$} if for all
  question pairs $(x, y) \in S$, we have $[A^x_a, B^y_b] = 0$ 
  for all $a
  \in \cal{A}, b \in \cal{B}$, where $[A, B] = AB - BA$ denotes the commutator.
\end{definition}

\begin{definition}[Consistent measurements]
  \label{def:consistent-measurement}
  Let $\cal{A}$ be a finite set, let $\ket{\psi} \in \mH \otimes \mH$ be a
  state, and let $\{ M_a \}_{a \in \cal{A}}$ be a projective measurement on
  $\mH$.
  We say that \emph{$\{M_a\}_{a \in \cal{A}}$ is consistent on $\ket{\psi}$} if
  and only if
  \begin{equation*}
    \forall a \in \mA\;,\quad M_a \otimes \Id_\bob\, \ket{\psi}
    = \Id_\alice \otimes M_a\, \ket{\psi}\;.
  \end{equation*}
	When the state $\ket{\psi}$ is clear from context we simply say that \emph{$\{M_a\}_{a \in \cal{A}}$ is consistent}.
\end{definition}

The terminology ``consistent'' arises from the fact that when the measurement $\{M_a\}$ is performed on both registers of $\ket{\psi}$, the probability of obtaining twice the same outcome is 
\[ \sum_a \bra{\psi} M_a \otimes M_a \ket{\psi}\,=\, \sum_a \bra{\psi} \Id_\alice \otimes (M_a)^2 \ket{\psi} \,=\, \sum_a \bra{\psi} \Id_\alice \otimes M_a \ket{\psi} \,=\,1\;,\]
where the first equality is by definition of consistency, the second uses that $\{M_a\}$ is projective, and the last uses $\sum_a M_a=\Id$. 

\begin{definition}[Consistent strategies]
  \label{def:consistent-strategy}
	Let $\strategy = (\ket{\psi}, A, B)$ be a projective strategy with state
  $\ket{\psi} \in \mH \otimes \mH$, for some Hilbert space~$\mH$, which is
  defined on question alphabets $\cal{X}$ and~$\cal{Y}$ and answer alphabets
  $\cal{A}$ and~$\cal{B}$, respectively.
  We say that the strategy $\strategy$ is \emph{consistent} if for all $x \in
  \cal{X}$, the measurement $\{A^x_a\}_{a \in \cal{A}}$ is consistent on
  $\ket{\psi}$ and if for all $y \in \cal{Y}$, the measurement $\{B^y_b\}_{b \in
    \cal{B}}$ is consistent on $\ket{\psi}$.
\end{definition}

We almost always restrict our attention to symmetric strategies. In this case, consistency implies that whenever both players are sent the same question, they provide the same answer. 

\begin{definition}
  \label{def:spcc}
  We say that a strategy $\strategy$ for a game $\game$ is \emph{PCC} if it is projective, consistent, and commuting for $\game$.
  Additionally, we say that a PCC strategy $\strategy$ is \emph{SPCC} if it is, furthermore, symmetric.
\end{definition}

Recall that for a (pure) state $\ket{\psi}\in \mH_\alice \otimes \mH_\bob$, the \emph{Schmidt rank} of $\ket{\psi}$ is the smallest integer $k$ such that 
\[ \ket{\psi} = \sum_{i=1}^k\, \alpha_i \ket{u_i}\otimes \ket{v_i}\;,\]
for some orthonormal families $\{\ket{u_i}\}\subset \mH_\alice$ and $\{\ket{v_i}\}\subset\mH_\bob$ and $\alpha_i>0$. The coefficients $\alpha_i$ are uniquely defined by $\ket{\psi}$ and called \emph{Schmidt coefficients}. The state $\ket{\psi}$ is called \emph{maximally entangled} if $\alpha_i=\frac{1}{\sqrt{k}}$ for all $i=1,\ldots,k$.

\begin{definition}[Entanglement requirements of a game]
  \label{def:ent}
	For all games $\game$ and $\nu \in [0, 1]$, let $\Ent(\game, \nu)$ denote the
  minimum integer $d$ such that there exists a finite-dimensional tensor product
  strategy $\strategy$ that achieves success probability at least $\nu$ in the
  game $\game$ with a state $\ket{\psi}$ whose Schmidt rank is at most $d$.
  If there is no finite-dimensional strategy that achieves success probability
  $\nu$, then define $\Ent(\game, \nu)$ to be $\infty$.
\end{definition}

\begin{remark}
  A strategy $\strategy = (\ket{\psi},A,B)$ for a symmetric game $\game =
  (\cal{X}, \cal{X}, \cal{A}, \cal{A}, \mu, D)$ is called \emph{synchronous} if
  it holds that for every $x\in \cal{X}$ and $a\neq b \in \cal{A}$, $\bra{\psi}
  A^x_a \otimes B^x_b \ket{\psi} = 0$; in other words,
  the players never return different answers when simultaneously asked the same
  question.
  As shown in~\cite{paulsen2016estimating} the condition for a
  finite-dimensional strategy of being synchronous is equivalent to the
  condition that it is projective, consistent, and, moreover, $\ket{\psi}$ is a
  maximally entangled state.
  (The equivalence is extended to infinite-dimensional strategies, as well as
  correlations induced by limits of finite-dimensional strategies,
  in~\cite{kim2018synchronous}.)
\end{remark}
\subsection{Distance measures}

We introduce several distance measures that are used throughout.

\begin{definition}[Distance between states]
  \label{def:state-distance}

	Let $\{\ket{\psi_n}\}_{n\in \N}$ and $\{\ket{\psi'_n}\}_{n\in \N}$ be two
  families of states in the same space $\mH$.
  For some function $\delta : \N \to [0,1]$ we say that $\{ \ket{\psi_n} \}$ and
  $\{ \ket{\psi'_n} \}$ are \emph{$\delta$-close}, denoted as $\ket{\psi}
  \approx_\delta \ket{\psi'}$, if $ \norm{\ket{\psi_n} - \ket{\psi'_n}}^2 =
  O(\delta(n))$.
  (For convenience we generally leave the dependence of the states and $\delta$
  on the indexing parameter $n$ implicit, writing e.g.,\ $\ket{\psi}$ for $\ket{\psi_n}$.)
\end{definition}

\begin{definition}[Consistency between POVMs]
  \label{def:consistency}
	Let $\mX$ be a finite set and $\mu$ a distribution on $\mX$.
  Let $\ket{\psi} \in \cal{H}_A \otimes \cal{H}_B$ be a quantum state, and for
  all $x\in \mX$, $\{A^x_a\}$ and $\{B^x_a\}$ POVMs.
  We write
  \begin{equation*}
    A^x_a \otimes I_\bob \simeq_\delta I_\alice \otimes B^x_a
  \end{equation*}
  on state $\ket{\psi}$ and distribution $\mu$ if
  \begin{equation*}
    \E_{x\sim \mu} \sum_{a\ne b} \bra{\psi} A^x_a \otimes B^x_b \ket{\psi}
    = O(\delta)\;.\footnote{Here the use of the $O(\cdot)$ notation refers to the fact that the notation will often be used to measure consistency for \emph{famillies} of POVM and states indexed by an implicit parameter $n\in \mathbb{N}$, which will always be clear from context but omitted for legibility; see Definition~\ref{def:state-distance}. The $O(\cdot)$ is taken as $n\to\infty$.}
  \end{equation*}
  In this case, we say that $\{A^x_a\}$ and $\{B^x_a\}$ are
  \emph{$\delta$-consistent} on $\ket{\psi}$.
\end{definition}

Note that a consistent measurement on $\ket{\psi}$ according to
Definition~\ref{def:consistent-measurement} is $0$-consistent with itself on the same $\ket{\psi}$ and under
the singleton distribution, according to Definition~\ref{def:consistency} (and
vice-versa). This is because 
\[ \sum_a \bra{\psi} M_a \otimes M_a \ket{\psi} =1 \,\implies \, \forall a\;,\;\Id\otimes M_a \ket{\psi} = M_a \otimes \Id \ket{\psi}\;,\]
because using that $\{M_a\}$ is a POVM there must be equality in the Cauchy-Schwarz inequality
\[ 1 = \sum_a \bra{\psi} M_a \otimes M_a \ket{\psi}\, \leq\, \Big(\sum_a \|\Id\otimes M_a \ket{\psi}\|^2 \Big)\Big(\sum_a \|M_a\otimes \Id \ket{\psi}\|^2 \Big)^{1/2} \,\leq\,1 \;.\]

\begin{definition}[Distance between POVMs]
  \label{def:povm-distance}
	Let $\mX$ be a finite set and $\mu$ a distribution on $\mX$. 
  Let $\ket{\psi} \in \cal{H}$ be a quantum state, and for all $x\in \mX$,
  $\{M^x_a\}$ and $\{N^x_a\}$ two POVMs on $\mH$.
  We say that \emph{$\{M^x_a\}$ and $\{N^x_a\}$ are $\delta$-close on state
    $\ket{\psi}$ and under distribution $\mu$} if
  \begin{equation*}
    \E_{x\sim \mu} \sum_a \norm{ (M^x_a - N^x_a) \ket{\psi} }^2 \leq O(\delta)\;,
  \end{equation*}
  and we write $M^x_a \approx_\delta N^x_a$ to denote this when the state
  $\ket{\psi}$ and distribution $\mu$ are clear from context.
  This distance is referred to as the \emph{state-dependent} distance.
\end{definition}

\begin{definition}[Distance between strategies]
  \label{def:strategy-distance}
  Let $\game = (\cal{X}, \cal{Y}, \cal{A}, \cal{B}, \mu, D)$ be a nonlocal game
  and let $\strategy = (\psi, A, B)$, $\strategy' = (\psi', A', B')$ be 
  strategies for $\game$.
  For $\delta\in[0,1]$ we say that \emph{$\strategy$ is $\delta$-close to
    $\strategy'$} if the following conditions hold.
  \begin{enumerate}
	\item The states $\ket{\psi}, \ket{\psi'}$ are states in the same
    Hilbert space $\mH_A \otimes \mH_B$ and are $\delta$-close.
	\item For all $x \in \cal{X}, y \in \cal{Y}$, we have $A^x_a \approx_\delta
    (A')^x_a$ and $B^y_b \approx_\delta (B')^y_b$, with the approximations
    holding under the distribution $\mu$, and on either $\ket{\psi}$ or
    $\ket{\psi'}$.
  \end{enumerate}
\end{definition}

We record several useful facts about the consistency measure and the
state-dependent distance without proof.
Readers are referred to Sections 4.4 and 4.5 in~\cite{NW19} for additional
discussion and proofs.

\begin{fact}[Fact 4.13 and Fact 4.14 in~\cite{NW19}]
  \label{fact:agreement}
  For POVMs $\{A^x_a\}$ and $\{B^x_a\}$, the following hold.
  \begin{enumerate}
  \item If $A^x_a \otimes I_\bob \simeq_\delta I_\alice \otimes B^x_a$ then
    $A^x_a \otimes I_\bob \approx_\delta I_\alice \otimes B^x_a$.
    \label{item:consistency-implies-approx}
  \item If $A^x_a \otimes I_\bob \approx_\delta I_\alice \otimes B^x_a$
    \emph{and} $\{A^x_a\}$ and $\{B^x_a\}$ are projective measurements, then
    $A^x_a \otimes I_\bob \simeq_\delta I_\alice \otimes B^x_a$.
    \label{item:both-projectors-implies-consistency}
  \item If $A^x_a \otimes I_\bob \approx_\delta I_\alice \otimes B^x_a$ and
    either $\{A^x_a\}$ or $\{B^x_a\}$ is a projective measurement, then
    $A^x_a \otimes I_\bob \simeq_{\delta^{1/2}} I_\alice \otimes B^x_a$.
    \label{item:one-projector-implies-consistency}
  \end{enumerate}
\end{fact}

\begin{fact}[Fact 4.20 in~\cite{NW19}]
  \label{fact:add-a-proj}
  Let $\cal{A},\cal{B},\cal{C}$ be finite sets, and let $\mu$ be a distribution
  over question pairs $(x,y)$.
  Let $\{A_{a,b}^x\}$ and $\{B_{a,b}^x\}$ be operators whose outcomes range over the
  product set $\cal{A} \times \cal{B}$.
  Suppose a set of operators $\{C_{a,c}^y\}$, whose outcomes range over the
  product set $\cal{A} \times \cal{C}$, satisfies the condition $\sum_{c}
  (C_{a,c}^y)^\dagger C_{a,c}^y \leq \Id$ for all $a$ and all $y$.
  If $A_{a,b}^x \approx_\delta B_{a,b}^x$ on average over $x$ sampled from the
  corresponding marginal of distribution $\mu$, then $C^y_{a,c} A^x_{a,b}
  \approx_\delta C^y_{a,c} B^x_{a,b}$ on average over $(x,y)$ sampled from $\mu$.
\end{fact}
\begin{proof}
  Fix questions $x,y$ and answers $a \in \cal{A},b \in \cal{B}$. We have then that
  \begin{align}
    &\sum_{c} \left \| (C^y_{a,c} A^x_{a,b} - C^y_{a,c} B^x_{a,b} ) \ket{\psi} \right \|^2
    \nonumber\\
    & = \sum_{c} \bra{\psi} (A^x_{a,b} - B^x_{a,b})^\dagger (C^y_{a,c})^\dagger (C^y_{a,c})
      (A^x_{a,b} - B^x_{a,b} ) \ket{\psi}\\
    & \leq \bra{\psi} (A^x_{a,b} - B^x_{a,b})^\dagger
      (A^x_{a,b} - B^x_{a,b} ) \ket{\psi} \\
    & = \left \| (A^x_{a,b} - B^x_{a,b})\ket{\psi} \right \|^2
  \end{align}
  where the inequality follows from the assumption that $\sum_c (C_{a,c}^y)^\dagger
  C_{a,c}^y \leq \Id$.
  Thus we obtain the desired conclusion
  \begin{equation*}
    \E_{(x,y) \sim \mu} \sum_{a,b,c} \left \| (C^y_{a,c} A^x_{a,b} - C^y_{a,c} B^x_{a,b} )
      \ket{\psi} \right \|^2 	\leq \E_{(x,y) \sim \mu} \sum_{a,b}
    \left \| (A^x_{a,b} - B^x_{a,b})\ket{\psi} \right \|^2 \leq \delta.
  \end{equation*}
\end{proof}

\begin{fact}
\label{fact:add-a-proj2}
Let $\mathcal{X},\mathcal{A}$ denote finite sets, and let $\mathcal{G}$ denote a set of functions $g: \mathcal{X} \to \mathcal{A}$. Let $\{A^x_a \}$, $\{B^x_a\}$ be POVMs indexed by $\mathcal{X}$ and outcomes in $\mathcal{A}$. Let $\{ S_g^x \}_{x\in \mathcal{X}, g\in\mathcal{G}}$ denote a set of operators such that for all $x \in \mathcal{X}$, $\sum_g (S_g^x)^\dagger S_g^x \leq \Id$. 
If $A^x_a \approx_\delta B^x_a$ on average over $x$, then $S_g^x A^x_{g(x)} \approx_\delta S_g^x B^x_{g(x)}$. 
\end{fact}
\begin{proof}
	We expand:
	\begin{align*}
		&\E_x \sum_g \norm{ S^x_g (A^x_{g(x)} - B^x_{g(x)}) \ket{\psi}}^2 \\
 &= \E_x \sum_g \bra{\psi} (A^x_{g(x)} - B^x_{g(x)})^\dagger  (S_g^x)^\dagger S_g^x (A^x_{g(x)} - B^x_{g(x)}) \ket{\psi} \\
		&= \E_x \sum_a \bra{\psi} (A^x_a - B^x_a)^\dagger  \Big(\sum_{g: g(x) = a} (S_g^x)^\dagger S_g^x \Big) (A^x_a - B^x_a) \ket{\psi} \\
		&\leq \E_x \sum_a \bra{\psi} (A^x_a - B^x_a)^\dagger (A^x_a - B^x_a) \ket{\psi} \\
		&= \E_x \sum_a \norm{ (A^x_a - B^x_a) \ket{\psi} }^2.
	\end{align*}
	The inequality follows from the fact that $\sum_{g: g(x) = a} (S_g^x)^\dagger S_g^x \leq \sum_g (S_g^x)^\dagger (S_g^x) \leq \Id$. The last line is at most $\delta$ by assumption, and we obtain the desired conclusion.
\end{proof}

\begin{lemma}
\label{lem:cool-closeness-fact}
Let $\mathcal{X}$ and $\mathcal{A}$ be finite sets and $\mu$ a distribution on $\mX$. Let $\{A^x_a\}_{a \in \mathcal{A}}$ be a projective measurement and let $\{B^x_a\}_{a \in \mathcal{A}}$ be a set of matrices. If $A^x_a \approx_\delta B^x_a$, then for all subsets $S \subseteq \mathcal{A}$, we have
\[
	\sum_{a \in S} A^x_a \approx_\delta \sum_{a \in S} A^x_a \cdot B^x_a.
\]
\end{lemma}
\begin{proof}
We expand:
\begin{align*}
	\E_x \norm{ \sum_{a \in S} \Big(A^x_a - A^x_a \cdot B^x_a\Big) \ket{\psi}}^2 &= \E_x \norm{ \sum_{a \in S} A^x_a \cdot \Big(A^x_a - B^x_a\Big) \ket{\psi}}^2 \\
	&= \E_x \sum_{a \in S} \bra{\psi} (A^x_a - B^x_a)^\dagger A^x_a (A^x_a - B^x_a) \ket{\psi} \\
	&\leq \E_x \sum_a \norm{ (A^x_a -  B^x_a) \ket{\psi}}^2\;,
\end{align*}
where all expectations are taken according to $\mu$. 
In the second line we used the projectivity of $\{ A^x_a \}$, and in the third line we used that $A^x_a \leq \Id$.
\end{proof}

\begin{fact}[Triangle inequality, Fact 4.28 in~\cite{NW19}]
  \label{fact:triangle}
  If $A_a^x \approx_\delta B_a^x$ and $B_a^x \approx_\epsilon C_a^x$, then
  $A_a^x \approx_{\delta + \epsilon} C_a^x$.
\end{fact}

The following triangle equality fact is an improvement of Proposition 4.29 in~\cite{ML20}
\begin{fact}[Triangle-like
  inequalities]\label{fact:triangle-for-simeq}
The following two facts are true.
\begin{enumerate}
\item Suppose $A^x_a \otimes I_{\bob} \simeq_\delta I_{\alice} \otimes B^x_a$,
	$I_{\alice} \otimes B^x_a \simeq_\delta C^x_a \otimes I_{\bob}$,
	and $C^x_a \otimes I_{\bob} \simeq_\delta I_{\alice} \otimes D^x_a$.
	Then $A^x_a \otimes I_{\bob} \simeq_\delta I_{\alice} \otimes D^x_a$.
\item Suppose $A^x_a \otimes I_{\bob} \approx_\delta I_{\alice} \otimes B^x_a$,
	$I_{\alice} \otimes B^x_a \approx_\delta C^x_a \otimes I_{\bob}$,
	and $C^x_a \otimes I_{\bob} \approx_\delta I_{\alice} \otimes D^x_a$.
	Then $A^x_a \otimes I_{\bob} \approx_\delta I_{\alice} \otimes D^x_a$.
\end{enumerate}
\end{fact}
\begin{proof}
The second fact follows from several applications of Fact~\ref{fact:triangle}.
As for the first fact, 
given POVM measurements $\{R^x_a\}$ and $\{S^x_a\}$, let us define
\begin{equation*}
    p(R, S) := \E_{x \sim  \cal{D}} \sum_a \bra{\psi} R^{x}_a \otimes S^{x}_a \ket{\psi}.
\end{equation*}
By assumption, we know that
\begin{equation*}
    p(A, B) \geq 1 -\delta,
    \qquad
    p(C, B) \geq 1 - \delta,
    \quad\text{and}\quad
    p(C, D) \geq 1 -\delta.
\end{equation*}
Now, let us introduce two new auxiliary registers corresponding to the questions $x$ and the answers $a$, and define the vectors
\begin{align*}
    &\ket{u_A} := \sum_{x, a} \sqrt{ \cal{D}(x)} \cdot \ket{x} \otimes \ket{a} \otimes (A_a \otimes I) \ket{\psi}, \\
    & 
    \ket{u_B} := \sum_{x, a} \sqrt{ \cal{D}(x)} \cdot \ket{x} \otimes \ket{a} \otimes (I \otimes B_a) \ket{\psi},\\
    &\ket{u_C} := \sum_{x, a} \sqrt{ \cal{D}(x)} \cdot \ket{x} \otimes \ket{a} \otimes (C_a \otimes I) \ket{\psi}, \\
    & 
    \ket{u_D} := \sum_{x, a} \sqrt{ \cal{D}(x)} \cdot \ket{x} \otimes \ket{a} \otimes (I \otimes D_a) \ket{\psi}.
\end{align*}
Then for any pair $(R, S) \in \{(A, B), (C, B), (C, D)\}$, we have that
\begin{equation*}
    \ip{u_R}{u_S} = p(R, S).
\end{equation*}
In addition, we have that
\begin{equation*}
    \ip{u_A}{u_A} = \E_{x \sim  \cal{D}}\sum_{a \in A} \bra{\psi} (A_a^{x})^2 \otimes I \ket{\psi}
    \leq \E_{x \sim  \cal{D}}\sum_{a \in A} \bra{\psi} A_a^{x} \otimes I \ket{\psi}
    = 1,
\end{equation*}
using the fact that $A_a^{x} \preceq I$ because $\{A^x_a\}$ is a POVM.
A similar argument shows that $\ket{u_B}, \ket{u_C}, \ket{u_D}$ all have length at most one.
To make these unit vectors,
let us enlarge our space by direct summing it with a dimension-4 subspace with an orthonormal basis labeled $\ket{e_A}, \ket{e_B}, \ket{e_C}, \ket{e_D}$.
Then for $R \in \{A, B, C, D\}$ we can define
\begin{equation*}
    \ket{\widehat{u}_R} := \ket{u_R} \oplus \sqrt{1 - \ip{u_R}{ u_R}} \cdot \ket{e_R}.
\end{equation*}
By construction, these are unit vectors, and for $(R,\allowbreak  S) \allowbreak \in\allowbreak  \{(A,\allowbreak  B),\allowbreak  (C,\allowbreak  B),\allowbreak  (C,\allowbreak  D)\}$ they satisfy
\begin{equation*}
    \ip{\widehat{u}_R}{\widehat{u}_S}
    = \ip{u_R}{u_S}
    = p(R, S).
\end{equation*}
Moreover,
\begin{equation*}
    \Vert \ket{\widehat{u}_R} - \ket{\widehat{u}_S} \Vert^2 = 2 - 2 p(R, S) \leq 2 \delta.
\end{equation*}
Thus, by the triangle inequality,
\begin{equation*}
    \Vert \ket{\widehat{u}_A} - \ket{\widehat{u}_D} \Vert
    \leq \Vert \ket{\widehat{u}_A} - \ket{\widehat{u}_B} \Vert + \Vert \ket{\widehat{u}_B} - \ket{\widehat{u}_C} \Vert + \Vert \ket{\widehat{u}_C} - \ket{\widehat{u}_D} \Vert \leq 3 \sqrt{2 \delta}.
\end{equation*}
Thus,
\begin{equation*}
    p(A, D) = 1 - \frac{1}{2} \Vert \ket{\widehat{u}_A} - \ket{\widehat{u}_B} \Vert^2 \geq 1 - 9 \delta.
\end{equation*}
This completes the proof.
\end{proof}

\begin{fact}[Data processing, Fact 4.26 in~\cite{NW19}]
  \label{fact:data-processing}
  Suppose $A^x_a \abc B^x_a$.
  Then $A^x_{[f(\cdot) = b]} \abc B^x_{[f(\cdot) = b]}$.
\end{fact}

The state-dependent distance is the right tool for reasoning about the closeness
of measurement operators in a strategy.
The following lemma ensures that, when two families of measurements are close on a
state, changing from one family of measurement to the other only introduces
a small error to the value of the strategy.

\begin{lemma}
  \label{lem:commutation-analysis}
  Let $\{A^x_{a,\, b}\}$, $\{B^x_{a,\, b,\, c}\}$, $\{C^x_{a,\, c}\}$ be POVMs.
  Suppose $\{B^x_{a,\, b,\, c}\}$ is projective, and
  \begin{align*}
    A^x_{a,\, b} \otimes \Id_\bob &\approx_\delta \Id_\alice \otimes B^x_{a,\, b}\;,\\
    C^x_{a,\, c} \otimes \Id_\bob &\approx_\delta \Id_\alice \otimes B^x_{a,\, c}\;,
  \end{align*}
	where recall that $B^x_{a,\, b} = \sum_c B^x_{a,b,c}$ and similarly $B^x_{a,\, c}=\sum_b B^x_{a,b,c}$. 
  Then the following approximate commutation relation holds:
  \begin{equation*}
    [A^x_{a,\, b}, C^x_{a,\, c}] \otimes I_\bob \approx_\delta 0\;.
  \end{equation*}
\end{lemma}

\begin{proof}
  Applying Fact~\ref{fact:add-a-proj} to $C^x_{a,\, c} \otimes \Id_\bob
  \approx_\delta \Id_\alice \otimes B^x_{a,\, c}$ and $\{A^x_{a,b} \otimes
  \Id_\bob \}$, we have
  \begin{equation}
    \label{eq:commutation-analysis-1}
    A^x_{a,\, b} C^x_{a,\, c} \otimes \Id_\bob
    \approx_\delta A^x_{a,\, b} \otimes B^x_{a,\, c}\,.
  \end{equation}
  Similarly, applying Fact~\ref{fact:add-a-proj} to $A^x_{a,\, b} \otimes
  \Id_\bob \approx_\delta \Id_\alice \otimes B^x_{a,\, b}$ and $\{\Id_\alice
  \otimes B^x_{a,c}\}$, and using the fact that $\{B^x_{a,b,c}\}$ is projective,
  we have
  \begin{equation}
    \label{eq:commutation-analysis-2}
    \begin{split}
      A^x_{a,\, b} \otimes B^x_{a,\, c}
      & \approx_\delta \Id_\alice \otimes B^x_{a,\, c} B^x_{a,\, b}\\
      & =_{\phantom{\delta}} \Id_\alice \otimes B^x_{a,\, b,\, c}\,.
    \end{split}
  \end{equation}
  Combining Equations~\eqref{eq:commutation-analysis-1}
  and~\eqref{eq:commutation-analysis-2}, we have
  \begin{equation}
    \label{eq:commutation-analysis-3}
    A^x_{a,\, b} C^x_{a,\, c} \otimes \Id_\bob \approx_\delta
    \Id_\alice \otimes B^x_{a,\, b,\, c}\,.
  \end{equation}
  A similar argument gives
  \begin{equation}
    \label{eq:commutation-analysis-4}
    C^x_{a,\, c} A^x_{a,\, b} \otimes \Id_\bob \approx_\delta
    \Id_\alice \otimes B^x_{a,\, b,\, c}\,.
  \end{equation}
  The claim follows from Equations~\eqref{eq:commutation-analysis-3}
  and~\eqref{eq:commutation-analysis-4}.
\end{proof}

The following lemma is a slightly modified version of~\cite[Fact~4.34]{NW19}.
\begin{lemma}\label{lem:ld-sandwich}
  Let $k \geq 0$ be an integer and let $\eps>0$.
  Let $\mX$ be a finite set and $\mu$ a distribution over $\mX$.
  For each $1 \leq i \leq k$ let $\mG_i$ be a finite set of functions $g_i : \mY
  \rightarrow \mR_i$ and for each $x \in \mX$ let $\{G^{i,\, x}_g\}_{g\in
    \mG_i}$ be a projective measurement.
  Suppose that for all $i\in \{1,\ldots,k\}$, $\mG_i$ satisfies the following
  property: for any two $g_i \neq g_i' \in \mG_i$, the probability that $g_i(y)
  = g_i'(y)$ over a uniformly random $y \in \mY$ is at most $\eps$.

  Let $\bigl\{ A^{x}_{g_1,\, g_2,\, \ldots\,,\, g_k} \bigr\}$ be a projective
  measurement with outcomes $(g_1,\allowbreak \ldots,\allowbreak g_k) \allowbreak \in\allowbreak  \mG_1\times\cdots\times
  \mG_k$.
  For each $1 \allowbreak \leq\allowbreak  i \allowbreak \leq\allowbreak  k$, suppose that on average over $x \sim \mu$ and
  $y\in \mY$ sampled uniformly at random and independently from $x$,
  \begin{equation}\label{eq:ld-sandwich-1}
    A^{x}_{[\eval_y(\cdot)_i = a_i]} \abc G^{i,\, x}_{[\eval_y(\cdot)=a_i]}\;.
  \end{equation}
  Define the POVM family $\{C^x_{g_1,\, g_2,\, \ldots\,,\, g_k}\}$, for $x\in \mX$, by
  \begin{equation*}
    C^x_{g_1,\, g_2,\, \ldots\,,\, g_k} = G^{k,\, x}_{g_1} \cdots
    G^{2,\, x}_{g_{k-1}} \, G^{1,\, x}_{g_k} \, G^{2,\, x}_{g_{k-1}} \cdots
    G^{k,\, x}_{g_1}\;.
  \end{equation*}
  Then on average over $x\sim \mu$ and $y\in \mY$ sampled uniformly at random and independently from $x$,
  \begin{equation}\label{eq:ld-sandwich-2}
    A^x_{[\eval_y(\cdot) = (a_1,\, a_2,\, \ldots,\, a_k)]} \abc[k(\delta+\eps)^{1/2}]
    C^x_{[\eval_y(\cdot) = (a_1,\, a_2,\, \ldots,\, a_k)]}\;.
\end{equation}
\end{lemma}

\begin{proof}
  The proof is identical to the one given in \cite[Fact~4.34]{NW19}, with the
  only modification needed to insert the dependence on $x$ for all measurements
  considered.
\end{proof}

The following lemma is almost identical to Fact 4.35 in~\cite{NW19}, except for the addition of the assumption~\eqref{eq:gonna-compare-like-a-pear-backwards}. The proof given in~\cite{NW19} contains an error, and we include a corrected proof here.

\begin{lemma}\label{lem:pasting-updated}
Let $k \geq 0$ be a constant.
Let $\mathcal{D}$ be a distribution on questions $(x, y_1, \ldots, y_k)$, where each $y_i \in \mathcal{Y}_i$.
For each $1 \leq i \leq k$, let $\mathcal{G}_i$ be a set of functions $g_i : \mathcal{Y}_i \rightarrow \mathcal{R}_i$.
and let $\{(G_i)^x_g\}$ be a projective measurement with outcomes from this set.
(For the $i=1$ case, we also allow this measurement to be a POVM.)
Suppose that the set $\mathcal{G}_i$ has the following distance property:
fix a question $z = (x, y_1, \ldots, y_{i-1}, y_{i+1}, \ldots, y_k)$, and let $\mathcal{D}_z$ be the distribution on~$y_i$ conditioned on the other outcomes~$z$.
Then for any two nonequal $g_i, g_i' \in \mathcal{G}_i$, the probability that $g_i(y_i) = g_i'(y_i)$, over a random $y_i \sim \mathcal{D}_z$, is at most $\eps$.

Let $\{A^{x, y_1, \ldots, y_k}_{a_1, \ldots, a_k} \}$ be a projective
measurement with outcomes $(a_1, \dots a_k) \in \mathcal{R}_1 \times
\dots \times \mathcal{R}_k$.
For each $1 \leq i \leq k$, suppose that
\begin{equation}\label{eq:gonna-compare-like-a-pear}
(A^{x, y_1, \ldots, y_k}_{a_i})_{\alice} \otimes I_{\bob} \simeq_{\delta} I_{\alice} \otimes ((G_i)^x_{[g_i(y_i)=a_i]})_{\bob}
\end{equation}
and
\begin{equation}\label{eq:gonna-compare-like-a-pear-backwards}
I_{\alice} \otimes (A^{x, y_1, \ldots, y_k}_{a_i})_{\bob}  \simeq_{\delta} ((G_i)^x_{[g_i(y_i)=a_i]})_{\alice} \otimes I_{\bob}.
\end{equation}
Suppose also that
\begin{equation}\label{eq:spend-some-time-reflecting}
(A^{x, y_1, \ldots, y_k}_{a_1, \ldots, a_k})_{\alice} \otimes I_{\bob}
	\simeq_{\delta} I_{\alice} \otimes (A^{x, y_1, \ldots, y_k}_{a_1, \ldots, a_k})_{\bob}.
\end{equation}
Define the POVM measurement $\{J^x_{g_1, \ldots, g_k}\}$ as
\begin{equation*}
J^x_{g_1, \ldots, g_k} := (G_k)^x_{g_k} \cdots (G_2)^x_{g_2} \cdot (G_1)^x_{g_1} \cdot (G_2)^x_{g_2} \cdots (G_k)^x_{g_k}.
\end{equation*}
Then there is a
\[ \delta_{pasting} = \delta_{pasting}(\eps, \delta) = \poly(\eps,
  \delta) \]
such that
\begin{equation*}
(A^{x, y_1, \ldots, y_k}_{a_1, \ldots, a_k})_{\alice} \otimes I_{\bob}
	\simeq_{\delta_{pasting}} I_{\alice} \otimes (J^x_{[g_1(y_1), \ldots, g_k(y_k) = a_1, \ldots, a_k]})_{\bob}.
\end{equation*}
\end{lemma}
\begin{proof}
First, we show how to reduce this to the $k = 2$ case. Then we prove it for that case.
Assume the fact holds when $k = 2$.  Define the POVM measurement $\{(J_i)^x_{g_1, \ldots, g_i}\}$ as
\begin{equation*}
(J_i)^x_{g_1, \ldots, g_i} := (G_i)^x_{g_i} \cdots (G_2)^x_{g_2} \cdot (G_1)^x_{g_1} \cdot (G_2)^x_{g_2} \cdots (G_i)^x_{g_i}.
\end{equation*}
We will show by induction that 
\begin{equation}\label{eq:hi-i-work-in-graphics-but-not-that-graphics}
(A^{x, y_1, \ldots, y_k}_{a_1, \ldots, a_i})_{\alice} \otimes I_{\bob}
	\simeq_{\poly(\delta, \eps)} I_{\alice} \otimes ((J_i)^x_{[g_1(y_1), \ldots, g_i(y_i) = a_1, \ldots, a_i]})_{\bob},
\end{equation}
the base case being trivial. Assume this holds for~$i$.
We apply the $k = 2$ case as follows:
consider the question tuple $(y_1, \ldots, y_i)$ as a single question
and consider functions of the form $(y_1, \ldots, y_i) \mapsto (g_1(y_1), \ldots, g_i(y_i))$.
Then the first POVM measurement is $J_i$, which satisfies \eqref{eq:gonna-compare-like-a-pear} due to \eqref{eq:hi-i-work-in-graphics-but-not-that-graphics}.
The second measurement is the projector $G_{i+1}$.
Then the $k = 2$ case immediately implies the $i+1$ case of \eqref{eq:hi-i-work-in-graphics-but-not-that-graphics}.

Now we prove the $k = 2$ case.
Our goal is to show that
\begin{equation}\label{eq:food-rakes}
\E_{x, y_1, y_2} \sum_{a_1, g_2} \bra{\psi} (A^{x, y_1, y_2}_{a_1, g_2(y_2)})_{\alice}
\otimes ((G_2)^{x}_{g_2} \cdot (G_1)^{x}_{[g_1(y_1) = a_1]} \cdot (G_2)^{x}_{g_2})_{\bob} \ket{\psi}
\end{equation}
is at least $1- \poly(\delta, \eps)$.
We will do this in two steps. First, we will show that the related
expression \begin{equation}
\E_{x, y_1, y_2} \sum_{a_1, g_2} \bra{\psi} (A^{x, y_1, y_2}_{a_1, g_2(y_2)})_{\alice}
\otimes ((G_2)^{x}_{g_2} \cdot (G_2)^{x}_{g_2} \cdot (G_1)^{x}_{[g_1(y_1) = a_1]})_{\bob} \ket{\psi} \label{eq:the-related-expression}
\end{equation}
satisfies the bound
\begin{equation}
  |\eqref{eq:the-related-expression} - 1|  \leq 2\sqrt{\delta}. \label{eq:the-related-exp-is-close-to-1}
 \end{equation}
 We will then show that
 \begin{equation}
   |\eqref{eq:the-related-expression} -  \eqref{eq:food-rakes}| \leq
   \poly(\delta,\eps) \label{eq:the-related-exp-is-actually-related}
 \end{equation}
 by showing that
\begin{equation}\label{eq:what-a-johnnie-wants}
I_{\alice} \otimes ((G_1)^x_{[g_1(y_1)=a_1]} \cdot (G_2)^x_{g_2})_{\bob}
\approx_{\poly(\delta, \eps)} I_{\alice} \otimes ((G_2)^x_{g_2} \cdot (G_1)^x_{[g_1(y_1)=a_1]})_{\bob}.
\end{equation}

To show \eqref{eq:the-related-exp-is-close-to-1}, note that because $G_2$ is a projector, we can rewrite \eqref{eq:the-related-expression} as
\begin{align}
    \eqref{eq:the-related-expression}&=\E_{x, y_1, y_2} \sum_{a_1, g_2} \bra{\psi} (A^{x, y_1, y_2}_{a_1, g_2(y_2)})_{\alice}
\otimes ((G_2)^{x}_{g_2} \cdot (G_1)^{x}_{[g_1(y_1) = a_1]})_{\bob} \ket{\psi}\nonumber\\
&= \E_{x, y_1, y_2} \sum_{a_1, g_2} \bra{\psi} (A^{x, y_1, y_2}_{a_1, g_2(y_2)})_{\alice}
\otimes ((G_2)^{x}_{g_2} \cdot I)_{\bob} \ket{\psi} \label{eq:label} \\
&\qquad - \E_{x, y_1, y_2} \sum_{a_1, g_2} \bra{\psi} (A^{x, y_1, y_2}_{a_1, g_2(y_2)})_{\alice}
\otimes \Big((G_2)^{x}_{g_2} \cdot (G_1)^{x}_{[g_1(y_1) \neq a_1]}\Big)_{\bob} \ket{\psi}, \label{eq:another-label}
\end{align}
where we write $(G_1)^{x}_{[g_1(y_1) \neq a_1]}$
as shorthand for $I - (G_1)^{x}_{[g_1(y_1) = a_1]}$.
Now, we bound the difference between this expression and $1$ as follows. For the first term, which appears in \eqref{eq:label}, it is equal to
\begin{equation*}
    \E_{x, y_1, y_2} \sum_{g_2} \bra{\psi} (A^{x, y_1, y_2}_{ g_2(y_2)})_{\alice}
\otimes ((G_2)^{x}_{g_2})_{\bob} \ket{\psi},
\end{equation*}
which is at least $1-\delta \geq 1 - \sqrt{\delta}$ due to \eqref{eq:gonna-compare-like-a-pear}.
For the second term, which appears in \eqref{eq:another-label}, let us first rewrite it as
\begin{align}
    \eqref{eq:another-label} &= \E_{x, y_1, y_2} \sum_{a_1, a_2} \bra{\psi} (A^{x, y_1, y_2}_{a_1, a_2})_{\alice}
\otimes \Big((G_2)^{x}_{[g_2(y_2) = a_2]} \cdot (G_1)^{x}_{[g_1(y_1) \neq a_1]})\Big)_{\bob} \ket{\psi} \label{eq:another-label-alt}
\end{align}
Cauchy-Schwarz allows us to bound its magnitude by
\begin{multline}
    |\eqref{eq:another-label-alt}| \leq 
   \sqrt{\E_{x, y_1, y_2} \sum_{a_1, a_2} \Vert(A^{x, y_1, y_2}_{a_1, a_2})_{\alice}
\otimes ((G_2)^{x}_{[g_2(y_2) = a_2]} )_{\bob} \ket{\psi} \Vert^2}\\
   \cdot \sqrt{\E_{x, y_1, y_2} \sum_{a_1, a_2} \Vert (A^{x, y_1, y_2}_{a_1, a_2})_{\alice}
\otimes  \big((G_1)^{x}_{[g_1(y_1) \neq a_1]}\big)_{\bob} \ket{\psi} \Vert^2}. \label{eq:second-square-root}
\end{multline}
The expression inside the first square root is always at most 1. For the second square root, we will expand out the squared norm and use the fact that for each $a_1$, $((G_1)^{x}_{[g_1(y_1) \neq a_1]})^2 \preceq (G_1)^{x}_{[g_1(y_1) \neq a_1]}$, to obtain
\begin{equation}
\eqref{eq:second-square-root} \leq \sqrt{\E_{x, y_1, y_2} \sum_{a_1} \bra{\psi} (A^{x, y_1, y_2}_{a_1})_{\alice}
\otimes  \big((G_1)^{x}_{[g_1(y_1) \neq a_1]}\big)_{\bob} \ket{\psi} } \leq \sqrt{\delta},
\end{equation}
where the final inequality is by \eqref{eq:gonna-compare-like-a-pear}.
This concludes the proof of \eqref{eq:the-related-exp-is-close-to-1}.

We now show \eqref{eq:the-related-exp-is-actually-related}, assuming
that \eqref{eq:what-a-johnnie-wants} is true. Write the left-hand side
of \eqref{eq:the-related-exp-is-actually-related} as 
\begin{multline*}
\Big| \E_{x, y_1, y_2} \sum_{a_1, g_2} \bra{\psi} (A^{x, y_1, y_2}_{a_1, g_2(y_2)})_{\alice}\\
\otimes ((G_2)^{x}_{g_2} \cdot ((G_1)^{x}_{[g_1(y_1) = a_1]} \cdot
(G_2)^{x}_{g_2}- (G_2)^{x}_{g_2} \cdot (G_1)^{x}_{[g_1(y_1) =
  a_1]}))_{\bob} \ket{\psi} \Big|. 
\end{multline*}
Cauchy-Schwarz allows us to bound this by
\begin{multline*}
\leq \E_{x, y_1, y_2} \sqrt{ \sum_{a_1, g_2}\Vert (A^{x, y_1, y_2}_{a_1, g_2(y_2)})_{\alice} \otimes ((G_2)^{x}_{g_2})_{\bob} \ket{\psi} \Vert^2}\\
	\cdot \sqrt{ \sum_{a_1, g_2}\Vert I_{\alice} \otimes ((G_1)^{x}_{[g_1(y_1) = a_1]} \cdot (G_2)^{x}_{g_2}- (G_2)^{x}_{g_2} \cdot (G_1)^{x}_{[g_1(y_1) = a_1]})_{\bob} \ket{\psi} \Vert^2}.
\end{multline*}
The expression inside the first square root is always at most~$1$.
This allows us to bring the expectation into the second square root by Jensen's inequality,
and the resulting expression we can bound due to \eqref{eq:what-a-johnnie-wants}.

Now, all that remains is to prove \eqref{eq:what-a-johnnie-wants}.
Showing this is equivalent to showing
\begin{equation*}
\E_{x, y_1, y_2}
\sum_{a_1, g_2}\Vert  I_{\alice} \otimes ((G_1)^{x}_{[g_1(y_1) = a_1]}
	\cdot (G_2)^{x}_{g_2}- (G_2)^{x}_{g_2} \cdot (G_1)^{x}_{[g_1(y_1) = a_1]})_{\bob}  \ket{\psi} \Vert^2
\end{equation*}
is small.  Expanding this, we get
\begin{align}
\E_{x, y_1, y_2} \sum_{a_1, g_2} \bra{\psi}\big(&I_{\alice} \otimes
	(G_2)^{x}_{g_2} \cdot (G_1)^{x}_{[g_1(y_1) = a_1]} \cdot (G_1)^{x}_{[g_1(y_1) = a_1]} \cdot (G_2)^{x}_{g_2}\nonumber\\
	+ &I_{\alice} \otimes(G_1)^{x}_{[g_1(y_1) = a_1]} \cdot (G_2)^{x}_{g_2} \cdot (G_2)^{x}_{g_2} \cdot (G_1)^{x}_{[g_1(y_1) = a_1]}\nonumber\\
	- &I_{\alice} \otimes(G_2)^{x}_{g_2} \cdot (G_1)^{x}_{[g_1(y_1) = a_1]} \cdot (G_2)^{x}_{g_2} \cdot (G_1)^{x}_{[g_1(y_1) = a_1]}\nonumber\\
	- &I_{\alice} \otimes(G_1)^{x}_{[g_1(y_1) = a_1]} \cdot (G_2)^{x}_{g_2} \cdot (G_1)^{x}_{[g_1(y_1) = a_1]} \cdot (G_2)^{x}_{g_2} \big)\ket{\psi}.\label{eq:burger-king}
\end{align}
We \emph{do} know that $G_1$ and $G_2$ satisfy some form of commutation.
Because they satisfy \eqref{eq:gonna-compare-like-a-pear} and~$A$ is a projector, 
we know that
\begin{equation*}
 I_{\alice}\otimes ((G_1)^x_{[g_1(y_1)=a_1]} \cdot (G_2)^x_{[g_2(y_2) = a_2]})_{\bob} 
\approx_{\delta}  I_{\alice}\otimes ((G_2)^x_{[g_2(y_2) = a_2]} \cdot (G_1)^x_{[g_1(y_1)=a_1]})_{\bob} .
\end{equation*}
Expanding this as above, we can bound the following expression by~$\delta$:
\begin{align}
&\E_{x, y_1, y_2} \sum_{a_1, a_2} \bra{\psi}\big(\nonumber\\
	&I_{\alice} \otimes (G_2)^x_{[g_2(y_2) = a_2]} \cdot (G_1)^{x}_{[g_1(y_1) = a_1]} \cdot (G_1)^{x}_{[g_1(y_1) = a_1]} \cdot (G_2)^x_{[g_2(y_2) = a_2]} \nonumber\\
	&+ I_{\alice} \otimes (G_1)^{x}_{[g_1(y_1) = a_1]} \cdot (G_2)^x_{[g_2(y_2) = a_2]} \cdot (G_2)^x_{[g_2(y_2) = a_2]} \cdot (G_1)^{x}_{[g_1(y_1) = a_1]} \nonumber\\
	&- I_{\alice} \otimes (G_2)^x_{[g_2(y_2) = a_2]} \cdot (G_1)^{x}_{[g_1(y_1) = a_1]} \cdot (G_2)^x_{[g_2(y_2) = a_2]} \cdot (G_1)^{x}_{[g_1(y_1) = a_1]} \nonumber\\
	&- I_{\alice} \otimes (G_1)^{x}_{[g_1(y_1) = a_1]} \cdot (G_2)^x_{[g_2(y_2) = a_2]} \cdot (G_1)^{x}_{[g_1(y_1) = a_1]} \cdot (G_2)^x_{[g_2(y_2) = a_2]} \big)\ket{\psi}.\label{eq:what-to-do-with-this}
\end{align}

We can, therefore, show \eqref{eq:burger-king} is small by
upper-bounding (\eqref{eq:burger-king}$ - $\eqref{eq:what-to-do-with-this}).
There are four terms in this difference;
write $\Delta_i$ for the $i$-th term in \eqref{eq:burger-king} minus the $i$-th term in \eqref{eq:what-to-do-with-this}.
We will bound each $\Delta_i$ one-by-one.

The first term in the difference, $\Delta_1$, is
\begin{multline*}
\E_{x, y_1, y_2} \sum_{a_1}\sum_{g_2} \bra{\psi}
	I_{\alice} \otimes(G_2)^{x}_{g_2}
	\\
\cdot (G_1)^{x}_{[g_1(y_1) = a_1]} \cdot (G_1)^{x}_{[g_1(y_1) = a_1]} \cdot (G_2)^{x}_{g_2} \ket{\psi}\\
-\E_{x, y_1, y_2} \sum_{a_1, a_2} \bra{\psi}
	I_{\alice} \otimes(G_2)^{x}_{[g_2(y_2) = a_2]}
	\cdot (G_1)^{x}_{[g_1(y_1) = a_1]} \cdot (G_1)^{x}_{[g_1(y_1) = a_1]}\\
\cdot (G_2)^{x}_{[g_2(y_2) = a_2]} \ket{\psi}.
\end{multline*}
The first of these terms is at most~$1$, and so we just have to show that the second term is close to~$1$ as well.
Note that by repeated applications of \eqref{eq:gonna-compare-like-a-pear}, we have that 
\begin{equation*}
I_{\alice} \otimes (G_2)^{x}_{[g_2(y_2) = a_2]}
	\cdot (G_1)^{x}_{[g_1(y_1) = a_1]} \cdot (G_1)^{x}_{[g_1(y_1) = a_1]} \cdot (G_2)^{x}_{[g_2(y_2) = a_2]} 
	\approx_{\delta} A^{x, y_1, y_2}_{a_1, a_2} \otimes  I_{\bob}.
\end{equation*}
But then by
Fact~4.31 from~\cite{NW19}, the expression we want to lower-bound is $O(\delta^{1/2})$-close to
\begin{equation*}
\E_{x, y_1, y_2} \sum_{a_1, a_2} \bra{\psi}  A^{x, y_1, y_2}_{a_1, a_2} \otimes I_{\bob}\ket{\psi},
\end{equation*}
which is exactly~$1$. As a result, $\Delta_1$ is at most $O(\delta^{1/2})$.

The second term in the difference, $\Delta_2$, can be written as
\begin{equation*}
-\E_{x, y_1, y_2} \sum_{a_1}\sum_{\substack{g_2\neq g_2',\\g_2(y_2) = g_2'(y_2)}} \bra{\psi}(
	I_{\alice}\otimes (G_1)^{x}_{[g_1(y_1) = a_1]} \cdot (G_2)^{x}_{g_2} \cdot (G_2)^{x}_{g_2'} \cdot (G_1)^{x}_{[g_1(y_1) = a_1]}  )\ket{\psi}.
\end{equation*}
This is zero because $G_2$ is a projector.

The third and fourth terms in \eqref{eq:burger-king} are complex conjugates of each other,
as are the third and fourth terms in \eqref{eq:what-to-do-with-this}.
As a result, it suffices to bound the magnitude of $\Delta_4$, and this will serve to bound $\Delta_3$ as well.
We begin by manipulating the fourth term in \eqref{eq:burger-king};
specifically, we will show that it is close to
\begin{equation}\label{eq:cloud}
-\E_{x, y_1, y_2} \sum_{a_1, g_2} \bra{\psi}
	 (G_2)^{x}_{g_2} 
     \otimes (G_1)^{x}_{[g_1(y_1) = a_1]} \cdot (G_2)^{x}_{g_2} \cdot (G_1)^{x}_{[g_1(y_1) = a_1]} \ket{\psi}.
\end{equation}
To do so, we take their difference:
\begin{align*}
&\E_{x, y_1, y_2} \sum_{a_1, g_2} \bra{\psi}
	 (I_{\alice} \otimes (G_1)^{x}_{[g_1(y_1) = a_1]} \cdot
                 (G_2)^{x}_{g_2} \cdot (G_1)^{x}_{[g_1(y_1) = a_1]}) \\
&\cdot (  (G_2)^{x}_{g_2} \otimes I_{\bob}- I_{\alice} \otimes (G_2)^{x}_{g_2}) \ket{\psi}.
\end{align*}
To bound the magnitude, we apply Cauchy-Schwarz:
\begin{multline*}
\E_{x, y_1, y_2}
	\sqrt{\sum_{a_1, g_2} \Vert  (I_{\alice} \otimes ((G_1)^{x}_{[g_1(y_1) = a_1]})^{1/2} \cdot (G_2)^{x}_{g_2} \cdot (G_1)^{x}_{[g_1(y_1) = a_1]}) \ket{\psi} \Vert^2}\\
	\cdot \sqrt{\sum_{a_1, g_2} \Vert (I_{\alice} \otimes ((G_1)^{x}_{[g_1(y_1) = a_1]})^{1/2})
	\cdot ((G_2)^{x}_{g_2} \otimes I_{\bob}- I_{\alice}\otimes (G_2)^{x}_{g_2}) \ket{\psi} \Vert^2}.
\end{multline*}
The expression inside the first square root is always at most~$1$.
This allows us to bring the expectation into the second square root by Jensen's inequality.
Because $G_1$ is a POVM, we can bound the resulting expectation by
\begin{equation}\label{eq:majora}
\E_{x, y_1, y_2} \sum_{g_2} \Vert ((G_2)^{x}_{g_2} \otimes I_{\bob}  - I_{\alice}  \otimes (G_2)^{x}_{g_2}) \ket{\psi} \Vert^2.
\end{equation}
To bound this, we note that \eqref{eq:gonna-compare-like-a-pear}, \eqref{eq:gonna-compare-like-a-pear-backwards}, and \eqref{eq:spend-some-time-reflecting} along with Fact~\ref{fact:triangle-for-simeq} imply that
\begin{equation*}
(G_2)^{x}_{[g_2(y_2) = a_2]} \otimes I_{\bob}
\simeq_{\delta} I_{\alice} \otimes (G_2)^{x}_{[g_2(y_2) = a_2]}.
\end{equation*}
Using the distance properties of $\mathcal{G}_2$, this implies that
\begin{equation*}
(G_2)^{x}_{g_2} \otimes I_{\bob}
\simeq_{\delta + \eps} I_{\alice} \otimes (G_2)^{x}_{g_2}.
\end{equation*}
Hence, \eqref{eq:majora} is at most $O((\delta+\eps)^{1/2})$.
A similar argument shows that the fourth term in \eqref{eq:what-to-do-with-this}
is $O(\delta^{1/2})$-close to
\begin{equation}\label{eq:strife}
-\E_{x, y_1, y_2} \sum_{a_1, a_2} \bra{\psi}(G_2)^x_{[g_2(y_2) = a_2]}  \otimes
(G_1)^{x}_{[g_1(y_1) = a_1]} \cdot (G_2)^x_{[g_2(y_2) = a_2]} \cdot (G_1)^{x}_{[g_1(y_1) = a_1]}\ket{\psi}.
\end{equation}

Now, we compute \eqref{eq:cloud} minus \eqref{eq:strife}:
\begin{align*}
&\E_{x, y_1, y_2} \sum_{a_1}\sum_{\substack{g_2\neq g_2'\\g_2(y_2) = g_2'(y_2)}} \bra{\psi}\\
& (G_2)^{x}_{g_2'} \otimes 
	 (G_1)^{x}_{[g_1(y_1) = a_1]} \cdot (G_2)^{x}_{g_2} \cdot (G_1)^{x}_{[g_1(y_1) = a_1]} \ket{\psi}\\
&=\E_{x, y_1, y_2} \sum_{a_1}\sum_{g_2, g_2'} \bra{\psi}\\
& (G_2)^{x}_{g_2'} \otimes
	 (G_1)^{x}_{[g_1(y_1) = a_1]} \cdot (G_2)^{x}_{g_2} \cdot (G_1)^{x}_{[g_1(y_1) = a_1]} \ket{\psi}
	 	\cdot 1(g_2, g_2', y_2),
\end{align*}
where $1(g_2, g_2', y_2)$ is the indicator that $g_2 \neq g_2'$ but $g_2(y_2) = g_2'(y_2)$.
This is the only part of the expression that depends on $y_2$, and by our distance assumption it is at most $\eps$ in expectation.
Since the rest of the expression is guaranteed to be positive, we can upper-bound this by
\begin{equation*}
\E_{x, y_1} \sum_{a_1}\sum_{g_2, g_2'} \bra{\psi}
	 (G_2)^{x}_{g_2'} \otimes (G_1)^{x}_{[g_1(y_1) = a_1]} \cdot (G_2)^{x}_{g_2} \cdot (G_1)^{x}_{[g_1(y_1) = a_1]} \ket{\psi} \cdot \eps.
\end{equation*}
But the remaining part of the expression is at most~$1$, and so in total we can upper-bound it by~$\eps$.
This completes the proof.
\end{proof}

\begin{lemma}[Fact~4.32 of \cite{NW19}]
  \label{lem:close-strategies-have-close-values}
  Let $\game$ be a nonlocal game, and let $\strategy, \strategy'$ be
  two strategies that are $\delta$-close (in the sense of
  Definition~\ref{def:strategy-distance}) for $\delta \in [0,1]$, and use the
  same state $\ket{\psi}$. If either $\strategy$ or $\strategy'$ is
  projective, then $|\val^*(\game,\strategy) - \val^*(\game,
  \strategy')| \leq O(\delta^{1/2})$.
\end{lemma}

\subsection{The class $\MIP^*$}

The complexity class $\MIP^*$ of \emph{multi-prover interactive proof
  systems with entangled provers} is the class of languages that can be
decided by a \emph{proof system} in which a polynomial-time classical
\emph{verifier} interacts with noncommunicating, computationally unbounded
 \emph{provers} who may share a \emph{finite-dimensional} entangled state. In general there may be polynomially many provers,
and the verifier may interact with them over polynomially many
rounds. In each round, the verifier generates a question for
each prover and receives a response from them. At the end of
the interaction the verifier 
decides to either accept or reject. A language $L$ is in $\MIP^*$ if for
every input $z \in L$, there exists a strategy that the provers can
use to convince the verifier to accept with high probability ($\geq 2/3$), and for every input $z \not\in L$, there is
\emph{no} strategy that the provers can use to convince the verifier
to accept with more than low probability ($\geq 1/3$). The former
property is called \emph{completeness} of the proof system, and the
latter \emph{soundness}. The
completeness and soundness probabilities $2/3$ and $1/3$ may be
amplified by sequentially repeating the protocol. 

For a formal definition of the class $\MIP^*$, see, e.g.,~\cite[Section
  6.1]{vidick2016quantum}. 
The main result of this paper is a lower bound, and for this purpose,
it suffices  to restrict our attention to proof
systems which involve only two provers, one round of interaction, and
completeness probability $1$. This class is often denoted $\MIP^*_{1,1/2}(2,1)$. A formal definition for this restricted setting follows. 

\begin{definition}\label{def:mipstar}
  A language $L$ is in $\MIP^*_{1,1/2}(2,1)$ if and only if there exist two randomized Turing machines
  $\sampler$ and $\decider$ with the following properties.
  \begin{enumerate}
    \item \textbf{Efficiency:} For every $z\in\{0,1\}^*$ there is
      a game $\game_z = (\cal{X}, \cal{Y}, \cal{A}, \cal{B}, \mu, D)$
      such that:
      \begin{enumerate}
      \item The Turing machine $\sampler$ given input $z$ runs in time $\poly(|z|)$ and returns a
      pair $(x,y) \in \cal{X} \times \cal{Y}$ such that the distribution of $(x,y)$, over the random choices of $\sampler$, is $\mu$.
    \item The Turing machine $\decider$ given as input $z$ and a tuple $(x,y,a,b)
      \in \cal{X} \times \cal{Y} \times \cal{A} \times \cal{B}$ runs in time $\poly(|z|)$ and returns $D(x,y,a,b)$.\footnote{Note that the running time of $\decider$ should be $\poly(|z|)$, even for long inputs $a,b$. This can be ensured by having $D$ return $0$ whenever $x,y,a,b$ are too long with respect to $|z|$.}
    \end{enumerate}
  \item \textbf{Completeness:} If $z \in L$, then $\val^*(\game_z) = 1$
  \item \textbf{Soundness:} If $z \not\in L$, then $\val^*(\game_z) \leq 1/2$.
  \end{enumerate}
  We say that the pair $(\sampler, \decider)$ form an $\MIP^*$
  \emph{verifier} for the language $L$, and the associated family of
  games $\game_z$ are an $\MIP^*$ protocol for $L$.
\end{definition}

  It is clear that $\MIP^*_{1,1/2}(2,1) \subseteq \MIP^*$ (the
  probability $1/2$ of acceptance of inputs not in the language may be
  reduced to $1/3$ by sequential repetition as noted above). The main result of this paper is that $\RE
  \subseteq \MIP^*_{1,1/2}(2,1)$. Since it is known that $\MIP*\subseteq \RE$ it follows
  that $\MIP^*_{1,1/2}(2,1) = \MIP^* = \RE$. 

 To show our lower bound it will suffice to consider an even more restricted class of $\MIP^*$
  verifiers, for which the Turing machines $\sampler$ and
  $\decider$ have a special structure which we refer to as
  \emph{normal form}. This is defined in the following subsection.
	
\subsection{Normal form verifiers}
\label{sec:normal-form}

We introduce a normal form for verifiers in nonlocal games.
The normal form uses Turing machines to specify the two actions performed by the
verifier in a game: the generation of questions and the verification of answers.
For the generation of questions, we use the formalism of samplers introduced in
Section~\ref{sec:cls}.
The normal form for verifiers gives a uniform method to specify an infinite
family of nonlocal games.

\begin{definition}[Decider]
  \label{def:decider}
  A \emph{decider}  $\decider$ is a $5$-input Turing machine. The input $5$-tuple
  of $\decider$ is interpreted as $(n,x,y,a,b)$ where $n$ is an
  integer and $x,y,a,b \in \{0,1\}^*$. When $\decider$ halts and outputs $1$ we say that it \emph{accepts};  if it halts and outputs $0$ (or enters the $\TIMEOUT$ state), we say that it \emph{rejects}. (If $\decider$ does not halt, then it neither accepts nor rejects.) 
  Furthermore, we call the input $n$ to a decider the \emph{index}.
\end{definition}

\hnote{added May 25, 2026:}  For a decider $\decider$ in standard timeout-counter form, let $B_\decider(n)$ denote its timeout bound as defined in Section~\ref{sec:tms}. Such a decider halts on every input of index $n$ in time $\poly(B_\decider(n))$, with $\TIMEOUT$ treated as rejection.


\begin{definition}
  \label{def:normal-ver}
  A \emph{normal form verifier} is a pair $\verifier = (\sampler, \decider)$
  where $\sampler$ is a sampler with field size $q(n) = 2$ and $\decider$ is a
  decider.


  The description length of $\verifier$ is defined to be $\abs{\verifier} =
  \max \{ \abs{\sampler} ,  \abs{\decider} \}$, the maximum of the description lengths of
  $\sampler$ and $\decider$.
  The number of levels of verifier $\verifier$ is defined to be the number of
  levels of its sampler $\sampler$.
\end{definition}


Next we introduce a notion of \emph{bounded normal form
  verifiers}, in which a single parameter (in this case, an integer $\lambda$)
specifies a bound on the time complexity of $\verifier$, as well as on the
description length of the verifier.

\begin{definition}[$\lambda$-bounded verifiers]
  \label{def:lambda}
  Let $\lambda \in \N$ be an integer.
  A normal form verifier $\verifier = (\sampler, \decider)$ is
  \emph{$\lambda$-bounded} if the following two conditions hold
  \begin{enumerate}
  \item The time complexity bounds of the verifier $\TIME_\sampler(n)$,
    $\TIME_\decider(n)$ are at most $n^\lambda$ for $n \ge 2$.\footnote{We do
      not require the bound to hold for $n=1$ as $n^\lambda$ is always $1$ and
      hence it is usually not satisfied.
    }
  \item The description length of the verifier $\abs{\verifier}$ is bounded by $\lambda$.
  \end{enumerate}
\end{definition}

Normal form verifiers specify an infinite family of nonlocal games indexed by
natural numbers in the following way.

\begin{definition}
  \label{def:normal-game}
    Let $\verifier = (\sampler, \decider)$ be a normal form verifier such that
  $\TIME_\decider(n)<\infty$ for all $n\in\N$.
  For $n \in \N$, we define the following nonlocal game $\verifier_{n}$ to be
  the \emph{$n$-th game corresponding to the verifier $\verifier$}.
  The question sets $\cal{X}$ and $\cal{Y}$ are $\{0,1\}^{s(n)}$, with $s(n)$ the
  dimension of the sampler.
  The answer sets $\cal{A}$ and $\cal{B}$ are the sets of binary strings of
  length at most $\TIME_\decider(n)$.
  The question distribution is the distribution $\mu_{\sampler,\, n}$ specified
  in Definition~\ref{def:sampler-sample}.
  The decision predicate is the total function $D_n(x,y,a,b)$ that is $1$ if and only if $\decider(n,x,y,a,b)$ halts and accepts. Otherwise, $D_n(x,y,a,b) = 0$. 
  The value of the game is denoted by $\val^*(\verifier_{n})$.
\end{definition}

We note that the game $\verifier_n$ is well-defined since for a normal form
verifier the distribution $\mu_{\sampler,\, n}$ is supported on
$\{0,1\}^{s(n)}\times \{0,1\}^{s(n)}$ (with the obvious identification between $\{0,1\}$ and $\mathbb{F}_2$).

\begin{definition}[Verifier with commuting strategy]
  \label{def:comm-with-checker}
  Let $\verifier = (\sampler, \decider)$ be a normal form verifier.
  For $v:\N\to [0,1]$ say that $\verifier$ has a \emph{value-$v$ commuting
    strategy} if for all $n \in \N$, the game $\verifier_n$ has a value-$v(n)$
  commuting strategy.
\end{definition}


\section{Types}
\label{sec:types}

We augment the definition of conditionally linear functions with a construct
we call \emph{types}.
A type~$\tvar$ is an element of a \emph{type set} $\type$, and a $\type$-typed
family of conditionally linear functions is a collection
$\{L_\tvar\}_{\tvar\in\type}$ containing a CL function $L_\tvar$ for each
type~$\tvar\in\type$.
The utility of this definition is that it allows us to define another object,
namely conditionally linear distributions parameterized by an undirected graph
$G = (\type, E)$ on the set of types known as a \emph{type graph}.
Given two $\type$-typed families of conditionally linear functions
$\{L_\lvar\}_{\lvar\in\type}, \{R_\rvar\}_{\rvar\in\type}$, the $(\type,
G)$-typed conditionally linear distribution corresponding to them is the
distribution which samples a pair of types $(\lvar, \rvar)$ uniformly at random
from the edges of~$G$ (with each endpoint having equal probability as being
chosen for $\lvar$ or $\rvar$, respectively) and then samples $(x, y)$ from
$\mu_{L_\lvar,\, R_\rvar}$.
The output is the pair $( (\lvar, x), (\rvar, y) )$.

The normal form verifiers we present in the paper frequently use typed CL
distributions to sample their questions, rather than untyped CL distributions.
Types allow us to model the parts of their question distributions which are
unstructured and unsuitable for being sampled from CL distributions.
A common use of types is to allow the verifier to use previously defined games
as subroutines.
Here, the type helps indicate which subroutine the verifier selects, and an edge
in the type graph between two different types allows us to introduce a test that
cross-checks the results of one subroutine with the results of another.

Finally, we show how to convert any typed CL distribution into an equivalent (in
the precise sense defined below) untyped CL distribution with two additional
levels, a technique we call \emph{detyping}.
This entails showing how to ``simulate'' the \emph{graph distribution} of $G =
(\type, E)$, i.e.,\ the uniform distribution on its edges, using an untyped CL
distribution.
The simulation we give is based on rejection sampling and is only approximate:
its quality degrades exponentially with the number of types in~$\type$.
As a result, we will ensure throughout the paper that all type sets we consider
are of a small, in fact generally constant, size.

This section is organized as follows.
In Section~\ref{sec:typed-samplers} we define typed variants of CL distributions,
samplers, deciders, and verifiers.
In Section~\ref{sec:graph-dist} we define a CL distribution which samples from the
graph distribution of a given graph $G = (\type, E)$.
In Section~\ref{sec:detype} we define a canonical way to detype typed samplers,
deciders, and verifiers using the graph sampler from Section~\ref{sec:graph-dist}.
We then prove the main result of the section, Lemma~\ref{lem:detyping-verifiers},
which relates the value of the detyped normal form verifier to the value of
the original typed verifier.

\subsection{Typed samplers, deciders, and verifiers}
\label{sec:typed-samplers}

\begin{definition}[Typed conditionally linear functions]
  Let $\type$ be a finite set and $V$ be $\F^n$ for some integer $n \geq 0$.
  A \emph{$\type$-typed family of $\ell$-level conditionally linear functions}
  (implicitly, \emph{on $V$}) is a collection $\{L_\tvar\}_{\tvar \in \type}$
  such that, for each $\tvar \in \type$, $L_\tvar$ is an $\ell$-level
  conditionally linear function on~$V$.
\end{definition}

\begin{definition}[Graph distribution]
  \label{def:graph-distribution}
  Let $G = (U, E)$ be an undirected graph with vertex set $U$ and edge set~$E$.
  Edges in~$E$ are written as multisets $\{u, v\}$ of two vertices; the case $u
  = v$ represents a self-loop.
  Suppose there are $m$ edges, $k$ of which are self-loops.
  Then the \emph{graph distribution $\mu_G$ of~$G$} is the distribution over
  $U\times U$ such that for every $(u, v)\in U\times U$,
  \begin{equation*}
    \mu_G(u, v) =
    \begin{cases}
      1/(2m-k) & \text{if $\{u, v\} \in E$},\\
      0 & \text{otherwise}.
    \end{cases}
  \end{equation*}
  This is identical to the uniform distribution over pairs $(u, v)\in U\times U$
  such that $\{u, v\} \in E$.
\end{definition}

\begin{definition}[Typed conditionally linear distributions]
  Let $\type$ be a type set and
  $L=\{L_\lvar\}_{\lvar\in\type},R=\{R_\rvar\}_{\rvar\in\type}$ be $\type$-typed
  families of conditionally linear functions on~$V$.
  Let $G = (\type, E)$ be a graph with vertex set $\type$.
  The \emph{$(\type, G)$-typed conditionally linear distribution $\mu_{L,\,
      R}^G$ corresponding to~$(L, R)$} is the distribution over pairs $( (\lvar,
  x), (\rvar, y) )$, where $(\lvar, \rvar)$ is drawn from $\mu_G$ and $(x,y)$ is
  drawn from $\mu_{L_\lvar,\, R_\rvar}$.
\end{definition}

\begin{definition}[Typed conditionally linear samplers]
  \label{def:typed-sampler}
	Let $q:\N \to \N$ be an admissible field size function and $s: \N \to \N$ be a
  function.
  Let $\type$ be a finite type set and let $G = (\type, E)$ be a graph with vertex set $\type$.
  A $7$-input Turing machine $\sampler$ is a \emph{$(\type, G)$-typed, $\ell$-level
    conditionally linear sampler with field size $q(n)$ and dimension $s(n)$} if
  for all $n \in \N$, letting $q = q(n)$ and $s = s(n)$, there exist
  $\type$-typed families of $\ell$-level conditionally linear functions
  $\{L^{\alice,\, n}_{\tvar}\}_{\tvar \in \type}$ and $\{L^{\bob,\,
    n}_{\tvar}\}_{\tvar \in \type}$ on $V = \F_q^s$ where $\tvar \in \type, w\in
  \AB$, the conditionally linear function $L^{w,\, n}_{\tvar}$ has marginal
  functions $\{L^{w,\, n}_{\tvar,\, \leq j}\}$ and factor spaces $\{V^{w,\,
    n}_{\tvar,\, j,\, u}\}$ satisfying the conditions of Lemma~\ref{lem:cl-kth},
  and for all $\tvar \in \type$, $w \in \ab$, $j \in \{1, \ldots, \ell\}$, and
  $z, u, y \in V$:
  \begin{itemize}
  \item On input $(n, \gamestyle{dimension})$, the sampler $\sampler$ returns
    the dimension $s(n)$.

  \item On input $(n,\gamestyle{field})$, output $q(n)$.
	
	\item On input $(n, w, \gamestyle{marginal}, j, z, \tvar)$, the sampler
    $\sampler$ returns the binary representation of $L^{w,\, n}_{\tvar,\, \le
      j}(z)$.
			
			\item On input $(n, w, \gamestyle{linear}, j, u, y, \tvar)$, the sampler $\sampler$
    outputs the binary representation of $L^{w,\, n}_{\tvar,\, j,\, u}(y)$,

	\item On input $(n, w, \gamestyle{factor}, j, u, \tvar)$, the sampler
    $\sampler$ returns the factor space $V^{w,\, n}_{\tvar,\, j,\, u}$ of
    $L^{w,\, n}_{\tvar}$, represented as an indicator vector in $\{0,1\}^s$.
  \end{itemize}
  Furthermore, the sampler is required to halt on every input of the form
  $(n,\ldots)$. On every such input, it first computes $s(n)$ and $q(n)$ and then performs
  the format and length checks prescribed by the query type before parsing any
  auxiliary input as an element of $\F_{q(n)}^{s(n)}$, an index, a player label,
  or a type. If any check fails, the sampler halts with a fixed default output.

  We call $\F_{q(n)}^{s(n)}$ the \emph{ambient space of $\sampler$}.
  We call $\{ L^{\alice,\, n}_{\tvar} \}, \{ L^{\bob,\, n}_{\tvar} \}$ the
  \emph{CL functions of $\sampler$ on index $n$}.
 The \emph{time complexity} of $\sampler$, denoted $\TIME_\sampler(n)$, is
defined as in Section~\ref{sec:tms}: it is the supremum of the running time of
$\sampler$ over all inputs of the form $(n,\ldots)$.  In particular,
malformed/excessively-long inputs are included in this supremum.
\end{definition}

We assume that types $\tvar \in \type$ are represented using binary strings of
length at most $\lceil \log |\type| \rceil$; if a type $\tvar$ is given as input
to the sampler $\sampler$ and is not an element of $\type$, then the sampler
returns $0$.
Furthermore, as described in Remark~\ref{rmk:sampler-inputs} for un-typed
samplers, we write typed samplers with different numbers of arguments depending
on the input.
We note that Definition~\ref{def:typed-sampler} includes the graph $G$ as a parameter, even though it is not explicitly used anywhere;
this is so we can define the following concept, i.e.,\ the \emph{distribution} of a typed sampler, which \emph{does} depend on $G$.

\begin{definition}[Distribution of a typed sampler]
  \label{def:typed-sampler-sample}
  Let $\sampler$ be a $(\type, G)$-typed sampler.
  Let $L^w = \{ L^{w}_{\tvar} \}_\tvar$ for $w\in \{\alice, \bob\}$ be the CL
  functions of $\sampler$ on index~$n$.
  The \emph{distribution of sampler $\sampler$ on index $n$},
  denoted $\mu^G_{\sampler,\, n}$, is the $(\type, G)$-typed conditionally
  linear distribution corresponding to $(L^\alice, L^\bob)$.
\end{definition}

\begin{definition}[Downsizing typed CL samplers]
  \label{def:downsize-typed-sampler}
  Let $\sampler$ be a typed sampler.
  The typed downsized sampler $\downsize(\sampler)$ is defined as in
  Definition~\ref{def:downsize_sampler} with the only difference that the type
  $\tvar$ is included as part of the input to the sampler, as in
  Definition~\ref{def:typed-sampler}.
  In particular, $\downsize(\sampler)$ performs the appropriate format and length
checks from Definition~\ref{def:downsize_sampler} for the requested query type.
It also checks that the type input has length at most
$\lceil \log |\type| \rceil$ and encodes an element of $\type$.  If any of these
checks fails, it halts with the same fixed default output.
\end{definition}

\begin{lemma}
  \label{lem:downsize_typed_sampler}
  Let $\sampler$ be a $(\type, G)$-typed $\ell$-level CL sampler, for some finite set
  $\type$, graph $G$,  and integer $\ell\geq 0$.
  Let $q(n)$ and $s(n)$ be as in Definition~\ref{def:typed-sampler}.
  Then $\downsize(\sampler)$ defined in
Definition~\ref{def:downsize-typed-sampler} is a $(\type, G)$-typed
$\ell$-level CL sampler with field size $2$, dimension $s(n)\log q(n)$, and
time complexity
\[
  \TIME_{\downsize(\sampler)}(n)
  =
  \poly(\TIME_\sampler(n),|\type|).
\]
Moreover, the description of $\downsize(\sampler)$ can be computed from the
description of $\sampler$ in time $\poly(|\sampler|,|\type|)$, and has length
$\poly(|\sampler|,|\type|)$.
  Furthermore, for every integer $n \geq 1$, the CL functions of
  $\downsize(\sampler)$ on index $n$ are
  $\{(L_\tvar^{w,n})^\downsize\}_{w\in\{\alice,\bob\}, \tvar\in\type}$, as
  defined in Definition~\ref{def:cl-downsize}.
\end{lemma}

\begin{proof}
  The proof is the typed analogue of Lemma~\ref{lem:downsize_sampler}.  The only
  additional input is the type string.  The downsized sampler performs the appropriate
  format and length checks from Definition~\ref{def:downsize-typed-sampler};
  in particular, it checks that the type string has length at most
  $\lceil \log |\type| \rceil$ and encodes an element of $\type$.  If any of
  these checks fails, it halts with the fixed default output.  After these
  checks, the extra work for handling the type input is polynomial in $|\type|$,
  and the remaining parsing, conversion, and padding work is the same as in
  Lemma~\ref{lem:downsize_sampler}.  
\end{proof}

Typed deciders are the typed analogue of the deciders from
Definition~\ref{def:decider}; the only difference is that the two question types
are given explicitly as inputs.

\begin{definition}[Typed decider]
  \label{def:typed-decider}
A \emph{typed decider} is a $7$-input Turing machine $\decider$.  The input
  $7$-tuple of $\decider$ is interpreted as
  $(n,\lvar,x,\rvar,y,a,b)$, where $n$ is an integer and
  $\lvar,x,\rvar,y,a,b\in\{0,1\}^*$.
  When $\decider$ halts and outputs $1$ we say that it \emph{accepts}; 
  if it halts and outputs $0$ (or enters the $\TIMEOUT$ state), we say that it \emph{rejects}. (If $\decider$ does not halt, then it neither accepts nor rejects.)
\end{definition}

\begin{definition}
  \label{def:typed-normal-ver}
  Let $\type$ be a finite set and let $G = (\type, E)$ be a graph.
  A $(\type, G)$-\emph{typed normal form verifier} is a pair $\verifier =
  (\sampler, \decider)$ where $\sampler$ is a $(\type,G)$-typed sampler with field
  size $q(n) = 2$ and $\decider$ is a typed decider.

\end{definition}

\begin{definition}
  \label{def:typed-normal-game}
 Let $\verifier = (\sampler, \decider)$ be a $(\type, G)$-typed normal form
  verifier such that $\TIME_\decider(n)<\infty$ for all $n\in\N$.
  For $n \in \N$, we define the following nonlocal game $\verifier_{n}$ to be
  the \emph{$n$-th game corresponding to the verifier $\verifier$}.
  The question sets $\cal{X}$ and $\cal{Y}$ are
  $\type \times \{0,1\}^{s(n)}$, where $s(n)$ is the dimension of $\sampler$.
  The answer sets $\cal{A}$ and $\cal{B}$ are the sets of binary strings of
  length at most $\TIME_\decider(n)$.
  The question distribution is the distribution $\mu^G_{\sampler,\, n}$
  specified in Definition~\ref{def:typed-sampler-sample}.
  The decision predicate is the total function that, on input
  $((\lvar,x),(\rvar,y),a,b)\in
  \cal{X}\times\cal{Y}\times\cal{A}\times\cal{B}$, runs
  $\decider(n,\lvar,x,\rvar,y,a,b)$ and returns $1$ if $\decider$ accepts, and returns $0$ otherwise. 
  The value of the game is denoted by $\val^*(\verifier_{n})$.
\end{definition}

For $w\in\{\alice,\bob\}$ and a question $(\lvar, x)$ to player $w$ we refer to
$\lvar$ as the \emph{question type} and $x$ as the \emph{question content}.

\subsection{Graph distributions}
\label{sec:graph-dist}

We describe a construction of conditionally linear distributions which sample
from the graph distribution (see Definition~\ref{def:graph-distribution}) of a
graph $G = (U, E)$.
We begin with a technical definition, followed by the definition of the
conditionally linear distribution.

\begin{definition}[Neighbor indicator]\label{def:type-neighbor}
  Given a graph $G = (U, E)$, the \emph{neighbor indicator} of a vertex $u \in
  U$ is the vector $\mathrm{neigh}_G(u) \in \F_2^U$ in which, for all $v \in U$,
  \begin{equation*}
    \mathrm{neigh}_G(u)_v =
    \begin{cases}
      1 & \text{ if } \{u, v\} \in E,\\
      0 & \text{ otherwise}.
    \end{cases}
  \end{equation*}
  In addition, the \emph{$\F_2$-encoding} of a vertex $u \in U$ is the vector
  $\mathrm{enc}_G(u) \in \F_2^U \times \F_2^U$ given by $\mathrm{enc}_G(u) =
  (e_u, \mathrm{neigh}_G(u))$, where $e_u$ is the standard basis vector with
  a~$1$ in the $u$-th position and $0$'s everywhere else.
\end{definition}

\begin{definition}[Graph sampler]\label{def:graph-sampler}
  Let $G = (U, E)$ be a graph with $n$ vertices.
  Then the \emph{conditionally linear functions corresponding to~$G$} are the
  pair of functions $L_G^\alice, L_G^\bob$ on linear space $V_G$ specified in
  Fig.~\ref{fig:graph-distribution} where $V_G = V_{\vertex{\alice}} \oplus
  V_{\edge{\alice}} \oplus V_{\vertex{\bob}} \oplus V_{\edge{\bob}}$.
\end{definition}

\begin{figure}[!htb]
  \begin{gamespec}
    \centering
    \newcolumntype{C}{>{\centering\arraybackslash $}X<{$}}
    \begin{tabularx}{\textwidth}{*{4}C}
      \multicolumn{4}{c}{\textbf{Subspaces}}\\
      \midrule
      V_{\vertex{\alice}} & V_{\edge{\alice}} & V_{\vertex{\bob}} &
       V_{\edge{\bob}} \\
      \midrule
      \F_2^U & \F_2^U & \F_2^U & \F_2^U\\
      \midrule\\
    \end{tabularx}

    \begin{tabularx}{\textwidth}{p{3.8cm}X}
      \multicolumn{2}{c}{\textbf{Conditionally linear function $L_G^\alice$}}\\
      \midrule
      1st factor subspace & $V_{\vertex{\alice}} \oplus V_{\edge{\alice}}$\\
      1st linear function & Identity function\\
      \midrule
      2nd factor subspace & $V_{\vertex{\bob}} \oplus V_{\edge{\bob}}$\\
      2nd linear functions & For all $x \in V_{\vertex{\alice}} \oplus
      V_{\edge{\alice}}$, suppose there exists a $u \in U$ such that $x =
      \mathrm{enc}_G(u)$.
      Then for all $y \in V_{\vertex{\bob}} \oplus V_{\edge{\bob}}$,
      $L^\alice_{G,\, 2,\, x}$ zeroes out all entries of~$y$ except for
      $(y^{V_{\edge{\bob}}})_u$.
      Otherwise, $L^\alice_{G,\, 2,\, x} = 0$.\\
      \midrule\\
    \end{tabularx}

    \begin{tabularx}{\textwidth}{p{3.8cm}X}
      \multicolumn{2}{c}{\textbf{Conditionally linear function $L^\bob_G$}}\\
      \midrule
      1st factor subspace & $V_{\vertex{\bob}} \oplus V_{\edge{\bob}}$\\
      1st linear function & Identity function\\
      \midrule
      2nd factor subspace & $V_{\vertex{\alice}} \oplus V_{\edge{\alice}}$\\
      2nd linear functions & Similarly defined as those for $L^\alice_G$ by swapping
      $V_{\vertex{\alice}}$ and $V_{\edge{\alice}}$ with $V_{\vertex{\bob}}$ and
      $V_{\edge{\bob}}$, respectively.\\
      \midrule
    \end{tabularx}
  \end{gamespec}
  \caption{Specification of the conditionally linear functions corresponding
    to~$G$.}
  \label{fig:graph-distribution}
\end{figure}

These conditionally linear functions do not simulate the graph distribution
in the sense of sampling directly from it.
The following proposition, however, does show a sense in which these functions
simulate the graph distribution, namely via rejection sampling.

\begin{proposition}[Simulating the graph distribution]
  \label{prop:simulating-graph}
  Let $G = (U, E)$ be a graph with $n$ vertices and $m$ edges, $k$ of which
  are self-loops.
  Let $L^\alice_G$, $L^\bob_G$ be the conditionally linear functions
  corresponding to $G$ (see Fig.~\ref{fig:graph-distribution} for the
  definition and associated notation).
  Let $(x, y) \sim \mu_{L^\alice_G,\, L^\bob_G}$.
  Consider the event $\mathcal{E}_G$ that there exists $u, v \in U$ such that
  the following two statements are true.
  \begin{enumerate}
  \item $x^{V_{\vertex{\alice}} \oplus V_{\edge{\alice}}} = \mathrm{enc}_G(u)$
   	and $y^{V_{\vertex{\bob}} \oplus V_{\edge{\bob}}} = \mathrm{enc}_G(v)$,
  \item  $(x^{V_{\edge{\bob}}})_u = (y^{V_{\edge{\alice}}})_v = 1$.
  \end{enumerate}
  Then
  \begin{enumerate}
  \item \label{item:graph-prob} $\Pr_{x, y} (\mathcal{E}_G) = (2m-k)/16^n$.
  \item \label{item:edge-dist} Conditioned on $\mathcal{E}_G$, $(u, v)$ are
    distributed as the graph distribution of~$G$ (see
    Definition~\ref{def:graph-distribution}).
  \end{enumerate}
  Note that $\mathcal{E}_G$ occurs if and only if both $x^{V_\edge{\bob}}$ and
  $y^{V_\edge{\alice}}$ are nonzero.
  In particular, if $x$ and $y$ are sampled from $\mu_{L^\alice_G,\, L^\bob_G}$
  and given to the respective players, then at least one of them knows when the
  event $\mathcal{E}_G$ does \emph{not} occur.
\end{proposition}

\begin{proof}
  Let $z$ be drawn uniformly at random from $V_{\vertex{\alice}}\oplus
  V_{\edge{\alice}}\oplus V_{\vertex{\bob}} \oplus V_{\edge{\bob}}$, and let $x
  = L^\alice_G(z)$ and $y = L^\bob_G(z)$.
  Then with probability $n^2/16^n$, there exist $u, v \in U$ such that
  \begin{equation*}
    \text{$x^{V_{\vertex{\alice}} \oplus V_{\edge{\alice}}} = \mathrm{enc}_G(u)$
      and $y^{V_{\vertex{\bob}} \oplus V_{\edge{\bob}}} = \mathrm{enc}_G(v)$}\;.
  \end{equation*}
  Conditioned on this occurring, $u$ and~$v$ are distributed as independent,
  uniformly random vertices in~$U$.
  If we further condition on $\{u, v\} \in E$, which occurs with probability
  $(2m-k)/n^2$, then by definition, $(u, v)$ is distributed as the graph
  distribution of~$G$.
  But this event is exactly the event that $\mathcal{E}_G$ holds on~$(x, y)$,
  establishing the proposition.
\end{proof}

\subsection{Detyping typed verifiers}
\label{sec:detype}  

We give a canonical method for taking a typed normal form verifier and producing
an untyped normal form verifier which simulates it.
Throughout this section, $\type$ denotes a finite set, $G = (\type, E)$ denotes
a graph, and $L_G^\alice, L_G^\bob$ denote the conditionally linear functions
corresponding to~$G$ acting on the vector space $V_G = V_{\vertex{\alice}}
\oplus V_{\edge{\alice}} \oplus V_{\vertex{\bob}} \oplus V_{\edge{\bob}}$ of
dimension $4\cdot |\type|$ over $\F_2$, as in
Definition~\ref{def:graph-sampler}.

\begin{definition}[Detyped CL functions]
  \label{def:detyped-CL}
  Let $L^\alice = \{L^{\alice}_{\tvar}\}, L^\bob = \{L^{\bob}_{\tvar}\}$ be
  $\type$-typed families of $\ell$-level conditionally linear functions on $V$.
  We define the \emph{detyped CL functions corresponding to $(L^\alice, L^\bob)$
    on~$G$} to be the pair of $(\ell+2)$-level CL functions $(R^\alice, R^\bob)
  = \detype_G(L^\alice, L^\bob)$ on linear space $V_{\gamestyle{detype}} = V_G
  \oplus V$ as follows.
  For $w \in \{\alice, \bob\}$, and $z \in L^w_G (V_G)$, define the family of
  $\ell$-level CL functions $\{L^w_z\}$ on $V$ as
  \begin{equation*}
    L^w_z =
    \begin{cases}
      0 & \text{ if } z^{V_{\edge{}\overline{w}}} = 0,\\
      L^w_\tvar & \text { otherwise, for } z^{V_{\vertex{}w}} = e_\tvar.
    \end{cases}
  \end{equation*}
  We note that when $z^{V_{\edge{}\overline{w}}}$ is nonzero, it is always the
  case that $z^{V_{\vertex{}w}} = e_\tvar$ for a unique type $\tvar$, by
  Definition~\ref{def:graph-sampler}.
  For $w\in \{\alice, \bob\}$, $R^w$ is the concatenation of $L^w_G$ and
  $\{L^w_z\}_z$ (cf.\ Lemma~\ref{lem:cl-concat}).
\end{definition}

\hnote{added: May 26, 2026} Before defining detyped samplers we fix one convention.  For
$w\in\{\alice,\bob\}$ and $g\in V_G$, say that $g$ \emph{selects} type $\tvar$
for player $w$ if
\[
  g^{V_{\edge{}\overline{w}}}\neq 0
  \qquad\text{and}\qquad
  g^{V_{\vertex{}w}}=e_\tvar.
\]
If no such $\tvar$ exists, then $g$ selects no type.  By
Definition~\ref{def:graph-sampler}, if a type is selected then it is unique.

\begin{definition}[Detyped samplers]
   \label{def:detyped-sampler}
 Let $\sampler$ be a $(\type,G)$-typed $\ell$-level CL sampler with field size
  $2$ and dimension $s(n)$.  We define $\detype(\sampler)$ to be the following
  (standard) sampler.

  On input with index $n$, the machine first calls
  $\sampler(n,\gamestyle{dimension})$ to compute $s(n)$.  Its ambient space on
  index $n$ is
  \[
    V_{\gamestyle{detype}} = V_G \oplus \F_2^{s(n)}.
  \]
  We write elements of this space as pairs $(g,v)$, where $g\in V_G$ and
  $v\in\F_2^{s(n)}$. Before parsing any input component as a vector in this space, it checks that
  the component has the prescribed length.  If any format or length check fails,
  it halts with a fixed default output.

  The machine is specified as follows.
  \begin{itemize}
  \item On input $(n,\gamestyle{dimension})$, output $4|\type|+s(n)$.

  \item On input $(n,\gamestyle{field})$, output $2$.
    
  \item On input $(n,w,\gamestyle{marginal},j,z)$, parse
    $z=(z_G,z_V)\in V_G\oplus \F_2^{s(n)}$.
    If $j\leq 2$, let $g$ be the $j$-th marginal output of the graph sampler
    $L_G^w$ on input $z_G$, and output $(g,0^{s(n)})$.

    If $j=2+r$ with $1\leq r\leq \ell$, compute $g=L_G^w(z_G)$.  If $g$
    selects type $\tvar$ for player $w$, let $v$ be the output of $\sampler(n,w,\gamestyle{marginal},r,z_V,\tvar)$, and output $(g,v)$.  If $g$ selects no type,
    output $(g,0^{s(n)})$.

  \item On input $(n,w,\gamestyle{linear},j,u,y)$, parse
    $u=(u_G,u_V)$ and $y=(y_G,y_V)$ in
    $V_G\oplus \F_2^{s(n)}$.
    If $j\leq 2$, let $g$ be the $j$-th linear-map output of the graph sampler
    $L_G^w$ on the $V_G$ components, and output
    $(g,0^{s(n)})$.

    If $j=2+r$ with $1\leq r\leq \ell$, use $u_G$ to determine whether a type is
    selected.  If $u_G$ selects type $\tvar$, let $v$ be the output of $\sampler(n,w,\gamestyle{linear},r,u_V,y_V,\tvar)$, 
    and output $(0^{4|\type|},v)$.  If $u_G$
    selects no type, output $(0^{4|\type|},0^{s(n)})$.

  \item On input $(n,w,\gamestyle{factor},j,u)$, parse
    $u=(u_G,u_V)\in V_G\oplus \F_2^{s(n)}$.
    If $j\leq 2$, let $g$ be the $j$-th factor-space indicator of the graph
    sampler $L_G^w$, and output of $(g,0^{s(n)})$.

    If $j=2+r$ with $1\leq r\leq \ell$, use $u_G$ to determine whether a type is
    selected.  If $u_G$ selects type $\tvar$, let $v$ be the output of $\sampler(n,w,\gamestyle{factor},r,u_V,\tvar)$, 
    and output $(0^{4|\type|},v)$.  If $u_G$
    selects no type and $r=1$, output
    $(0^{4|\type|},1^{s(n)})$.  If $u_G$ selects no type and $r>1$, output $(0^{4|\type|},0^{s(n)})$.
  \end{itemize}

\end{definition}


The preceding definition specifies $\detype(\sampler)$ as a Turing machine.  We
now check that this machine is in fact a conditionally linear sampler in the
sense of Definition~\ref{def:sampler}, and identify its associated CL functions.

\begin{proposition}
  \label{prop:detyped-sampler-functions}
  Let $\sampler$ be a $(\type,G)$-typed $\ell$-level CL sampler with field size
  $2$ and dimension $s(n)$.  Then $\detype(\sampler)$ is an $(\ell+2)$-level
  CL sampler with field size $2$ and dimension $4|\type|+s(n)$.

  More precisely, fix $n\in\N$, and let
  \[
    L^{\alice,n}=\{L^{\alice,n}_\tvar\}_{\tvar\in\type},
    \qquad
    L^{\bob,n}=\{L^{\bob,n}_\tvar\}_{\tvar\in\type}
  \]
  be the typed CL functions of $\sampler$ on index $n$.  If
  \[
    (R^{\alice,n},R^{\bob,n})
    =
    \detype_G(L^{\alice,n},L^{\bob,n})
  \]
  as in Definition~\ref{def:detyped-CL}, then $R^{\alice,n},R^{\bob,n}$ are the
  CL functions of $\detype(\sampler)$ on index $n$.
\end{proposition}

\begin{proof}
  Fix $n\in\N$ and let
  $(R^{\alice,n},R^{\bob,n})
  = \detype_G(L^{\alice,n},L^{\bob,n})$
  as in Definition~\ref{def:detyped-CL}.  We show that the CL functions of
  $\detype(\sampler)$ on index $n$ are $R^{\alice,n}$ and $R^{\bob,n}$.

  Fix $w\in\{\alice,\bob\}$.  By construction, the first two levels of
  $\detype(\sampler)$ are the graph sampler $L_G^w$ on the $V_G$ component.  If
  the resulting graph output $g$ selects type $\tvar$, the remaining levels call
  the typed sampler for $L^{w,n}_\tvar$ on the $\F_2^{s(n)}$ component; otherwise
  they use the zero CL function.  This is precisely the concatenation defining
  $R^{w,n}$ in Definition~\ref{def:detyped-CL}.

  Thus the CL functions of $\detype(\sampler)$ are
  $R^{\alice,n},R^{\bob,n}$.  By Lemma~\ref{lem:cl-concat}, these are
  $(\ell+2)$-level CL functions on $V_G\oplus\F_2^{s(n)}$, so
  $\detype(\sampler)$ is an $(\ell+2)$-level CL sampler with field size $2$ and
  dimension $4|\type|+s(n)$.
\end{proof}

\begin{definition}[Detyped deciders]
  Let $\decider$ be a typed decider in the standard timeout-counter form of
  Section~\ref{sec:tms}, and let $B_\decider(n)$ denote its timeout bound.
  We define the \emph{detyped decider} $\detype_G(\decider)$ to be the
  (standard) decider, in standard timeout-counter form, that behaves as follows on input $(n,x,y,a,b)$.
  
  It attempts to parse $x = (x', x'')$, $y =(y', y'') \in V_G \times \{0,1\}^*$.
  If it cannot, it rejects. 
  Otherwise, suppose that there exists $\{\lvar, \rvar\} \in E_G$ such that,
  using notation from Definition~\ref{def:type-neighbor},
  \begin{equation*}
\begin{aligned}
&x' = (e_\lvar, \mathrm{neigh}_G(\lvar), 0, e_\lvar)\;,\\
 &
    y' = (0, e_\rvar, e_\rvar, \mathrm{neigh}_G(\rvar))
    \, \in V_{\vertex{\alice}} \oplus V_{\edge{\alice}}
    \oplus V_{\vertex{\bob}} \oplus V_{\edge{\bob}}\;.
\end{aligned}
\end{equation*}
  If such an edge exists, the machine simulates $\decider$ on
$(n,\allowbreak \lvar,\allowbreak x'',\allowbreak \rvar,\allowbreak y'',\allowbreak a,\allowbreak b)$. In preparing this input to $\decider$, it reads only
prefixes of $x'',\allowbreak y'',\allowbreak a,\allowbreak b$ of length \[\textstyle \poly(B_\decider(n)),\] with the polynomial $\poly(\cdot)$ chosen large enough to include every input-tape cell that $\decider$ can inspect as well as $\TIME_\decider(n)$. If any of $x'',\allowbreak y'',\allowbreak a,\allowbreak b$ are longer than this, $\detype_G(\decider)$ will reject. If no such edge exists, the
machine accepts.

  The detyped decider $\detype_G(\decider)$ is equipped with a timeout counter with
  timeout bound 
  \[
    B_{\detype_G(\decider)}(n) = \poly(|\type|,B_\decider(n))
  \]
  where $\poly(\cdot)$ denotes a universal polynomial such that $B_{\detype_G(\decider)}(n)$ exceeds the time taken by the above procedure on any input of the form $(n,x,y,a,b)$.
\end{definition}

We note that we include the subscript ``$G$'' in $\detype_G(\decider)$
because we do not specify a graph when defining a typed decider (cf.\ Definition~\ref{def:typed-decider}).
This is different than the situation for samplers $\sampler$,
where we do not include a subscript in $\detype(\sampler)$
because the graph $G$ is already specified when defining $\sampler$.

\begin{definition}[Detyped verifiers]
\label{def:detyped-verifier}
  Let $\verifier = (\sampler, \decider)$ be a $(\type, G)$-typed normal form
  verifier whose decider $\decider$ is in standard timeout-counter form as in Section~\ref{sec:tms}.
  We define the \emph{detyped verifier}, denoted by $\detype(\verifier)$, to be
  the (standard) normal form verifier
  $(\detype(\sampler), \detype_G(\decider))$.
\end{definition}

\begin{lemma}[Typed verifiers to detyped verifiers]
  \label{lem:detyping-verifiers}
  Let $\verifier = (\sampler,\decider)$ be a $(\type,G)$-typed normal form
  verifier whose decider $\decider$ is in the standard timeout-counter form of
  Section~\ref{sec:tms}.  Let $B_\decider(n)$ denote the timeout bound of
  $\decider$. Let $s(n)$ denote the dimension of $\sampler$, and suppose that $s(n) \leq B_\decider(n)$.
  The detyped verifier $\detype(\verifier) =
  (\detype(\sampler),\detype_G(\decider))$ satisfies the following properties:
  for all $n \in \N$,
  \begin{enumerate}
	\item (\textbf{Completeness}) If $\verifier_n$ has a value-$1$ PCC strategy,
    then $\detype(\verifier)_n$ has a value-$1$ PCC strategy.
	\item (\textbf{Soundness}) If $\val^*(\detype(\verifier)_n) \geq 1 - \eps$, 
    then $\val^*(\verifier_n) \geq 1 - 16^{|\type|} \cdot \eps$.
    Furthermore,
    \[
      \Ent(\detype(\verifier)_n,1 - \eps) \geq
      \Ent(\verifier_n,1 - 16^{|\type|} \cdot\eps).
    \]
	\item (\textbf{Sampler parameters}) If $\sampler$ is an $\ell$-level sampler,
  then $\detype(\sampler)$ is an $(\ell+2)$-level sampler.
  The sampler $\detype(\sampler)$ has time complexity
  \[
    \TIME_{\detype(\sampler)}(n)
    =
    \poly(|\type|,\TIME_\sampler(n)).
  \]
\item (\textbf{Decider complexity}) The decider $\detype_G(\decider)$ is in the
  standard time\-out-counter form of Section~\ref{sec:tms}.  Its timeout bound
  satisfies
  \[
    B_{\detype_G(\decider)}(n)
    =
    \poly(|\type|,B_\decider(n)).
  \]
  In particular,
  \[
    \TIME_{\detype_G(\decider)}(n)
    =
    \poly(|\type|,B_\decider(n)).
  \]
  Moroever, we have 
  \[
    s_{\detype(\sampler)}(n)
  \leq
  B_{\detype_G(\decider)}(n)~.
  \]

\item (\textbf{Efficient computability}) The descriptions of
  $\detype(\sampler)$ and $\detype_G(\decider)$ are polynomial time computable
  from the description of~$G$ and the descriptions of $\sampler$ and $\decider$,
  respectively.
  \end{enumerate}
\end{lemma}

\begin{proof}
  Throughout this proof, we fix an index~$n$.
  Let $s = s(n)$ be the dimension of $\sampler$.
  The ambient space of~$\sampler$ is $V = \F_2^s$ and the ambient space
  of~$\detype(\sampler)$ is $V_{\gamestyle{detype}} = V_G \oplus V$.
  Let $\lvar, \rvar \in \type$.
  For this proof, we introduce the notation
  \begin{equation*}
\begin{aligned}
&\mathrm{view}^\alice(\lvar) =
    (e_\lvar, \mathrm{neigh}_G(\lvar), 0, e_\lvar),\;
    \\
 &\mathrm{view}^\bob(\rvar) =
    (0, e_\rvar, e_\rvar, \mathrm{neigh}_G(\rvar))\\
 &\in V_{\vertex{\alice}} \oplus V_{\edge{\alice}}
    \oplus V_{\vertex{\bob}} \oplus V_{\edge{\bob}}\;.
\end{aligned}
\end{equation*}
Supposing that players~\alice and~\bob receive~$x$ and~$y$ in
$V_{\gamestyle{detype}}$, and supposing that $(x, y)$ satisfies event
$\mathcal{E}_G$ from Proposition~\ref{prop:simulating-graph}, then $x^{V_G} =
\mathrm{view}^\alice(\lvar)$ and $y^{V_G} = \mathrm{view}^\bob(\rvar)$ for a unique
$\{\lvar, \rvar\} \in E$.

We assume that the dimension of the typed sampler $s(n)$ is at most the timeout bound of the typed decider $B_\decider(n)$. This assumption is used so that the detyped decider does not reject validly-sampled input components $x'',y''$ (which will be of length $s(n)$ each).

\paragraph{Completeness.}
Let $\strategy = (\ket{\psi},A, B)$ be a value-$1$ PCC strategy for
$\verifier_n$.
We construct a PCC strategy $\strategy^{\gamestyle{DETYPE}}$ for
$\detype(\verifier)_n$ with value~$1$.
This strategy also uses the state $\ket{\psi}$. 
When a player receives a question, they perform measurements described as
follows.
\begin{description}
\item Player \alice: given $x \in V_{\gamestyle{detype}}$ the player checks if
  for some $\lvar \in \type$, $x^{V_G} = \mathrm{view}^\alice(\lvar)$.
	If so, they perform the
  measurement
  \begin{equation*}
    \Big\{ A^{(\lvar,\, x^V)}_{\, a} \Big\}
  \end{equation*}
  to obtain an outcome~$a$, which they use as their answer.
	If not, they reply with the empty string.
	(This entails performing the measurement whose POVM element corresponding to
  the empty string is the identity matrix.)
\item Player \bob: given $y \in V_{\gamestyle{detype}}$, the player checks if
  for some $\rvar \in \type$, $y^{V_G} = \mathrm{view}^\bob(\rvar)$.
	If so, they perform the measurement
  \begin{equation*}
    \Big\{ B^{(\rvar,\, y^V)}_{\, b} \Big\}
  \end{equation*}
  to obtain an outcome~$b$, which they use as their answer.
	If not, they reply with the empty string.
\end{description}
This strategy is projective and consistent because the only measurements it uses
are those in $\strategy$ and ``trivial" measurements containing the identity
matrix.
Suppose the players receive questions~$x$ and~$y$ such that both $x^{V_G} =
\mathrm{view}^\alice(\lvar)$ and $y^{V_G} = \mathrm{view}^\bob(\rvar)$.
In this case, the questions $(\lvar, x^{V})$ and $(\rvar, y^{V})$ are in the
support of the question distribution of $\verifier$.
As a result, the players succeed with probability~$1$ on these questions, and
their measurements always commute.
For the remaining pairs of questions, the decider $\detype_G(\decider)$ always
accepts, and the measurements always commute by virtue of the fact that at least
one is trivial, i.e.,\ containing the identity matrix as a POVM element.

\paragraph{Soundness.}
Let $\strategy = (\ket{\psi}, A, B)$ be a strategy for $\detype(\verifier)_n$
with value $1-\eps$.
Suppose $G$ has~$m$ edges, $k$ of which are self-loops.
For any $(x, y)$ drawn from $\mu_{\detype(\sampler),\, n}$, the decider
$\detype_G(\decider)$ automatically accepts unless $(x^{V_G},y^{V_G})$ satisfies
event $\mathcal{E}_G$ from Proposition~\ref{prop:simulating-graph}, which occurs
with probability $(2m-k)/16^{|\type|}$.
When this happens, $x^{V_G}$ and $y^{V_G}$ are distributed as
$\mathrm{view}^\alice(\lvar)$ and $\mathrm{view}^\bob(\rvar)$, where $(\lvar,
\rvar)$ are distributed as the graph distribution on~$G$.
As a result, conditioned on $\mathcal{E}_G$, the probability that $\strategy$
succeeds on $\detype(\verifier)_n$ is equal to the probability that the strategy
$\strategy' = (\ket{\psi}, A', B')$ succeeds on $\verifier_n$, where
\begin{equation*}
  (A')^{\lvar,\, x^V}_a = A^{(\mathrm{view}^\alice(\lvar), x^V)}_a\;, \quad
  (B')^{\rvar,\, y^V}_b = B^{(\mathrm{view}^\bob(\rvar), y^V)}_b\;.
\end{equation*}
This means that
\begin{align*}
  \val^*(\detype(\verifier)_n, \strategy)
  & = \left(1 - \frac{2m-k}{16^{|\type|}}\right) +
    \frac{2m-k}{16^{|\type|}} \cdot \val^*(\verifier_n, \strategy')\\
  & \leq \left(1 - \frac{1}{16^{|\type|}}\right) +
    \frac{1}{16^{|\type|}} \cdot \val^*(\verifier_n, \strategy')\;.
\end{align*}
Thus, $\strategy'$ has value at least $1-16^{|\type|} \cdot \epsilon$.
This proves the first statement in the soundness.
As for the second, $\strategy$ and $\strategy'$ use the same state $\ket{\psi}$,
and, therefore, both strategies have the same Schmidt rank, which by definition is
at least $\Ent(\verifier_n,1 - 16^{|\type|} \cdot\eps)$.

\paragraph{Complexity.}
The sampler-level statement follows from
Proposition~\ref{prop:detyped-sampler-functions}: if $\sampler$ is
$\ell$-level, then $\detype(\sampler)$ is $(\ell+2)$-level and has dimension
$4|\type|+s(n)$.

The machine $\detype(\sampler)$ contains the description of $\sampler$ as a
subroutine in its finite description.  By the subroutine convention of
Section~\ref{sec:tms}, a call to $\sampler$ contributes
$\poly(\TIME_\sampler(n))$ to the running time of $\detype(\sampler)$.  The
description length of this subroutine is part of the fixed description of the
constructed machine, not an additional runtime parameter.

We also use that $s(n)$ is absorbed into $\TIME_\sampler(n)$: on any valid
factor-space input, $\sampler$ outputs an indicator vector in
$\{0,1\}^{s(n)}$, and writing this output already takes $s(n)$ steps up to a
constant factor.

Apart from subroutine calls, $\detype(\sampler)$ performs the fixed graph
sampler computations, the format and length checks from
Definition~\ref{def:detyped-sampler}, and the parsing needed to split vectors
into their $V_G$ and $\F_2^{s(n)}$ components.  This takes
$\poly(|\type|,s(n))$ time, hence $\poly(|\type|,\TIME_\sampler(n))$ time.

The parsing and graph-view checks take $\poly(|\type|)$ time.  If no
edge of $G$ is encoded, the machine accepts immediately.  Otherwise, the
subroutine simulation of $\decider$ takes $\poly(B_\decider(n))$ time.  Choosing the preceding standard timeout bound sufficiently large to dominate these computations gives
\[
  B_{\detype_G(\decider)}(n)=\poly(|\type|,B_\decider(n)),
\]
for some universal polynomial $\poly(\cdot)$, and hence the claimed time bound follows.

Assume that $s(n) \leq B_\decider(n)$. The dimension of the detyped sampler is $4|\type|+s(n)$, which is at most $B_{\detype_G(\decider)}(n)$ if we choose the preceding polynomial $\poly(\cdot)$ to be sufficiently large.

Finally, the descriptions of $\detype(\sampler)$ and $\detype_G(\decider)$ are
computed by copying the descriptions of $\sampler$, $\decider$, and $G$, and
adding the fixed detyping instructions.  Hence the construction time and output
description lengths are polynomial in the input description lengths.

\end{proof}


\section{Classical and quantum low-degree tests}
\label{sec:ldt}

In this section we introduce the classical and quantum low individual degree tests, that form the basis of our classical, quantum-sound and quantum PCP constructions. For quantum soundness of the classical test we refer to~\cite{ML20}. For the quantum test, we refer to~\cite{natarajan2018low} and its adaptation to the low individual degree test given in Appendix~\ref{sec:qld-analysis}. 
The protocol in this work combines both of these tests, using the quantum
low individual degree test for question reduction and the classical low individual degree test for
answer reduction.
Section~\ref{sec:ld-verifier} below introduces the classical test,
and Section~\ref{sec:pauli-verifier} does the same for the quantum 
test.
Prior to doing this, we introduce the \emph{Magic Square game} in
Section~\ref{sec:ms}, a key subroutine in the quantum test.

\subsection{The classical low-degree test}
\label{sec:ld-verifier}

We begin with a generalization of the classical low-degree test known as the
``simultaneous individual low-degree test.''
We sometimes refer to this as the ``classical low-degree test'' for short.
The low-degree test is used as a subroutine in the Pauli Basis test (see
Section~\ref{sec:pauli-verifier}) as well as the answer-reduction normal form
verifier (see Section~\ref{sec:ans}).
We describe the test as a nonlocal game in Section~\ref{sec:ld-game} and show
that the question distribution can be implemented as a (typed) conditionally
linear distribution (see Section~\ref{sec:types} for the definition of typed CL
distributions).

\subsubsection{The game}
\label{sec:ld-game}

The game $\game^\ld$ is parametrized by a tuple $\ldparams = (q, m, d, \ldc)$
where $m, d, \ldc \in \N$ are integers, $q \in \N$ is an admissible field size,
and $m$ divides $q$.
We sometimes write $\game^\ld_\ldparams$ to emphasize the dependence of the
classical low-degree test on the parameter tuple $\ldparams$.

We first provide a high-level description of the game for $k=1$.
It is based on the low-individual degree variant of the multilinearity test
of~\cite{babai1991non}, where one player (called the ``points player'') receives
a random point $u \in \F_q^m$, and the other player (called the ``lines
player'') receives a random axis-parallel line $\line$ that contains $u$, i.e.,\
the line parallel to the $i$-th axis which goes through point $u$, for uniformly
random $i\in\{1,\ldots,m\}$.
The points player is supposed to respond with a value $a \in \F_q$, and the
lines player is supposed to respond with a univariate polynomial $f_\line$ of
degree at most $d$ such that $f_\line(u_i) = a$, where $u_i$ is the $i$-th
coordinate of $u$.

Suppose the players agree beforehand on a global polynomial $g: \F_q^m \to \F_q$
such that every variable has degree at most $d$.
Then a winning strategy is the following: the points player responds with
$g(u)$, and the lines player responds with the univariate polynomial $f_\line$
that is the restriction of $g$ to line $\line$; formally, for $t \in \F_q$,
\[f_\line(t)=g(u_1,u_2,\ldots,u_{i-1},t,u_{i+1},\ldots,u_m)\;.\] It is easy to
see that $f_\line(t)$ has degree at most $d$ in the variable $t$.
Thus in this case the players will pass the low-degree test with probability
$1$.
The low-degree testing theorem states that the converse approximately holds: if
the players pass the low-degree test with probability close to $1$, then their
responses must be approximately consistent (in some sense) with a global
polynomial $g$ with individual degree $d$.

When $k > 1$, the players are supposed to respond with a \emph{tuple} of
answers; and the test is intended to check that the players' responses are
consistent with a $k$-tuple of functions $g_i: \F_q^m \to \F_q$ that are
polynomials where every variable has degree at most $d$.

The game that we use is slightly more elaborate than the ``axis-parallel
lines-versus-points'' test just described.
In addition, we have two other subtests.
One is a ``diagonal line-vs-points'' test: here, the points player receives a
point $u \in \F_q^m$ as before, but the lines player receives a line $\line$
that goes through $u$ but is not necessarily axis-aligned.
Instead, it's chosen by first picking an index $i \in \{1,2,\ldots,m\}$
uniformly at random, picking $v \in \F_q^m$ uniformly at random, and letting $v'
= \pi_{i-1}(v) = (0,0,\ldots,0,v_i,v_{i+1},\ldots,v_m)$ (i.e., $\pi_{i-1}$
zeroes out the first $i-1$ coordinates of its input).
Then, the line is defined as the set
\[
	\line = \{ u + tv' : t \in \F_q \}\;,
\]
which is a uniformly random line in the affine subspace of $\F^m_q$
corresponding to all points $v$ whose first $(i-1)$ coordinates match those of
$u$.
The points player responds with a value $a \in \F_q$, the lines player responds
with a univariate polynomial $f_\line$ of degree $md$,\footnote{Since $\line$ is
  no longer axis-parallel, the restriction of an individual degree-$d$
  polynomial $g$ to a line $\line$ may yield a degree $md$ polynomial.}
and the verifier checks that $f_\line(t_u) = a$ (where $t_u \in \F_q$ is the
projection of $u$ onto the line $\line$ with respect to a canonical
parameterization of $\line$).

The other additional test is a ``consistency test,'' where both players receive
either the same points question, the same axis-parallel lines question, or the
same diagonal lines question.
In all cases they are expected to respond with the same answer.


\paragraph{Axis-parallel lines and diagonal lines.}
Before presenting a formal description of the game $\game^\ld$, we first define
what we mean by lines, and their properties.

\begin{definition}[Lines]
  \label{def:line}
	A \emph{line} $\line$ in $\F^m_q$ specified by a pair $(u,v) \in (\F_q^m)^2$
  is the subset
  \begin{equation}
    \label{eq:line-set}
    \line(u,v) = \bigl\{\, u + tv : t \in \F_q \,\bigr\} \subseteq \F_q^m \;.
  \end{equation}
  We call the vector $v$ a \emph{direction} of line
  $\line(u,v)$.\footnote{For convenience we explicitly allow
    $v=0$, in which case the line is reduced to a singleton.}
  
  A line $\line(u,v)$ is \emph{axis-parallel} if there is a coordinate $i \in
  \{1,2,\ldots,m\}$ for which $v$ is equal to $e_i$, the $i$-th elementary basis
  vector in $\F_q^m$.
  We say that such a line $\line$ is \emph{parallel to the $i$-th direction}.
  A line $\line(u,v)$ is \emph{diagonal} if there exists an $i \in
  \{0,1,\ldots,m-1\}$ for which $v_1 = v_2 = \cdots = v_i = 0$.\footnote{Every line is diagonal. The terminology enables us to emphasize when we refer to a ``general'' line, as opposed to an axis-parallel line.}
\end{definition}

The following proposition establishes that, for a fixed direction, a line
$\line$ is an equivalence class of points on the line.
\begin{proposition}
  \label{prop:line-equiv}
  Let $u,v \in \F_q^m$.
  Then for all $u' \in \line(u,v)$, we have that $\line(u,v) = \line(u',v)$.
\end{proposition}
\begin{proof}
	Fix $u' \in \line(u,v)$.
  We have that $u' = u + tv$ for some $t \in \F_q$.
  Let $a \in \line(u,v)$.
  This means that $a = u + sv$ for some $s\in \F_q$.
  We also have that $a = u + (s - t + t)v = u + tv + (s - t)v = u' + (s - t)v$,
  so $\line(u,v) \subseteq \line(u',v)$.
  A similar argument shows that $\line(u',v) \subseteq \line(u,v)$.
\end{proof}

We now specify a canonical representation of lines $\line = \line(u,v)$ as pairs $(u_0,v)$
where $u_0$ is a \emph{canonical representative} of $\line$ specified as
follows.
Let $L_v^{\lnf}: \F_q^m \to \F_q^m$ denote the canonical linear map with the
one-dimensional kernel basis $\{v\}$ (see Definition~\ref{def:cl-canonical} for a
definition of canonical linear map).
Note that for all $u \in \line$, we have that $L_v^{\lnf}(u)$ is a point in
$\line$.
This is because since the image and kernel of $L_v^{\lnf}$ are complementary
subspaces by definition (see Definition~\ref{def:cl-canonical}), any $u \in \F_q^m$ has a
unique decomposition $u = u_0 + u_1$ where $u_0$ is in the image of $L_v^{\lnf}$
and $u_1 \in \ker(L_v^{\lnf})$, which by definition is the subspace spanned by
$v$.
Thus $u_1 = tv$ for some $t \in \F_q$ and we have that $u_0 = u - tv \in \line$.
In particular, if $u\in \line$ then $u_0 = L_v^{\lnf}(u) \in \line$ as well. 
Furthermore, since $L_v^{\lnf}$ is a projection onto its image by definition, we
have $L_v^{\lnf}(u_0) = u_0$.

Next, note that for any $u,u' \in \line$, since $u-u' = sv$ for some $s\in \F_q$
it follows that $L_v^\lnf(u) = L_v^\lnf(u')$.
Thus $u_0 = u_0'$ and this shows that $L_v^\lnf$ gives a means to compute a
canonical representative of lines.

\begin{definition}[Canonical representative of a line]
  \label{def:line-representative}
  Let $\line$ be a line in direction $v$.
  A canonical representative of $\line$ is the point $u_0 \in \F_q^m$ such that
  \[
    u_0 = L_v^\lnf(u)
  \]
  for all $u \in \line$.
\end{definition}

%
%

\paragraph{Question distribution.}
We now present the question distribution of the game $\game^\ld$ as a typed CL
distribution (see Section~\ref{sec:types} for the definition of typed CL functions and
typed CL distributions).
The type set $\type^\ld$ of the game $\game^\ld$ we use is
\[
	\type^\ld = \{ \Point, \ALine, \DLine \}\;.
\]
The distribution over question types is uniform over $\type^\ld \times \type^\ld$. 

We introduce the corresponding typed CL functions $L_\Point^\alice=L_\Point^\bob=L_\Point$, $L_\ALine^\alice=L_\ALine^\bob=L_\ALine$,
$L_\DLine^\alice=L_\DLine^\bob=L_\DLine$ where the corresponding CL distributions $\mu_{L_\ALine,L_\Point}$
and $\mu_{L_\DLine,L_\Point}$ implement the axis line-versus-point and diagonal
line-versus-point distributions, respectively.
The functions are parametrized by a field size $q$, a dimension $m$, and three
complementary register subspaces $V_\xpt,V_{\coord}, V_{\dir{}}$ of some ambient
space.
The register $V_\xpt$ (which is isomorphic to $\F_q^m$) is called the
\emph{point register}, $V_\coord$ (which is isomorphic to $\F_q$) is called the
\emph{coordinate register}, and $V_{\dir{}}$ (which is isomorphic to $\F_q^m$)
is called the \emph{direction register}, respectively.
We let $V$ denote the direct sum $V_\xpt \oplus V_{\coord} \oplus V_{\dir{}}$.
We identify elements of $V$ as triples $(u,s,v) \in \F_q^m \times \F_q \times
\F_q^m$.

Define $L_\Point$ to be the $1$-level CL function that projects onto $V_\xpt$: for every $(u,s,v) \in V$,
\begin{equation}
  \label{eq:cl-ptf}
	L_\Point(u,s,v) = (u,0,0) \;.
\end{equation}
Define $L_\ALine$ to be the following $2$-level CL function:
\begin{equation}
  \label{eq:cl-alnf}
	L_\ALine(u,s,v) = (L^\lnf_{e_i}(u), s, 0)\ \text{for all } (u,s,v) \in V,
\end{equation}
where $L^\lnf$ is the linear function used to compute the canonical
representative of lines (see Definition~\ref{def:line-representative}), and $i = \chi(s)$
where we define $\chi(s)$ to be the unique integer such that
\begin{equation}
\label{eq:chi-func}
	s = (\chi(s) - 1) \frac{q}{m} + r
\end{equation}
for some $0 \leq r < q/m$, and we fix a bijection between field elements $s \in \F_q$ and
integers between $0$ and $q - 1$ as in e.g.,\ Remark~\ref{rmk:tm_fields}.
The function $L_\ALine$ is the concatenation of the $1$-level CL function that
projects the subspaces $V_{\coord} \oplus V_{\dir{}}$ down to $V_\coord$ (i.e.,
zeroes out the entire $V_{\dir{}}$ register), and the family of $1$-level CL
functions $\{L^\lnf_{e_i} \}$ acting on $V_\xpt$, indexed by $i \in
\{1,2,\ldots,m\}$.
(See Lemma~\ref{lem:cl-concat} for definition of CL function concatenation.)
Thus $L_\ALine$ is a $2$-level CL function.

Define $L_\DLine$ to be the following $3$-level CL function: for all $(u,s,v) \in
V$,
\begin{equation}
\label{eq:cl-dlnf}
	L_\DLine(u,s,v) = (L^\lnf_{\pi_{i-1}(v)}(u),s,\pi_{i-1}(v))
\end{equation}
where $i = \chi(s)$.
The function $L_\DLine$ is the concatenation of the following CL functions (again, see Lemma~\ref{lem:cl-concat} for definition of CL function concatenation): (i) the identity function on $V_{\coord}$ (which is a $1$-level CL function); (ii) for every $i \in \{1,2,\ldots,m\}$ the $1$-level CL function on $V_{\dir{}}$ that maps $v \mapsto \pi_{i-1}(v)$ (i.e.,~it zeroes out the first $i-1$ coordinates of $v$), and (iii) for every pair $(s,v') \in V_\coord \oplus V_{\dir{}}$, the $1$-level CL function $L^\lnf_{v'}$ acting on $V_\xpt$. Thus $L_\DLine$ is a $3$-level CL function.

The following two lemmas establish that the CL functions
$L_\Point,\allowbreak L_\ALine,\allowbreak L_\DLine$ give rise to the axis-parallel line-versus-point and
diagonal line-versus-point distributions.
\begin{lemma}
	\label{lem:alnf}
	Consider the pair $(\line, u)$ sampled in the following way: first, sample the
  tuple $((u_0,s,0),(u,0,0))$ from the CL distribution $\mu_{L_\ALine,L_\Point}$,
  and let $\line = \line(u_0,e_i)$ where $i = \chi(s)$.
  Then $u$ is a uniformly random point in $\F_q^m$ and $\line$ is a line that
  goes through $u$ and is parallel to a uniformly random axis $i \in
  \{1,2,\ldots,m\}$.
\end{lemma}
\begin{proof}
	Since $s$ is uniformly random in $\F_q$ and $m$ divides $q$ by assumption, we
  have that $i$ is uniformly random between $\{1,2,\ldots,m\}$ and, therefore,
  $(u_0,s)$ specifies a uniformly random axis-parallel line $\line(u_0,e_i)$.
  Since $u_0 = L^{\lnf}_{e_i}(u)$, we have that $u \in \line(u_0,e_i)$.
\end{proof}


\begin{lemma}
	\label{lem:dlnf}
	Consider the pair $(\line,u)$ sampled in the following way: first, sample the
  tuple $((u_0,s,v),(u,0,0))$ from the CL distribution $\mu_{L_\DLine,L_\Point}$,
  and let $\line = \line(u_0,v)$ where $i = \chi(s)$.
  Then $u$ is a uniformly random point in $\F_q^m$ and $\line$ is a random
  diagonal line that goes through $u$ and shares the first $i-1$ coordinates
  with $u$, for uniformly random $i$.
\end{lemma}
\begin{proof}
	The proof of this follows identically to Lemma~\ref{lem:alnf}.
\end{proof}

For later convenience we make the following definition. 

\begin{definition}[Line-point distributions]
        \label{def:line-point-dist}
 The \emph{axis line-point distribution} $D_{\ALine}$ is the
 distribution $\mu_{L_\ALine,L_\Point}$ over pairs $(\ell=(u_0,e_i),u)$
 (we omit the $0$ elements for simplicity). The \emph{diagonal
   line-point distribution} $D_{\DLine}$ is the distribution
 $\mu_{L_\DLine,L_\Point}$  over $(\ell=(u_0,s,v),u)$. The
 \emph{line-point distribution} $D_{\Line}$ is the equal mixture of the axis-parallel line-point distribution and the diagonal line-point distribution.
 In other words, a sample from $D_{\Line}$ is distributed as follows:
 with probability $\tfrac{1}{2}$, output a sample from $D_{\ALine}$,
 and with probability $\tfrac{1}{2}$, output a sample from $D_{\DLine}$.
\end{definition}

\paragraph{Decision procedure} The decision procedure $\decider^\ld$ for the
game $\game^\ld$ is presented in Figure~\ref{fig:ld-decider}.
The table at the top specifies a parsing scheme for the questions and answers,
depending on the type of question.
For example, when a player receives a question with type $\Point$, the question
content $x$, a bit string of length $m \log q$, should be interpreted by the
decision procedure and the players as an element of the vector space $\F_q^m$,
as indicated in Section~\ref{sec:ff-representations}.
Similarly the answer to a question with type $\Point$ is expected to be a bit
string of length $\ldc \log q$, and is interpreted as an element of $\F_q^\ldc$.
For questions with type $\ALine$, the question content is a $(m \log q + \log
q)$-bit string, which is parsed as a pair $(u_0, s) \in \F_q^m \times \F_q$,
which in turn can be interpreted as a specification for an axis-parallel line
$\line(u_0,e_i)$ in $\F_q^m$ where $i = \chi(s)$, as described in
Lemma~\ref{lem:alnf}.
The answer is interpreted as the description of $\ldc$ degree-$d$ univariate
polynomials defined on the line $\line(u_0,e_i)$, each polynomial given by a
list of $d+1$ coefficients.
For questions with type $\DLine$, the question content is a $(2m \log q + \log
q)$-bit string, which is parsed as a triple $(u_0, s, v) \in \F_q^m \times \F_q
\times \F_q^m$ that specifies a diagonal line $\line(u_0,v')$ where $v' =
\pi_{i-1}(v)$ as described in Lemma~\ref{lem:dlnf}.
The answer is interpreted as the description of $\ldc$ degree-$md$ univariate
polynomials defined on the line $\line(u_0,v')$.
If the answers returned by the players do not fit this format the decision
procedure rejects.

\begin{figure}[!htbp]
  \centering
  \begin{gamespec}
    \begin{tabularx}{\textwidth}{ l   l   X   }
      \toprule
      Type & Question Content & Answer Format \\
      \midrule
      $\Point$ & $ u \in \F_q^m$ & Element of $\F_q^\ldc$  \\
      $\ALine$ & $ (u_0, s) \in \F_q^m \times \F_q$ &
      $\ldc$ degree-$d$ polynomials $f_j: \F_q \to \F_q$\\
      $\DLine$ & $(u_0, s, v) \in \F_q^m \times \F_q \times \F_q^m$ & 
      $\ldc$ degree-$md$ polynomials $f_j: \F_q \to \F_q$\\
      \bottomrule
    \end{tabularx}

    Input to $\decider^\ld$: $(\tvar_\alice,x_\alice, \tvar_\bob, x_\bob,
    a_\alice, a_\bob)$. \tnote{Added this in response to reviewer comment. We need to make sure that $\decider$ has bounded runtime. As described below,\allowbreak  it is not clear if it might take infinite time just to parse some super-long dummy input}Check that $(\tvar_\alice,x_\alice)$ and $(\tvar_\bob,x_\bob)$ are correctly formatted --- the types correspond to valid types,\allowbreak  and the questions have the correct length --- and that $a_\alice$ and $a_\bob$ also have the correct length, as can be inferred from the table above.
		    For $w \in \AB$,

    \begin{enumerate}
    \item (\textbf{Consistency test}) If $\tvar_\alice = \tvar_\bob$, accept iff
      $a_\alice = a_\bob$.
    \item (\textbf{Axis-parallel line-versus-point test}) If $\tvar_w = \ALine$
      and $\tvar_{\overline{w}} = \Point$, accept iff $f_j(t) =
      (a_{\overline{w}})_j$ for all $j \in \{1,2,\ldots,\ldc\}$ where $t \in
      \F_q$ is such that $x_{\overline{w}} = u_0 + te_i$ where $i = \chi(s)$.
      
    \item (\textbf{Diagonal line-versus-point test}) If $\tvar_w = \DLine$ and
      $\tvar_{\overline{w}} = \Point$, accept iff $f_j(t) =
      (a_{\overline{w}})_j$ for all $j \in \{1,2,\ldots,\ldc\}$ where $t \in
      \F_q$ is such that $x_{\overline{w}} = u_0 + tv'$, where $v' =
      \pi_{i-1}(v)$ with $i = \chi(s)$.
    \end{enumerate}
		    In all cases where no action is indicated, accept.
  \end{gamespec}
  \caption{The decision procedure $\decider^\ld$ for the simultaneous low-degree
    test, parameterized by $\ldparams = (q,m,d,\ldc)$.
    The function $\chi(s)$ is defined in \eqref{eq:chi-func}, and $\pi_{i-1}(v)$
    zeroes out the first $i-1$ coordinates of $v$.}
  \label{fig:ld-decider}
\end{figure}

We define a special class of measurements that are relevant to the soundness
properties of the low-degree test.

\begin{definition}[Low-degree polynomial measurements]
  \label{def:ld-meas}
  \leavevmode\par Define $\polymeas{m}{d}{q}$ to be the set of POVM measurements whose outcomes
  correspond to (individual) degree-$d$ polynomials of $m$ variables over
  $\F_q$.
  More generally, for an integer $\ldc$ and tuples $m=(m_1,\allowbreak  m_2,\allowbreak  \ldots,\allowbreak 
  m_\ldc)$, $d=(d_1,\allowbreak  d_2,\allowbreak  \ldots,\allowbreak  d_\ldc)$ and $q=(q_1,\allowbreak  q_2,\allowbreak  \ldots,\allowbreak  q_\ldc)$,
  we let $\simulpolymeas{m}{d}{q}{\ldc}$ be the set of measurements $G =
  \{G_{g_1,\, g_2,\, \ldots\,,\, g_\ldc}\}$ such that for $i\in\{1,\allowbreak  2,\allowbreak  \ldots,\allowbreak 
  \ldc\}$, $g_i$ is a polynomial $g_i:\F_{q_i}^{m_i} \rightarrow
  \F_{q_i}$ with individual degree $d_i$ (see Section~\ref{sec:ld-encoding} for a definition of individual degree). 
\end{definition}

Quantum soundness of a version of the classical low-degree test for polynomials of low \emph{total} degree was claimed
in~\cite{natarajan2018low}, building on an analysis of a three-prover version of the test in~\cite{vidick2016three}. Unfortunately there is a gap in the soundness analysis of this test. For our construction we use the low individual degree test, whose quantum soundness is shown in~\cite{ML20} and generalized in~\cite{ji2021quantum}. We state the result in the form that is most directly useful for us, and show how the stated result follows from~\cite{ji2021quantum}.

\begin{theorem}[Quantum soundness of the simultaneous classical low-degree test]
  \label{lem:ld-soundness}
  There exists a function $\delta_{\ld}(\eps,\allowbreak  q,\allowbreak  m,\allowbreak  d,\allowbreak  \ldc) = a (dm \ldc)^a (\eps^b + q^{-b} + 2^{-bmd})$ for universal constants $a\geq 1$ and $0<b\leq 1$ 
  such that the following
  holds.
  For all $\eps > 0$ and parameter tuple $\ldparams = (q, m, d, \ldc)$, for all
  projective strategies $(\psi,A,B)$ that succeed with probability at least
  $1-\eps$ in the game $\game^\ld_\ldparams$, there exists \tnote{added:}projective measurements
  \[G^w \in \simulpolymeas{m}{d}{q}{\ldc}\] on $\mH_w$, for
  $w\in\{\alice,\bob\}$, such that
  \begin{equation*}
    \begin{gathered}
      A^{\Point,\, u}_{a_1,\, a_2,\, \ldots\,,\, a_\ldc}
      \abc[\delta_{\ld}]
      G^\bob_{[(g_1(u),g_2(u),\ldots,g_\ldc(u)) = (a_1,\, a_2,\, \ldots \,,\, a_\ldc)]}\;,\\
      G^\alice_{[(g_1(u),g_2(u),\ldots,g_\ldc(u)) = (a_1,\, a_2,\, \ldots \,,\, a_\ldc)]}
      \abc[\delta_{\ld}]
			 B^{\Point,\, u}_{a_1,\, a_2,\, \ldots\,,\, a_\ldc}\;,\\
      G^\alice_{g_1,\, g_2,\, \ldots \,,\, g_\ldc}
      \abc[\delta_{\ld}]
      G^\bob_{g_1,\, g_2,\, \ldots \,,\, g_\ldc}\;,
    \end{gathered}
  \end{equation*}
  where $\delta_{\ld} = \delta_{\ld}(\eps, q, m, d, \ldc)$, 
  and all three statements are with respect to the state $\ket{\psi}$.
	In addition, the first two approximations hold under the uniform distribution over $u\in \F_q^m$,
	whereas there is no question distribution associated with the third approximation.
\end{theorem}

\begin{proof}
Let $\ldparams = (q, m, d, \ldc)$
where $m, d, \ldc \in \N$ are integers, $q \in \N$ is an admissible field size,
and $m$ divides $q$. Let $\Sigma=\F_q$. Let $\code \subseteq \Sigma^q$ be the Reed-Solomon code with degree $d$. Explicitly, $\code$ is the set of all $(p(x_1),\ldots,p(x_n))$ where $p:\F_q\to\F_q$ is a univariate polynomial of degree at most $d$ and $x_1,\ldots,x_n$ are an enumeration of $\F_q$. Then in the (standard) notation from~\cite{ji2021quantum} $\code$ is a linear $[q,d+1,n-d+1]_\Sigma$ code.

Observe that the two-prover tensor code test, for the code $\code^{\otimes m}$, as described in~\cite[Figure 1 and Figure 2]{ji2021quantum}, is identical to the game $\game^\ld$ with parameters $\ldparams$ and $\ldc=1$, whose decider is presented in Figure~\ref{fig:ld-decider}. This is because 
\begin{enumerate}
\item The \emph{axis-parallel line-versus-point test} from Figure~\ref{fig:ld-decider} is identical to the \emph{axis-parallel lines test} from~\cite[Figure 1]{ji2021quantum}, with the only difference that in Figure~\ref{fig:ld-decider} the roles of A and B are chosen uniformly at random. This is consistent with the two-prover variant of the tensor code test from~\cite[Section 4.1]{ji2021quantum}.
\item The \emph{diagonal line-versus-point test} from Figure~\ref{fig:ld-decider} is identical to the \emph{subcube commutation test} from~\cite[Figure 1]{ji2021quantum}, with again the added symmetrization between players. 
\item The \emph{consistency test} from Figure~\ref{fig:ld-decider} is identical to the \emph{synchronicity test} from~\cite[Figure 2]{ji2021quantum}.
\end{enumerate}

As a result,~\cite[Theorem 4.7]{ji2021quantum}, in which we choose the parameter $k$ (which is a distinct parameter to the $k$ which appears in the present proof) as $k=m^3d$,  immediately yields Lemma~\ref{lem:ld-soundness}.

To conclude it remains to extend the result to the case of general $\ldc$. This is done 
via a standard reduction, following exactly the same steps as the derivation of Theorem 4.43 from Theorem 4.40 in~\cite{NW19}.
\end{proof}

%

%

\subsubsection{Complexity of the classical low-degree test.}
The CL functions and decision procedure of the low-degree test are incorporated
as subroutines in some of the normal form verifiers constructed in subsequent
sections.
The next lemma establishes the time complexity of these procedures as a function
of the parameter tuple $\ldparams = (q,m,d,\ldc)$.
The lemma also establishes the time complexity of computing the {description} of
the decision procedure $\decider^\ld$ as a Turing machine, given the parameter
tuple $\ldparams$ as input.


\begin{lemma}[Complexity of the classical low-degree test]
  \label{lem:ld-complexity}
  Let $\ldparams = (q,m,d,\ldc)$ denote a parameter tuple.
  \begin{enumerate}
	\item The time complexity of the decision procedure $\decider^\ld$
    parametrized by $\ldparams$ is $\poly(m, d, \ldc, \log q)$.
	\item The time complexity of evaluating marginals of the CL functions
    $L_\Point$, $L_\ALine$, and $L_\DLine$ at a given input point is $\poly(m, \log
    q)$.
	\item The Turing machine description of the decision procedure $\decider^\ld$
    parametrized by $\ldparams$ can be computed from $\ldparams$ in $\polylog
    (q,m,d,\ldc)$ time.
  \end{enumerate}
\end{lemma}

\begin{proof}
The input check performed by $\decider^\ld$ ensures that its entire input has length $O(m\log q)$, and, furthermore, the questions and answers are formatted as indicated in Figure~\ref{fig:ld-decider}. Performing these formatting checks takes linear time, $O(m\log q)$. We now evaluate the complexity of each of the tests, assuming the inputs are correctly formatted.	Finite field arithmetic over $\F_q$ can be performed in time $\polylog q$, by
  Lemma~\ref{lem:efficient_arithmetic}.
	The most expensive step in $\decider^\ld$ is to evaluate a univariate
  polynomial $f: \F_q \to \F_q^\ldc$ at some point $t_u \in \F_q$ that corresponds
  to a point $u \in \F_q^m$, which takes time $\poly(m,d,\ldc,\log q)$.
  The function $L_\Point$ is a projection onto $V_\xpt$, which takes time
  $\poly(m,\log q)$ to compute (it just involves ``zeroing out'' the registers
  outside of $V_\xpt$).
  The functions $L_\ALine$ and $L_\DLine$ require computing a canonical linear
  map, which requires performing Gaussian elimination and can be done in time
  $\poly(m,\log q)$.

\hnote{added May 30, 2026:} Due to the input length check and the time complexity of all the tests described, the decider
$\decider^\ld$ halts in time $\poly(m,d,\ldc,\log q)$ as desired. 

  The Turing-machine description of the decision procedure $\decider^\ld$ can be
  uniformly computed from the integers $(q,m,d,\ldc)$ expressed in binary; the
  complexity of computing the description comes from describing the parameter
  tuple $\ldparams$, which takes time that is at most polynomial in the bit
  length of $(q,m,d,\ldc)$.
\end{proof}

\subsection{The magic square game}
\label{sec:ms}

We recall the \emph{Magic Square game} of Mermin and
Peres~\cite{mermin1990simple,peres1990incompatible,aravind2002simple}.
The Magic Square game is a simple self-test for EPR pairs (it tests for two of
them). 
In addition, it allows one to test that a pair of observables
\emph{anticommutes}.
Here we use it as a building block to construct the quantum low-degree test. 

There are several formulations of the Magic Square game; here we present it as a
\emph{binary constraint satisfaction} game~\cite{cleve2014characterization}.
In this formulation of the game (denoted by $\game_\MS$) there are $6$ linear
equations defined over $9$ variables that take values in $\F_2$.
The variables correspond to the cells of a $3 \times 3$ grid, as depicted in
Figure~\ref{fig:ms}.
Five of the equations correspond to the constraint that the sum of the variables
in each row and the first two columns must be equal to $0$, and the last
equation requires that the sum of the variables in the last column must be equal
to $1$.

\begin{figure}[ht!]
	\begin{center}
		\begin{tabular}{|c|c|c|}
			\hline
			$x_1$ & $x_2$ & $x_3$ \\
			\hline 
			$x_4$ & $x_5$ & $x_6$ \\	
			\hline
			$x_7$ & $x_8$ & $x_9$ \\
			\hline
		\end{tabular}
		\caption{The Magic Square game}
		\label{fig:ms}
		\end{center}
\end{figure}

The question set $\type^\ms$ of the Magic Square game is the following:
\begin{equation*}
  \begin{split}
    \type^\msc & = \{ \Constraint_i : i = 1, 2, \ldots, 6\}\;,\\
    \type^\msv & = \{ \Variable_j : j = 1, 2, \ldots, 9\}\;,\\
    \type^\ms & = \type^\msc \cup \type^\msv\;.
  \end{split}
\end{equation*}
The questions $\Constraint_i$ for $i \in \{1,2,3\}$ correspond to the three row
constraints, the questions $\Constraint_4$, $\Constraint_5$ correspond to the first
two column constraints, and question $\Constraint_6$ corresponds to the third
column constraint.

In the Magic Square game, the verifier first samples a constraint\newline $\Constraint_i
\in \type^\msc$ uniformly at random, and then samples $\Variable_j$, one of the
three variables in the row or column corresponding to $\Constraint_i$, uniformly
at random.
One player is randomly assigned to be the $\Constraint$ player, and the other is
assigned to be the $\Variable$ player.
The $\Constraint$ player is sent the question $\Constraint_i$ and is expected to
respond with three bits $(\beta_{v_1},\allowbreak  \beta_{v_2},\allowbreak  \beta_{v_3}) \in \F_2^3$,
where $(v_1,\allowbreak v_2,\allowbreak v_3)$ are the indices of the three variables corresponding to
$\Constraint_i$.
The $\Variable$ player is given question $\Variable_j$ and is expected to
respond with a single bit $\gamma \in \F_2$.
The players win if the $\Constraint$ player's answers satisfy the equation
associated with $\Constraint_i$, and $\gamma = \beta_j$.
More precisely, the verifier samples an edge of the type graph (see Section~\ref{sec:types}) $G^\ms$
in Fig.~\ref{fig:type-graph-ms}, sends one endpoint to a random player, and the
other endpoint to the other player.

\begin{figure}[!htbp]
  \centering
  \begin{tikzpicture}[scale=.8]

    \tikzset{type/.style args={[#1]#2}{
        draw,circle,fill,scale=0.25,
        label={[font=\scriptsize, label distance=1pt]#1:#2}
      }}


    \foreach \i in {1,...,6} \draw (0,8-9/8*\i)
    coordinate (Constraint-\i)
    node[type={[180]$\Constraint_\i$}] {};

    \foreach \i in {1,...,9} \draw (2.5,9-\i)
    coordinate (Variable-\i)
    node[type={[0]$\Variable_\i$}] {};

    \foreach \i in {1,...,3} \foreach \j in {1,...,3}
    \pgfmathsetmacro{\k}{(\i-1)*3+\j}
    \draw (Constraint-\i) -- (Variable-\k);

    \foreach \i in {4,...,6} \foreach \j in {1,...,3}
    \pgfmathsetmacro{\k}{\i-3+(\j-1)*3}
    \draw (Constraint-\i) -- (Variable-\k);
  \end{tikzpicture}
  \caption{Type graph $G^\ms$ for the Magic Square game.  }
  \label{fig:type-graph-ms}
\end{figure}

The following theorem records the self-testing (also known as \emph{rigidity})
properties of the Magic Square game.
Its self-testing properties
are crucial to the Pauli basis test.
In particular, it is used to enforce anticommutation relations between certain
pairs of operators.
We will refer to this theorem in the proof of the soundness of the Pauli basis test (Theorem~\ref{thm:pauli} below) in Section~\ref{sec:qld-analysis}.

\begin{theorem}[Rigidity of the Magic Square game]
  \label{thm:ms-rigidity}

Let $\strategy = (\ket{\psi},A,B)$ be a strategy that succeeds in the Magic Square game $\game^{\MS}$ with probability $1 - \eps$, where $\ket{\psi} \in \mH_A \otimes \mH_B$ is a state. Then there exist local isometries $\phi_A: \mH_A \to \mH_{A'} \otimes \mH_{A''},\allowbreak \phi_B: \mH_B \to \mH_{B'} \otimes \mH_{B''}$ (where $\mH_{A'},\mH_{B'} \cong (\C^2)^{\otimes 2}$ and $\mH_{A''},\mH_{B''}$ are finite dimensional) and a state $\ket{\aux} \in \mH_{A''} \otimes \mH_{B''}$ such that 
\begin{enumerate}
    \item $\left \| \phi_A \otimes \phi_B \ket{\psi} - \ket{\EPR_2}^{\otimes 2} \otimes \ket{\aux} \right \| \leq O(\sqrt{\eps})$,
    \item Letting $\tilde{A}^x_a = \phi_A\, A^x_a \, \phi_A^\dagger$ and $\tilde{B}^y_b = \phi_B B^y_b \phi_B^\dagger$, we have that
    \begin{gather*}
        \tilde{A}^{\Variable_1}_b \otimes I_{B'B''} \approx_{\sqrt{\eps}} (\sigma^X_b)_{A'} \otimes I_{A'' B' B''} \\
        \tilde{A}^{\Variable_5}_b \otimes I_{B'B''} \approx_{\sqrt{\eps}} (\sigma^Z_b)_{A'} \otimes I_{A'' B' B''} \\
        I_{A'A''} \otimes \tilde{B}^{\Variable_1}_b    \approx_{\sqrt{\eps}} I_{A' A'' B''} \otimes (\sigma^X_b)_{B'} \\
        I_{A'A''} \otimes \tilde{B}^{\Variable_5}_b    \approx_{\sqrt{\eps}} I_{A' A'' B''} \otimes (\sigma^Z_b)_{B'}
    \end{gather*}
    where the $\approx_{\sqrt{\eps}}$ statement holds with respect to the state $\ket{\EPR_2}^{\otimes 2}_{A'B'} \otimes \ket{\aux}_{A''B''}$ and the answer summation is over $b \in \{0,1\}$. As a consequence, letting 
    \begin{gather*}
    \tilde{A}^{\Variable_1} = \tilde{A}^{\Variable_1}_0 - \tilde{A}^{\Variable_1}_1 \\
    \tilde{A}^{\Variable_5} = \tilde{A}^{\Variable_5}_0 - \tilde{A}^{\Variable_5}_1 \\
    \end{gather*}
    it holds that
    \begin{gather*}
        \tilde{A}^{\Variable_1} \tilde{A}^{\Variable_5} \otimes I_{B' B''} \approx_{\sqrt{\eps}} - \tilde{A}^{\Variable_5}  \tilde{A}^{\Variable_1} \otimes I_{B' B''} \\
        I_{A' A''} \otimes \tilde{B}^{\Variable_1} \tilde{B}^{\Variable_5}  \approx_{\sqrt{\eps}} - I_{A' A''} \otimes \tilde{B}^{\Variable_5}  \tilde{B}^{\Variable_1} \;,
      \end{gather*}
      where the notation $M \approx_{\delta} N$ for $M, N$ that are not POVM
      elements means
      \[  \bra{\psi} (M-N)^\dagger (M-N) \ket{\psi}  \leq  O(\delta). \]
\end{enumerate}

\end{theorem}
\begin{proof}
 	A proof of the rigidity of the Magic Square game can be found in~\cite[Theorem 6.9]{coladangelo2017robust}. There are a couple minor differences between their rigidity statement and the one stated here. First, they state that the robustness of the Magic Square self-test is $O(\eps)$; however, they specify the closeness between the actual and ideal states in terms of the trace norm, whereas we specify the closeness between $\phi_\alice \otimes \phi_\bob \ket{\psi}$ and $\ket{\EPR_2}^{\otimes 2}\otimes \ket{\aux}$ in terms of the Euclidean distance, which translates to $O(\sqrt{\eps})$ instead of $O(\eps)$. Second, their choice of ideal strategy specifies that the observables corresponding to $\Variable_1$ and $\Variable_5$ questions are $I \otimes \sigma^Z$ and $\sigma^{Z} \sigma^{X} \otimes \sigma^Z \sigma^X$; however, under a local change of basis these are equivalent to $\sigma^X \otimes I$ and $\sigma^Z \otimes I$, respectively.
\end{proof}

We will need the following theorem,
which shows that any pair of anticommuting observables 
can be used to form a value-$1$ strategy for the Magic Square game.

\begin{theorem}
  \label{thm:ms-from-ac}
  Let $A = \{A_b\}_{b \in \F_2}$ and $B = \{B_b\}_{b \in \F_2}$ be two-outcome
  projective measurements acting on $(\C^q)^{\otimes n}$ which are consistent on
  $\ket{\EPR_q}^{\otimes n}$, and let $\mathcal{O}_A = A_0 - A_1$ and
  $\mathcal{O}_B = B_0 - B_1$ be the corresponding observables.
  Suppose that $\mathcal{O}_A \mathcal{O}_B = -\mathcal{O}_B \mathcal{O}_A$.
  Then there exists a symmetric strategy $\strategy = (\psi, M)$ for the Magic
  Square game with the following properties.
  \begin{enumerate}
  \item $\strategy$ is an SPCC strategy of value $1$.
  \item The state $\ket{\psi}$ has the form $\ket{\psi} = \ket{\EPR_q}^{\otimes
      n} \otimes \ket{\EPR_2}$.
  \item For $b \in \{0,1\}$, we have $M^{\Variable_1}_b = A_b \otimes I$ and
    $M^{\Variable_5}_b = B_b \otimes I$.\label{item:embeds}
  \end{enumerate}
\end{theorem}

\begin{proof}
  The strategy~$\strategy$ is based on the canonical two-qubit strategy for the
  Magic Square game as described in, for example,~\cite{aravind2002simple}.
  The state is $\ket{\psi} = \ket{\EPR_q}^{\otimes n} \otimes \ket{\EPR_2}$.
  We specify the measurements of~$\strategy$ in
  Figure~\ref{fig:ms-operator-soln} as an \emph{operator solution} for the Magic
  Square game, meant to be read as follows: each cell contains a two-outcome
  projective measurement $\{E_0, E_1\}$ on $(\C^q)^{\otimes n} \otimes \C^2$
  written as its \emph{$\pm 1$-valued observable} $E_0 - E_1$.
  When Player~$\alice$ or~$\bob$ receives the question $\Variable_j$ for $j \in
  \{1, \ldots, 9\}$, they measure their share of~$\ket{\psi}$ using the
  measurement specified by the cell corresponding to~$\Variable_j$ and receive a
  single-bit measurement.
  When they receive the question $\Constraint_i$ for $i \in \{1,2,\ldots,6\}$,
  they simultaneously perform the three measurements in the corresponding row or
  column on $\ket{\psi}$ to obtain three bits.
  For example, if Player~$\alice$ receives question~$\Variable_1$, they
  measure~$\ket{\psi}$ using the measurement $\{A_0 \otimes I, A_1 \otimes I\}$
  corresponding to the observable $\cal{O}_A \otimes I$ (where the first
  operator acts on $\ket{\EPR_q}^{\otimes n}$ and the second acts on
  $\ket{\EPR_2}$).
  Similarly, on question~$\Variable_5$, they use the measurement $\{B_0 \otimes
  I, B_1 \otimes I\}$.
  This establishes Item~\ref{item:embeds} of the theorem.
  \begin{figure}[ht!]
    \begin{center}
      \begin{tabular}{|c|c|c|}
      \hline
      $ \cal{O}_A \otimes I$
      & $ I \otimes \sigma^X$
      & $ \cal{O}_A \otimes \sigma^X$ \\
      \hline
      $ I \otimes \sigma^Z$
      & $ \cal{O}_B \otimes I$
      & $ \cal{O}_B \otimes \sigma^Z$ \\
      \hline
      $ \cal{O}_A \otimes \sigma^Z$
      & $ \cal{O}_B \otimes \sigma^X$
      & $ \cal{O}_A \cal{O}_B \otimes \sigma^Z \sigma^X$ \\
      \hline
      \end{tabular}
      \caption{Observables for Magic Square strategy}
      \label{fig:ms-operator-soln}
		\end{center}
  \end{figure}
  
  First, we show that this gives a well-defined strategy.
  The $\Variable_j$ measurements are well-defined because each cell contains a
  $\pm 1$-valued observable.
  This is obvious for all $j \neq 9$; when $j = 9$, the bottom-right cell
  contains $\cal{O}_A \cal{O}_B \otimes \sigma^Z \sigma^X$.
  Because~$\cal{O}_A$ and~$\cal{O}_B$ anti-commute,
  \begin{equation}\label{eq:am-i-hermitian}
  \cal{O}_A \cal{O}_B \otimes \sigma^Z \sigma^X
  = - \cal{O}_A \cal{O}_B \otimes \sigma^X \sigma^Z
  = \cal{O}_B \cal{O}_A \otimes \sigma^X \sigma^Z
  = (\cal{O}_A \cal{O}_B \otimes \sigma^Z \sigma^X)^\dagger.
  \end{equation}
  As a result, this matrix is Hermitian. In addition, 
  \begin{equation*}
\begin{aligned}
&(\cal{O}_A \cal{O}_B \otimes \sigma^Z \sigma^X)^2\\
 &= (\cal{O}_A \cal{O}_B \otimes \sigma^Z \sigma^X) \cdot (\cal{O}_B \cal{O}_A \otimes \sigma^X \sigma^Z)\\
 &= (\cal{O}_A \cal{O}_B \cdot \cal{O}_B \cal{O}_A) \otimes (\sigma^Z \sigma^X \cdot \sigma^X \sigma^Z)\\
 &= I,
\end{aligned}
\end{equation*}
  where the first step uses Equation~\eqref{eq:am-i-hermitian} and the final
  step uses the fact that $\cal{O}_A, \cal{O}_B, \sigma^X, \sigma^Z$ are $\pm
  1$-valued observables and hence square to the identity.
  As a result, this matrix is Hermitian and squares to the identity.
  Therefore, it is a $\pm 1$-valued observable.
  
  As for the $\Constraint_i$ measurements,
  we must show that the three measurements in each row and column are simultaneously measurable.
  This is equivalent to the three $\pm 1$-valued observables being simultaneously diagonalizable,
  which is equivalent to them being pairwise commuting. This can be easily verified for the cases of $i = 1, 2, 4, 5$
  (i.e.,\ the first two rows and columns).
  In the case of $i = 3$, commutativity of $\cal{O}_A \otimes \sigma^Z$ and $\cal{O}_B \otimes \sigma^X$ 
  follows from Equation~\eqref{eq:am-i-hermitian}.
  Since these two matrices commute, they also commute with their product 
  $(\cal{O}_A \otimes \sigma^Z)(\cal{O}_B \otimes \sigma^X) = \cal{O}_A \cal{O}_B \otimes \sigma^Z \sigma^X$.
  The case of $i = 6$ is similar.
  
  By construction, $\strategy$ is symmetric, and we have already shown that it
  is projective.
  It remains to show that it is commuting, consistent, and value~$1$.
  To show that it is commuting, it suffices to show that the measurement for
  each cell is simultaneously measurable with all three measurements in its row
  or column, which was already proved above.
  Now we show consistency of the measurements.
  We will show this by instead showing consistency of their observables.
  By this we mean the following: if $E = \{E_0, E_1\}$ is a two-outcome projective measurement on $\cal{H}$
  then its corresponding $\pm 1$-valued observable $\cal{O} = E_0 - E_1$
  is consistent on a state $\ket{\phi} \in \cal{H} \otimes \cal{H}$ if
  \begin{equation*}
  \cal{O}\otimes I_{\bob} \cdot \ket{\phi} = I_{\alice} \otimes \cal{O} \cdot \ket{\phi}.
  \end{equation*}
  This is in fact equivalent to the notion of $E$ being consistent on $\ket{\phi}$ from Definition~\ref{def:consistent-measurement}.
  To see this, if $E$ is consistent on $\ket{\phi}$, then so is $\cal{O}$, because
  \begin{equation*}
  \cal{O}\otimes I_{\bob} \cdot \ket{\phi}
  = (E_0 - E_1) \otimes I_{\bob} \cdot \ket{\phi}
  = I_{\alice} \otimes (E_0 - E_1) \cdot \ket{\phi}
  = I_{\alice} \otimes \cal{O} \cdot \ket{\phi},
  \end{equation*}
  where the second equality is by consistency of~$E$.
  On the other hand, suppose $\cal{O}$ is consistent on $\ket{\phi}$.
  Then because $E_0 + E_1 = I$, we can write $E_0 = \tfrac{1}{2} \cdot (I + \cal{O})$ and $E_1 = \tfrac{1}{2} \cdot (I - \cal{O})$.
  As a result,
  \begin{equation*}
  E_0 \otimes I_{\bob} \cdot \ket{\phi}
  = \tfrac{1}{2} \cdot (I + \cal{O}) \otimes I_{\bob} \cdot \ket{\phi}
  = I_{\alice} \otimes \tfrac{1}{2} \cdot (I + \cal{O}) \cdot \ket{\phi}
  = I_{\alice} \otimes E_0 \cdot \ket{\phi},
  \end{equation*}
  and similarly for $E_0$. Thus, $E$ is consistent on $\ket{\phi}$.
  This proves the equivalence.
  
  Now we use this to prove consistency of $\strategy$.
  To begin, we note that because $A$ and $B$ are consistent on $\ket{\EPR_q}^{\otimes n}$ by assumption,
  so to are their $\pm 1$-valued observables $\cal{O}_A$ and $\cal{O}_B$.
  Next, we note that $\sigma^X$ and $\sigma^Z$ are consistent on $\ket{\mathrm{EPR}_2}$;
  to verify this, we recall that $\sigma^X \ket{0} = \ket{1}$ and $\sigma^X \ket{1} = \ket{0}$, and so
  \begin{align*}
  \sigma^X \otimes I_{\bob} \cdot \ket{\mathrm{EPR}_2}
&  =   \sigma^X \otimes I_{\bob} \cdot \tfrac{1}{\sqrt{2}} (\ket{00} + \ket{11})\\
&  = \tfrac{1}{\sqrt{2}} (\ket{10} + \ket{01})
 \\
&=    I_{\alice} \otimes \sigma^X \cdot \tfrac{1}{\sqrt{2}} (\ket{11} + \ket{00})
 \\
&=    I_{\alice} \otimes \sigma^X \cdot \ket{\mathrm{EPR}_2}.
  \end{align*}
  This shows $\sigma^X$ is consistent on $\ket{\mathrm{EPR}_2}$,
  and a similar argument shows the same for $\sigma^Z$.
  We now use these two facts to show that the observable in each cell of
  Figure~\ref{fig:ms-operator-soln} is consistent on $\ket{\psi}$.
  To see why, consider the $j = 9$ case:
  \begin{align*}
&  (\cal{O}_A \cal{O}_B \otimes \sigma^Z \sigma^X)_\alice \otimes I_\bob
		\cdot \ket{\EPR_q}^{\otimes n} \otimes \ket{\EPR_2} \\
  ={}& (\cal{O}_A \cal{O}_B \otimes \sigma^Z)_\alice \otimes (I \otimes \sigma^X)_\bob
  		\cdot \ket{\EPR_q}^{\otimes n} \otimes \ket{\EPR_2}\tag{consistency of $\sigma^X$}\\
  ={}& (\cal{O}_A \cal{O}_B \otimes I)_\alice \otimes (I \otimes \sigma^X\sigma^Z)_\bob
  		\cdot \ket{\EPR_q}^{\otimes n} \otimes \ket{\EPR_2}\tag{consistency of $\sigma^Z$}\\
  ={}& (\cal{O}_A \otimes I)_\alice \otimes ( \cal{O}_B \otimes \sigma^X\sigma^Z)_\bob
  		\cdot \ket{\EPR_q}^{\otimes n} \otimes \ket{\EPR_2}\tag{consistency of $\cal{O}_B$}\\
  ={}& I_\alice \otimes (\cal{O}_B \cal{O}_A \otimes \sigma^X\sigma^Z)_\bob
  		\cdot \ket{\EPR_q}^{\otimes n} \otimes \ket{\EPR_2}\tag{consistency of $\cal{O}_A$}\\
  ={}& I_\alice \otimes (\cal{O}_A \cal{O}_B \otimes \sigma^Z \sigma^X)_\bob
  		\cdot \ket{\EPR_q}^{\otimes n} \otimes \ket{\EPR_2}. \tag{by Equation~\eqref{eq:am-i-hermitian}}
  \end{align*}
  The remaining cases of $j \in\{1, \ldots, 8\}$ are similar, and we omit them.
  This shows that the observable in each cell is consistent on $\ket{\psi}$,
  and as a result the corresponding two-outcome projective measurements are consistent as well.
  As a result, the $\Variable_j$ measurements are consistent.
  
  As for the $\Constraint_i$ measurements, each such measurement \[\textstyle \{F_{a, b,
    c}\}_{a, b, c \in \{0, 1\}}\] is of the form $F_{a, b, c} = E^1_a \cdot E^2_b
  \cdot E^3_c$, where $E^1$, $E^2$, and $E^3$ are $\Variable$ measurements.
  But then consistency of~$F$ follows from the~$\Variable$ consistencies:
  \begin{align*}
  F_{a, b, c} \otimes I_{\bob} \ket{\psi}
 & = (E^1_a \cdot E^2_b \cdot E^3_c)_{\alice} \otimes I_{\bob} \ket{\psi}\\
 & = (E^1_a \cdot E^2_b)_{\alice} \otimes (E^3_c)_{\bob} \ket{\psi} \tag{consistency of $E^3$}\\
 &   = (E^1_a)_{\alice} \otimes (E^3_c \cdot E^2_b)_{\bob} \ket{\psi}\tag{consistency of $E^2$}\\
 &       = I_{\alice} \otimes (E^3_c \cdot E^2_b \cdot E^1_a)_{\bob} \ket{\psi}\tag{consistency of $E^1$}\\
 &       = I_{\alice} \otimes (E^1_a \cdot E^2_b \cdot E^2_c)_{\bob} \ket{\psi}\tag{commutativity of $E^1$, $E^2$, and $E^3$}\\
 &               = I_{\alice} \otimes F_{a, b,c} \ket{\psi}.
  \end{align*}
  Hence, all measurements are consistent.
  
  Since all measurements are consistent, this implies that the answer bit of the
  player receiving a $\Variable$ question is always consistent with the
  corresponding answer bit of the player receiving the $\Constraint$ question.
  Similarly, the answers of the player receiving the $\Constraint$ question
  always satisfy the given constraint; observe that in all rows and the first
  two columns, the observables multiply to $I$, whereas the observables in the
  last column multiply to $-I$.
  This implies that the strategy is value-$1$.
  To see why, 
  let us again consider a $\Constraint_i$ measurement, which we write as $F=\{F_{a, b, c}\}_{a, b, c \in \{0, 1\}}$.
  As above, it can be written as  $F_{a, b, c} = E^1_a \cdot E^2_b \cdot E^3_c$, where $E^1$, $E^2$, and $E^3$ are $\Variable$ measurements.
  Consider the measurement $S=\{S_0, S_1\}$ that corresponds to measuring $\{F_{a, b, c}\}$ and then outputting the sum of measured values of~$a$, $b$, and~$c$.
  In other words,
  \begin{equation*}
  S_0 = F_{0, 0, 0} + F_{0, 1, 1} + F_{1, 0, 1} + F_{1, 1, 0},
  \qquad
  S_1 = F_{0, 0, 1} + F_{0, 1, 0} + F_{1, 0, 0} + F_{1, 1, 1}.
  \end{equation*}
  The following expression gives a convenient formula for the $\pm 1$-valued observable $S_0 - S_1$:
  \begin{equation*}
\begin{aligned}
&(E_0^1 - E_1^1) \cdot (E_0^2 - E_1^2) \cdot (E_0^3 - E_1^3)\\
 &= E_0^1 E_0^2 E_0^3 - E_0^1 E_0^2 E_1^3 - E_0^1 E_1^2 E_0^3 + E_0^1 E_1^2 E_1^3 + \cdots\\
 &= S_0 - S_1.
\end{aligned}
\end{equation*}
  When we are looking at $\Constraint_i$ for $i \in \{1, 2, 3, 4, 5\}$,
  the left-hand side of this equation is equal to $I$, and so $I = S_0- S_1$.
  Because $S_0$ and $S_1$ are both positive semidefinite,
  this is true only if $S_0 = I$ and $S_1 = 0$, which implies that the three bits output by $F$ always sum to~$0$.
  As a result, this strategy always satisfies the $\Constraint_i$ question.
  On the other hand, when $i = 6$, then the left-hand side of this equation is equal to $-I$,
  and so $S_1 = I$, and the three bits sum to~$1$.
  Thus, the strategy always satisfies the $\Constraint_6$ question as well.
  This concludes the proof.
\end{proof}

\subsection{The Pauli basis test}
\label{sec:pauli-verifier}

We introduce the quantum low-degree test of \cite{natarajan2018low} in the form
of a slight modification to it by~\cite{NW19} known as the \emph{Pauli basis
  test}.
Informally, the quantum low-degree test asks the players to measure a large
number of qubits and return a highly compressed version of the measurement
outcome.
The Pauli basis test simply asks that the players return their uncompressed
measurement outcomes, and it is designed by direct reduction to the quantum
low-degree test.
In Section~\ref{sec:qld-game} we describe the Pauli basis test as a nonlocal
game $\game^\pauli$ and show that its question distribution is implementable via
a CL distribution, as we did with the classical low-degree test in
Section~\ref{sec:ld-verifier}.
In Section~\ref{sec:qld-complexity} we exhibit canonical parameters for the
Pauli basis test and give bounds on the time complexity of executing the test.

\subsubsection{The game}
\label{sec:qld-game}

We start by discussing parameter settings.
The game $\game^\pauli$ is parametrized by a tuple
\[\qldparams = (q,m,d)\;,\] where $q,m,d \in \N$ are integers. 
We sometimes write $\game^\pauli_\qldparams$ to emphasize the dependence of the
Pauli basis test on the parameters.

Informally, the test is meant to certify that the players share a state of the
form $\ket{\EPR_q}^{\otimes M}$, where $M=2^m$.
Its question set includes questions that are axis-parallel/diagonal lines and points in $\F_q^m$,
which are meant to correspond to questions in the classical low-degree test, and questions
of the form $(\Pauli, W)$, for $W \in \{X, Z\}$.
Upon receipt of a question of the latter form, the players are expected to
perform the POVM $\{\tau_h^W\}_{h \in \F_q^{M}}$ and report the outcome $h$ as
their answer.

The key idea behind the test
is for the provers to encode their outcomes using the \emph{low-degree encoding} from Section~\ref{sec:ld-encoding}.
Given an outcome $h \in \F_q^M$, the low-degree encoding of~$h$ is the polynomial $g_h:\F_q^m \rightarrow \F_q$.
Rather than asking the provers to always return the entire string~$h$,
many of the subtests in the game $\game^\pauli$ will ``probe'' limited information about~$h$ as follows:
the verifier provides the provers with a string $u_W \in \F_q^m$,
and using this they are expected to evaluate $g_h(u_W) \in \F_q$ and return it to the verifier.
For a fixed value of~$W$, then, the prover's responses to these ``probes'' should be consistent with some low-degree polynomial in the input~$u_W$.
To check this, the verifier can perform the classical low-degree test from Section~\ref{sec:ld-verifier}.
This ensures that the prover's responses for the $W = X$ basis are consistent with each other,
as well as their responses for the $W = Z$ basis, although it does not test consistency between the $X$ and $Z$ responses.

The next set of subtests the verifier performs ensure that the provers' $X$ and $Z$ measurements are consistent with each other.
On the $X$ side, consider the measurement in which the prover performs the $\{\tau_h^X\}$ POVM,
receives outcome $h_X$, and outputs $g_{h_X}(u_X) \in \F_q$.
For technical reasons that will become clear shortly,
it will be convenient for the prover's measurement to have two outcomes rather than~$q$ outcomes.
To accomplish this, the verifier also provides the prover with an element $r_X \in \F_q$;
given this, the prover should output not $g_{h_X}(u_X) \in \F_q$ but $\tr(g_{h_X}(u_X) \cdot r_X)$, which is an element of $\F_2$.
Similarly, on the $Z$ side, suppose the prover is provided $r_Z \in \F_q$,
performs $\{\tau_h^Z\}$ to receive $h_Z$, and outputs $\tr(g_{h_Z}(u_Z) \cdot r_Z) \in \F_2$.
Together, these two are a pair of two-outcome measurements, one in the~$X$ basis and one in the~$Z$ basis,
and as it turns out, this pair of measurements either commutes with each other or anticommutes with each other
(this is why we modified the measurements to have only two outcomes).
The key quantity to determine which of these is the case is the following complicated-looking expression:
\begin{equation*}
\gamma = \tr \bigl( (\ind_{m}(u_\xpt) \cdot r_\xpt) \cdot
      (\ind_{m}(u_\zpt) \cdot r_\zpt) \bigr)\;.
\end{equation*}
If $\gamma = 0$, then these two measurements commute, and if $\gamma = 1$, then these two measurements anticommute
(this fact is established in the proof of Lemma~\ref{lem:pauli-completeness} below).
In the $\gamma = 0$ case, the verifier can check whether these two measurements commute by performing the ``commutation check,''
which asks the provers to simultaneously measure both $\tr(g_{h_X}(u_X) \cdot r_X)$ and $\tr(g_{h_Z}(u_Z) \cdot r_Z)$ and report their values.
In the $\gamma = 1$ case, on the other hand, the measurements anticommute, which implies that they cannot simultaneously be measured.
In spite of this, the verifier can still check that the measurements anticommute by using the Magic Square game from Section~\ref{sec:ms} above.
As was shown in Theorem~\ref{thm:ms-from-ac}, any pair of anticommuting measurements
can be used to form a perfect strategy for the Magic Square game
(and indeed the reverse is true as well: any perfect strategy for the Magic Square game entails a pair of anticommuting measurements).

Together, these give the main subtests for the game $\game^\pauli$.

\begin{definition}[Admissible parameters]\label{def:admissible}
  \leavevmode\par We say that the tuple $\qldparams=(q,\allowbreak  m,\allowbreak  d)$ is \emph{admissible} if $q$ is
  an admissible field size (Definition~\ref{def:admissible-size}) and $m | q$.
\end{definition}

\paragraph{Question distribution.}
We now describe the question distribution of the game $\game^\pauli$, and show
that it is a (typed) CL distribution.
The question types in the game $\game^\pauli$ are
\begin{equation}
  \label{eq:pauli-type}
  \type^\pauli= \bigl( \{\Point, \ALine, \DLine, \Pauli, \Pair \} \times \{X,Z\} \bigr)
  \cup \type^\ms \cup \{\Pair\}\;,
\end{equation}
where $\type^\ms$ is the question type set of the Magic Square game defined in
Section~\ref{sec:ms}.


Before presenting the CL functions of the Pauli basis test, we first give an
intuitive description of the question distribution of the Pauli basis test: a
pair of questions $((\tvar_\alice, x_\alice), (\tvar_\bob, x_\bob))$ in
$\game^\pauli$ can be sampled via the following procedure:

\begin{enumerate}
\item Sample a pair of types by sampling an edge $(\tvar_\alice, \tvar_\bob)$ of
  the graph $G^\pauli$ given in Figure~\ref{fig:type-graph-pauli} uniformly at
  random (including the self-loops).
\item Sample the following uniformly at random:
	\begin{enumerate}
	\item (\emph{Points}) $u_\xpt,u_\zpt \in \F_q^m$,
	\item (\emph{Directions}) $s \in \F_q$, $v \in \F_q^m$,
  \item (\emph{Qubit basis for (anti-)commutation}) $r_\xpt,r_\zpt \in \F_q$.
	\end{enumerate}
\item For $w \in \AB$ and $W\in\{X,Z\}$,
	\begin{enumerate}
  \item If $\tvar_w = (\Point, W)$, then set $x_w = u_W$,
  \item If $\tvar_w = (\ALine, W)$, then set $x_w = (u_0,s)$, where $u_0 =
    L^\lnf_{e_i}(u_W)$, with $i = \chi(s)$ (see Section~\ref{sec:ld-game} for
    definition of $L^\lnf$ and $\chi(s)$),
  \item If $\tvar_w = (\DLine,W)$, then set $x_w = (u_0,s,v')$, where $i =
    \chi(s)$, $v' = \pi_{i-1}(v)$, and $u_0 = L^\lnf_{v'}(u_W)$,
  \item If $\tvar_w = \Constraint_i$ for some $i\in\{1,\allowbreak \ldots,\allowbreak 6\}$, then set
    $x_w = (u_\xpt,\allowbreak  u_\zpt,\allowbreak  r_\xpt,\allowbreak  r_\zpt)$,
  \item If $\tvar_w = \Variable_j$ for some $j\in\{1,\allowbreak \ldots,\allowbreak 9\}$, then set $x_w
    = (u_\xpt,\allowbreak  u_\zpt,\allowbreak  r_\xpt,\allowbreak  r_\zpt)$,
  \item If $\tvar_w = \Pair$, then set $x_w = (u_\xpt,\allowbreak  u_\zpt,\allowbreak  r_\xpt,\allowbreak  r_\zpt)$,
  \item If $\tvar_w = (\Pair, W)$, then set $x_w = (u_\xpt, u_\zpt, r_\xpt, r_\zpt)$,
  \item If $\tvar_w = (\Pauli, W)$, then set $x_w = 0$.
	\end{enumerate}
\end{enumerate}

\begin{figure}[!htbp]
  \centering
  \begin{tikzpicture}[scale=.8]

    \tikzset{type/.style args={[#1]#2}{
        draw,circle,fill,scale=0.25,
        label={[font=\scriptsize, label distance=1pt]#1:#2}
      }}


    \foreach \i in {1,...,6} \draw (0,8-9/8*\i)
    coordinate (Constraint-\i)
    node[type={[180]$\Constraint_\i$}] {};

    \foreach \i in {1,...,9} \draw (2.5,9-\i)
    coordinate (Variable-\i)
    node[type={[330]$\Variable_\i$}] {};

    \foreach \i in {1,...,3} \foreach \j in {1,...,3}
    \pgfmathsetmacro{\k}{(\i-1)*3+\j}
    \draw (Constraint-\i) -- (Variable-\k);

    \foreach \i in {4,...,6} \foreach \j in {1,...,3}
    \pgfmathsetmacro{\k}{\i-3+(\j-1)*3}
    \draw (Constraint-\i) -- (Variable-\k);


  	\draw (6,9) coordinate (DLine-X) node[type={[90]$(\DLine, X)$}] {};
    \draw (8,9) coordinate (ALine-X) node[type={[90]$(\ALine, X)$}] {};
    \draw (7,8) coordinate (Point-X) node[type={[315]$(\Point, X)$}] {};
    \draw (9.5,8) coordinate (Pauli-X) node[type={[0]$(\Pauli, X)$}] {};
    \draw (6,3) coordinate (DLine-Z) node[type={[270]$(\DLine, Z)$}] {};
    \draw (8,3) coordinate (ALine-Z) node[type={[270]$(\ALine, Z)$}] {};
    \draw (7,4) coordinate (Point-Z) node[type={[45]$(\Point, Z)$}] {};
    \draw (9.5,4) coordinate (Pauli-Z) node[type={[0]$(\Pauli, Z)$}] {};
    \draw (8,6) coordinate (Pair) node[type={[0]$\Pair$}] {};
    \draw (7.3,6.8) coordinate (Pair-X) node[type={[0]$(\Pair,X)$}] {};
    \draw (7.3,5.2) coordinate (Pair-Z) node[type={[0]$(\Pair,Z)$}] {};


    \foreach \from/\to in {ALine-X/Point-X, DLine-X/Point-X, Point-X/Pauli-X,
      DLine-Z/Point-Z, ALine-Z/Point-Z, Point-Z/Pauli-Z,
      Point-X/Variable-1, Point-Z/Variable-5}
    \draw (\from) -- (\to);

    \draw (Point-X) to [out=275,in=120] (Pair-X) to [out=300,in=140] (Pair);
    \draw (Point-Z) to [out=85,in=240] (Pair-Z) to [out=60,in=220] (Pair);

  \end{tikzpicture}
  \caption{Graph $G^\pauli$ for the Pauli basis test.
    Each vertex also has a self-loop which is not drawn on the figure for
    clarity.}
  \label{fig:type-graph-pauli}
\end{figure}

We now specify the corresponding CL functions for each of the question types in
the Pauli basis test.

\paragraph{Conditional linear functions for the Pauli basis test.}
The CL functions corresponding to the Pauli basis test question distribution are
parameterized by the parameter tuple $\qldparams = (q,m,d)$.
Let $V^\pauli$ denote the linear space $(\F_q^m)^2 \times \F_q \times \F_q^m
\times (\F_q)^2$.
The space $V^\pauli$ is decomposed into a direct sum of the following register
subspaces: $V_\xpt,V_\zpt$ (which are $m$-dimensional), $V_\coord$ (which is
$1$-dimensional), $V_{\dir{}}$ (which is $m$-dimensional), and
$V_{\rxpt},V_{\rzpt}$ (which are $1$-dimensional).
We identify elements of $V^\pauli$ as tuples $(u_\xpt,u_\zpt,s,v,r_\xpt,r_\zpt)
\in (\F_q^m)^2 \times \F_q \times \F_q^m \times (\F_q)^2$.
We define CL functions $L_\tvar: V^\pauli \to V^\pauli$ for every type $\tvar
\in \type^\pauli$:
\begin{enumerate}
\item For $W \in \{X,Z\}$, define $L_{\Point,W}(u_\xpt,u_\zpt,s,v,r_\xpt,r_\zpt)
  = L_\Point(u_W,s,v)$ where $L_\Point$ is the $1$-level CL function defined in
  \eqref{eq:cl-ptf}.\footnote{The range $V_{\xpt} \oplus V_\coord \oplus
    V_{\dir{}}$ or $V_{\zpt} \oplus V_\coord \oplus V_{\dir{}}$ of $L_\Point$ is
    embedded in $V^\pauli$ in the natural way.
    The same convention is used also in Item 2 and 3 for $L_\ALine$ and
    $L_\DLine$.}
\item For $W \in \{X,Z\}$, define $L_{\ALine,W}(u_\xpt,u_\zpt,s,v,r_\xpt,r_\zpt)
  = L_\ALine(u_W,s,v)$ where $L_\ALine$ is the $2$-level CL function defined in
  \eqref{eq:cl-alnf}.
\item For $W \in \{X,Z\}$, define $L_{\DLine,W}(u_\xpt,u_\zpt,s,v,r_\xpt,r_\zpt)
  = L_\DLine(u_W,s,v)$ where $L_\DLine$ is the $3$-level CL function defined in
  \eqref{eq:cl-dlnf}.
\item For $\tvar \in \type^\ms \cup \{\Pair,\allowbreak  (\Pair,\allowbreak X),\allowbreak  (\Pair,\allowbreak Z) \}$,
  $L_{\tvar}(u_\xpt,\allowbreak u_\zpt,\allowbreak s,\allowbreak v,\allowbreak r_\xpt,\allowbreak r_\zpt) =
  (u_\xpt,\allowbreak u_\zpt,\allowbreak 0,\allowbreak 0,\allowbreak r_\xpt,\allowbreak r_\zpt)$, i.e., it projects onto $V_\xpt \oplus
  V_\zpt \oplus V_{\rxpt} \oplus V_{\rzpt}$.
  Thus $L_{\tvar}$ is a $1$-level CL function.
  
\item For $W \in \{X,Z\}$, define $L_{\Pauli,W} = 0$ as the identically $0$
  function.
  Thus it is a $0$-level CL function.

\end{enumerate}
We note that the maps $L_{\Point,W}, L_{\ALine,W}, L_{\DLine,W}$ act as the zero function
on $V_{\overline{W}} \oplus V_{\rxpt} \oplus V_{\rzpt}$ where $\overline{W} = X$
if $W = Z$ and $\overline{W} = Z$ if $W = X$.

The question distribution of $\game^\pauli$ is thus a typed CL distribution on
$\type^\pauli \times V^\pauli \times \type^\pauli \times V^\pauli$ where
$(\tvar_\alice,x_\alice,\tvar_\bob,x_\bob)$ is sampled by first uniformly
sampling an edge $(\tvar_\alice,\tvar_\bob)$ from the graph $G^\pauli$ defined
in Figure~\ref{fig:type-graph-pauli}, sampling a uniformly random $z \in
V^\pauli$, and then setting $x_w = L_{\tvar_w}(z)$ for $w \in \AB$.
%
%
%


\paragraph{Decision procedure.}
The decision procedure for $\game^\pauli$ is presented in
Figure~\ref{fig:decider_pauli}.
Similarly to Figure~\ref{fig:ld-decider}, we provide a table that summarizes a
parsing scheme for the questions and answers, depending on the type of question.
The answers are bit strings that are interpreted as more structured objects such
as elements over $\F_q$, vectors, or polynomials, depending on the question.
In the ``low-degree check,'' the decision procedure $\decider^\pauli$ calls the
classical low-degree decision procedure $\decider^\ld$ parametrized by the tuple
$\ldparams = (q,m,d,1)$ (defined in Section~\ref{sec:ld-verifier}) as a
subroutine.

\begingroup
  \begin{gamespec}
    
    \begin{center}
      \begin{tabularx}{.92\textwidth}{ l X l }
        \toprule
        Type & Question Content & Answer Format \\
        \midrule
        (\Point, $W$) & $y \in  \F_q^m$ & Element of $\F_q$  \\
        (\ALine, $W$) & $(u_W, s) \in \F_q^m \times \F_q$
        & Polynomial $f: \F_q \to \F_q$\\
				&& of degree at most $d$\\
        (\DLine, $W$) & $(u_W,\allowbreak  s,\allowbreak  v) \in \F_q^m \times \F_q \times \F_q^m$ 
        & Polynomial $f: \F_q \to \F_q$ \\
				&& of degree at most $md$\\
        \Pair & $(u_\xpt,\allowbreak  u_\zpt,\allowbreak  r_\xpt,\allowbreak  r_\zpt) \in
        (\F_q^m)^2 \times \F_q^2 $ & $(\beta_\xpt,\allowbreak \beta_\zpt) \in \F_2^2$ \\
        (\Pair, $W$) & $(u_\xpt,\allowbreak  u_\zpt,\allowbreak  r_\xpt,\allowbreak  r_\zpt) \in
        (\F_q^m)^2 \times \F_q^2 $ & Element of $\F_2$ \\
        $\Constraint_i$ & $(u_\xpt,\allowbreak  u_\zpt,\allowbreak  r_\xpt,\allowbreak  r_\zpt) \in (\F_q^m)^2
        \times \F_q^2$ &
        $(\alpha_{v_1},\allowbreak  \alpha_{v_2},\allowbreak  \alpha_{v_3}) \in \F_2^3$ \\
        $\Variable_j$ & $(u_\xpt,\allowbreak  u_\zpt,\allowbreak  r_\xpt,\allowbreak  r_\zpt) \in
        (\F_q^m)^2 \times \F_q^2$ & Element of $\F_2$ \\
        (\Pauli, $W$) & $0$ & Element of $\F_q^{M}$ \\
        \bottomrule \multicolumn{3}{c}{Table: Question and answer format of the
          Pauli basis game.}
      \end{tabularx}
    \end{center}

    On input $(\tvar_\alice,\allowbreak x_\alice,\allowbreak  \tvar_\bob,\allowbreak  x_\bob,\allowbreak  a_\alice,\allowbreak  a_\bob)$,
    the decision procedure $\decider^\pauli$ \tnote{added:}first checks that $(\tvar_\alice,x_\alice)$ and $(\tvar_\bob,x_\bob)$ are correctly formatted --- the types correspond to valid types, and the questions have the correct length --- and then that $a_\alice$ and $a_\bob$ also have the correct length, as can be inferred from the table above. If not, it rejects. It then 
		performs the following checks for
    $w \in \AB$:
    \begin{enumerate}
    \item (\textbf{Consistency check}).
      \label{item:same-type}
      If $\tvar_\alice = \tvar_\bob$, accept iff $a_\alice = a_\bob$.

    \item (\textbf{Low-degree}).
      \label{item:low-degree}
      Let $\ldparams = (q,m,d,1)$. 
      \begin{enumerate}
      	\item If $\tvar_w = (\Point,\allowbreak  W),\allowbreak  \tvar_{\overline{w}} = (\ALine,\allowbreak  W)$, accept if
      $\decider^\ld_{\ldparams}$ accepts $(\Point,\allowbreak  x_w,\allowbreak  \ALine,\allowbreak  x_{\overline{w}},\allowbreak  a_w,\allowbreak 
      a_{\overline{w}})$.
      	\item If $\tvar_w = (\Point,\allowbreak  W),\allowbreak  \tvar_{\overline{w}} = (\DLine,\allowbreak  W)$, accept if
      $\decider^\ld_{\ldparams}$ accepts $(\Point,\allowbreak  x_w,\allowbreak  \DLine,\allowbreak  x_{\overline{w}},\allowbreak  a_w,\allowbreak 
      a_{\overline{w}})$.
		\end{enumerate}
    \item (\textbf{Consistency check}).
      \label{item:low-degree-consistency}
      If $\tvar_w = (\Point,\allowbreak  W),\allowbreak  \tvar_{\overline{w}} = (\Pauli,\allowbreak  W)$, accept if
      $g_{a_{\overline{w}}}(x_w) = a_w$, where $g_{a_{\overline{w}}}$ is
      the low-degree encoding of $a_{\overline{w}} \in \F_q^M$ defined in
      Section~\ref{sec:ld-encoding}.
    \end{enumerate}
    In the remaining four cases, the decision procedure first computes the
    number
    \begin{equation}
      \label{eq:gamma-value}
      \gamma = \tr \bigl( (\ind_{m}(u_\xpt) \cdot r_\xpt) \cdot
      (\ind_{m}(u_\zpt) \cdot r_\zpt) \bigr)\;,
    \end{equation}
    where we recall the $\ind_{m}(\cdot)$ vector from Section~\ref{sec:ld-encoding}.
    \begin{enumerate}
      \setcounter{enumi}{3}
    \item (\textbf{Commutation check}).
      \label{item:commutation} If $\tvar_w = (\Pair,\allowbreak  W),\allowbreak  \tvar_{\overline{w}} =
      \Pair$, accept if $a_w = \beta_W$ or $\gamma \neq 0$.
    \item (\textbf{Consistency check}).
      \label{item:commutation-consistency} If $\tvar_w = (\Point,\allowbreak W),\allowbreak  \tvar_{\overline{w}} =
      (\Pair,\allowbreak W)$, accept if $\tr (a_w r_W) = a_{\overline{w}}$ or $\gamma \neq 0$.
      
    \item (\textbf{Magic square check}).
      \label{item:magic-square} If $\tvar_w = \Constraint_i,\allowbreak 
      \tvar_{\overline{w}} = \Variable_j$, accept if $\gamma = 0$, or $a_w$
      satisfies constraint $\Constraint_i$ and $ \alpha_j = a_{\overline{w}}$.

    \item (\textbf{Consistency check}).
      \label{item:magic-square-consistency}
      If $\tvar_w = (\Point,\allowbreak W),\allowbreak  \tvar_{\overline{w}} = \Variable_j$, accept if
      $\gamma = 0$ or if $j = 1$, $W = X$, and $\tr(a_w r_\xpt) =
      a_{\overline{w}}$ or if $j = 5$, $W = Z$, and $\tr(a_w r_\zpt) =
      a_{\overline{w}}$.
    \end{enumerate}
  \end{gamespec}
  \captionof{figure}{Specification of the decision procedure $\decider^\pauli_\qldparams$.}
  \label{fig:decider_pauli}
\endgroup

We describe an honest, value-$1$ PCC strategy for the Pauli basis test.
%

\begin{lemma}\label{lem:pauli-completeness}
  The Pauli basis test $\game^\pauli_\qldparams$ has a value-$1$ SPCC strategy.
\end{lemma}

\begin{proof}
  We begin by specifying the value-$1$ strategy $\strategy = (\ket{\psi}, E)$.
  The state is
  \begin{equation*}
    \ket{\psi} = \ket{\EPR_q}^{\otimes M} \otimes \ket{\EPR_2}\;.
  \end{equation*}
  Now we specify the measurements.
  We start with measurements associated with questions of type $\Point$,
  $\ALine$, $\DLine$, and $\Pauli$.
  Using notation introduced in Section~\ref{sec:ld-encoding}, for $W \in \{X,
  Z\}$,
  \begin{align*}
    E^{(\Point,W),\, y}_{\, a}
    & = \tau^W_{[g_{\cdot}(y) = a]} \otimes I\;,\\
    E^{(\ALine,W),\, \line}_{\, f}
    & = \tau^W_{[(g_{\cdot})|_\line = f]} \otimes I\;,\\
    E^{(\DLine,W),\, \line}_{\, f}
    & = \tau^W_{[(g_{\cdot})|_\line = f]} \otimes I\;,\\    
    E^{(\Pauli,W)}_{\, a}
    & = \tau^W_{a} \otimes I\;.
  \end{align*}
  Here, the bracket notation used to post-process measurement outcomes is
  defined in Definition~\ref{def:bracket}.
  Thus, for example, the measurement on question $((\Point,W),y)$ corresponds to
  first performing the measurement $\{\tau^W_h\}_{h \in \F_q^M}$, receiving an
  outcome $h \in \F_q^M$, and then outputting the value $a = g_{h}(y)$.

  For $\ALine$ and $\DLine$ questions, the question content $\line$ denotes
  either an axis-parallel line $\line(u_0,e_i)$ for some $i \in
  \{1,2,\ldots,m\}$, or a diagonal line $\line(u_0,v')$ for some $v' \in
  \F_q^m$.
  As described at the beginning of Section~\ref{sec:qld-game}, axis-parallel lines are
  specified by pairs $(u_0,s) \in \F_q^m \times \F_q$ and diagonal lines are
  specified by triples $(u_0,s,v) \in \F_q^m \times \F_q \times \F_q^m$.
  The measurement on question $((\ALine,W),\line)$, for example, corresponds to
  first performing the measurement $\{\tau^W_h\}_{h \in \F_q^M}$ to get an
  outcome $h \in \F_q^M$, and reporting the univariate polynomial $f_\line(t) =
  g_{h}(u_0 + te_i)$ where $\line$ is specified by $(u_0,s)$ and $i = \chi(s)$.
  
  
  Next we specify the POVMs associated with questions of type\newline $\Constraint$,
  $\Variable$, and $\Pair$.
  Questions with these question types have a question content that is a tuple
  $\omega= (u_\xpt,\allowbreak  u_\zpt,\allowbreak  r_\xpt,\allowbreak  r_\zpt)\in (\F_q^m)^2 \times \F_q^2$.
  Given such a tuple, consider the two $\F_2$-valued POVMs $A=\{A_b\}_{b \in
    \F_2}$ and $B=\{B_b\}_{b \in \F_2}$ defined as
  \begin{align}
    A_b & = \tau^X_{[\tr(g_{\cdot}(u_X)\, r_\xpt) = b]} \otimes I\;,\label{eq:gonna-expand-A}\\
    B_b & = \tau^Z_{[\tr(g_{\cdot}(u_Z)\, r_\zpt) = b]} \otimes I\;.\label{eq:gonna-expand-B}
  \end{align}
  We would like to determine when the two observables $\mathcal{O}_A = A_0 -
  A_1$ and $\mathcal{O}_B = B_0 - B_1$ commute or anti-commute.
  Towards this we derive alternative expressions for these observables from
  which their commutativity becomes plain from inspection.
  We begin by inspecting the first matrices on the right-hand sides of
  Equations~\eqref{eq:gonna-expand-A} and~\eqref{eq:gonna-expand-B}:
  \begin{align}
  \tau^W_{[\tr(g_{\cdot}(u_W)\, r_W) = b]}
  &= \sum_{h : \tr(g_{h}(u_W)\, r_W) = b} \tau^W_h\nonumber\\
  &= \sum_{h : \tr((h\, \cdot\, \ind_{m}(u_W)) \cdot r_W ) = b} \tau^W_h\;,\label{eq:expand-with-dot-product}
  \end{align}
  where~\eqref{eq:expand-with-dot-product}
  follows by  Definition~\ref{eq:low-degree-encoding-definition}.
  As a result,
  \begin{align}
&  \tau^W_{[\tr(g_{\cdot}(u_W)\, r_W) = 0]}
  -   \tau^W_{[\tr(g_{\cdot}(u_W)\, r_W) = 1]} \notag \\
  & = \sum_{h : \tr((h\, \cdot\, \ind_{m}(u_W)) \cdot r_W ) = 0} \tau^W_h
  	- \sum_{h : \tr((h\, \cdot\, \ind_{m}(u_W)) \cdot r_W ) = 1} \tau^W_h\nonumber\\
  & = \sum_{h} (-1)^{\tr((h\, \cdot\, \ind_{m}(u_W)) \cdot r_W )} \tau^W_h\nonumber\\
  & = \tau^W(\ind_{m}(u_W)r_W)\;,\label{eq:finally-got-an-observable}
  \end{align}
  where the last step uses~\eqref{eq:pauli-obs-proj}.
  As a result, Equation~\eqref{eq:finally-got-an-observable}
  and Equations~\eqref{eq:gonna-expand-A} and~\eqref{eq:gonna-expand-B} imply that
  \begin{equation*}
  \mathcal{O}_A = \tau^X(\ind_{m}(u_X)r_X) \otimes I,\quad
  \mathcal{O}_B = \tau^Z(\ind_{m}(u_Z)r_Z) \otimes I\;.
  \end{equation*}
  Now, let $\gamma = \tr((\ind_{m}(u_X)r_X) \cdot (\ind_{m}(u_Z)r_Z)) \in \F_2$,
  as in Equation~\eqref{eq:gamma-value}.
  Then by Equation~\eqref{eq:twisted-fq},
  \begin{equation*}
  \mathcal{O}_A \mathcal{O}_B = (-1)^{\gamma} \mathcal{O}_B \mathcal{O}_A\;.
  \end{equation*}
  As a result, $\gamma$ quantifies whether the observables $\mathcal{O}_A$ and
  $\mathcal{O}_B$ commute, and, therefore, whether the measurements~$A$ and~$B$
  commute.
  If $\gamma = 0$, they commute.
  If $\gamma = 1$, they anti-commute.
  We now specify the POVMs associated with questions of type $\Constraint$,
  $\Variable$, and $\Pair$, considering separately the cases when $\gamma = 0$
  or $\gamma = 1$.
  \begin{enumerate}
  \item If $\gamma = 0$, 
     for each $\beta_X, \beta_Z \in \F_2$ define
    \begin{equation}\label{eq:pair-definition}
      E^{\Pair,\, \omega}_{\beta_X,\, \beta_Z} = A_{\beta_X} \cdot B_{\beta_Z}\;,
    \end{equation}
    \begin{equation*}
      E^{(\Pair,X),\, \omega}_{a} = A_{a} \;,\qquad
      E^{(\Pair,Z),\, \omega}_{a} = B_{a}\;.
    \end{equation*}
    The POVM associated with questions of type $\Constraint$ and $\Variable$ are
    defined to be trivial.
    In particular, we define
    \begin{equation*}
      E^{\Constraint_i,\, \omega}_{\, 0,\, 0,\, 0}
      = E^{\Variable_j,\, \omega}_{\, 0} = I\;,
    \end{equation*}
    and the remaining POVM elements in these measurements are set to be zero.
  \item If $\gamma = 1$, then $\mathcal{O}_A$ and $\mathcal{O}_B$ anti-commute.
    In this case we define measurements $E^{\Constraint_i,\, \omega}$ and
    $E^{\Variable_j,\, \omega}$, for all $i\in\{1,\ldots,6\}$ and $j
    \in\{1,\ldots,9\}$ to be those guaranteed by Theorem~\ref{thm:ms-from-ac}.
    Measurements associated with inputs of type $\Pair$ are defined to be
    trivial.
    In particular, we define
    \begin{equation*}
      E^{\Pair,\, \omega}_{0,\, 0} = E^{(\Pair,X),\, \omega}_{0} = E^{(\Pair,Z),\, \omega}_{0} = I\;,
    \end{equation*}
    and the remaining matrices in these measurements are set to be zero.
  \end{enumerate}
  This completes the specification of the strategy.

  Now, we show that $\strategy$ is a value-$1$ SPCC strategy.
  It is clearly symmetric and projective. 
  To show that it is consistent, we note that all measurements are Pauli basis
  measurements, which are consistent, or measurements produced by
  Theorem~\ref{thm:ms-from-ac}, which are also consistent.
  The only exception is the $\Pair$ measurement in the $\gamma = 0$ case, which
  by Equation~\eqref{eq:pair-definition} is a product of two commuting,
  consistent measurements, and so it is, therefore, also consistent.
  To show that it is commuting, we note that for each $W \in \{X, Z\}$, all
  $(\Point, W)$, $(\ALine, W)$, $(\DLine,W)$ and $(\Pauli, W)$ measurements
  commute as they are all measurements in the Pauli~$W$ basis.
  Next, if $\gamma = 0$ then the $\Constraint$ and $\Variable$ measurements
  commute trivially, and for $W \in \{X, Z\}$, the $(\Point, W)$ measurement
  commutes with the $(\Pair, W)$ measurement, as they are both $W$ basis
  measurements, and the measurement $E^{\Pair, \omega}_{\beta_X, \, \beta_Z} =
  A_{\beta_X}\cdot B_{\beta_Z}$ commutes with $E^{(\Pair, W), \omega}_{a}$
  because $A$ and~$B$ commute.
  On the other hand, if $\gamma \neq 0$, then the $\Pair$ measurements commute
  trivially, and the $\Constraint$ and $\Variable$ measurements commute by
  Theorem~\ref{thm:ms-from-ac}.
  Finally, the $(\Point, X)$ measurement commutes with $\Variable_1$ because
  both are $X$-basis measurements, and likewise both $(\Point, Z)$ and
  $\Variable_5$ are $Z$-basis measurements.
  
  It remains to show that $\strategy$ is value-$1$.
	Consider first the first three tests executed by the decision procedure in
  Figure~\ref{fig:decider_pauli}.
  The strategy passes the consistency checks with probability~$1$ because it is
  projective and consistent.
  It passes the low-degree checks because it answers those consistently with an
  honest strategy in the classical low-degree test.
		
  Next, consider the remaining four tests.
  Fix an $\omega= (u_\xpt, u_\zpt, r_\xpt, r_\zpt)$ and~$\gamma$ as
  in~\eqref{eq:gamma-value}.
  If $\gamma = 0$, then the strategy passes the commutation check with
  probability~$1$ by construction.
  As for the consistency check in Item~\ref{item:commutation-consistency},
  we can write the $(\Point, W)$ measurement as follows: 
  \begin{equation*}
  E^{(\Point,W),\, y}_{[\tr(\,\cdot\, r_W) = a_{\overline{w}}]}
  = \tau^W_{[\tr(g_{\cdot}(y)r_W) = a_{\overline{w}}]} \otimes I
  = E^{(\Pair, W), \omega}_{a_{\overline{w}}} \otimes I.
  \end{equation*}
  As a result, due to the consistency of the $(\Pair, W)$ measurement, the
  consistency check is passed with probability~$1$.
  On the other hand, if $\gamma \neq 0$, then the strategy passes the Magic
  Square check by Theorem~\ref{thm:ms-from-ac}.
  As for the consistency check in Item~\ref{item:magic-square-consistency}, we
  can write the $(\Point, X)$ measurement as follows:
  \begin{equation*}
  E^{(\Point,X),\, y}_{[\tr(\,\cdot\, r_X) = a_{\overline{w}}]}
  = \tau^X_{[\tr(g_{\cdot}(y)r_X) = a_{\overline{w}}]} \otimes I
  = E^{\Variable_1,\, \omega}_{a_{\overline{w}}} \otimes I,
  \end{equation*}
  where the last step is by Theorem~\ref{thm:ms-from-ac}.
  As these measurements are consistent, this test is passed with
  probability~$1$.
\end{proof}

\paragraph{Soundness of the Pauli basis test.} 
We state the soundness properties of the Pauli basis test.
The following is an adaptation of the self-testing statement in~\cite[Theorem
6.4]{NW19}.

\begin{theorem}
  \label{thm:pauli}
There exists a function 
\[\delta_{\qld}(\eps,m,d,q) = a(md)^a(\eps^b + q^{-b} + 2^{-bmd})\]
 for universal constants $a \geq 1$ and $0 < b < 1$ such that the following holds. For all admissible parameter tuples $\qldparams = (q,m,d)$ and for all strategies $\strategy = (\ket{\psi},A,B)$ for the game $\game^\pauli_\qldparams$ that succeed with probability at least $1 - \eps$, there exist local isometries $\phi_\alice: \mH_\alice \to \mH_{\alice'} \otimes \mH_{\alice''},\phi_\bob: \mH_\bob \to \mH_{\bob'} \otimes \mH_{\bob''}$ (where $\ket{\psi} \in \mH_\alice \otimes \mH_\bob$ and $\mH_{\alice''},\mH_{\bob''} \cong (\C^q)^{\otimes M}$ with $M = 2^m$) and a state $\ket{\aux} \in \mH_{\alice'} \otimes \mH_{\bob'}$ such that 
\begin{enumerate}
    \item $\left \| \phi_\alice \otimes \phi_B \ket{\psi} - \ket{\aux} \otimes \ket{\EPR_q}^{\otimes M}  \right \| \leq \delta_{\qld}(\eps,m,d,q)$,
    \item Letting $\tilde{A}^x_a = \phi_A\, A^x_a \, \phi_A^\dagger$ and $\tilde{B}^y_b = \phi_B B^y_b \phi_B^\dagger$, we have for $W \in \{X,Z\}$
    \begin{gather*}
        \tilde{A}^{(\pauli,W)}_u \otimes I_{\bob'\bob''} \approx_{\delta_{\qld}} (\tau^W_u)_{\alice''} \otimes I_{\alice' \bob' \bob''} \\
        I_{\alice'\alice''} \otimes \tilde{B}^{(\pauli,W)}_u    \approx_{\delta_{\qld}} I_{\alice' \alice'' \bob'} \otimes (\tau^W_u)_{\bob''}\;,
    \end{gather*}
    where the $\approx_{\delta_{\qld}}$ statement holds with respect to the state $\ket{\aux}_{\alice'\bob'} \otimes \ket{\EPR_q}^{\otimes M}_{\alice''\bob''}$ and the answer summation is over $u \in \F_q^M$.
\end{enumerate}
\end{theorem}

We provide a proof of Theorem~\ref{thm:pauli} in Section~\ref{sec:qld-analysis}.

The strategies described in the proof of Lemma~\ref{lem:pauli-completeness} and in the conclusion of Theorem~\ref{thm:pauli} are described in terms of qudit Pauli measurements and maximally entangled states defined over $q$-dimensional qudits. 
However, it is more convenient for our application of the Pauli basis test to
instead deal with \emph{qubits}.
In particular, we use the Pauli basis test in the ``introspection game'' of
Section~\ref{sec:introspection}, where the players are commanded to sample
questions according to a sampler $\sampler$ of a normal form verifier.
By definition of normal form verifier, $\sampler$ is a sampler over $\F_2$, and,
therefore, it is natural to use a test for qubit Pauli measurements.

Since we use field sizes $q$ that are powers of $2$, Lemma~\ref{lem:pauli-binary}
shows that the conclusions of Theorem~\ref{thm:pauli} can also be described in terms of
testing for maximally entangled states over qubits and qubit Pauli measurements.



\begin{corollary}
\label{cor:pauli-binary}  
There exists a function
\[\delta_{\qld}(\eps,q,m,d) = a(md)^a(\eps^b + q^{-b} + 2^{-bmd})\]
for universal constants $a \geq 1$ and $0 < b < 1$ such that the following holds. For all admissible parameter tuples $\qldparams = (q,m,d)$ and for all strategies $\strategy = (\ket{\psi},A,B)$ for the game $\game^\pauli_\qldparams$ that succeeds with probability at least $1 - \eps$, there exist local isometries $\phi_\alice: \mH_\alice \to \mH_{\alice'} \otimes \mH_{\alice''},\allowbreak \phi_\bob: \mH_\bob \to \mH_{\bob'} \otimes \mH_{\bob''}$ (where $\ket{\psi} \in \mH_\alice \otimes \mH_\bob$ and $\mH_{\alice''},\mH_{\bob''} \cong (\C^2)^{\otimes M \log q}$ with $M = 2^m$) and a state $\ket{\aux} \in \mH_{\alice'} \otimes \mH_{\bob'}$ such that 
\begin{enumerate}
    \item $\left \| \phi_\alice \otimes \phi_B \ket{\psi} - \ket{\aux} \otimes \ket{\EPR_2}^{\otimes M \log q}  \right \| \leq \delta_{\qld}(\eps,\allowbreak m,\allowbreak d,\allowbreak q)$,
    \item Letting $\tilde{A}^x_a = \phi_A\, A^x_a \, \phi_A^\dagger$ and $\tilde{B}^y_b = \phi_B B^y_b \phi_B^\dagger$, we have for $W \in \{X,Z\}$
    \begin{gather*}
        \tilde{A}^{(\pauli,W)}_u \otimes I_{\bob'\bob''} \approx_{\delta_{\qld}} (\sigma^W_u)_{\alice''} \otimes I_{\alice' \bob' \bob''} \\
        I_{\alice'\alice''} \otimes \tilde{B}^{(\pauli,W)}_u    \approx_{\delta_{\qld}} I_{\alice' \alice'' \bob'} \otimes (\sigma^W_u)_{\bob''}
    \end{gather*}
    where the $\approx_{\delta_{\qld}}$ statement holds with respect to the state $\ket{\aux}_{\alice'\bob'} \otimes \ket{\EPR_2}^{\otimes M \log q}_{\alice''\bob''}$ and the answer summation is over $u \in \F_q^M$. Here, $\sigma^W_u$ denotes the tensor product of single-qubit Pauli measurements
    \[
    	\bigotimes_{i=1}^{M} \bigotimes_{j=1}^{\log q} \sigma^W_{u_{ij}}
    \]
    where $\kappa(u_i) = (u_{ij})_{1 \leq j \leq \log q}$ (using the bijection $\kappa: \F_q \to \F_2^{\log q}$ from Section~\ref{sec:finite-fields}).
\end{enumerate}  
\end{corollary}

\begin{proof}
Lemma~\ref{lem:pauli-binary} implies that there are local isometries that map \[\textstyle \ket{\EPR_q}^{\otimes M}\] to $\ket{\EPR_2}^{\otimes M\log q}$, and map the generalized Pauli measurements $\tau^W_u$ to \[\textstyle \bigotimes_{i=1}^M \bigotimes_{j=1}^{\log q} \sigma^W_{u_{ij}}\] where $\kappa(u_i) = (u_{i1},\allowbreak \ldots,\allowbreak u_{i\log q})$. Combining this with the isometries guaranteed by Theorem~\ref{thm:pauli}, we obtain the statement of the Corollary.
\end{proof}
\subsubsection{Canonical parameters and complexity of the Pauli basis test}
\label{sec:qld-complexity}

We specify a canonical setting of the parameter tuple $\qldparams$ as a function
of an integer $R$.
We call this canonical setting $\introparams$ because we use these in the
``introspection'' section (Section~\ref{sec:introspection}).
We then give bounds on the complexity of computing the decision procedure and CL
functions of the Pauli basis test corresponding to the parameter tuple
$\qldparams$, as a function of $R$.

\begin{definition}[Canonical parameters of the Pauli basis test]
  \label{def:introparams}
  Let $a,b$ be the universal constants specified in Theorem~\ref{thm:pauli}.
  Let $c$ denote the smallest even integer that is at least $(b + a)/b$.
  For all integers $R \geq 4$, define the tuple $\introparams(R) = (q,m,d)$
  where
  \begin{itemize}
  	\item $q = 2^k$ for $k = c \lceil  \log \log R  \rceil + 1$.
	\item $m = 2^j$ where $2^j \leq c \lceil \log R \rceil + 1 < 2^{j+1}$. 
	\item $d = 1$. 
  \end{itemize}
\end{definition}

Intuitively, this choice of parameters is such that the Pauli basis test
certifies the presence of an $M=2^m$-qudit EPR state (of local dimension $q$), or
equivalently an $M\log q$-qubit EPR state.

\begin{lemma}
  \label{lem:delta-bound}
 For all integers $R \geq 4$, the parameter tuple $\introparams(R) \allowbreak =\allowbreak  (q,\allowbreak m,\allowbreak d)$ is
  admissible (see Definition~\ref{def:admissible}), satisfies $2^m \allowbreak \geq\allowbreak  R$, and, furthermore, there exist
  universal constants $a'\geq 1$ and $0 < b' \allowbreak \leq\allowbreak  1$ such that the function
  $\delta_\qld(\eps,\allowbreak  m,\allowbreak  d,\allowbreak  q)$ from Theorem~\ref{thm:pauli} is at most
   \[
   	a' ( \log(R)^{a'} \cdot \eps^{b'} + \log(R)^{-b'}) \;.
	\]
\end{lemma}

\begin{proof}
  We verify the admissibility of $\introparams(R)$ first: the field size $q =
  2^k$ is admissible because $k$ is odd. We have $m | q$ because for all $R \geq 4$ we have  
  \[
  	2^j \leq 2 c \lceil \log R \rceil \leq 2 \log^c R \leq 2^k~.
  \]
	Next, we show that $R \leq 2^m$. Since $m = 2^j \geq (c/2) \log R$, we have that
	\[
		2^m \geq R^{c/2} \geq R
	\]
	where we used that $c \geq 2$. 

Finally, we show the desired bound on the error function $\delta_\qld(\eps,m,d,q)$. First, observe that $2^{-bmd} \leq q^{-b}$ because $k \leq m$.
  Note that for $R \geq 2$, we have $m \leq 2c \log R$, and, furthermore, $q =
  2^k \geq \log^c R$.
Then, we get
\begin{align*}
  	\delta_\qld(\eps,m,d,q)
    & = a (md)^a (\eps^b + q^{-b} + 2^{-bmd}) \tag{by Theorem~\ref{thm:pauli}}  \\
    & \leq a (md)^a (\eps^b + 2q^{-b}) \tag{because  $2^{-bmd} \leq q^{-b}$}  \\
    & = a (m)^a (\eps^b + 2q^{-b}) \tag{because  $d=1$}  \\
    & \leq a (2c)^a \log^{a} (R) ( \eps^{b} + 2q^{-b}) \tag{because  $m \leq 2c \log R$}\\
    & \leq a (2c)^a \log^{a} (R) ( \eps^{b} + 2(\log R)^{-cb}) \tag{because  $q \geq \log^c R$}\\
    & \leq 2a (2c)^a \log^{a} (R) ( \eps^{b} + (\log R)^{-cb}) \\
    & = 2a (2c)^a ( (\log R)^{a}\eps^{b} + (\log R)^{a-cb}) \\
    & \leq 2a(2c)^{a} \left ( (\log R)^{a} \eps^{b}
      + (\log R)^{-b} \right) \tag{because  $cb - a \geq b \geq 0$}\\
    & \leq a' ( (\log R)^{a'} \eps^{b} + (\log R)^{-b}),
  \end{align*}
  where $a'$ is defined as the quantity $2a(2c)^a$.
%
%
  This completes the proof of the lemma.
\end{proof}

\begin{lemma}
	\label{lem:introparams-complexity}
	Fix an integer $R \geq 4$, let $\introparams(R) = (q,m,d)$.
	The integers $(q,m,d) = \introparams(R)$ can be computed in time $\polylog
  R$ when given the binary representation of $R$ as input.
%
\end{lemma}
\begin{proof}
  This follows from the definitions of the parameters in
  Definition \ref{def:introparams}.
  
\end{proof}

\begin{lemma}
  \label{lem:qld-complexity}
  Let $R \geq 4$ be an integer, and let $\introparams(R)$ denote the parameter tuple
  specified by Definition~\ref{def:introparams}.
  \begin{enumerate}
	\item The time complexity of the decision procedure $\decider^\pauli$
    parameterized by $\introparams(R)$ is $\poly(R)$.
	\item The time complexity of computing marginals of the CL functions $L_\tvar$
    as well as the associated factor spaces, for $\tvar \in \type^\pauli$, is
    $\polylog R$.
	\item The Turing machine description of the decision procedure
    $\decider^\pauli$ parameterized by $\introparams$ can be computed from the
    binary presentation of $R$ in $\polylog (R)$ time.
  \end{enumerate}
\end{lemma}

\begin{proof}
	Finite field arithmetic over $\F_q$ can be performed in time $\polylog q$, by
  Lemma~\ref{lem:efficient_arithmetic}.
  The parameters of $\introparams(R)$, which the decision procedure
  $\decider^\pauli$ implicitly computes given $R$, can be computed in time
  $\polylog(R)$, by Lemma~\ref{lem:introparams-complexity}. 
  
  The decider $\decider^\pauli$ attempts to parse the questions and answers
according to the formats prescribed in Figure~\ref{fig:decider_pauli}.  If the
parsing fails, it rejects.  For inputs that are parsed successfully, all
question and answer strings have length at most $\poly(R)$ for the canonical
parameters $\introparams(R)$.

  The most expensive step in $\decider^\pauli$ is to evaluate the low-degree
  encoding $g_{a}(y)$ where $a \in \F_q^M$ and $y \in \F_q^m$, which
  takes time $\poly(M,\log q) = \poly(R)$ by Lemma~\ref{lem:ld-encoding-complexity}.
	
  The complexity of computing the CL functions $L_\tvar$ for types $\tvar \in
  \type^\pauli$ is dominated by the complexity of computing the CL functions
  $L_\ALine$, $L_\DLine$, and $L_\Point$ from the classical low-degree test, which takes time
  $\poly(m, \log q) = \polylog(R)$.

  The factor spaces of $L_\tvar$ depend only on the question type $\tvar$ (of
  which there are only constantly many), and outputting the length-$m$ indicator
  vectors of the factor spaces takes $O(m) = \polylog (R)$ time.

  Together with the input parsing checks above, all checks performed by
$\decider^\pauli$, therefore, run in time $\poly(R)$.

  Finally, the time complexity of computing the description of $\decider^\pauli$
  from the binary representation of $R$ requires $\polylog R$ time, because the
  checks performed in the decision procedure $\decider^\pauli$ depend on
  $\introparams$ which ultimately depends on $R$; we assume that the decision
  procedure computes the parameter tuple $\introparams$ based on $R$.
  Thus the time complexity is dominated by the time to write $R$ in binary.
\end{proof}


\section{Introspection games}
\label{sec:introspection}

\newcommand{\ai}[1][y,\, a]{A^\Intro_{#1}}
\newcommand{\bs}[1][z,\, a]{B^\Sample_{#1}}
\newcommand{\br}[1][y,y^\perp, a]{B^\Read_{#1}}
\newcommand{\ahk}[1][y_{<k}]{A^{\Hide{k}}_{#1}}
\newcommand{\ahp}[1][y_{\le k}]{A^{\Hide{k+1}}_{#1}}
\newcommand{\bhk}[1][y_{<k}]{B^{\Hide{k}}_{#1}}
\newcommand{\bhp}[1][y_{\le k}]{B^{\Hide{k+1}}_{#1}}
\newcommand{\z}[1][z]{\sigma^Z_{#1}}
\newcommand{\x}[1][x]{\sigma^X_{#1}}

\newcommand{\La}{[L(\cdot) = y]}
\newcommand{\Lk}[1][k]{[L_{#1,\, y_{<#1}}(\cdot) = y_{#1}]}
\newcommand{\Llk}[1][k]{[L_{<#1}(\cdot) = y_{<#1}]}
\newcommand{\Llek}[1][k]{[L_{\le #1}(\cdot) = y_{\le #1}]}
\newcommand{\Lperpk}[1][k]{[L_{#1,\, y_{<#1}}^\perp (\cdot) = y_{#1}^\perp]}

\newcommand{\aigek}[1][y_{\ge k},\, a]{A^{\Intro,\, y_{<k}}_{#1}}
\newcommand{\aigk}[1][y_{> k},\, a]{A^{\Intro,\, y_{\le k}}_{#1}}
\newcommand{\bsL}{\bs[\La,\, a]}
\newcommand{\zL}{\z[\La]}
\newcommand{\zLk}[1][k]{\sigma^{Z}_{\Lk[#1]}}
\newcommand{\zLlk}[1][k]{\z[{\Llk[#1]}]}
\newcommand{\zLlek}{\z[\Llek]}
\newcommand{\xLperpk}[1][k]{\sigma^{X}_{\Lperpk[#1]}}

\newcommand{\aizk}[1][y,\, a]{A^{\Intro,\, Z_{<k}}_{#1}}

\subsection{Overview}
\label{sec:intro-overview}

Consider a normal form verifier $\verifier = (\sampler, \decider)$ (see
Definition~\ref{def:normal-game}).
In this section we design a normal form verifier $\verifier^\intro =
(\sampler^\intro, \decider^\intro)$ (called the \emph{introspective verifier})
such that in the $n$-th game $\verifier^\intro_n$ (called the
\emph{introspection game}; see Definition~\ref{def:normal-game} for the
definition of the $n$-th game associated with a normal form verifier) the
verifier expects the players to sample for themselves questions $x$ and $y$
distributed as their questions in the game $\verifier_N$ for index $N =
2^n$---this is the ``introspection'' step.
Note the exponential separation of the indices of the introspection game versus
the original game!
Our construction of the introspection game generalizes the introspection
technique of~\cite{NW19}.

Recall the execution of the $N$-th game corresponding to the ``original''
verifier $\verifier$ (see Definition~\ref{def:normal-game}).
Let $n\geq 1$, $N=2^n$, and suppose that the CL functions of $\sampler$ on index
$N$ are $L^\alice,L^\bob$ acting on an ambient space $V$.
In the game $\verifier_N$, the verifier first samples $z \in V$ uniformly at
random and gives each player $w\in\AB$ the question $x_w = L^w(z)$.
In this context, the string $z$ is referred to as the ``seed.''
The players respond with answers $a_\alice$ and $a_\bob$, respectively, and the
verifier accepts or rejects according to the output of $\decider(N, x_\alice,
x_\bob, a_\alice, a_\bob)$.

In the introspection game, with some constant probability independent of $n$,
the verifier $\verifier^\intro_n$ sends the question pair $(\Introspect,\alice)$
to player $w$ and $(\Introspect,\bob)$ to the other player $\overline{w}$, where
$w \in \AB$ and recall that $\overline{w}=\bob$ if $w=\alice$ and
$\overline{w}=\alice$ otherwise.
The verifier expects player $w$ to measure their share of the state
$\ket{\EPR}_V$ using a coarse-grained $Z$-basis measurement whose outcomes range
over $L^\alice(V)$, and similarly player $\overline{w}$ measures the state
$\ket{\EPR}_V$ using a coarse-grained $Z$-basis measurement with outcomes that
range over $L^\bob(V)$.
If the players perform these measurements honestly, then the outcomes
$(x_\alice,x_\bob)$ are distributed exactly according to $\mu_{\sampler,\, N}$, the
question distribution of the game $\verifier_N$.
Players $w$ and $\overline{w}$ are then expected to respond with the question
$x_w$ and $x_{\ol{w}}$ that they each sampled, together with strings $a_w$ and
$a_{\overline{w}}$, respectively, which are intended to be the answers for the
question pair $(x_\alice,x_\bob)$ in the game $\verifier_N$.
In other words, the players introspectively sample the question pair
$(x_\alice,x_\bob)$ and then respond with the question itself and an answer for
it.

To facilitate comprehension, we call the players that interact with the
introspective verifier $\verifier^\intro$ the ``introspecting players,'' and the
players that interact with the ``original'' verifier $\verifier$ the ``original
players.''
To ensure that the introspecting players follow the above procedure honestly,
the introspective verifier $\verifier^\intro_n$ first uses the (binary) Pauli
basis test described in Section~\ref{sec:pauli-verifier} to force the
introspecting players to share the state $\ket{\EPR}_{V}$.
The Pauli basis test also ensures that the players measure $\sigma^W$ and report
the outcome honestly when they receive questions $(\Pauli, W)$ for $W \in \{X,
Z\}$.
For $v\in\{\alice,\bob\}$ the verifier cross-checks the question pairs
$(\Introspect, \trole)$ and $(\Pauli, Z)$ to enforce that the honest measurement
is performed for question $(\Introspect, \trole)$.

The main difficulty in the soundness analysis is to ensure that the answer of
player $w$ who received question $(\Introspect, \trole)$ depends only on
$L^\trole(z)$ and not on any other information about the string $z$.
First assume for simplicity that $L^\trole$ is a linear function.
As shown below (based on Lemma~\ref{lem:why-didnt-i-think-of-this-before}),
$L^\trole(z)$ can be obtained by measuring a specific collection of~$\sigma^Z$
observables; namely, the set
\begin{equation}
  \label{eq:intro-overview-z}
  \{\sigma^Z(u) : u \in \ker(L^\trole)^\perp\}.
\end{equation}
To prevent the player from obtaining any additional information the verifier
needs to enforce that the player does \emph{not} additionally measure any
$\sigma^Z(u)$ for $u \notin \ker(L^\trole)^\perp$.
(We refer to any such $\sigma^Z$ as a ``prohibited'' $\sigma^Z$.)

The introspective verifier achieves this by sometimes sending the $\Read$ type
question $(\Read, \trole)$ or $\Hide{k}$ type questions $(\Hide{k}, \trole)$ to the
players.
When receiving the $\Read$ question, the players are required to also measure
observables from the set
\begin{equation}
  \label{eq:intro-overview-x}
  \{\sigma^X(r) : r \in \ker(L^\trole)\},
\end{equation}
which (as shown in Lemma~\ref{lem:commute} below) commute with every $Z$-basis
measurement in~\eqref{eq:intro-overview-z}.
On the other hand, any prohibited $\sigma^Z(u)$ must anticommute with at least
one of the $\sigma^X(r)$, as otherwise $u$ would be in $\ker(L^\trole)^\perp$.
As a result, honestly measuring the~$\sigma^X$ observables
of~\eqref{eq:intro-overview-x} has the effect of preventing the player from
measuring any of the prohibited~$\sigma^Z$ observables, so that the answer $a$
can effectively only depend on the question $L^\trole(z)$.
(In the protocol, the verifier asks the player to measure the function
$(L^\trole)^\perp(z)$ in the $X$ basis, rather than all of the~$\sigma^X$
observables.
By Lemma~\ref{lem:why-didnt-i-think-of-this-before} the two are equivalent.)
Similarly to how questions $(\Pauli, Z)$ and $(\Introspect, \trole)$ are
cross-checked, the questions $(\Pauli, X)$, $(\Hide{k}, \trole)$ and $(\Read,
\trole)$ are cross-checked in order to ensure that the honest $X$-basis
measurements are performed for the $\Hide{k}$ questions and the $\Read$ questions.

When the CL functions $L^\trole$ are $\ell$-level for $\ell > 1$, the
introspective verifier sends one of $O(\ell)$ different hiding questions to the
players, chosen at random; together these hiding checks ensure that each of the
constituent linear maps of $L^\trole$ are honestly measured.
Intuitively, these hiding questions ``interpolate'' between questions
$(\Pauli,Z)$, $(\Introspect,\trole)$ and $(\Pauli,X)$ in a way that, for all
pairs of questions asked by the verifier, the honest measurements commute.
(See Figure~\ref{fig:intro-honest} for an overview of the honest measurements.)
This strong commuting property of the strategy is essential for the
oracularization transformation in Section~\ref{sec:oracle} to be possible.

A key property of the introspection game is that the distribution of questions
(which include the Pauli basis test questions as well as the introspection
questions and the hiding questions) is also conditionally linear.
This means that the introspection game can be ultimately specified by a normal
form verifier $\verifier^\intro = (\sampler^\intro,\decider^\intro)$, which is
crucial for recursive compression of games.
Moreover, while the time complexity of the introspective verifier's decider
$\decider^\intro$ remains polynomially related to that of $\decider$ on index
$N$, the time complexity of the sampler $\sampler^\intro$ is $\polylog(N)$
(exponentially smaller), due to the efficiency of the Pauli basis test.
Finally, $\sampler^\intro$ only depends on $\verifier$ through the number of
levels $\ell$ of $\sampler$ as well as upper bounds on the time complexities of
$\sampler$ and $\decider$.

\subsection{The introspective verifier}
\label{sec:intro-verifier}

\def\tsint{\hat{\sampler}^\intro}
\def\tdint{\hat{\decider}^\intro}
\def\tvint{\hat{\verifier}^\intro}
\def\sint{\sampler^\intro}
\def\dint{\decider^\intro}
\def\vint{\verifier^\intro}

Let $\verifier = (\sampler, \decider)$ be an $\ell$-level normal form verifier.
We call $\verifier$, $\sampler$, and $\decider$ the ``original'' verifier,
sampler, and decider, respectively.
For all original verifier $\verifier$, we define the introspective verifier
$\vint$ in this section.
We use $N = 2^n$ to denote the index of the original verifier $\verifier$ that
is simulated by the introspective verifier $\vint$ on index $n$.



The \emph{introspective verifier corresponding to $\verifier$ and parameters
  $(\lambda,\ell)$} is a \emph{typed} normal form verifier $\tvint
=(\tsint, \tdint)$, sketched in
Section~\ref{sec:intro-overview} and specified in detail in the present section
(see Section~\ref{sec:types} for the definition of typed normal form verifiers).
In the following descriptions of the sampler $\tsint$ and decider
$\tdint$, all parameters are functions of the index $n$, the number of
levels $\ell$ of the sampler $\sampler$, and the parameter $\lambda$; we often
leave this dependence implicit.

Let $R = N^\lambda$, and let $\introparams(R) =
(q,m,d)$ denote the parameter tuple specified in
Section~\ref{sec:qld-complexity}.
Note that $\introparams$ is implicitly a function of $n$ (since $R$ is a
function of $n$), and that the field size function $q(n)$ is admissible according to definition~\ref{def:admissible-size}. 
The associated parameter tuple $\introparams$ is intended to parametrize a Pauli
basis test that certifies the presence of $Q = 2^m \log q$-qubit EPR states; intuitively, these qubits are meant to serve as the source of randomness for the CL functions of the
original sampler $\sampler$. Since we will assume that the original verifier $\verifier$ is $\lambda$-bounded, the original sampler $\sampler$ on index $N$ has time complexity at most $R$, and, therefore, the amount of randomness needed is at most $R$. By Lemma~\ref{lem:delta-bound}, the number of qubits $Q$ in the EPR states is at least $R$, and thus the EPR qubits can be used to properly introspect the original game. 

Recall that the players in the introspection game are referred to as
``introspecting players'' and the players in the original game are referred to
as ``original players.''
We use the following notation in order to distinguish between questions and
answers meant for the introspecting players versus the original players.
The questions and answers of the introspecting players are denoted by hatted
variables (e.g., $\hat{x}$ and $\hat{a}$).
Similarly, the associated question types are denoted using hatted variables
$\hat{\tvar}$.
The questions and answers of the original players in the original game
$\verifier_N$ are denoted using nonhatted variables (e.g.,\ $x$, $a$, and so
on).

\paragraph{Types and type graph.}
The type set $\type^\intro$ for the introspective verifier $\tvint$ is
\begin{align*}
  &\type^\intro = \type^\pauli \cup \Big( \Big(\big\{ \Introspect, \Sample,
  \Read \big\}\\
 &\cup \Big(\, \bigcup_{k=1}^{\ell} \bigl\{ \Hide{k} \bigr\} \,
  \Big) \Big) \times \ab \Big) \;,
\end{align*}
where $\type^\pauli$ is the type set of the Pauli basis test, defined in
Section~\ref{sec:pauli-verifier}.
The type graph $G^\intro$ is specified in Figure~\ref{fig:type-graph-intro}.

\begin{figure}[!htbp]
  \centering
  \begin{tikzpicture}[x=1pt,y=1pt]

    \def\typegraphlabelsep{1pt}
    \tikzset{type/.style args={[#1]#2}{
        draw,circle,fill,scale=0.25,
        label={[font=\scriptsize, label distance=\typegraphlabelsep]#1:#2}
      }}


    \foreach \i in {1,...,6} \draw ({298-54*(\i-1)},0)
    coordinate (Constraint-\i)
    node[type={[90]$\Constraint_\i$}] {};

    \foreach \i in {1,...,9} {
      \pgfmathsetlengthmacro{\typegraphlabelsep}{1+10*mod(\i+1,2)}
      \def\typegraphlabelangle{270}
      \ifnum\i=1 \def\typegraphlabelangle{0}\fi
      \ifnum\i=5 \def\typegraphlabelangle{315}\fi
      \draw ({310-36*(\i-1)},-64)
      coordinate (Variable-\i)
      node[type={[\typegraphlabelangle]$\Variable_\i$}] {};
    }

    \foreach \i in {1,...,3} \foreach \j in {1,...,3}
    \pgfmathsetmacro{\k}{(\i-1)*3+\j}
    \draw (Constraint-\i) -- (Variable-\k);

    \foreach \i in {4,...,6} \foreach \j in {1,...,3}
    \pgfmathsetmacro{\k}{\i-3+(\j-1)*3}
    \draw (Constraint-\i) -- (Variable-\k);


    \draw (300,-113) coordinate (DLine-X) node[type={[0]$(\DLine, X)$}] {};
    \draw (300,-193) coordinate (Plane-X) node[type={[0]$(\ALine, X)$}] {};
    \draw (260,-153) coordinate (Point-X) node[type={[0]$(\Point, X)$}] {};
    \draw (260,-248) coordinate (Pauli-X) node[type={[0]$(\Pauli, X)$}] {};
    \draw (60,-113) coordinate (DLine-Z) node[type={[180]$(\DLine, Z)$}] {};    
    \draw (60,-193) coordinate (Plane-Z) node[type={[180]$(\ALine, Z)$}] {};
    \draw (100,-153) coordinate (Point-Z) node[type={[180]$(\Point, Z)$}] {};
    \draw (100,-248) coordinate (Pauli-Z) node[type={[0]$(\Pauli, Z)$}] {};
    \draw (180,-213) coordinate (Pair) node[type={[270]$\Pair$}] {};
    \draw (210,-178) coordinate (Pair-X) node[type={[0]$(\Pair,X)$}] {};
    \draw (150,-178) coordinate (Pair-Z) node[type={[180]$(\Pair,Z)$}] {};


    \draw (300,-292) coordinate (Hide-1-A) node[type={[0]$(\Hide{1}, \alice)$}] {};
    \draw (300,-336) coordinate (Hide-2-A) node[type={[0]$(\Hide{2}, \alice)$}] {};
    \draw (300,-362) node {$\vdots$};
    \draw (300,-388) coordinate (Hide-l-A) node[type={[0]$(\Hide{\ell}, \alice)$}] {};

    \draw (220,-292) coordinate (Hide-1-B) node[type={[0]$(\Hide{1}, \bob)$}] {};
    \draw (220,-336) coordinate (Hide-2-B) node[type={[0]$(\Hide{2}, \bob)$}] {};
    \draw (220,-362) node {$\vdots$};
    \draw (220,-388) coordinate (Hide-l-B) node[type={[0]$(\Hide{\ell}, \bob)$}] {};

    \draw (60,-292) coordinate (Sample-A) node[type={[0]$(\Sample, \alice)$}] {};
    \draw (60,-336) coordinate (Intro-A) node[type={[180]$(\Introspect, \alice)$}] {};
    \draw (60,-388) coordinate (Read-A) node[type={[0]$(\Read, \alice)$}] {};

    \draw (140,-292) coordinate (Sample-B) node[type={[0]$(\Sample, \bob)$}] {};
    \draw (140,-336) coordinate (Intro-B) node[type={[0]$(\Introspect, \bob)$}] {};
    \draw (140,-388) coordinate (Read-B) node[type={[0]$(\Read, \bob)$}] {};


    \foreach \from/\to in {DLine-X/Point-X, Plane-X/Point-X, Point-X/Pauli-X,
      DLine-Z/Point-Z, Plane-Z/Point-Z, Point-Z/Pauli-Z,
      Point-X/Variable-1,
      Hide-2-A/Hide-1-A,
      Hide-2-B/Hide-1-B,
      Sample-A/Intro-A, Intro-A/Read-A,
      Sample-B/Intro-B, Intro-B/Read-B,
      Intro-A/Intro-B}
    \draw (\from) -- (\to);

    \draw (Point-Z) to [out=90,in=270] (Variable-5);

    \draw (Pauli-X) to [out=270,in=90] (Hide-1-A);
    \draw (Pauli-X) to [out=270,in=90] (Hide-1-B);
    \draw (Pauli-Z) to [out=270,in=90] (Sample-B);
    \draw (Pauli-Z) to [out=270,in=90] (Sample-A);
    \draw [rounded corners=8] (Hide-l-A) -- (300,-442) -- (60,-442) -- (Read-A);
    \draw [rounded corners=8] (Hide-l-B) -- (220,-420) -- (140,-420) -- (Read-B);
    \draw (Point-X) to [out=180,in=90] (Pair-X) to [out=270,in=0] (Pair);
    \draw (Point-Z) to [out=0,in=90] (Pair-Z) to [out=270,in=180] (Pair);

  \end{tikzpicture}
  \caption{Type graph $G^\intro$ for the introspection game.
    Each vertex also has a self-loop which is not drawn in the figure for
    clarity. }
  \label{fig:type-graph-intro}
\end{figure}

\paragraph{Sampler.}
We first define a $3$-level $(\type^\intro,G^\intro)$-typed sampler
$\tilde{\sampler}^\intro$, which has field size $q(n)$ and dimension $3m(n) +
3$, where $q(n)$ and $m(n)$ are specified by $\introparams(R)$.
 Note that the dimension matches that of the space $V^\pauli$ of the CL
functions of the Pauli basis test parameterized by $\introparams(R)$, specified in
Section~\ref{sec:qld-game}. 

Fix $n \in \N$.
We specify the CL functions of $\tilde{\sampler}^\intro$ on index $n$.
Since the functions $L_{\hat{\tvar}}^{\alice,\, n}$ and
$L_{\hat{\tvar}}^{\bob,\, n}$ are identical for all $n$ and $\hat{\tvar}$, we
omit the superscripts $\alice$ and $\bob$.
For types $\hat{\tvar} \in \type^\pauli$, the CL functions $L_{\hat{\tvar}}^n$
are given by those specified in Section~\ref{sec:qld-game}, parameterized by
$\introparams(R)$.
For types $\hat{\tvar} \in \type^\intro \setminus \type^\pauli$, the associated
CL functions are defined to be $0$-level CL functions (i.e., they are the $0$
map).
This means that for question types such as $\hat{\tvar}=(\Introspect, \trole)$
or $\hat{\tvar}=(\Sample, \trole)$ for $\trole\in\{\alice,\bob\}$ the associated question is solely comprised
of the question type label. The sampler $\tilde{\sampler}^\intro$ then behaves as in Definition~\ref{def:typed-sampler} (and returns $0$ if its input is not properly formatted or too long). 

Finally, we define the typed sampler $\tsint$ to be
$\downsize(\tilde{\sampler}^\intro)$, the downsized sampler
(Definition~\ref{def:downsize-typed-sampler}) corresponding to $\tilde{\sampler}^\intro$.
The typed CL functions associated with
the Pauli basis test are $3$-level; using Remark~\ref{rk:higher-level}
and Lemma~\ref{lem:cl-downsize}
it follows that $\tsint$ is a $3$-level typed sampler.

The following lemma establishes the complexity of the sampler $\tsint$
as well as the complexity of computing a {description} of it from the parameters
$(\lambda,\ell)$.
\begin{lemma}
  \label{lem:intro-sampler-complexity}
  There exists a $2$-input Turing machine $\ComputeIntroSampler$ that
  on input $(\lambda,\ell)$ outputs a description of the sampler
  $\tsint$ (corresponding to parameters $\lambda,\ell$) in time $\polylog(\lambda,\ell)$.
  Furthermore, 
  \begin{enumerate}
  \item $\TIME_{\tsint}(n) = \poly(n, \lambda, \ell)$, and
	\item $\tsint$  is a $3$-level typed sampler.
  \end{enumerate}
\end{lemma}

\begin{proof}
  \hnote{rewrote this (May 30, 2026):} 
  By the hardwiring convention of Section~\ref{sec:tms}, it suffices to describe
the sampler obtained after the parameters $\lambda$ and $\ell$ have been
hardwired.  On input $(\lambda,\ell)$, the machine $\ComputeIntroSampler$
outputs the description of the following sampler.

Let $\tsint_{\lambda,\ell}$ denote the sampler with $\lambda$ and $\ell$
hardwired.  On input $(n,\allowbreak x_1,\allowbreak \ldots,\allowbreak x_6)$, it first computes
$R = 2^{\lambda n}$ and the corresponding Pauli-test parameters
$\introparams(R)$.  By Lemma~\ref{lem:introparams-complexity}, this takes time
$\polylog R=\poly(n,\allowbreak \lambda)$.

The sampler $\tsint_{\lambda,\ell}$ implements the downsized typed sampler
$\downsize(\tilde{\sampler}^{\intro})$ from
Definition~\ref{def:downsize-typed-sampler}.  It first performs the formatting
checks required by that construction.  In particular, it checks that the type
input is a valid element of $\type^\intro$, and before parsing any input string
as an element of the relevant ambient space, it checks that the string has the
prescribed length determined by $\introparams(R)$.  If any formatting check
fails, the sampler halts immediately with a fixed default output.

If the input requests the dimension, the sampler outputs the dimension of the
downsized ambient space.  If the input requests a marginal, linear map, or
factor space for a type $\tvar\in\type^\pauli$, the sampler executes the
corresponding downsized Pauli-basis sampler routine with parameters
$\introparams(R)$.  By Lemma~\ref{lem:qld-complexity} and
Lemma~\ref{lem:downsize_typed_sampler}, this takes time $\poly(n,\lambda,\ell)$.

If $\tvar\in\type^\intro\setminus\type^\pauli$, the corresponding CL function
of $\tilde{\sampler}^{\intro}$ is the zero map, as specified above.  Thus the
sampler answers marginal, linear, and factor-space inputs for these types by
outputting the corresponding fixed zero vector or fixed indicator vector.  This
requires only checking and parsing the type label, and hence takes time
$\poly(\log \ell)$.

Putting everything together, we thus have  $\TIME_{\tsint_{\lambda,\ell}}(n)=\poly(n,\lambda,\ell)$.

The description-length and construction-time bounds follow from the hardwiring
convention of Section~\ref{sec:tms}.  The parameters hardwired into
$\tsint_{\lambda,\ell}$ are $\lambda$ and $\ell$.  Hence
$\ComputeIntroSampler$ outputs the description of $\tsint_{\lambda,\ell}$ in
time $\polylog(\lambda,\ell)$.

By construction, $\tsint_{\lambda,\ell}$ is the sampler $\tsint$ corresponding
to parameters $(\lambda,\ell)$, and it is a $3$-level typed sampler.

\end{proof}

\paragraph{Decider.}
The typed decider $\tdint$ is specified in
Figure~\ref{fig:intro-decider}.
We explain how to interpret the figure, including the notation.
(It may also be helpful to review the description of the intended strategy for
the players in the game, described in Section~\ref{sec:intro-completeness}.)

\begingroup
  \begin{gamespec}
    \begingroup
      \centering
      \begin{tabularx}{.7\textwidth}{l X}
        \toprule
        Type  & Answer format\\
        \midrule
        $(\Introspect, \trole)$ & $(y,a) \in V \times \{0,1\}^*$ \\
        $(\Sample, \trole)$ &  $(z,a) \in V \times \{0,1\}^*$ \\
        $(\Read, \trole)$ & $(y,\allowbreak y^{\perp},\allowbreak a) \in V \times V \times \{0,\allowbreak 1\}^*$\\
        $(\Hide{k}, \trole)$ & $(y,\allowbreak y^{\perp},\allowbreak x) \in V \times V \times V$\\
        \bottomrule
      \end{tabularx}
    \par\endgroup
    In the following, whenever $\sampler$ or $\decider$ is called as a subroutine,
    $\tdint$ aborts and rejects if the subroutine takes more than
    $N^\lambda$ time steps.
    On input $(n,\allowbreak  \hat{\tvar}_\alice,\allowbreak  \hat{x}_\alice,\allowbreak  \hat{\tvar}_\bob,\allowbreak 
    \hat{x}_\bob,\allowbreak  \hat{a}_\alice,\allowbreak  \hat{a}_\bob)$, the decider $\tdint$
    first computes the dimension $s(N)$ of $V$ by calling the original sampler
    $\sampler$ on input $(N,\gamestyle{dimension})$.
    The decider rejects if $s(N) > N^\lambda$, if \tnote{edited:}$\hat{\tvar}_\alice$ or $\hat{\tvar}_\bob$ is not a valid type from $\type^\intro$, or if $\max \bigl\{ |\hat{x}_\alice|,\allowbreak  |\hat{x}_\bob|,\allowbreak  \abs{\hat{a}_\alice},\allowbreak 
    \abs{\hat{a}_\bob} \bigr\} \geq 3 \cdot 2^m \cdot \log q$ where $\introparams(N^\lambda) = (q,\allowbreak m,\allowbreak d)$. Otherwise, it performs the following tests for all $w,\trole \in \AB$.
    (If no test applies, the decider accepts.)

\begin{enumerate}
    \item (\textbf{Pauli test}).
      \label{enu:pauli}
      If $\hat{\tvar}_\alice,\allowbreak  \hat{\tvar}_\bob \in \type^\pauli$, the decider
      accepts if $\decider^\pauli_\introparams$ accepts the input
      $(\hat{\tvar}_\alice,\allowbreak  \hat{x}_\alice,\allowbreak  \hat{\tvar}_\bob,\allowbreak  \hat{x}_\bob,\allowbreak 
      \hat{a}_\alice,\allowbreak  \hat{a}_\bob)$.
    \item (\textbf{Sampling tests}).
      \label{enu:sampling}
      \begin{enumerate}
      \item If $\hat{\tvar}_w = (\Pauli,Z)$ and $\hat{\tvar}_{\overline{w}} =
        (\Sample,\allowbreak  \trole)$, accept if $\hat{a}_w^V = z_{\overline{w}}$.
        \label{enu:sampling-pauli}

      \item If $\hat{\tvar}_w = (\Introspect,\allowbreak  \trole)$,
        $\hat{\tvar}_{\overline{w}} = (\Sample,\allowbreak  \trole)$, accept if $y_{w} =
        L^\trole(z_{\overline{w}})$ and $a_w=a_{\overline{w}}$.
        \label{enu:sampling-intro}
      \end{enumerate}

    \item (\textbf{Hiding tests}).
      \label{enu:hiding}
      \begin{enumerate}

      \item If $\hat{\tvar}_w = (\Introspect,\allowbreak  \trole)$,
        $\hat{\tvar}_{\overline{w}} = (\Read,\allowbreak  \trole)$, accept if $y_{w} =
        y_{\overline{w}}$ and $a_w = a_{\overline{w}}$.
        \label{enu:hiding-intro}

      \item If $\hat{\tvar}_w = (\Hide{\ell},\allowbreak  \trole)$,
        $\hat{\tvar}_{\overline{w}} = (\Read,\allowbreak  \trole)$, accept if $y_{w,\,
          <\ell} = y_{\overline{w},\, <\ell}$, and $y^{\perp}_w =
        y^{\perp}_{\overline{w}}$.
        \label{enu:hiding-read}
        
      \item \label{enu:hiding-same} If $\hat{\tvar}_w = (\Hide{k},\allowbreak  \trole)$,
        $\hat{\tvar}_{\overline{w}} = (\Hide{k+1},\allowbreak  \trole)$ for some $k \in \{ 1,\allowbreak 
        2,\allowbreak  \ldots,\allowbreak  \ell-1 \}$, accept if
        \begin{equation*}
          y_{w,\, <k} = y_{\overline{w},\, <k}\;, \quad
          y^{\perp}_{w,\, \le k} = y^{\perp}_{\overline{w},\, \le k}\;, \quad
          x_{w,\, >k+1} = x_{\overline{w},\, >k+1}\;,
        \end{equation*}
        and $y^{\perp}_{\overline{w},\, k+1} = \big( L^{\trole}_{k+1,\, u} \big)^\perp
        (x_{w,\, k+1})$ where $u = y_{\overline{w},\, \le k}$.
      \item If $\hat{\tvar}_w = (\Pauli,X)$, $\hat{\tvar}_{\overline{w}} =
        (\Hide{1},\allowbreak  \trole)$, accept if $y^\perp_{\overline{w},\, 1} =
        \bigl( L^{\trole}_1 \bigr)^\perp (\hat{a}_w^{V_1})$ and $\hat{a}_w^{V_{> 1}} =
        x_{\overline{w},\, >1}$.
        \label{enu:hiding-pauli}
      \end{enumerate}

    \item (\textbf{Game test}).
      \label{enu:intro-game}
      If $\hat{\tvar}_w = (\Introspect,\allowbreak  \alice)$ and $\hat{\tvar}_{\overline{w}}
      = (\Introspect,\allowbreak  \bob)$, accept if $\decider$ accepts $(N,\allowbreak  y_w,\allowbreak 
      y_{\overline{w}},\allowbreak  a_w,\allowbreak  a_{\overline{w}})$.

    \item (\textbf{Consistency test}).
      \label{enu:intro-consistency}
      If $\hat{\tvar}_\alice = \hat{\tvar}_\bob$, accept if and only if
      $\hat{a}_\alice = \hat{a}_\bob$.
    \end{enumerate}
  \end{gamespec}
  \captionof{figure}{The typed decider $\tdint$ for the introspective verifier,
    parameterized by integers $\lambda, \ell$, on index $n$.
    $N$ denotes $2^n$, $V$ is the ambient space for $\sampler$, and
    $\{L^\trole\}_{\trole\in\{\alice,\bob\}}$ the associated CL functions on
    index $N$.}
  \label{fig:intro-decider}
\endgroup

\emph{Question and answer format.}
The decider takes as input a tuple $(n,\allowbreak  \hat{\tvar}_\alice,\allowbreak  \hat{x}_\alice,\allowbreak 
\hat{\tvar}_\bob,\allowbreak  \hat{x}_\bob,\allowbreak  \hat{a}_\alice,\allowbreak  \hat{a}_\bob)$ where
$(\hat{\tvar}_w,\allowbreak \hat{x}_w)$ denotes the question for introspecting player $w \in
\AB$, and $\hat{a}_w$ denotes their answer.
As in the specification of the Pauli basis test, in
Figure~\ref{fig:intro-decider} we include an ``answer key'' indicating how the
players' answers are parsed, depending on the question type.
When the question type is from $\type^\pauli$, the question and answer format
are as described in Figure~\ref{fig:decider_pauli}.
When the question type is in $\type^\intro \setminus \type^\pauli$, the answer
format is described in the table at the top of
Figure~\ref{fig:intro-decider}.\footnote{There is no question format
  specification for the question types in $\type^\intro \setminus \type^\pauli$,
  because the question is solely comprised of the question type label.}
For each such question, there is an associated variable $\trole \in \AB$ that
indicates to the introspecting player which {original player} it is supposed to
impersonate in the introspection game.

In the figure, $V$ denotes the ambient space of the original sampler $\sampler$
on index $N = 2^n$. We assume that the original verifier $\verifier$ is $\lambda$-bounded and, in particular, the running time of the original sampler $\sampler$ (and, therefore, its dimension) is at most $R = N^\lambda$. If the dimension were larger than $N^\lambda$, the decider $\hat{\decider}^\intro$ would always reject in the beginning. 

Since $V$ is isomorphic to $\F_2^{s(N)}$, where $s(N)$ is the dimension of
$\sampler$, the space $V$ is identified in a canonical way as the register
subspace of $\F_2^Q$ spanned by $e_1,\ldots,e_{s(N)}$ where $e_i$ is the $i$-th
elementary basis vector.
For example, if $\hat{\tvar}_w = (\Read,\trole)$, then syntactically the
player's answer is a triple $(y,y^\perp,a)$ in $\F_2^Q \times \F_2^Q
\times \{0,1\}^*$.
We assume that the decider $\hat{\decider}^\intro$ computes the dimension $s(N)$ of the
subspace $V$ by calling $\sampler$ on input $(N,\gamestyle{dimension})$, and
 if $y,y^\perp$ are not presented as vectors in the subspace $V$,
then the decider rejects. Thus in the analysis we directly consider $y,y^\perp$ as vectors in $V$.
In Fig.~\ref{fig:intro-decider}, the components of the players' answers are
subscripted by the player index.
For example, if player $w$ receives question $(\Hide{k},\trole)$, then their
answers are denoted by $(y_w,y_w^\perp,x_w)$.

The notation used in the ``answer key'' is meant to give an indication of the
intended meaning of the players' answers.
We use $y$ to denote a vector that is supposed to be the result of measuring a
CL function $L^\trole$; $y^\perp$ is supposed to be the result of measuring
``dual'' linear maps $L^\perp$ (as used in Step~\ref{enu:hiding} in
Figure~\ref{fig:intro-decider}); $x$ is supposed to be the result of $\sigma^X$
measurements, and $z$ is supposed to be the result of $\sigma^Z$ measurements.
We use $a$ to denote the introspected answers meant for the original decider
$\decider$.

\emph{CL functions and factor spaces.}
For $\trole \in \AB$, let $L^\trole = L^{\trole,\, N}$ denote the CL function
for original player $\trole$ specified by $\sampler$ on index $N = 2^n$.
For $z \in V$, the decider $\hat{\decider}^\intro$ computes $L^\trole(z)$ by calling
$\sampler$ on input $(N, \trole, \gamestyle{marginal}, \ell, z)$.

\tnote{edited:}
For $y=L^\trole(z) \in V$ and $k \in \{1,\ldots,\ell\}$ we define register subspaces
$V_{k}^\trole (y)$ by induction on $k$.
For $k=1$, $V_1^\trole(y)$ is the first factor space
of $L^\trole$ and is independent of $y$.\footnote{See
  Lemma~\ref{lem:cl-kth} for the definitions of factor spaces of a CL function.}
Suppose $V_j^\trole(y)$ has been defined for all $j < k$.
Then we define the marginal space $V_{< k}^\trole(y) = \bigoplus_{j=1}^{k-1}
V_j^\trole(y)$, and define $V_k^\trole(y)$ as the $k$-th factor space $V_{k,\,
  u}^\trole$ of $L^\trole$ with prefix $u = L_{<k}^\trole(z)$.
We also define $V^\trole_{\leq k}(y) = V^\trole_{< k+1}(y)$, and $V^\trole_{>
  k}(y)$ to be the complementary register subspace to $V^\trole_{\leq k}(y)$
within $V$.

The decider $\tdint$ computes factor spaces $V_j^\trole(y)$ from $y \in
V$ in the following sequential manner: first, the indicator vector for the
factor space $V^\trole_1$ is computed by calling the original sampler $\sampler$
on input $(N,\trole,\gamestyle{factor},1,0)$.
Let $y_1$ denote $L_1^\trole(z)$. Note that this indeed only depends on $y$, not on $z$, and is in fact the projection of $y$ on $V_1^v(y)$. 
Then, for $j \in \{2,\ldots,\ell\}$, the factor space $V^\trole_j(y)$ is
computed by calling $\sampler$ on input $(N,\trole,\gamestyle{factor},j,y_{\le
  j-1})$, where $y_{\le j-1} = L^\trole_{\leq j-1}(z)$, which again only depends on $y$, not $z$.

Finally, Lemma~\ref{lem:cl-kth} implies that the CL function $L^\trole$ gives rise to functions $\{ L_{k,u}^\trole \}$ where for all $k$ and prefixes $u$, $L_{k,u}^\trole$ is a map from $V_{k,u}^\trole$ to $V_{k,u}^\trole$.

\emph{Detailed explanation of the steps of $\tdint$.}
We give more details on the implementation of decider $\tdint$
specified in Figure~\ref{fig:intro-decider}.

\begin{enumerate}
\item The decider $\tdint$ first checks that the answers are not too
  long;  the maximum-length answer should be a tuple $(y, y^\perp, a)$ in response to question $(\Read,v)$ 
  where $y, y^\perp \in V$ and $a$ is an answer intended for the original
  decider $\decider$ on index $N$ (which we assume runs in time at most
  $N^\lambda$), or a tuple $(y,y^\perp,x) \in V \times V \times V$ in response to question $(\Hide{k},v)$. Either way, the maximum answer length should be $3Q = 3 \cdot 2^m \cdot \log q = \poly(R)$ bits long. 
  This check is necessary in order to ensure that the decider $\tdint$
  halts in time that is a fixed polynomial of $R = N^\lambda$.

\item If the question types for the players are both in $\type^\pauli$,
  $\tdint$ executes the decision procedure $\decider^\pauli$ for the
  Pauli basis test parametrized by $\introparams(R) = (q,m,d)$ (see
  Section~\ref{sec:qld-game} for the definition of $\decider^\pauli$).
 The Pauli basis test is intended to ensure that the players share at least $Q \geq R$ EPR pairs, where $Q = 2^m \log q$. This number of EPR pairs is chosen
  so that, under the assumption that the original verifier $\verifier$ is $\lambda$-bounded (see Definition~\ref{def:lambda}), 
  the time complexity of $\sampler$ on index $N$ (and, therefore, its dimension) is at most $R = N^\lambda$. 
	When analyzing the Completeness, Soundness, and Entanglement properties of the introspective verifier $\verifier^\intro$,
	we assume that the original verifier $\verifier$ is $\lambda$-bounded (see Theorem~\ref{thm:introspection}). 
   

\item In Step~\ref{enu:sampling-pauli} of $\tdint$, player
  $w\in\{\alice,\bob\}$ receives question $(\Pauli,Z)$ and player $\overline{w}$
  receives question $(\Sample,\trole)$, for some $v\in\{\alice,\bob\}$.
  According to the answer key, player $w$ is expected to return an answer
  $\hat{a}_w\in \F_2^Q$ and player $\overline{w}$ is expected to respond with a
  pair $(y_{\overline{w}},a_{\overline{w}}) \in V \times \{0,1\}^*$.
  Thus, the dimension of answer $\hat{a}_w$ may be larger than that of
  $y_{\overline{w}}$; this is the reason that Step~\ref{enu:sampling-pauli}
  checks consistency between $y_{\overline{w}}$ and the projection of
  $\hat{a}_w$ to $V$.

\item In Step~\ref{enu:sampling-intro} of $\tdint$, player $w \in \ab$
  receives question $(\Introspect,\trole)$ and player $\overline{w}$ receives
  question $(\Sample,\trole)$.
  As specified by the ``answer key,'' player $w$ responds with $(y_w,a_w) \in V
  \times \{0,1\}^*$ and player $\overline{w}$ responds with
  $(z_{\overline{w}},a_{\overline{w}}) \in V \times \{0,1\}^*$.
  The decider $\tdint$ checks that $a_w = a_{\overline{w}}$ and $y_w =
  L^\trole(z_{\overline{w}})$ where $L^\trole$ denotes the CL function for
  player $\trole$ specified by $\sampler$ for index $N = 2^n$.
  The CL function is computed by calling $\sampler$ on input
  $(N, \trole, \gamestyle{marginal}, \ell, z)$.

\item In Step~\ref{enu:hiding-read}, $\tdint$ checks that the answer of
  player $w$ who received $(\Hide{\ell},\allowbreak \trole)$ is consistent with the answer
  of player $\overline{w}$ who received $(\Read,\allowbreak \trole)$. One of the checks is
  that $y_{w,< \ell} = y_{\overline{w},<\ell}$~; these are, respectively, 
  $L^\trole_{< \ell}(y_w)$ and $L^\trole_{< \ell}(y_{\overline{w}})$.
  

\item\label{enu:decider-perp}
  \def\Lfunc{L^{\trole}_{k+1,\, y_{\overline{w},\, \le k}}}

  In Steps~\ref{enu:hiding-same}, the vectors $x_{w,\, >k+1}$ and
  $x_{\overline{w},\, >k+1}$ denote the projections of $x_w$ and
  $x_{\overline{w}}$ to the factor spaces $V^\trole_{> k+1}(y_w)$ and $V^\trole_{>
    k+1}(y_{\overline{w}})$, respectively.
  Similarly, $y_{\overline{w},\, k+1}^\perp$ denotes the projection of
  $y_{\overline{w}}^\perp$ to the factor space $V^\trole_{k+1} (y_{\overline{w}})$ and
  $x_{w,\, k+1}$ denotes the projection of $x_{w}$ to
 the factor space $V^\trole_{k+1}(y_{\overline{w}})$. We emphasize that the latter
 factor space depends on $y_{\overline{w}}$ and not $y_w$.\footnote{ This is because it is the $\overline{w}$ player
 who receives the $(\Hide{k+1},\trole)$ question, and in the ``honest'' strategy (described in Section~\ref{sec:intro-completeness}) 
 they are supposed to measure $k$ registers to obtain a sequence of values $(y_{\overline{w},1},y_{\overline{w},2},\ldots,y_{\overline{w},k})$ 
 which specify the $(k+1)$-st factor subspace, whereas the $w$ player is only supposed to measure $k-1$ registers.
 }

  The decider also has to compute $( \Lfunc )^\perp (x_{w,\, k+1})$.
  According to Definition~\ref{def:Lperp}, this requires specifying a basis for
  $\ker( \Lfunc )^\perp$.
  To compute the value, the decider performs the following steps:
	\begin{enumerate}
  \item Call $\sampler$ on input $(N, \trole, \gamestyle{factor}, k+1, y_{\overline{w},\, \le k})$ to
    obtain a subset $H = \{h_1,\ldots,h_m\}$ of the canonical basis for
    $\F_2^{s(N)}$ that is a basis of the register subspace $V^\trole_{k+1}(y_{\overline{w}})$.
  \item For $i \in \{1, 2, \ldots, m\}$ compute $c_i = \sampler(N, \trole,
    \gamestyle{linear}, k+1, y_{\overline{w},\, \le k}, h_i)$.
    Compute a matrix representation $M$ for $ \Lfunc $ in the basis
    $H$, whose columns are the vectors $c_1, \ldots, c_m$ as elements of
    $V^\trole_{k+1}(y_{\overline{w}})$.

  \item Using a canonical algorithm for Gaussian elimination, compute a basis
    $F$ for $\ker(M)$.

  \item Compute the canonical complement $S$ of $F$, as in
    Definition~\ref{def:canonical-complement}.
    $S$ is a basis for $\ker( \Lfunc )^\perp$.

  \item To compute $(\Lfunc)^\perp$ on input
    $x_{w,\, k+1}$, compute the canonical linear map with kernel basis
    $S$ (see Definition~\ref{def:cl-canonical}) on input $x_{w,\,
      k+1}$.
  \end{enumerate}

\item In Step~\ref{enu:hiding-pauli}, the player $w$ that receives $(\Pauli,X)$
  is expected to return an answer $\hat{a}_w$ in $\F_2^Q$.
  Part of this step checks that the projection of $\hat{a}_w$ to $V_{>1}$ is
  equal to $x_{\overline{w},\, >1}$ (which is the projection to $V_{>1}$ of the
  third component of the answer triple of player $\overline{w}$ that receives
  question $(\Hide{1},\trole)$).
\end{enumerate}

\subsubsection{Complexity of the introspective verifier}
\label{sec:intro-complexity}

In this section we determine the complexity of the introspective verifier.
The following lemma establishes the complexity of the decider $\tdint$
as well as the complexity of computing a {description} of it from the parameters
$(\lambda,\ell)$ and the description of the original verifier $\verifier =
(\sampler,\decider)$.

\begin{lemma}
  \label{lem:intro-decider-complexity}
  \leavevmode\par There exists a polynomial time Turing machine $\ComputeIntroDecider$ that
  on input $(\verifier,\allowbreak  \lambda,\allowbreak  \ell)$ outputs a description of a typed decider
$\tdint$ in standard timeout counter form (as described in Section~\ref{sec:tms}) such that:
  \begin{enumerate}
  		\item Its description length is at most $\poly(\lambda, \ell)$.
		\item It has timeout counter bound \[\textstyle B_{\tdint}(n) = \poly(2^{\lambda n},\allowbreak \ell)\] satisfying $s_{\tsint}(n) \leq B_{\tdint}(n)$ (where $s_{\tsint}(n)$ is the dimension of the sampler $\tsint$). In particular, its time complexity is at most \[\textstyle \TIME_{\tdint}(n) = \poly(2^{\lambda n},\allowbreak \ell).\]
  \end{enumerate}
  Furthermore, if $|\verifier| \leq \lambda$, then the output of $\ComputeIntroDecider$ is the introspective decider 
  $\tdint$ corresponding to verifier $\verifier$ and parameters $(\lambda,\ell)$. 
\end{lemma}

\begin{proof}
  \hnote{rewrote this (May 30, 2026):}
On input $(\verifier,\lambda,\ell)$, the machine $\ComputeIntroDecider$ first
checks whether $\verifier$ can be parsed as a pair $(\sampler,\decider)$ and
whether $|\verifier|\leq \lambda$.  
If either check fails, $\ComputeIntroDecider$ outputs an always-rejecting typed
decider in standard timeout-counter form. Its timeout parameters are chosen so
that
\[
  B_{\tdint}(n)=\poly(2^{\lambda n},\ell)
  \qquad\text{and}\qquad
  s_{\tsint}(n)\leq B_{\tdint}(n)
\]
for every $n\in\N$. Its description has length $\poly(\lambda,\ell)$.

Now assume that $\verifier=(\sampler,\decider)$ and $|\verifier|\leq\lambda$.  We describe the decider $\tdint$ whose description is computed
by $\ComputeIntroDecider$.  By the hardwiring convention of
Section~\ref{sec:tms}, $\ComputeIntroDecider$ hardwires the descriptions of
$\sampler$ and $\decider$, together with the binary representations of
$\lambda$ and $\ell$, into the decider $\tdint$ specified in Figure~\ref{fig:intro-decider}.

On input $(n,\hat{\tvar}_\alice,\hat{x}_\alice,
    \hat{\tvar}_\bob,\hat{x}_\bob,\hat{a}_\alice,\hat{a}_\bob)$,
the decider $\tdint$ first computes $N=2^n$, $R=N^\lambda=2^{\lambda n}$, and
$\introparams(R)=(q,m,d)$.  By Lemma~\ref{lem:introparams-complexity}, this
takes time $\polylog R$.

It then computes the dimension $s(N)$ of the ambient space $V$ by simulating the
hardwired sampler $\sampler$ on input $(N,\gamestyle{dimension})$ for at most
$R$ steps.  If the simulation does not halt within $R$ steps, or if the output
is not a valid integer at most $R$, the decider rejects.  It then performs the
tests described in Figure~\ref{fig:intro-decider}.  The initial type and length
checks are implemented by reading only up to the prescribed bound
$3\cdot 2^m\log q$; if an input string exceeds this bound, the decider rejects.

The running time of $\tdint$ is bounded as follows.  For inputs that pass the type and
length checks, all question and answer strings have length at most
$3\cdot 2^m\log q=\poly(R)$.  The Pauli test takes time $\poly(R)$ by
Lemma~\ref{lem:qld-complexity}.  Any call to the hardwired original decider
$\decider$ on index $N$ is simulated for at most $R$ steps.  Any call to the
hardwired original sampler $\sampler$ on index $N$ is also simulated for at most
$R$ steps.  Since $s(N)\leq R$ after the dimension check, the sampler is called
at most $\poly(R,\ell)$ times while computing the needed marginals, factor
spaces, and linear maps.  Finally, the computations of
$(L^\trole_{k,u})^\perp(x_{\overline{w},k})$ use Gaussian elimination and
finite-field operations on matrices of size at most $\poly(R)$.

Thus all tests in Figure~\ref{fig:intro-decider} run in time
$\poly(R,\ell)=\poly(2^{\lambda n},\ell)$. If any invocation of $\sampler$ or $\decider$ does not halt within $R$ steps,
the decider rejects.  Thus $\tdint$ has time complexity $\poly(2^{\lambda n},\ell)$, as desired.

We equip $\tdint$ with a standard timeout counter with bound of the form $B_{\tdint}(n) = c_0 (c_1 2^n)^{c_2}$, where the constants are chosen large enough (possibly depending on $\lambda$ and $\ell$) so that 
 $B_{\tdint}(n)$ dominates the $\poly(2^{\lambda n},\ell)$ running time bound just computed above as well as the dimension $s_{\tsint}(n)$ of the sampler $\tsint$; note that $s_{\tsint}(n) = \poly(n,\lambda,\ell)$ by Lemma~\ref{lem:intro-sampler-complexity}. Thus the timeout counter satisfies the conditions specified by Lemma~\ref{lem:intro-decider-complexity}. 

Finally, we bound the time needed to compute the description of $\tdint$.  The
hardwired data are the descriptions of $\sampler$ and $\decider$ and the binary
representations of $\lambda$ and $\ell$.  Since $|\verifier|\leq\lambda$, the
description length of the resulting decider is at most $\poly(\lambda,\ell)$.
Moreover, producing this description only requires copying the hardwired data and
adding the fixed instructions of Figure~\ref{fig:intro-decider}, so
$\ComputeIntroDecider$ runs in polynomial time.

If $|\verifier|\leq\lambda$, the output is exactly the introspective decider
corresponding to $\verifier$ and parameters $(\lambda,\ell)$.  
\end{proof}

\subsubsection{Introspection theorem}
\label{sec:intro-theorem}

We are now ready to state the introspection theorem.

\begin{theorem}[Introspection theorem]
  \label{thm:introspection}
  There is a polynomial time Turing machine $\ComputeIntroVerifier$ which takes
  as input a tuple $(\verifier, \lambda, \ell)$ for $\lambda, \ell \in \N$ and
  returns the description of the introspective verifier $\vint=(\sint,\dint)$,
where the decider $\dint$ is in the standard timeout-counter form of
Section~\ref{sec:tms}. 
  For all $\ell \in \N$, $\vint$ is a $5$-level normal-form verifier satisfying
  \begin{equation*}
    \begin{split}
      \TIME_{\sint}(n) & = \poly \big (n, \lambda, \ell \big),\\
      B_{\dint}(n) & = \poly \big (2^{\lambda n}, \ell \big),\\
      |\dint| &= \poly(\lambda,\ell)~.
    \end{split}
  \end{equation*}
  Moreover, for all $\ell$, there exist constants $a>0$, \, $0 < b < 1$ (depending on $\ell$),
  and a function $\delta(\eps, n) = a \bigl( (\lambda n)^a\eps^b + (\lambda
  n)^{-b} \bigr)$ such that for all $n \in \N$ and $\eps \ge 0$ the following
  statements hold if $\verifier$ is a $\lambda$-bounded $\ell$-level verifier.
  \begin{enumerate}
  \item (\textbf{Completeness}) If $\verifier_{2^n}$ has a projective, consistent,
    and commuting (PCC) strategy with value $1$ then $\vint_n$ also
    has a PCC strategy with value $1$.

  \item (\textbf{Soundness}) If $\val^*(\vint_n) > 1 - \eps$, then
    $\val^*(\verifier_{2^n}) \geq 1 - \delta(\eps,n)$.
  \item (\textbf{Entanglement}) The entanglement bound $\Ent$ defined
    in Definition~\ref{def:ent} satisfies
    \begin{equation*}
      \Ent(\vint_n, 1 - \eps) \geq \max
      \left \{ \Ent(\verifier_{2^n}, 1 - \delta(\eps,n)),
        (1 - \delta(\eps,n))\, 2^{2^{\lambda n}} \right \}\;.
    \end{equation*}
  \end{enumerate}
\end{theorem}

\begin{proof}
  The Turing machine $\ComputeIntroVerifier$ does the following on input
  $(\verifier, \lambda, \ell)$: it first computes
  \begin{align*}
    \tsint & = \ComputeIntroSampler(\lambda,\ell)\;,\\
    \tdint & = \ComputeIntroDecider(\verifier, \lambda, \ell)\;,
  \end{align*}
  using Lemmas~\ref{lem:intro-sampler-complexity} and~\ref{lem:intro-decider-complexity}, runs
  the detyping procedure from Definition~\ref{def:detyped-verifier} to compute
  \begin{equation*}
    \vint = \detype(\tvint)
  \end{equation*}
  where $\tvint = (\tsint,\allowbreak \tdint)$, and output the resulting verifier. 
  This takes time $\poly(\abs{\verifier},\allowbreak  \log\lambda,\allowbreak  \log\ell)$, which is polynomial
  in the input length of\newline $\ComputeIntroVerifier$. 
  The complexity parameters of the typed sampler $\tsint$ and typed decider
  $\tdint$ follow from
  Lemmas~\ref{lem:intro-sampler-complexity} and~\ref{lem:intro-decider-complexity}.
  The complexity parameters of the detyped verifier $\vint$ then follow from
  Lemma~\ref{lem:detyping-verifiers}. 
  
  By Lemmas~\ref{lem:intro-sampler-complexity} and
\ref{lem:intro-decider-complexity}, and by the choice of the timeout bound,
\[
  s_{\tsint}(n)
  \leq
  B_{\tdint}(n)
  =
  \poly(2^{\lambda n},\ell).
\]
Therefore, the completeness and soundness conclusions of
Lemma~\ref{lem:detyping-verifiers} apply to $\tvint$, and the detyped decider
$\dint=\detype_G(\tdint)$ is in standard timeout-counter form with timeout bound $B_{\dint}(n)=\poly(|\type^\intro|,\allowbreak B_{\tdint}(n))$ such that
\[
  s_{\sint}(n) \leq B_{\dint}(n) =\poly(2^{\lambda n},\ell).
\]
In particular,
\[
  \TIME_{\dint}(n)=\poly(2^{\lambda n},\ell).
\]

  As $\tvint$ is a $3$-level verifier, $\vint$ is $5$-level as detyping will
  increase the number of levels by at most $2$.

  The completeness part of the theorem is proven in Section~\ref{sec:intro-completeness} and
  the soundness and entanglement properties are proven in Section~\ref{sec:intro-soundness}.
\end{proof}

\subsection{Completeness of the introspective verifier}
\label{sec:intro-completeness}

In this section we establish the completeness property of the introspection
game: if for $N = 2^n$, $\verifier_N$ has a PCC strategy (see
Definition~\ref{def:spcc}) with value $1$, then so does $\vint_n$.
For this we describe the actions that are expected of the players in the
introspection game (i.e.,\ the ``honest strategy'').
We first prove several preliminary lemmas that will be used in both the
completeness and soundness analysis.

\subsubsection{Preliminary lemmas}

The lemmas in this section are stated for general fields $\F$ and generalized
Pauli observables and projectors, although in the application to introspection
games we use $\F = \F_2$, $\omega = -1$, and qubit Pauli observables and
projectors.
Furthermore, the Pauli observables $\tau^W(v)$ and projectors $\tau^W_u$ act on
$\C^{\F^k}$ for some integer $k$ (in our application, we set $k=R$).

The following lemma generalizes Eq.~\eqref{eq:pauli-inversion-0}.

\begin{lemma}\label{lem:why-didnt-i-think-of-this-before}
Let $L: \F^k \to \F^k$ be a linear map, and let $W \in \{X, Z\}$.
For each $a$ in the range of $L$, let $u_a \in \F^k$ be such that $L(u_a) = a$. 
\begin{enumerate}
\item For each $v \in \ker(L)^\perp$, 
\begin{equation*}
\tau^W(v) = \sum_{a \in \F^k} \omega^{\tr(u_a \cdot v)} \tau^W_{[L(\cdot)=a]}\;.
\end{equation*}
\item For all $a$ in the range of $L$,
\begin{equation*}
\tau^W_{[L(\cdot)=a]} = \E_{v \sim \ker(L)^\perp} \omega^{-\tr(v \cdot u_a)} \tau^W(v)\;.
\end{equation*}
\end{enumerate}
\end{lemma}

\begin{proof}
  Let $V$ denote the image of $\F^k$ under $L$.
  Let $a\in V$, $v \in \ker(L)^\perp$.
  For all $u, u' \in L^{-1}(a)$, we have that $u-u'\in \ker(L)$ and thus $\tr(u
  \cdot v) = \tr(u' \cdot v)$.
  As a result, using~\eqref{eq:pauli-obs-proj}, for any $v\in\ker(L)^\perp$ it
  holds that
  \begin{equation*}
\begin{aligned}
&\tau^W(v)\\
 &= \sum_{u \in \F^k} \omega^{\tr(u \cdot v)} \tau^W_u\\
 &= \sum_{a \in V} \sum_{u \in L^{-1}(a)} \omega^{\tr(u \cdot v)} \tau^W_u\\
 &= \sum_{a \in V} \omega^{\tr(u_a \cdot v)} \tau^W_{[L(\cdot)=a]}\\
 &= \sum_{a \in \F^k} \omega^{\tr(u_a \cdot v)} \tau^W_{[L(\cdot)=a]}\;,
\end{aligned}
\end{equation*}
  where in the last equality we used that for $a \notin V$, the projector
  $\tau^W_{[L(\cdot)=a]}$ vanishes.
  This shows the first item.
  For the second item,
  \begin{align*}
    \tau^W_{[L(\cdot)=a]}
    & = \sum_{u  \in L^{-1}(a)} \tau^W_u\\
    & =  \sum_{u  \in L^{-1}(a)} \E_{v \sim \F^k} \Bigl(
      \omega^{-\tr(u \cdot v)} \tau^W(v) \Bigr) && \\
    & =  \sum_{u \in \ker(L)} \E_{v \sim \F^k} \Bigl(
      \omega^{-\tr((u_a + u) \cdot v)} \tau^W(v) \Bigr)\\
    & = \frac{|\ker(L)|}{|\F^k|} \sum_{v \in \F^k} \biggl( \Big(
      \E_{u \sim \ker(L)} \omega^{-\tr(u \cdot v)} \Big)\,
      \omega^{-\tr(u_a \cdot v)}  \tau^W(v) \biggr)\\
    & = \E_{v \in \ker(L)^\perp} \omega^{-\tr(v \cdot u_a)} \tau^W(v)\;,
\end{align*}
where the second equality follows from~\eqref{eq:pauli-inversion-0} and the last
uses the fact that $|\F^k|=|\ker(L)||\ker(L)^\perp|$, as shown in
Lemma~\ref{lem:perp_perp}, and Lemma~\ref{lem:cancellation} applied to the
expectation over~$u$.
\end{proof}

\begin{lemma}[Commuting~$X$ and~$Z$ measurements]
  \label{lem:commute} 
	For all linear maps $L,R: \F^k \to \F^k$ such that
  \begin{equation*}
    \ker (R)^\perp \subseteq \ker (L)\;,
  \end{equation*}
  the measurements
  $\big\{ \tau^Z_{[L(\cdot)=b]} \big\}_{b\in \F^k}$ and $\big\{ \tau^X_{[R(\cdot) = d]}
  \big\}_{d\in \F^k}$ commute.
\end{lemma}
\begin{proof}
  Let $b,d\in \F_k$.
  If either $b$ is not in the range of $L$, or $d$ is not in the range of $R$,
  then at least one of $\tau^Z_{[L(\cdot) = b]}$ or $\tau^X_{[R(\cdot) = d]}$ is
  $0$, and thus the operators trivially commute.
  Otherwise, both $b$ and $d$ are in the range of $L$ and $R$, respectively.
  Let $a_0 \in L^{-1}(b)$, $c_0 \in R^{-1}(d)$.
  By Lemma~\ref{lem:why-didnt-i-think-of-this-before},
  \begin{equation*}
    \tau^Z_{[L(\cdot)=b]} = \E_{u \in \ker(L)^\perp}
    \omega^{-\tr(u \cdot a_0)} \tau^Z(u)\;,
  \end{equation*}
  \begin{equation*}
    \tau^X_{[R(\cdot)=d]} = \E_{v \in \ker(R)^\perp}
    \omega^{-\tr(v \cdot c_0)} \tau^X(v)\;.
  \end{equation*}
  For any $v \in \ker(R)^\perp$, by assumption $v\in \ker(L)$, so for $u \in
  \ker(L)^\perp$ it holds that $u\cdot v=0$.
  Thus $\tau^Z(u)$ and $\tau^X(v)$ commute, and it follows that
  $\tau^Z_{[L(\cdot)=b]}$ and $\tau^X_{[R(\cdot)=d]}$ commute as well.
\end{proof}

\subsubsection{Completeness of the introspective verifier}
\label{sec:intro-completeness-sub}

\begin{proof}[Proof of the completeness part of Theorem~\ref{thm:introspection}]
  We analyze the completeness of the \emph{typed}
  verifier $\tvint$; the corresponding completeness of the
  \emph{detyped} verifier $\vint = \detype(\tvint)$ follows from
  Lemma~\ref{lem:detyping-verifiers}, and the fact that the type set
  $\type^\intro$ has size $O(\ell)$.

  \paragraph{Completeness.}
  We first show completeness.
  Let $n\geq 1$ be an index for $\tvint$ and $N = 2^n$ be the corresponding
  index for $\verifier$.
  Since we assume that the verifier $\verifier$ is $\lambda$-bounded, we have
  that
  \begin{equation}
    \label{eq:intro-complexity-assump}
    \max \big\{ \TIME_\sampler(N), \, \TIME_\decider(N) \big\} \leq N^\lambda = R\;.
  \end{equation}
  This assumption on the time complexity of $\verifier$ ensures that
  $\tdint$ never aborts due to a timeout when simulating $\sampler$ and $\decider$ for more than $R$ steps.
	Let $L^\trole = L^{\trole,\, N}$ denote the CL function of the original
  sampler $\sampler$ on index $N$ corresponding to player $\trole \in \AB$.
  Let $R = N^\lambda$, and let $\introparams(R) =
  (q,m,d)$, as in Section~\ref{sec:intro-verifier}.
  Set $Q = M\log q$ where $M=2^m$\,; this represents the number of qubits that are
  certified by the Pauli basis test parameterized by $\introparams(R)$.
  
  Let $\strategy = (\ket{\aux}, A, B)$ be a PCC strategy for $\verifier_N$ with
  value~$1$.
  We first construct a PCC strategy $\strategy^\intro_n$ for the typed verifier
  $\tvint_n$ with value $1$.
  We then conclude using Lemma~\ref{lem:detyping-verifiers}.

  \begin{figure}[htb!]
    \centering
    \begin{gamespec}
      \newcolumntype{M}{>{\arraybackslash $}X<{$}}
      \begin{tabularx}{\textwidth}{*{6}M}
        \toprule
        & \phantom{I1} V^\trole_1
        & \phantom{I1} V^\trole_2 & \phantom{I1} V^\trole_3 & \aux \\
        \midrule
        (\Introspect, \trole) & \sigma^Z_{L_1} & \sigma^Z_{L_2} &
        \sigma^Z_{L_3} & A^x/B^x \\
        (\Sample, \trole) & \sigma^Z & \sigma^Z & \sigma^Z & A^x/B^x \\
        (\Read, \trole) & \sigma^Z_{L_1^{\phantom{\perp}}}\;
        \sigma^X_{L_1^\perp} &
        \sigma^Z_{L_2^{\phantom{\perp}}}\;
        \sigma^X_{L_2^\perp} &
        \sigma^Z_{L_3^{\phantom{\perp}}}\;
        \sigma^X_{L_3^\perp} & A^x/B^x \\
        \midrule
        (\Hide{3}, \trole) & \sigma^Z_{L_1^{\phantom{\perp}}}\;
        \sigma^X_{L_1^\perp} &
        \sigma^Z_{L_2^{\phantom{\perp}}}\;
        \sigma^X_{L_2^\perp} &
        \phantom{\sigma^Z_{L_3^{\phantom{\perp}}}\;}
        \sigma^X_{L_3^\perp} & I \\
        (\Hide{2}, \trole) & \sigma^Z_{L_1^{\phantom{\perp}}}\;
        \sigma^X_{L_1^\perp}
        & \phantom{\sigma^Z_{L_1^{\phantom{\perp}}}\;}
        \sigma^X_{L_2^\perp}
        & \sigma^X & I \\
        (\Hide{1}, \trole) & \phantom{\sigma^Z_{L_1^{\phantom{\perp}}}\;}
        \sigma^X_{L_1^\perp} & \sigma^X & \sigma^X & I \\
        \midrule
      \end{tabularx}
      The left-most column denotes the introspection/hiding questions that a
      player may receive.
      The top row denotes the registers corresponding to the factor spaces of a
      CL function $L^\trole$ (we note that the partition of the registers depends
      on the prefixes), as well as the register corresponding to the state
      $\ket{\aux}$ coming from the original PCC strategy $\strategy$.
      We use $\sigma^Z_{L_j}$ as shorthand for $\sigma^Z_{[ L^\trole_{j,\,
          x_{<j}} (\cdot) = x_j ]}$ and similarly $\sigma^X_{L^\perp_j}$ for $
      \sigma^X_{[ (L^\trole_{j,\, x_{<j}})^\perp (\cdot)=x_j^\perp]}$.
      A symbol $I$ means that the register is left unmeasured.
    \end{gamespec}
    \caption{Summary of the honest strategy $\strategy_n^\intro$ for
      $\hat{\verifier}_n^\intro$ for a $3$-level sampler.}
    \label{fig:intro-honest}
  \end{figure}

  \begin{remark}
    \label{rk:downsizing}
    Note that by definition in $\verifier_n^\intro$ the players receive
    questions $(x,y)$ that are sampled according to the distribution
    $\mu_{\tsint,\, n}^{G^\intro}$ associated with the downsized typed
    sampler $\downsize(\tilde{\sampler}^\intro)$.
    Using the definition of the downsized typed sampler,
    Definition~\ref{def:downsize-typed-sampler}, and
    Lemma~\ref{lem:downsize-cl-dist} the distribution is identical to the
    distribution $\mu_{\tilde{\sampler}^\intro,\, n}^{G^\intro}$, up to the
    bijective mapping $\downsize$.
    This mapping can be computed by the players themselves.
    Therefore, we construct a strategy for players that receive questions from
    $\mu_{\tilde{\sampler}^\intro,\, n}^{G^\intro}$, and a strategy for
    questions from $\mu_{\tsint,\, n}^{G^\intro}$ follows immediately.
  \end{remark}

  The strategy $\strategy^\intro_n$ uses the state $\ket{\EPR_2}^{\otimes (Q+1)}
  \otimes \ket{\aux}$, where recall that $\ket{\EPR_2} =
  \frac{1}{\sqrt{2}}(\ket{00}+\ket{11})$.
  For all register subspaces $K \subseteq \F_2^{Q}$ (see
  Definition~\ref{def:cl-func} for the definition of register subspace),
  whenever we refer to ``register $K$,'' we mean the qubits of
  $\ket{\EPR_2}^{\otimes Q}$ corresponding to $K$ (see
  Section~\ref{sec:lin-reg}).
  The most frequent register subspace we consider is~$V$, spanned by $e_1,
  \ldots, e_{s(N)}$.
  We write $\overline{V}$ for the complement of~$V$, i.e.,
  the register subspace spanned by $e_{s(N)+1}, \ldots, e_{Q}$.
  (Note that $s(N) \leq R$ by
  assumption~\eqref{eq:intro-complexity-assump}, and by Lemma~\ref{lem:delta-bound} we have $R \leq M \leq Q$.)
  Then $\ket{\EPR_2}^{\otimes Q} = \ket{\EPR}_V \otimes
  \ket{\EPR}_{\overline{V}}$.

  Let $\strategy^\pauli_n$ be the honest Pauli strategy with respect to
  $\introparams$ defined in the proof of Lemma~\ref{lem:pauli-completeness}. Using the 
  isomorphism between $\C^q$ and $(\C^2)^{\otimes \log q}$ specified by Lemma~\ref{lem:pauli-binary},
 the measurements   of $\strategy^\pauli_n$ can be treated as qubit Pauli measurements acting on the
  maximally entangled state over qubits. Furthermore, the questions and answer labels
  of the measurements can be viewed as binary strings
 using the bijection $\kappa: \F_q \to \F_2^{\log q}$ specified in Section~\ref{sec:finite-fields}.
  For a question with a type from $\type^\pauli$ the player measures the shared state
  $\ket{\EPR_2}^{\otimes (Q+1)}$ using the measurements specified by
  $\strategy^\pauli_n$, and reports the measurement outcomes.
  When a player receives questions of type $\hat{\tvar} \in \type^\intro
  \setminus \type^\pauli$, they perform measurements described as follows.
  (Below, whenever we write a Pauli operator $\sigma^W_a$ the register on which
  the operator acts should always be clear from context, and is implicit from
  the space in which the outcome $a$ ranges.)
  The reader may find it helpful to consult Figure~\ref{fig:intro-honest} to see
  a summary of the honest strategy $\strategy^\intro_n$ for the special case
  when $\ell = 3$.

  \begin{description}
  \item $(\Introspect, \trole)$: The player performs the measurement
    \begin{equation}
      \label{eq:main-measurement}
      \bigl\{ \sigma^Z_{[L^\trole(\cdot) = y]} \bigr\}
    \end{equation}
    to obtain an $y\in V$.
    Intuitively, the player has now introspectively sampled the question $y$ for
    original player $\trole$ in game~$\verifier_N$.
    The player then measures $\ket{\aux}$ using player $\alice$'s
    measurement~$\{A^y_a\}$ from $\strategy$ if $v = \alice$ and using player
    $\bob$'s measurement~$\{B^y_a\}$ if $\trole = \bob$ to obtain an answer~$a$.
    The player replies with $(y,a)$.
  \item $(\Sample, \trole)$: The player measures their share of $\ket{\EPR}_V$
    in the $Z$ basis to obtain $z \in V$.
    Using this, they compute the question $y = L^\trole(z)$.
    The player then uses player $v$'s strategy and question $y$ to measure
    $\ket{\aux}$ and obtain outcome $a$.
    The player replies with $(z, a)$.
  \item $(\Read, \trole)$: The player first performs all measurements as in the
    $(\Introspect, \trole)$ question for player $\trole$ and records the
    outcomes as $y\in L(V)$ and $a\in\{0,1\}^*$.
    For $j\in \{1, 2, \ldots, \ell\}$, the player measures their share of the state
    $\ket{\EPR}_V$ with the measurement
    \begin{equation}
      \label{eq:perp-measurement}
      \big\{\sigma^X_{[L_j^\perp (\cdot) = y_j^\perp]} \big\}_{y_j^\perp}
    \end{equation}
    to obtain $y^\perp = y_1^\perp + \cdots + y_\ell^\perp$.
    Here $L_j^\perp$ is shorthand for the function $(L^{\trole}_{j,\,
      y_{<j}})^\perp$ defined in Item~\ref{enu:decider-perp} of the decider
    description in Section~\ref{sec:intro-verifier}.
    (That these operators are simultaneously measurable with the measurement
    in~\eqref{eq:main-measurement} follows from Lemma~\ref{lem:commute} and the
    fact that $\ker(L^\perp_{j})^\perp = \ker(L_{j})$ by
    Lemma~\ref{lem:perp_perp} and the definition of $L^\perp_{j}$ in
    Section~\ref{sec:intro-verifier}.)
    
    The player measures $\ket{\aux}$ with player $\trole$'s measurement strategy
    in $\strategy$ for question $y$ to obtain $a$ and replies with
    $(y,y^\perp,a)$.

  \item $(\Hide{k}, \trole)$:
    The player performs the following sequence of measurements: first measure
    $\{ \sigma^Z_{[L^\trole_{1}(\cdot) = y_{1}]} \}$ on register $V_1^v$ to
    obtain $y_1$.
    Then, use $y_1$ to specify the second linear function
    $L^\trole_{2,\, y_1}(\cdot)$ and measure register $V^\trole_{2,\,y_1}$ using
    $\{ \sigma^Z_{[L^\trole_{2,\, y_1}(\cdot) = y_{2}]} \}$ to obtain $y_2$.
    This process continues until the $(k-1)$-th linear map $L_{k-1,\, y_{<
        k-1}}^\trole(\cdot)$ has been measured to obtain $y_{k-1}$ in factor
    space $V_{k-1,\, y_{< k-1}}^\trole$.
    Let $y = y_1 + y_2 + \cdots + y_{k-1}$.
    
    Next, for $j \in \{1, 2, \ldots, k\}$ the player measures
    \begin{equation*}
      \Big \{ \sigma^X_{[L_{j}^\perp(\cdot) = y^\perp_{j}]} \Big \}_{y^\perp_j}\;,
    \end{equation*}
    where $L_j^\perp$ denotes the linear map $(L^\trole_{j,\, y_{<j}})^\perp$ as in
    the case $(\Read, \trole)$.
    Let $y^\perp = y_1^\perp + y_2^\perp + \cdots + y_k^\perp$, where each
    $y_j^\perp$ is a vector in the factor space $V^\trole_{j,\, y_{<j}}$.
    Finally, the player measures register $V_{> k}^\trole(y)$ using
    $\{\sigma^X_{x_{>k}}\}$ to obtain outcome $x_{> k}$.
    Let $x = x_{> k}$.
    The player replies with $(y,y^\perp,x)$.
\end{description}

By definition, when player $w$ performs the honest measurement for question
$(\Introspect, \alice)$ and player $\overline{w}$ performs the honest
measurement for question $(\Introspect, \bob)$, the joint outcome $(y, y')$ has
distribution $\mu_{\sampler,\, N}$.
In this case, the players play according to strategy~$\strategy$ and succeed
with probability~$1$.
In all other cases, it is straightforward to verify that the players succeed in
all tests performed by $\tdint$ (Figure~\ref{fig:intro-decider}) with
probability~$1$.
As a result, the value of this strategy is~$1$. 

The strategy $\strategy^\intro$ is projective by construction.
It is also consistent because of the assumed consistency of the strategy
$\strategy$ as well as consistency of the honest Pauli strategy $\strategy^\pauli$.
Furthermore, note that $\strategy^\intro$ only calls $\strategy$ for both
players on question pairs such that both types are in
$\{\Introspect,\Sample,\Read\}$.
In all these cases, $\strategy$ is called on a pair of questions $(y,y')$
distributed as $(L^\trole(z),L^{\trole'}(z))$ for
$\trole,\trole'\in\{\alice,\bob\}$ and $z$ uniform in $V$.
When $v \neq v'$, any such pair by definition has positive probability under
$\mu_{\sampler,\, n}$, and so by assumption the associated measurements from
$\strategy$ commute.
On the other hand, when $v = v'$, then the players apply the same measurements
from $\strategy$, and because $\strategy$ only uses projective measurements,
their measurements commute as well.
Examining all other cases, it follows by direct inspection that the strategy
commutes on all question pairs whose corresponding types appear as an edge in
the graph $G^\intro$.
Thus the strategy commutes with respect to the support of the distribution
$\mu_{\tsint,\, n}$.
\end{proof}

\begin{remark}\label{rk:intro-normalform}
  For future reference, we note that on any input $(\verifier,\allowbreak  \lambda,\allowbreak  \ell)$,
the Turing machine $\ComputeIntroVerifier$ \emph{always} returns a normal form verifier $\vint \allowbreak =\allowbreak 
  (\sint,\allowbreak \dint)$ --- even if $\verifier$ itself is not the description of a normal form verifier. 
  This is because, for any two integer $\lambda,\allowbreak \ell$,
  $\ComputeIntroSampler(\lambda,\allowbreak \ell)$ returns a sampler with field size $2$,
  and for any $\verifier \allowbreak =\allowbreak  (\sampler,\allowbreak \decider),\allowbreak  \lambda$ and $\ell$ the typed decider $\tdint$ specified in
  Figure~\ref{fig:intro-decider} is a $7$-input Turing machine.\footnote{We note that, even though it is not required in the definition of a normal form verifier, it is the case that $\tdint$ always halts with a
  single-bit output, even if $\sampler$ or $\decider$ themselves do not halt.}
\end{remark}

\subsection{Soundness of the introspective verifier}
\label{sec:intro-soundness}

The main result of this section is the soundness part of
Theorem~\ref{thm:introspection}.



\subsubsection{The Pauli twirl}

A key tool in the proof of soundness is the \emph{Pauli twirl}.
In this section we introduce the Pauli twirl and establish several of its
properties. 

The section closely follows Sections~$8$ and~$10$ of~\cite{NW19}. 

To begin, we define the twirl with respect to an arbitrary distribution over
unitaries.

\begin{definition}[Twirl]
  \label{def:mixing}
  Let $\mu$ be a probability distribution over a finite set of unitary matrices.
  Then for any matrix $A$, the \emph{twirl of~$A$ with respect to~$\mu$},
  denoted $\mathscr{T}_\mu(A)$, is defined as
  \begin{equation*}
    \mathscr{T}_\mu (A) = \E_{U \sim \mu} \bigl( U A U^\dagger \bigr)\;.
  \end{equation*}
\end{definition}

\tnote{added:}The twirling operation is most natural when the underlying distribution is uniform over a subgroup of unitaries, in which case fixed points of $\mathscr{T}_\mu$ have a simple structure that can be evidenced through Schur's lemma from representation theory; see, e.g.,~\cite{gross2007evenly}. Here, due to the need to work with approximation errors and the somewhat ad-hoc state dependent norm, we develop special-purpose tools that adapt the general theory; see, in particular, Lemma~\ref{lem:W} below. 

In the next two lemmas we consider the Pauli twirl, in which the distribution
$\mu$ is over subsets of Pauli observables.
First, we show how the Pauli twirl acts on Pauli matrices.
Then, using this, we derive an expression for the Pauli twirl applied to general
matrices.

\begin{lemma}[Pauli twirl of Pauli matrices]
  \label{lem:twirl-pauli}
  Let $V$ be a subspace of $\F^k$, let $W \in \{X, Z\}$, and let $\mu$ be the
  uniform distribution over~$\{\tau^W(w) : w \in V\}$.
  Let $W'\neq W$ and $u\in \F^k$.
  Then
  \begin{equation*}
    \mathscr{T}_\mu(\tau^{W'}(u)) =
    \begin{cases}
      \tau^{W'}(u)&\text{ if }u \in V^{\perp}\;,\\
      0 &\text{ if }u \notin V^{\perp}\;.
    \end{cases}
  \end{equation*}
\end{lemma}

\begin{proof}
  Let $u\in \F^k$.
  Let $c = 1$ if $W = Z$ and let $c = -1$ if $W = X$.
  Then
  \begin{align*}
    \mathscr{T}_\mu(\tau^{W'}(u))
    &= \E_{z \sim V} (\tau^W(z)  \tau^{W'}(u) \tau^W(z)^\dagger)\\
    &= \E_{z \sim V} (\omega^{c\cdot \tr(u \cdot z)} \tau^{W'}(u) \tau^W(z)
      \tau^W(z)^\dagger)\\
    &= \Big(\E_{z \sim V} \omega^{c\cdot \tr(u \cdot z)}\Big) \tau^{W'}(u)\;.
  \end{align*}
  The lemma now follows from Lemma~\ref{lem:cancellation}
  and the fact that $c \cdot u \in V^\perp$ if and only if~$u\in V^\perp$.
\end{proof}

\begin{lemma}[Pauli twirl of general matrices]
  \label{lem:mixing-L}
  Let $V= \F^k$, and let $L : V \rightarrow V$ be a linear map.
  Let $\zeta$ be the uniform distribution over $\{\tau^Z(z) \mid z \in V\}$ and
  $\chi$ the uniform distribution over $\{\tau^X(x) \mid x \in \ker(L)\}$.
  Let~$M$ be a matrix acting on $\C^V \otimes \mH_\alice$, where $\mH_\alice$ is
  a finite-dimensional Hilbert space.
  Then there exist matrices $\{M^y\}_{y\in L(V)}$ acting on $\mH_\alice$ such
  that the twirl of~$M$ with respect to~$\zeta$ and~$\chi$ is given by
  \begin{equation}
    \label{eq:mixing-L-0}
    (\mathscr{T}_{\chi} \circ \mathscr{T}_{\zeta} \otimes \Id_A)(M) = \sum_{y\in
      V} \tau^Z_{[L(\cdot)= y]} \otimes M^y\;.
  \end{equation}
	Moreover, if we apply~\eqref{eq:mixing-L-0} to each element of a POVM
  measurement $\{M_a\}$, then for each $y \in V$, the set $\{M^y_a\}$ also forms
  a POVM measurement.
\end{lemma}

\begin{proof}
  The collection $\{\tau^X(x)\tau^Z(z)\}_{x, z \in V}$ forms a basis for the
  complex linear space of matrices acting on $\C^V$.
  As a result, we can write
  \begin{equation*}
    M = \sum_{x, z \in V} \tau^X(x) \tau^Z(z) \otimes M_{x, z}\;,
  \end{equation*}
  for matrices $M_{x,z}$ on $\mH_\alice$.
  We now use Lemma~\ref{lem:twirl-pauli} to compute the twirl first with respect
  to $\zeta$ and then with respect to $\zeta$ and $\chi$:
  \begin{align*}
    (\mathscr{T}_{\zeta} \otimes \Id_{A})(M)
    &= \sum_{x, z \in V} \mathscr{T}_{\zeta}(\tau^X(x)) \tau^Z(z) \otimes M_{x, z}
      = \sum_{z \in V} \tau^Z(z) \otimes M_{0, z}\;,\\
    (\mathscr{T}_{\chi} \circ \mathscr{T}_{\zeta} \otimes \Id_{A})(M)
    &= \sum_{z \in V} \mathscr{T}_{\chi}(\tau^Z(z)) \otimes M_{0, z}
      =  \sum_{z \in \ker(L)^\perp} \tau^Z(z) \otimes M_{0, z}\;.
  \end{align*}
  For all $y \in V$, let $u_y$ denote an arbitrary element of $L^{-1}(y)$ if $y$
  is in the image of $L$; otherwise, set $u_y = 0$.
  Expanding $\tau^Z(z)$ using the first part of
  Lemma~\ref{lem:why-didnt-i-think-of-this-before},
  \begin{align}
    \label{eq:post-twirl-matrix}
    \sum_{z \in \ker(L)^\perp} \tau^Z(z) \otimes M_{0, z}
    & = \sum_{z \in \ker(L)^\perp} \,\, \sum_{y} \omega^{\tr(u_y \cdot z)}
      \tau^Z_{[L(\cdot)=y]} \otimes M_{0, z} \notag \\
    & =  \sum_{y} \tau^Z_{[L(\cdot)=y]} \otimes \Big(\sum_{z \in \ker(L)^\perp}
      \omega^{\tr(u_y \cdot z)}  M_{0, z}\Big)\;.
  \end{align}
  Equation~\eqref{eq:mixing-L-0} follows by setting $M^y = \sum_{z \in
    \ker(L)^\perp} \omega^{\tr(u_y \cdot z)} M_{0, z}$.

  For the ``moreover'' part, note first that whenever $M \geq 0$ it holds that
  any twirl satisfies $0 \leq \mathscr{T}_{\mu}(M)$.
  As a result, each matrix $M^y$ must be positive semi-definite due to
  Equation~\eqref{eq:post-twirl-matrix} and the fact that the
  $\{\tau^Z_{[L(\cdot)=y]}\}_y$ matrices are orthogonal projections.
  Next, suppose $\{M_a\}$ is a POVM measurement, and write $N = \sum_a M_a$ for
  the identity matrix.
  Then by linearity, for each $y \in V$, $\sum_a M_a^y = N^y$.
  In addition, $N_{0, 0} = \Id$, and $N_{x, z} = 0$ otherwise.
  As a result, $N^y = N_{0, 0} = \Id$, and so $\{M_a^y\}$ also forms a POVM.
\end{proof}

In the next few lemmas we derive a sufficient condition for a measurement to be
close to its own Pauli twirl, namely that it satisfies certain commutation
relations with the Pauli basis measurements.

\begin{lemma}[Commuting with Pauli basis implies commuting with Pauli observables]
  \label{lem:W} 
  Let~$\cal{X},\cal{A}$ be finite sets and~$D$ be a distribution over $\cal{X}$.
  For each~$x \in \cal{X}$, let $V_x$ be a register subspace of $V = \F^k$, and
  let $L_x:V_x \to V_x$ be a linear map.

  Consider a state $\ket{\psi} = \ket{\EPR}_V \otimes \ket{\aux}$, where
  $\ket{\EPR}_{V} \in \mH_\alice \otimes \mH_\bob$, for $\mH_\alice,\mH_\bob
  \cong \C^{V}$, is defined in Definition~\ref{def:EPR} and $\ket{\aux}\in
  \mH_{\alice'} \otimes \mH_{\bob'}$ is arbitrary.
  For each~$x \in \cal{X}$, let $\{M^{x}_a\}_{a \in \cal{A}}$ be a measurement
  on $\mH_{\alice} \otimes \mH_{\alice'}$.
  Let $W\in\{X,Z\}$.
  Then the following are equivalent on the state $\ket{\psi}$:
	\begin{itemize}
	\item On average over $x\sim D$, 
    \begin{equation*}
      \big[ M^x_a \,,\, \big(\tau^W_{[L_x(\cdot)=y]} \otimes \Id_{\overline{V_x}}
      \otimes \Id_{\alice'}\big) \big]  \otimes \Id_\bob \approx_\eps 0\;.
    \end{equation*}
  \item On average over $x \sim D$ and $v$ drawn uniformly from $\ker
    (L_x)^\perp$,
    \begin{equation*}
      \big[ M^x_a \,,\, \big( \tau^W(v) \otimes \Id_{\overline{V_x}} \otimes
      \Id_{\alice'} \big) \big]  \otimes \Id_\bob \approx_\eps 0\;.
    \end{equation*}
	\end{itemize}
	In the equations above, for each $x \in \cal{X}$ we decompose $\mH_\alice = \C^{V_x} \otimes \C^{\overline{V_x}}$ and interpret $\tau^W_{[L_x(\cdot)=y]}$ and $\tau^W(v)$ for $v \in \ker(L_x)^\perp$ as acting on $\C^{V_x}$. 
\end{lemma}

\begin{proof}
  For $x\in \cal{X}$, $a\in \cal{A}$, $y$ in the range of $L_x$, and $v\in \F^k$
  define
  \begin{equation*}
    \Delta^x_{a,y} = \big[ M^x_a \,,\, \big(\tau^W_{[L_x(\cdot)=y]} \otimes
    \Id_{\overline{V_x}} \otimes \Id_{\alice'}\big) \big]\otimes \Id_\bob\:,
  \end{equation*}
  \begin{equation*}
    \Delta^x_a(v) = \big[ M^x_a \,,\, \big( \tau^W(v) \otimes
    \Id_{\overline{V_x}} \otimes \Id_{\alice'} \big) \big]  \otimes \Id_\bob\:.
  \end{equation*}
  By the first item of Lemma~\ref{lem:why-didnt-i-think-of-this-before}, for
  each $v \in \ker(L_x)^\perp$,
  \begin{equation*}
    \Delta^x_a(v) = \sum_{y \in V_x} \omega^{\tr(u_y \cdot v)} \Delta^x_{a,y}\;,
  \end{equation*}
	where for every $y$ in the range of $L_x$, $u_y$ is a fixed element in
  $L_x^{-1}(y)$.
  The expression for the closeness of $\Delta^x_a$ to $0$ on average over $x
  \sim D$ and $v \sim \ker(L_x)^\perp$ (i.e.,\ the second quantity of the Lemma
  statement), is equal to
  \begin{equation*}
    \E_{x\sim D} \E_{v \sim \ker(L_x)^\perp} \sum_a\, \bigl \| \Delta^x_a (v)
  \ket{\psi}\bigr \|^2,
  \end{equation*}
  and can be expanded as
  \begin{equation}\label{eq:lemW-1}
    \E_{x\sim D} \E_{v \sim \ker(L_x)^\perp} \sum_a\, \bra{\psi} \sum_{y,y'}
    \omega^{\tr ((u_y'-u_y) \cdot v)} \bigl( \Delta^x_{a,y} \bigr)^\dagger
    \Delta^x_{a,y'} \ket{\psi}\;.
  \end{equation}
  If $y \neq y'$, then by definition $L_x(u_y) \neq L_x(u_{y'})$, and so $u_y' -
  u_y$ is not in $\ker(L_x)$.
  Lemma~\ref{lem:cancellation} (with $V = \ker(L_x)^\perp$) thus implies
  that~\eqref{eq:lemW-1} equals
  \begin{equation*}
    \E_{x\sim D} \sum_a\, \bra{\psi} \sum_{y} \bigl( \bigl( \Delta^x_{a,y}
    \bigr)^\dagger \Delta^x_{a,y} \bigr) \ket{\psi} = \E_{x\sim D} \sum_{a, y}
    \, \bigl \|\Delta^x_{a,y} \ket{\psi} \bigr \|^2
  \end{equation*}
  which is the closeness of $\Delta^x_{a,y}$ to $0$ on average over $x \sim D$.
  Thus, $\Delta^x_a(v) \approx_\eps 0$ on average over $x$ and $v$ if and only
  if $\Delta^x_{a,y} \approx_\eps 0$ on average over $x$.
\end{proof}

\begin{lemma}[Commuting implies twirl]
  \label{lem:mixing-U}
	Let $\cal{X}$ be a finite set and $D$ a distribution on $\cal{X}$. 
  For each~$x\in \cal{X}$, let $\{ M^x_a \}$ be a POVM on $\mH_\alice$, and let
  $\mu_x$ be a distribution over unitary matrices acting on $\mH_\alice$.
  Suppose that on average over $x \sim D$ and $U \sim \mu_x$ it holds that
  \begin{equation*}
    \bigl[ M^x_a , U^\dagger \bigr] \otimes \Id_\bob \approx_{\eps} 0\;,
  \end{equation*}
  where the commutator is evaluated on a state $\ket{\psi} \in \mH_\alice
  \otimes \mH_\bob$.
  Then on average over $x\sim D$,
  \begin{equation*}
    M^x_a \otimes \Id_\bob \approx_\eps \mathscr{T}_{\mu_x} ( M^x_a ) \otimes
    \Id_\bob \;.
  \end{equation*}
\end{lemma}

\begin{proof}
  Observe that
  \begin{equation}
    \E_{x\sim D} \sum_a \, \bigl\| \bigl( \mathscr{T}_{\mu_x} (M^x_a) - M^x_a
    \bigr) \otimes \Id_\bob \ket{\psi} \bigr\|^2 =
    \E_{x\sim D} \sum_a \, \bigl\| \E_{U\sim \mu_x} \bigl( U \big[ M^x_a, U^\dagger
    \bigr] \bigr) \otimes \Id_\bob \ket{\psi} \bigr\|^2.
    \label{eq:twirl2}
  \end{equation}
	Applying Jensen's inequality, the right-hand side of~\eqref{eq:twirl2} is at
  most
  \begin{equation*}
    \E_{x\sim D} \E_{U\sim \mu_x} \sum_a \, \bigl\| \bigl( U \big[ M^x_a, U^\dagger
    \bigr] \bigr) \otimes \Id_\bob \ket{\psi} \bigr\|^2
    = \E_{x\sim D} \E_{U\sim \mu_x} \sum_a \, \bigl\| \big[ M^x_a, U^\dagger
    \bigr]  \otimes \Id_\bob \ket{\psi} \bigr\|^2,
  \end{equation*}
  using the unitary invariance of the Euclidean norm.
  This last quantity is $O(\eps)$, by assumption.
\end{proof}

\begin{lemma}[Commuting with each implies commuting with both]
  \label{lem:commute-with-both}
  Let $\cal{X}$ be a finite set and $D$ a distribution on $\cal{X}$.
  For each~$x\in \cal{X}$, let $\{ M^x_a \}$ be a POVM on $\mH_\alice$ and let
  $\mu_{x,1}, \mu_{x,2}$ be two distributions over unitary matrices acting on
  $\mH_\alice$.
  Suppose that for each $i \in \{1, 2\}$, on average over $x \sim D$ and $U_i
  \sim \mu_{x,i}$,
  \begin{equation}\label{eq:to-be-tagged}
    \bigl[ M^x_a , U_i^\dagger \bigr] \otimes \Id_\bob \approx_{\eps} 0,
  \end{equation}
  where the expression is evaluated on some state $\ket{\psi} \in
  \mH_\alice\otimes\mH_\bob$, where $\mH_\bob \cong \mH_\alice$.
  Suppose further that on average over $x \sim D$ and $U_2 \sim \mu_{x, 2}$,
  \begin{equation}\label{eq:to-be-tagged-dos}
    U_2^\dagger \otimes \Id_\bob \approx_{\eps} \Id_A \otimes U_2\;.
  \end{equation}
  (The corresponding statement for $i = 1$ is not needed.)
  Then on average over $x \sim D$, $U_1 \sim \mu_{x, 1}$, and $U_2 \sim \mu_{x,
    2}$,
  \begin{equation*}
    \bigl[ M^x_a , U_1^\dagger U_2^\dagger \bigr] \otimes \Id_\bob
    \approx_{\eps} 0\;.
  \end{equation*} 
\end{lemma}

\begin{proof}
  The claim follows from the following sequence of approximations:
  \begin{align*}
    M^x_a U_1^\dagger U_2^\dagger \otimes \Id_\bob
    & \approx_{\eps} M^x_a U_1^\dagger \otimes U_2
      \tag{by~\eqref{eq:to-be-tagged-dos}}\\
    & \approx_{\eps} U_1^\dagger M^x_a  \otimes U_2
      \tag{by~\eqref{eq:to-be-tagged} for $i = 1$}\\
    & \approx_{\eps} U_1^\dagger M^x_a U_2^\dagger \otimes \Id_\bob
      \tag{by~\eqref{eq:to-be-tagged-dos}}\\
    & \approx_{\eps} U_1^\dagger U_2^\dagger M^x_a \otimes \Id_\bob
      \tag{by~\eqref{eq:to-be-tagged} for $i = 2$}\;,
  \end{align*}
  where each step also uses Fact~\ref{fact:add-a-proj}.
  This is equivalent to $\bigl[ M^x_a , U_1^\dagger U_2^\dagger \bigr] \otimes
  \Id_\bob \approx_{\eps} 0$, completing the proof.
\end{proof}

The following is a slight generalization of~\cite[Fact~4.25]{NW19},
and we give a similar proof.

\begin{lemma}[Close to sub-measurement implies close to measurement]
  \label{lem:replace-with-measurement}
  Let $\cal{X},\cal{A}$ be finite sets and $D$ be a distribution on $\cal{X}$.
  Suppose that for each $x\in \cal{X}$, $\{A^x_a\}_{a\in \cal{A}}$ is a
  projective measurement and $\{B^x_a\}_{a\in \cal{A}}$ is a set of matrices
  such that each~$B^x_a$ is positive semidefinite and $\sum_{a}B^x_a \leq \Id$.
  Suppose $\{C^x_a\}_{a \in \cal{A}}$ is a POVM such that $C^x_a \geq B^x_a$ for
  all~$x$ and~$a$.
  Then if, on average over $x\sim D$, $A^x_a \approx_{\eps} B^x_a$, then, on
  average over $x\sim D$, $A^x_a \approx_{\eps^{1/2}} C^x_a$.
\end{lemma}

\begin{proof}
  By the triangle inequality,
  \begin{equation*}
    \E_{x} \sum_a \Vert (A^{x}_a - C^{x}_a) \ket{\psi} \Vert^2
    \leq 2\E_{x} \sum_a \Vert (A^{x}_a - B^{x}_a) \ket{\psi} \Vert^2
    + 2 \E_{x}\ \sum_a \Vert (C^x_a - B^x_a) \ket{\psi}\Vert^2.
\end{equation*}
The first term on the right-hand side is $O(\eps)$ by assumption.
For the second,
\begin{multline*}
  \E_{x}\ \sum_a \Vert (C^x_a - B^x_a) \ket{\psi}\Vert^2
  = \E_{x}\ \sum_a \bra{\psi} (C^x_a - B^x_a)^2 \ket{\psi}
  \leq \E_{x}\ \sum_a \bra{\psi} (C^x_a - B^x_a) \ket{\psi}\\
  = 1- \E_{x} \sum_a\bra{\psi}  B^{x}_a \ket{\psi}
  \leq1- \E_{x} \sum_a\bra{\psi}  (B^{x}_a)^2 \ket{\psi}\;,
\end{multline*}
where the middle inequality uses $0\leq C^x_a-B^x_a\leq \Id$ for all $x,a$. 
Write $1 = \E_{x} \sum_a \bra{\psi} (A^{x}_a)^2 \ket{\psi}$, which holds
because~$A$ is a projective measurement.
Then
\begin{multline*}
  \E_{x} \sum_a \bra{\psi} ((A^{x}_a)^2  - (B^{x}_a)^2)\ket{\psi}
  = \Re\Big(\E_{x} \sum_a \bra{\psi} (A^{x}_a  + B^{x}_a)(A^{x}_a - B^{x}_a)
  \ket{\psi}\Big)\\
  \leq \E_{x} \sqrt{\sum_a \Vert (A^{x}_a + B^{x}_a) \ket{\psi}\Vert^2}
	\cdot \sqrt{\sum_a \Vert (A^{x}_a - B^{x}_a)\ket{\psi} \Vert^2}
\end{multline*}
where the first equality follows from the fact that $A^x_a$ and $B^x_a$ are
Hermitian.
For each~$x\in X$ the first square root is $O(1)$.
This allows us to move the expectation into the second square root by Jensen's
inequality.
The result is $O(\eps^{1/2})$ by assumption. 
\end{proof}

Now we put everything together to show the main result of this section. 

\begin{lemma}
  \label{lem:mixing}
  Let~$\cal{X},\cal{A}$ be a finite sets and~$D$ be a distribution over
  $\cal{X}$.
  For each~$x \in \cal{X}$, let $V_x$ be a register subspace of $V = \F^k$, let
  $U_x$ be a register subspace of~$V_x$, and let $L_x:U_x \to U_x$ be a linear
  map.

  Consider a state $\ket{\psi} = \ket{\EPR}_V \otimes \ket{\aux}$, where
  $\ket{\EPR}_{V} \in \mH_\alice \otimes \mH_\bob$, for $\mH_\alice,\mH_\bob
  \cong \C^{V}$, is defined in Definition~\ref{def:EPR} and $\ket{\aux}\in
  \mH_{\alice'} \otimes \mH_{\bob'}$ is arbitrary.
  For each~$x \in \cal{X}$, let $\{M^{x}_{y,a}\}_{y \in U_x,\, a\in \cal{A}}$ be
  a projective measurement on $\mH_{V_x} \otimes \mH_{\alice'}$.
  Suppose that on average over~$x \sim D$ the following conditions hold.
  \begin{alignat*}{2}
    \text{(Consistency):} && \quad
    \bigl( M^{x}_{y} \otimes \Id_{\overline{V_x}} - \tau^Z_{[L_x(\cdot) = y]}
    \otimes \Id_{\overline{U_x}} \otimes \Id_{\alice'}\bigr) \otimes \Id_\bob
    \approx_\eps 0\;, &\\
    \text{(Commutation):} && \quad
      \bigl[ M^{x}_{y,\, a} \otimes \Id_{\overline{V_x}}\,,\, \tau^Z_{z} \otimes
      \Id_{\overline{U_x}}   \otimes \Id_{\alice'} \bigr] \otimes
      \Id_\bob\approx_\eps 0\;, &\\
      && \quad
      \bigl[ M^{x}_{y,\, a}\otimes \Id_{\overline{V_x}}\,,\,
      \tau^X_{[L_x^\perp(\cdot) = y^\perp]}  \otimes \Id_{\overline{U_x}}
      \otimes \Id_{\alice'} \bigr]  \otimes \Id_\bob  \approx_\eps 0\;.&
  \end{alignat*}
  Here, the projector $\tau^Z_{[L_x(\cdot) = y]}$ acts on the register subspace
  $\mH_{U_x}$, and $\overline{U_x}$ and $\overline{V_x}$ denote the
  complementary register subspaces of $U_x$ and $V_x$, respectively, within $V$. Furthermore, the answer summations are over all $y, z \in U_x$ and $a \in \cal{A}$.

  Then for each $x\in \cal{X}$ and $y \in U_x$, there exists a POVM measurement
  $\bigl\{ M^{x,\, y}_a \bigr\}_{a \in \cal{A}} $ on $\mH_{V_x \setminus
    U_x}\otimes \mH_{\alice'}$ such that on average over $x \sim D$,
  \begin{equation*}
    (M^{x}_{y,\, a} \otimes \Id_{\overline{V_x}}) \otimes \Id_{B}
    \approx_{\eps^{1/2}} \Bigl(\tau^Z_{[L_x(\cdot) = y]} \otimes M^{x,\, y}_a
    \otimes \Id_{\overline{V_x}}\Bigr) \otimes \Id_\bob\;.
  \end{equation*}
\end{lemma}

\begin{proof}
  For each~$x \in X$, let $\zeta_{x}$ be the uniform distribution over $U_x$ and
  $\chi_{x}$ be the uniform distribution over $\ker(L_x)$.
   Applying Lemma~\ref{lem:W} to the first commutation assumption (letting ``$L_x$'' in Lemma~\ref{lem:W} be the identity map on $U_x$, and letting ``$D$'' be the uniform distribution over $\ker(L_x)^\perp = U_x$), we get that on average over $x \sim \cal{X}$ and $v \sim \zeta_x$, we have
  \begin{equation*}
    \bigl[ M^{x}_{y,\, a} \otimes \Id_{\overline{V_x}} \,,\, \tau^Z(v) \otimes
    \Id_{\overline{U_x}} \otimes \Id_{\alice'} \bigr]  \otimes \Id_\bob
    \approx_\eps 0\;.
  \end{equation*}
  Applying Lemma~\ref{lem:W} to the second commutation assumption (letting ``$L_x$'' in Lemma~\ref{lem:W} be the $L_x^\perp$ in the statement of Lemma~\ref{lem:mixing} and letting $D$ be the uniform distribution over $\ker(L_x^\perp)^\perp$, which is $\ker(L_x)$ by Lemma~\ref{lem:L_perp_perp}), we get 
on average over $x \sim \cal{X}$ and $v \sim \xi_x$, we have
  \begin{equation*}
    \bigl[ M^{x}_{y,\, a} \otimes \Id_{\overline{V_x}} \,,\, \tau^X(v) \otimes
    \Id_{\overline{U_x}} \otimes \Id_{\alice'} \bigr]  \otimes \Id_\bob
    \approx_\eps 0\;.
  \end{equation*}    
  By Lemma~\ref{lem:commute-with-both}, this implies that on average over $x
  \sim D$, $u \sim \zeta_{x}$, and~$v \sim \chi_{x}$,
  \begin{equation*}
    \bigl[ M^{x}_{y,\, a} \otimes \Id_{\overline{V_x}} \,,\, \tau^Z(u) \tau^X(v)
    \otimes \Id_{\overline{U_x}} \otimes \Id_{\alice'} \bigr] \otimes \Id_\bob
    \approx_\eps 0\;.
  \end{equation*}
  By Lemma~\ref{lem:mixing-U} and Lemma~\ref{lem:mixing-L},
  this implies that on average over $x\sim D$,
  \begin{align}
    M^{x}_{y,\, a} \otimes \Id_{\overline{V_x}} \otimes \Id_\bob
    & \approx_\eps \bigl( \mathscr{T}_{\chi_{x}} \circ \mathscr{T}_{\zeta_{x}} (
      M^{x}_{y,\, a}) \bigr) \otimes \Id_{\overline{V_x}} \otimes
      \Id_\bob\notag\\
    & = \Big(\sum_{y' \in V_x} \tau^Z_{[L_x(\cdot) = y']} \otimes  M^{x,\,
      y'}_{y,\, a} \Big) \otimes \Id_{\overline{V_x}} \otimes
      \Id_\bob\;,\label{eq:mixing-eq-1}
  \end{align}
	for some POVM measurement $\{M^{x,\, y'}_{y,\, a} \}$ on $\mH_{V_x \setminus
    U_x} \otimes \mH_{\alice'}$.
	
	In the following sequence of equations, whenever an operator does not act on a
  subsystem it should be assumed that it is appropriately tensored with the
  identity.
  For clarity, we explicitly indicate using a subscript $\alice$ or $\bob$
  whether a Pauli operator acts on $\mH_\alice$ or $\mH_\bob$.
  Then on average over $x\sim D$ we have
  \begin{align*}
    M^{x}_{y, \, a}
    & =  M^{x}_{y, \, a} \cdot M^{x}_y  \tag{$M^x$ is projective}\\
    & \approx_\eps  M^{x}_{y, \, a} \cdot \big(
      \tau^Z_{[L_x(\cdot)=y]}\big)_\alice
      \tag{Consistency assumption}\\
    & \approx_0 M^x_{y, \, a} \otimes \big(\tau^Z_{[L_x(\cdot)=y]}\big)_\bob
      \tag{Paulis are self-consistent}\\
    & \approx_\eps \Big( \sum_{y'} \big(\tau^Z_{[L_x(\cdot) = y']}\big)_\alice
      \otimes  M^{x,\, y'}_{y,\, a} \Big) \otimes \big(\tau^Z_{[L_x(\cdot)=y]}
      \big)_\bob \tag{Equation~\eqref{eq:mixing-eq-1}}\\
    & \approx_0 \sum_{y'} \big(\tau^Z_{[L_x(\cdot) =
      y']}\tau^Z_{[L_x(\cdot)=y]}\big)_\alice \otimes  M^{x,\, y'}_{y,\, a}
      \tag{Paulis are self-consistent} \\
    & = \big( \tau^Z_{[L_x(\cdot) = y]} \big)_\alice
      \otimes M^{x,\, y}_{y,\, a} \;,
  \end{align*}
  where $\approx_0$ indicates equality with respect to the state $\ket{\EPR}_{V}
  \otimes \ket{\aux}$, and we have repeatedly used Fact~\ref{fact:add-a-proj}.
  This is essentially the statement promised by the lemma, except that
  $\{M^{x,\, y}_{y,\, a}\}_a$ does not necessarily sum to identity (since we
  only sum over~$a$).
  To remedy this, define $M^{x,\, y}_a = \sum_{y'} M^{x,\, y}_{y',\, a}$ and note
  that $M^{x,\, y}_{a} \geq M^{x,\, y}_{y,\, a}$, so
  by~Lemma~\ref{lem:replace-with-measurement} and the fact that~$M^x$ is
  projective,
  \begin{equation}\label{eq:mixing-eq-2}
    (M^{x}_{y,\, a} \otimes \Id_{\overline{V_x}}) \otimes \Id_{B} \approx_{\eps^{1/2}}
    \Bigl(\tau^Z_{[L_x(\cdot) = y]} \otimes M^{x,\, y}_a \otimes \Id_{\overline{V_x}}\Bigr)
    \otimes \Id_\bob\;.
  \end{equation}
  $\{M^{x\,y}_a\}$ is the POVM measurement guaranteed in the lemma statement,
  which concludes the proof.
\end{proof}

\subsubsection{Preliminary lemmas}

We show a few simple lemmas that allow us to argue about measurements that
have a decomposition across a tensor product of two Hilbert spaces, within the
space of a single player.

\def\ha{\mH_{\alice}}
\def\hb{\mH_{\bob}}
\def\haq{\mH_{\alice,\, a}}
\def\haa{\mH_{\alice',\, a}}
\def\hbq{\mH_{\bob,\, a}}
\def\hba{\mH_{\bob',\, a}}
\def\sq{\ket{\psi_{\textsc{que},\, a}}}
\def\sa{\ket{\psi_{\textsc{ans},\, a}}}

\begin{lemma}
  \label{lem:conditional}
	Let $\cal{A}, \cal{B}$ be finite sets.
	Let $\ket{\psi} \in \ha \otimes \hb$ be a state.
  Consider the following: for all $a \in \cal{A}$,
  \begin{enumerate}
  \item Let $\haq$, $\hbq$, $\haa$, and $\hba$ be Hilbert spaces such that
    \begin{equation*}
  		\ha = \haq \otimes \haa\quad
      \text{and}\quad
      \hb = \hbq \otimes \hba\;,
		\end{equation*}
    and let $\sq \in \haq \otimes \hbq$ be a ``question state'' and $\sa \in
    \haa \otimes \hba$ be an ``answer state'' such that $\ket{\psi} = \sq
    \otimes \sa$.
  \item Let $Q_a$ be projectors on $\haq$ such that $\{Q_a \otimes
    \Id_{\haa}\}$ forms a projective measurement on $\ha$ and let
    $\{A^{a}_b\}_{b \in \cal{B}}$ and $\{B^{a}_b\}_{b \in \cal{B}}$ be matrices
    acting on $\haa$.
  \item Let $D$ be the distribution on~$\cal{A}$ obtained by measuring
    $\ket{\psi}$ using $\{ Q_a \otimes I_{\haa}\}_{a \in \cal{A}}$.
  \end{enumerate}
  Then the following are equivalent:
	\begin{itemize}
	\item On average over $a\sim D$ and with respect to state $\ket{\psi}$,
    \begin{equation*}
      (\Id_{\haq} \otimes A^{a}_b) \otimes \Id_\bob \approx_\eps
      (\Id_{\haq} \otimes B^{a}_b) \otimes \Id_\bob\;.
    \end{equation*}
  \item $\bigl( Q_a \otimes A^{a}_b \bigr) \otimes \Id_{\bob} \approx_\eps
    \bigl( Q_a \otimes B^{a}_b \bigr) \otimes \Id_{\bob}$ on state $\ket{\psi}$.
	\end{itemize}
\end{lemma}

\begin{proof}
 Expand
  \begin{align*}
    & \sum_{a,b} \norm{ (Q_a \otimes A^a_b - Q_a \otimes B^a_b) \otimes
      \Id_{\bob} \cdot \sq \otimes \sa }^2\\
    = & \sum_{a,b} \norm{ Q_a \otimes (A^a_b -  B^a_b) \otimes
        \Id_{\bob} \cdot \sq \otimes \sa}^2\\
    = & \sum_{a, b} \norm{ Q_a \otimes I_{\hbq} \sq}^2 \cdot
        \norm{  (A^a_b -  B^a_b) \otimes \Id_{\hba} \sa }^2\\
	= & \sum_{a, b} \norm{ Q_a \otimes \Id_{\bob} \ket{\psi}}^2 \cdot
      \norm{  (A^a_b -  B^a_b) \otimes \Id_{\hba} \sa }^2\\
    = & \E_{a \sim D} \sum_{b}
        \norm{  (A^a_b -  B^a_b) \otimes \Id_{\hba} \sa }^2\\
    = & \E_{a \sim D} \sum_{b}
      \norm{  \Id_{\haq} \otimes (A^a_b -  B^a_b) \otimes
        \Id_{\bob} \cdot \sq \otimes \sa }^2\;.
  \end{align*}
  In going from the third to the fourth line we used the fact that
  \[
\begin{aligned}
&\norm{ Q_a \otimes I_{\hbq} \sq }^2 \\
 &=
    \norm{ Q_a \otimes \Id_{\haa} \otimes I_{\bob} \sq \otimes \sa }^2\\
 &= \norm{ Q_a \otimes \Id_{\haa} \otimes \Id_{\bob} \ket{\psi} }^2\;.
\end{aligned}
\]
	\tnote{added:}The last equality above is obtained in a similar manner. 
  Hence, the first line is $O(\eps)$ if and only if the last one is.
\end{proof}

\begin{lemma}\label{lem:commutation-simplification}
  Let $\mA,\mB,\mC$ be finite sets.
  Let $\ket{\psi} \in \ha \otimes \hb$ be a state, and let $\{A_{a,\, b}\}_{a
    \in \mA, b \in \cal{B}}$ and $\{B_{a,\, c}\}_{a \in \mA, c \in \cal{C}}$ be
  POVMs acting on $\ha$.
  Suppose further that:
  \begin{enumerate}
  \item The measurements approximately commute,
    i.e.,\, \label{item:almost-compute}
    \begin{equation*}
      \bigl[ A_{a,\, b}, B_{a,\, c} \bigr] \otimes \Id_\bob \approx_\delta 0\;,
    \end{equation*}
    where the approximation holds with respect to the state $\ket{\psi}$. 
  \item For all $a \in \cal{A}$, there exist Hilbert spaces $\haq$, $\haa$,
    $\hbq$, $\hba$, and states $\sq \in \haq \otimes
    \hbq$, $\sa \in \haa \otimes \hba$ such that
    \begin{align*}
      \ha & = \haq \otimes \haa\;,\\
      \hb & = \hbq \otimes \hba\;,\\
      \ket{\psi} & = \sq \otimes \sa\;.
    \end{align*}
  \item \label{item:weird-decomposition} For all $a \in \cal{A}$, there exist
    projectors $Q_a$ on $\haq$ and matrices $\{A^{a}_b\}_{b \in
      \cal{B}}$, $\{B^{a}_c\}_{c \in \cal{C}}$ acting on $\haa$
    such that $\{Q_a \otimes \Id_{\haa}\}$ is a projective
    measurement on $\ha$ and
    \begin{equation*}
      A_{a, \, b} = Q_a \otimes A^a_b\;, \quad
      B_{a, \, c} = Q_a \otimes B^a_c\;.
    \end{equation*}
  \end{enumerate}
  Then
  \begin{equation*}
    \bigl[ \Id_{\haq} \otimes A^{a}_b \,,\, \Id_{\haq}
    \otimes B^{a}_c \bigr] \otimes \Id_\bob \approx_\delta 0\;,
  \end{equation*}
  on average over $a \sim D$ where $D$ is the distribution on~$\cal{A}$ obtained
  by measuring $\ket{\psi}$ using $\{ Q_a \otimes I_{\haa} \}_{a \in \cal{A}}$.
\end{lemma}
 
\begin{proof}
  The assumptions of the lemma imply that
  \begin{align*}
    (Q_a \otimes A^a_b B^a_c) \otimes \Id_\bob
    & = ((Q_a \otimes A^a_b) \cdot (Q_a \otimes B^a_c)) \otimes \Id_\bob
      \tag{$Q_a$ is a projector}\\
    & = (A_{a, b} \cdot B_{a, c}) \otimes \Id_\bob
      \tag{Item~\ref{item:weird-decomposition}}\\
    & \approx_\delta (B_{a, c} \cdot A_{a, b}) \otimes \Id_\bob
      \tag{Item~\ref{item:almost-compute}}\\
    & = ((Q_a \otimes B^a_c) \cdot (Q_a \otimes A^a_b)) \otimes \Id_\bob
      \tag{Item~\ref{item:weird-decomposition}}\\
    & = (Q_a \otimes B^a_c A^a_b) \otimes \Id_\bob\;.
      \tag{$Q_a$ is a projector}
  \end{align*}
  We apply Lemma~\ref{lem:conditional} as follows.
  The set ``$\mA$'' in Lemma~\ref{lem:conditional} is the same as $\cal{A}$
  here, and the set ``$\mB$'' is the product set $\mB \times \mC$ here.
  The matrices ``$\{A^a_b\}$'' are $\{A^a_{b}B^a_c\}$ here and ``$\{B^{a}_b\}$''
  are $\{B^a_c A^a_{b}\}$ here.
  We then obtain, on average over $a \sim
  D$,
  \begin{equation*}
    (\Id_{\mH_{\alice, a}}\otimes A^{a}_b B^a_c) \otimes \Id_\bob
    \approx_\delta (\Id_{\mH_{\alice, a}} \otimes B^{a}_c A^a_b) \otimes \Id_\bob\;.
  \end{equation*}
  This implies the conclusion of the lemma.
\end{proof}

\begin{lemma}
  \label{lem:conditional-consistency}
	Let $ \cal{Y}$ be a finite set and for all $y\in \cal{Y}$ let $\{A^y_{x,\, z}\}$ be a POVM on $\mH_\alice$. Let 
  $\{B_{x,\, y,\, z}\}$ be a projective measurement on $\mH_\bob$.
  Suppose that
  \begin{equation}\label{eq:cc-1}
    \sum_{x,\, y,\, z} \bra{\psi} A^y_{x,\, z} \otimes B_{x,\, y,\, z}
    \ket{\psi} \ge 1 - \delta\;.
  \end{equation}
  Then with respect to state $\ket{\psi}$,
  \begin{equation*}
    \ia \otimes B_{x,\, y,\, z} \approx_\delta A^y_{x,\, z} \otimes B_{x,\,
      y}\;.
  \end{equation*}
\end{lemma}

\begin{proof}
  Using the fact that $\{B_{x,\, y,\, z}\}$ is projective, we have $B_{x,\, y,\, z} =
  B_{x,\, y} B_z$ for all $x,y,z$, so that~\eqref{eq:cc-1} implies
  \begin{equation*}
    \sum_{x,\, y,\, z} \bra{\psi} A^y_{x,\, z} \otimes B_{x,\, y} B_{z}
    \ket{\psi} \ge 1 - \delta\;.
  \end{equation*}
  Define $\hat{A}_z = \sum_{x,\, y} A^y_{x,\, z} \otimes B_{x,\, y}$ and
  $\hat{B}_z = \ia \otimes B_z$.
  The above equation simplifies to
  \begin{equation*}
    \sum_{y,\, z} \bra{\psi} \hat{A}_z \hat{B}_z \ket{\psi} \ge 1 - \delta\;.
  \end{equation*}
  This implies that $\hat{A}_z \approx_\delta \hat{B}_z$ as
  \begin{equation*}
    \sum_z \norm{(\hat{A}_z - \hat{B}_z) \ket{\psi}}^2 = \sum_z \bra{\psi}
    \hat{A}_z^2 + \hat{B}_z^2 \ket{\psi} - 2 \bra{\psi} \hat{A}_z \hat{B}_z
    \ket{\psi} \le 2 \delta\;,
  \end{equation*}
  where the equality uses the fact that $\hat{A}_z$ and $\hat{B}_z$ commute.
  To conclude the proof, we have
  \begin{equation*}
    \ia \otimes B_{x,\, y,\, z} = \ia \otimes B_{x,\, y} B_z \approx_\delta (\ia
    \otimes B_{x,\, y}) \sum_{x',\, y'} A^{y'}_{x',\, z} \otimes B_{x',\, y'} =
    A^y_{x,\, z} \otimes B_{x,\, y},
  \end{equation*}
  where the approximation follows from $\hat{A}_z \approx_\delta \hat{B}_z$ and
  Fact~\ref{fact:add-a-proj}.

\end{proof}

\subsubsection{Soundness proof}

We analyze the soundness of the introspective verifier.

\begin{proof}[Proof of the soundness part of Theorem~\ref{thm:introspection}]
  Let $\verifier=(\sampler, \decider)$ be a normal form verifier such that
  $\sampler$ is an $\ell$-level sampler.
  Recall the definition of the introspective verifier $\tvint =
  (\tsint, \tdint)$ corresponding to $\verifier$ from
  Section~\ref{sec:intro-verifier}.
  Fix an index $n\geq 1$ and let $N = 2^n$.
  Let $R = N^\lambda$ and $\introparams(R) = (q,m,d)$,
  as in Section~\ref{sec:intro-verifier}.
 
  As in the proof of completeness part of Theorem~\ref{thm:introspection}, we make the
  following simplifications.
  First, we analyze the soundness of the typed introspective verifier
  $\tvint$; the soundness of the detyped verifier
  $\vint = \detype(\tvint)$ follows from Lemma~\ref{lem:detyping-verifiers}
  and the fact that the type set $\type^\intro$ has size $O(\ell)$.
  Second, analogously to Remark~\ref{rk:downsizing} without loss of generality
  for notational simplicity we consider strategies for questions sampled from
  $\tilde{\sampler}^\intro$ rather than the downsized sampler
  ${\sampler}^\intro$.

  Suppose that $\val^*(\verifier_n^\intro) > 1-\eps$ for some $0 < \eps < 1$,
  and let $\strategy = (\ket{\hat{\psi}}, \hat{A}, \hat{B})$ be a strategy for
  $\tvint_n$ with value at least~$1-\eps$.
  Since $\strategy$ has success probability that is strictly positive, the
  decider $\tdint$ does not automatically reject, which means that
	\begin{equation}\label{eq:intro-sound-lambda}
	s(N) \,\leq\, N^\lambda\;.
	\end{equation}
	We analyze each of the tests performed by $\tdint$ (see
  Figure~\ref{fig:intro-decider}) in sequence, and state consequences of each
  test.
  We start with Item~\ref{enu:pauli}, the Pauli test.
	
  \begin{lemma}\label{lem:intro-pauli-strat}
    There exists a function $\delta_1(\eps,R) = a( (\log R)^a \eps^{b} + (\log R)^{-b})$ for universal constants $a > 1$ and $0 < b < 1$ such that the following holds.
    Let $Q = M \log q$, where $M=2^m$.
    There is a projective strategy $\strategy'=(\ket{\psi},A,B)$ for
    $\tvint_n$ that succeeds with probability at least $1-\delta_1$
    and, furthermore,
    \begin{equation}
      \label{eq:intro-state-a}
      \ket{\psi}_{\alice\bob} =  \ket{\EPR_2}_{\alice'\bob'}^{\otimes Q} \otimes
      \ket{\aux}_{\alice''\bob''}
    \end{equation}
    for some  bipartite state $\ket{\aux}$, and for all $W\in\{X,Z\}$,
    \begin{equation}\label{eq:intro-sound-paulia}
      A^{\Pauli,W}_x = \sigma^W_x \;,\qquad B^{\Pauli,W}_x = \sigma^W_x \;,
    \end{equation}
    where $\sigma^W_x$ acts on the first $s(N)$ qubits of player $\alice$'s
    share (resp.
    $\bob$'s share) of $\ket{\EPR_2}^{\otimes Q}$.
  \end{lemma}
	
  \begin{proof}
    Given the definition of the type graph $G^\intro$, for
    $((\hat{t}_\alice,\allowbreak \hat{x}_\alice),\allowbreak (\hat{t}_\bob,\allowbreak \hat{x}_\bob))$ sampled
    according to $\mu_{\tsint,n}$ it holds that
    $(\hat{t}_\alice,\allowbreak \hat{t}_\bob) \allowbreak \in\allowbreak  \type^\pauli \times \type^\pauli$ with
    constant probability.
    Therefore, conditioned on the Pauli test, Item~\ref{enu:pauli}, being
    executed, $\strategy$ must succeed in the test with probability $1-O(\eps)$.

    Observe that conditioned on the test being executed, the distribution of
    $((\hat{t}_\alice,\hat{x}_\alice),(\hat{t}_\bob,\hat{x}_\bob))$ is, by
    definition, exactly the distribution of questions in the Pauli basis game
    with parameters $\qldparams$, as described in Section~\ref{sec:qld-game}.
    By Corollary~\ref{cor:pauli-binary} it follows that there exists a local isometry
    $\phi = \phi_\alice \otimes \phi_\bob$ and a state $\ket{\aux}\in
    \mH_{\alice''}\otimes \mH_{\bob''}$ such that
    \begin{equation}\label{eq:intro-sound-0}
      \Vert \phi (\ket{\hat{\psi}}) - \ket{\EPR_2}^{\otimes Q} \otimes
      \ket{\aux} \Vert^2 \leq \delta'(\eps,R)\;,
    \end{equation}
    where $\delta'(\eps,R)$ is an upper bound on $\delta_\qld(O(\eps),q,m,d)$
    that only depends on $\eps$ and $R$, as stated in
    Lemma~\ref{lem:delta-bound}.
    In addition, defining $A^{\hat{x}}_{\hat{a}} = \phi_\alice~
    \hat{A}^{\hat{x}}_{\hat{a}}~\phi_\alice^\dagger$ for all questions $\hat{x}$ and answers
    $\hat{a}$, for $W \in \{X, Z\}$ it holds that
    \begin{equation}
      \label{eq:intro-sound-pauli}
      A^{\Pauli,W}_x \otimes \Id_\bob \approx_{\delta'(\eps,R)}
      \sigma^W_x \otimes \Id_\bob\;,
    \end{equation}
    and a similar set of equations hold for operators associated with the second
    player.
    Using Naimark's theorem as formulated in~\cite[Theorem 5.1]{ML20}, at the
    cost of extending the state $\ket{\aux}$ we may assume that the measurements
    are projective without loss of generality.
    Define the strategy $\strategy'$ which uses the state $\ket{\EPR_2}^{\otimes
      Q} \otimes \ket{\aux}$ and measurement operators $\{ A^{\hat{x}}_{\hat{a}}
    \}$ and $\{ B^{\hat{x}}_{\hat{a}} \}$ for all questions $\hat{x}$, except
    for $(\Pauli,W)$-type questions where instead the Pauli measurements
    $\sigma^W_{\hat{a}}$ are used.
    Using~\eqref{eq:intro-sound-0} and~\eqref{eq:intro-sound-pauli} the strategy
    $\strategy'$ succeeds in $\tvint_n$ with probability at least
    $1-\delta_1(\eps,R)$, where the claimed bound on $\delta_1$ follows from the bound on $\delta'$ given in    Lemma~\ref{lem:delta-bound}.
  \end{proof}
	
  In the remainder of the proof we analyze the strategy $\strategy'$ from
  Lemma~\ref{lem:intro-pauli-strat}.
  We use the following notation conventions:
  \begin{enumerate}
  \item 
    We use indices $\alice$ and $\bob$ to label each player's Hilbert space
    after application of the isometry $\phi$ from
    Lemma~\ref{lem:intro-pauli-strat}.

  \item We write~$V$ for the register subspace of $\F_2^{Q}$ spanned by $e_1,
    \ldots, e_{s(N)}$ and $\overline{V}$ for its complement.
    (Note that by definition of $\introparams$ in
    Section~\ref{sec:qld-complexity} it holds that $s(N)\leq R \leq Q$, where
    the first inequality follows from~\eqref{eq:intro-sound-lambda}.)

  \item Whenever we write a Pauli operator $\sigma^W_a$ the register on which
    the operator acts should always be clear from context, and is implicit from
    the space in which the outcome $a$ ranges.
  
  \item For measurement operators in the introspection game, the variables for
    the measurement outcomes follow the specification of the ``answer key'' in
    Figure~\ref{fig:decider_pauli} (for $\type^\Pauli$-type questions) and
    Figure~\ref{fig:intro-decider} (for all other question types).
    For example, the measurement operators $\{A_x^{\Pauli,W} \}$ corresponding
    to question type $(\Pauli,W)$ are indexed by vectors $x \in \F_q^Q$ where $Q
    = M \log q$ where $M = 2^m$.
    The measurement operators corresponding to question type $(\Intro,\trole)$
    for $\trole \in \AB$ are indexed by pairs $(y,a) \in V \times \{0,1\}^{\leq
      3Q}$.\footnote{Technically the answer $a$ may be a binary string of any
      length; however, if $a$ is too long the decider rejects due to the answer
      length check.
      Thus we assume without loss of generality that the answer $a$ is a binary
      string of length at most $3Q = 3\cdot 2^m \cdot \log q$.}
    We often refer to marginalized measurement operators, e.g., the operator
    $A^{\Intro,\trole}_{y}$ denotes marginalizing $A^{\Intro,\trole}_{y,\, a}$ over
    all $a$.
    In these cases, the part of the answer that is marginalized over will be
    clear from context.
		
  \item We use the notation $\delta$ to denote a function which is polynomial in
    $\delta_1$, although the exact expression may differ from occurrence to
    occurrence.
    The polynomial itself may depend on $\ell$, but we leave this dependence
    implicit; due to the use of inductive steps that involve taking the square
    root of the error $\ell$ times in sequence (e.g.,
    Lemma~\ref{lem:intro-sound-induction}) the exponent generally depends on
    $\ell$.
  \end{enumerate}

  The next two lemma derive conditions implied by Items~\ref{enu:sampling}
  and~\ref{enu:hiding} of the checks performed by the decider $\tdint$
  described in Figure~\ref{fig:intro-decider}.
  As these two parts are performed independently for the two possible values of
  $\trole\in\{\alice,\bob\}$, we only discuss the case where $\trole = \alice$.
  For notational simplicity, whenever possible we omit $v$ when referring to the
  measurement operators.
  For example $L$, $\ai$ and $\bs$ are used as shorthand notation
  for $L^\alice$, $A^{\Intro,\, \alice}_{y,\, a}$, and $B^{\Sample,\,
    \alice}_{z,\, a}$, respectively.

  \begin{lemma}[Sampling test, Item~\ref{enu:sampling} of
    Figure~\ref{fig:intro-decider}]
    \label{lem:intro-sound-1}
    For each $k\in\{ 1,\allowbreak  2,\allowbreak  \ldots,\allowbreak  \ell\}$,
    \begin{align}
      \ia \otimes \bs[z]
      & \simeq_\delta \sigma^Z_z \otimes \ib\;,
        \label{eq:intro-sound-1} \\
      \ai[y_{\leq k},\, a] \otimes \ib
      & \simeq_\delta \ia \otimes \bs[\Llek,\, a]\;,
        \label{eq:intro-sound-3b}
    \end{align}
    where $z$ ranges over $V$ and $y_{\leq k}$ ranges over $L_{\leq k}(V)$.
    Moreover, analogous equations hold with operators acting on the other side
    of the tensor product.
  \end{lemma}
	
  \begin{proof}
    When $((\hat{t}_\alice,\hat{x}_\alice),(\hat{t}_\bob,\hat{x}_\bob))$ is
    sampled according to $\mu_{\tsint,n}$, each check in
    Item~\ref{enu:sampling} of Figure~\ref{fig:intro-decider} is executed with
    probability $\Omega(1/\ell)$ (this is due to the number of types in
    $\type^\intro$ and the structure of the type graph $G^\intro$).
    Therefore, in each of the checks specified by Items~\ref{enu:sampling-pauli}
    and~\ref{enu:sampling-intro}, the strategy $\strategy'$ succeeds with
    probability at least $1 - O(\ell \delta)$, conditioned on the test being
    executed.
    Item~\ref{enu:sampling-pauli} for $w=\alice$ combined
    with~\eqref{eq:intro-sound-paulia} implies~\eqref{eq:intro-sound-1}.
    Item~\ref{enu:sampling-intro} for $w=\alice$, combined with
    Fact~\ref{fact:data-processing}, implies~\eqref{eq:intro-sound-3b}.
    The lemma follows from repeating the same argument with the tensor factors
    interchanged.
  \end{proof}

  \begin{lemma}[Hiding test, Item~\ref{enu:hiding} of
    Figure~\ref{fig:intro-decider}]\label{lem:intro-sound-2}
    For each $k \in \{1, \ldots, \ell+1\}$,
    \begin{equation}\label{eq:intro-sound-4letter-before-a}
      \ai[y_{< k},\, a] \abc \br[y_{< k},\, a]\;,
    \end{equation}
    and if $k\leq \ell$,
    \begin{equation}\label{eq:intro-sound-4a-2}
      \ia \otimes \bhk  \approx_\delta \zLlk \otimes \ib\;.
    \end{equation}
    Furthermore, for all $j,k \in \{1,\ldots,\ell -1 \}$ such that $j \leq k$,
    we have
    \begin{equation}
      \label{eq:hide-chain}
      \ahk[y_{<j},\, y_{\leq j}^\perp] \otimes \ib \approx_\delta
      \ahp[y_{<j},\, y_{\leq j}^\perp] \otimes \ib\;.
    \end{equation}
    Analogous equations
    to~\eqref{eq:intro-sound-4letter-before-a},~\eqref{eq:intro-sound-4a-2},
    and~\eqref{eq:hide-chain} hold with operators acting on the other side of
    the tensor product.
  \end{lemma}
	
  \begin{proof}
    When $((\hat{t}_\alice,\hat{x}_\alice),(\hat{t}_\bob,\hat{x}_\bob))$ is
    sampled according to $\mu_{\tsint,n}$, each check in
    Item~\ref{enu:hiding} of Figure~\ref{fig:intro-decider} is executed with
    probability $\Omega(1/\ell)$.
    Therefore, in each of the checks specified by
    Items~\ref{enu:hiding-intro},~\ref{enu:hiding-read}, and~\ref{enu:hiding-same}
    (conditioned on the right types) the strategy
    $\strategy'$ succeeds with probability at least $1 - O(\ell \delta)$.

    Item~\ref{enu:hiding-intro} for $w=\alice$ combined with
    Fact~\ref{fact:data-processing}
    implies~\eqref{eq:intro-sound-4letter-before-a}.
    Combining \eqref{eq:intro-sound-1} and~\eqref{eq:intro-sound-3b} with
    \eqref{eq:intro-sound-4letter-before-a} yields
    \begin{equation}
      \label{eq:intro-sound-4a}
      \ia \otimes \br[y_{< k}] \approx_\delta \zLlk \otimes \ib\;.
    \end{equation}
    The fact that $\strategy'$ is projective and succeeds in
    Items~\ref{enu:hiding-read} and~\ref{enu:hiding-same} of
    Figure~\ref{fig:intro-decider} with probability $1 - O(\ell \delta)$,
    along with Item~\ref{item:consistency-implies-approx} of
    Fact~\ref{fact:agreement}, imply~\eqref{eq:intro-sound-4a-2}.
  
    We now establish the ``Furthermore'' part of the lemma statement.
    Let $1 \leq j \leq k \leq \ell-1$.
    Item~\ref{enu:hiding-same}, Item~\ref{item:consistency-implies-approx} of
    Fact~\ref{fact:agreement}, and Fact~\ref{fact:data-processing} imply that
    \begin{equation}
      \ahk[y_{<j},\, y_{\leq j}^\perp] \otimes \ib \approx_\delta
      \ia \otimes \bhp[y_{<j},\, y_{\leq j}^\perp].
    \end{equation}
    Item~\ref{enu:intro-consistency} and Fact~\ref{fact:data-processing} imply
    \begin{equation}
      \ahp[y_{<j},\, y_{\leq j}^\perp] \otimes \ib \simeq_\delta
      \ia \otimes \bhp[y_{<j},\, y_{\leq j}^\perp]\;.
    \end{equation}
    This proves~\eqref{eq:hide-chain}.

    The lemma follows from repeating the same arguments with the tensor factors
    interchanged.
	\end{proof}
	
	We exploit the tests performed in Item~\ref{enu:hiding} further to show the
  following lemma.

  \begin{lemma}
    \label{lem:hide-rigidity}
    For all $k \in \{1, 2, \ldots, \ell\}$,
    \begin{equation*}
      \ia \otimes \bhk[y_{<k},\, y^\perp_k,\, x_{>k}] \approx_\delta
      \bigl( \zLlk \otimes \xLperpk \otimes \x[x_{>k}] \bigr) \otimes \ib\;,
    \end{equation*}
    and an analogous equation holds with operators acting on the other side of
    the tensor product.
  \end{lemma}

  \begin{proof}
    The proof is by induction on $k$. We first show the case $k=1$.
    Under the distribution $\mu_{\tsint,n}$, the check in
    Item~\ref{enu:hiding-pauli} of Fig.~\ref{fig:intro-decider} is executed with
    probablity $\Omega(1/\ell)$. That part of the check for $w=\alice$ together
    with \eqref{eq:intro-sound-paulia} implies that
    \begin{equation}
      \label{eq:base-case-hide}
      \ia \otimes B^{\Hide{1}}_{y_1^\perp,\, x_{>1}} \approx_\delta \bigl(
      \xLperpk[1] \otimes \x[x_{>1}] \bigr) \otimes \ib\;.
    \end{equation}
    This proves the case for $k=1$.
    Next we perform the induction step.
    Assume that the lemma holds for some $k \in \{1, 2, \ldots, \ell-1\}$.
    The check of Item~\ref{enu:hiding-same} is executed with probablity
    $\Omega(1/\ell)$; using that $\strategy'$ succeeds in Item~\ref{enu:hiding-same}
    (conditioned on the right types having been sampled) with probablity at least $1 - O(\ell
    \delta)$, we have
    \begin{equation*}
      \sum_{y_{\le k},\, y^\perp_{k+1},\, x_{>k+1}}
      \bra{\psi} \ahk[{y_{<k},\, [L^\perp_{k+1,\, y_{\le k}}
      (\cdot ) = y^\perp_{k+1}],\, x_{>k+1}}] \otimes \bhp[y_{\le k},\,
      y^\perp_{k+1},\, x_{>k+1}] \ket{\psi} \ge 1 - O(\ell \delta)\;.
    \end{equation*}

    We now apply Lemma~\ref{lem:conditional-consistency}, choosing the measurements
    $A$, $B$ and outcomes $x$, $y$, $z$ in the lemma as follows:
    \begin{gather*}
      \text{``$A$''}: \ahk[{y_{<k},\, [L^\perp_{k+1,\, y_{\le k}}(\cdot) =
          y^\perp_{k+1}],\, x_{>k+1}}]\;, \quad
      \text{``$B$''}: \bhp[y_{\le k},\, y^\perp_{k+1},\, x_{>k+1}]\;, \\
      \text{``$x$''}: y_{<k}\;, \quad
      \text{``$y$''}: y_k\;, \quad
      \text{``$z$''}: (y^\perp_{k+1},\, x_{>k+1})\;.
    \end{gather*}
		Note that here $A$ does not depend on $y$, so we use the same $A$ for all values of $y$. 
    Lemma~\ref{lem:conditional-consistency} with the above choices of parameters
    implies that
    \begin{equation*}
      \begin{split}
        &\ia \otimes \bhp[y_{\le k},\, y^\perp_{k+1},\, x_{>k+1}]
        \\
        & \approx_\delta \ahk[{y_{<k},\, [L^\perp_{k+1,\, y_{\le k}}(\cdot) =
          y^\perp_{k+1}],\, x_{>k+1}}] \otimes \bhp[y_{\le k}]\\
        & \approx_\delta  \bigl( \zLlk \otimes \xLperpk[k+1] \otimes \x[x_{>k+1}]
        \bigr) \otimes \bhp[y_{\le k}]\\
        & \approx_\delta \bigl( \zLlk \zLlek \otimes \xLperpk[k+1] \otimes \x[x_{>k+1}]
        \bigr) \otimes \ib,\\
        & \approx_0 \bigl( \zLlk[k+1] \otimes \xLperpk[k+1] \otimes \x[x_{>k+1}]
        \bigr) \otimes \ib,
      \end{split}
    \end{equation*}
    where the input to $L^\perp_{k+1,\, y_{<k+1}}(\cdot)$ is $x_{k+1}$.
    The second approximation uses the induction hypothesis,
    Fact~\ref{fact:data-processing}, Fact~\ref{fact:agreement}, and
    Fact~\ref{fact:add-a-proj}.
    The third approximation follows from \eqref{eq:intro-sound-4a-2} and
    Fact~\ref{fact:add-a-proj}.
    The fourth approximation follows from the definition of CL functions.
    This completes the induction.
  \end{proof}

  \begin{lemma}
    For all $k \in \{1, \ldots, \ell\}$,
	  \begin{equation}
      \label{eq:intro-sound-4}
      \ia \otimes \br[y_{<k},\, y_k^\perp] \approx_\delta
      \bigl( \zLlk \otimes \xLperpk \bigr) \otimes \ib \;.
    \end{equation}
    Moreover, analogous equations hold with operators acting on the other side
    of the tensor product.
  \end{lemma}

  \begin{proof}
    Lemmas~\ref{lem:hide-rigidity} and~\ref{fact:data-processing} imply that
    \begin{equation}\label{eq:intro-sound-3d}
      \ahk[y_{<k},\, y^\perp_k] \otimes \ib \approx_\delta
      \Bigl( \zLlk \otimes \xLperpk \Bigr) \otimes \ib\:.
    \end{equation}
    Since the strategy $\strategy'$ succeeds in Item~\ref{enu:hiding-read} with
    probability at least $1 - O(\ell \delta)$ it follows from
    Fact~\ref{fact:data-processing} that
    \begin{equation}\label{eq:intro-sound-3e}
      A^{\Hide{\ell}}_{y_{<\ell},\, y^\perp_{\leq \ell}} \otimes \ib
      \approx_\delta \ia \otimes \br[y_{<\ell},\, y^\perp_{\leq
          \ell}].
    \end{equation}
    An inductive argument applied to~\eqref{eq:hide-chain} of
    Lemma~\ref{lem:intro-sound-2}, combined with Fact~\ref{fact:data-processing},
    implies that for all $1 \leq k \leq \ell$ we have
    \begin{equation}\label{eq:intro-sound-3f}
      A^{\Hide{k}}_{y_{<k},\, y^\perp_{k}} \otimes \ib \approx_\delta A^{\Hide{
          \ell}}_{y_{<k},\, y^\perp_{k}} \otimes \ib.
    \end{equation}
    Equations~\eqref{eq:intro-sound-3d},~\eqref{eq:intro-sound-3e},
    and~\eqref{eq:intro-sound-3f}, combined with Fact~\ref{fact:data-processing},
    then establishes the lemma statement.
  \end{proof}

  \begin{lemma}
    \label{lem:intro-sound-induction}
    For each $k\in \{1, 2, \ldots, \ell+1\}$, there exists a product state
    $\ket{\ancilla_k} = \ket{\ancilla_{k,\alice}} \otimes
    \ket{\ancilla_{k,\bob}} \in \mH_{\alice'_k} \otimes \mH_{\bob'_k}$ and, for
    each $y_{<k} \in L_{<k}(V)$, a projective measurement $\bigl\{\aigek
    \bigr\}$ that acts on $\mH_{\alice} \otimes \mH_{\alice'_k}$ such that the
    following holds.
		First, for all $y \in V$, the operator $\aigek$ acts as identity on the
    register subspace spanned by basis vectors for the subspace $V_{<
      k}(y_{<k})$, and as a consequence the operator
    \begin{equation}\label{eq:intro-sound-4aa}
      \aizk = \zLlk \otimes \aigek\;
    \end{equation}
    is well-defined.
    Second, let $\strategy''_k$ be the strategy defined as follows.
    The state is $\ket{\EPR_2}^{\otimes Q} \otimes \ket{\aux} \otimes
    \ket{\ancilla_k}$.
    The measurements are identical to those in $\strategy'$ defined in
    Lemma~\ref{lem:intro-pauli-strat}, except that $\bigl\{ \ai \bigr\}$ is
    replaced with $\bigl\{ \aizk \bigr\}$.
    Then $\strategy''_k$ succeeds with probability at least $1-\delta$ in the
    game $\verifier_n^\intro$, \tnote{added:}for some $\delta=\poly(\delta_1)$ where $\delta_1$ is defined in Lemma~\ref{lem:intro-pauli-strat} and the implicit constants in the $\poly(\cdot)$ notation may depend on $k$ only. 
  \end{lemma}
  
  \begin{proof}

    The proof is by induction on $k$ from $1$ to $\ell+1$. \tnote{added:}For the proof, recall that we use the notation ``$\delta$'' to denote any $\poly(\delta_1)$ where $\delta_1$ is defined in Lemma~\ref{lem:intro-pauli-strat} and the implicit constants in the $\poly(\cdot)$ notation may depend on $k$ only. Thus, the exact value of ``$\delta$'' may vary each time it is used. 
		
    The case $k=1$ is trivial by setting $A_{y_{\geq 1},\, a}^{\Intro,\, y_{<1}}
    = \ai[y,\, a]$ for all $y,a$.
    Assume that for some $k \in \{1, 2, \ldots, \ell\}$ there exist projective
    measurements $\bigl\{\aigek \bigr\}$ for every $y_{< k}$ and a strategy
    $\strategy''_k$ satisfying the conditions of the lemma statement.
    We show the statement of the lemma for $k+1$.

    \paragraph{Commutation with $Z$-basis measurements.}
    We first prove that on average over $y_{< k}$, the measurement operator
    $\aigek$, which comes from the inductive assumption, commutes with the
    projective measurement $\{\sigma^Z_{z_k}\}$ where the outcomes $z_k$ range
    over the factor space $V_k(y_{<k})$.

    To do so, we first apply Lemma~\ref{lem:commutation-analysis} where we
    choose the measurements ``$A$,'' ``$B$,'' and~``$C$'' and outcomes ``$a$,''
    ``$b$,'' and~``$c$'' in the lemma as follows:
    \begin{gather*}
      \text{``$A$''}: \{\aizk\}\;,\quad
      \text{``$B$''}: \{\bs[z, \, a]\}\;,\quad
      \text{``$C$''}: \{\z  \}\;, \\
      \text{``$a$''}: y_{< k} \;, \quad
      \text{``$b$''}: (y_{\geq k},a) \;, \quad
      \text{``$c$''}: z \;.
    \end{gather*}
    To make sense of how the ``$B$'' POVM is indexed by ``$a$,'' ``$b$,'' and
    ``$c$'' as described above, we use the following relabelling: for all
    $(z,a)$, identify $\bs[z,\, a]$ with $\bs[y,\, a,\, z]$ where $y =
    L(z)$.
    Similarly, for the ``$C$'' POVM, we identify $\sigma^Z_z$ with the operator
    $\sigma^Z_{y_{<k},\, z}$ where $y_{<k} = L_{<k}(z)$.
    By applying Lemma~\ref{lem:intro-sound-1} to $\strategy_k''$ (the strategy
    given by the inductive hypothesis) with ``$k$'' in
    Lemma~\ref{lem:intro-sound-1} set to $\ell$, we have that
    \begin{equation}
      \label{eq:intro-sound-bsL2}
      \aizk \otimes \ib \approx_\delta \ia \otimes \bs[y,\, a]\;
    \end{equation}
    where $\bs[y,\, a] = \sum_{z: L(z) = y} \bs[z,\, a]$.
    Equations~\eqref{eq:intro-sound-1} and~\eqref{eq:intro-sound-bsL2} imply that the
    conditions of Lemma~\ref{lem:commutation-analysis} are satisfied, and thus
    we obtain
    \begin{equation}\label{eq:commute-with-Z}
      \bigl[\aizk \,,\, \z \bigr] \otimes \ib \approx_\delta 0\;
    \end{equation}
    where in the answer summation, $y$ is a deterministic function of $z$.
    We now apply Lemma~\ref{lem:commutation-simplification}, choosing the
    measurements ``$A$,'' ``$B$,'' ``$Q$'' and outcomes ``$a$,'' ``$b$,''
    ``$c$'' in the lemma as follows:
    \begin{gather*}
      \text{``$A$''}: \{\aizk\}\;,\quad
      \text{``$B$''}: \{\z[\Llk] \otimes \z[z_k] \}\;, \quad
      \text{``$Q$''}: \{\z[\Llk] \}\;, \\
      \text{``$a$''}: y_{< k} \;, \quad
      \text{``$b$''}: (y_{\geq k}, a) \;, \quad
      \text{``$c$''}: z_k \;.
    \end{gather*}
    Here, we write $\z[\Llk] \otimes \z[z_k]$ to denote
    \[
    	\sum_{\substack{z' \in V: \\ L_{<k}(z') = y_{<k} \\ (z')^{V_k(y_{<k})} = z_k}} \sigma^Z_{z'}.
    \]
    We choose the Hilbert spaces ``$\mH_{\alice,a}$'' and ``$\mH_{\bob,a}$'' as
    the register subspace $ V_{<k}(y_{<k})$, and ``$\mH_{\alice',a}$'' and
    ``$\mH_{\bob',a}$'' as the register subspace $V_{\geq k}(y_{<k})$ tensored
    with $\mH_{\alice''} \otimes \mH_{\bob''}$, the Hilbert space of the state
    $\ket{\aux}$.
    Thus for every $y_{<k}$, the state $\ket{\EPR}^{\otimes Q} \otimes
    \ket{\aux}$ of the strategy $\strategy_k''$ can be decomposed into a tensor
    product of a ``question state'' and an ``answer state'' as follows:
    \[
      \Big ( \ket{\EPR_2}_{V_{<k}(y_{<k})} \Big)_{\mH_{\alice,a} \mH_{\bob,a}}
      \otimes \Big ( \ket{\EPR_2}_{V_{\geq k}(y_{<k})} \otimes \ket{\aux}
      \Big)_{\mH_{\alice',a} \mH_{\bob',a}}\;.
    \]

    Let $\mu_{L_{<k}}$ denote the distribution over outcomes $y_{< k}$ generated
    by performing the ``$Q$'' measurement on the state $\ket{\EPR}^{\otimes Q}$,
    which is equivalent to the distribution generated by the following
    procedure: (1) sample a uniformly random $z \in V$; (2) compute $y = L(z)$;
    (3) return $y_{<k}$.
    Then, since Equation~\eqref{eq:commute-with-Z} and the inductive hypothesis
    about the structure of $A^{\Intro\,,Z_{<k}}_{y_{\leq k},\,a}$ satisfy the
    conditions of Lemma~\ref{lem:commutation-simplification}, we obtain on
    average over $y_{< k} \sim \mu_{L_{< k}}$
    \begin{equation}
      \label{eq:intro-sound-5}
      \bigl[ \aigek \,,\, \z[z_k] \bigr] \otimes \ib \approx_\delta 0 \;.
    \end{equation}
    Here, the measurement outcomes $z_k$ range over $V_k(y_{<k})$.

    \paragraph{Commutation with $X$-basis measurements.}
    Next, we first prove that on average over $y_{< k}$, the measurement
    operator $\aigek$ commutes with the projective measurement
    $\bigl\{ \xLperpk \bigr\}$ where the
    outcomes $y_k^\perp$ are elements of the factor space $V_k(y_{<k})$.
  
    We again apply Lemma~\ref{lem:commutation-analysis}, choosing the
    measurements and outcomes in the lemma as follows:
    \begin{gather*}
      \text{``$A$''}: \{\aizk\}\;,\\
      \text{``$B$''}: \{ \br[y,\, y^\perp_k,\, a] \}\;,\\
      \text{``$C$''}: \{ \z[\Llk] \otimes \xLperpk \}\;, \\
      \text{``$a$''}: y_{< k} \;, \quad
      \text{``$b$''}: (y_{\geq k},a) \;, \quad
      \text{``$c$''}: y_k^\perp \;.
    \end{gather*}
    Equations~\eqref{eq:intro-sound-4letter-before-a} (with ``$k$'' in
    Lemma~\ref{lem:intro-sound-2} chosen to be $\ell+1$)
    and~\eqref{eq:intro-sound-4} imply the conditions of
    Lemma~\ref{lem:commutation-analysis}, so we obtain
    \begin{equation}\label{eq:commute-with-some-X}
      \Bigl[ \aizk \,,\,  \zLlk \otimes \xLperpk \Bigr] \otimes
      \ib \approx_\delta 0\;.
    \end{equation}
    We then apply Lemma~\ref{lem:commutation-simplification} with the following
    choice of measurements and outcomes:
    \begin{gather*}
      \text{``$A$''}: \{\aizk\}\;,\\
      \text{``$B$''}: \{\zLlk \otimes \xLperpk \}\;,\\
      \text{``$Q$''}: \{\zLlk \}\;, \\
      \text{``$a$''}: y_{< k} \;, \quad
      \text{``$b$''}: (y_{\geq k},a) \;, \quad
      \text{``$c$''}: y_k^\perp \;.
    \end{gather*}
    The Hilbert space and state decomposition are the same as in the previous
    invocation of Lemma~\ref{lem:commutation-simplification}.
    Equation~\eqref{eq:commute-with-some-X} and the inductive hypothesis satisfy
    the conditions of Lemma~\ref{lem:commutation-simplification}, and we
    similarly obtain that on average over $y_{< k} \sim \mu_{L_{<k}}$,
    \begin{equation}
      \label{eq:intro-sound-6}
      \bigl[ \aigek \,,\, \xLperpk \bigr] \otimes \ib
      \approx_\delta 0\;,
    \end{equation}

    \paragraph{Applying the Pauli twirl.}
    The last step is to apply the Pauli twirl to decompose the family of
    measurements $\{\aigek \}$ into a tensor product measurement, with the first
    part of the tensor product measuring the $k$-th linear map of $L$.

    Again applying Lemma~\ref{lem:intro-sound-1} to $\strategy_k''$, and using
    Facts~\ref{fact:data-processing} and~\ref{fact:agreement}, we obtain that
    \begin{equation*}
      \aizk[y_{\le k}] \otimes \ib \approx_\delta \zLlek \otimes \ib\;
    \end{equation*}
    which is equivalent to, by the inductive hypothesis,
    \begin{equation}\label{eq:intro-sound-8}
      \zLlk \otimes \aigek[y_k] \otimes \ib \approx_\delta \zLlk \otimes \zLk
      \otimes \ib\;.
    \end{equation}
    Applying Lemma~\ref{lem:conditional} to Equation~\eqref{eq:intro-sound-8},
    we conclude that
    \begin{equation}\label{eq:consistency-equation}
      \aigek[y_k] \otimes \ib \approx_\delta
      \zLk \otimes \ib\;,
    \end{equation}
    on average over $y_{<k} \sim \mu_{L_{<k}}$.
  
    Now we apply Lemma~\ref{lem:mixing} with the following identification:
    \begin{gather*}
      \text{``$x$''}: y_{< k} \;, \quad
      \text{``$y$''}: y_{k} \;, \quad
      \text{``$a$''}: (y_{> k}, a) \;, \quad
      \text{``$L_x$''}: L_{k,\, y_{<k}} \;,\\
      \text{``$U_x$''}: V_k(y_{< k}) \;, \quad
      \text{``$M^{x}_{y,\, a}$''}: \aigek \;, \quad
      \text{``$M^{x,\, y}_a$''}: \aigk \;.
    \end{gather*}
    The ``Consistency'' condition is implied by
    Equation~\eqref{eq:consistency-equation} and the ``Commutation'' conditions
    are implied by Equations~\eqref{eq:intro-sound-5}
    and~\eqref{eq:intro-sound-6}.
    We obtain that for all $y_{\leq k}$ there exists POVM measurements $\bigl\{
    \aigk \bigr\}$ that act as identity on the register subspace spanned by
    basis vectors for the subspace $V_{< k}(y_{<k})$ and, on average over
    $y_{<k} \sim \mu_{L_{<k}}$, we have
    \begin{equation}\label{eq:intro-sound-7}
      \aigek \otimes \ib
      \approx_\delta \Bigl( \zLk \otimes \aigk  \Bigr)
      \otimes \ib\;.
    \end{equation}
    Using the inductive assumption on the structure of $A^{\intro,\,
      Z_{<k}}_{y,\, a} $ from~\eqref{eq:intro-sound-4aa}, we get
    \begin{align}
      \aizk \otimes \ib
      & = \Bigl( \zLlk \otimes \aigek \Bigr) \otimes \ib \\
      & \approx_\delta \bigl( \zLlek \otimes \aigk \bigr) \otimes \ib\;.
        \label{eq:whatever}
    \end{align}
    where the second line follows from~\eqref{eq:intro-sound-7} and
    Lemma~\ref{lem:conditional}.
    By~\eqref{eq:whatever}, replacing the projective measurement $\bigl\{\aizk
    \bigr\}$ with the POVM
    \begin{equation*}
      \bigl\{\zLlek \otimes \aigk \bigr\}
    \end{equation*}
    in the strategy $\strategy''_k$ results in a strategy that succeeds with
    probability at least $1 - \delta$.
    To show that $\{\aigk\}$ can, furthermore, be turned into a projective
    measurement we use Naimark's theorem as formulated in~\cite[Theorem
    5.1]{ML20}. In particular, we see that the state
    $\ket{\ancilla'}$ it produces is a product state with no entanglement.
    This yields a strategy $\strategy_{k+1}''$ with ancilla state
    $\ket{\ancilla_{k+1}} = \ket{\ancilla_k}\ket{\ancilla'}$, which establishes
    the induction hypothesis for $k+1$.
    This completes the proof of Lemma~\ref{lem:intro-sound-induction}.
  \end{proof}

  Taking $k = \ell+1$ in Lemma~\ref{lem:intro-sound-induction} we obtain a
  strategy $\strategy''=\strategy''_{\ell+1}$ with value~$1-\delta$ in which,
  given question $(\Introspect, \alice)$, player~$\alice$ performs the
  measurement
  \begin{equation}
    \label{eq:intro-sound-final}
    \bigl\{ \z[{[L^\alice(\cdot)=y]}] \otimes A^{\Intro,\, y}_a \bigr\} \;,
  \end{equation}
  for a family of measurements $\bigl\{ A^{\Intro,\, y}_a \bigr\}$ acting on the
  state $\ket{\EPR}_{\overline{V}} \otimes \ket{\aux} \otimes \ket{\ancilla}$
  where $\ket{\ancilla}=\ket{\ancilla_{\ell+1}}$ is unentangled.
  An analogous argument for Player~$\bob$'s measurements shows that we may
  additionally assume Player~$\bob$ responds to the question~$(\Introspect,
  \bob)$ using the measurement
  \begin{equation}
    \label{eq:intro-sound-final-bob}
    \bigl\{\sigma^Z_{[L^\bob(\cdot)=y]} \otimes
    B^{\Intro,\, y}_a\bigr\}\;,  
  \end{equation}
  for a family of measurements $\bigl\{ B^{\Intro,\, y}_a \bigr\}$ acting on the
  state $\ket{\EPR}_{\overline{V}} \otimes\ket{\aux} \otimes \ket{\ancilla'}$,
  where $\ket{\ancilla'}$ is unentangled.
  Summarizing, the strategy $\strategy''$ uses the state
  \begin{equation*}
    \ket{\theta} = \ket{\EPR}^{\otimes Q} \otimes \ket{\aux} \otimes
    \ket{\ancilla} \otimes \ket{\ancilla'}\;,
  \end{equation*}
  and the measurements given by Equations~\eqref{eq:intro-sound-final}
  and~\eqref{eq:intro-sound-final-bob}.
  
	To conclude the proof of the soundness part of the theorem we analyze
  Item~\ref{enu:intro-game} in Figure~\ref{fig:intro-decider}.
  The test in Item~\ref{enu:intro-game} is executed with probability
  $\Omega(1/\ell)$, so the strategy $\strategy''$ succeeds with probability at
  least $1 - \delta$ in that test, conditioned on the right types (here we
  absorb factors of $O(\ell)$ into $\delta$).
  Using~\eqref{eq:intro-sound-final} and~\eqref{eq:intro-sound-final-bob},
  conditioned on the test being executed the distribution of the
  part~$(y_\alice, y_\bob)$ of the players' answers in the test is exactly the
  distribution $\mu_{\sampler, N}$ associated with game~$\verifier_N$.
  As a result, the strategy which uses the state $\ket{\EPR}_{\overline{V}}
  \otimes \ket{\aux} \otimes \ket{\ancilla} \otimes \ket{\ancilla'}$ and
  measurements $ \{ A^{\Intro,\, y}_a \},\{ B^{\Intro,\, y}_a \}$ succeeds with
  probability at least~$1- \delta$ in the game~$\verifier_N$.
  Thus, $\val^*(\verifier_N) \geq 1- \delta$.
  It remains to show that $\delta$ has the required form. 
  As $\delta = \poly(\delta')$, $\delta'(\eps, R) = a' \bigl( (\log
  R)^{a'}\eps^{b'} + (\log R)^{-b'} \bigr)$, and $R = (2^n)^\lambda$, there is a
  constant $C \ge 1$ such that
  \begin{equation*}
    \begin{split}
      \delta(\eps, n) & \le C \Bigl( a' \bigl( (\lambda n)^{a'} \eps^{b'} +
      (\lambda n)^{-b'} \bigr) \Bigr)^{\frac{1}{C}}\\
      & \le C (a')^{1/C} \bigl(  (\lambda n)^{a'/C} \cdot \eps^{b'/C} + (\lambda n)^{-b'/C} \bigr)\\
      & \le a \bigl( (\lambda n)^a \cdot \eps^b + (\lambda n)^{-b} \bigr)\;,
    \end{split}
  \end{equation*}
  for $a = \max \bigl\{ C (a')^{1/C}, a'/C \bigr\}$, and $b = b'/C$, where the
  second line uses the inequality $(x+y)^{1/C} \le x^{1/C} + y^{1/C}$ for
  $x,y>0$ and $C\ge 1$.

  This establishes the soundness part of the theorem.

  To show the entanglement bound, we observe that local isometries do not change
  the Schmidt rank of a state.
  Define $\ket{\psi'} = \phi(\ket{\psi}) \otimes \ket{\ancilla} \otimes
  \ket{\ancilla'}$.
  Since the strategy $(\ket{\psi'},\allowbreak  \{ A^{\Intro,\, y}_a \},\allowbreak \{ B^{\Intro, y}_a
  \})$ is $\delta$-close to $(\ket{\theta},\allowbreak  \{ A^{\Intro,\, y}_a \},\allowbreak \{
  B^{\Intro,\, y}_a \})$ which has value $1 - \delta$, the strategy
  $(\ket{\psi'},\allowbreak  \{ A^{\Intro,\, y}_a \},\allowbreak \{ B^{\Intro,\, y}_a \})$ has value at
  least $1 - 2\delta$ in the game $\verifier_N$, and, therefore, the Schmidt rank
  of $\phi(\ket{\psi})$ (and thus of $\ket{\psi}$) must be at least
  $\Ent(\verifier_N,\allowbreak 1 - 2\delta)$.
  Here, we use the fact that ancilla states are product states and, therefore,
  have Schmidt rank~$1$.
  
  Moreover, recall that $\phi(\ket{\psi})$ is $\delta$-close to
  $\ket{\EPR_2}^{\otimes Q} \otimes \ket{\aux}$ whose Schmidt coefficients are
  all at most $2^{-Q/2}$.
  For any bipartite state $\ket{a}$ with Schmidt rank at most $R$ and $\ket{b}$
  whose Schmidt coefficients are all at most $\beta$, it follows from the
  Cauchy-Schwarz inequality that $|\langle a | b \rangle|^2 \leq R \beta^2$.
  Therefore, the Schmidt rank of $\phi(\ket{\psi})$ (and thus of $\ket{\psi}$) is
  at least
  \[
  	\left (1 - \delta \right)^2 \cdot 2^{Q} \geq \left
      (1 - \delta \right)^2 \cdot 2^{N^\lambda}\;,
  \]
  where we used that $Q = M \log q \geq R$ (using the canonical parameter
  settings of Definition~\ref{def:introparams}) and $\delta \geq \|
  \phi(\ket{\psi}) - \ket{\theta} \|^2 = 2 - 2\Re \bra{\theta}
  \phi(\ket{\psi})$.
  Combining the two lower bounds on the Schmidt rank of $\ket{\psi}$ shows the
  desired lower bound on $\Ent(\tvint_n,1 - \eps)$.
\end{proof}


\section{Oracularization}
\label{sec:oracle}

\newcommand{\tvora}{\hat{\verifier}^\ora}
\newcommand{\tsora}{\hat{\sampler}^\ora}
\newcommand{\tdora}{\hat{\decider}^\ora}

\subsection{Overview}

In this section we introduce the \emph{oracularization} transformation.
At a high level, the oracularization $\game^\ora$ of a nonlocal game $\game$ is
intended to implement the following: one player (called the \emph{oracle
  player}) is supposed to receive questions $(x,y)$ meant for both players in
the original game $\game$, and the other player (called the \emph{isolated
  player}) only receives either $x$ or $y$ (but not both), along with a label
indicating which player in the original game the question is associated with (we
refer to such players as the \emph{original} players, e.g.,\ ``original $\alice$
player'' and ``original $\bob$ player'').
The oracle player is supposed to respond with an answer pair $(a,b)$, and the
other player is supposed to respond with an answer $c$.
The oracle and isolated player win the oracularized game if $(x,y,a,b)$
satisfies the predicate of the original game $\game$ and the isolated player's
answer is consistent with the oracle player's answer.

The oracularization step is needed in preparation to the next section, in which
we perform answer reduction on the introspection game.
To implement answer reduction we need at least one player to be able to compute
a proof, in the form of a PCP, that the decider of the original game would have
accepted the questions $(x,y)$ and answers $(a,b)$.
This requires the player to have access to both questions, and be able to
compute both answers.

\subsection{Oracularizing normal form verifiers}
\label{sec:orac-def}

Let $\verifier = (\sampler,\decider)$ be a normal form verifier such that
$\sampler$ is an $\ell$-level sampler.  Assume, moreover,
that $\decider$ is in the standard timeout-counter form of Section~\ref{sec:tms}.
We specify the \emph{typed oracularized verifier} $\tvora =
(\tsora,\tdora)$ associated with $\verifier$ as follows.

\paragraph{Sampler.}
Define the type set $\type^\ora = \{\oracle,\alice,\bob\}$.
(In the remainder of this section we refer to the types in $\type^\ora$ as
\emph{roles}.)
Define the type graph $G^\ora$ that is the complete graph on vertex set
$\type^\ora$ (including self-loops on all vertices).
Define the $\type^\ora$-type sampler $\tsora$ as follows.
For a fixed index $n \in \N$, let $V$ be the ambient space of $\sampler$ and
$L^w$ for $w \in \ab $ be the pair of CL functions of $\sampler$ on index $n$.

Define two $\type^\ora$-typed families of CL functions $\{ L^{w}_\tvar: V \to V
\}$, for $w \in \{\alice,\bob\}$ and $\tvar \in \type^\ora$, as follows:
\begin{equation*}
	L^{w}_\tvar =
  \begin{cases}
		L^\tvar & \text { if $\tvar \in \{\alice,\bob\}$ }, \\
		\mathrm{Id} & \text { if $\tvar = \oracle$ }.
	\end{cases}
\end{equation*}
In other words, if a player gets the type $\tvar \in \{\alice,\bob\}$, then they
get the question that original player $\tvar$ would have received in the game
played by $\verifier_n$.
If they get type $\tvar = \oracle$, then they get the entire seed $z$ that is
used by the sampler $\sampler$, from which they can compute both $L^\alice(z)$
and $L^\bob(z)$, the pair of questions sampled for the players in game
$\verifier_n$.

On all inputs beginning with $(n,\ldots$), the sampler $\tsora$ first computes the dimension $s(n)$ of the ambient space by calling $\sampler(n,\gamestyle{dimension})$, as well as the field size $q(n)$ by calling $\sampler(n,\gamestyle{field})$.  It then performs the appropriate format
checks, and before parsing any input string as a vector in the ambient space, it
checks that the string has the prescribed length determined by $s(n)$ and $q(n)$.  If any
check fails, $\tsora$ halts immediately with a fixed default output.

By definition, the sampler distribution $\mu_{\tsora,\, n}$ has the
following properties.
\begin{enumerate}
\item Conditioned on both players receiving the $\oracle$ role, both players
  receive $z$ for a uniformly random $z \in V$.
		
\item Conditioned on both players receiving the isolated player role, the
  player(s) with role $\alice$ (respectively,~$\bob$) receives $L^\alice(z)$
  (respectively, $L^\bob(z)$) for a uniformly random $z \in V$.
		
\item Conditioned on player $w\in\{\alice,\bob\}$ receiving the $\oracle$ role
  and player $\overline{w}$ receiving the isolated player role, their question
  tuple is distributed according to $\bigl( (\oracle,z), (v,L^v(z)) \bigr)$ if
  $w=\alice$ and $\bigl( (v, L^v(z)), (\oracle,z) \bigr)$ if $w=\bob$, where $z
  \in V$ is uniformly random and $v$ indicates the role of player
  $\overline{w}$.
\end{enumerate}

\paragraph{Decider.}
The typed decider $\tdora$ is specified in Figure~\ref{fig:oracle-decider}.

\begin{figure}[!htb]
  \centering
  \begin{gamespec}
    Input to decider $\tdora$: $(n, \tvar_\alice, x_\alice , \tvar_\bob,
    x_\bob ,a_\alice, a_\bob)$. 
    \hnote{added (May 31, 2026):} 
    First, compute the ambient dimension $s(n)$ of the sampler $\sampler$ by running $\sampler(n,\gamestyle{dimension})$. Then, check that $\tvar_\alice$ and $\tvar_\bob$ correspond to valid types from $\type^\ora$, and that $x_\alice ,  x_\bob$ each have length $s(n)$. If not, reject.

    Next, compute the timeout bound $B_{\decider}(n)$ from the description of $\decider$, as described in Section~\ref{sec:tms}.  For $w \in \{\alice,\bob\}$, if $\tvar_w = \oracle$, then attempt to parse $a_w$ as a pair
$(a_{w,\alice},a_{w,\bob})$ whose components each have length at most
$B_{\decider}(n)$; if this bounded parsing check fails, reject. If $\tvar_w \in \{\alice,\bob\}$, then check that $a_w$ has length at most $B_{\decider}(n)$; if not, reject.

    Perform the following steps sequentially. If any subroutine call to $\sampler$ or $\decider$ rejects or enters $\TIMEOUT$, then $\tdora$ rejects. 

    \begin{enumerate}
    \item (\textbf{Game check}).
      \label{enu:oracle-game}
      For all $w \in \{\alice,\bob\}$, if $\tvar_w = \oracle$, then compute
      $x_{w,v} = L^v(x_w)$ for $v \in \{\alice,\bob\}$.
      If $\decider$ rejects $(n,
      x_{w,\alice},x_{w,\bob},a_{w,\alice},a_{w,\bob})$, then reject.

    \item (\textbf{Consistency checks}).
      \label{enu:oracle-full-consistency}
      \begin{enumerate}
      \item \label{enu:check-same} If $\tvar_\alice = \tvar_\bob$ and $a_\alice
        \neq a_\bob$, then reject.
      \item \label{enu:oracle-versus-player} If for some $w \in
        \{\alice,\bob\}$, $\tvar_w = \oracle$, $\tvar_{\overline{w}} \in
        \{\alice,\bob\}$, and $a_{w,\, \tvar_{\overline{w}}} \neq
        a_{\overline{w}}$, then reject.
      \end{enumerate}
    \item \label{enu:oracle-acc} Accept if none of the preceding steps rejects.
    \end{enumerate}
  \end{gamespec}
  \caption{Specification of the typed decider $\tdora$.}
  \label{fig:oracle-decider}
\end{figure}

\subsection{Completeness and complexity of the oracularized verifier}

We determine the complexity of the oracularized verifier and
establish the completeness property.

\begin{theorem}[Completeness and complexity of the oracularized verifier]
  \label{thm:oracle-completeness}
  Let $\verifier = (\sampler, \decider)$ be a normal form verifier such that
$\sampler$ is an $\ell$-level sampler and $\decider$ is in the standard
timeout-counter form of Section~\ref{sec:tms}.  Let $B_{\decider}(n)$ denote
the timeout bound of $\decider$, and let $\tvora = (\tsora,\tdora)$ be the
typed oracularized verifier associated with $\verifier$. Then the following hold.
  \begin{itemize}
  \item (\textbf{Completeness}) For all $n \in \N$, if $\verifier_n$ has a PCC
    strategy of value $1$, then $\tvora_n$ has an
    SPCC strategy of value $1$.
  
  \item (\textbf{Sampler complexity}) The sampler $\tsora$ depends only
    on $\sampler$ (and not on $\decider$).
    Moreover, the time complexity of $\tsora$ satisfies
    \begin{equation*}
      \TIME_{\tsora}(n) = \poly(\TIME_{\sampler}(n))~.
    \end{equation*}
    Furthermore, if $\sampler$ is an $\ell$-level sampler, then
    $\tsora$ is a $\ell$-level typed sampler.
\item (\textbf{Decider complexity}) The typed decider $\tdora$ has time
  complexity
  \begin{equation*}
    \TIME_{\tdora}(n)
    =
    \poly(\TIME_{\sampler}(n), B_{\decider}(n)).
  \end{equation*}

	\item (\textbf{Efficient computability}) There is a Turing machine
    $\ComputeOracleVerifier$ which takes as input $\verifier =
    (\sampler,\decider)$ and returns $\tvora = ( \tsora,
    \tdora)$ in time $\poly(\abs{\verifier})$.
  \end{itemize}
\end{theorem}


\begin{remark}
Unlike with the Introspection transformation, we do not detype the oracularized verifier $\tvora$; this is because the analysis of the Answer Reduction transformation in Section~\ref{sec:ans} \emph{directly} reduces to the analysis of the typed oracularized verifier $\tvora$ defined here.
\end{remark}

\begin{proof}
  We analyze the completeness and complexity properties of the typed verifier
  $\tvora$.

  \paragraph{Completeness.}
  For $n \in \N$, let $\strategy = (\ket{\psi},A,B)$ be a PCC strategy for
  $\verifier_n$ with value $1$. Since $\decider$ by assumption has a timeout counter bound, it can inspect at most $B_{\decider}(n)$ symbols of each answer string.
Thus truncating every answer outcome of $\strategy$ to its first
$B_{\decider}(n)$ symbols does not change the outcome of $\decider$ on any input tuple that it accepts. Thus we may assume without loss of generality that all
answers produced by $\strategy$ have length at most $B_{\decider}(n)$.

  Consider the following symmetric strategy $\strategy^\ora = (\ket{\psi}, M)$
  for $\tvora_n$.
  Depending on the role received, each player performs the following:
  \begin{enumerate}
	\item Suppose the player receives role $v \in \{\alice,\bob\}$ and question
    $x$.
    Then the player performs the measurement that player $v$ would on question
    $x$ according to strategy $\strategy$ to obtain outcome $a$ (either
    $\{A^{x}_a\}$ or $\{B^x_a\}$, depending on $v$).
    The player replies with $a$.
		\label{enu:oracle-honest-isolated}
		
	\item Suppose the player receives role $v = \oracle$ and question $x$.
    The player first computes $y_w = L^w(x)$ for $w \in \{\alice,\bob\}$ where
    for $w\in \ab$, $L^w$ is the CL functions of $\sampler$ corresponding to
    player $w$.
    Then, the player measures using the POVM $\{ M^{\Oracle,\, x}_{a_\alice,\,
      a_\bob}\}$ where
    \begin{equation}
      \label{eq:oracle-honest}
      M^{\oracle,\, x}_{a_\alice,\, a_\bob} = B^{y_\bob}_{a_\bob}\,
      A^{y_\alice}_{a_\alice}\;.
    \end{equation}
    The projectors $A^{y_\alice}_{a_\alice}$ and $B^{y_\bob}_{a_\bob}$
    commute because $(y_\alice, y_\bob)$ is distributed according to
    $\mu_{\sampler,\, n}$ (over the choice of $x$) and $\strategy$ is a
    commuting strategy for $\verifier_n$.
    Thus $M^{\oracle,\, x}_{a_\alice,\, a_\bob}$ is a projector.
    The player replies with $(a_\alice, a_\bob)$.
		\label{enu:oracle-honest-oracle}
  \end{enumerate}

  The strategy $\strategy^\ora$ is symmetric and projective by construction, and
  consistency follows from the consistency of $\strategy$.
  We now argue that the strategy is commuting and has value $1$ in the game
  $\tvora_n$.
  We consider all possible pairs of roles.

  \begin{enumerate}
  \item ({Oracle, isolated}) Suppose without loss of generality that player
    $w=\alice$ gets the $\oracle$ role and player $\overline{w}=\bob$ gets the
    isolated player $\bob$ role.
    Then player $w$ gets question $x$ and player $\overline{w}$ gets question
    $L^\bob(x)$, where $x$ is uniformly sampled from $V$.
    The oracle player computes $y_v = L^v(x)$ for all $v \in \ab$.
    Notice that $(y_\alice, y_\bob)$ is distributed according to
    $\mu_{\sampler,\, n}$.
    The two players return $\bigl( (a_\alice, a_\bob), a'_\bob \bigr)$ with
    probability
    \begin{align*}
      \bra{\psi} M^{\oracle,\, x}_{a_\alice,\, a_\bob} \otimes
      B^{y_\bob}_{a'_\bob} \ket{\psi}
      & = \bra{\psi} B^{y_\bob}_{a_\bob} A^{y_\alice}_{a_\alice} \otimes
        B^{y_\bob}_{a'_\bob} \ket{\psi}\\
      & = \bra{\psi}  A^{y_\alice}_{a_\alice} \otimes
        B^{y_\bob}_{a_\bob} B^{y_\bob}_{a'_\bob} \ket{\psi}\\
      & = \delta_{a_\bob,\, a'_\bob}\, \bra{\psi} A^{y_\alice}_{a_\alice} \otimes
        B^{y_\bob}_{a_\bob} \ket{\psi}\;,
    \end{align*}
    where the first equality uses the definition of $M^{\oracle,\,
      x}_{a_\alice,\, a_\bob}$ from Eq.~\eqref{eq:oracle-honest}, the second
    equality uses the consistency of $\strategy$, and the third equality uses
    the projectivity of $\strategy$.
    Notice that when $a_\bob = a'_\bob$, this is exactly the probability of
    obtaining answers $(a_\alice,a_\bob)$ when player $\alice$ and player $\bob$
    get question pair $(y_\alice,y_\bob)$ in the game $\verifier_n$.
Since $\strategy$ is value-$1$, the answers are accepted by $\decider$ with
probability $1$; in particular, the simulated decider neither rejects nor enters
$\TIMEOUT$ on these transcripts.
    Thus the oracle's answers pass the ``Game check'' of the oracularized
    decider with certainty, and, furthermore, the oracle's answers are consistent
    with the isolated player's answers and thus pass the ``Consistency check''
    with certainty as well.
	
    Commutativity of $M^{\Oracle,\, x}_{a_\alice,\, a_\bob}$ and
    $B^{y_\bob}_{a_\bob}$ follows from the commutativity of $\strategy$ for the
    game $\verifier_n$.

  \item ({Both oracle}) If both players get the $\oracle$ role, then both
    players receive the same question $x \in V$.
    Using a similar analysis as for the previous item, the players return the
    same answer pair (thus passing the ``Consistency check'') and pass the
    ``Game check.''
    Both players' measurements commute because they are identical.

  \item ({Both isolated}) Suppose that both players receive the same isolated
    player role (e.g., they both receive the isolated player role $\alice$).
    They then perform the same measurements, which produce the same outcomes due
    to the consistency of the strategy $\strategy$, and thus they pass the
    ``Consistency check.''
    Otherwise, suppose that one player receives the $\alice$ role and the other
    player receives the $\bob$ role.
    Then the decider $\tdora$ automatically accepts.
    Furthermore, their measurements commute because their questions are
    distributed according to $\mu_{\sampler,\, n}$, and $\strategy$ is a
    commuting strategy with respect to $\mu_{\sampler,\, n}$.
  \end{enumerate}

  \paragraph{Complexity.}
It is clear from the definition that $\tsora$ depends only on $\sampler$.

We first bound the time complexity of $\tsora$.  On a dimension query,
$\tsora$ runs $\sampler(n,\gamestyle{dimension})$.  On every other input,
$\tsora$ first computes the dimension $s(n)$ by running $\sampler(n,\gamestyle{dimension})$.  It then performs
the format and length checks described above, reading only up to the prescribed
length before either rejecting or calling the corresponding routine of
$\sampler$.  Since $\sampler$ outputs factor-space indicator vectors of length
$s(n)$, we have $s(n)\leq \TIME_{\sampler}(n)$. Hence $\TIME_{\tsora}(n)=\poly(\TIME_{\sampler}(n))$.


We next bound the time complexity of $\tdora$.  The machine first computes
$B_{\decider}(n)$ from the description of
$\decider$.  It only attempts to parse answer components of length at most
$B_{\decider}(n)$, and whenever it applies a CL function of $\sampler$ it first
checks that the relevant question string has the prescribed length $s(n)$ (which it can compute by calling $\sampler(n,\gamestyle{dimension})$).  The decider $\tdora$ calls $\decider$ only on answer components of length at most
$B_{\decider}(n)$; if that call enters $\TIMEOUT$, then $\tdora$ rejects.  The
remaining consistency checks compare strings of length at most
$B_{\decider}(n)$.  Therefore, $\TIME_{\tdora}(n)
  =
  \poly(\TIME_{\sampler}(n),B_{\decider}(n))$.

The sampler $\tsora$ is an $\ell$-level typed sampler because $\sampler$ is an
$\ell$-level sampler and the new CL functions for $\tvar=\oracle$ are identity
maps, which can be viewed as $\ell$-level CL functions by
Remark~\ref{rk:higher-level}.

  \paragraph{Efficient computability.}
  The description of $\tsora$ can be computed, in polynomial time, from
  the description of $\sampler$ alone.
  The description of $\tdora$ can be computed in polynomial time from the
  descriptions of $\sampler$ and $\decider$.
  Moreover, in each case the computation amounts to copying the description of
  $\sampler$ and $\decider$, respectively, and adding constant-sized additional
  instructions.
\end{proof}

\subsection{Soundness of the oracularized verifier}

\begin{theorem}[Soundness of the oracularized verifier]
  \label{thm:oracle-soundness}
Let $\verifier = (\sampler, \decider)$ be a normal form verifier such that
$\decider$ is in the standard timeout-counter form of Section~\ref{sec:tms},
and let $\tvora = (\tsora,\tdora)$ be the corresponding typed oracularized
verifier.
  Then there exists a function $\delta(\eps) = \poly(\eps)$ such that for all $n
  \in \N$ the following hold.
  \begin{enumerate}
  \item If $\val^*(\tvora_n) > 1 - \eps$, then $\val^*(\verifier_n) \geq
    1 - \delta(\eps)$.
	\item For all $\eps > 0$, we have that
    \begin{equation*}
      \Ent \bigl( \tvora_n, 1 - \eps \bigr) \geq
      \Ent(\verifier_{n}, 1 - \delta(\eps))
    \end{equation*}
    where $\Ent(\cdot)$ is as in Definition~\ref{def:ent}.
  \end{enumerate}
\end{theorem}

\begin{proof}
	Fix $n \in \N$.
  Let $\strategy^\ora = \bigl( \ket{\psi}, A, B \bigr)$ be a strategy
  for $\tvora_n$ with value $1 - \eps$ for some $0 < \eps \leq 1$. Using Lemma~\ref{lem:symmetric-strat} we may without loss of generality assume that $\strategy^\ora$ is projective. 
  Let $(\tvar,x) \in \type^\ora \times V$ be a question to player $w=\alice$.
  In the event that $\tvar = \oracle$ (which occurs with probability $1/3$), let
  $y_v = L^v(x)$ for each $v \in \{\alice,\bob\}$.
  From the consistency check performed by $\tdora$ and
  item~\ref{item:consistency-implies-approx} of Fact~\ref{fact:agreement}, we
  have that for all $v \in \{\alice,\bob\}$ and on average over $x$ sampled by
  $\tsora$,
  \begin{equation}
    \label{eq:oracle-1}
    A^{\oracle,\, x}_{a_v} \otimes \Id_\bob  \approx_\eps
    \Id_\alice \otimes B^{v,\, y_v}_{a_v}\;.
  \end{equation}     
  Here, we used that with probability $1/9$ player $w=\alice$ gets the $\oracle$
  role and player $\overline{w}=\bob$ gets the isolated player $v$ role;
  conditioned on this, player $\bob$ gets question $y_v$.
	
	Using 
	Fact~\ref{fact:add-a-proj} on~\eqref{eq:oracle-1}, with the $C$ operators from  Fact~\ref{fact:add-a-proj} chosen as $\{A^{\oracle,\, x}_{a_\alice,\,
    a_\bob}\}$ here we get
		  \begin{align}
   A^{\oracle,\, x}_{a_\alice,a_\bob} \otimes \Id_\bob  &=  A^{\oracle,\, x}_{a_\alice,a_\bob} \cdot\big( A^{\oracle,\, x}_{a_\bob} \otimes \Id_\bob \big) \notag\\
	&\approx_\eps
   A^{\oracle,\, x}_{a_\alice,a_\bob}\cdot\big( \Id_\alice \otimes B^{\bob,\, y_\bob}_{a_\bob}\big) \notag\\
	&= A^{\oracle,\, x}_{a_\alice,a_\bob}\otimes B^{\bob,\, y_\bob}_{a_\bob}\;,   \label{eq:oracle-1b}
  \end{align}   
		where for the first equality we used that the POVM elements $\{A^{\oracle,\, x}_{a_\alice,\,
    a_\bob}\}$ are projective. Using a similar calculation for the second and third approximations below we get
  \begin{equation}
    \label{eq:oracle-2}
    \begin{split}
      A^{\oracle,\, x}_{a_\alice,\, a_\bob} \otimes \Id_\bob
      & \approx_{\eps} A^{\oracle,\, x}_{a_\alice,\, a_\bob}  \otimes
      B^{\bob,\, y_\bob}_{a_\bob} \\
      &\approx_{\eps} A^{\oracle,\, x}_{a_\alice} \otimes
      B^{\bob,\, y_\bob}_{a_\bob} \\
      & \approx_\eps \Id_\alice \otimes
      B^{\bob,\, y_\bob}_{a_\bob} B^{\alice,\, y_\alice}_{a_\alice} \\
      & \approx_\eps \Id_\alice \otimes
      B^{\alice,\, y_\alice}_{a_\alice} B^{\bob,\, y_\bob}_{a_\bob}\;,
    \end{split}
  \end{equation}
	where the last line uses that starting from $A^{\oracle,\, x}_{a_\alice,\, a_\bob} \otimes \Id_\bob$ we may perform all operations leading to the before-last line hwile reversing the order in which $B^\bob$ and $B^\alice$ are introduced. 
	
  Using Item~\ref{enu:check-same} of the consistency check, we have that on
  average over a random $y = L^\alice(x) \in L^\alice(V)$,
  \begin{equation}
    \label{eq:oracle-3}
    A^{\alice,\, y}_{a} \otimes I_\bob \approx_\eps
    \Id_\alice \otimes B^{\alice,\, y}_{a}\;.
  \end{equation}
  Define, for all $y \in L^\alice(V)$, measurement operators $\{C^y_a\}_a$ where
  $C^y_a = A^{\alice,\, y}_a$.
  Similarly, define $D^y_a = B^{\bob,\, y}_a$.
  This defines a strategy $\strategy = (\ket{\psi},C,D)$ for the game
  $\verifier_n$ that we now argue succeeds with high probability.
    
  Let $x \in V$ be uniformly random.
  Let $y_\alice = L^\alice(x)$ and $y_\bob =L^\bob(x)$.
  \begin{align*}
    A^{\oracle,\, x}_{a_\alice,\, a_\bob} \otimes \Id_\bob
    & \approx_\eps \Id_\alice \otimes
      B^{\bob,\, y_\bob}_{a_\bob} B^{\alice,\, y_\alice}_{a_\alice} \\
    & \approx_{\eps} A^{\alice,\, y_\alice}_{a_\alice} \otimes
      B^{\bob,\, y_\bob}_{a_\bob} \\
    & = C^{y_\alice}_{a_\alice} \otimes D^{y_\bob}_{a_\bob}\;.
  \end{align*}
	The first approximation follows from Equation~\eqref{eq:oracle-2}.
  The second approximation follows from Equation~\eqref{eq:oracle-3} and
  Fact~\ref{fact:add-a-proj} (where we let $C^y_b$ in Fact~\ref{fact:add-a-proj}
  represent $\Id\ot B^{\bob,\, y_\bob}_{a_\bob}$).
  The last equality follows from definition of $C^{y_\alice}_{a_\alice}$ and
  $D^{y_\bob}_{a_\bob}$.
  The pair of questions $(y_\alice, y_\bob) \in V \times V$ is distributed
  according to $\mu_{\sampler,\, n}$.

  The game check part of $\tdora$ succeeds with probability $1 -
  O(\eps)$, which implies that the answer pair $(a_\alice,\allowbreak  a_\bob)$ that arises
  from the measurement \[\textstyle A^{\oracle,\, x}_{a_\alice,\, a_\bob} \otimes \Id_\bob\]
  is accepted by the decider $\decider$ on question pair $(y_\alice,\allowbreak y_\bob)$
  with probability $1 - O(\eps)$.
  This in turn implies that the strategy $\strategy = (\ket{\psi},\allowbreak C,\allowbreak D)$ succeeds
  with probability $1 - O(\sqrt{\eps})$ in the game $\verifier_n$.
  As an additional consequence, the Schmidt rank of $\ket{\psi}$ must be at
  least $\Ent(\verifier_n,\allowbreak  1 - O(\sqrt{\eps}))$.
\end{proof}


\section{Answer reduction}
\label{sec:ans}

In this section we show how to transform a normal form verifier $\verifier =
(\sampler, \decider)$ into an ``answer reduced'' normal form verifier
$\verifier^\ar = (\sampler^\ar, \decider^\ar)$ such that the values of the
associated nonlocal games are directly related, yet the answer-reduced
verifier's decision runtime is only polylogarithmic in the answer length of the
original verifier (the answer-reduced verifier's sampling runtime remains
polynomially related to the sampling runtime of $\sampler$).
The polylogarithmic dependence is achieved by composing a probabilistically
checkable proof (PCP) with the oracularized verifier given in Section~\ref{sec:oracle}.
This step generalizes the answer reduction technique of~\cite[Part V]{NW19}.

Given index~$n \in \N$, the answer reduced verifier $\verifier^\ar$ simulates
the oracularization~$\verifier^\ora$ of $\verifier$ on index $n$.
To do so, it first samples questions~$x$ and~$y$ using the oracularized sampler
$\sampler^\ora$ and distributes them to the players, who compute answers~$a$
and~$b$.
Let us suppose that the first player is assigned the $\oracle$ role, and parse
their question and answer as pairs $x = (x_\alice, x_\bob)$ and $a = (a_\alice,
a_\bob)$, while the second player is an ``isolated'' player receiving the
question $x_\alice$ and responding with answer $b$.
Instead of executing the decider $\decider^\ora$ on the answers $(a, b)$, the
verifier $\verifier^\ar$ asks the first player to compute a PCP $\Pi$ of
$\decider(n, x_\alice, x_\bob, a_\alice, a_\bob) = 1$, and the second player to
compute an encoding $g_b$ of answer $b$.
$\verifier^\ar$ then requests randomly chosen locations of the proof $\Pi$ and
the encoding $g_b$, and executes the PCP verifier on the players' answers.
By the soundness of the PCP, $\verifier^\ar$ accepts with high probability only
if the player's answers satisfy $\decider(n, x_\alice, x_\bob, a_\alice, a_\bob)
= 1$ and $b = a_\alice$.

There are several challenges that arise when implementing answer reduction.
One challenge, already encountered in~\cite{NW19}, is that we need to ensure the
PCP $\Pi$ computed by the first player can be cross-tested against the encoding
$g_b$ computed by the second player (who doesn't know the entire structure of
the PCP $\Pi$).
This was handled in~\cite{NW19} by using a special type of PCP called a
\emph{probabilistically checkable proof of proximity (PCPP)}, which allows one
to efficiently check that a \emph{specific} string $x$ is a satisfying
assignment to a Boolean formula $\varphi$, as opposed to simply checking that
$\varphi$ is satisfiable.
In a PCPP, an encoding of the specific string $x$ is provided separately from
the proof of satisfiability.
The answer reduction scheme of~\cite{NW19} was able to use an ``off-the-shelf''
PCPP in a relatively black-box fashion to handle this.

In our answer reduction scheme, however, there is a further requirement: we need
the question distribution of the answer reduced verifier to be conditionally
linear.
This is necessary to maintain the invariant that the verifier after each step of
the compression procedure (introspection, answer reduction, parallel repetition)
is a normal form verifier.
Unfortunately, simulating the question distributions of off-the-shelf PCPPs with
conditionally linear distributions can be quite cumbersome.
Instead, we design a bespoke PCP verifier for the protocol whose question
distribution is more easily seen to be conditionally linear.

This section is organized as follows.
We start with some preliminaries on formulas and encodings in
Section~\ref{sec:ar-tms}.
In Section~\ref{sec:cook-levin} we show how to use the Cook-Levin reduction to
reduce the Bounded Halting problem for deciders to a succinct satisfiability
problem called Succinct-3SAT.
Following this, in Section~\ref{sec:succinct-deciders}, we reduce the
Succinct-3SAT instance to an instance of a related problem called Succinct
Decoupled 5SAT, which is easier to use in our answer reduction step.
Then in Section~\ref{sec:pcp-cktval-new} we introduce a PCP for Succinct
Decoupled 5SAT.
The verifier for the PCP expects a proof consisting of the evaluation tables of
low-degree polynomials, including the low-degree encodings of the players'
answers $a$ and $b$.
In Section~\ref{sec:ld-compiler} we provide the definition of a normal-form
verifier $\verifier^\ar$ that executes the composition of $\verifier^\ora$ with
the PCP verifier from Section~\ref{sec:pcp-cktval-new}.
In Section~\ref{sec:ar-completeness} we show completeness of the construction
and analyze its complexity.
In Section~\ref{sec:ar-soundness} we prove soundness.

\subsection{Circuit preliminaries}
\label{sec:ar-tms}
\newcommand{\true}{\mathrm{true}}
\newcommand{\false}{\mathrm{false}}
\newcommand{\out}{\mathrm{out}}

Recall the definitions pertaining to Turing machines from Section~\ref{sec:tms}. In this section it is useful to model
computation using \emph{Boolean circuits}. A Boolean circuit is a directed
acyclic graph, where each node is assigned a label from the set
$\{\true,\false,\wedge,\vee,\neg,\out\}\cup\{x_1,x_2,\ldots\}$. A
directed edge is also referred to as a \emph{wire}. Nodes with a label
in $\{x_1,x_2,\ldots\}$ are referred to as input nodes. There is a
unique node with the label $\out$, which encodes the
single-bit output. The
in-degree of each node with label in $\{\vee,\wedge\}$ is restricted
to be $2$, the in-degree of $\neg$ and $\out$ is $1$, and the
in-degree of nodes with all other labels is zero. \tnote{was:There is no restriction on
the out-degree (the ``fan-out'')}Here we restrict the out-degree of all nodes to be at most $2$.\footnote{Eventually, this is without loss of generality as one is always able to simulate a gate with fan-out $k$ using $(k-2)$ OR gates with fan-out $2$ and $(k-2)$ $\false$ inputs with fan-out $1$, as explained in e.g.,~\cite[p.37]{Pap94}.} The \emph{size} of a circuit is the total
number of nodes and wires that it contains.\anote{To self: check that
  this fan-out business is OK.}

To a circuit with $n$ input nodes we can associate a function
$f:\{0,1\}^n\to\{0,1\}$ in a straightforward way, and we will often
refer to the circuit as ``computing'' that function. For the sake of
concreteness, we sketch the process of computing the function $f$ on a
given input $x \in \{0,1\}^n$. Fix a topological
order on the nodes in the circuit, and processing nodes in this order,
assign a boolean value in $\{0,1\}$ to each node as follows: every node labeled
$x_i$ is assigned the value of the $i$th bit of the input $x$, every
node labeled $\true$ is assigned $1$, every node labeled $\false$ is
labeled $0$, every node labeled $\vee$ is assigned the boolean OR of
the values of its in-neighbors, every node labeled $\wedge$ is assigned the boolean AND of
the values of its in-neighbors, every node labeled $\neg$ is assigned the boolean
NOT of the value of its in-neighbor, and the node labeled $\out$ is
assigned the same value as its in-neighbor. The value assigned to the
node labeled $\out$ is the value $f(x)$ of the function computed by
the circuit. For a more detailed description, see Section 4.3 of~\cite{Pap94}.

\begin{remark}[Plugging integers into circuits]\label{rem:plugging-in-integers}
  Let $\circuit$ be a circuit with $n$ input nodes.
  Inputs to~$\circuit$ are identified with strings $x \in \{0, 1\}^n$.
  In this section, we will also allow~$\circuit$ to receive inputs $a \in \{0,
  1, \ldots, 2^n-1\}$.
  In doing so, we use the convention that a number~$a$ between~$0$ and~$2^{n}-1$
  is interpreted as its $n$-digit binary encoding $\binary{n}(a)$
when provided as input to a set of
  $n$ single-bit wires.
  In other words, $\circuit(a) = \circuit(x)$, where $x = \binary{n}(a)$.

We sometimes group the $n$ input bits to a circuit, for semantic convenience. If we refer to a circuit $\circuit$ with~$k$ different inputs of length
  $n_1, \ldots, n_k$, then we mean that the circuit can be evaluated on inputs $a_1 \in \{0, 1, \ldots,
  2^{n_1}-1\}, \ldots, a_k \in \{0, 1, \ldots, 2^{n_k}-1\}$ as follows:
  \begin{equation*}
    \circuit(a_1, \ldots, a_k) = \circuit(x_1, \ldots, x_k)\;,
  \end{equation*}
  where $x_1 = \binary{n_1}(a_1), \ldots, x_k = \binary{n_k}(a_k)$ and the $n$ input wires to $\circuit$ are set to the concatenation of $x_1,\ldots,x_k$.
\end{remark}

A 3SAT formula is a Boolean formula in conjunctive normal form in which at most
three literals appear in each clause.
More precisely, $\varphi$ is a 3SAT formula on $N$ variables $x_1, x_2, \ldots,
x_N$ if it has the form $\bigwedge_{j=1}^m C_j$ and each clause $C_j$ is the
disjunction of at most three literals, where a literal is either a variable
$x_i$ or its negation $\neg x_i$.
We use $x_i^o$ to denote the literal $x_i$ if $o = 1$ and $\neg x_i$ if $o = 0$.

\begin{definition}[Succinct description of 3SAT formulas]
  \label{def:succinct-formulas}
  Let $N = 2^n$, and let $\varphi$ be a 3SAT formula on~$N$ variables named
  $x_0, \ldots, x_{N-1}$.
  Let $\circuit$ be a Boolean circuit with 3 inputs of length~$n$ and three
  single-bit inputs.
  Then $\circuit$ is a \emph{succinct description of $\varphi$} if for each
  $i_1, i_2, i_3 \in \{0, 1, \ldots, N-1\}$ and $o_1, o_2, o_3 \in \{0, 1\}$,
  \begin{equation}\label{eq:circuit-val-is-one}
    \circuit(i_1, i_2, i_3, o_1, o_2, o_3) = 1
  \end{equation}
  if and only if $x_{i_1}^{o_1} \lor x_{i_2}^{o_2} \lor x_{i_3}^{o_3}$ is a
  clause in $\varphi$.
  In Equation~\eqref{eq:circuit-val-is-one}, we use the notation from
  Remark~\ref{rem:plugging-in-integers}.
\end{definition}
  
\begin{definition}[Succinct-3SAT problem]
  The \emph{Succinct-3SAT} problem is the language containing encodings of
  circuits~$\circuit$ in which $\circuit$ is a succinct description of a
  satisfiable 3SAT formula~$\varphi$.
\end{definition}

%

We make use of the \emph{Tseitin transformation}, which is a mapping from circuits to Boolean
formulas, and \emph{arithmetization}, which is a mapping from Boolean formulas to functions defined over a finite field. Both of these are standard constructs, introduced in detail in~\cite[Section 3.8]{NW19}. Here we briefly recall some relevant properties. 

\begin{lemma}
  \label{lem:tseitin}
  Let $\circuit$ be a Boolean circuit with $n$ inputs and size $s$. Then the corresponding \emph{Tseitin formula} $\formula$ is a
  Boolean formula on $n+s$ variables with the property that for all $x
  \in \{0,1\}^n$, $\circuit(x) = 1$ if and only if there exists $w \in
\{0,1\}^s$ such that $\formula(x, w) = 1$. The formula $\formula$ has size
$O(s)$ and a description of it is computable from $\circuit$ in time
$O(s \log s)$.
\end{lemma}
We remark that \cite{NW19} claims that the Tseitin formula can be
constructed in time $O(s)$ but this is not correct: the construction
given there runs in time linear in the bit length of the description
of $\circuit$, which is $O(s \log s)$ bits long.

\begin{lemma}
  \label{lem:formula-arithmetization}
  Let $\formula$ be a Boolean formula over $m$ variables. An arithmetization
  $\formula_{\mathrm{arith}}$ over $\F_q$ is a function
  $\formula_{\mathrm{arith}}:\F_q^{m}\rightarrow \F_q$  such that
  \begin{equation}
    \label{eq:farith-general}
    \forall x \in \{0, 1\}^{m}\;,
    \quad \formula_{\mathrm{arith}}(x) = \formula(x)\;.
  \end{equation}
  
\end{lemma}

\begin{proposition}
  \label{prop:tseitin-arith-degree}
  Let $\circuit$ be a circuit on $n$ inputs with size $s$ and
  out-degree at most $2$. Then there exists an arithmetization $\formula_{\mathrm{arith}}$ of its Tseitin formula
  which is a polynomial over $\F_q$ on $n +s$ variables with individual
  degree at most $6$ in each variable\tnote{fixed; was degree $2$}. For admissible field sizes $q$,
  a description of the polynomial
  $\formula_{\mathrm{arith}}$ can be computed in $\poly(s, \log q)$ time
  and it can be evaluated at a specific point $z \in \F_q^{n+s}$ in time
  $\poly(s, \log q)$.
\end{proposition}

\begin{proof}
    The properties follow by inspecting the construction in Definition~3.27 and~3.28 of~\cite{NW19}, together with
    Lemma~\ref{lem:efficient_arithmetic}. \tnote{edited:}In particular, the degree bound
    follows by observing that every variable in
    $\formula$ that is associated to an input occurs at most $2d$ times, if the gate associated with this input has out-degree $d$, and the variable associated to a wire occurs $2(1+d)$ times, with $d$ the out-degree of the gate that this wire enters. Since all out-degrees are at most $2$, the arithmetization
    $\formula_{\mathrm{arith}}$ has individual degree at most $6$.
\end{proof}

\subsection{A Cook-Levin theorem for bounded deciders}
\label{sec:cook-levin}



\begin{definition}[Bounded Halting problem]
	The \emph{$k$-input Bounded Halting} problem is the language
  $\BoundedHalting_k$ containing the set of tuples $(\overline{\cal{M}}, T, z_1,\ldots,z_k)$
  where $\overline{\cal{M}}$ is the description of a $k$-input Turing
  machine $\cal{M}$,  $T \in \N$,
  $z_1, \ldots, z_k \in \{0,1\}^*$, and $\cal{M}$ accepts input $(z_1,
  \ldots, z_k)$ (i.e., halts and outputs $1$ given this input) in at most $T$ time steps.
\end{definition}

We begin by defining natural encodings of a decider's tape alphabet and set of
states.

\begin{definition}[Decider encodings]
  Let~$\decider$ be a decider with tape alphabet $\Gamma = \{0, 1, \sqcup\}$ and
  set of states~$K$.
  We will write $\mathrm{enc}_{\Gamma}: \Gamma \cup \{\square\} \rightarrow \{0,
  1\}^2$ for the function which encodes the elements of~$\Gamma$, and a special
  ``$\square$'' symbol described below, as length-two binary strings in the
  following manner:
  \begin{alignat*}{2}
    & \mathrm{enc}_{\Gamma}(0) = 00\;,\quad
    && \mathrm{enc}_{\Gamma}(1) = 01\;,\\
    & \mathrm{enc}_{\Gamma}(\sqcup) = 10\;,\quad
    && \mathrm{enc}_{\Gamma}(\square) = 11\;.
  \end{alignat*}
  In addition, we write $\mathrm{enc}_K : K \rightarrow \{0, 1\}^\kappa$ for
  some arbitrary fixed $\kappa$-bit encoding of the states in~$K$, where
  $\kappa = \lceil \log(|K|) \rceil$.
\end{definition}

Now, we give the main result of this section.
It states that any decider~$\decider$ can be converted into a circuit~$\circuit$
which succinctly represents a 3SAT formula~$\varphi_{\mathrm{3SAT}}$ that
carries out the time~$T$ computation of~$\decider$.
In addition, $\circuit$ is extremely small---size $\poly \log(T)$ rather than
$\poly(T)$.

\begin{proposition}[Succinct representation of deciders]
  \label{prop:standard-succinct-sat}
  There is an algorithm with the following properties.
  Let $\decider$ be a decider, let $n$, $T$,~$\qlen$, and $\sigma$ be integers
  with $\max \{ \qlen, 2 \lceil \log n \rceil \} \leq T$ \hnote{june 14: added $\log n$ lower bound} and $|\decider| \leq \sigma$, and let~$x$ and~$y$ be
  strings of length at most~$\qlen$.
  Then on input $(\sigma,\decider, n, T, \qlen, x, y)$, the algorithm outputs
  a circuit~$\circuit$ on $3m + 3$ inputs which succinctly describes a 3SAT
  formula $\varphi_{\mathrm{3SAT}}$ on $M = 2^m$ variables.
  Furthermore, the algorithm and the corresponding 3SAT formula
  $\varphi_{\mathrm{3SAT}}$ satisfy the following properties:
  \begin{enumerate}
    \item For all $a, b \in \{0, 1\}^{2T}$, there exists a
          $c \in \{0, 1\}^{M-4T}$ such that $w = (a, b, c)$ satisfies
          $\varphi_{\mathrm{3SAT}}$ if and only if there exist
          $a_{\mathrm{prefix}}, b_{\mathrm{prefix}} \in \{0, 1\}^*$ of
          lengths~$\ell_a, \ell_b \leq T$, respectively, such that
          \begin{equation*}
            a = \mathrm{enc}_\Gamma(a_{\mathrm{prefix}}, \sqcup^{T - \ell_a})
            \quad
            \text{and}
            \quad
            b = \mathrm{enc}_\Gamma(b_{\mathrm{prefix}}, \sqcup^{T - \ell_b})
          \end{equation*}
          and $\decider$ accepts
          $(n, x, y, a_{\mathrm{prefix}}, b_{\mathrm{prefix}})$ in time~$T$.
    \item $m = O(\log(T) + \log(\sigma))$ is a power of two, and
    \item $\circuit$ has at most
          $s(n,T,\qlen,\sigma)=\poly(\log(n), \log(T), \qlen, \sigma)$ gates,
    \item On any input, the algorithm runs in time
          $\poly(\log(n), \log(T), \qlen,\sigma)$,
    \item Explicit values for $m(T, \sigma)$ and $s(n, T, \qlen, \sigma)$ can be
          computed in time polynomial in $\log(n)$, $\log(T)$, $\log(\qlen)$,
          and $\log(\sigma)$.
  \end{enumerate}
\end{proposition}

Proposition~\ref{prop:standard-succinct-sat} is essentially the standard fact
that Succinct-3SAT is an $\mathsf{NEXP}$-complete language, i.e.,\ that every
nondeterministic computation which takes time~$2^n$ can be represented as a
Succinct-3SAT instance of size only~$\poly(n)$, and our proof will follow the standard
proof that Succinct-3SAT is $\mathsf{NEXP}$-complete as presented
in~\cite{Pap94}.
This proof observes that the Cook-Levin reduction, which is used to show that
3SAT is $\mathsf{NP}$-complete, produces a 3SAT instance whose clauses follow
such a simple pattern that they can be described succinctly using an
exponentially-smaller circuit.

However, Proposition~\ref{prop:standard-succinct-sat} makes a few additional
stipulations on the Succinct-3SAT instance and reduction, which will require us
to prove it from scratch.
Firstly, Item~1 of the Proposition requires that the variables $w$ of the
formula $\varphi_{\mathrm{3SAT}}$ should split into three blocks $a,b, c$ such
that the $a$ variables correspond to the input $a$ of $\decider$, and the $b$
variables correspond to the input $b$ of $\decider$.
\anote{Old: First, we require that the coordinates of~$a$ and~$b$ embed into~$w$
  not randomly but as its lexicographically first coordinates (for reasons that
  are explained below in Section~\ref{sec:succinct-deciders}).}
Secondly, we need explicit bounds on how quantities such as the size
of~$\circuit$ relate to quantities such as $\sigma$, an upper bound on the
description length of $\decider$ in bits.

\begin{proof}[Proof of Proposition~\ref{prop:standard-succinct-sat}]
  \tnote{added this here. It's a bit inelegant, but the alternative is to rewrite the whole proof for a $3$-tape decider, which feels quite cumbersome}

  \znote{Use the pack TM idea to make sure we can argue about the exact run time of the decider.}

  \leavevmode\par First, algorithm checks that the description length of $\decider$ is at most $\sigma$; if not, then it rejects. Otherwise it proceeds as follows.

  The proof follows the Cook-Levin reduction, with one modification to address a
  technical issue concerning how we refer to variable-length components of the
  input $(n,x,y,a,b)$ to the decider $\decider$, as required in the first item
  of Proposition~\ref{prop:standard-succinct-sat}.
  Specifically, we introduce a $7$-tape \emph{packing} machine $\PACK$
  whose sole purpose is to take the five inputs $n$, $x$, $y$, $a$, and $b$
  appearing on its first five input tapes and write them in the format of a
  single $5$-tuple input to $\decider$, which is a $3$-tape Turing machine as
  specified in Section~\ref{sec:tms}.
  We then apply the Cook-Levin construction to $\PACK$ and $\decider$ separately
  and add constraints ensuring that the input provided to $\decider$ matches the
  output produced by $\PACK$.
  This special treatment resolves the difficulty of referring to variable-length
  components of the input $(n,x,y,a,b)$ to $\decider$, as required by
  Proposition~\ref{prop:standard-succinct-sat}, while simultaneously allowing us
  to analyze the running time of $\decider$ independently of the reformatting
  preprocessing.

  \begin{definition}[Packing machine]
    Let $\textsc{PACK}$ be the following $7$-tape Turing machine over tape alphabet
    $\Gamma=\{0,1,\sqcup\}$ that has five read-only input tapes, one working
    tape, and one output tape.
    On input $n,x,y,a,b$ placed on its first five tapes (each followed by
    blanks), $\textsc{PACK}$ writes to tape~$7$ the concatenation
    \begin{equation}\label{eq:pack-string}
      \mathrm{pack}(n,x,y,a,b) \;:=\; n\,\sqcup\,x\,\sqcup\,y\,\sqcup\,a\,\sqcup\,b
    \end{equation}
    where the concatenation is over $\Gamma$ and $\sqcup$ is the blank symbol. Once this has been completed 
    it then enters a designated accepting state in which it makes no further
    changes to any tape.
    Let $K'$ be the state set of $\PACK$. We write
    $\mathrm{enc}_{K'} : K' \rightarrow {\{0, 1\}}^{\kappa'}$ for some arbitrary
    fixed $\kappa'$-bit encoding of the states in $K'$ where
    $\kappa' = \lceil \log(|K'|) \rceil$.
  \end{definition}

  Whenever $\abs{x}, \abs{y} \le \qlen \le T$, and $\abs{a}, \abs{b} \le T$, the
  string $\mathrm{pack}(n,x,y,a,b)$ over $\Gamma$ has length at most $5T+4$.
  For $S = 5T+5$, $\textsc{PACK}$ halts (and thereafter stays put) within $S$
  steps 
	on all inputs that will arise in the statement of the proposition, and
  it never needs to access beyond the first $S$ tape cells of any tape.

  We will construct a single 3SAT instance as the conjunction of: (i) a
  Cook-Levin tableau for $\textsc{PACK}$ with time bound~$S$, (ii) a Cook-Levin
  tableau for the original $3$-tape machine~$\decider$ with time bound~$T$, and
  (iii) a short ``linking'' formula forcing the output written by
  $\textsc{PACK}$ to be the input read by~$\decider$.
  Crucially, the tableau for $\decider$ uses the \emph{original} time bound~$T$
  which is required by Proposition~\ref{prop:standard-succinct-sat}.

  \paragraph{Cook-Levin for $\PACK$.}
  We start by defining an execution tableau for $\textsc{PACK}$ up to time~$S$, restricted
  to tape positions $\{0,1,\ldots,S+1\}$, exactly as in the standard Cook-Levin
  construction.
  The execution tableau contains, for each time $t \in \{1, \ldots, S\}$,
  variables describing the state of~$\PACK$ and the contents of each of its
  tape cells at time~$t$.
  More formally, it consists of the following three sets of variables.
  \begin{enumerate}
  \item \label{item:cell-variables} For each time $t \in \{1, \ldots, S\}$, tape
    $i \in \{1, \ldots, 7\}$, and tape position $j \in \{0, 1,\ldots, S+1\}$,
    the tableau contains two Boolean-valued variables
    \begin{equation*}
      c^{\PK}_{t, i, j} = (c^{\PK}_{t, i, j, 1}, c^{\PK}_{t, i, j, 2}) \in \{0, 1\}^2
    \end{equation*}
    which are supposed to correspond to the contents of the $j$-th tape cell on
    tape~$i$ at time~$t$ according to $\mathrm{enc}_{\Gamma}(\cdot)$.
    The variables with $j \in \{0, S+1\}$ do not correspond to any cell on the
    tape; rather, the $j=0$ variables correspond to the left-boundary of the
    tape, and the $j=S+1$ variables correspond to the right-boundary of the
    first~$S$ cells on the tape.
    These are expected to always contain the special boundary symbol
    ``$\square$,'' i.e.,\ $c^{\PK}_{t, i, j}$ should be equal to
    $\mathrm{enc}_{\Gamma}(\square)$ whenever $j \in \{0, S+1\}$.
    As we will see below, it is convenient to define these so that for each~$t
    \in \{1, \ldots, S\}$, $i\in \{1, \ldots, 7\}$, and $j \in \{1, \ldots,
    S\}$, the variable $c^{\PK}_{t, i, j}$ also has a variable to its left $c^{\PK}_{t, i,
      j-1}$ and to its right $c^{\PK}_{t, i, j+1}$.
  \item \label{item:head-variables} For each time $t \in \{1, \ldots, S\}$, tape
    $i \in \{1, \ldots, 7\}$, and tape position $j \in \{0, 1,\ldots, S+1\}$,
    the tableau contains Boolean-valued variables $h^{\PK}_{t, i, j} \in \{0, 1\}$
    which are supposed to indicate whether the $i$-th tape head is in cell~$j$
    at time~$t$.
    For the boundary cells $j \in \{0, S+1\}$, we expect that $h^{\PK}_{t, i, j} = 0$
    for all $t \in \{1, \ldots, S\}$ and $i \in \{1, \ldots, 7\}$.
  \item For each time $t\in \{1, \ldots, S\}$, the tableau contains $\kappa'$
    Boolean-valued variables
    \begin{equation*}
      s^{\PK}_t = (s^{\PK}_{t, 1}, \ldots, s^{\PK}_{t, \kappa}) \in \{0, 1\}^{\kappa'}
    \end{equation*}
    which are supposed to correspond to the state of $\PACK$ at time~$t$
    according to $\mathrm{enc}_{K'}(\cdot)$.
  \end{enumerate}
  Finally, we let $\mathcal{V}^{\PK}$ denote the set of all of these variables.
  In other words,
  \begin{equation*}
    \mathcal{V}^{\PK} = \{c^{\PK}_{t, i, j, k}\}_{t, i, j, k} \cup \{h^{\PK}_{t, i, j}\}_{t, i, j}
    \cup \{s^{\PK}_{t, k}\}_{t, k}.
  \end{equation*}
  In total, the number of variables in the execution tableau for $\PACK$ is given by
  \begin{equation}\label{eq:pack-tableau-variable-count}
    |\mathcal{V}^{\PK}|
    = O\big(S^2 + S \cdot \log(|K'|)\big)
    = O\big(T^{2}\big)\;.
  \end{equation}

  As stated above, we expect the variables in~$\mathcal{V}^{\PK}$ to correspond to
  some execution of the machine~$\PACK$ of at most $S$ time steps.
  However, in general these are just arbitrary $\{0, 1\}$-valued variables.
  We now describe a set of constraints placed on these variables which, if
  satisfied, ensure they \emph{do} indeed correspond to some execution
  of~$\PACK$ of at most $S$ timesteps.
  These constraints will be split into two categories: (i) the constraints
  corresponding to the boundary, which ensure that the $t=1$ variables are
  initialized to a valid starting configuration and the $j \in \{0, S+1\}$
  variables are set according to items~\ref{item:cell-variables}
  and~\ref{item:head-variables} above, and (ii) the constraints corresponding to the
  execution of~$\PACK$, which ensure that the variables at each time~$(t+1)$
  follow from the variables at time~$t$ according to the computation
  of~$\PACK$.
  We start with the boundary constraints, which are simple enough to be
  described with a 3SAT formula.

  \begin{definition}
    The \emph{boundary formula} $\varphi^{\PK}_{\mathrm{Boundary}}$ is the 3SAT
    formula on the variables $\mathcal{V}^{\PK}$ described as follows.
    Let $o = \mathrm{enc}_{K'}({\mathsf{start}}) \in \{0, 1\}^{\kappa'}$, where
    ${\mathsf{start}} \in K'$ is the start state of~$\PACK$.
    For the $t= 1$ boundary, $\varphi^{\PK}_{\mathrm{Boundary}}$ contains the
    following set of clauses.
    \begin{alignat}{2}
      & \text{Indices} & \qquad & \text{Clauses}\nonumber\\
      & i \in \{1, \ldots, 7\} && h^{\PK}_{1, i, 1}
      \label{eq:pack-tape-heads-start}\\
      & i \in \{1, \ldots, 7\},~ j \neq 1 && \neg h^{\PK}_{1, i, j}
      \label{eq:pack-no-more-tape-heads}\\
      & k \in \{1, \ldots, \kappa'\} && \bigl(s^{\PK}_{1, k}\bigr)^{o_k}
      \label{eq:pack-start-state}\\
      & i \in \{1, \ldots, 7\},~ j \in \{1, \ldots, S\} &&
      \neg c^{\PK}_{1, i, j, 1} \lor \neg c^{\PK}_{1, i, j, 2}
      \label{eq:pack-not-square}\\
      & i \in \{1, \ldots, 5\},~ j < j' \in \{1, \ldots, S\} &&
      \neg c^{\PK}_{1, i, j, 1} \lor c^{\PK}_{1, i, j', 1}
      \label{eq:pack-trailing-empty}\\
      & i \in \{6, 7\},~ j \in \{1, \ldots, S\}
      && c^{\PK}_{1, i, j, 1} \text{ and } \neg c^{\PK}_{1, i, j, 2}
      \label{eq:pack-all-empty}
    \end{alignat}
    This is meant to be read as follows: for each row, the ``Indices'' column
    specifies the range of the indices that the clauses in the ``Clauses'' column
    are quantified over.
    For example, row~\eqref{eq:pack-tape-heads-start} specifies that for all $i \in
    \{1, \ldots, 7\}$, $\varphi^{\PK}_{\mathrm{Boundary}}$ contains the clause $h^{\PK}_{1,
      i, 1}$.
    For the $j \in \{0, S+1\}$ boundary, $\varphi^{\PK}_{\mathrm{Boundary}}$ contains
    the following set of clauses.
    \begin{alignat}{2}
      & \text{Indices} & \qquad & \text{Clauses}\nonumber\\
      & t \in \{1, \ldots, S\},~i \in \{1, \ldots, 7\},
      j \in \{0, S+1\},~k \in \{1, 2\}
      && c^{\PK}_{t, i, j, k} \label{eq:pack-square}
    \end{alignat}
  \end{definition}

  Rows~\eqref{eq:pack-tape-heads-start} and~\eqref{eq:pack-no-more-tape-heads}
  ensure that at time~$t = 1$, each tape has exactly one tape head, and it is
  located on cell $j = 1$.
  Row~\eqref{eq:pack-start-state} ensures that at time~$t=1$, the state is the
  start state $\mathsf{start}$.
  (Recall the notation $x_i^o$ to denote the literal $x_i$ if $o = 1$ and
  $\neg x_i$ if $o = 0$.)
  For the remaining rows, we recall that under the encoding of the tape
  alphabet, $\mathrm{enc}_\Gamma(\sqcup) = 10$ and
  $\mathrm{enc}_\Gamma(\square) = 11$.
  As a result, (i) row~\eqref{eq:pack-square} ensures that for all times and
  tapes, the cells $j \in \{0, S+1\}$ contain the~$\square$ symbol, (ii)
  row~\eqref{eq:pack-not-square} ensures that for time~$t=1$, no cell
  $j \notin \{0, S+1\}$ contains the~$\square$ symbol, and (iii)
  row~\eqref{eq:pack-all-empty} ensures that for time~$t=1$ and tapes~$6$
  and~$7$, all cells $j \in \{1, \ldots, S\}$ contain the $\sqcup$ symbol.
  Finally, row~\eqref{eq:pack-trailing-empty} says that for tapes
  $i \in \{1, \ldots, 5\}$, if cell~$j$ contains~$\sqcup$, then every cell
  $j' > j$ must contain~$\sqcup$ as well.
  This means that the five strings encoded by
  $c^{\PK}_{1, 1}, \ldots, c^{\PK}_{1,5}$ each consist of a string of $0$'s and
  $1$s followed by a string of $\sqcup$'s.
  In short, suppose we write $n$, $x$, $y$, $a$, and~$b$ for the prefixes of
  these strings with no $\sqcup$'s.
  If $\varphi^{\PK}_{\mathrm{Boundary}}$ is satisfied, then the execution
  tableau correctly encodes that the tapes of $\PACK$ contain inputs $n$, $x$,
  $y$, $a$, and~$b$ at time~$t=1$.

  Next, we describe the execution constraints.
  These are more complicated than the boundary constraints, and so we will begin
  by describing them in terms of a general Boolean circuit known as the
  \emph{local check circuit}.
  For any time~$t\in \{1, \ldots, S-1\}$ and tape positions $j_1, \ldots, j_7
  \in \{1, \ldots, S\}$, the local check circuit can check that the execution
  tableau properly encodes these tape positions at time~$t+1$ by looking only at
  the encodings of these tape positions and their neighbors (i.e.,\ the tape
  positions $j_i \pm 1$ for each~$i \in \{1, \ldots, 7\}$) at time~$t$.

  \begin{definition}\label{def:local-check-circuit}
    In this definition, we will define the \emph{local check circuit}
    $\circuit^{\PK}_{\mathrm{Check}}$.
    It has the following inputs.
    \begin{itemize}
      \item For each $i \in \{1, \ldots, 7\}$, it has the eight inputs
      \begin{equation}
        \label{eq:pack-local-check-inputs}
        \begin{matrix}
          c^{\PK}_{\mathrm{Check}, 0, i, -1} & c^{\PK}_{\mathrm{Check}, 0, i, 0} &
          c^{\PK}_{\mathrm{Check}, 0, i, 1} & \qquad &
          h^{\PK}_{\mathrm{Check}, 0, i, -1} & h^{\PK}_{\mathrm{Check}, 0, i, 0} &
          h^{\PK}_{\mathrm{Check}, 0, i, 1} \\
          & c^{\PK}_{\mathrm{Check}, 1, i, 0} & & & &
          h^{\PK}_{\mathrm{Check}, 1, i, 0} &
        \end{matrix}
      \end{equation}
      where the $c$-inputs are in $\{0, 1\}^2$ and the $h$-inputs are in
      $\{0, 1\}$.
      \item It has two inputs
            $s^{\PK}_{\mathrm{Check}, 0},\allowbreak  s^{\PK}_{\mathrm{Check}, 1} \in \{0,\allowbreak  1\}^{\kappa'}$.
    \end{itemize}
    In addition, for each time $t \in \{1, \ldots, S-1\}$ and tape positions
    $j_1, \ldots, j_7 \in \{1, \ldots, S\}$, we will define a circuit
    $\circuit_{\mathrm{Check}, t, j_1, \ldots, j_7}$ that has the same number of
    inputs as $\circuit^{\PK}_{\mathrm{Check}}$.
    For convenience we label the inputs of
    $\circuit_{\mathrm{Check}, t, j_1, \ldots, j_7}$ using names of variables
    in~$\mathcal{V}^{\PK}$.
    We perform this labeling by associating the following eight labels
    from~$\mathcal{V}^{\PK}$ with the corresponding input variables as described
    in Equation~\eqref{eq:pack-local-check-inputs}:
    \begin{equation*}
      \begin{matrix}
        c^{\PK}_{t, i, j_i-1} & c^{\PK}_{t, i, j_i} & c^{\PK}_{t, i, j_i+1} & \qquad\qquad &
        h^{\PK}_{t, i, j_i-1} & h^{\PK}_{t, i, j_i} & h^{\PK}_{t, i, j_i+1} \\
        & c^{\PK}_{t+1, i, j_i} & & & & h^{\PK}_{t+1, i, j_i} &
      \end{matrix}
    \end{equation*}
    as well as by associating $s^{\PK}_t$ and $s^{\PK}_{t+1}$ from~$\mathcal{V}^{\PK}$ with
    $s^{\PK}_{\mathrm{Check}, 0}$ and $s^{\PK}_{\mathrm{Check},1}$, respectively.

    We first define~$\circuit^{\PK}_{\mathrm{Check}}$ as a function.
    In Lemma~\ref{lem:check-size} below we give an implementation of this
    function as a circuit.
    To do so it is convenient to define a function
    $\circuit^{\PK}_{\mathrm{Check}, t, j_1, \ldots, j_7}$ for each value of
    $t, j_1, \ldots, j_7$.
    It will be clear that each of these is in fact the same function applied to
    different inputs, and hence this will
    define~$\circuit^{\PK}_{\mathrm{Check}}$ as well.
    Now, the functions~$\circuit^{\PK}_{\mathrm{Check}, t, j_1, \ldots, j_7}$
    perform the following computation.
    \begin{itemize}
      \item Suppose for each $i \in \{1, \ldots, 7\}$, exactly one of the tape
            positions $j_i-1$, $j_i$, and $j_i+1$ at time~$t$ contains a tape
            head.
            First, $\circuit^{\PK}_{\mathrm{Check}, t, j_1, \ldots, j_7}$
            computes the transition function of the Turing machine applied to
            the contents of these~$7$ tape positions, when the state of~$\PACK$
            at time~$t$ is $s^{\PK}_{\mathrm{Check}, 0}$.
            This transition function produces the state of~$\PACK$ and contents
            of these tape positions at time~$t+1$, as well as directions to move
            the~$7$ tape heads in.
            Then $\circuit^{\PK}_{\mathrm{Check}, t, j_1, \ldots, j_7}$ checks
            that the variables for the tape positions $j_1, \ldots, j_7$ and the
            state of~$\PACK$ at time~$t+1$ match what they should be according
            to the computed transition function.
            In addition, if a tape head is marked as moving into a tape cell
            containing a~$\square$ symbol, that tape head remains in place
            instead.
      \item Suppose for each $i \in \{1, \ldots, 7\}$, none of the tape
            positions $j_i -1$, $j_i$, or $j_i+1$ contain a tape head at
            time~$t$.
            Then $\circuit^{\PK}_{\mathrm{Check}, t, j_1, \ldots, j_7}$ checks
            that the variables for the tape positions $j_1, \ldots, j_7$ at
            time~$t+1$ match the variables at time $t$.
      \item Otherwise, $\circuit^{\PK}_{\mathrm{Check}, t, j_1, \ldots, j_7}$
            accepts.
    \end{itemize}
    This defines $\circuit^{\PK}_{\mathrm{Check}, t, j_1, \ldots, j_7}$, and, therefore,
    $\circuit^{\PK}_{\mathrm{Check}}$.
  \end{definition}

  Definition~\ref{def:local-check-circuit} shows the utility of introducing the
  boundary variables $c^{\PK}_{t, i, 0}$ and $c^{\PK}_{t, i, S+1}$.
  The function $\circuit^{\PK}_{\mathrm{Check}}$ checks the contents of a cell $c^{\PK}_{t,
    i, j}$ at time $t+1$ by looking at the cell and its neighbors $c^{\PK}_{t, i,
    j-1}, c^{\PK}_{t, i, j+1}$ at time $t$.
  However, those cells with $j \in \{1, S\}$ only have either a left neighbor or
  a right neighbor, and so without the boundary variables we'd have to introduce
  two other local check circuits designed just for these boundary cases.
  The boundary variables then allow us to use the same local check circuit for
  all cells.

  \begin{lemma}\label{lem:pack-check-size}
    A circuit~$\circuit^{\PK}_{\mathrm{Check}}$ that computes the function described
    above can be computed in constant time (depending only on the description length of the
    fixed Turing machine $\PACK$).
  \end{lemma}

  \begin{proof}
    The circuit has $7 \cdot 12 + 2\cdot \kappa' = 84 + 2\kappa'$ total Boolean
    inputs, giving a total of $2^{84} \cdot 4^{\kappa'} = O(|K'|^2)$ possible input
    strings.
    For each possible fixed input string, the circuit
    $\circuit^{\PK}_{\mathrm{Check}}$ checks if the actual input is equal to the
    fixed input, which takes $O(\kappa')$ gates, and then it accepts if the
    fixed input should be accepting, i.e.,\ for each fixed input the correct
    output is hard-coded in the circuit.
    Doing so takes $O(|K'|^2 \cdot \kappa')$ gates.
    Computing $\circuit^{\PK}_{\mathrm{Check}}$ requires looping over all
    possible input strings and checking which ones are accepting or rejecting.
    This requires computing the transition function of~$\PACK$, which takes time
    $O(1)$.
  \end{proof}

  \paragraph{Cook-Levin for $\decider$.}
  We now perform the Cook-Levin construction for the decider $\decider$.
  Although it may be a significantly more complex machine, its Cook-Levin
  construction is very similar to that for $\PACK$.
  We will, therefore, be brief and only highlight the necessary changes for
  $\decider$.

  First, we define an execution tableau for the given $3$-tape decider
  $\decider$ up to time~$T$, restricted to tape positions $\{0,1,\ldots,T+1\}$.
  For each time $t \in \{1, \ldots, T\}$, tape $i \in \{1,2,3\}$, and tape
  position $j \in \{0, 1,\ldots, T+1\}$, introduce cell variables
  \begin{equation*}
    c^{\decider}_{t,i,j}=(c^{\decider}_{t,i,j,1},c^{\decider}_{t,i,j,2})\in\{0,1\}^2,
  \end{equation*}
  head variables $h^{\decider}_{t,i,j}\in\{0,1\}$, and state variables
  $s^{\decider}_{t}\in\{0,1\}^{\kappa}$ encoding the states of~$\decider$ (via
  $\mathrm{enc}_K$ from Definition ``Decider encodings'').
  Let $\mathcal{V}^{\decider}$ denote the collection of all tableau variables
  for $\decider$.
  We have
  \begin{equation*}
    \mathcal{V}^{\decider} = \{c^{\decider}_{t, i, j, k}\}_{t, i, j, k} \cup \{h^{\decider}_{t, i, j}\}_{t, i, j}
    \cup \{s^{\decider}_{t, k}\}_{t, k}.
  \end{equation*}
  In total, the number of variables in the execution tableau for $\decider$ is given by
  \begin{equation}\label{eq:tableau-variable-count}
    |\mathcal{V}^{\decider}|
    = O\big(T^2 + T \cdot \log(|K|)\big)
    = O\big(T^2 + T \cdot \log(|\decider|)\big)\;.
  \end{equation}

  \begin{definition}
    The \emph{boundary formula} $\varphi^{\decider}_{\mathrm{Boundary}}$ is the 3SAT
    formula on the variables $\mathcal{V}^{\decider}$ described as follows.
    Let $o = \mathrm{enc}_K({\mathsf{start}}) \in \{0, 1\}^\kappa$, where
    ${\mathsf{start}} \in K$ is the start state of~$\decider$.
    For the $t= 1$ boundary, $\varphi^{\decider}_{\mathrm{Boundary}}$ contains the
    following set of clauses.
    \begin{alignat}{2}
      & \text{Indices} & \qquad & \text{Clauses}\nonumber\\
      & i \in \{1, 2, 3\} && h^{\decider}_{1, i, 1}
      \label{eq:tape-heads-start}\\
      & i \in \{1, 2, 3\}, j \neq 1 && \neg h^{\decider}_{1, i, j}
      \label{eq:no-more-tape-heads}\\
      & k \in \{1, \ldots, \kappa\} && \bigl(s^{\decider}_{1, k}\bigr)^{o_k}
      \label{eq:start-state}\\
      & i \in \{1, 2, 3\},~ j \in \{1, \ldots, T\} &&
      \neg c^{\decider}_{1, i, j, 1} \lor \neg c^{\decider}_{1, i, j, 2}
      \label{eq:not-square}\\
      & i \in \{2, 3 \},~ j \in \{1, \ldots, T\}
      && c^{\decider}_{1, i, j, 1} \text{ and } \neg c^{\decider}_{1, i, j, 2}
      \label{eq:all-empty}
    \end{alignat}
    Rows~\eqref{eq:tape-heads-start} and~\eqref{eq:no-more-tape-heads} ensure
    that the head is at the first cell before the computation begins and
    row~\eqref{eq:start-state} checks the correct initial state.
    Rows~\eqref{eq:not-square} and~\eqref{eq:all-empty} make sure that no square
    symbol ``$\square$'' is appearing for cells indexed by $j\in {1, \ldots, T}$ and that the
    working and write-only tape is initialized with blanks, respectively.
    Note that no further restriction is necessary to be placed on the content of
    the input tape at time $t=1$; it will be determined by the output of $\PACK$
    and the linking formula below.

    For the $j \in \{0, T+1\}$, $\varphi^{\decider}_{\mathrm{Boundary}}$
    contains the following set of clauses to mark the boundary of the tableau.
    \begin{alignat}{2}
      & \text{Indices} & \qquad & \text{Clauses}\nonumber\\
      & t \in \{1, \ldots, T\}, i \in \{1, 2, 3\},
      j \in \{0, T+1\}, k \in \{1, 2\}
      && c^{\decider}_{t, i, j, k} \label{eq:square}
    \end{alignat}
    We also need to check that the machine's state at time $T$ is accepting,
    which can be expressed as clauses in
    $\varphi^{\decider}_{\mathrm{Boundary}}$ as follows.
    \begin{alignat}{2}
      & \text{Indices} & \qquad & \text{Clauses}\nonumber\\
      & k \in \{1, 2, \ldots, \kappa\}
      && \bigl(s^{\decider}_{T, k}\bigr)^{o'_k}\label{eq:accepts-at-time-T}
    \end{alignat}
    where $o' = \mathrm{enc}_{K}(\mathsf{accept})$ for $\mathsf{accept} \in K$
    representing the only accept state of $\decider$.
    \znote{I think we also need to check that the final configuration of the tm
      accepts. so I added the above.}
  \end{definition}

  Similar to the construction of $\circuit^{\PK}_{\mathrm{Check}}$ and
  $\circuit^{\PK}_{\mathrm{Check}, t, j_{1}, \ldots, j_{7}}$, we can construct
  circuits $\circuit^{\decider}_{\mathrm{Check}}$ and
  $\circuit^{\decider}_{\mathrm{Check}, t, j_{1}, j_{2}, j_{3}}$ that check the
  propagation of computation locally using the transition function of machine
  $\decider$.
  They satisfy the following claim.

  \begin{lemma}\label{lem:check-size}
    A circuit~$\circuit^{\decider}_{\mathrm{Check}}$ of size at
    most~$\poly(|\decider|)$ can be computed in time~$\poly(|\decider|)$.
  \end{lemma}
	
  \begin{proof}
    The circuit has a total of $3 \cdot 12 + 2\cdot \kappa = 36 + 2\kappa$
    Boolean input bits, and, therefore, $2^{36} \cdot 4^{\kappa} = O(|K|^2)$
    possible input strings.
    For each fixed input string, the circuit
    $\circuit^{\decider}_{\mathrm{Check}}$ verifies whether the actual input
    matches this fixed string, which can be implemented using $O(\kappa)$ gates.
    The circuit then accepts if and only if the fixed input is accepting; that
    is, the correct output for each fixed input is hard-coded into the circuit.
    This construction requires a total of $O(|K|^2 \cdot \kappa)$ gates.

    Constructing $\circuit^{\decider}_{\mathrm{Check}}$ involves iterating over
    all possible input strings and determining whether each should be accepted
    or rejected.
    This requires evaluating the transition function of $\decider$, which takes
    time $O(|\decider|)$, since the canonical encoding $\ol{\decider}$
    explicitly contains a lookup table for the transition function.
    The proposition follows from the observation that $|K| \leq |\decider|$.
\end{proof}

\begin{lemma}\label{lem:correct-tableau}
  \leavevmode\par Suppose that the execution tableau for $\PACK$ satisfies
  $\varphi^{\PK}_{\mathrm{Boundary}}$ and the circuit
  $\circuit^{\PK}_{\mathrm{Check}, t, j_1, \ldots, j_7}$ for each
  $t \in \{1,\allowbreak  \ldots,\allowbreak  S-1\}$ and $j_1,\allowbreak  \ldots,\allowbreak  j_7 \in \{1,\allowbreak  \ldots,\allowbreak  S\}$.
  Suppose also that the tableau for $\decider$ satisfies
  $\varphi^{\decider}_{\mathrm{Boundary}}$ and the circuit
  $\circuit^{\decider}_{\mathrm{Check}, t, j_1, \ldots, j_3}$ for each
  $t \in \{1,\allowbreak  \ldots,\allowbreak  T-1\}$ and $j_1,\allowbreak  j_2,\allowbreak  j_3 \in \{1,\allowbreak  \ldots,\allowbreak  T\}$.
  Then the execution tableau  for $\PACK$ correctly encodes the execution of
  $\PACK$ on inputs $n,\allowbreak  x,\allowbreak  y,\allowbreak  a,\allowbreak  b$ when run for time~$S$.
  Similarly, the execution tableau for $\decider$ encodes the execution of
  $\decider$ for time~$T$.
\end{lemma}

\begin{proof}
  We show this for the tableau of $\PACK$; that for $\decider$ follows similarly.
  We will show for each time $t \in \{1, \ldots S\}$ that the
  variables in the execution tableau corresponding to time~$t$ correctly encode
  the state of~$\PACK$ and the contents of the seven tapes at time~$t$.
  The proof is by induction on~$t$.
  The base case of~$t=1$ follows from the tableau satisfying
  $\varphi^{\PK}_{\mathrm{Boundary}}$.

  Next we perform the induction step.
  Assuming the statement holds for time $t \in \{1, \ldots, S-1\}$, we will show
  it holds for time~$t+1$ as well.
  Let $i^* \in \{1, \ldots, 7\}$ be a tape, and consider a tape position
  $j_{i^*} \in \{1, \ldots, S\}$.
  We will show that the variables correctly encode the contents of this tape
  position at time $t+1$.
  Suppose one of the tape positions $j_{i^*}-1$, $j_{i^*}$, or $j_{i^*}+1$ at
  time $t$ has a tape head.
  For each of the other tapes $i \neq i^*$, we select a tape position $j_{i}$
  such that either $j_{i}-1$, $j_{i}$, or $j_{i}+1$ has a tape head at time $t$.
  (These positions are guaranteed to exist since each tape has exactly one tape
  head.)
  
  By the induction hypothesis, for each tape $i \in \{1, \ldots, 7\}$ the
  variables corresponding to tape cells $j_i-1$, $j_i$, and $j_i+1$ correctly
  encode the contents of these cells at time~$t$.
  By assumption, $\circuit^{\PK}_{\mathrm{Check}, t, j_1, \ldots, j_7}$ evaluates
  to~$1$.
  In this case, it calculates the transition function of~$\PACK$ to compute
  the contents of the tape cells $j_1, \ldots, j_7$ at time~$t+1$ and checks
  that the corresponding variables encode these contents.
  As a result, the tape position $j_{i^*}$ on tape~$i^*$ is correctly encoded.
  In addition, it computes the state of~$\PACK$ at time~$t+1$ and checks that
  the corresponding variables encode this state.
  This completes the induction step.
  The case when none of the tape positions $j_{i^*}-1$, $j_{i^*}$, and
  $j_{i^*}+1$ at time $t$ contain a tape head follows similarly.
  Finally, the variables for all tape positions $j \in \{0, S+1\}$ are correctly
  encoded due to $\varphi^{\PK}_{\mathrm{Boundary}}$ being satisfied.
\end{proof}

Lemma~\ref{lem:correct-tableau} gives a set of constraints that ensure the
execution tableau properly encodes the execution of~$\PACK$ and $\decider$.
Our next step will be to convert these constraints into a single 3SAT formula.
This entails transforming each circuit
$\circuit^{\PK}_{\mathrm{Check}, t, i_1, \ldots, i_7}$ and
$\circuit^{\decider}_{\mathrm{Check}, t, j_{1}, j_{2}, j_{3}}$ into a 3SAT
formula.
We do so using the following reduction.

\begin{lemma}[Circuit-to-3SAT]\label{lem:circuit-to-sat-reduction}
  There is an algorithm which, on input a size-$r$ circuit~$\circuit$ on
  variables $x \in \{0, 1\}^n$, runs in time $\poly(r)$ and outputs a 3SAT
  formula $\varphi$ on variables $x \in \{0, 1\}^n$ and $y \in \{0, 1\}^r$ with
  $O(r)$ clauses such that for all $x$, $\circuit(x) = 1$ if and only if there
  exists a $y$ such that $x$ and~$y$ satisfy $\varphi$.
\end{lemma}
\begin{proof}
  This is the textbook circuit-to-3SAT reduction.
  For each gate $i \in \{1, \ldots, r\}$, the algorithm introduces a variable
  $y_i \in \{0, 1\}$, so that the total collection of variables is
  $\mathcal{V}_{\mathrm{total}} = \{x_i\}_{i \in \{1, \ldots, n\}} \cup
  \{y_i\}_{i \in \{1, \ldots, r\}}$.
  Consider gate~$i \in \{1, \ldots, r\}$, and let $z_1, z_2 \in
  \mathcal{V}_{\mathrm{total}}$ be the pair of variables feeding into it.
  If gate~$i$ is an AND gate, then~$\varphi$ includes the constraints
  \begin{equation}\label{eq:AND-gate}
    (\neg z_1 \lor \neg z_2 \lor y_i)\;,
    \quad
    (z_1 \lor  z_2 \lor \neg y_i)\;,
    \quad
    (z_1 \lor  \neg z_2 \lor \neg y_i)\;,
    \quad
    (\neg z_1 \lor  z_2 \lor \neg y_i)\;.
  \end{equation}
  These constraints are satisfied if and only if $y_i = z_1 \land z_2$.
  The case of gate~$i$ being an OR gate follows similarly.
  Finally, $\varphi$ includes the constraint $(y_{i^*})$, where $i^* \in \{1,
  \ldots, r\}$ is the output gate.
  It follows that $x, y$ satisfy $\varphi$ if and only if $\circuit(x) = 1$ and
  for each $i \in \{1, \ldots, r\}$, $y_i$ is the value computed by gate~$i$ in
  circuit~$\circuit$ on input~$x$.
  In total, $\varphi$ has $4r+1$ clauses and is computable in time $\poly(r)$.
\end{proof}

We now apply the algorithm from Lemma~\ref{lem:circuit-to-sat-reduction} to
$\circuit^{\PK}_{\mathrm{Check}}$ and $\circuit^{\decider}_{\mathrm{Check}}$,
which by Lemmas~\ref{lem:pack-check-size} and~\ref{lem:check-size} have size
$O(1)$ and $\poly(|\decider|)$, respectively.
Let $r$ be the maximum of the two sizes.
This produces a 3SAT formula $\varphi^{\PK}_{\mathrm{Check}}$ on the variables
in $\circuit^{\PK}_{\mathrm{Check}}$ plus auxiliary variables $a^{\PK}_k$ for
$k \in \{1, \ldots, r\}$ added by the reduction.
Now, for each $t \in \{1, \ldots, S-1\}$ and
$j_1, \ldots, j_7 \in \{1, \ldots, S\}$, we define a 3SAT formula
$\varphi^{\PK}_{\mathrm{Check}, t, j_1, \ldots, j_7}$ analogously to
$\circuit^{\PK}_{\mathrm{Check}, t, j_1, \ldots, j_7}$.
We begin by associating those variables in $\varphi_{\mathrm{Check}}$ which come
from $\circuit^{\PK}_{\mathrm{Check}}$'s inputs with the variables
in~$\mathcal{V}^{\PK}$ as in Definition~\ref{def:local-check-circuit}.
Next, for each $k \in \{1, \ldots, r\}$, we introduce a new variable
$a^{\PK}_{t, j_1, \ldots, j_7, k} \in \{0, 1\}$ and associate it with the variable
$a^{\PK}_k$.
This defines $\varphi^{\PK}_{\mathrm{Check}, t, j_1, \ldots, j_7}$
and we define $\varphi^{\decider}_{\mathrm{Check}, t, j_{1}, j_{2}, j_{3}}$ similarly
using auxiliary variables $a^{\decider}_{t, j_1, j_2, j_3}$.

\paragraph{Linking $\PACK$ and $\decider$.} 
To relate the output of $\PACK$ and the input of $\decider$, we introduce a formula
$\varphi_{\mathrm{Link}}$ to be the conjunction, over all
$j\in\{1,\ldots,T\}$ and $k\in\{1,2\}$, of the two-literal clauses enforcing
bitwise equality
\begin{equation}\label{eq:linking}
  \neg c^{\PK}_{S,7,j,k} \lor c^{\decider}_{1,1,j,k} \quad \text{and} \quad
  c^{\PK}_{S,7,j,k} \lor \neg c^{\decider}_{1,1,j,k}.
\end{equation}
(Note that these constraints can easily be turned into a 3SAT formula by e.g., rewriting $x\lor y$ as $x\lor y\lor y$.)

In summary, the final 3SAT instance produced by the Cook-Levin reduction is
\begin{equation}\label{eq:cook-levin-3sat}
\begin{aligned}
&\varphi:= \varphi^{\PK}_{\mathrm{Boundary}} \land
  \varphi^{\decider}_{\mathrm{Boundary}} \land\\
 &\bigg(\bigwedge_{t, j_1, \ldots, j_7}
  \varphi^{\PK}_{\mathrm{Check}, t, j_1, \ldots, j_7}\bigg) \land\\
 &\bigg(\bigwedge_{t, j_1, j_2, j_3}
  \varphi^{\decider}_{\mathrm{Check}, t, j_1, j_2, j_3}\bigg) \land
  \varphi_{\mathrm{Link}}\;.
\end{aligned}
\end{equation}
Note that it is a CNF formula on variable set
$\mathcal{V} = \mathcal{V}^{\PK} \cup \mathcal{V}^{\decider}$.
By Lemma~\ref{lem:circuit-to-sat-reduction} and the definition of
$\varphi_{\mathrm{Link}}$, the execution tableau properly encode the execution
of $\decider$ when given a five-tuple $(n,x,y,a,b)$ as the input for time $T$ if
and only if there exists a setting to the auxiliary variables satisfying the
3SAT formula $\varphi$.
In total, $\varphi$ contains
\begin{equation}\label{eq:variable-count}
|\mathcal{V}^{\PK}| + |\mathcal{V}^{\decider}| + O(T^8) \cdot r
\leq O\big(T^2 + T \cdot \log(|\decider|)
    + T^8\cdot \poly(|\decider|)\big)
 = \poly(T, |\decider|)
\end{equation}
variables.
The third term on the left-hand side of Equation~\eqref{eq:variable-count}
corresponds to the auxiliary variables in each of the $O(T^8)$ copies of
$\varphi^{\PK}_{\mathrm{Check}}$ and $\varphi^{\decider}_{\mathrm{Check}}$.
The inequality follows by Equation~\eqref{eq:tableau-variable-count} and the
fact that $r \leq \poly(|\decider|)$.

Our next task is to represent the 3SAT formula~$\varphi$ succinctly.
To do this, we will provide a circuit~$\circuit$ which succinctly describes a
3SAT formula~$\varphi_{\mathrm{3SAT}}$ which, while not literally equal
to~$\varphi$, will be ``isomorphic'' to it.
Essentially, the issue is that our definition of succinctly represented 3SAT
formulas requires the set of variable labels to be the set of all bitstrings of
some length, which means that the number of variables must be a power of $2$.
Since $\varphi$ need not have a number of variables that is a power of 2, our
circuit $\circuit$ will represent a formula that is essentially $\varphi$
``padded'' by extra dummy variables that do not appear in any clause.
Specifically, $\varphi_{\mathrm{3SAT}}$ will contain at least as many variables
as $\varphi$, with $2^m$ variables labeled by bitstrings in $\{0,1\}^m$ for some
parameter $m$, and will contain exactly the same number of clauses as $\varphi$.
Every variable in~$\varphi$ will correspond to a distinct variable
in~$\varphi_{\mathrm{3SAT}}$, and every clause in $\varphi$ will correspond to a
clause in $\varphi_{\mathrm{3SAT}}$ according to the correspondence on
variables.
All the remaining variables in~$\varphi_{\mathrm{3SAT}}$ will be ``dummy''
variables that do not appear in any clauses.
This circuit is constructed as follows.

\begin{definition}
  In this definition we construct the circuit~$\circuit$.
  It has three inputs~$z_1, z_2, z_3$ of length~$m$, which we specify below in
  Equation~\eqref{eq:length-of-m}, and three inputs~$o_1, o_2, o_3 \in \{0,
  1\}$.
	For each~$\nu \in \{1, 2, 3\}$, each $z_\nu$ is supposed to specify a variable
  in $\varphi$ according to a format we will now specify.
	If~$z_\nu$ is not properly formatted, then it does not correspond to a
  variable in~$\varphi$; if any of~$z_1, z_2, z_3$ is not properly formatted,
  then~$\circuit$ automatically outputs~$0$.
	Below, we will often write substrings of the $z_\nu$'s as though they are
  integers from some specified range, i.e.,\ $a \in \{b, \ldots, c\}$.
	This means that $a$ is represented as a binary string of length $\lceil
  \log(c+1) \rceil$, which is to be interpreted as the binary encoding of an
  integer between~$b$ and~$c$.

  The input $z_\nu$ is formatted as a string
  $(\omega, \alpha, \beta_1, \ldots, \beta_8)$.
	The first substring $\omega$ has length
  $m - (|\alpha| + |\beta_1| + \cdots + |\beta_8|)$ bits and is formatted to be
  the all-zeroes string.
	Its purpose is to pad the inputs to have the length~$m$ we specify below.
	Next, $\alpha$ is formatted as an integer $\alpha \in \{1, 2, \ldots, 8\}$.
	The variable encoded by~$z_\nu$ is specified by $\beta_\alpha$, and the other
  three $\beta$'s should be the all-zeros string of the appropriate length
  (matching the maximum possible binary length of a nonzero value for the
  corresponding $\beta$, as indicated below).
	We now specify the encoding of $\beta_\alpha$, conditioned on the value
  of~$\alpha$.
	\begin{enumerate}
	\item $\beta_1$ is formatted as $(t, i, j, k)$, where\label{item:pack-a=1}
		\begin{equation*}
      t \in \{1, \ldots, S\}, \quad i \in \{1, \ldots, 7\},
      \quad j \in \{0, 1, \ldots, S+1\}, \quad k \in \{1, 2\}.
		\end{equation*}
		This corresponds to the variable $c^{\PK}_{t, i, j, k}$.
	\item $\beta_2$ is formatted as $(t, i, j)$, where\label{item:pack-a=2}
		\begin{equation*}
      t \in \{1, \ldots, S\}, \quad i \in \{1, \ldots, 7\},
      \quad j \in \{0, 1, \ldots, S+1\}.
		\end{equation*}
		This corresponds to the variable $h^{\PK}_{t, i, j}$.
	\item $\beta_3$ is formatted as $(t, k)$, where\label{item:pack-a=3}
		\begin{equation*}
      t \in \{1, \ldots, S\}, \quad k \in \{1, \ldots, \kappa\}.
		\end{equation*}
		This corresponds to the variable $s^{\PK}_{t, k}$.
	\item $\beta_4$ is formatted as $(t, j_1, \ldots, j_7, k)$, where\label{item:pack-a=4}
		\begin{equation*}
      t \in \{1, \ldots, S-1\}, \quad j_1, \ldots, j_7 \in \{1, 2, \ldots, S\},
      \quad k \in \{1, \ldots, r\}.
		\end{equation*}
		This corresponds to the variable $a^{\PK}_{t, j_1, \ldots, j_7, k}$.
	\item $\beta_5$ is formatted as $(t, i, j, k)$, where\label{item:a=1}
		\begin{equation*}
      t \in \{1, \ldots, T\}, \quad i \in \{1, 2, 3\},
      \quad j \in \{0, 1, \ldots, T+1\}, \quad k \in \{1, 2\}.
		\end{equation*}
		This corresponds to the variable $c^{\decider}_{t, i, j, k}$.
	\item $\beta_6$ is formatted as $(t, i, j)$, where\label{item:a=2}
		\begin{equation*}
      t \in \{1, \ldots, T\}, \quad i \in \{1, 2, 3\},
      \quad j \in \{0, 1, \ldots, T+1\}.
		\end{equation*}
		This corresponds to the variable $h^{\decider}_{t, i, j}$.
	\item $\beta_7$ is formatted as $(t, k)$, where\label{item:a=3}
		\begin{equation*}
      t \in \{1, \ldots, T\}, \quad k \in \{1, 2, \kappa\}.
		\end{equation*}
		This corresponds to the variable $s^{\decider}_{t, k}$.
	\item $\beta_8$ is formatted as $(t, j_1, j_2, j_3, k)$, where\label{item:a=4}
		\begin{equation*}
      t \in \{1, \ldots, T-1\}, \quad j_1, \ldots, j_7 \in \{1, 2, \ldots, T\},
      \quad k \in \{1, \ldots, r\}.
		\end{equation*}
		This corresponds to the variable $a^{\decider}_{t, j_1, \ldots, j_7, k}$.
	\end{enumerate}
	In total, the length of these substrings is
	\begin{equation*}
    |\alpha| + |\beta_1| + \cdots + |\beta_8| = O(\log(T) + \log(\kappa) +
    \log(r)) = O(\log(T) + \log(|\decider|))\;.
	\end{equation*}
	As a result, because $|\decider| \leq \sigma$, the length of the
  ``padding''~$\omega$ can be chosen so that each~$z_i$ has length at most
	\begin{equation}\label{eq:length-of-m}
    m = O(\log(T) + \log(\sigma))
	\end{equation}
  and, furthermore, $m$ is a power of $2$. 
	Checking that each $z_\nu$ is properly formatted can be done by checking that
  certain substrings of $z_1$, $z_2$, and $z_3$ encode integers which fall
  within specified ranges.
	This can be done using $O(m)$ gates.

  Having specified the inputs, we can now specify the execution of the circuit,
  and we may assume that $z_1, z_2, z_3$ are properly formatted.
  Implementing the clauses from $\varphi^{\PK}_{\mathrm{Boundary}}$ and $\varphi^{\decider}_{\mathrm{Boundary}}$  is simple; we
  specify how to implement the clause from Equation~\eqref{eq:tape-heads-start}.
  \begin{itemize}
  \item Suppose for input $z_1$, $\alpha = 6$.
    Then $\beta_6$ can be parsed as $(t, i, j)$.
    The circuit accepts if $t=j = 1$ and $o_1 = 1$, regardless of~$z_2$ and~$z_3$.
    This ensures that $\varphi_{\mathrm{3SAT}}$ includes $x_{z_1} \lor
    x_{z_2}^{o_2} \lor x_{z_3}^{o_3}$ for any $z_2, z_3, o_2, o_3$, which is
    equivalent to including the arity-one clause $x_{z_1}$.
  \end{itemize}
  This can be implemented with $O(m)$ gates.
  Similar arguments can be used to implement the clauses from
  Equations~\eqref{eq:no-more-tape-heads}-\eqref{eq:accepts-at-time-T} and~\eqref{eq:linking} using $O(m)$ gates
  apiece.

  Implementing the clauses from the formulas
  \[\textstyle \varphi^{\PK}_{\mathrm{Check}, t, j_1, \ldots, j_7}\] and
  $\varphi^{\decider}_{\mathrm{Check}, t, j_1, j_2, j_3}$ is more challenging.
  From Equation~\eqref{eq:AND-gate}, we can see that any constraint in
  $\varphi^{\decider}_{\mathrm{Check}, t, j_1, j_2, j_3}$ always involves a
  variable of the form $a^{\decider}_{t, j_1, j_2, j_3, k}$ for some
  $k \in \{1,\allowbreak  \ldots,\allowbreak  r\}$.
  \begin{enumerate}
  \item First check if one of its inputs $z_1, z_2, z_3$ corresponds to such a
    variable.
    This can be done with $O(1)$ gates simply by checking if for any of the
    $z_i$'s, $\alpha = 8$.
    If so, this specifies the values of $t, j_1, j_2, j_3$.
  \item Each variable in $\varphi^{\decider}_{\mathrm{Check}, t, j_1, j_2, j_3}$ is
    associated with a variable in $\varphi^{\decider}_{\mathrm{Check}}$.
    The circuit checks if all the $z_i$'s are contained in
    $\varphi^{\decider}_{\mathrm{Check}, t, j_1, j_2, j_3}$ and then it computes which
    variable in $\varphi^{\decider}_{\mathrm{Check}}$ they are associated with.
    We include below the example of checking whether~$z_1$ is associated with
    the variable $c^{\decider}_{\mathrm{Check},1, 1, 0, 0}$.
    \begin{itemize}
    \item The circuit~$\circuit$ first computes $t+1$, which takes $O(m)$ gates.
      It then looks at~$z_1$ and checks if~$\alpha=5$.
      If so, then $\beta_5 = (t', i', j', k')$, and so it tests the equalities
      $t' = t+1$, $i' = 1$, $j' = j_1$, and $k' = 0$, each of which takes $O(m)$
      gates to test.
      If so, then~$z_1$ is associated with $c^{\decider}_{\mathrm{Check},1, 1, 0, 0}$.
    \end{itemize}
    There are $\poly(|\decider|)$ variables in $\varphi^{\decider}_{\mathrm{Check}}$ and,
    for each variable $z_i$, it takes $O(m)$ gates to determine whether $z_i$ is
    associated with this variable.
    As a result, because $|\decider| \leq \sigma$, this takes $\poly(\sigma,
    \log(T))$ gates to compute.
  \item Whether $z_1$, $z_2$, and $z_3$ share a clause in
    $\varphi^{\decider}_{\mathrm{Check}, t, j_1, \ldots, j_7}$ depends only on which
    variables in $\varphi^{\decider}_{\mathrm{Check}}$ they are associated with.
    As a result, after having computed these variables, the algorithm can
    hard-code whether the circuit~$\circuit$ should accept.
  \end{enumerate}

  Finally, we show how to implement the clauses in $\varphi_{\mathrm{Link}}$.
  If for $z_{1}$ the $\alpha$ component equals $1$ and $\beta_{1} = (S,7,j,k)$
  and if for $z_{2}$ the $\alpha$ component equals $5$ and
  $\beta_{5} = (1,1,j,k)$, $\circuit$ accepts if $o_{1} \oplus o_{2} =1$.
  This completes the description of~$\circuit$.
  In total, it contains $\poly(\sigma, \log(T))$ gates.
  Computing $\circuit$ first requires computing $\varphi_{\mathrm{Check}}$,
  which takes time $\poly(|\decider|) \leq \poly(\sigma)$ by
  Lemmas~\ref{lem:check-size} and~\ref{lem:circuit-to-sat-reduction}.
  After that, the steps outlined above for construction~$\circuit$ take
  time~$\poly(\sigma, \log(T))$.
\end{definition}

The circuit~$\circuit$ succinctly describes~$\varphi$ in the loose sense
described above.
Recall that~$\varphi$ accepts if and only if it encodes the execution
of~$\decider$ up to time~$T$.
We now modify~$\circuit$ to hard code~$n$, $x$, and~$y$ onto~$\PACK$'s 
input tapes. 
We demonstrate how to do so by hard coding~$n$ as an example.
\begin{itemize}
\item The input~$n$ is described by some string $\nu$ of length~$\ell =
  O(\log(n))$.
	We would like to hard-code the string $(\nu, \sqcup^{T - \ell})$ into the
  first tape of~$\PACK$. ( \hnote{june 14: added $\log n$ lower bound} This is where we use the condition that $T \geq 2 \lceil \log n \rceil$, as it takes $2 \lceil \log n \rceil$ bits to encode the integer $n$.)
	To do this, we first check if~$z_1$ corresponds to the variable $c^\PK_{t, i, j,
    k}$ for some values of~$t, i, j, k$.
	Then, we check if $t = 1$, the time where the inputs appear on the tapes, and
  $i = 1$, corresponding to the first tape.
	If so, we branch on whether $j \leq \ell$.
	If it is, then if $k = 1$ the circuit accepts if $o_1 = 0$, and if $k = 2$ the
  circuit accepts if $o_1 = \nu_j$.
	Otherwise, if $j > \ell$, then if $k = 1$ the circuit accepts if $o_1 = 1$,
  and if $k = 2$ the circuit accepts if $o_1 = 0$.
	This can be done with $\poly(\log(n), \log(T))$ gates.
\end{itemize}
This modifies~$\varphi$ so that it only accepts if~$n$ is written on its first
tape at input.
We can similarly hard-code~$x$ and~$y$ onto the second and third input tapes. 
 In total, this takes $\poly(\log(n), |x|, |y|, \log(T), \sigma)$ gates, which is at most
\[s(n,T,\qlen,\sigma)=\poly(\log(n), \qlen, \log(T), \sigma)\]
 because~$x$ and~$y$ have length at
most~$\qlen$.

It remains to ensure that the variables corresponding to the~$4$th and~$5$th input tapes of $\PACK$ at
time~$t = 1$ are the lexicographically-first named variables in~$\varphi$.
However, this is simple and can be done using $\poly(\log(n), \qlen, \log(T),
\sigma)$ gates.
This concludes the construction.
\end{proof}

\subsection{A succinct 5SAT description for deciders}
\label{sec:succinct-deciders}

Proposition~\ref{prop:standard-succinct-sat} allows us to convert any decider
$\decider$ and inputs $n, x, y$ into a 3SAT formula $\varphi_{\mathrm{3SAT}}$,
succinctly described by a circuit~$\circuit$, which represents it.
However, there are two undesirable properties of this construction, which we
describe below.
\begin{enumerate}
\item First, evaluating any clause $(w_{i_1}^{o_{i_1}} \lor
  w_{i_2}^{o_{i_2}}\lor w_{i_3}^{o_{i_3}})$ of $\varphi_{\mathrm{3SAT}}$
  requires evaluating the \emph{same} assignment~$w$ at three separate points.
	While this is fine when the assignment~$w$ is provided in full to the
  verifier, it can be a problem when the verifier is only able to query the
  points in~$w$ by interacting with a prover.
	In this case, the verifier might send the prover the values $i_1, i_2, i_3$,
  who responds with three bits $b_1, b_2, b_3 \in \{0, 1\}$, purported to be the
  values $w_{i_1}, w_{i_2}, w_{i_3}$ for some assignment $w \in \{0, 1\}^M$.
	As it will turn out, in the answer reduced protocol below, the verifier will
  actually be able to force the prover to reply using \emph{three} different
  assignments.
	In other words, the prover will have three assignments $w_1, w_2, w_3 \in \{0,
  1\}^M$ such that, for any $i_1, i_2, i_3$ provided to it by the verifier, it
  will respond with $b_1 = w_{1, i_1}, b_2 = w_{2, i_2}, b_3 = w_{3, i_3}$.
	However, even given this there is no guarantee that the three assignments are
  the \emph{same} assignments (i.e.,\ that $w_1 = w_2 = w_3$).
	In the work of~\cite{NW19}, this was accomplished by an additional subroutine
  called the \emph{intersecting lines test}, which would enforce consistency
  between $w_1$, $w_2$, and $w_3$.
	In this work, on the hand, we would like to relax the assumption that $w_1$,
  $w_2$, and $w_3$ must be the same.
	This will allow us to not use the intersecting lines test, simplifying the
  answer reduction protocol.
	(In fact, the answer reduced verifier will not be querying the assignments
  directly, but rather low-degree encodings of these assignments.)
\item Second, we are guaranteed that $w = (a, b, c)$ satisfies
  $\varphi_{\mathrm{3SAT}}$ if and only if $\decider$ accepts $(n, x, y, a, b)$.
	However, it is inconvenient that~$a$ and~$b$ are contained as substrings
  of~$w$.
	To see why, recall from Section~\ref{sec:oracle} that the oracularized
  verifier sometimes gives one prover a pair of questions $(x, y)$ and another
  prover just one of the questions---say, $x$.
	The answer reduced verifier will sample its questions similarly; as for its
  answers, it might expect the first prover to respond with a string $w = (a, b,
  c)$ that satisfies $\varphi_{\mathrm{3SAT}}$ and the second prover to respond
  with a string $a'$ such that $a = a'$.
	Verifying that $a = a'$ requires the verifier to sample a uniformly random
  point from~$w$, restricted to the coordinates in~$a$.
	As it turns out, generating a uniform point from a substring is extremely
  cumbersome, though not impossible, to do when the verifier's questions are
  expected to be sampled from conditional linear functions.
	To remove this complication, though, we would instead prefer if the provers'
  answers were formatted in a way that gave the verifier direct access to the
  strings~$a$ and~$b$.
\end{enumerate}

In the remainder of this section, we will show how to modify the Succinct-3SAT
circuits produced by Proposition~\ref{prop:standard-succinct-sat} in order to
ameliorate these two difficulties.
Doing so entails modifying the $\varphi_{\mathrm{3SAT}}$ formula from
Proposition~\ref{prop:standard-succinct-sat} to produce a \emph{5SAT formula}
$\varphi_{\mathrm{5SAT}}$.
The clauses of $\varphi_{\mathrm{5SAT}}$ will be of the form
\begin{equation*}
  a_{i_1}^{o_1} \lor b_{i_2}^{o_2} \lor w_{1, i_3}^{o_3}
  \lor w_{2, i_4}^{o_4} \lor w_{3, i_5}^{o_5}\;,
\end{equation*}
where $a$, $b$, $w_1$, $w_2$, and $w_3$ are five separate assignments which are
not assumed to be equal.
The guarantee is that $(a, b, w_1, w_2, w_3)$ satisfies
$\varphi_{\mathrm{5SAT}}$ if and only if $\decider$ accepts $(n, x, y, a, b)$.
This addresses the two concerns from above: each clause is totally
\emph{decoupled}, meaning it samples~$5$ variables from~$5$ different
assignments, and so no consistency check must be performed between the
assignments.
In addition, the first two strings exactly correspond to~$a$ and~$b$, addressing
the second item.

We now formally define decoupled 5SAT instances and how they succinctly
represent bounded deciders.
Following that, we show how the succinct 3SAT instance $\varphi_{\mathrm{3SAT}}$
produced by Proposition~\ref{prop:standard-succinct-sat} can be modified to
produce a succinct 5SAT instance $\varphi_{\mathrm{5SAT}}$ which represents
$\decider$.

\begin{definition}[Decoupled 5SAT and its succinct descriptions]
  A \emph{block} of variables~$x_i$ is a tuple $x_i = (x_{i,0}, \ldots, x_{i,
    N_i-1})$.
  A formula $\varphi$ on~$5$ blocks $x_1, x_2, \ldots, x_5$ of variables is
  called a \emph{decoupled 5SAT formula} if every clause is of the form
  \begin{equation}\label{eq:5sat}
    x_{1,\, i_1}^{o_1}
    \lor x_{2,\, i_2}^{o_2}
    \lor x_{3,\, i_3}^{o_3}
    \lor x_{4,\, i_4}^{o_4}
    \lor x_{5,\, i_5}^{o_5}\;,
  \end{equation}
  for $i_j \in \{0, 1, \ldots, N_j-1\}$ and $o_1, \ldots, o_5 \in \{0, 1\}$.
  (Recall from Definition~\ref{def:succinct-formulas} that the notation $x^{o}$
  means $x$ if $o = 1$ and $\neg x$ if $o = 0$.)

  For each $i\in\{1, 2, \ldots, 5\}$, suppose each $N_i$ is a power of two, and write
  it as $N_i = 2^{n_i}$.
  Let $\circuit$ be a circuit with five inputs of length $n_1, n_2, \ldots, n_5$ and
  five single-bit inputs.
  Then $\circuit$ \emph{succinctly describes decoupled $\varphi$} if, for all
  $i_j \in \{0, 1, \ldots, N_j-1\}$ and $o_1, o_2, \ldots, o_5 \in \{0, 1\}$,
  \begin{equation}\label{eq:put-in-numbers}
    \circuit(i_1, i_2, \ldots, i_5,
    o_1, o_2, \ldots, o_5) = 1
  \end{equation}
  if and only if the clause in~\eqref{eq:5sat} is included in~$\varphi$.
  As in Definition~\ref{def:succinct-formulas}, we slightly abuse notation and
  use the convention that a number~$a$ between~$0$ and~$2^{n_i}-1$ is
  interpreted as its binary encoding $\binary{n_i}(a)$ when provided as input to
  a set of $n_i$ single-bit wires.
\end{definition}

\begin{definition}[Succinct descriptions for bounded deciders]
  Let $\decider$ be a decider.
  Fix an index $n \in \N$ and a time $T \in \N$.
  Let $L = 2^\ell$ be at least~$2T$.
  Let~$x$ and~$y$ be strings, $r\in\N$ and $R=2^r$.

  Consider a circuit~$\circuit$ with two inputs of length~$\ell$, three inputs
  of length~$r$, and $5$ single-bit inputs.
  Let $\varphi_C$ be the decoupled 5SAT instance with two blocks of variables of
  size~$L$ and three blocks of size~$R$ which~$\circuit$ succinctly describes.
  Then we say that $\circuit$ \emph{succinctly describes $\decider$ (on
    inputs~$n$, $x$, and~$y$ and time~$T$)} if, for all $a, b \in \{0, 1\}^{L}$,
  there exists $w_1, w_2, w_3 \in \{0, 1\}^{R}$ such that $a, b, w_1, w_2, w_3$
  satisfy $\varphi_{\circuit}$ if and only if there exist $a_{\mathrm{prefix}},
  b_{\mathrm{prefix}} \in \{0, 1\}^*$ of lengths~$\ell_a, \ell_b \leq T$,
  respectively, such that
  \begin{equation*}
    a = \mathrm{enc}_\Gamma(a_{\mathrm{prefix}}\,,\, \sqcup^{L/2 - \ell_a})
  \quad
  \text{and}
  \quad
  b = \mathrm{enc}_\Gamma(b_{\mathrm{prefix}}\,,\, \sqcup^{L/2 - \ell_b})
  \end{equation*}
  and $\decider$ accepts $(n, x, y, a_{\mathrm{prefix}}, b_{\mathrm{prefix}})$
  in time~$T$.
\end{definition}

In this definition of succinct descriptions, the answers~$a$ and~$b$ are
isolated, in that the first input of~$\circuit$ of length~$\ell$ indexes
into~$a$ and the second input of length~$\ell$ indexes into~$b$.
The next proposition shows how to construct such descriptions.

\begin{proposition}[Explicit succinct descriptions]
  \label{prop:explicit-succinct-deciders}
  There is a Turing machine $\succinctdecider$ with the following properties.
  Let $\decider$ be a decider, let $n$, $T$,~$\qlen$, and $\sigma$ be integers
  with $\max \{ \qlen, 2 \lceil \log n \rceil \} \leq T$  \hnote{june 14: added $\log n$ lower bound}, and $|\decider| \leq \sigma$, and let~$x$ and~$y$ be
  strings of length at most~$\qlen$.
  Then on input $(\sigma,\decider, n, T, \qlen, x, y)$, $\succinctdecider$
  outputs a circuit~$\circuit$ with two inputs of length~$\ell_0(T)$, three of
  length $r_0(T, \sigma)$, and five single-bit inputs which succinctly
  describes $\decider$ on inputs~$n$, $x$, and~$y$ and time~$T$.
  Moreover, the following hold.
  \begin{enumerate}
  \item $\ell_0(T) = \lceil \log(2T) \rceil$.
  \item $r_0(T,\sigma) = O(\log(T)+\log(\sigma))$,
  \item $\circuit$ has at most $s_0(n,T,\qlen,\sigma) = \poly(\log(T), \log(n),
    \qlen,\sigma)$ gates,
  \item $\succinctdecider$ runs in time $\poly(\log(T),\allowbreak  \log(n),\allowbreak 
    \qlen,\allowbreak \sigma)$, and the parameters $\ell_0,\allowbreak  r_0,\allowbreak  s_0$ can be computed
    from $n,\allowbreak  T,\allowbreak  \qlen,\allowbreak \sigma$ in time $\poly(\log(T),\allowbreak  \log(n),\allowbreak 
    \log(\qlen),\allowbreak \log(\sigma))$.
  \end{enumerate}
\end{proposition}

\begin{proof}
  The Turing machine $\succinctdecider$ begins by running the algorithm in
  Proposition~\ref{prop:standard-succinct-sat} on input
  $(\sigma,\decider,n,T,\qlen,x,y)$ to produce a
  circuit~$\circuit_{\mathrm{3SAT}}$ on~$3r_0+3$ inputs, where $r_0 =
  O(\log(T)+\log(\sigma))$ is the parameter $m$ from the proposition.
  Set $\ell_0 = \lceil \log(2T) \rceil$, $L = 2^{\ell_0}$ and $R = 2^{r_0}$.
  Given this, $\succinctdecider$ returns the circuit~$\circuit$ with inputs
  $i_1, i_2 \in \{0, 1, \ldots, L-1\}$, $i_3, i_4, i_5 \in \{0, 1, \ldots,
  R-1\}$, $o_1, o_2, \ldots, o_5 \in \{0, 1\}$, and
  \[
    \circuit(i_1, i_2, \ldots, i_5, o_1, o_2, \ldots, o_5) = 1,
  \]
  if one of the following conditions hold.
  \begin{align*}
    & \circuit_{\mathrm{3SAT}}(i_3, i_4, i_5, o_3, o_4, o_5) = 1\;,\\
    & (i_1 < 2T) \land (i_1 = i_3) \land (o_1 \neq o_3)\;,\\
    & (i_2 < 2T)\land (i_2 = i_3 - 2T) \land (o_2 \neq o_3)\;,\\
    & (i_1 \geq 2T) \land (\text{$i_1$ is odd}) \land (o_1 = 1)\;,\\
    & (i_1 \geq 2T) \land (\text{$i_1$ is even}) \land (o_1 = 0)\;,\\
    & (i_2 \geq 2T) \land (\text{$i_2$ is odd}) \land (o_2 = 1)\;,\\
    & (i_2 \geq 2T) \land (\text{$i_2$ is even}) \land (o_2 = 0)\;,\\
    & (i_3 = i_4) \land (o_3 \neq o_4)\;,\\
    & (i_4 = i_5) \land (o_4 \neq o_5)\;.
  \end{align*}
  It is not hard to verify that testing ``$(i_1 < 2T)$'' can be done with
  $O(\ell_0)$ AND and OR gates, and testing $(i_2 = i_3 - 2T)$ can be done with
  $O(m)$ AND and OR gates.
  Using similar estimates for the remaining sub-circuits, we compute
  \begin{equation*}
    \begin{split}
      &\mathrm{size}(\circuit)\\
      &= \mathrm{size}(\circuit_{\mathrm{3SAT}}) +
      O(r_0 + \ell_0) \\
      & = \mathrm{size}(\circuit_{\mathrm{3SAT}}) + O(r_0) \\
      & \leq \poly(\log(T), \log(n), \qlen,\sigma) + O(\log(T) + \log(\sigma))\;.
    \end{split}
  \end{equation*}
  In addition, due to the simplicity of these modifications, we conclude that
  the runtime of $\succinctdecider$ is dominated by the runtime of the algorithm
  from Proposition~\ref{prop:standard-succinct-sat}, which is $\poly(\log(T),
  \log(n), \qlen,\sigma)$.

  Now we show that $\circuit$ succinctly describes $\decider$ on inputs~$n$,
  $x$, and~$y$ and time~$T$.
  To begin, we describe the decoupled 5SAT formula $\varphi_{\circuit}$.
  Let us first consider the constraints in $\varphi_{\circuit}$ which are
  implied by the final constraint, i.e.,\ those of the form
  \begin{equation*}
    a_{i_1}^{o_1} \lor b_{i_2}^{o_2} \lor (w_1)_{i_3}^{o_3} \lor (w_2)_{i_4}^{o_4}
    \lor (w_3)_{i_5}^{o_5}
  \end{equation*}
  whenever $i_4 = i_5$ and $o_4 \neq o_5$.
  For any fixed $i_1, i_2, i_3$, the negations $o_1, o_2, o_3$ can take any
  values, and as a result, the first three bits in the constraint vary over all
  assignments in $\{0, 1\}^3$.
  This means that these constraints are satisfied if and only if
  $(w_2)_{i_4}^{o_4} \lor (w_3)_{i_5}^{o_5}$ is satisfied whenever $i_4 = i_5$
  and $o_4 \neq o_5$.
  This, in turn, is equivalent to the constraint $w_2 = w_3$.
  Carrying out similar arguments for the entire circuit, we can express the
  formula~$\varphi_{\circuit}$ as follows.
  \begin{multline*}
    \varphi_{\circuit}(a, b, w_1, w_2, w_3)
    =
    \varphi_{\mathrm{3SAT}}(w_1, w_2, w_3)
    \land (w_{1, 1} = a_1) \land (w_{1, 2} = b_1)\\
    \land (a_2= (10)^{L/2 -T}) \land (b_2 = (10)^{L/2-T})
    \land (w_1 = w_2)
    \land (w_2 = w_3).
  \end{multline*}
  Here, we write $\varphi_{\mathrm{3SAT}}(w_1, w_2, w_3)$ for the formula in
  which, for each constraint in $\varphi_{\mathrm{3SAT}}$, the first variable is
  taken from $w_1$, the second from $w_2$, and the third from $w_3$.
  In addition, we write $a = (a_1, a_2)$, where $a_1$ is the first~$2T$ bits
  in~$a$ and $a_2$ is the remaining $L-2T$ bits, and similarly for~$b = (b_1,
  b_2)$.
  We also write $w_1 = (w_{1,1}, w_{1,2}, w_{1,3})$, where $w_{1, 1}$ contains
  the first~$2T$ bits in~$w_1$, $w_{1,2}$ contains the second~$2T$ bits,
  and~$w_{1,3}$ contains the remaining $R-4T$ bits.
  As a result, $\varphi_{\circuit}$ is satisfied only if $w_1 = w_2 = w_3 =
  (a_1, b_1, c)$ for some string $c \in \{0, 1\}^{R-4T}$.
  In this case, calling $w = (a_1, b_1, c)$, $\varphi_{\circuit}$ is satisfied
  only if $\varphi_{\mathrm{3SAT}}(w)$ is.
  By Proposition~\ref{prop:standard-succinct-sat}, this implies that there exists a
  string~$a_{\mathrm{prefix}}$ of length $\ell_a \leq T$ such that $ a_1 =
  \mathrm{enc}_\Gamma(a_{\mathrm{prefix}},\,\sqcup^{T-\ell_a})$.
  This, in turn, implies that
  \begin{equation*}
    a = (a_1, a_2) = (\mathrm{enc}_\Gamma(a_{\mathrm{prefix}},\,
    \sqcup^{T-\ell_a}), (10)^{L-2T})
    = \mathrm{enc}_\Gamma(a_{\mathrm{prefix}},\,\sqcup^{L/2-\ell_a})\;,
  \end{equation*}
  using the fact that $\mathrm{enc}_\Gamma(\sqcup) = 10$, and similarly for~$b$.
  Finally, Proposition~\ref{prop:standard-succinct-sat} implies that $\decider$
  accepts $(n, x, y, a_{\mathrm{prefix}}, b_{\mathrm{prefix}})$ in time~$T$.
  This completes the proof.
\end{proof}

As stated above, moving from 3SAT to 5SAT allows us to devote the first two
inputs to~$a$ and~$b$.
In addition, we have added extra constraints into $\varphi_{\circuit}$ which
enforce that $w_1 = w_2 = w_3$, which means that we can relax this assumption on
these assignments.

We now show a simple transformation that takes in a succinct circuit~$\circuit$
and outputs another succinct circuit~$\circuit'$ whose~$5$ inputs are ``padded''
to contain more input bits.


\begin{proposition}[Padding]\label{prop:exciting-padding-prop}
  There is an algorithm which satisfies the following.
	Let $\circuit$ be a circuit of size~$s$ with two inputs of length~$\ell$,
  three inputs of length~$r$, and~$5$ single-bit inputs.
  Suppose $\circuit$ succinctly describes $\decider$ on inputs $n$, $x$, and $y$
  and time~$T$.
  Then the algorithm takes as input $(\circuit, \ell, r, \ell',
  r')$, with $\ell' \geq \ell$ and $r' \geq r$, and in time~$\poly(s, \ell',
  r')$ outputs a circuit~$\circuit'$ with the following properties.
  First, $\circuit'$ has two inputs of length~$\ell'$, three inputs of
  length~$r'$, and~$5$ single-bit inputs, and its size is $s + \poly(\ell',
  r')$.
  Second, it succinctly describes $\decider$ on inputs~$n$, $x$, and~$y$ and
  time~$T$.
\end{proposition}
\begin{proof}
  Write $L = 2^\ell, R = 2^r$ and $L' = 2^{\ell'}, R' = 2^{r'}$.
  The algorithm constructs the circuit~$\circuit'$ which on inputs $i_1, i_2 \in
  \{0, \ldots, L'-1\}$, $i_3, i_4, i_5 \in \{0, \ldots, R'-1\}$, and $o_1,
  \ldots, o_5 \in \{0, 1\}$, outputs~$1$ if and only if one of the following
  conditions hold:
  \begin{align*}
    & (i_1, i_2 < L) \land (i_3, i_4, i_5 < R) \land
      \circuit(i_1, i_2, i_3, i_4, i_5, o_1, o_2, o_3, o_4, o_5) = 1\;,\\
    & (i_1 \geq L) \land (\text{$i_1$ is odd}) \land (o_1 = 1)\;,\\
    & (i_1 \geq L) \land (\text{$i_1$ is even}) \land (o_1 = 0)\;,\\
    & (i_2 \geq L) \land (\text{$i_2$ is odd}) \land (o_2 = 1)\;,\\
    & (i_2 \geq L) \land (\text{$i_2$ is even}) \land (o_2 = 0)\;.
  \end{align*}
  Now we show that $\circuit'$ succinctly describes $\decider$ on inputs~$n$,
  $x$, and~$y$ and time~$T$.
  To begin, we describe the decoupled 5SAT formula $\varphi_{\circuit'}$.
  The second and third constraints imply that for each $i_1 \geq L$,
  $\varphi_{\circuit'}$ contains the constraint $(a_{i_1})$ if $i_1$ is odd and
  $(\neg a_{i_1})$ if $i_1$ is even.
  Likewise, the fourth and fifth constraints imply that for each $i_2 \geq L$,
  $\varphi_{\circuit'}$ contains the constraint $(b_{i_2})$ if $i_2$ is odd and
  $(\neg a_{i_2})$ if $i_2$ is even.
  Thus, we can express the formula $\varphi_{\circuit'}$ as follows.
  \begin{equation}\label{eq:write-formula}
\begin{aligned}
&\varphi_{\circuit'}(a, b, w_1, w_2, w_3)
    = \varphi_{\circuit}(a_1, b_1, w_{1, 1}, w_{2, 1} ,w_{3, 1})\\
 &\land (a_2 = (10)^{(L'-L)/2}) \land (b_2 = (10)^{(L'-L)/2}).
\end{aligned}
\end{equation}
  Here, we write $a = (a_1, a_2)$ and $b = (b_1, b_2)$, where $a_1, b_1$ have
  length $L$ and $a_2, b_2$ have length $L' - L$, and for each $i \in \{1, 2,
  3\}$, we write $w_i = (w_{i, 1}, w_{i, 2})$, where $w_{i, 1}$ has length~$R$
  and $w_{i, 2}$ has length~$R'-R$.

  Now, suppose there exist $w_1, w_2, w_3$ such that $a, b, w_1, w_2, w_3$
  satisfy $\varphi_{\circuit'}$.
  Then because $\circuit$ succinctly represents~$\decider$, there exists
  $a_{\mathrm{prefix}}, b_{\mathrm{prefix}}$ of lengths $\ell_a, \ell_b \leq T$
  such that $\decider$ accepts $(n, x, y, a_{\mathrm{prefix}},
  b_{\mathrm{prefix}})$.
  In addition,
  \begin{equation*}
    a_1 = \mathrm{enc}_\Gamma(a_{\mathrm{prefix}}, \sqcup^{L/2-\ell_a}),
  \end{equation*}
  and likewise for~$b_1$.
  Equation~\eqref{eq:write-formula} then implies that
  \begin{align*}
    a = (a_1, a_2)
    &= (\mathrm{enc}_\Gamma(a_{\mathrm{prefix}}, \sqcup^{L/2-\ell_a}),
      (10)^{(L' - L)/2})\\
    &= (\mathrm{enc}_\Gamma(a_{\mathrm{prefix}}, \sqcup^{L/2-\ell_a}),
      \mathrm{enc}_\Gamma(\sqcup)^{(L' - L)/2})\\
    &= \mathrm{enc}_\Gamma(a_{\mathrm{prefix}}, \sqcup^{L'/2-\ell_a}),
  \end{align*}
  where the third step used the fact that $\mathrm{enc}_\Gamma(\sqcup) = 10$.
  As a similar statement holds for~$b$, this establishes that $\circuit'$
  succinctly describes $\decider$ on inputs~$n$, $x$, and~$y$ and time~$T$.
\end{proof}

By combining Proposition~\ref{prop:explicit-succinct-deciders} and
Proposition~\ref{prop:exciting-padding-prop} (choosing $\ell' = r' = r_0$ where $r_0$ is
from Proposition~\ref{prop:explicit-succinct-deciders}) \tnote{added:}and artificially introducing input wires and gates if necessary, we get the following:

\newcommand{\paddedsuccinctdecider}{\mathsf{PaddedSuccinctDecider}}

\begin{proposition}[Explicit padded succinct descriptions]
  \label{prop:explicit-padded-succinct-deciders}
  There is a Turing machine $\paddedsuccinctdecider$ with the following
  properties.
  Let $\decider$ be a decider, let $n$, $T$,~$\qlen$, and $\sigma$ be integers
  with $\max\{\qlen, 2 \lceil \log n \rceil \} \leq T$  \hnote{june 14: added $\log n$ lower bound}, and $|\decider| \leq \sigma$, and let~$x$ and~$y$ be
  strings of length at most~$\qlen$.
  Then on input $(\sigma,\decider, n, T, \qlen, x, y)$,
  $\paddedsuccinctdecider$ outputs a circuit~$\circuit$, with five inputs of
  length~$m(T,\sigma)$ and five single-bit inputs, and $\circuit$ succinctly describes
  $\decider$ on inputs~$n$, $x$, and~$y$ and time~$T$.
  Moreover, the following hold.
  \begin{enumerate}
  \item $m(T,\sigma) = O(\log(T)+ \log(\sigma))$ is a power of two, and $2^m \geq 2T$.
  \item $\circuit$ has at most $s(n,T,\qlen,\sigma)$ gates, where $s$
    is such that $s\geq m$, $s(n,T,\qlen, \sigma) = \poly(\log(T), \log(n),
    \qlen,\sigma)$ and $5m(T, \sigma) + 5 + s(n, T, \qlen,\sigma)$ is
    a power of $2$.
  \item $\paddedsuccinctdecider$ runs in time $\poly(\log(T), \log(n),
    \qlen,\sigma)$, and the parameters $m(T,\sigma), s(n, T, \qlen, \sigma)$ can
    be computed from $n, T, \qlen,\sigma$ in time $\poly(\log(T), \log(n),
    \log(\qlen),\log(\sigma))$.
  \end{enumerate}
\end{proposition}

\subsection{A PCP for normal form deciders}
\label{sec:pcp-cktval-new}

We give a probabilistically checkable proof (PCP) for the Bounded Halting
problem specialized to the case of normal form deciders.
Our PCP will use standard techniques from the algebraic, low-degree-code-based
PCP literature.
In particular, we slightly modify the PCP for Succinct-3SAT described
in~\cite[Section 11]{NW19} (which itself is based on the proof of the PCP
theorem in~\cite{harsha2004robust}) to apply it to the decoupled Succinct-5SAT
instances described in Section~\ref{sec:succinct-deciders}.
We follow their treatment closely.
As the PCPs constructed in this section are only an intermediate object towards
the normal form verifier introduced in the next section we do not include
standard definitions on PCPs, and refer to these references (in
particular~\cite{harsha2004robust}) for background.
We begin with some preliminaries.

\subsubsection{Preliminaries}

A key part of the PCP will be to design a function $f:\F^n \rightarrow \F$ which
is zero on the subcube $H_{\mathrm{subcube}} = \{0,1\}^n$.
Our next proposition shows that given such a function~$f$, there is a way of
writing it so that the fact that it is zero on $H_{\mathrm{subcube}}$ is
self-evidently true.
Doing so involves showing that $f$ can be written in a simple basis of
polynomials which are constructed to be zero on the subcube.
This is one of the components of Alon's combinatorial
Nullstellensatz~\cite[Theorem~$1.1$]{alon1999combinatorial}, and is standard in the PCP literature (see, for example,
\cite[Lemma~$4.11$]{ben2008short}). We include its proof for
completeness.

\begin{proposition}[Polynomial basis of zero functions]\label{prop:zero-basis}
  Let $\F$ be a field.
  Define $\mathrm{zero}: \F \to \F$ as the univariate polynomial $x \mapsto
  x(1-x)$.
  Define $H_{\mathrm{subcube}} = \{0,1\}^n$.
  Suppose $f:\F^n \rightarrow \F$ is an individual degree $d$ polynomial such
  that $f(x) = 0$ for all $x \in H_{\mathrm{subcube}}$.
  Then there exist polynomials $c_1, \ldots, c_n:\F^n \rightarrow \F$ such that
  for all $x \in \F^n$,
  \begin{equation*}
  f(x) = \sum_{i=1}^n c_i(x) \cdot \mathrm{zero}(x_i)\;.
  \end{equation*}
  In addition, for each $i \in \{1, \ldots, n\}$, $c_i$ has individual
  degree $d$.
  \end{proposition}

  \begin{proof}
    To prove this, we first prove the following statement for each $k \in \{0,
    1, \ldots, n\}$: there exists an individual degree-$d$ polynomial $r_k:\F^{n}
    \rightarrow \F$ and polynomials $c_1, \ldots, c_k:\F^n \rightarrow \F$ such
    that
  \begin{equation}\label{eq:k-levels-of-division}
  f(x) = \sum_{i=1}^k c_i(x) \cdot \mathrm{zero}(x_i) + r_k(x)\;.
  \end{equation}
  In addition, for each $i \in \{1, \ldots, k\}$, $c_i$ has individual degree
  $d$ and the degree of~$x_i$ in $r_k$ is at most $1$.

  The proof is by induction on~$k$, the base case of $k = 0$ being trivial.
  Now, we perform the induction step.
  Assuming that Equation~\eqref{eq:k-levels-of-division} holds for $k$, we will
  show that it holds for~$k+1$ as well.
  Let~$r_k$ be the polynomial guaranteed by the inductive hypothesis.
  We now divide~$r_k$ by $\mathrm{zero}(x_{k+1})$ using polynomial
  division.
  This guarantees a polynomial $c_{k+1}$ and an individual degree-$d$
  polynomial~$r_{k+1}(x)$ such that
  \begin{equation*}
  r_k(x) = c_{k+1}(x) \cdot \mathrm{zero}(x_{k+1}) + r_{k+1}(x)\;.
  \end{equation*}
  In addition, $c_{k+1}$ still has individual degree $d$ (and in fact the degree
  of $x_{k+1}$ in $c_{k+1}$ is at most $d - 2$).
  Furthermore, for each $i \in \{1, \ldots, k+1\}$, $r_{k+1}$ has degree at most
  $1$ in~$x_i$.
  Plugging this into Equation~\eqref{eq:k-levels-of-division}, we see that
  \begin{equation}\label{eq:almost-there}
  f(x) = \sum_{i=1}^{k+1} c_i(x) \cdot \mathrm{zero}(x_i) + r_{k+1}(x)\;.
  \end{equation}
  This completes the induction.

  Applying the $k = n$ case of Equation~\eqref{eq:k-levels-of-division}, we see
  that
  \begin{equation}\label{eq:case-n-of-induction}
  f(x) = \sum_{i=1}^n c_i(x) \cdot \mathrm{zero}(x_i) + r(x)\;,
  \end{equation}
  where the degree of each variables in $r$ is at most $1$.
  Since $f(x)$ vanishes on the subcube $H_{\mathrm{subcube}}$, it must also be
  that $r(x)$ vanishes on $H_{\mathrm{subcube}}$ as well.
  We claim that $r$ must, therefore, be the zero polynomial.

  We prove this by showing the following statement: for every polynomial
  $s(x_1,\ldots,x_n)$ with individual degree $1$ (i.e.,
  a multilinear polynomial) that vanishes on all points $x \in \{0,1\}^n$, $s$
  must be the zero polynomial.
  We show this by induction on the number of variables.

  Consider the base case $n = 1$.
  Then $s$ is a univariate polynomial with degree at most $1$.
  However, it is zero on two points, so, therefore, it must be the zero polynomial.
  Now for the inductive step: assuming the proposition holds for some $n \geq
  1$, we show that it holds for $n+1$ as well.
  For any multilinear polynomial$s$, we can write
	\[
		s(x_1,\ldots,x_{n+1}) = x_{n+1} s_1(x_1,\ldots,x_n) + s_2(x_1,\ldots,x_n)
	\]
	where both $s_1$ and $s_2$ are $n$-variate multilinear polynomials.
  Fix $x_{n+1} = 0$.
  Then $s(x_1,\ldots,x_n,0) = s_2(x_1,\ldots,x_n)$.
  Since $s$ vanishes on $\{0,1\}^{n+1}$, this implies that $s_2$ vanishes on
  $\{0,1\}^n$ as well.
  By the inductive hypothesis, $s_2$ is the identically zero polynomial.
  Now fix $x_{n+1} = 1$.
	Then $s(x_1,\ldots,x_n,1) = s_1(x_1,\ldots,x_n)$.
  Again, since $s$ vanishes on $\{0,1\}^{n+1}$, this implies that $s_1$ vanishes
  on $\{0,1\}^n$ as well.
  Again by the inductive hypothesis, $s_1$ is the identically zero polynomial.
  This shows that $s$ is the identically zero polynomial, completing the
  induction.

  Thus, $r$ is the zero polynomial.
  Applying this fact to Equation~\eqref{eq:case-n-of-induction},
  we arrive at the statement in the proposition.
  \end{proof}

\subsubsection{The PCP}
\label{sec:ar-pcp}

\paragraph{The problem.}
The input to the PCP verifier is a tuple $(\sigma,\decider, n, T, \qlen, x,
y)$.
Here, $\decider$ is a decider, $n$, $T$, $\qlen$, $\sigma$ are integers
with $\qlen \leq T$ and $|\decider| \leq \sigma$, and~$x$ and~$y$ are a pair of
strings of length at most~$\qlen$ each.
The goal of the verifier is to check whether there exist two
strings~$a_{\mathrm{prefix}}$ and~$b_{\mathrm{prefix}}$ of length at most~$T$
such that $\decider$ halts and accepts on input
$(n,x,y,a_{\mathrm{prefix}},b_{\mathrm{prefix}})$ in time at most~$T$.
To do that the verifier makes random queries to a \emph{PCP
  proof} $\Pi$ encoded using a low-degree code, and decides whether to accept or reject based on the parts of
$\Pi$ that it reads. 
We first set the parameters used in the PCP construction.

\begin{definition}[Parameters for the PCP]\label{def:pcpparams}
  For all integers $n,T,\qlen,\sigma\in\N$ such that
\[
  \max\{\qlen,2\lceil\log n\rceil\}\leq T,
\]
define the tuple
\[
  \pcpparams(n,T,\qlen,\sigma)=(q,m,d,m',s)
\]
as follows. Let $m=m(T,\sigma)$ and
$s=s(n,T,\qlen,\sigma)$ be as in
Proposition~\ref{prop:explicit-padded-succinct-deciders}.
  Let $a' > 1$ and $0 < b' < 1$ be the universal constants in the error function $\delta_\ld$ from
  Theorem~\ref{lem:ld-soundness}.
  Define the following integers.
  \hnote{streamlined the following parameter settings, especially simplified the condition for $q$}
  \begin{enumerate}
	\item Let $m' = 5m + 5 + s$ (recall that by
          Proposition~\ref{prop:explicit-padded-succinct-deciders}, $s$ is chosen
          so that $m'$ is a power of 2). The parameter $m'$
          denotes the number of variables appearing in the polynomials
          comprising the PCP proof.
  \item Let $d = 7$. The parameter $d$ is an upper bound on the
    individual degree of the polynomials comprising the PCP proof
    (this bound will be attained by the polynomial $c_0$
    in Equation~\eqref{eq:pcp-c0}).

  \item Let $q=2^k$ where $k$ is the smallest odd integer satisfying
\begin{equation}\label{eq:pcp-q-choice}
  q=2^k \geq
  \bigl(\qlen\, a'(dm'(m'+6))^{a'}\bigr)^{1/b'}~.
\end{equation} 
By construction, $m$ and $m'$ are powers of $2$, and $m \leq m' \leq q$. Consequently, $m | q$ and $m' | q$. 
        The parameter $q$ denotes the size of the field over which
        polynomials comprising the PCP proof will be defined. The constraint on $q$ is chosen so that the error $\delta_\ld$ is controlled in the analysis of the soundness of the answer-reduced verifier (see Section~\ref{sec:ar-soundness}).

  \end{enumerate}

\end{definition}

Given $n,T,\qlen,\sigma$ represented in binary, the tuple
$\pcpparams(n,T,\qlen,\sigma)$ can be computed in time
\[
  \poly(\log n,\log T,\log\qlen,\log\sigma).
\]
Indeed, Proposition~\ref{prop:explicit-padded-succinct-deciders} gives
algorithms for computing $m(T,\allowbreak \sigma)$ and $s(n,\allowbreak T,\allowbreak \qlen,\allowbreak \sigma)$ within this
time bound. The value $m'=5m+5+s$ is computed by binary addition, and $d=7$ is
constant. Since $a'$, $b'$, and $d$ are universal constants, there is a
universal constant $C_q$ such that \eqref{eq:pcp-q-choice} is satisfied whenever
\[
  k \geq
  C_q\bigl(\lceil\log\qlen\rceil+\lceil\log(m'+6)\rceil+1\bigr).
\]
Hence the smallest odd such $k$ is at most this quantity plus $1$. It can,
therefore, be found by checking the odd integers $1,3,5,\ldots$ up to this value
and testing whether $2^k$ satisfies \eqref{eq:pcp-q-choice}. Once $k$ is found,
$q=2^k$ is represented in binary as a $1$ followed by $k$ zeros.

Next, we define the format of a valid PCP proof, which for our construction
consists of evaluation tables of low-degree polynomials.

\begin{definition}\label{def:pcp-proof}
  Given $n,\allowbreak  T,\allowbreak  \qlen,\allowbreak  \sigma \in \N$ and $(q,\allowbreak m,\allowbreak d,\allowbreak m',\allowbreak s)=
  \pcpparams(n,\allowbreak  T,\allowbreak  \qlen,\allowbreak  \sigma)$, a \emph{low-degree PCP proof} is a tuple
  $\Pi$ of evaluation tables of polynomials $g_1,\allowbreak  \dots,\allowbreak  g_5: \F_q^{m} \to \F_q$
  and $c_0,\allowbreak  \dots,\allowbreak  c_{m'}: \F_q^{m'} \to \F_q$ with all polynomials having
  individual degree at most $d$.
  We divide the $m'$ input variables of $c_0,\allowbreak  \dots,\allowbreak  c_{m'}$ into blocks as
  follows:
  \[
    \F_q^{m'} \ni z = (\underbrace{x_1}_{\F_q^m}, \dots,
    \underbrace{x_5}_{\F_q^m}, \underbrace{o}_{\F_q^5},
    \underbrace{w}_{\F_q^s})\;.
  \]
\end{definition}

\begin{definition}\label{def:pcp-eval}
  Given a low-degree PCP proof $\Pi$ and a point $z = (x_1, \dots, x_5, o, w)
  \in \F_q^{m'}$, where $x_1, \dots, x_5 \in \F_q^{m}$, $o \in \F_q^{5}$, and $w
  \in \F_q^{s}$, the \emph{evaluation} of $\Pi$ at $z$ is given by
  \[
    \ev_z(\Pi) = (\alpha_1, \dots, \alpha_5, \beta_0, \dots, \beta_{m'}) \in
    \F_{q}^{6 + m'}\;,
  \]
  where $\alpha_i = g_i(x_i)$ and $\beta_j = c_j(z)$.
\end{definition}

We use the following encoding convention for \emph{decider specifications}. A decider
specification is a tuple $(\sigma,\decider,n,T,\qlen,x,y)$. Such a
specification is \emph{valid} if $\decider$ is a description of a Turing machine,
$|\decider|\leq\sigma$, $|x|,|y|\leq\qlen$, and
$\max\{\qlen,2\lceil\log n\rceil\}\leq T$. The Turing machine
$\pcpverifier$, defined next, rejects any input whose decider specification is not valid.

\begin{theorem}\label{thm:pcp-decider}
  There exists a Turing machine $\pcpverifier$ with the following properties.
  \begin{enumerate}
  \item \textbf{(Input format)} The input to $\pcpverifier$ consists of two parts:
    a ``decider specification'' and a ``PCP view.''
    \begin{enumerate}
    \item \textbf{(Decider specification)} The decider specification $(\sigma,\allowbreak \decider,\allowbreak n,\allowbreak T,\allowbreak \qlen,\allowbreak x,\allowbreak y)$ is valid in the sense
      described above.  Let $(q,\allowbreak m,\allowbreak d,\allowbreak m',\allowbreak s)=\pcpparams(n,\allowbreak T,\allowbreak \qlen,\allowbreak \sigma)$ be as in
      Definition~\ref{def:pcpparams}.
    \item \textbf{(PCP view)} Let $z \in \F_{q}^{m'}$ and let $\pcpeval \in
      \F_q^{6 + m'}$.
      Then the PCP view is the pair $(z, \pcpeval)$.
    \end{enumerate}
    The Turing machine $\pcpverifier$ returns either $1$ (accept) or $0$
    (reject). It returns $0$ on any input whose decider specification is invalid
    or whose PCP view is not formatted correctly.
  \item
    \textbf{(Complexity):}
    $\pcpverifier$ runs in time at most~$\poly(\log(T), \log(n), \qlen, \sigma)$.

  \end{enumerate}
  For the remaining items, fix a decider specification $(\sigma,\decider, n, T, \qlen,x, y)$; we think of $\pcpverifier$ as a function of the PCP view
  input only.

  \begin{enumerate}
  \setcounter{enumi}{2}
  \item
    \textbf{(Completeness):} Suppose $a_{\mathrm{prefix}}, b_{\mathrm{prefix}}
    \in \{0, 1\}^*$ are two strings of length $\ell_a, \ell_b \leq T$,
    respectively, such that $\decider$ halts and accepts in time~$T$ on input $(n, x, y,
    a_{\mathrm{prefix}}, b_{\mathrm{prefix}})$.
    Let $M = 2^m$ and write
    \begin{equation*}
      a = \mathrm{enc}_\Gamma(a_{\mathrm{prefix}}\,,\, \sqcup^{M/2 - \ell_a})
      \quad
      \text{and}
      \quad
      b = \mathrm{enc}_\Gamma(b_{\mathrm{prefix}}\,,\, \sqcup^{M/2 - \ell_b})\;,
    \end{equation*}
    where recall that by
    Proposition~\ref{prop:explicit-padded-succinct-deciders}, $m$ is chosen so that $2T\leq M$, and $\sqcup$ is a special symbol that
    is encoded using two bits; thus $a$ and $b$ are each an $M$-bit string.
    Then there exists a low-degree PCP proof (Definition~\ref{def:pcp-proof})
    $\Pi = (g_1, \ldots, g_5, c_0, \ldots, c_{m'})$ with $g_1 = g_a$ and $g_2 =
    g_b$, the low-degree encodings of $a$ and $b$, respectively, (see
    Section~\ref{sec:ld-encoding}), which causes $\pcpverifier$ to accept with
    probability~$1$ over a uniformly random $z \in \F_q^{m'}$:
    \[ \Pr_{z \in \F_q^{m'}}\big(\pcpverifier(z, \ev_{z}(\Pi)) = 1\big) = 1\;. \]
  \item
    \textbf{(Soundness):} Let $\Pi = (g_1, \ldots, g_5, c_0, \ldots, c_{m'})$ be a
    low-degree PCP proof such that $\pcpverifier$ at a uniformly random $z$
    accepts with probability larger than $p_{\soundness} =\frac{1}{2}$:
    \[ \Pr_{z \in \F_q^{m'}}\big( \pcpverifier(z, \ev_{z}(\Pi)) = 1\big) >
      p_{\soundness}\;. \] Then there exist strings $a, b \in \{0, 1\}^{M}$ with the
    following properties.
    \begin{enumerate}
    \item There exist strings $a_{\mathrm{prefix}}, b_{\mathrm{prefix}} \in \{0,
      1\}^*$ of length $\ell_a, \ell_b\leq T$, respectively, such that
      \begin{equation*}
        a = \mathrm{enc}_\Gamma(a_{\mathrm{prefix}}\,,\, \sqcup^{M/2 - \ell_a})
        \quad
        \text{and}
        \quad
        b = \mathrm{enc}_\Gamma(b_{\mathrm{prefix}}\,,\, \sqcup^{M/2 - \ell_b})\;.
      \end{equation*}
    \item $\decider$ halts in time~$T$ and accepts on input $(n, x, y, a_{\mathrm{prefix}},
      b_{\mathrm{prefix}})$.
    \item $a = \coded(g_1)$ and $b = \coded(g_2)$, where $\coded(\cdot)$ is the
      Boolean decoding map of the low-degree encoding defined in
      Section~\ref{sec:ld-encoding}.
%
    \end{enumerate}

  \end{enumerate}
\end{theorem}

\begin{proof}[Proof of Theorem~\ref{thm:pcp-decider}]

  We present the description of the Turing machine $\pcpverifier$ in
  Figure~\ref{fig:pcpverifier}.

\begin{figure}[!htb]
  \begin{gamespec}
  The Turing machine $\pcpverifier$ takes the following as input:
	\begin{itemize}
		\item Decider specification $(\sigma,\decider, n, T, \qlen,x, y)$
      	\item PCP view $(z,\Xi)$
	\end{itemize}
	The Turing machine first checks that its input is properly formatted, according to the conditions stated in Theorem~\ref{thm:pcp-decider} 1(a) and (b). If not, it rejects. If so, it performs the following steps sequentially:
  \begin{enumerate}
  \item \label{enu:compute-circuit} Compute $\circuit =
    \paddedsuccinctdecider(\sigma,\decider, n, T, \qlen, x, y)$.
    The circuit $\circuit$ has five $m$-bit inputs and five single-bit inputs,
    and contains at most~$s$ AND and OR gates.


  \item \label{enu:compute-tseitin-formula} Compute the Tseitin formula
    $\formula$ (Lemma~\ref{lem:tseitin}) corresponding to $\circuit$, which is a boolean formula on $m' =
    5m + 5 + s$ variables.
    Let $\formula_{\mathrm{arith}}: \F_q^{m'} \to \F_q$ denote the
    arithmetization of $\formula$ (Lemma~\ref{lem:formula-arithmetization}).
  \item \label{enu:parse-input} Parse $z = (x,o,w) \in \F_q^{m'}$ where
    $x_1,\ldots,x_5 \in \F_q^m$, $o \in \F_q^5$, and $w \in \F_q^s$. (Here $x$ is a variable which should not be confused with an input $x$ to $\decider$.)
    Parse $\Xi = (\alpha_1,\ldots,\alpha_5,\beta_0,\ldots,\beta_{m'}) \in
    \F_q^{6 + m'}$.

  \item (\textbf{Formula test}) \label{enu:formula-test} Reject if $\beta_0 \neq
    \formula_{\mathrm{arith}}(x,o,w) \cdot (\alpha_1 - o_1) \cdots (\alpha_5 -
    o_5)$.
    Otherwise, continue.
  \item (\textbf{Zero on subcube test}) \label{enu:zero-test} Reject if $\beta_0
    \neq \sum_{i=1}^{m'} \beta_i \cdot \mathrm{zero}_i(z)$.
    Otherwise, accept.
  \end{enumerate}
  \end{gamespec}
\caption{The decision procedure $\pcpverifier$.}
\label{fig:pcpverifier}
\end{figure}

Before establishing the Complexity, Completeness, and Soundness properties of
$\pcpverifier$, we point out several important items.
Recall that $M = 2^m$.
\begin{enumerate}
\item We recall what it means for the circuit $\circuit$ computed in
  Step~\ref{enu:compute-circuit} of Figure~\ref{fig:pcpverifier} to succinctly
  describe the 5SAT formula $\varphi_\circuit$.
  Let the inputs of the formula $\varphi_\circuit$ be strings $a,b,u_3,u_4,u_5
  \in \{0,1\}^M$.
  Then for all $x_1,\ldots,x_5 \in \{0,1\}^m$ and $o \in \{0,1\}^5$, we have
  that $\circuit(x_1,\ldots,x_5,o) = 1$ if and only if $a_{x_1}^{o_1} \lor
  b_{x_2}^{o_2} \lor u_{3, x_3}^{o_3} \lor u_{4, x_4}^{o_4} \lor u_{5,
    x_5}^{o_5}$ is a clause in~$\varphi_\circuit$, where the coordinates of
  $a,b,u_3,u_4,u_5$ are indexed by strings of length $m$.

  Furthermore, the formula $\varphi_\circuit$ is related to the decider
  $\decider$ in the following way.
  For all $a, b\in \{0, 1\}^M$, there exist $u_3, u_4, u_5 \in \{0, 1\}^M$ such
  that $a, b, u_3, u_4, u_5$ satisfy $\varphi_{\circuit}$ if and only if there
  exist $a_{\mathrm{prefix}}, b_{\mathrm{prefix}} \in \{0, 1\}^*$ of
  lengths~$\ell_a, \ell_b \leq T$, respectively, such that
  \begin{equation*}
    a = \mathrm{enc}_\Gamma(a_{\mathrm{prefix}}, \sqcup^{M/2 - \ell_a})
    \quad \text{and} \quad
    b = \mathrm{enc}_\Gamma(b_{\mathrm{prefix}}, \sqcup^{M/2 - \ell_b})
  \end{equation*}
  and $\decider$ accepts $(n, x, y, a_{\mathrm{prefix}}, b_{\mathrm{prefix}})$
  in time~$T$.

\item We recall from Lemma~\ref{lem:tseitin} that the boolean formula $\formula$ is related to $\circuit$ in the following
  way.
  For all $x_1, \ldots,x_5 \in \{0, 1\}^m$ and $o \in \{0, 1\}^5$,
  $\circuit(x_1,\ldots,x_5, o) = 1$ if and only if there exists a $w \in \{0,
  1\}^s$ such that $\formula(x_1,\ldots,x_5, o, w) = 1$.
  The size of the formula $\formula$ is linear in the size of the circuit
  $\circuit$, which is $s$.

\item We recall from Lemma~\ref{lem:formula-arithmetization} that the arithmetization $\formula_{\mathrm{arith}}:=
  \mathrm{arith}_q(\formula)$ is a function
  $\formula_{\mathrm{arith}}:\F_q^{m'}\rightarrow \F_q$ such that
  \begin{equation}
    \label{eq:farith}
    \forall (x,o,w) \in \{0, 1\}^{m'}\;,
    \quad \formula_{\mathrm{arith}}(x,o,w) = \formula(x,o,w)\;.
  \end{equation}
  By Proposition~\ref{prop:tseitin-arith-degree}, $\formula_{\mathrm{arith}}$ is an
  individual degree $6$ polynomial.
\end{enumerate}

\paragraph{Complexity.}
We bound the complexity of the Turing machine $\pcpverifier$. First,
$\pcpverifier$ checks the validity of the decider specification
$(\sigma,\decider,n,T,\qlen,x,y)$ and computes
$(q,m,d,m',s)=\pcpparams(n,T,\qlen,\sigma)$. By the computability discussion
following Definition~\ref{def:pcpparams}, this parameter computation takes time
$\poly(\log n,\log T,\log\qlen,\log\sigma)$. The validity checks
$|\decider|\leq\sigma$, $|x|,|y|\leq\qlen$, and
$\max\{\qlen,2\lceil\log n\rceil\}\leq T$ take time
$\poly(\log T,\log n,\qlen,\sigma)$. The PCP view check parses
$z\in\F_q^{m'}$ and $\Xi\in\F_q^{6+m'}$, and, therefore, takes time
$\poly(m',\log q)$. By Definition~\ref{def:pcpparams},
$m'$ and $\log q$ are bounded by
$\poly(\log T,\log n,\qlen,\sigma)$.

From Proposition~\ref{prop:explicit-padded-succinct-deciders},
Step~\ref{enu:compute-circuit} takes time
$\poly(\log T,\log n,\qlen,\sigma)$.
Computing the Tseitin formula in Step~\ref{enu:compute-tseitin-formula} takes
time polynomial in the size of the circuit $\circuit$, which is
$\poly(s)=\poly(\log T,\log n,\qlen,\sigma)$.
The Formula Test, Step~\ref{enu:formula-test}, requires computing the
arithmetization $\formula_{\mathrm{arith}}$ at a point $z\in\F_q^{m'}$, which
takes time $\poly(s,\log q)$.
The Zero on Subcube Test, Step~\ref{enu:zero-test}, takes time
$\poly(m',\log q)$.

Using the parameter bounds from Definition~\ref{def:pcpparams}, the total
runtime of $\pcpverifier$ is, therefore,
\[
  \poly(\log T,\log n,\qlen,\sigma).
\]

\paragraph{Completeness.}
We now establish the Completeness property of $\pcpverifier$ by concocting a PCP
proof $\Pi$ that is accepted with probability $1$.

Let $a,b \in \{0,1\}^M$ be as specified in Item 3 of Theorem~\ref{thm:pcp-decider}.
Then by definition of the formula $\varphi_\circuit$ and the assumption that
 $\decider$ halts and accepts in time~$T$ on input $(n, x, y,
    a_{\mathrm{prefix}}, b_{\mathrm{prefix}})$, there exist strings
$u_3,u_4,u_5 \in \{0,1\}^M$ such that $(a,b,u_3,u_4,u_5)$ satisfies
$\varphi_\circuit$.

Let $g_1,g_2$ denote the low-degree encodings of $a$ and $b$, respectively, (see
Section~\ref{sec:ld-encoding} for definition of low-degree encoding), and let
$g_3, g_4, g_5$ denote the low-degree encodings of $u_3, u_4$, and
$u_5$, respectively.
In particular, $g_1,\ldots,g_5$ are $m$-variate multilinear polynomials, and for
all $x_1,\ldots,x_5 \in \{0,1\}^m$,
\begin{equation*}
\begin{aligned}
&g_1(x_1) = a_{x_1}\;, \quad
  g_2(x_2) = b_{x_2}\;,\\
  &g_3(x_3) = u_{3,x_3}\;, \quad
  g_4(x_4) = u_{4,x_4}\;,\\
  &g_5(x_5) = u_{5,x_5}\;,
\end{aligned}
\end{equation*}
where we index the coordinates of $a,b,u_3,u_4,u_5$ by $m$-bit strings in the
natural way.

Next, define the polynomial $c_0:\F_q^{m'} \rightarrow \F_q$ as
\begin{equation}
  c_0(x, o, w) = \formula_{\mathrm{arith}} (x, o, w)
  \cdot (g_1(x_1) - o_1) \cdots (g_5(x_5)- o_5)\;. \label{eq:pcp-c0}
\end{equation}
Note that $c_0$ has individual degree $7$. This is because, as recalled in item 3 above,
$\formula_{\mathrm{arith}} (x, o, w)$ has individual degree $6$, each of the
$g_i$'s are multilinear, and, moreover, the $g_i$s act on disjoint
collections of variables.

We now show that $c_0(x,o,w) = 0$ for all $(x,o,w) \in \{0,1\}^{m'}$.
By construction, the arithmetization $\formula_{\mathrm{arith}} (x, o, w)$ is
either $0$ or $1$.
If it is $0$, then we are done.
If it is $1$, then this means that $\formula(x,o,w) = 1$, which by definition
means that $\circuit(x,o) = 1$, which means that $a_{x_1}^{o_1} \lor
b_{x_2}^{o_2} \lor u_{3, x_3}^{o_3} \lor u_{4, x_4}^{o_4} \lor u_{5, x_5}^{o_5}$
is a clause in the formula $\varphi_\circuit$.
But since $(a,b,u_3,u_4,u_5)$ satisfies $\varphi_\circuit$, this means that at
least one of $a_{x_1} = o_1$, $b_{x_2} = o_2$, $u_{3,x_3} = o_3$, $u_{4,x_4} =
o_4$, or $u_{5,x_5} = o_5$.
But this means that the product $(g_1(x_1) - o_1) \cdots (g_5(x_5)- o_5) = 0$,
by definition of the $g_i$'s.
Thus $c_0(x,o,w) = 0$.

Thus, by Proposition~\ref{prop:zero-basis}, there exist polynomials $c_1,\ldots,c_{m'}:
\F_q^{m'} \to \F_q$ of individual degree at most $7$ that certify that $c_0$
vanishes on $\{0,1\}^{m'}$.
In other words, for all $z \in \F_q^m$ (not just over the boolean cube), we have
\begin{equation}
	\label{eq:restricted-to-subcube}
  c_0(z) = \sum_{i=1}^{m'} c_i(z) \cdot \mathrm{zero}(z_i)
\end{equation}
where $\mathrm{zero}(x) = x(1-x)$.

Let the PCP proof $\Pi$ be the collection of polynomials
$(g_1,\allowbreak \ldots,\allowbreak g_5,\allowbreak c_0,\allowbreak c_1,\allowbreak \ldots,\allowbreak c_{m'})$, where the $g_i$'s act on disjoint
variables and the $c_i$'s act on all of them.
Let $z \in \F_q^{m'}$ and let $\Xi =
(\alpha_1,\allowbreak \ldots,\allowbreak \alpha_5,\allowbreak \beta_0,\allowbreak \beta_1,\allowbreak \ldots,\allowbreak \beta_{m'}) = \eval_z(\Pi)$.
The PCP view $(z,\allowbreak \Xi)$ passes the Formula Test always, because $\beta_0 =
c_0(z)$ and $\alpha_i = g_i(z)$ for $i \in \{1,\allowbreak 2,\allowbreak \ldots,\allowbreak 5\}$.
The PCP view also passes the Zero on Subcube Test, because of~\eqref{eq:restricted-to-subcube}.

Thus the PCP view is accepted by the Turing machine $\pcpverifier$ with
probability $1$.
This establishes the Completeness property.

\paragraph{Soundness.}
Fix a low-degree PCP proof $\Pi = (g_1,\ldots,g_5,c_0,\ldots,c_{m'})$ such that
the PCP view $(z,\Xi)$ for $\Xi = \eval_z(\Pi)$ causes $\pcpverifier$ to accept
with probability greater than $p_\soundness$.
This, in particular, implies that the Formula Test is satisfied with high
probability.
With probability greater than $p_\soundness$ over the choice of $z \sim
\F_q^{m'}$, we have
\[
  c_0(z) = \formula_{\mathrm{arith}} (z) \cdot (g_1(x_1) - o_1) \cdots (g_5(x_5)- o_5)
\]
Note that $c_0(z)$, by definition of low-degree PCP proof, has individual degree
at most $d$.
The right-hand side has individual degree at most $6 + 5d$, because
$\formula_{\mathrm{arith}}$ has individual degree $6$ and the $g_i$'s have
individual degree at most $d$.
Thus both sides have total degree at most $(6 + 5d)m'$.
Suppose $c_0(z)$ was not identical to $\formula_{\mathrm{arith}} (z) \cdot
(g_1(x_1) - o_1) \cdots (g_5(x_5)- o_5)$.
Then the Schwartz-Zippel lemma implies that the probability they agree on a
randomly chosen $z$ is at most $(6 + 5d)m'/q$.
By our choice of $q$, $m'$, and $d$ in Definition~\ref{def:pcpparams}, this is less than
$p_\soundness$, which is a contradiction.
Thus $c_0(z)$ is equal to $\formula_{\mathrm{arith}} (z) \cdot (g_1(x_1) - o_1)
\cdots (g_5(x_5)- o_5)$ for all $z\in \F_q^{m'}$.

Next, we consider the Zero on Subcube Test.
With probability greater than $p_\soundness$ over the choice of $z \sim
\F_q^{m'}$, we have
\[
  c_0(z) = \sum_{i=1}^{m'} c_i(z) \cdot \mathrm{zero}_i(z).
\]
Again, the polynomials on both sides of the equation have individual degree at
most $d+2$, and, therefore, total degree at most $(6 + d)m'$.
By the Schwartz-Zippel lemma, the probability that both sides would agree on the
evaluation of a random $z$, if they weren't equal polynomials, would be at most
$(6 + d)m'/q$, which is also less than $p_\soundness$.
Thus $c_0(z) = \sum_{i=1}^{m'} c_i(z) \cdot \mathrm{zero}_i(z)$ for all $z \in
\F_q^{m'}$.
This, in particular, implies that $c_0(z)$ vanishes on the subcube $\{0,1\}^{m'}$.

We can now decode assignments $a,b,u_3,u_4,u_5 \in \{0,1\}^M$ that satisfy the
formula $\varphi_\circuit$.
Let $a = \coded(g_1), b = \coded(g_2), u_3 = \coded(g_3), u_4 = \coded(g_4), u_5
= \coded(g_5)$ where $\coded(\cdot)$ is the decoding map defined in
Section~\ref{sec:ld-encoding}. Recall that by the definition of $\coded$, we
are guaranteed that $a,b, u_3, u_4, u_5$ are all boolean strings.
The fact that $c_0(z)$ vanishes on the subcube $\{0,1\}^{m'}$ implies that all
clauses of $\varphi_\circuit$ are satisfied by $a,b,u_3,u_4,u_5$.
By construction of $\varphi_\circuit$, this implies that there exists
$a_{\mathrm{prefix}}, b_{\mathrm{prefix}}$ such that
$\decider(n,x,y,a_{\mathrm{prefix}},b_{\mathrm{prefix}})$ accepts in time $T$.
This completes the proof of the Soundness property, and the proof of the
Theorem.
\end{proof}

\subsection{A normal form verifier for the PCP}
\label{sec:ld-compiler}

\def\tvans{\hat{\verifier}^\ar}
\def\tsans{\hat{\sampler}^\ar}
\def\tdans{\hat{\decider}^\ar}
\def\vans{\verifier^\ar}
\def\sans{\sampler^\ar}
\def\dans{\decider^\ar}

In this section we show how to convert the PCP from Section~\ref{sec:pcp-cktval-new}
into a normal form verifier.
This results in an ``answer reduction'' scheme: a way to map a verifier
$\verifier$ into a new (typed) verifier $\tvans$ with a smaller answer size.
	
Let $\verifier = (\sampler, \decider)$ be a normal form verifier and
$(\lambda,\mu,\sigma)$ a tuple of integers.
Assume that $\decider$ is in the standard timeout-counter form of
Section~\ref{sec:tms}, and let $B_{\decider}(n)$ denote its timeout bound.
In the rest of this section we define the (typed) answer-reduced verifier $\tvans =
(\tsans, \tdans)$ associated with $\verifier$ and parameters $(\lambda, \mu,\sigma)$.

\subsubsection{Parameters and notation}\label{sec:ar-params}

We establish some notation and parameters that are used throughout Section~\ref{sec:ld-compiler}. 
First, define the functions
\begin{equation}
  \label{eq:ar-params-1}
  T(n)=(2^{\lambda n})^\mu \quad\text{and} \quad \qlen(n) = (\lambda n)^{\mu}~.
\end{equation} 
 \hnote{june 14: added $\log n$ lower bound mention} We note that by definition, since $\lambda, \mu$ are positive integers it holds that $T(n) \geq \max \{ \qlen(n) , 2 \lceil \log n\rceil \}$ as needed by Proposition~\ref{prop:explicit-padded-succinct-deciders}.
In the main theorem of this section, Theorem~\ref{thm:ar}, we assume that the input
verifier $\verifier$ satisfies \tnote{added condition on size}
\begin{equation}
  \label{eq:ar-time-assumption}
|\decider|\,\leq\,\sigma\;,\quad \TIME_\decider(n) \leq T(n)\quad
  \text{and}\quad \TIME_\sampler(n) \leq \qlen(n) \;.
\end{equation}
In what follows we write $T$ and $\qlen$ as free parameters (though they are
implicitly functions of the index $n$).

Next, for all integers $n \in \N$ define the PCP parameters $(q,m,d,m',s) =
\pcpparams(n, T, \qlen, \sigma)$ where \tnote{edit, was $\sigma=|\decider|$, but actually we will only ultimately get an upper bound on this:}$\sigma$ is a parameter (that we will ultimately set to an easily computable upper bound on $|\decider|$).
Note that the parameters $(q,m,d,m',s)$ are all implicitly functions of $n$.

Recall from Definition~\ref{def:sampler-sample} the notation $\mu_{\sampler}$
denoting the distribution over pairs of questions $(x_\alice,x_\bob)$ generated by
$\sampler$.
We use $\mu_{\sampler, \alice}$ to indicate the marginal distribution of
$\mu_\sampler$ on the first question $x_\alice$, and $\supp(\mu_{\sampler})$ to
indicate the set of question pairs that have nonzero probability under
$\mu_\sampler$.

\begin{remark}\label{rk:ab-01}
  Throughout Section~\ref{sec:ld-compiler}, for convenience we often identify the label
  $\alice$ with $1$ and $\bob$ with $2$.
  For example, $g_\alice$ is another label for the polynomial $g_1$.
\end{remark}

\subsubsection{The answer-reduced verifier}
\label{sec:ar-verifier}

Let $\verifier = (\sampler, \decider)$ be a normal form verifier and
$(\lambda,\mu,\sigma)$ integers.
All other required parameters and notation are introduced in
Section~\ref{sec:ar-params}.

\newcommand{\tspcp}{\hat{\sampler}^\pcp}

\paragraph{Sampler.}
In this section we define the (typed) sampler $\tsans$ for the (typed) answer reduced
verifier $\tvans$. 

We first give an intuitive description of the sampler, and then proceed to give
a formal definition.
The sampler $\tsans$ is a \emph{product} of the oracularized sampler $\tsora$
corresponding to $\sampler$, and a typed PCP sampler $\tspcp$.
The corresponding question distribution $\mu_{\tsans}$ is then a product
distribution $\mu_{\tsora} \times \mu_{\tspcp}$.

The PCP distribution $\mu_{\tspcp}$ corresponds to six copies of the classical
low-degree test question distribution (see Section~\ref{sec:ld-game}).
Recall that the classical low-degree test is a nonlocal game that checks whether
the players' answers are consistent with low-degree polynomials.
The answer-reduced verifier will use the classical low-degree test to check that
the players respond with evaluations of low-degree polynomials
$g_1,\ldots,g_5,c_0,\ldots,c_{m'}$, and then process the evaluations using the
PCP verifier $\pcpverifier$ from Theorem~\ref{thm:pcp-decider}.
Five of the six copies are meant to certify the ``low-degreeness'' of
$g_1,\ldots,g_5$, and the sixth copy is used to certify the ``low-degreeness''
of the entire bundle of polynomials $g_1,\ldots,g_5,c_0,\ldots,c_{m'}$.
The reason that the $g_i$'s are checked individually is to ensure that they only
depend on certain blocks of input variables, because ultimately the soundness of
the answer-reduced verifier reduces to the soundness of the PCP verifier
$\pcpverifier$, and Theorem~\ref{thm:pcp-decider} assumes that the polynomials
$g_1,\ldots,g_5$ of the PCP proof only depend on certain subsets of variables.

We now formally define the typed PCP sampler $\tspcp$. The type set is
\begin{equation*}
\begin{aligned}
&\type^\pcp = \{ \Point_1, \ldots,\Point_6\} \cup \{\ALine_1,
  \dots, \ALine_6\} \\
 &\cup \{\DLine_1,
  \dots, \DLine_6\}\;,
\end{aligned}
\end{equation*}
and the type graph $G^\pcp=(\type^\pcp,E^\pcp)$ uses the complete edge set
$E^\pcp = \type^\pcp \times \type^\pcp$.
The ambient vector space for the sampler is
\begin{equation}
  \label{eq:V-pcp}
  V^\pcp = \Bigl( \bigoplus_{i=1}^{5} V_{i, \xpt} \oplus V_{i, \coord} \oplus
  V_{i, \dir{}}\Bigr) \oplus V_{\aux, \xpt} \oplus V_{6, \coord} \oplus
  V_{\aux, \dir{}}\;,
\end{equation}
where the spaces $V_{i,\xpt}, V_{i, \dir{}}$ are each isomorphic to $\F_q^m$,
the space $V_{i, \coord}$ is isomorphic to $\F_{q}$, the spaces $V_{\aux, \xpt},
V_{\aux, \dir{}}$ are each isomorphic to $\F_q^{5+s}$, and the space $V_{6,
  \coord}$ is isomorphic to $\F_q$.
In addition define the following direct sums:
\begin{align*}
  V_{6, \xpt} & = \Big( \bigoplus_{i=1}^{5} V_{i, \xpt} \Big) \oplus V_{\aux, \xpt}\;,\\
  V_{6, \dir{}} & = \Big( \bigoplus_{i=1}^{5} V_{i, \dir{}} \Big) \oplus
                   V_{\aux, \dir{}}\;.
\end{align*}
For all $i \in \{1,2,\ldots,6\}$, we call the space $V_{i,\xpt}$ the
\emph{$i$-th point register}, the space $V_{i,\coord}$ the \emph{$i$-th
  coordinate register}, and the space $V_{i,\dir{}}$ the \emph{$i$-th direction
  register} (compare this with the subspaces defined in
Section~\ref{sec:ld-game}).

We remark that the purpose of this ``overlapping'' definition of spaces $V_{6,\xpt},
V_{6,\dir{}}$ is to allow for consistency tests where one player is
queried for the evaluation of a tuple of multiple polynomials on a
collection of points, and the other is queried
for the value of a single one of these polynomials on the associated
point. Since the $V_{i, \coord}$ space is not used to sample points in
the low degree test, we can take $V_{6,\coord}$ to be a separate space
independent of $V_{1, \coord}, \dots, V_{5, \coord}$.

For every type $\tvar \in \type^\pcp$, we define the following CL functions on
$V^\pcp$:
\begin{itemize}
\item For the types $\tvar = \Point_i$ for $i \in \{1, \dots, 6\}$, the
  corresponding $1$-level CL function $L_{\Point_i}$ is identical to the CL
  function $L_\Point$ defined in \eqref{eq:cl-ptf}, but acts on the
  subspace $V_{i,\xpt} \oplus V_{i, \coord} \oplus V_{i,
    \dir{}}$ (and zeroes out all the other registers).

\item For the types $\tvar = \ALine_i$ for $i \in \{ 1, \dots, 6\}$, the
  corresponding $2$-level CL function $L_{\ALine_i}$ is identical to the CL
  function $L_\ALine$ defined in \eqref{eq:cl-alnf}, but acts on the subspaces
  $V_{i,\xpt} \oplus V_{i,\coord} \oplus V_{i,\dir{}}$.
  
\item For the types $\tvar = \DLine_i$ for $i \in \{ 1, \dots, 6\}$, the
  corresponding $3$-level CL function $L_{\DLine_i}$ is identical to the CL
  function $L_\DLine$ defined in \eqref{eq:cl-dlnf}, but acts on the subspaces
  $V_{i,\xpt} \oplus V_{i,\coord} \oplus V_{i,\dir{}}$.
\end{itemize}

Observe that for $i \in \{1,\ldots,5\}$, the CL distributions
$\mu_{L_{\ALine_i},L_{\Point_i}}$ and $\mu_{L_{\DLine_i},L_{\Point_i}}$ correspond to
the question distributions of the classical low-degree test parameterized by $q$
and $m$ (see Section~\ref{sec:ld-game} for the definition of the low-degree test
distributions, which only depend on the first two arguments of the $4$-tuple
$\ldparams$), and the CL distributions $\mu_{L_{\ALine_6},L_{\Point_6}}$ and
$\mu_{L_{\DLine_6},L_{\Point_6}}$ correspond to the classical low-degree test
parameterized by $q$ and $m'$.

We finally give the formal definition of the sampler $\tsans$.
The type set of $\tsans$ is $\type^\ar = \type^\ora \times \type^\pcp$ and the
type graph is $G^\ar = G^\ora \times G^\pcp$ with edge set
\begin{equation*}
  E^\ar = \bigl\{\{(\lvar,\rvar),(\lvar',\rvar')\}:\,
  \{\lvar, \lvar'\} \in E^\ora \text{ and } \{\rvar, \rvar'\} \in
  E^\pcp \bigr\}\;.
\end{equation*}
Let $V^\ora$ denote the ambient space of the oracularized sampler $\tsora$.
The ambient space of $\tsans$ is then the direct sum $V^\ora \oplus V^\pcp$.
For all type pairs $(\tvar_\ora,\tvar_\pcp) \in \type^\ar$, the corresponding CL
function $L_{\tvar_\ora,\tvar_\pcp}$ is simply the direct sum of the CL functions $L_{\tvar_\ora}$ (coming from $\tsora$) and
$L_{\tvar_\pcp}$ (coming from $\tspcp$).
Thus one can see that the distribution corresponding to $\tsans$ is the product
distribution $\mu_{\tsora} \times \mu_{\tspcp}$.

Note that $\tsans$ is an $\max\{ \ell, 3\}$-level sampler, where $\ell$ is the
number of levels of the sampler $\sampler$.




\paragraph{Decider.}
The decider $\tdans$ is described in Figure~\ref{fig:decider-pcp}.

\begingroup
  \begin{gamespec}
  {
    \begingroup
      \centering
      \begin{tabularx}{\textwidth}{ l l X }
        \toprule
        Type & Question Format & Answer Format \\
        \midrule
        $\Point_i$ for $i=1, \ldots, 5$ & $y_i \in \F_q^m$ &
        $\alpha_i \in \F_q$ \\
        $\ALine_i$ for $i=1, \ldots, 5$ & $v_i \in \F_q^m \times \F_q$ &
        $h_i: \F_q \to \F_q$ \\
        $\DLine_i$ for $i=1, \ldots, 5$ & $v_i \in \F_q^m \times \F_q \times \F_q^m$ &
        $h_i: \F_q \to \F_q$ \\        
        $\Point_6$ & $z = (y, o, w) \in \F_q^{m'}$ & $(\alpha'_1,\allowbreak  \ldots,\allowbreak 
        \alpha'_5,\allowbreak  \beta_0,\allowbreak  \ldots,\allowbreak  \beta_{m'}) \in \F_q^{m'+6}$\\
        $\ALine_6$ & $v \in \F_q^{m'} \times \F_q$ & $(h'_1,\allowbreak  \dots,\allowbreak 
        h'_5,\allowbreak  f_0,\allowbreak  \dots,\allowbreak  f_{m'}): \F_q \to \F_q^{m' + 6}$ \\
        $\DLine_6$ & $v \in \F_q^{m'} \times \F_q \times \F_q^{m'}$ & $(h'_1,\allowbreak  \dots,\allowbreak 
        h'_5,\allowbreak  f_0,\allowbreak  \dots,\allowbreak  f_{m'}): \F_q \to \F_q^{m' + 6}$ \\        
        \bottomrule
      \end{tabularx}
      \captionof{table}{Question and answer formats for types in $\type^\pcp$.}\label{table:tpcp}
    \par\endgroup

    On input $(n,\allowbreak \tvar_\alice,\allowbreak x_\alice,\allowbreak \tvar_\bob,\allowbreak x_\bob,\allowbreak a_\alice,\allowbreak a_\bob)$,
the decider $\tdans$ first computes $T=T(n)$, $\qlen=\qlen(n)$, and
$(q,\allowbreak m,\allowbreak d,\allowbreak m',\allowbreak s)=\pcpparams(n,\allowbreak T,\allowbreak \qlen,\allowbreak \sigma)$. For each
$w\in\{\alice,\bob\}$, it attempts to parse
$\tvar_w=(\tvar_{Q,w},\allowbreak \tvar_{\Pi,w})\in\type^\ora\times\type^\pcp$ and
$x_w=(x_{Q,w},x_{\Pi,w})$. The component $x_{Q,w}$ is checked against the
question length for the oracularized sampler $\tsora$, and the components
$x_{\Pi,w}$ and $a_w$ are checked against the question and answer formats for
$\tvar_{\Pi,w}$ in Table~\ref{table:tpcp}. In particular, these format checks
include the prescribed length bounds: if, while parsing a component, the
component exceeds the length allowed by its claimed type, then the parsing fails
and $\tdans$ rejects.

    The decider performs the following steps sequentially, for all
    $w\in\{\alice,\bob\}$:
    \begin{enumerate}
    \item (\textbf{Global consistency check}): If $\tvar_{\alice} =
      \tvar_{\bob}$, reject if $a_\alice\neq a_\bob$.
      \label{enu:ar-global-consistency}

    \item (\textbf{Input consistency check}): If $\tvar_{Q,w} = \oracle$ and
      $\tvar_{Q, \ol{w}} = v \in \{\alice,\allowbreak \bob\}$, and if $(\tvar_{\Pi,w},\allowbreak 
      \tvar_{\Pi, \ol{w}}) = (\Point_6,\allowbreak  \Point_{v})$, reject if $\alpha_{v} \neq
      \alpha'_{v}$ (where $\alice \leftrightarrow 1$ and $\bob \leftrightarrow
      2$, as per Remark~\ref{rk:ab-01}).
      \label{enu:ar-input-consistency}

    \item (\textbf{Input low degree test}) If $\tvar_{Q,w} = \tvar_{Q,\ol{w}} =
      v \in \{\alice,\allowbreak \bob\}$, and if $(\tvar_{\Pi,w},\allowbreak  \tvar_{\Pi, \ol{w}}) =
      (\Point_{v},\allowbreak  \ALine_{v})$ (resp.
      $\DLine_v$ instead of $\ALine_v$), execute $\decider^\ld_\ldparams$ on
      input $(\Point,\allowbreak  x_{\Pi,w} ,\allowbreak \ALine,\allowbreak  x_{\Pi,\overline{w}},\allowbreak  a_w,\allowbreak  a_{\ol{w}})$
      (resp.\ $\DLine$ instead of $\ALine$), where $\ldparams = (q,m,d,1)$.
      Reject if $\decider^\ld_{\ldparams}$ rejects.
      \label{enu:ar-input-ld}

    \item (\textbf{Proof encoding checks}): If $\tvar_{Q,w} = \tvar_{Q,\ol{w}} =
      \oracle$,
      \begin{enumerate}
			\item (Consistency test) If $(\tvar_{\Pi,w},\allowbreak  \tvar_{\Pi, \ol{w}}) =
        (\Point_i,\allowbreak  \Point_6)$ for some $i \in \{3, 4, 5\}$, reject if
        $\alpha_i \neq \alpha'_i$.
        \label{enu:ar-proof-cons-ld}

      \item (Individual low degree test) If $(\tvar_{\Pi,w},\allowbreak  \tvar_{\Pi,
          \ol{w}}) = (\Point_i,\allowbreak  \ALine_i)$ (resp.\ $\DLine_i$) for some $i \in
        \{3, 4, 5\}$, execute $\decider^\ld_{\ldparams}$ on input $(\Point,\allowbreak 
        x_{\Pi,w} ,\allowbreak \ALine,\allowbreak x_{\Pi,\overline{w}},\allowbreak  a_w,\allowbreak a_{\ol{w}})$ (resp.\ $\DLine$).
        Reject if $\decider^\ld_{\ldparams}$ rejects.
        \label{enu:ar-proof-id-ld}

      \item (Simultaneous low degree test) If $(\tvar_{\Pi, w},\allowbreak  \tvar_{\Pi,
          \ol{w}}) = (\Point_6,\allowbreak  \ALine_6)$ (resp.\ $\DLine_6$), execute
        $\decider^\ld_{\ldparams'}$ on input $(\Point,\allowbreak  x_{\Pi,w},\allowbreak  \ALine,\allowbreak 
        x_{\Pi,\overline{w}},\allowbreak  a_w,\allowbreak  a_{\ol{w}})$ (resp.\ $\DLine$), where
        $\ldparams'=(q, m', d, m'+6)$.
        Reject if $\decider^\ld_{\ldparams'}$ rejects.
        \label{enu:ar-proof-sim-ld}
      \end{enumerate}
      \label{enu:ar-proof-encoding}

    \item (\textbf{Game check}): If $\tvar_{Q,w} = \oracle$, then for $v \in
      \{\alice,\bob\}$, compute $x_{w,v} = L^{v}(x_{Q,w})$.
      If $\tvar_{\Pi,w} = \Point_6$, reject if $\pcpverifier((\sigma,\allowbreak \decider,\allowbreak  n ,\allowbreak  T,\allowbreak 
      \qlen,\allowbreak  x_{w,\alice},\allowbreak  x_{w,\bob}),\allowbreak  (z,\allowbreak  a_{w} ))$ rejects.
      \label{enu:ar-game}
    \end{enumerate}
    If none of the preceding checks rejects, then $\tdans$ accepts.
  }
  \end{gamespec}
  
  \captionof{figure}{The decision procedure $\tdans$.
    Parameters $T,\qlen,q,m,m',d$ are defined in
    Section~\ref{sec:ar-params}.}
  \label{fig:decider-pcp}
  
\endgroup

The typed decider $\tdans$ is implemented in the standard timeout-counter form of
Section~\ref{sec:tms}, with timeout counter bound $B_{\tdans}(n) = c_0 (c_1 n)^{c_2} $ where the constants are chosen (possibly depending on $\sigma$, $\lambda$ and $\mu$) so that the bound dominates both the running
time of the procedure in Figure~\ref{fig:decider-pcp} and the dimension
$s_{\tsans}(n)$ of the typed sampler $\tsans$. Both quantities are bounded by
\[
  \poly(\sigma,(\lambda n)^\mu)~.
\]

\subsubsection{Main theorem for answer reduction}

\begin{theorem}\label{thm:ar}
\hnote{edited this heavily: got rid of $\gamma$, simplified the error function $\delta$, and added the condition that properties hold for all $n \geq C_\ar$}
  \leavevmode\par There exists a polynomial time Turing machine $\ComputeAnsVerifier$ that, on
  input $(\verifier,\allowbreak  \lambda,\allowbreak  \mu,\allowbreak \sigma)$ where
  $\verifier = (\sampler,\allowbreak \decider)$ is an $\ell$-level normal form verifier whose
  decider is in the standard timeout-counter form of Section~\ref{sec:tms}, and
  $\lambda,\allowbreak  \mu,\allowbreak  \sigma \in \N$, outputs the description of the answer reduced
  verifier $\vans = (\sans,\allowbreak  \dans)$.
  Suppose that $\verifier$ satisfies the complexity conditions
\[
|\decider|\,\leq\,\sigma\;,\quad\TIME_\sampler(n) \leq (\lambda n)^\mu \quad
  \text{and}\quad  \TIME_\decider(n) \leq (2^{\lambda n})^\mu
 \]
for all $n \geq 2$.
  Then $\vans$ is $\max\{\ell+2,5\}$-level, the decider $\dans$ is in
standard timeout-counter form of Section~\ref{sec:tms} with timeout bound
$B_{\dans}(n)=\poly(\sigma,(\lambda n)^\mu)$, and $\sans,\dans$ have time
complexity bounds
  \begin{align*}
    \TIME_{\sans}(n) = \poly\bigl( (\lambda n)^\mu, \sigma \bigr)\;, \quad \text{and} \quad
    \TIME_{\dans}(n)
    =  \poly\bigl( (\lambda n)^\mu, \sigma \bigr)~.
  \end{align*}
  Furthermore, the sampler $\sans$ only depends on
  $\sampler$ and the parameter tuple $(\lambda, \mu, \sigma)$, and its dimension $s_{\sans}(n)$ satisfies
  \[
    s_{\sans}(n) \leq B_{\dans}(n)~.
  \]
	Moreover, there exists a universal constant $C_{\ar} \in \N$, and a function
  \begin{equation*}
    \delta(\eps, n) = \sigma^a \cdot \bigl( (\lambda \cdot n)^{\mu a}
    \cdot \eps^b + (\lambda \cdot n)^{- \mu b} \bigr)
  \end{equation*}
  for universal constants $a>0$, $0<b<1$, such that the following hold for all $n \geq C_{\ar}$:
  \begin{enumerate}
  \item (\textbf{Completeness}) If $\verifier_n$ has a projective, consistent,
    and commuting (PCC) strategy of value $1$, then $\vans_n$
    has an SPCC strategy with value $1$.
    \label{enu:ar-completeness}
	\item (\textbf{Soundness}) If $\val^*(\vans_n) > 1 - \eps$ for some $\eps > 0$
    then $\val^*(\verifier_n) \geq 1 - \delta(\eps, n)$.\label{enu:ar-soundness}
  \item (\textbf{Entanglement}) Let $\Ent(\cdot)$ be as defined in Definition~\ref{def:ent}.
    Then for all $\eps\geq 0$,
    \begin{equation*}
      \Ent(\vans_n, 1 - \eps) \geq \frac{1}{2} \Ent(\verifier_{n}, 1 - \delta(\eps, n))\;.
    \end{equation*}
  \end{enumerate}
\end{theorem}
Observe that $\TIME_{\dans}(n)$ is polynomial in the \emph{logarithm}
of $T(n)$, the runtime of the decider of $\verifier$, achieving the
desired reduction in runtime and answer complexity.

\begin{proof}
The Turing machine $\ComputeAnsVerifier$ can be described as follows: 
\begin{enumerate}
\item From the description of $\verifier$, compute the description of the
  (typed) oracularized verifier $\tvora = (\tsora,\allowbreak \tdora)$ using the Turing
  machine\newline $\ComputeOracleVerifier$ from Theorem~\ref{thm:oracle-completeness}.
\item From the description of the typed sampler $\tsora$ and the parameter tuple
  $(\lambda,\mu,\sigma)$, compute the descriptions of the typed sampler $\tsans$
  as described in Section~\ref{sec:ar-verifier}.
\item From the descriptions of the decider $\decider$ and the parameter tuple
  $(\lambda,\mu,\sigma)$, compute the description of the typed decider $\tdans$
  in standard timeout-counter form, as described in Figure~\ref{fig:decider-pcp}.
\item Compute the detyped verifier $\vans=(\sans,\dans)=\detype(\tvans)$.
\end{enumerate}
The Turing machine $\ComputeAnsVerifier$ takes time that is
$\poly(|\verifier|,\allowbreak \log \lambda,\allowbreak \log \mu,\allowbreak \log\sigma)$, because each step,
including the detyping procedure (with its timeout counter), runs in time that is polynomial in the length
of the input $(\verifier,\allowbreak \lambda,\allowbreak \mu,\allowbreak \sigma)$.

\paragraph{Properties of the sampler $\sans$.}
It can be verified via inspection that the typed sampler $\tsans$ depends on
$\tsora$ (which itself only depends on $\sampler$; see
Theorem~\ref{thm:oracle-completeness}), and $\tspcp$ (which only depends on the
parameter tuple $(\lambda,\mu,\sigma)$).
Using Theorem~\ref{thm:oracle-completeness}, $\tsora$ is a $\ell$-level typed
sampler and, therefore, $\tsans$ is a $\max\{\ell, 3\}$-level typed sampler as
$\tspcp$ is a typed $3$-level sampler.
Using Lemma~\ref{lem:detyping-verifiers} for the detyping it follows that
$\sans$ is a $\max\{\ell+2, 5\}$-level sampler.
  
\paragraph{Complexity.}
In addition to the running time of $\tsora$, the time complexity of $\tsans$
also includes the running time of $\tspcp$, which is dominated by the complexity
of computing the CL functions $L_{\Point_i}$, $L_{\ALine_i}$, and $L_{\DLine_i}$.
From Lemma~\ref{lem:ld-complexity}, this takes time 
\begin{equation}\label{eq:t-c-tt} \poly(m', \log q) =
\poly(s) = \poly(\log T, \log n, \qlen , \sigma) = \poly((\lambda n)^\mu,\sigma)\;,
\end{equation}
where the last equality uses the formulas~\eqref{eq:ar-params-1}. Therefore, the time complexity of $\tsans$ is
\[
	\poly \left ( \TIME_{\tsora}(n), \poly((\lambda n)^\mu,\sigma) \right)\;.
\] 
The complexity of $\sans = \detype(\tsans)$ then follows by
Lemma~\ref{lem:detyping-verifiers} and \eqref{eq:ar-time-assumption}.


The decider $\tdans$ executes subroutines $\decider^\ld$ and $\pcpverifier$.
The runtime of $\decider^\ld$ is $\poly(m,d,m',\log q)$.
The runtime of $\pcpverifier$ is given in Theorem~\ref{thm:pcp-decider}; for
$T$ and $\qlen$ as in~\eqref{eq:ar-params-1} it is
$\poly(\log T,\log n,\qlen,\sigma)$.

Theorem~\ref{thm:oracle-completeness} gives that the relevant question length
and CL-function evaluations for $\tsora$ are bounded by a polynomial in
$\TIME_\sampler(n)$, hence by a polynomial in $\qlen(n)=(\lambda n)^\mu$.
On the PCP side, the formats in Table~\ref{table:tpcp} have lengths polynomial
in $m'$ and $\log q$, and Definition~\ref{def:pcpparams} gives
$m',\allowbreak \log q=\poly(\log T,\allowbreak \log n,\allowbreak \qlen,\allowbreak \sigma)=\poly((\lambda n)^\mu,\allowbreak \sigma)$.
Therefore, the parsing checks, the CL-function evaluations, tuple manipulations,
comparisons, and finite-field arithmetic outside the subroutine calls all take
time $\poly((\lambda n)^\mu,\sigma)$.

Thus
\[
  \TIME_{\tdans}(n)=\poly((\lambda n)^\mu,\sigma).
\]
We choose the timeout bound $B_{\tdans}(n)$ to dominate
$s_{\tsans}(n)$, the dimension of the sampler $\tsans$, as well as the running time of $\tdans$. 
Since $s_{\tsans}(n) \leq \TIME_{\tsans}(n) = \poly((\lambda n)^\mu,\sigma)$, we have
\[
  s_{\tsans}(n)
  \leq
  B_{\tdans}(n)
  =
  \poly((\lambda n)^\mu,\sigma).
\]
Consequently, the completeness and soundness conclusions of
Lemma~\ref{lem:detyping-verifiers} apply to $\tvans$, thus giving
\[
  s_{\sans}(n)\leq B_{\dans}(n) = \poly((\lambda n)^\mu,\sigma)~.
\]

In particular,
\[
  \TIME_{\dans}(n)=\poly((\lambda n)^\mu,\sigma).
\]
\paragraph{Completeness, Soundness, and Entanglement.}
The Completeness property is established in Section~\ref{sec:ar-completeness} and the
Soundness and Entanglement properties are proven in Section~\ref{sec:ar-soundness}.
\end{proof}

\subsection{Completeness of the answer-reduced verifier}
\label{sec:ar-completeness}

We establish the Completeness property of the answer reduced verifier as stated
in Theorem~\ref{thm:ar}.


\begin{proof}[Proof of the Completeness part of Theorem~\ref{thm:ar}]
  We establish the Completeness property of $\tvans$, which by the Detyping
  Lemma (Lemma~\ref{lem:detyping-verifiers}) establishes the Completeness property of
  $\vans$.

  Let $n\geq 2$ be an index for $\tvans$.
  Let $\strategy$ be a PCC strategy for $\verifier_n$ with value $1$.
  By Theorem~\ref{thm:oracle-completeness} it follows that there exists a
  symmetric PCC strategy $\strategy^\ora$ using a state $\ket{\psi}$ with value
  $1$ for $\tvora_n$.
  We define a strategy $\strategy^\ar$ for the typed verifier $\tvans_n$ as
  follows.
  The shared state is $\ket{\psi}$.
  Given the index $n$ and $(\lambda,\mu)$, each player can compute
  $(q,m,d,m',s) = \pcpparams(n, T, \qlen,\sigma)$
  (see~Definition~\ref{def:pcpparams}), where $T=T(n)$ and $\qlen=\qlen(n)$ are as
  in~\eqref{eq:ar-params-1}, and $\sigma \geq \abs{\decider}$.
  Let $M = 2^m$.

  \begin{enumerate}
  \item On receipt of a question $((\tvar_Q,\tvar_\Pi),(x_Q, x_\Pi))$ a player
    first measures their share of $\ket{\psi}$ using the projective measurement
    for $\strategy^\ora$ for the typed question $(\tvar_Q,x_Q)$ to obtain an
    outcome $a_Q$.
    The player then computes an answer, depending on $\tvar_Q, a_Q$ and
    $\tvar_\Pi,x_\Pi$, as follows:
    \begin{enumerate}
    \item Suppose $\tvar_Q = v \in \{\alice,\bob\}$ and $\tvar_\Pi \in
      \{\Point_v, \ALine_v \}$ (resp. $\DLine_v$ instead of $\ALine_v$).
      Let $a_Q' = a_Q$ if $a_Q$ has length at most~$T$, and let $a_Q'$ be the
      truncation of~$a_Q$ to its first~$T$ symbols otherwise.
      Let $\ell_a \leq T$ be the length of~$a_Q'$, and set 
      \begin{equation*}
      a_Q'' = \mathrm{enc}_\Gamma(a_Q', \sqcup^{M/2 - \ell_a}).
      \end{equation*}
      Next, the player computes the low-degree encoding $g_{a_Q''}$ of $a_Q''$
      using the low-degree encoding described in Section~\ref{sec:ld-encoding}, in the
      same way as described in Section~\ref{sec:ar-pcp}.
      The player then returns the restriction of $g_{a_Q''}$ to the
      axis-parallel line (resp.
      the diagonal line) specified by $x_\Pi$.
    \item
      If $\tvar_Q=\oracle$, for $v \in \{\alice, \bob\}$ the player computes
      questions $x_v = L^{v}(x_Q)$, as in Step \ref{enu:oracle-game} of
      $\tdora$.
      The player parses $a_Q$ as a pair $(a_\alice, a_\bob)$.
      Let $a_\alice' = a_\alice$ if $a_\alice$ has length at most~$T$, and let
      $a_\alice'$ be the truncation of~$a_\alice$ to its first~$T$ symbols
      otherwise.
      Let $\ell_{\alice} \leq T$ be the length of~$a_\alice'$, and set
      \begin{equation*}
        a_\alice'' = \mathrm{enc}_\Gamma(a_\alice', \sqcup^{M/2 - \ell_{\alice}}).
      \end{equation*}
      Define $a_\bob''$ similarly.
      The player computes a PCP proof $\Pi=(g_1,\allowbreak  \dots,\allowbreak  g_5,\allowbreak  c_0,\allowbreak  \dots,\allowbreak  c_{m'})$ as
      described in the completeness case of Theorem~\ref{thm:pcp-decider} for the tuple
      $(\decider,\allowbreak  n,\allowbreak  T,\allowbreak  x_\alice,\allowbreak  x_\bob)$, where the polynomials $g_1,\allowbreak  g_2$ are
      low-degree encodings of $a_\alice''$ and $a_\bob''$, respectively.
      \begin{enumerate}
      \item If $\tvar_\Pi \in \{\Point_i, \ALine_i\}$ (resp.
        $\DLine_i$ instead of $\ALine_i$) for $i \in \{1, \dots, 5\}$, the
        player returns the restriction of $g_i$ to the axis-parallel line (resp.
        diagonal line) of $\F_q^{m}$ specified by $x_\Pi$.
      \item If $\tvar_\Pi \allowbreak \in\allowbreak  \{\Point_6,\allowbreak  \ALine_6\}$ (resp.
        $\DLine_6$ instead of $\ALine_6$), the player returns the restriction of
        all the polynomials $g_1,\allowbreak  \dots,\allowbreak  g_5,\allowbreak $ $c_0,\allowbreak  \dots,\allowbreak  c_{m'}$ to the
        axis-parallel line of $\F_q^{m'}$ (resp.
        the diagonal line) specified by $x_\Pi$.
      \end{enumerate}
    \item In all other cases the player returns $0$.
    \end{enumerate}
  \end{enumerate}
  The strategy $\strategy^\ar$ is projective and consistent because $\strategy^\ora$
  is.
  To show that it has value $1$, we first observe that by definition it
  satisfies the input formatting checks as well as all consistency checks.
  Moreover, the strategy passes all low-degree tests with certainty because it
  always returns restrictions of consistent polynomials.
  Finally, it also passes the game check with probability $1$.
  This follows from the completeness statement of the PCP made in
  Theorem~\ref{thm:pcp-decider} and the fact that, if $\decider$ accepts the
  input $(n, x_\alice,x_\bob, a_\alice,a_\bob)$ in time at most $T$ then it also
  accepts $(n, x_\alice,x_\bob, a'_\alice,a'_\bob)$ in time at most $T$, where
  $a'_\alice$ and $a'_\bob$ are obtained from $a_\alice$ and $a_\bob$ by
  truncating them to strings of length~$T$ if their lengths exceed~$T$.

	To show that $\strategy^\ar$ is commuting, note that using the product structure
  of $\tsans$ every typed question pair with positive probability consists
  of a pair of questions $((\tvar_{Q,\alice},\allowbreak x_{Q, \alice}),\allowbreak (\tvar_{Q,\bob},\allowbreak 
  x_{Q,\bob})) $ with positive probability for $\tsora$, together with an
  arbitrary pair $((\tvar_{\Pi,\alice},\allowbreak x_{\Pi, \alice}),\allowbreak 
  (\tvar_{\Pi,\bob},\allowbreak  x_{\Pi,\bob}))$.
  Using that $\strategy^\ora$ is commuting and that the additional operations
  associated with $((\tvar_{\Pi,\alice},\allowbreak x_{\Pi, \alice}),\allowbreak 
  (\tvar_{\Pi,\bob},\allowbreak  x_{\Pi,\bob}))$ amount to classical post-processing it
  follows that $\strategy^\ar$ is commuting.

  This establishes the existence of a symmetric PCC strategy for
  $\tvans_n$ with value $1$.
	By Lemma~\ref{lem:detyping-verifiers} it follows that there exists a symmetric
  PCC strategy for $\vans_n = \detype(\tvans)_n$ with value $1$.

\qedhere

\end{proof}

\subsection{Soundness of the answer-reduced verifier}
\label{sec:ar-soundness}


\begin{proof}[Proof of the soundness part of Theorem~\ref{thm:ar}]
  We first show the soundness for the typed verifier $\tvans$.
  Soundness for the detyped verifier $\vans$ follows from
  Lemma~\ref{lem:detyping-verifiers}, with a constant-factor loss using that the
  type set $\type^\ar$ for $\tvans$ has constant size.
	
  We proceed in two steps.
  Fix an index $n\geq 2$ and suppose that $\val^*(\tvans_n) > 1-\eps$ for
  some $\eps > 0$.
  Observe that $\tsora$ and $\tspcp$ both sample distributions that
  are invariant under permutation of the two players; therefore, the same holds
  for $\tsans$.
  Moreover, the decider $\tdans$ treats both players symmetrically.
  Therefore, the game played by $\tvans_n$ is a symmetric game.
  Applying Lemma~\ref{lem:symmetric-strat} it follows that there exists a
  symmetric projective strategy $\strategy = (\ket{\psi},M)$ for
  $\tvans_n$ with value greater than $1 - \eps$.
	
  We use the following shorthand notation.
  A pair of questions to the players is
  $((\tvar_\alice,x'_\alice),(\tvar_\bob,x'_\bob))$ where for
  $w\in\{\alice,\bob\}$, $\tvar_w=(\tvar_{Q,w},\tvar_{\Pi,w})$ and
  $x'_w=(x'_{Q,w}, x'_{\Pi,w})$.
  When $w$ is clear from context we omit it from the subscript.
  Fixing a $w$, whenever $\tvar_Q = \oracle$ we introduce $x_\alice =
  L^\alice(x'_Q)$ and $x_\bob =L^\bob(x'_Q)$ and often write directly the player's
  question as $x_Q = (x_\alice,x_\bob)$.
  Whenever $\tvar_Q=v \in \{\alice,\bob\}$ we slightly abuse notation and write
  the question as $x_Q = (x_{v},v)$, explicitly including the type to clarify
  which player it points to.
	
  We denote the measurements used by both players in strategy $\strategy$ by
  $\{M({x}_Q)^{ {x}_{\Pi}}_{a}\}$, where for the sake of clarity we have
  notationally separated the two parts ${x}_Q$ and $x_\Pi$ of the question and
  omitted explicit mention of the associated types $\tvar_Q$ and $\tvar_\Pi$ (we
  include the type and write $M({x}_Q)^{\tvar_\Pi, {x}_{\Pi}}_{a}$ when it is
  needed for clarity).
  First we show that the strategy $\strategy$ is close to a strategy
  $\strategy'$ that performs ``low-degree'' measurements: upon receipt of a
  typed question $(\tvar,x)=((\tvar_Q,\tvar_\Pi),({x}_Q,x_\Pi))$ a player first
  performs a measurement depending on ${x}_Q$ to obtain a tuple of low-degree
  polynomials, and then returns evaluations of those polynomials on the
  subspaces (either axis-parallel or diagonal lines) specified by $x_\Pi$.
  This step of the argument uses the quantum soundness of the low-degree test
  performed in Steps~\ref{enu:ar-input-ld}
  and~\ref{enu:ar-proof-encoding} of Figure~\ref{fig:decider-pcp}.
  Next, we ``decode'' this strategy to produce a strategy $\strategy''$ for
  $\tvora$ with a high value.
  This step makes use of the classical soundness of the underlying PCP shown in
  Section~\ref{sec:pcp-cktval-new}.
  The conclusion of the theorem then follows from the soundness of
  $\tvora$ (Theorem~\ref{thm:oracle-soundness}).
  We proceed with the details.

	We start by showing a sequence of claims that establish approximations implied
  by the assumption that $\strategy^\ar$ succeeds with probability greater than 
  $1-\eps$ in the decision procedure implemented by the decider in
  Figure~\ref{fig:decider-pcp}.

  \begin{claim}[Global consistency check, Step~\ref{enu:ar-global-consistency}]
    \label{claim:ar-1}
    On average over questions
    $(\tvar_\alice,x_\alice)=((\tvar_Q,\tvar_\Pi),({x}_Q, x_{\Pi}))$ sampled
    from the marginal distribution of $ \mu_{\tsans}$ on the first player
    it holds that
    \begin{equation}
      \label{eq:global-consistency}
      M({x}_Q)^{x_{\Pi}}_a \ot I \simeq_{\eps} I \ot M({x}_Q)^{x_{\Pi}}_a\;.
    \end{equation}
  \end{claim}

  \begin{proof}
    First we observe that the condition $\tvar_\alice = \tvar_\bob$ for the
    global consistency check, Step~\ref{enu:ar-global-consistency} in
    Figure~\ref{fig:decider-pcp}, holds with constant probability over the
    choice of a pair of questions $(\tvar_\alice,x_\alice),(\tvar_\bob,x_\bob)$
    sampled according to $\mu_{\tsans}$.
    Thus $\strategy$ must succeed in this test with probability $1-O(\eps)$,
    conditioned on the test being executed: this is because each of
    $\tsora$ and $\tspcp$ have a constant probability of returning
    a pair of questions of the same type.

    Moreover, observe that conditioned on $\tvar_\alice = \tvar_\bob$ a pair of
    questions $((\tvar_\alice,\allowbreak x_\alice),\allowbreak (\tvar_\alice,\allowbreak x_\bob))\sim
    \mu_{\tsans}$ is such that $x_\alice = x_\bob = L_{\tvar_\alice}(z)$,
    where $z$ is the sampler seed and $L_{\tvar_\alice}$ the CL function of type
    $\tvar_\alice$ associated with $\tsans$.
    The claim then follows directly from the test and the definition of
    approximate consistency (Definition~\ref{def:consistency}).
  \end{proof}

  \begin{claim}[Input consistency check, Step~\ref{enu:ar-input-consistency}]
    \label{claim:ar-2}
    For all $v \in \{\alice,\bob\}$, on average over question pairs
    $({x}_\alice, {x}_\bob) \sim \mu_{\sampler}$ and $z=(y_1,\ldots,y_5,o,w)\in
    \F_q^{m'}$ sampled uniformly at random,
    \begin{equation}
      M({x}_\alice, {x}_\bob)^{\Point_6, z}_{\alpha_v} \ot I \simeq_{\eps} I \ot
      M(x_v,v)^{ \Point_v, y_v}_{\alpha_v}\;, \label{eq:input-consistency}
    \end{equation}
    where as in Remark~\ref{rk:ab-01} we made the identification
    $1\leftrightarrow \alice$ and $2\leftrightarrow \bob$.
    Moreover, an analogous relation holds for operators acting on opposite sides
    of the tensor product.
  \end{claim}

	\begin{proof}
    For $w=\alice$ and fixed $v\in\{\alice,\bob\}$ there is a constant
    probability that $\tvar_{Q,w} = \oracle$, $\tvar_{Q,\ol{w}}=v$, and
    $(\tvar_{\Pi,w},\tvar_{\Pi, \ol{w}}) = (\Point_6, \Point_{v})$.
    Therefore, the input consistency check in Step~\ref{enu:ar-input-consistency} is
    executed with constant probability, and $\strategy$ must pass it with
    probability $1-O(\eps)$, conditioned on the test being executed.

    Moreover, conditioned on $\tvar_{Q,w} = \oracle$, $\tvar_{Q,\ol{w}}=v$, and
    $(\tvar_{\Pi,w},\allowbreak \tvar_{\Pi, \ol{w}}) = (\Point_6,\allowbreak  \Point_{v})$, the
    distribution of $(x_{Q,w},\allowbreak x_{Q,\overline{w}})$ is $((x_\alice,\allowbreak x_\bob),\allowbreak x_v)$
    for $(x_\alice,\allowbreak x_\bob)\sim \mu_\sampler$ and the distribution of
    $(x_{\Pi,w},\allowbreak x_{\Pi,\overline{w}})$ is $(z,\allowbreak y_v)$ for a uniformly random $z
    \in \F_q^{m'}$.
    Eq.~\eqref{eq:input-consistency} then follows directly from the
    specification of the test and the definition of approximate consistency.
    The ``moreover'' part follows from the case $w=\bob$.
  \end{proof}
	
	\begin{claim}[Input low degree test, Step~\ref{enu:ar-input-ld}]\label{claim:ar-3}
    For each $v \in \{\alice,\bob\}$ and for each $x$ in the support of the
    marginal of $\mu_S$ on player $v$ there exists a \tnote{added:}projective measurement
    $\{G^{{x},v}_{g}\} \in \polymeas{m}{d}{q}$ such that the following hold for
    some $\delta_1 = O(\delta_{\ld}(O({\eps}), q, m, d, 1))$, where
    $\delta_{\ld}$ is defined in Theorem~\ref{lem:ld-soundness}.
    For all $v\in\{\alice,\bob\}$, on average over ${x}$ chosen from the
    marginal of $\mu_{\sampler}$ on player $v$ and $y_v\in \F_q^m$ sampled
    uniformly at random,
    \begin{align}
      M({x}, v)^{\Point_v,y_v}_{\alpha}\ot I
      & \simeq_{\delta_1} I \ot G^{{x}, v}_{[\eval_{y_v}(\cdot) =
        \alpha]}\;, \label{eq:ld-single-input} \\
      I\ot  M({x}, v)^{\Point_v,y_v}_{\alpha}
      & \simeq_{\delta_1}  G^{{x}, v}_{[\eval_{y_v}(\cdot) =
        \alpha]} \ot I\;, \label{eq:ld-single-input-b} \\
      G^{{x}, v}_{g} \ot I
      & \simeq_{\delta_1} I \ot G^{{x},
        v}_{g}\;,\label{eq:ld-self-input}
    \end{align}
    where we used the notation $\eval_{y_v}(g)=g(y_v)$ for the evaluation map.
	\end{claim}
	
	\begin{proof}
    Fix $v\in\{\alice,\bob\}$.
    For any $w\in\{\alice,\bob\}$ there is a constant probability that
    $\tvar_{Q,w} = \tvar_{Q,\ol{w}} = v$ and $(\tvar_{\Pi,w}, \tvar_{\Pi,
      \ol{w}}) = (\Point_{v}, \ALine_{v})$ (resp.\ $\DLine_v$).
    Therefore, the input low degree test in Step~\ref{enu:ar-input-ld} of
    Figure~\ref{fig:decider-pcp} is executed with constant probability, and
    $\strategy$ must pass it with probability $1-O(\eps)$, conditioned on the
    test being executed.
			
    Observe that by definition the distribution of $(x_{\Pi,\alice},
    x_{\Pi,\bob})$ conditioned on $\tvar_{Q,w} = \tvar_{Q,\ol{w}} = v$,
    uniformly random $x_Q = (x_v,v)$, and $(\tvar_{\Pi,w}, \tvar_{\Pi, \ol{w}})
    = (\Point_{v}, \ALine_{v})$ (resp.
    $\DLine_v$), where $w\in\{\alice,\bob\}$ is uniformly random, is exactly the
    distribution of questions in the game $\game^\ld$ described in
    Section~\ref{sec:ld-game}, parametrized by $\ldparams = (q, m, d, 1)$.
			
    For every $v\in\{\alice,\bob\}$ and question $x=L^v(z)$ in the support of
    the marginal distribution of $\mu_\sampler$ on player $v$ let $\eps_{x,v}$
    be the probability that $\strategy$ is rejected in Step~\ref{enu:ar-input-ld},
    conditioned on the test being executed and on average over
    $w\in\{\alice,\bob\}$.
    Then $\E[\eps_{x,v}] = O(\eps)$, where the expectation is taken over a
    uniformly random $v\in\{\alice,\bob\}$ and $x=L^v(z)$ for uniformly random
    $z$.

    By definition it follows that the strategy $\tsans$ conditioned on the
    first part of the players' questions being $\tvar_{Q,w} = \tvar_{Q,\ol{w}} =
    v$ and $x_{Q,\alice}=x_{Q,\bob}=x$ is a projective strategy that succeeds
    with probability $1-\eps_{x,v}$ in the low-degree test
    $\decider^\ld_\ldparams$ executed in Item~\ref{enu:ar-input-ld}.
	
    We may thus apply Theorem~\ref{lem:ld-soundness} to obtain projective measurements \[\textstyle \{G^{{x},v}_{g}\}
    \in \polymeas{m}{d}{q}\] such
    that~\eqref{eq:ld-single-input},~\eqref{eq:ld-single-input-b}
    and~\eqref{eq:ld-self-input} each hold with approximation error
    $O(\delta_{\ld}(\eps_{x,v},\allowbreak q,\allowbreak m,\allowbreak d,\allowbreak 1))$.
    Using that for fixed $q,\allowbreak m,\allowbreak d$ the function
    $\eps\mapsto\delta_{\ld}(\eps,\allowbreak q,\allowbreak m,\allowbreak d,\allowbreak 1)$ is concave, the claim follows from
    Jensen's inequality.
  \end{proof}

  \begin{claim}[Proof encoding checks, Step~\ref{enu:ar-proof-encoding}]
    \label{claim:ar-4}
    For each $x_Q=(x_\alice,x_\bob)$ in the support of $\mu_\sampler$ there
    exist measurements $\{G^{(x_\alice,x_\bob),i}_{g}\} \in \polymeas{m}{d}{q}$
    for each $i\in\{3,4,5\}$ and
    \begin{equation*}\{J^{({x}_\alice, {x}_\bob)}_{f_1, \dots, f_5, c_0,
      \dots, c_{m'}}\} \in \simulpolymeas{m'}{d}{q}{m'+6}
      \end{equation*}
      such that
    the following hold for some
    \begin{equation*}
      \delta_2 = O\big(\delta_{\ld}(O({\eps}), q, m, d, 1)+
      \delta_{\ld}(O({\eps}), q, m', d, m'+6)\big)\;.
    \end{equation*}
    First, for all $i\in \{3,\allowbreak 4,\allowbreak 5\}$, on average over $({x}_\alice,\allowbreak  {x}_\bob)
    \sim \mu_{\sampler}$ and $z=(y_1,\allowbreak \ldots,\allowbreak y_5,\allowbreak o,\allowbreak w)$ of type $\Point_6$ sampled
    uniformly at random,
    \begin{equation}
      I \ot M({x}_\alice, {x}_\bob)^{ \Point_i, y_i}_{\alpha_i} \simeq_{\eps}
      M({x}_\alice, {x}_\bob)^{\Point_6, z}_{\alpha_i} \ot I\;.
      \label{eq:ld-cons-pf}
    \end{equation}
    Second, for all $i\in \{3,4,5\}$ and on average over $({x}_\alice, {x}_\bob)
    \sim \mu_{\sampler}$ and $y_i\in\F_q^m$ sampled uniformly at random,
    \begin{align}
      M({x}_\alice, {x}_\bob)^{\Point_i,  y_i}_{\alpha}\ot I
      &\simeq_{\delta_2} I \ot G^{({x}_\alice, {x}_\bob), i}_{[\eval_{y_i}(\cdot) =
        \alpha]}\;, \label{eq:ld-single-pf} \\
      G^{({x}_\alice, {x}_\bob), i}_{g} \ot I
      &\simeq_{\delta_2} I \ot G^{({x}_\alice, {x}_\bob),
        i}_{g}\;. \label{eq:ld-self-pf}
    \end{align}
    Third, on average
    over $({x}_\alice, {x}_\bob) \sim \mu_{\sampler}$ and $z\in\F_q^{m'}$
    sampled uniformly at random,
    \begin{align}
      M({x}_\alice, {x}_\bob)^{\Point_6,  z}_{\alpha_1, \ldots, \alpha_5,
      \beta_0, \ldots, \beta_{m'}} \ot I
      & \simeq_{\delta_2} I \ot J^{({x}_\alice, {x}_\bob) }_{[\eval_{z}(\cdot) =
        (\alpha_1,\ldots,\alpha_5,\beta_0,\ldots,\beta_{m'})]}\;,
        \label{eq:ld-simul-a}\\
      J^{({x}_\alice,{x}_\bob)}_{f_1, \dots, f_5, c_0, \dots, c_{m'}} \ot I
      &\simeq_{\delta_2} I \ot J^{({x}_\alice, {x}_\bob)}_{f_1, \dots, f_5,
        c_0, \dots, c_{m'}}\;.\label{eq:ld-self-simul}
    \end{align}
    Moreover, analogous equations
    to~\eqref{eq:ld-cons-pf},~\eqref{eq:ld-single-pf} and~\eqref{eq:ld-simul-a}
    hold with operators acting on opposite sides of the tensor product.
    Here, recall the definition of $\polymeas{\cdot}{\cdot}{\cdot}$ from
    Definition~\ref{def:ld-meas}.

\end{claim}
	
	\begin{proof} 
    The proof of the first item is similar to the proof of
    Claim~\ref{claim:ar-2}, and we omit it.

    The proof of the second and third items is similar to the proof of
    Claim \ref{claim:ar-3}, and we include more details.
    Fix an $i\in\{3,\allowbreak 4,\allowbreak 5\}$.
    For any $w\in\{\alice,\allowbreak \bob\}$ there is a constant probability that
    $\tvar_{Q,w} = \tvar_{Q,\ol{w}} = \oracle$ and $(\tvar_{\Pi,w},\allowbreak  \tvar_{\Pi,
      \ol{w}}) = (\Point_i,\allowbreak  \ALine_i)$ (or $\DLine_i$ instead of $\ALine_i$), in
    which case the individual low-degree test in~Step~\ref{enu:ar-proof-id-ld}
    is executed.
    Therefore, $\strategy$ must succeed in that part of the test with
    probability $1-O(\eps)$ conditioned on the test being executed.
					
    Furthermore, for fixed $i\in \{3,4,5\}$ and uniformly random
    $w\in\{\alice,\bob\}$, conditioned on the test being executed for that $i$
    and $w$ the distribution of $(x_{\Pi,\alice},x_{\Pi,\bob})$ is exactly the
    distribution of questions in the game $\game^\ld$ described in
    Section~\ref{sec:ld-game}, parametrized by $\ldparams = (q, m, d, 1)$.
			
    For every $i\in\{3,4,5\}$ and $x=(x_\alice,x_\bob)$ in the support of
    $\mu_\sampler$ let $\eps_{x,i}$ be the probability that $\strategy$ is
    rejected in~Step~\ref{enu:ar-proof-id-ld}, conditioned on the test being
    executed for that $i$ and on average over $w\in\{\alice,\bob\}$.
    Then for each $i$, $\E[\eps_{x,i}] = O(\eps)$, where the expectation is
    taken over a uniformly random $x\sim \mu_\sampler$.

    By definition of the individual low-degree test it follows
    from Lemma~\ref{lem:ld-soundness} that for every $x=(x_\alice,x_\bob)$ in the
    support of $\mu_\sampler$ and $i\in\{3,4,5\}$ there is a projective measurement
    $\{G^{({x}_\alice, {x}_\bob), i}_{g}\} \in \polymeas{m}{d}{q}$ such that on
    average over $y_i\in \F_q^{m'}$ of type $\Point_i$ sampled uniformly at
    random, Eq.~\eqref{eq:ld-single-pf} and~\eqref{eq:ld-self-pf} both hold with
    approximation error $O(\delta_{\ld}(\eps_{x,i},q,m,d,1))$.
    Eq~\eqref{eq:ld-single-pf} and~\eqref{eq:ld-self-pf} follow using the concavity
    of $\delta_{\ld}$ as a function of $\eps$.

    Finally we consider the simultaneous low-degree
    test,~Step~\ref{enu:ar-proof-sim-ld}.
    Here as well, using that there is a constant probability that $\tvar_{Q,w} =
    \tvar_{Q,\ol{w}} = \oracle$ and $(\tvar_{\Pi,w}, \tvar_{\Pi, \ol{w}}) =
    (\Point_6, \ALine_6)$ (resp.
    $\DLine_6$) it follows that $\strategy^\ar$ must succeed in that part of the
    test with probability $1-O(\eps)$.
    Using a similar argument as before it follows from Lemma~\ref{lem:ld-soundness}
    (this time for parameters $(q,m',d,m'+6)$) that for every
    $(x_\alice,x_\bob)$ there is a family of projective measurements $\{J^{({x}_\alice,
      {x}_\bob)}_{f_1, \dots, f_5, c_0, \dots, c_{m'}}\} \in
    \simulpolymeas{m'}{d}{q}{m'+6}$ such that on average over $z\in \F_q^{m'}$
    sampled uniformly at random, Eq.~\eqref{eq:ld-simul-a}
    and~\eqref{eq:ld-self-simul} both hold with approximation error
    $O(\delta_{\ld}(O({\eps}), q, m', d, m'+6))$.
  \end{proof}

  The families of projective measurements $\{G^{x_Q,i}_g\}$ and
  $\{J^{x_Q}_{f_1,\ldots,f_5,c_0,\ldots,c_{m'}}\}$, for $x_Q$ in the support of
  $\mu_\sampler$ and $i\in\{1,\ldots,5\}$, whose existence follows from
  Claim~\ref{claim:ar-3} and Claim~\ref{claim:ar-4} have outcomes that are low
  (individual) degree polynomials: for the first family, individual degree $d$
  polynomials $g:\F_q^m\to \F_q$, and for the second, tuples of individual
  degree $d$ polynomials $f_i,c_j:\F_q^{m'} \to \F_q$.
  Recall that $m' = 5m + 5 +s$ and that an element $z\in \F_q^{m'}$ is written
  as a triple $(y,o,w)$ with $x=(y_1,\ldots,y_5)\in \F_q^{5m}$, $o\in\F_q^5$ and
  $w\in\F_q^s$.
  The following claim, whose proof is based on Lemma~\ref{lem:ld-sandwich}, shows
  that we can reduce to a situation where the polynomials $f_1,\ldots,f_5$
  returned by $J$ are such that for each $i\in \{1,\ldots,5\}$, $f_i$ only
  depends on the $y_i$, and not on the entire variable $z$.
	
	\begin{claim}\label{claim:ar-5}
    For all $(x_\alice,x_\bob)$ in the support of $\mu_S$ and individual degree $d$
    polynomials $g_1,\ldots,g_5:\F_q^m\to \F_q$ and $c_0,\ldots,c_{m'}:\F_q^{m'}
    \to \F_q$ define
    \begin{equation}
\begin{aligned}
&\Lambda^{{x}_\alice, {x}_\bob}_{ g_1, \dots, g_5, c_0, \dots, c_{m'}} =\\
      &      G^{{x}_\alice, 1}_{g_1} G^{{x}_\bob, 2}_{g_2} G^{({x}_\alice, {x}_\bob),
        3}_{g_3} G^{({x}_\alice,{x}_\bob),4}_{g_4} G^{({x}_\alice, {x}_\bob),5}_{g_5} J^{({x}_\alice, {x}_\bob)}_{c_0,
        \dots, c_{m'}} \\
      &G^{({x}_\alice, {x}_\bob), 5}_{g_5}
      G^{({x}_\alice,{x}_\bob),4}_{g_4}
      G^{({x}_\alice, {x}_\bob), 3}_{g_3} G^{x_\bob,2}_{g_2} G^{{x}_\alice,1}_{g_1}\;,
      \label{eq:lambda-giant-sandwich}
\end{aligned}
\end{equation}
    where the outcomes $(f_1,\ldots,f_5)$ of the $J$ operator in the middle have
    been marginalized over.
    Then there is a
    \[\delta_3 = O\Big(\Big(\delta_{\ld}({\eps}, q, m', d,
      m'+6)^{1/2}+ \frac{m'd}{q}\Big)^{1/2}\Big)\]
    such that
    \[ \delta_3 \,\geq\, \max\big\{ \eps, \,\delta_1,\,\delta_2\big\}\]
    and on average over $({x}_\alice, {x}_\bob) \sim
    \mu_{\sampler}$ and $z\in \F_q^{m'}$ sampled uniformly at random,
    \begin{equation}
      \Lambda^{{x}_\alice, {x}_\bob}_{[\eval_{z}(\cdot) = ( \alpha, \beta)]} \ot I
      \simeq_{\delta_3} I \ot J^{{x}_\alice, {x}_\bob}_{[\eval_{z}(\cdot)
        = ( \alpha, \beta)]}\;.
    \label{eq:lambda-consistent}
  \end{equation}
	Moreover, a similar equation holds with the operators acting on opposite sides
  of the tensor product.

\end{claim}
	\begin{proof}
    We apply Lemma~\ref{lem:ld-sandwich} with the following setting of parameters.
    The number of sets of functions $k$ is set to $6$.
    The question set $\mathcal{X}$ is set to the support of $\mu_\sampler$, and
    the distribution $\mu$ on it is the distribution $\mu_\sampler$.
    The sets $\mathcal{G}_i$ for $i \in \{1, \dots, 5\}$ consist of individual
    degree $d$ polynomials over $\F_q^{m'}$ that depend only on the $i$-th block
    of $m$ variables.
    The set $\mathcal{G}_6$ consists of $(m' + 1)$-tuples of individual degree
    $d$ polynomials over $\F_q^{m'}$.

    We first verify the assumption on the sets of functions.
    Since all polynomials have individual degree at most $d$ (and, therefore,
    total degree at most $m'd$), by Lemma~\ref{lem:schwartz-zippel} the parameter
    $\eps$ in Lemma~\ref{lem:ld-sandwich} can be set to $m'd/q$.

    The family of measurements $\{A^x_{g_1,\ldots,g_6}\}$ in
    Lemma~\ref{lem:ld-sandwich} is the family of measurements
    $\{J^{(x_\alice,x_\bob)}_{f_1,\ldots,f_5,c_0,\ldots,c_{m'}}\}$ here, where
    we set $g_i = f_i$ for $i\in \{1,\ldots,5\}$ and $g_6 =
    (c_0,\ldots,c_{m'})$.
    The measurements $\{G^{i,x}_g\}$ in Lemma~\ref{lem:ld-sandwich} are
    $\{G^{x_\alice,i}_{g}\}$ for $i\in \{1,2\}$,
    $\{G^{(x_\alice,x_\bob),i}_{g}\}$ for $i\in \{3,4,5\}$, and
    $\{J^{(x_\alice,x_\bob)}_g\}$ for $i=6$.
    To ensure that all polynomials are defined over the same range, we treat
    $g:\F_q^m \to \F_q$ that is an outcome of some $\{G^{i,x}_g\}$ as a
    polynomial $g':\F_q^{m'} \to \F_q$, where the role of the $m$ variables of
    $g$ is taken by the $i$-th block of $m$ variables of $g'$.

    We verify that assumption~\eqref{eq:ld-sandwich-1} in the lemma
    holds with an error $\delta = O((\max\{\eps, \delta_1, \delta_2\})^{1/2})$.
    For $i\in\{1,2\}$ the assumption follows by
    combining~\eqref{eq:input-consistency} and~\eqref{eq:ld-single-input-b}
    with~\eqref{eq:ld-simul-a} and Fact~\ref{fact:data-processing}.
    For $i\in \{3,4,5\}$ we use~\eqref{eq:ld-cons-pf} instead
    of~\eqref{eq:input-consistency} and~\eqref{eq:ld-single-pf} instead of
    ~\eqref{eq:ld-single-input-b}.
    Finally, for $i=6$ we use~\eqref{eq:ld-self-simul} and
    Fact~\ref{fact:data-processing}.
    In these derivations, we use Fact~\ref{fact:triangle-for-simeq}, the
    triangle inequality for ``$\simeq$.''

    The conclusion follows from Lemma~\ref{lem:ld-sandwich}, with an error
    $\delta_3 = O((\delta + m'd/q)^{1/2})$. Observing that
    $\eps,\delta_1,\delta_2 = O(\delta_{\ld}(\eps,q,m',d,m'+6))$, as can be
    verified from the definition of $\delta_{\ld}$ given in
    Theorem~\ref{lem:ld-soundness}, and, therefore, $\delta =
    O(\delta_{\ld}(\eps, q,m',d,m'+6)^{1/2})$, the claimed bound
    \[ \delta_3 = O\Big(\Big( \delta_{\ld}(\eps, q, m', d, m'+6)^{1/2} +
      \frac{m'd}{q} \Big)^{1/2} \Big) \]
    follows.

\end{proof}

At this point, we have constructed measurements $G$ and $\Lambda$ that return
low-degree polynomials in a similar way as is expected from the honest strategy
in $\tvans_n$, as described in the proof of Theorem~\ref{thm:ar}.
These measurements can be used to specify a new strategy $\strategy'$ for the
game $\tvans_n$ as follows.
The shared state remains the state $\ket{\psi}$ used in $\strategy$.
For $w\in \{\alice,\bob\}$, upon reception of a question $(\tvar_w,x_w)$ player
$w$ performs the following.
If $\tvar_w = (\tvar_{Q,w},\tvar_{\Pi,w})$ is such that $\tvar_{Q,w} = \oracle$,
the player measures their share of $\ket{\psi}$ using the measurement
$\Lambda^{x_{Q,w}}$ defined in Claim~\ref{claim:ar-5} to obtain a tuple
$(g_1,\ldots,g_5,c_0,\ldots,c_{m'})$ of polynomials.
The player then answers exactly as in the strategy described in the
``completeness'' part of the proof of Theorem~\ref{thm:ar}.
Similarly, if $\tvar_{Q,w} = v \in \{\alice,\bob\}$ the player first measures
their share of $\ket{\psi}$ using the measurement $G^{x_{Q,w},v}$ from
Lemma~\ref{claim:ar-3} to obtain a polynomial $g$ as outcome; the player then
answers according to the same honest strategy.
	
\begin{lemma}\label{lem:ar-ar}
	The strategy $\strategy'$ succeeds with probability $1-O(\delta_3^{1/2})$ in the
  game $\tvans_n$.
\end{lemma}

\begin{proof}
  First we establish useful consistency relations.
  By combining \eqref{eq:lambda-consistent} and \eqref{eq:ld-simul-a} and applying
  Fact~\ref{fact:data-processing} we obtain that for all $i\in \{1,\ldots,5\}$,
  on average over $({x}_\alice, {x}_\bob) \sim \mu_{\sampler}$ and $z\in
  \F_q^{m'}$ sampled uniformly at random,
  \begin{equation}
    \label{eq:m-lambda-con}
    M({x}_\alice, {x}_\bob)^{\Point_6, z}_{\alpha_i} \ot I
    \simeq_{\delta_3} I \ot  \Lambda^{({x}_\alice, {x}_\bob)}_{[\eval_z(\cdot)_i =
      \alpha_i]}\;,
  \end{equation}
	and a similar equation holds with the operators acting on opposite sides of
  the tensor product.
  Next, combining~\eqref{eq:m-lambda-con} together with
  \eqref{eq:input-consistency} and \eqref{eq:ld-single-input} it follows that for
  each $v \in \{\alice,\bob\}$, on average over $({x}_\alice, {x}_\bob) \sim
  \mu_{\sampler}$ and $z=(y_1,\ldots,y_5,o,w) \in \F_q^{m'}$ sampled uniformly
  at random,
  \begin{equation}\label{eq:m-lambda-con-b}
    G^{{x}_v, v}_{[\eval_{y_v}(\cdot) = \alpha_v]} \ot I \simeq_{\delta_3} I \ot
    \Lambda^{({x}_\alice, {x}_\bob)}_{[\eval_z(\cdot)_v = \alpha_v]}\;.
  \end{equation}
  From the Schwartz-Zippel lemma (Lemma~\ref{lem:schwartz-zippel}) it follows that
  the probability that any two distinct individual degree $d$ polynomials $g_v$
  (an outcome of $G^{{x}_v, v}$) and $g'_v$ (an outcome of
  $\Lambda^{({x}_\alice, {x}_\bob)}$) agree at a uniformly random point $y_v\in
  \F_q^m$ is at most $m'd/q$.
  It thus follows from~\eqref{eq:m-lambda-con-b} that for all
  $v\in\{\alice,\bob\}$, on average over $({x}_\alice, {x}_\bob) \sim
  \mu_{\sampler}$ and $y_v \in\F_q^M$ sampled uniformly at random,
  \begin{equation}
    \label{eq:ld-strat-con-def}
    G^{{x}_v, v}_{g_v} \ot I \simeq_{\delta_3+m'd/q} I \ot \Lambda^{({x}_\alice,
      x_\bob)}_{g_v}\;.
  \end{equation}
  We now show that $\strategy'$ is accepted by $\tvans_n$ with high
  probability.
  We bound the probability of succeeding in each subtest.

  First note that the strategy is accepted in Step 1 of Figure~\ref{fig:decider-pcp}.
  For the $G$ measurements, consistency follows from~\eqref{eq:ld-self-input}.
  For the $\Lambda$ measurements, note first that
  by~\eqref{eq:lambda-consistent} and~\eqref{eq:ld-self-simul} it follows that
  on average over $(x_\alice,x_\bob)\sim \mu_\sampler$,
  \begin{equation}
    \Lambda^{{x}_\alice, {x}_\bob}_{[\eval_{z}(\cdot) = ( \alpha, \beta)]} \ot I
    \simeq_{\delta_3} I \ot \Lambda^{{x}_\alice, {x}_\bob}_{[\eval_{z}(\cdot)
      = ( \alpha, \beta)]} \;.
    \label{eq:lambda-consistent-b}
  \end{equation}
	Using that all outcomes of $\Lambda^{{x}_\alice, {x}_\bob}$ are individual degree $d$
  polynomials and the Schwartz-Zippel lemma (Lemma~\ref{lem:schwartz-zippel}) it
  follows that whenever a measurement of $\Lambda^{{x}_\alice, {x}_\bob}\ot
  \Lambda^{{x}_\alice, {x}_\bob}$ returns distinct outcomes, the outcomes take a
  different value at a uniformly random $z$ with probability at least $1-m'd/q$.
  It then follows from~\eqref{eq:lambda-consistent-b} and the fact that
  $\delta_3 \geq m'd/q$ by definition that the strategy $\strategy'$ is accepted
  in Step 1 with probability $1 - O(\delta_3)$.

  Next, the strategy is also accepted in the consistency check performed in item
  2 due to \eqref{eq:ld-strat-con-def}, and the consistency check in item 4(a)
  for the same reasons as for item 1.
  Finally, for the low-degree tests performed in item 3 and items 4(b) and 4(c),
  the strategy succeeds due to consistency and the fact that, as long as both
  players obtain the same polynomial outcomes, they pass the low-degree tests
  with probability 1.

  It remains to analyze the strategy's success probability
  in Step ~\ref{enu:ar-game}, the game check. Recall that this check
  is nontrivial only when $\tvar_{Q,w} = \oracle$ and $\tvar_{\Pi,w} =
  \Point_6$. By assumption the original strategy $\strategy$ succeeds with
  probability $1-O(\eps)$ in the game check. Thus, it will suffice to
  show that the answer distribution from $\strategy'$ is close to the
  answer distribtion from $\strategy$ whenever $\tvar_{Q,w} = \oracle$
  and $\tvar_{\Pi,w} = \Point_6$ for either or both players $w \in
  \{\alice, \bob\}$. For these question types, the measurement
  operator used by $\strategy$ is $M(x_\alice, x_\bob)^{\Point_6,
    z}_{(\alpha, \beta)}$, and
  the operator used by $\strategy'$ is $\Lambda^{(x_\alice,
    x_\bob)}_{[\eval_z(\cdot) = (\alpha, \beta)]}$. By converting
  \eqref{eq:m-lambda-con} and \eqref{eq:lambda-consistent-b} to
  $\approx$ bounds using Fact~\ref{fact:agreement}, and then applying the
  triangle inequality (Fact~\ref{fact:triangle}) we obtain
  that on average over $(x_\alice, x_\bob) \sim \mu_\sampler$,
  \[ \Lambda^{x_\alice, x_\bob}_{[\eval_{z}(\cdot) = (\alpha, \beta)]}
    \ot I \approx_{\delta_3} M(x_\alice,
    x_\bob)^{\Point_6, z}_{(\alpha, \beta)} \ot I_{\bob}. \]
  Thus, restricting to the game check, $\strategy$ and $\strategy'$
  are $\delta_3$-close in the sense of
  Definition~\ref{def:strategy-distance}. They both use the same state, and,
  moreover, $\strategy$ is projective. Therefore, by Lemma~\ref{lem:close-strategies-have-close-values}, this implies that $\strategy'$ must
  succeed in the game check with probability $1 - \eps -
  O(\delta_3^{1/2}) = 1 - O(\delta_3^{1/2})$.  
  Thus, we have shown that $\strategy'$ succeeds in all subtests of the
  game with probability $1 - O(\delta_3^{1/2})$, and thus the lemma follows.
\end{proof}

We now complete the proof by a reduction to the game $\verifier_n^\ora$: from
the strategy $\strategy'$ we construct a symmetric strategy $\strategy'' =
(\ket{\psi}, A)$ for $\tvora$ by ``decoding'' the low-degree
measurements $G$ and $\Lambda$.
The state $\ket{\psi}$ in $\strategy''$ is identical to the state used in
$\strategy'$ (which is identical to the state used in $\strategy$).
To begin, we define a decoding map $\Delta(\cdot)$, which takes in a polynomial
$g:\F_q^m \to \F_q$ and outputs a string in $\{0,1\}^*$.
This map is computed as follows:
\begin{itemize}
\item First, compute $a = \coded(g) \in \{0,1\}^M$, where $\coded$ is the
  (boolean) decoding of the low-degree code defined in Section~\ref{sec:ld-encoding}.
\item If there exists an $a_{\mathrm{prefix}} \in \{0,\allowbreak  1\}^*$ of length $\ell_a \leq T$
  such that $a = \mathrm{enc}_\Gamma(a_{\mathrm{prefix}},\allowbreak  \sqcup^{M/2 - \ell_a})$,
  then $\Delta(g) = a_{\mathrm{prefix}}$.
  Otherwise, $\Delta(g)$ is allowed to be arbitrary.
\end{itemize}
We can now define the ``decoded'' measurements $\{A^{{x}_v,v}\}$ and
$\{{A}^{{x}_\alice,{x}_\bob}\}$ as follows:
\begin{equation}
  \label{eq:decoded-meas}
  A^{{x}_v,v}_a  = G^{{x}_v, v}_{[\Delta(\cdot)=a]} \;,\quad
  A^{{x}_\alice,{x}_\bob}_{a_\alice, a_\bob}  =
  \Lambda^{{x}_\alice,{x}_\bob}_{[\Delta(\cdot)_{\alice,\bob}=(a_\alice,a_\bob)]}\;.
\end{equation}
  
	
\begin{lemma}\label{lem:ar-ora}
  The strategy $\strategy''$ succeeds with probability $1-O(\delta_3^{1/2})$ in the
  game $\tvora_n$.
\end{lemma}
	
\begin{proof}
  We consider the different subtests executed by $\tdora$ (see
  Figure~\ref{fig:oracle-decider}).
  We start with Step~\ref{enu:oracle-full-consistency}, the consistency checks.
  Success in the first check, Step~\ref{enu:check-same}, follows from the success of
  $\strategy'$ in the global consistency check,
  Step~\ref{enu:ar-global-consistency} of $\tdans$, the
  definition~\eqref{eq:decoded-meas}, and the fact that conditioned on
  $\tvar_{Q,\alice}=\tvar_{Q,\bob}=\oracle$, the distribution of
  $(x_{Q,\alice},x_{Q,\bob})$ in $\tvans_n$ is the same as the
  distribution of $(x_{\alice},x_{\bob})$ in $\verifier_n^\ora$, conditioned on
  $\tvar_\alice = \tvar_\bob = \oracle$.
  Similarly, success in the second check, Step~\ref{enu:oracle-versus-player},
  follows from success of $\strategy'$ in the input consistency check,
  Step~\ref{enu:ar-input-consistency} of $\tdans$.

  Next we consider the game check of $\tdora$,
  Step~\ref{enu:oracle-game}.
  To analyze the success probability of $\strategy''$ we use that $\strategy'$
  succeeds in the game check of $\tdans$, Step~\ref{enu:ar-game}, and the
  soundness of the PCP, as shown in Theorem~\ref{thm:pcp-decider}.
  Let $p_{\soundness}$ be as in Theorem~\ref{thm:pcp-decider}.

  Let $w\in\{\alice,\allowbreak \bob\}$, $(x_\alice,\allowbreak x_\bob)$ be in the support of
  $\mu_\sampler$, and $\Pi = (g_1,\allowbreak \ldots,\allowbreak g_5,\allowbreak c_0,\allowbreak \ldots,\allowbreak c_{m'})$ an outcome of
  $\Lambda^{x_\alice,x_\bob}$ such that conditioned on that outcome being
  obtained by player $w$ in the game check of $\tdans$, $\pcpverifier$
  accepts the pair of inputs $(\sigma,\allowbreak \decider,\allowbreak  n ,\allowbreak  T,\allowbreak  \qlen,\allowbreak  x_{\alice},\allowbreak  x_{\bob})$
  and $(z,\allowbreak  a_{w} )$ with probability at least $p_{\soundness}$ over the choice of
  a uniformly random $z\in \F_q^{m'}$ and $a_w = \eval_z(\Pi)$.
			
  For any such $\Pi$, the soundness statement of Theorem~\ref{thm:pcp-decider}
  states that there exist $a_\alice, a_\bob \in \{0,1\}^*$ such that
  $\decider(n, {x}_\alice, {x}_\bob, a_\alice, a_\bob) = 1$ and, furthermore,
  $a_v = \Delta(g_v)$ for $v \in \{\alice,\bob\}$.   
  It follows that for any proof $\Pi$ returned by
  $\{\Lambda^{{x}_\alice,{x}_\bob}_\Pi\}$ which is accepted with probability
  greater than $p_{\soundness}$ in the game check of $\tdans$ it holds
  that $\decider(n, {x}_\alice, {x}_\bob, \Delta(g_\alice), \Delta(g_\bob)) =
  1$.
  Using this observation we evaluate the probability $q_g''$ that the strategy
  $\strategy''$ succeeds in the game check of $\tvora$.
  Let $q_g'$ be the probability that $\strategy'$ succeeds in the game check of
  $\tdans$.
  \begin{align*}
    &q_g'' \\
 &= \Es{({x}_\alice,{x}_\bob) \sim \mu_{\sampler}} \sum_{a_\alice,
            a_\bob: \decider(n,{x}_\alice,{x}_\bob, a_\alice, a_\bob) = 1}
            \\
          &\bra{\psi} A^{({x}_\alice,{x}_\bob)}_{a_\alice, a_\bob} \ot I
            \ket{\psi}    \\
          &= \Es{({x}_\alice,{x}_\bob) \sim \mu_{\sampler}} \sum_{\Pi
            : \decider(n,{x}_\alice,{x}_\bob, \Delta(g_1), \Delta(g_2)) = 1}
            \\
          &\bra{\psi} \Lambda^{({x}_\alice,{x}_\bob)}_{\Pi} \ot I
            \ket{\psi} \\
          &\geq \Es{({x}_\alice,{x}_\bob) \sim \mu_{\sampler}} \sum_{\Pi
            : \decider(n,{x}_\alice,{x}_\bob, \Delta(g_1), \Delta(g_2)) = 1}
            \\
          &\bra{\psi} \Lambda^{({x}_\alice,{x}_\bob)}_{\Pi} \ot I
            \ket{\psi} \cdot \Pr_{z \sim \F_q^{m'}} [
            \pcpverifier(z,\eval_z(\Pi)) = 1]\\
          &= q_g' - \\
          &\Es{({x}_\alice,{x}_\bob) \sim \mu_{\sampler}} \sum_{\Pi
            : \decider(n,{x}_\alice,{x}_\bob, \Delta(g_1), \Delta(g_2)) = 0}
            \\
          &\bra{\psi} \Lambda^{({x}_\alice,{x}_\bob)}_{\Pi} \ot I
            \ket{\psi} \cdot \Pr_{z \sim \F_q^{m'}} [
            \pcpverifier(z,\eval_z(\Pi))= 1] \\
          &\geq q_g' - ( 1- q_g'')\cdot
            p_{\soundness} \;.
  \end{align*}
  Rearranging terms,
  \begin{equation}
    q_g'' \geq \frac{q_g' - p_{\soundness}}{1 - p_{\soundness}} 
    = 1 - \frac{1-q_g'}{1 - p_{\soundness}}\;. \label{eq:pgameora}
  \end{equation}
  Altogether, using Lemma~\ref{lem:ar-ar} we have shown that $\strategy''$ is
  accepted in each subtest performed by $\tdora$ with probability at
  least $1 - O(\delta_3^{1/2})$.
  Since every subtest occurs with constant probability, the lemma follows.
  \end{proof}
	
	To conclude the proof of the soundness, we appeal to the soundness statement
  for $\tvora$, given in Theorem~\ref{thm:oracle-soundness}.
  Now we establish the form of the error function $\delta$ as stated in the
  theorem. 
  
\hnote{june 14: simplified the calculations below.}  Let $a' \geq 1, 0 < b' < 1$ be the universal constants from Lemma~\ref{lem:ld-soundness} such that
  \[
    \delta_\ld(\eps,q,m',d,t) = a' (dm't)^{a'} (\eps^{b'} + q^{-b'} + 2^{-b' m' d}).
 	\]
(Note that we use $m'$ as the number of variables). 
	Let $m'$ be as set in the PCP parameters from Definition~\ref{def:pcpparams}. We set $t = m' + 6$. 
By definition of $q$ in Definition~\ref{def:pcpparams} and $\qlen$ in \eqref{eq:ar-params-1} we have that
	\[
\frac{a'(dm't)^{a'}}{q^{b'}} \leq \frac{1}{(\lambda n)^{\mu}}~.
	\]
Since $m \leq s$  from
Proposition~\ref{prop:explicit-padded-succinct-deciders}, $m' = 5m + 5
+ s$ from Definition~\ref{def:pcpparams}, and
Proposition~\ref{prop:explicit-padded-succinct-deciders} and \eqref{eq:ar-time-assumption} imply $s = \poly( (\lambda n)^\mu , \sigma)$, we have $a' (dm' t)^{a'} = \poly((\lambda n)^\mu,\sigma)$.
Since $2^{m'} \geq T = (2^{\lambda n})^\mu$ by Proposition~\ref{prop:explicit-padded-succinct-deciders} and \eqref{eq:ar-time-assumption}, we have 
	\begin{equation}
	\frac{a'(dm't)^{a'}}{2^{b' m' d}} \leq \frac{\poly((\lambda n)^\mu,\sigma)}{(2^{\Theta(\lambda n)})^\mu} \leq \frac{\poly(\sigma)}{(\lambda n)^{\mu}}
	\label{eq:ar-sound-3}
	\end{equation}
	for all $n$ larger than some universal constant. This means that we can bound
	\[
	\delta_\ld(\eps,q,m',d,t) \leq \poly((\lambda n)^\mu, \sigma) \eps^{b'} + \frac{\poly(\sigma)}{(\lambda n)^\mu}
	\]
	for all sufficiently large $n$. Finally, we have 
	\[
		\frac{m'd}{q} \leq \frac{a'(dm't)^{a'}}{q^{b'}} \leq \frac{1}{(\lambda n)^{\mu}}
	\]
	because $a',t \geq 1$, and $0 < b' < 1$. 
	Putting everything together we have
there exist universal constants $C_\ar \in \N$, $a > 1$
  and $0 < b < 1$ such that for all $n \geq C_\ar$,
  \begin{align*}
    \delta_3^{1/2}
    & = O \left ( \Big (\delta_\ld(\eps, q, m', d, m'+6)^{1/2} +
      \frac{m'd}{q} \Big)^{1/4} \right ) \\
    & \leq \sigma^a \Big ( (\lambda \cdot n)^{\mu a} \cdot \eps^{b}
      +  (\lambda \cdot n)^{- \mu b } \Big )~.
  \end{align*}

To prove the entanglement bound, let $\strategy_0$ be any strategy for
$\vans_n$ of value at least $1-\eps$, and let $r$ be the Schmidt rank of its
state.  The soundness construction in
Lemma~\ref{lem:detyping-verifiers} produces a strategy for $\tvans_n$ using
the same state; the constant-factor change in the error is absorbed into the
definition of $\delta(\eps,n)$ because $|\type^\ar|$ is constant.

Applying Lemma~\ref{lem:symmetric-strat} to this strategy produces the
symmetric projective strategy $\strategy$ used above, whose state has Schmidt
rank at most $2r$.  All subsequent constructions, including the strategies
$\strategy'$ and $\strategy''$, use this same symmetrized state.  The resulting
strategy for $\verifier_n$, therefore, has Schmidt rank at most $2r$ and value at
least $1-\delta(\eps,n)$.  Hence
\[
  \Ent(\verifier_n,1-\delta(\eps,n))\leq 2r.
\]
Minimizing over all such strategies $\strategy_0$ gives
\[
  \Ent(\vans_n,1-\eps)
  \geq
  \frac{1}{2} \Ent(\verifier_n,1-\delta(\eps,n)),
\]
as claimed.
\end{proof}



\section{Parallel repetition}
\label{sec:parrep}

\def\tvanch{\hat{\verifier}^\anch}
\def\tdanch{\hat{\decider}^\anch}
\def\tsanch{\hat{\sampler}^\anch}
\def\vanch{\verifier^\anch}
\def\danch{\decider^\anch}
\def\sanch{\sampler^\anch}

In each of the transformations on verifiers presented so far (introspection,
oracularization, and answer reduction), the \emph{soundness gap} of the
resulting verifier is slightly degraded: while the completeness property (i.e.,\
the property of having a PCC strategy with success probability $1$) is
preserved, if the starting game $\verifier_n$ has value at most $1 - \eps$, the
resulting game $\verifier_n'$ has value at most $1 - \delta(\eps, n)$ for some
function $\delta$.
In order to apply the compression procedure recursively we need a way to restore
the soundness gap after a sequence of transformations.
We accomplish this using (a modification of) the technique of \emph{parallel
  repetition}.
  This is a transformation of a two-player game $\game$ into another two-player game $\game^k$
  where the verifier samples $k$ independent pairs $(x_1,y_1),\ldots,(x_k,y_k)$ from the
  question distribution $\mu$ of $\game$, sends the tuple $(x_1,\ldots,x_k)$ to the first player
  to obtain an answer tuple $(a_1,\ldots,a_k)$, and sends the tuple $(y_1,\ldots,y_k)$ to the second
  player to obtain an answer tuple $(b_1,\ldots,b_k)$. The players win if and only if $D(x_i,y_i,a_i,b_i) = 1$ 
  for all $i$.  

Intuitively, if the value of $\game$ is $v < 1$, then one would expect the value
of $\game^k$ to decay exponentially with the number of repetitions $k$.
It is not true in general that the value of $\game^k$ is $v^k$, but exponential
decay bounds on the (tensor product) value of parallel-repeated games are known
for specific classes of
games~\cite{jain2014parallel,dinur2015parallel,bavarian2017hardness}.
In particular, it was shown in~\cite{bavarian2017hardness} that the class of
\emph{anchored games} satisfies exponential-decay parallel repetition, and,
furthermore, every game can be efficiently transformed into an equivalent
anchored game.
Put together, this gives a general soundness amplification procedure called
``anchored parallel repetition,'' which we use in our compression procedure to
reset the soundness gap to a fixed constant.

We describe the ``anchoring'' transformation of a game $\game$ in more detail. The anchoring of $\game$ is another game $\game_\perp$ where the verifier behaves as follows: it samples a question pair $(x,y)$ from the question distribution $\mu$ of
$\game$, and then for each player independently changes their question to a special symbol $\perp$ (not part of the original game $\game$) with probability $1/2$ to
obtain new questions $(x',y')$. The first player receives question $x'$ and responds with answer $a$; the second player receives question $y'$ and responds with answer $b$. If neither player receives the question $\perp$, then they win if and only if $D(x',y',a,b) = 1$ where $D$ is the decision predicate of $\game$. Otherwise, the players win if and only if the players receiving the question $\perp$ output a specific, fixed answer (e.g., $0$). 

It is easy to see that $\val^*(\game_\perp) = \frac{3}{4} + \frac{1}{4} \val^*(\game)$. In particular, we have that $\val^*(\game_\perp) = 1$ if and only if $\val^*(\game) = 1$. Bavarian, Vidick and Yuen proved the following bound on the entanglement requirements of $\game_\perp^k$, which is the $k$-fold repetition of $\game_\perp$. 

\begin{theorem}[Parallel repetition of anchored games, Theorem 6.1 of \cite{bavarian2017hardness}]
\label{thm:bvy}
There exists a universal constant $c > 0$ such that for all two-player games $\game$ and for all positive integers $k$, for all $0 < \eps \leq 1$, for all $p$ satisfying
\[
 p > \frac{4}{\eps}\, \exp \left ( - \frac{c \, \eps^{17} \, k}{s} \right)
\]
where $s$ denotes the bit-length of the players' answers in the game $\game$, we have
\[
	\Ent(\game_\perp^k,p) \geq \Ent(\game,1-\eps)~.
\]
\end{theorem}
We note that a corollary of~Theorem~\ref{thm:bvy} is that if $\val^*(\game) < 1 - \eps$, then $\val^*(\game_\perp^k) \leq p$ for $p = \frac{4}{\eps}\, \exp \left ( - \frac{c \, \eps^{17} \, k}{s} \right)$. This is because if $\val^*(\game_\perp^k)$ were larger than $p$, there would be a finite upper bound on $\Ent(\game_\perp^k,p)$, but on the other hand $\Ent(\game,1-\eps)$ is by definition infinite, a contradiction. 

In this section, we present the anchoring and parallel repetition transformations on normal form verifiers; starting with a normal form verifier $\verifier = (\sampler,\decider)$, we first apply the anchoring transformation to obtain a normal form verifier $\vanch = (\sanch,\danch)$, then apply parallel repetition to obtain a normal form verifier $\verifier^\rep = (\sampler^\rep,\decider^\rep)$. The main theorem of the section, Theorem~\ref{thm:repetition}, establishes the properties of the verifier $\verifier^\rep$.


\begin{remark}
The parallel repetition theorems of~\cite{dinur2015parallel,yuen2016parallel}
are also applicable for the purpose of amplification of the soundness gap, but are not
sufficient for us. This is because the (proofs of the) parallel repetition theorems
of~\cite{dinur2015parallel,yuen2016parallel} only imply that $\Ent(\game^k,v)
\geq \log \Ent(\game,1 - \eps)$, which is not sufficient for the recursive compression application; we need a larger entanglement lower bound 
on $\game^k$. The weaker entanglement bound is due to the use of the
  ``quantum correlated sampling lemma'' of~\cite{dinur2015parallel}; it is an open question whether there is
  an improved analysis that obtains a better entanglement lower bound.
%
\end{remark}

\subsection{The anchoring transformation} 
\label{sec:anchoring}

We present the anchoring transformation on normal form verifiers
$\verifier=(\sampler,\decider)$ whose decider is in the standard
timeout-counter form of Section~\ref{sec:tms}.  Let $B_{\decider}(n)$ denote
the timeout bound of $\decider$.  The transformation produces another normal
form verifier $\vanch=(\sanch,\danch)$, also in standard timeout-counter form.
%
We first define a typed verifier $\tvanch = (\tsanch,
\tdanch)$, and then detype $\tvanch$ using
Lemma~\ref{lem:detyping-verifiers} to obtain $\vanch$.

Define the type set $\type^\anch = \{\Game, \Anchor\}$ and type graph
$G^\anch$ to be the complete graph over $\type^\anch$ along with self-loops at
each vertex. The Turing machine $\tsanch$ is defined as follows:
\begin{enumerate}
  \item On input $(n,\gamestyle{dimension})$, it returns
  $\sampler(n,\gamestyle{dimension})$.

  \item On input $(n,\gamestyle{field})$, it returns
  $2$.

  \item On input $(n,w,\gamestyle{marginal},j,z,\tvar)$, it first computes
  $s=\sampler(n,\gamestyle{dimension})$. It checks that $\tvar\in\{\Game,\Anchor\}$
  and that $z$ has length exactly $s$. If either check fails, it halts with a
  fixed default output. If $\tvar=\Game$, it returns
  $\sampler(n,w,\gamestyle{marginal},j,z)$; if $\tvar=\Anchor$, it returns the
  binary representation of the zero vector in $\F_2^s$.

  \item On input $(n,w,\gamestyle{linear},j,u,y,\tvar)$, it first computes
  $s=\sampler(n,\gamestyle{dimension})$. It checks that $\tvar\in\{\Game,\Anchor\}$
  and that $u$ and $y$ each have length exactly $s$. If any check fails, it
  halts with a fixed default output. If $\tvar=\Game$, it returns
  $\sampler(n,w,\gamestyle{linear},j,u,y)$; if $\tvar=\Anchor$, it returns the
  binary representation of the zero vector in $\F_2^s$.

  \item On input $(n,w,\gamestyle{factor},j,u,\tvar)$, it first computes
  $s=\sampler(n,\gamestyle{dimension})$. It checks that $\tvar\in\{\Game,\Anchor\}$
  and that $u$ has length exactly $s$. If either check fails, it halts with a
  fixed default output. If $\tvar=\Game$, it returns
  $\sampler(n,w,\gamestyle{factor},j,u)$; if $\tvar=\Anchor$, it returns the
  binary representation of the zero vector in $\F_2^s$.
\end{enumerate}

Define the Turing machine $\tdanch$ that, on input
$(n,\tvar_\alice,x_\alice,\tvar_\bob,x_\bob,a_\alice,a_\bob)$, if either
$\tvar_\alice$ or $\tvar_\bob$ is equal to the type $\Anchor$, then the decider
accepts as long as the players receiving the $\Anchor$-type answer with $0$ (a player who receives the $\Game$ type can respond with any answer). 
Otherwise, it accepts only if $\decider(n,x_\alice,x_\bob,a_\alice,a_\bob)$
accepts. In preparing the input to $\decider$, it checks that each of the strings $x_\alice,x_\bob,a_\alice,a_\bob$ has length at most $\poly(B_\decider(n))$. If any 
string is longer, $\tdanch$ rejects. 

The typed decider $\tdanch$ is implemented in standard timeout-counter form with
\[
  B_{\tdanch}(n)=\poly(B_\decider(n)).
\]
The polynomial is chosen to dominate the runtime of $\tdanch$ and to satisfy
$B_{\tdanch}(n)\geq B_\decider(n)$. Let $s(n)$ denote the dimension of the sampler $\sampler$. Assuming $s(n)\leq B_\decider(n)$, we have
\[
  s_{\tsanch}(n)=s(n)\leq B_{\tdanch}(n).
\]

Finally, define the Turing machines $(\sanch,\allowbreak \danch)$ to be the result of the detyping transformation $\detype(\tvanch)$ on the pair $\tvanch \allowbreak =\allowbreak  (\tsanch,\allowbreak \tdanch)$. By Lemma~\ref{lem:detyping-verifiers}, the detyped decider $\danch$ is also in the standard timeout-counter form of Section~\ref{sec:tms}, with timeout bound $B_{\danch}(n) \allowbreak =\allowbreak  \poly(B_{\tdanch}(n)) \allowbreak =\allowbreak  \poly(B_{\decider}(n))$, and, furthermore, $s_{\sanch}(n) \allowbreak \leq\allowbreak  B_{\danch}(n)$. 

The transformation above is defined for normal form verifiers whose decider is in
the standard timeout-counter form of Section~\ref{sec:tms}.  The descriptions of
$\sanch$ and $\danch$ are computable in time polynomial in the descriptions of
$\sampler$ and $\decider$. The following proposition records the completeness, soundness, and complexity
properties of the resulting verifier.




\begin{proposition}
  \label{prop:anchoring}
  If $\verifier=(\sampler,\decider)$ is a normal form verifier whose decider is in
the standard timeout-counter form of Section~\ref{sec:tms}, with timeout bound
$B_{\decider}(n)$ satisfying $s_\sampler(n) \leq B_\decider(n)$ where $s_\sampler(n)$ is the dimension of the sampler $\sampler$, then $\vanch=(\sanch,\danch)$ is a normal form verifier satisfying the following
   for all $n \in \N$.  
  \begin{enumerate}
  \item \label{enu:anch-completeness} (\textbf{Completeness}) If there is a
    value-$1$ PCC strategy for $\verifier_n$, then there is a value-$1$ PCC
    strategy for $\vanch_n$.
  \item \label{enu:anch-soundness} (\textbf{Soundness}) For all $\eps > 0$, we have \[\textstyle \Ent(\vanch_n,\allowbreak 1 - \eps) \geq \Ent \left (\verifier_n,\allowbreak  1 - (4
        \cdot 16^2) \eps \right).\]
%
  \item \label{enu:anch-complexity} (\textbf{Complexity}) The decider $\danch$ is
  in the standard timeout-counter form of Section~\ref{sec:tms}, with timeout
  bound \tnote{This is not true, is it, because of the dependence on $k$? Say $B_{\decider}(n)$ is zero or log } \hnote{There's no $k$ here, this is just the anchoring transformation.}
  \[
    B_{\danch}(n)=\poly(B_{\decider}(n)).
  \]
  The time complexities satisfy
  \begin{align*}
    \TIME_{\sanch}(n) & = \poly(\TIME_\sampler(n)),\\
    \TIME_{\danch}(n) & = \poly(B_{\decider}(n)).
  \end{align*}
    Furthermore, the number of levels of $\sanch$ is $\ell + 2$ and the dimension is $s_{\sanch}(n) =s(n)+8$, where the number of levels and dimension of sampler $\sampler$ is $\ell$ and $s(n)$, respectively. Moreover, we have that the dimension of $\sanch$ satisfies $s_{\sanch}(n) \leq B_{\danch}(n)$. 
  \item \label{enu:anch-efficient-computability} (\textbf{Efficient
      computability}) The descriptions of $\sanch$ and $\danch$
    can be computed in polynomial time from the descriptions of $\sampler$ and
    $\decider$.  In particular, the sampler $\sanch$ only depends on the sampler
    $\sampler$.
  \end{enumerate}
\end{proposition}

\begin{proof}
If $\sampler$ is an $\ell$-level sampler with dimension $s(n)$, then by construction $\tsanch$ is a $(\type^\anch, G^\anch)$-typed sampler, with
field size $q(n) = 2$ and the same dimension $s(n)$. It has the following CL functions. Fix an integer $n \in \N$.
Let $V = \F_2^{s(n)}$ denote the ambient space of $\sampler$ on index $n$.
Let $L^\alice, L^\bob: V \to V$ denote the CL functions of $\sampler$ on index
$n$.
For $w\in\{\alice,\bob\}$ the associated CL functions $\{ L^{\anch,\, w}_\tvar
\}$ of $\tsanch$ are
\begin{equation*}
	L^{\anch,\, w}_\tvar =
  \begin{cases}
		 \; L^w & \quad \text{ if } \tvar = \Game \;, \\
		 \; 0 & \quad \text{ if } \tvar = \Anchor \;.
	\end{cases}
\end{equation*}
Intuitively, when the type $\tvar$ sampled for player $w$ is $\Game$, then they
receive a question $L^w(z)$ as they would according to $\sampler$.
Otherwise if $\tvar = \Anchor$, then their question is the zero string.
Thus if $L^w$ is an $\ell$-level CL function, then $L^{\anch,\, w}_\Game$ is
also an $\ell$-level CL function, and $L^{\anch,\, w}_\Anchor$ is a $0$-level CL
function.

  We analyze the completeness, soundness, and complexity properties of the typed
  verifier $\tvanch$; the corresponding properties of the
  detyped verifier $\vanch$ follow from
  Lemma~\ref{lem:detyping-verifiers} and the fact that the type set
  $\type^\anch$ has size $2$.

  Fix an index $n \in \N$.
  For the completeness property, let $\strategy$ be a value-$1$ PCC strategy for
  $\verifier_n$.
  We define a value-$1$ PCC strategy $\strategy^\anch$ for
  $\tvanch_n$: whenever a player receives the $\Anchor$ type as
  a question type, they perform the trivial measurement (i.e.,\ measure the
  identity operator), and output $0$.
  Otherwise, the player performs the same measurement as in $\strategy$.
  This is clearly value-$1$ and PCC.
  Item~\ref{enu:anch-completeness} follows from this and
  Lemma~\ref{lem:detyping-verifiers}.

  For the soundness property, we observe that if a strategy $\strategy^\anch$
  has value $1 - \eps$ in $\tvanch_n$, then
  \[ 1 - \eps = \frac{3}{4} + \frac{p}{4} \]
  where $p$ is the value of $\strategy^\anch$ in the game $\verifier_n$, and
  $1/4$ is the probability that neither player receives the question
  type $\Anchor$; this follows from the distribution associated with the typed
  sampler $\tsanch$.
  Solving for $p$, this implies that $\Ent(\tvanch_n,1 - \eps) \geq \Ent(\verifier_n,1 - 4\eps)$.
  The soundness property (Item~\ref{enu:anch-soundness}) then follows from Lemma~\ref{lem:detyping-verifiers}.
  

  Items~\ref{enu:anch-complexity} and~\ref{enu:anch-efficient-computability} are
  straightforward and also follow from Lemma~\ref{lem:detyping-verifiers}.
\end{proof}

\subsection{The parallel repetition transformation} 
\label{sec:anchored-repetition}
Starting with the anchored verifier $\vanch = (\sanch,\danch)$, we 
then perform the parallel repetition transformation.

Fix integers $\lambda,\tau\in\N$, and let
$k(n)=(\lambda n)^{\tau}$. Define the Turing machine $\sampler^\rep$ depending on $\lambda,\tau$ as
follows.
\begin{enumerate}
  \item On input $(n,\gamestyle{dimension})$, it computes
  $s'=\sanch(n,\gamestyle{dimension})$ and $k=k(n)$, and outputs $k\cdot s'$.

  \item On input $(n,\gamestyle{field})$, it returns
  $2$.

  \item On input $(n,w,\gamestyle{marginal},j,z)$, it computes
  $s'=\sanch(n,\gamestyle{dimension})$ and $k=k(n)$. It checks that $z$ has
  length exactly $k\cdot s'$. If not, it halts with a fixed default output.
  Otherwise, it parses $z=(z_1,\ldots,z_k)$ with each $z_i\in\F_2^{s'}$, and
  outputs
  \[
    \Big(
      \sanch(n,w,\gamestyle{marginal},j,z_1),\ldots,
      \sanch(n,w,\gamestyle{marginal},j,z_k)
    \Big).
  \]

  \item On input $(n,w,\gamestyle{linear},j,u,y)$, it computes
  $s'=\sanch(n,\gamestyle{dimension})$ and $k=k(n)$. It checks that $u$ and $y$
  each have length exactly $k\cdot s'$. If not, it halts with a fixed default
  output. Otherwise, it parses
  $u=(u_1,\ldots,u_k)$ and $y=(y_1,\ldots,y_k)$ with each component in
  $\F_2^{s'}$, and outputs
  \[
    \Big(
      \sanch(n,w,\gamestyle{linear},j,u_1,y_1),\ldots,
      \sanch(n,w,\gamestyle{linear},j,u_k,y_k)
    \Big).
  \]

  \item On input $(n,w,\gamestyle{factor},j,u)$, it computes
  $s'=\sanch(n,\gamestyle{dimension})$ and $k=k(n)$. It checks that $u$ has
  length exactly $k\cdot s'$. If not, it halts with a fixed default output.
  Otherwise, it parses $u=(u_1,\ldots,u_k)$ with each $u_i\in\F_2^{s'}$, and
  outputs
  \[
    \Big(
      \sanch(n,w,\gamestyle{factor},j,u_1),\ldots,
      \sanch(n,w,\gamestyle{factor},j,u_k)
    \Big).
  \]
\end{enumerate}
Define $\decider^\rep$ as follows. On input $(n,x,y,a,b)$, it first computes
$k=k(n)$. It checks that $|x|=|y|$ and that this common length is divisible
by $k$, rejecting if either check fails. It then splits $x$ and $y$ into
$k$ equal consecutive blocks,
\[
  x=x_1\cdots x_k,\qquad y=y_1\cdots y_k.
\]
It attempts to parse only $a$ and $b$ as canonically encoded $k$-tuples,
\[
  a=(a_1,\ldots,a_k),\qquad b=(b_1,\ldots,b_k),
\]
and rejects if either parsing fails. Otherwise, it runs
$\danch(n,x_i,y_i,a_i,b_i)$ for all $i\in\{1,\ldots,k\}$ and accepts if and
only if all these calls accept.

The decider $\decider^\rep$ is equipped with a timeout counter as in Section~\ref{sec:tms}. Its
timeout bound $B_{\rep}(n)$ is chosen to be of the same form (polynomial or exponential) as $B_{\decider}(n)$, with constants chosen in such a way that 
 $B_{\rep}(n)$ dominates the time to perform the
tuple parsing, bookkeeping, and the $k(n)$ calls to $\danch$, which is of the form $\poly(k(n),B_{\decider}(n))$. If the timeout counter
expires, $\decider^\rep$ enters $\TIMEOUT$, which is treated as rejection.

The sampler construction above is defined for any anchored sampler $\sanch$.
The next theorem records
the completeness, soundness, and complexity guarantees when
$(\sanch,\danch)$ is obtained by anchoring a normal form verifier
$\verifier=(\sampler,\decider)$ whose decider is in standard timeout-counter
form.



\begin{theorem}[Anchored parallel repetition of normal form verifiers]
  \label{thm:repetition}
    There exist universal constants $c,c'>0$ and a polynomial-time Turing machine
  $\ComputeParrepVerifier$ with the following property.  On input
$(\verifier,\lambda,\tau)$, where $\verifier=(\sampler,\decider)$ is a pair of
Turing machines, $\lambda,\tau\in\N$, and the description of $\decider$ is in
the standard timeout-counter form of Section~\ref{sec:tms}, the machine outputs
a pair of Turing machines
$\verifier^\rep=(\sampler^\rep,\decider^\rep)$.

  Suppose further that $\verifier=(\sampler,\decider)$ is an $\ell$-level normal
form verifier with sampler dimension $s(n)$. Let $B_{\decider}(n)$ denote the timeout bound of $\decider$. Suppose that $s(n) \leq B_{\decider}(n)$. 

Then $\verifier^\rep$ is a normal form verifier satisfying the following
properties for all $n\in\N$, where $k(n)=(\lambda n)^\tau$.
  \begin{enumerate}
	\item \label{enu:pr-completeness} (\textbf{Completeness}) If $\verifier_n$ has
    a value-$1$ PCC strategy, then $\verifier^{\rep}_n$ has a value-$1$ PCC
    strategy.

  \item \label{enu:pr-soundness} (\textbf{Soundness}) For all $\eps>0$ and all
\[
  p>
  \frac{4}{\eps}
  \exp\left(
    -c\eps^{17}\frac{k(n)}{ (c' B_{\decider}(n))^{c'}}
  \right),
\]
we have
\[
  \Ent(\verifier^\rep_n,p)\geq \Ent(\verifier_n,1-\eps).
\]

  \item \label{enu:pr-complexity} (\textbf{Complexity}) The repeated verifier
$\verifier^\rep$ is $(\ell+2)$-level and its decider $\decider^\rep$ is in the
standard timeout-counter form of Section~\ref{sec:tms}, with timeout bound
\[
  B_{\decider^\rep}(n)=\poly(k(n),B_{\decider}(n)).
\]
The time complexities of $\sampler^\rep$ and $\decider^\rep$ satisfy
\begin{align*}
  \TIME_{\sampler^\rep}(n)
    & = \poly\bigl(k(n),\TIME_\sampler(n)\bigr),\\
  \TIME_{\decider^\rep}(n)
    & = \poly\bigl(k(n),B_{\decider}(n)\bigr).
\end{align*}
\end{enumerate}
  Furthermore, the repeated sampler $\sampler^\rep$ only depends on $\sampler$ and the parameters $\lambda,\tau$. Moreover, the dimension $s_{\sampler^\rep}(n)$ of $\sampler^\rep$ satisfies $s_{\sampler^\rep}(n) \leq B_{\decider^\rep}(n)$.
\end{theorem}

\begin{proof}

%

Suppose that $\sampler$ is an $\ell$-level sampler with dimension $s(n)$. Then by Proposition~\ref{prop:anchoring}, the number of levels of $\sanch$ is $\ell' = \ell+2$, and its dimension is $s'(n) = s(n)+8$. Then by construction the Turing machine $\sampler^\rep$ is a sampler with the following properties. It has field size $q(n) = 2$ 
and it has dimension $s^\rep(n) = k(n) s'(n)$.
We treat the ambient space $V^\rep$ of $\sampler^\rep$ on index $n$ as the
$k(n)$-fold direct sum of the ambient space $V^\anch$ of $\sanch$ on index $n$.
For all integers $n \in \N$, $w \in \AB$, the CL functions $L^{\rep,\,
  w}: V^\rep \to V^\rep$ of $\sampler^\rep$ are as follows:
\begin{equation*}
	L^{\rep,\, w} = \bigoplus_{i = 1}^{k(n)} L^w\;,
\end{equation*}
where $L^\alice,L^\bob:V^\anch \to V^\anch$ are the CL functions of the sampler $\sampler^\anch$
on index $n$. The CL functions $L^{\rep,\, w}$ are $\ell'$-level; the $j$-th
factor spaces $V^{\rep,\, w}_{j,\, u}$ of $L^{\rep,\, w}$ are defined as
\begin{equation*}
	V^{\rep,\, w}_{j,\, u} = \bigoplus_{i = 1}^{k(n)} V^w_{j,\, u_i}
\end{equation*}
for all $u = (u_1,\ldots,u_{k(n)})$ where $V^w_{j,\, u_i}$ is the $j$-th factor
space of $L^w$ with prefix $u_i \in V^\anch$.

%
%
%
%



The number of levels of $\sampler^\rep$ is the same as $\sanch$, which is $\ell' = \ell + 2$. 

Thus, assuming that $\verifier = (\sampler,\decider)$ is a normal form verifier, we obtain that $\verifier^\rep = (\sampler^\rep,\decider^\rep)$ is a normal form verifier, and thus defines an infinite family of games $(\verifier_n^\rep)_{n \in \N}$. 


	Item~\ref{enu:pr-completeness} follows from
  Item~\ref{enu:anch-completeness} of Proposition~\ref{prop:anchoring}.  Let
  $\strategy^\anch$ be the resulting value-$1$ PCC strategy for $\vanch_n$.
  In the repeated game $\verifier^\rep_n$, the players share $k(n)$ independent
  copies of the state of $\strategy^\anch$.  On question tuples
  $x=(x_1,\ldots,x_{k(n)})$ and $y=(y_1,\ldots,y_{k(n)})$, they apply
  $\strategy^\anch$ independently to each coordinate and return the tuple of
  answers.  The tuple answers have length bounded by the timeout bound chosen
  for $\decider^\rep$, since this bound dominates the $k(n)$ calls to $\danch$
  and the tuple-encoding overhead.  The resulting strategy is PCC, and
  $\decider^\rep$ accepts exactly when every anchored coordinate accepts, so the
  repeated strategy has value $1$.

To show Item~\ref{enu:pr-soundness}, we apply Theorem~\ref{thm:bvy} to the
anchored game $\vanch_n$. The exponential decay bound in Theorem~\ref{thm:bvy}
depends on the answer length in $\vanch_n$. By Proposition~\ref{prop:anchoring},
the anchored decider $\danch$ has runtime bounded by a universal polynomial in
$\TIME_\decider(n)$. Since $\decider$ is in standard timeout-counter form with
timeout bound $B_{\decider}(n)$, increasing the universal constant $c'$ if
necessary gives
\[
  \TIME_{\danch}(n)\leq (c'B_{\decider}(n))^{c'}.
\]
By Definition~\ref{def:normal-game}, the answer alphabet of the anchored game
$\vanch_n$ consists of strings of length at most $\TIME_{\danch}(n)$.  Hence the
answer length parameter in Theorem~\ref{thm:bvy} is bounded by
$(c'B_{\decider}(n))^{c'}$.  Theorem~\ref{thm:bvy} then gives the displayed
soundness bound.

Item~\ref{enu:pr-complexity} follows as follows. For the sampler, on every
input of the form $(n,\ldots)$, $\sampler^\rep$ first computes
$s'=\sanch(n,\gamestyle{dimension})$ and $k=k(n)$. It checks that every input
string interpreted as an element of the repeated ambient space has length
exactly $k\cdot s'$, and returns a fixed default output otherwise. If these
checks pass, it evaluates the corresponding routine of $\sanch$ on each of the
$k$ components. Using Proposition~\ref{prop:anchoring}, this gives
\[
  \TIME_{\sampler^\rep}(n)=\poly(k(n),\TIME_\sampler(n)).
\]

Note that the dimension $s_{\sampler^\rep}(n)$ of the repeated sampler $\sampler^\rep$ satisfies
\[
  s_{\sampler^\rep}(n) = k(n) s_{\sanch}(n) \leq k(n) B_{\danch}(n) = \poly(k(n),B_{\decider}(n))
\]
where we used the fact that $s_{\sanch}(n) \leq B_{\danch}(n)$ from Proposition~\ref{prop:anchoring} (this uses the assumption that $s(n) \leq B_\decider(n)$).

For the decider, the preceding bound
$\TIME_{\danch}(n)\leq(c'B_{\decider}(n))^{c'}$ implies that, on inputs for
which the tuple parsing and length checks succeed, the procedure defining $\decider^\rep$ runs in
time polynomial in $k(n)$ and $B_{\decider}(n)$. There exists a universal constant polynomial $\poly(\cdot)$ such that the timeout bound
\[
  B_{\rep}(n)=\poly(k(n), B_{\decider}(n))
\]
dominates this computation as well as $s_{\sampler^\rep}(n)$.  Therefore, $\decider^\rep$ is in standard
timeout-counter form and has the claimed timeout bound.

  Since the CL functions of the sampler $\sanch$ are $(\ell+2)$-level
  (by Proposition~\ref{prop:anchoring}), and taking the direct sum of CL
  functions does not increase the number of levels (by
  Lemma~\ref{lem:cl-func-prod}), the CL functions $L^{\rep,\, w}$ are
  $(\ell+2)$-level.
  
  The ``Furthermore'' part of the theorem can be easily verified by inspecting the construction of the samplers $\sampler^\rep$ and $\sanch$. 

  Finally, we discuss the efficient computability of the description of the repeated verifier. The Turing machine $\ComputeParrepVerifier$, given input
$(\verifier,\lambda,\tau)$, first computes the description of the Turing
machines $(\tsanch,\tdanch)$ and then their detyping $(\sanch,\danch)$ as
described in Section~\ref{sec:anchoring}. Then, it computes the pair of Turing
machines $(\sampler^\rep,\decider^\rep)$ as described above. 

To construct $\decider^\rep$, the machine $\ComputeParrepVerifier$ applies\newline
$\IdentifyTimeoutCounter$ to the description of $\decider$ and recovers its
timeout mode (polynomial or exponential) and parameters. From these parameters and the binary
representations of $\lambda$ and $\tau$, it computes parameters of the same
timeout mode defining a standard bound
\[
  B_{\rep}(n)=\poly(k(n),B_\decider(n)).
\]
The new parameters have bit length
polynomial in the input description length. The constructor then applies
$\AddPolynomialTimeoutCounter$ or $\AddExponentialTimeoutCounter$, as
appropriate, to the underlying repeated computation. Therefore,\newline
$\ComputeParrepVerifier$ runs in polynomial time.
\end{proof}


\section{Gap-preserving compression}
\label{sec:compression}

We combine the transformations from the previous sections (the introspection
games, answer reduction, and parallel repetition) to obtain our
main technical result, a \emph{gap-preserving compression theorem}
(Theorem~\ref{thm:compression}) for normal form verifiers.

Recall from Definition~\ref{def:lambda} that a verifier $\verifier = (\sampler,
\decider)$ is $\lambda$-bounded if the description length $\abs{\overline{\verifier}}= \max \{ |\sampler|,|\decider| \}$ is at most
$\lambda$ and the time complexity bounds $\TIME_\sampler(n)$ and
$\TIME_\decider(n)$ are at most $n^\lambda$ for all $n \ge 2$.
The compression theorem states the existence of an efficient ``compression
procedure'' $\Compress$ that achieves the following.
Given as input a $\lambda$-bounded verifier $\verifier$ and the parameter
$\lambda$, the procedure returns a ``compressed verifier'' $\verifier^\compr$
such that the $n$-th game $\verifier^\compr_n$ simulates (in a sense to be made precise shortly) the $N$-th game
$\verifier_N$ for $N = 2^n$, and, furthermore, both the sampler time complexity
and the decider time complexity of $\verifier^\compr_n$ can be exponentially
smaller than those of $\verifier_N$, which can be as large as $N^\lambda$.
We note that the time complexity bounds of the compressed verifier hold without
the assumption that the input verifier is $\lambda$-bounded.
The completeness and soundness bounds of the theorem require the
$\lambda$-boundedness of the input verifier $\verifier$.

\begin{theorem}[Compression theorem]
  \label{thm:compression}
There exists a universal constant $C_0 > 0$ and a polynomial-time Turing machine $\Compress$ that takes as input a
tuple $(\verifier, \lambda)$ where $\verifier = (\sampler,\decider)$ is a pair of Turing machines and $\lambda > 0$ is an integer, and outputs a $9$-level normal form verifier $\verifier^\compr = (\sampler^\compr,\decider^\compr)$ whose decider $\decider^\compr$ is in the standard timeout-counter form of Section~\ref{sec:tms} with timeout counter $B_{\decider^\compr}(n)$. 
The time complexities of $\sampler^\compr,\decider^\compr$ satisfy
  \begin{equation*}
    \begin{split}
      \TIME_{\sampler^\compr}(n) & = \poly (n, \lambda),\\
      \TIME_{\decider^\compr}(n) & = \poly (n, \lambda).
    \end{split}
  \end{equation*}
  Moreover, letting $s_{\sampler^\compr}(n)$ denote the dimension of the sampler $\sampler^\compr$, we have $s_{\sampler^\compr}(n) \leq B_{\decider^\compr}(n)$. 
  The description of $\sampler^\compr$ is independent of $\verifier$ and is
  computable from the binary representation of $\lambda$ in time $\polylog
  (\lambda)$.	
  If, furthermore, $\verifier$ is a $\lambda$-bounded, $9$-level normal form verifier, then 
  $\verifier^\compr$ satisfies the following: for all $n \geq C_0$ and $N = 2^n$,
%

%
	\begin{enumerate}
  \item \label{enu:compr-completeness} (\textbf{Completeness}) If $\verifier_{N}$
    has a value-$1$ PCC strategy, then $\verifier^\compr_n$ has a value-$1$ PCC
    strategy.
  \item \label{enu:compr-soundness} (\textbf{Soundness})
    $\Ent(\verifier_{n}^\compr,\frac{1}{2}) \geq \max \frac12\left \{
      \Ent(\verifier_{N}, \frac{1}{2}), 2^{N^{\lambda}-1}
    \right \}$.
  \end{enumerate}
\end{theorem}

\subsection{Proof of theorem~\ref{thm:compression}}

Recall the following Turing machines. 
\begin{enumerate}
\item $\ComputeIntroVerifier$ takes as input a tuple $(\verifier, \lambda,
  \ell)$ and returns in polynomial time a description of the
  introspective verifier $\verifier^\intro$ corresponding to the verifier
  $\verifier$ and parameters $(\lambda, \ell)$. (See Theorem~\ref{thm:introspection}.)

\item $\ComputeAnsVerifier$ takes input $(\verifier, \lambda, \mu, \sigma)$ and returns
  in polynomial time a description of the answer reduced verifier
  $\verifier^\ar$ corresponding to $\verifier$ and parameters $(\lambda, \mu,\sigma)$. (See Theorem~\ref{thm:ar}.)

\item $\ComputeParrepVerifier$ takes input $(\verifier, \lambda, \tau)$ and returns in
  polynomial time a description of the anchored repeated verifier $\verifier^\rep$
  corresponding to $\verifier$ and a number of repetitions $k(n) = (\lambda n)^\tau$. (See Theorem~\ref{thm:repetition}.)
\end{enumerate}

We specify the Turing machine $\Compress$ in Fig.~\ref{fig:compress}. The
universal constants $\mu$ and $\tau$, and the parameter $\sigma=\sigma(\lambda)$,
are specified in \eqref{eq:mu-sigma} and \eqref{eq:c_rep}.

\begin{figure}[!htbp]
  \centering
  \begin{gamespec}
  	Input: $(\verifier, \lambda)$ where $\verifier = (\sampler,\decider)$ is a pair of Turing machines 
	and $\lambda$ is an integer.
    \begin{enumerate}
	
		\item \label{enu:compress-intro} Compute $\verifier^{(1)} =
      \ComputeIntroVerifier(\verifier, \lambda, 9)$.

		\item \label{enu:compress-ans} Compute $\verifier^{(2)} =
      \ComputeAnsVerifier(\verifier^{(1)}, \lambda, \mu, \sigma)$.


		\item \label{enu:compress-rep} Compute $\verifier^{(3)} =
      \ComputeParrepVerifier(\verifier^{(2)}, \lambda, \tau)$.

		\item Return $\verifier^\compr = \verifier^{(3)}$.
	\end{enumerate}
  \end{gamespec}
\caption{The definition of the Turing machine $\Compress$, where the universal
constant $\mu$ and the parameter $\sigma=\sigma(\lambda)$ are specified in
\eqref{eq:mu-sigma}, and the universal constant $\tau$ is specified in
\eqref{eq:c_rep}.}
  \label{fig:compress}
\end{figure}

\begin{lemma}
  \label{lem:compress-independent-samplers}
  Let $\Compress$ be the Turing machine specified in Fig.~\ref{fig:compress}.
  For all pairs of Turing machines $\verifier = (\sampler,\decider)$ and integer $\lambda$, the output of
  $\Compress(\verifier, \lambda)$ is a normal form verifier $\verifier^\compr =
  (\sampler^\compr, \decider^\compr)$ such that $\sampler^\compr$ does not
  depend on $\verifier$ but only on the parameter $\lambda$.
  Furthermore, there is a Turing machine $\ComputeSampler$ that on input
  $\lambda$ outputs the description of the compressed sampler $\sampler^\compr$
  in time $\polylog(\lambda)$.
\end{lemma}

\begin{proof}
  The verifier $\verifier^\compr = \verifier^{(3)}$ is a normal form verifier
  because the introspective verifier $\verifier^{(1)}$ is a normal form verifier (even if
  the input $\verifier$ is not; see Remark~\ref{rk:intro-normalform}), and 
  the verifier transformations $\ComputeAnsVerifier$, $\ComputeParrepVerifier$ preserve 
  the normal form (by Theorems~\ref{thm:ar} and \ref{thm:repetition}).  
  
  The introspective verifier computed in Step~\ref{enu:compress-intro} of
  Fig.~\ref{fig:compress} has a sampler $\sampler^{(1)}$ that only depends on
  the parameter $\lambda$, as implied by
  Lemma~\ref{lem:intro-sampler-complexity}.
  The answer reduced verifier from Step~\ref{enu:compress-ans} has a sampler $\sampler^{(2)}$ 
  that, according to Theorem~\ref{thm:ar}, only depends on the sampler $\sampler^{(1)}$ and the parameter tuple $(\lambda,\mu,\sigma)$. Note that $\sampler^{(1)}$ only depends on $\lambda$, the parameters $\mu$ is a universal constant, and the parameter $\sigma$ itself depends only on $\lambda$ (and is polynomial in it)\tnote{Edited this part to refer to $\sigma$}.
	Thus $\sampler^{(2)}$ only depends on $\lambda$. Similarly, the sampler $\sampler^{(3)}$ of the repeated verifier from Step~\ref{enu:compress-rep} only depends on the sampler $\sampler^{(2)}$, the parameter $\lambda$, and the universal constant $\tau$, as implied by Theorem~\ref{thm:repetition}. Therefore, the final sampler $\sampler^\compr = \sampler^{(3)}$ only depends on $\lambda$.


  Let $\verifier^*$ be the pair of trivial Turing machines $(0,\allowbreak 0)$. 
  Define $\ComputeSampler$ as the Turing machine that on input $\lambda$
  computes $(\sampler,\allowbreak  \decider^{\gamestyle{Dummy}}) =
  \Compress(\verifier^*,\allowbreak  \lambda)$ and returns the description of $\sampler$ as
  the output. By the above discussion, the description of $\sampler$ is the same
  as that of $\sampler^\compr$ and the time complexity of $\ComputeSampler$ is 
  $\polylog(\lambda)$ as $\Compress$ runs in polynomial time: this 
  follows from
  the fact that the Turing machines $\ComputeIntroVerifier$,\newline
  $\ComputeAnsVerifier$, and $\ComputeParrepVerifier$ all run in polynomial time.
\end{proof}

\begin{proof}[Proof of Theorem~\ref{thm:compression}]
  We evaluate the parameters of the verifiers generated in each step of the
  $\Compress$ procedure. 
  These parameters are summarized in Table~\ref{tab:params} and are explained in the
  following. Let $(\verifier, \lambda)$ be the input to the Turing machine $\Compress$. We write $N$ to denote $2^n$. 

  \begin{table}[!htb]
    \centering
    \begin{tabularx}{\linewidth}{l X X c X}
    \toprule & \multicolumn{2}{c}{Time complexity} & Level & Completeness\\
    \cmidrule(lr){2-3}
    & Sampler & Decider & & \\
    \midrule $\verifier^{(1)}$ & $\poly(n, \lambda)$ & $\poly(N^\lambda)$ & $5$ & value-$1$ PCC\\
    \midrule $\verifier^{(2)}$ & $\poly(n,\lambda)$ & $\poly(n,\lambda)$ & $7$ & value-$1$ PCC\\
    \midrule $\verifier^{(3)}$ & $\poly(n,\lambda)$ & $\poly(n,\lambda)$ & $9$ & value-$1$ PCC\\
    \bottomrule
    \end{tabularx}
    \begin{tabularx}{\linewidth}{l X}
    \toprule & Soundness\\
    \midrule $\verifier^{(1)}$ & $\begin{aligned}[t]
      &\Ent \big(\verifier_n^{(1)}, 1 - \eps_1 \big)\\
      &\quad\geq \max \big\{\Ent(\verifier_N, \frac{1}{2}), 2^{N^\lambda - 1} \big\}
      \end{aligned}$\\
    \midrule $\verifier^{(2)}$ & $\begin{aligned}[t]
      &\Ent \bigl(\verifier_n^{(2)}, 1 - \eps_2 \bigr)\\
      &\quad\geq \frac12\Ent \bigl(\verifier_n^{(1)}, 1 - \eps_1 \bigr)
      \end{aligned}$\\
    \midrule $\verifier^{(3)}$ & $\begin{aligned}[t]
      &\Ent \bigl(\verifier_n^{(3)}, \frac{1}{2} \bigr)\\
      &\quad\geq \Ent \bigl(\verifier_n^{(2)}, 1 - \eps_2 \bigr)
      \end{aligned}$\\
    \bottomrule
    \end{tabularx}
\caption{Parameters and properties of the input verifier $\verifier$,
  introspective verifier $\verifier^{(1)}$, answer reduced verifier
  $\verifier^{(2)}$, and parallel repeated verifier $\verifier^{(3)}$
  computed by $\Compress$.  For deciders in standard timeout-counter form, the
  decider time-complexity entry also bounds the timeout counter.  The quantities
  $\eps_1,\eps_2$ (which are functions of $n$) are specified below in the main
  text.}
    \label{tab:params}
  \end{table}

  \paragraph{Introspection.}
  The verifier $\verifier^{(1)} = (\sampler^{(1)}, \decider^{(1)})$ is the
  introspective verifier corresponding to $\verifier$ and parameters $\lambda$
  and $\ell = 9$ computed by $\ComputeIntroVerifier$.
  We state its parameters and properties in the second row of Table~\ref{tab:params}
  in terms of the index $n \in \N$.

  The complexity bounds and the number of levels of the introspective verifier are given by
  Theorem~\ref{thm:introspection}. Using that the argument $\ell$ in $\ComputeIntroVerifier$ is set to $9$, 
  there exists a universal constant $C_\intro \geq 2$ such that for all $\lambda\in \N$ and for all $n \geq 2$, we have
  \begin{equation}
    \label{eq:c_intro}
    \begin{split}
    \TIME_{\sampler^{(1)}}(n) & \le (\lambda \cdot n)^{C_\intro},\\
    \TIME_{\decider^{(1)}}(n) & \le 2^{C_\intro \lambda\,n}~, \\
    |\decider^{(1)}| &\le C_\intro \, \lambda^{C_\intro}~.
    \end{split}
  \end{equation}
  By Theorem~\ref{thm:introspection} these bounds hold for all inputs $(\verifier,
  \lambda, 9)$ (in particular, these bounds do not assume that the input $\verifier$ is $\lambda$-bounded or is even a normal form verifier).

  The statements in the completeness and soundness entries of the row assume that the input
  verifier $\verifier$ is $9$-level and $\lambda$-bounded. 
  The completeness entry follows from Theorem~\ref{thm:introspection} because in the completeness case we assume the verifier $\verifier$ has a value-$1$ PCC strategy, and thus the resulting introspective verifier has a value-$1$ PCC strategy. We deduce the soundness entry as follows. 
  Let the function $\delta(\eps,n)$ from Theorem~\ref{thm:introspection} be denoted by $\delta_1(\eps_1,n)$. Let $a_1 > 0, 0 < b_1 < 1$ be the universal constants such that
   \begin{equation}
    \label{eq:re-delta-1}
    \delta_1(\eps_1, n) = a_1 \Bigl((\lambda n)^{a_1} \cdot \eps_1^{b_1} +
    ( \lambda n )^{-b_1} \Bigr)
  \end{equation}
  By setting
  \begin{equation}
  \label{eq:re-eps-1}
  	\eps_1 = \Bigl( \frac{1}{8 a_1 (\lambda n)^{a_1}} \Bigr)^{1/b_1}, \qquad \qquad C_1 = ( 4 a_1 )^{1/b_1}
  \end{equation}
  we get for all $n \geq C_1$, 
  \begin{equation*}
    \begin{split}
      \delta_1(\eps_1,n) &= a_1 \Bigl( (\lambda n)^{a_1} \cdot (\eps_1)^{b_1}+
      ( \lambda n)^{-b_1} \Bigr) \\
      & \le a_1 \Bigl( (\lambda n)^{a_1} \cdot (\eps_1)^{b_1}
      \Bigr) + \frac{1}{4} \\
      & \le a_1 \Bigl( \frac{1}{8 a_1} \Bigr) + \frac{1}{4}\\
      & < \frac{1}{2} \;,
    \end{split}
  \end{equation*}
  where we used the fact that $\lambda \ge 1$ in the second line. Thus by the soundness guarantee of Theorem~\ref{thm:introspection} we get that
  \[
  	\Ent \big(
    \verifier_n^{(1)}, 1 - \eps_1 \big) \geq \max \big
    \{\Ent(\verifier_N, \frac{1}{2}), 2^{N^\lambda - 1} \big \}
  \]
  for all $n \geq C_1$ as desired.

  

  

  \paragraph{Answer reduction.} 
  
  	First we define the parameters $(\mu,\sigma)$ that are passed to
$\ComputeAnsVerifier$. Let $\delta_2(\eps_2,n)$ denote the function
$\delta(\eps,n)$ from Theorem~\ref{thm:ar}, so
\begin{equation}\label{eq:lambda2-15}
  \delta_2(\eps_2, n) = \sigma^{a_2} \cdot
  \Bigl( (\lambda \cdot n)^{a_2 \mu} \cdot \eps_2^{b_2}
  +(\lambda \cdot n)^{-b_2 \mu} \Bigr)
\end{equation}
for universal constants $a_2>0$ and $0<b_2<1$. Define
\begin{equation}
\label{eq:mu-sigma}
\sigma = \lceil C_\intro \, \lambda^{C_\intro} \rceil
\quad \text{and} \quad
\mu = \left\lceil
  \max\left\{
    C_\intro,
    \frac{9a_1 + 2a_2 C_\intro}{b_1 b_2}
  \right\}
\right\rceil~.
\end{equation}
Here $\mu$ is a universal constant, while $\sigma$ is a parameter depending on
$\lambda$. The choice $\mu\geq C_\intro$ ensures that the sampler-time
hypothesis of Theorem~\ref{thm:ar} holds for $\verifier^{(1)}$, and $\sigma$
upper bounds $|\decider^{(1)}|$ by \eqref{eq:c_intro}. The parameter $\mu$ is chosen so that $(\lambda n)^{-b_2 \mu} < \frac{\eps_1}{2\sigma^{a_2}}$ for all $n \geq 2$. 
	
  Let $\verifier^{(2)} = (\sampler^{(2)}, \decider^{(2)})$ denote the answer
  reduced verifier corresponding to $\verifier^{(1)}$ and parameters $(\lambda,
\mu,\sigma)$ computed by $\ComputeAnsVerifier$. 
%
  The third row in Table~\ref{tab:params} gives the properties of
  the answer reduced verifier.

The bounds on the number of levels, the time complexities, and the timeout
bound for the decider follow from Theorem~\ref{thm:ar}.  The introspective
verifier $\verifier^{(1)}$ is $5$-level, so the answer-reduced verifier
$\verifier^{(2)}$ is $7$-level.  The timeout-counter hypothesis of
Theorem~\ref{thm:ar} holds because Theorem~\ref{thm:introspection} outputs
$\decider^{(1)}$ in standard timeout-counter form.  Using the bounds on
$\sampler^{(1)}$ and $\decider^{(1)}$ from \eqref{eq:c_intro}, and using that
$\sigma$ upper bounds $|\decider^{(1)}|$, the remaining hypotheses of
Theorem~\ref{thm:ar} are satisfied with $\qlen(n)=(\lambda n)^\mu$.
Moreover, Theorem~\ref{thm:ar} gives that $\decider^{(2)}$ is in the standard
timeout-counter form of Section~\ref{sec:tms}.  Let
$B_{\decider^{(2)}}(n)$ denote its timeout bound.  Since $\mu$ is universal and
$\sigma=\poly(\lambda)$, there exists a universal constant $C_2\geq 1$ such
that for all $n\geq C_2$,
\begin{equation}
  \label{eq:c_2}
    \max \Bigl\{
      \TIME_{\sampler^{(2)}}(n),
      \TIME_{\decider^{(2)}}(n),
      B_{\decider^{(2)}}(n)
    \Bigr\} \le (\lambda \cdot n)^{C_2}~.
\end{equation}
  The constant $C_2$ depends on the universal constant $\mu$. Again, these time complexities do not depend on whether the starting verifier $\verifier$ is $\lambda$-bounded. Moreover, Theorem~\ref{thm:ar} gives that 
  \begin{equation}
  \label{eq:sampler_2_dimension}
  s_{\sampler^{(2)}}(n)\leq B_{\decider^{(2)}}(n),
  \end{equation}
where $s_{\sampler^{(2)}}(n)$ is the dimension of $\sampler^{(2)}$. 

  For the soundness and completeness entries, assume that the starting verifier
$\verifier$ is $9$-level and $\lambda$-bounded.  For completeness, fix
$n$ and set $N=2^n$.  If $\verifier_N$ has a value-$1$ PCC strategy, then
the completeness guarantee of Theorem~\ref{thm:introspection} gives a
value-$1$ PCC strategy for $\verifier^{(1)}_n$.  The completeness guarantee of
Theorem~\ref{thm:ar} then gives a value-$1$ PCC strategy for
$\verifier^{(2)}_n$, and Item~\ref{enu:pr-completeness} of
Theorem~\ref{thm:repetition} gives a value-$1$ PCC strategy for
$\verifier^{(3)}_n$.
  
  The soundness entry is argued as follows. Set
  \begin{equation}
  \label{eq:re-eps-2}
  	\eps_2 = \Bigl(\frac{\eps_1}{3 \sigma^{a_2} (\lambda n)^{a_2 \mu}}\Bigr)^{1/b_2}~.
  \end{equation}   
By our choice of parameters $\mu,\sigma,\eps_2$, we have that $\delta_2(\eps_2, n) < \eps_1$.

%
Let $C_\ar$ be the constant from Theorem~\ref{thm:ar}. By the soundness guarantee of Theorem~\ref{thm:ar} we get that
  \[
  	\Ent \big(
    \verifier_n^{(2)}, 1 - \eps_2 \big) \geq \frac12 \Ent(\verifier_n^{(1)}, 1 - \delta_2(\eps_2,n)) \geq \frac12\Ent(\verifier_n^{(1)}, 1 - \eps_1)
  \]
  for all $n \geq \max \{C_1, C_2, C_\ar \}$ as desired.  
  	

  \paragraph{Parallel repetition.}
  Let $\verifier^{(3)} = (\sampler^{(3)},\decider^{(3)})$ denote the anchored
  repetition (see Section~\ref{sec:anchored-repetition}) of $\verifier^{(2)}$
  computed by $\ComputeParrepVerifier$, with parameters $\lambda$ and $\tau$ defined below in \eqref{eq:c_rep}. 
  We state the parameters and properties of $\verifier^{(3)}$ in the fourth row
  of Table~\ref{tab:params}. 

  Equation~\eqref{eq:sampler_2_dimension} implies that the sampler $\sampler^{(2)}$ satisfies the dimension hypothesis of Theorem~\ref{thm:repetition}.
  
  The number of levels of $\sampler^{(3)}$ is $9$ by
Item~\ref{enu:pr-complexity} of Theorem~\ref{thm:repetition}, because
$\verifier^{(2)}$ is $7$-level.  The same item also gives that
$\decider^{(3)}$ is in the standard timeout-counter form of
Section~\ref{sec:tms}.  Using the complexity bounds for $\sampler^{(2)}$ and
the timeout bound for $\decider^{(2)}$ from \eqref{eq:c_2}, and using
$k(n)=(\lambda n)^\tau$, we get
\[
  \TIME_{\sampler^{(3)}}(n)
    = \poly(k(n),\TIME_{\sampler^{(2)}}(n))
    = \poly(n,\lambda)
\]
and
\[
  \TIME_{\decider^{(3)}}(n)
    = \poly(k(n),B_{\decider^{(2)}}(n))
    = \poly(n,\lambda).
\]
These bounds only depend on the bounds stated for $\verifier^{(2)}$, and again
do not depend on whether the input verifier $\verifier$ is $\lambda$-bounded.


The completeness entry for $\verifier^{(3)}$ follows from
Item~\ref{enu:pr-completeness} of Theorem~\ref{thm:repetition}: for each $n$,
if $\verifier^{(2)}_n$ has a value-$1$ PCC strategy, then
$\verifier^{(3)}_n$ has a value-$1$ PCC strategy.  
  
  The soundness entry is argued as follows. Let $c_3,c_3'>0$ denote the universal
constants $c,c'$ from Item~\ref{enu:pr-soundness} of
Theorem~\ref{thm:repetition}. Recalling that $\eps_2$ from
\eqref{eq:re-eps-2} is a function of $n$ and $\lambda$, define $\tau$ to be the
minimum integer satisfying
\[
  \tau \geq \max\{C_\ar,C_2\}
\]
and
\begin{equation}
\label{eq:c_rep}
  \frac{4}{\eps_2}
  \exp\left(
    -c_3\eps_2^{17}
    \frac{(\lambda n)^\tau}{(c_3'(\lambda n)^{C_2})^{c_3'}}
  \right)
  < \frac{1}{2}
\end{equation}
for all $n\geq\tau$ and all integers $\lambda\geq1$. Such a universal
$\tau$ exists because the reciprocal of $\eps_2$ is bounded by a fixed
polynomial in $\lambda n$, while the exponent in \eqref{eq:c_rep} can be made
an arbitrarily large fixed power of $\lambda n$ by increasing $\tau$.

For this choice of $\tau$, Theorem~\ref{thm:repetition} uses
$k(n)=(\lambda n)^\tau$. By \eqref{eq:c_2}, for all $n\geq\max\{C_2,\tau\}$,
we have $B_{\decider^{(2)}}(n)\leq(\lambda n)^{C_2}$. Hence
\[
  \frac{4}{\eps_2}
  \exp\left(
    -c_3\eps_2^{17}
    \frac{k(n)}{(c_3'B_{\decider^{(2)}}(n))^{c_3'}}
  \right)
  \leq
  \frac{4}{\eps_2}
  \exp\left(
    -c_3\eps_2^{17}
    \frac{(\lambda n)^\tau}{(c_3'(\lambda n)^{C_2})^{c_3'}}
  \right)
  < \frac{1}{2}.
\]
Thus by the soundness guarantee of Theorem~\ref{thm:repetition} we get that
\[
  \Ent \big(
  \verifier_n^{(3)}, \frac{1}{2} \big) \geq \Ent(\verifier_n^{(2)}, 1 - \eps_2)
\]
for all $n \geq \max \{ C_1, C_2, C_\ar, \tau \}$ as desired.

The sampler dimension conclusion of Theorem~\ref{thm:repetition} also gives
  \begin{equation}
  \label{eq:sampler_3_dimension}
  s_{\sampler^{(3)}}(n)\leq B_{\decider^{(3)}}(n)~.
  \end{equation}

  \paragraph{Putting everything together.}
  The verifier $\verifier^\compr$ is $\verifier^{(3)}$.
  We now put together the bounds and parameters from Table~\ref{tab:params} to obtain
  the conclusions stated in Theorem~\ref{thm:compression}.

  The claim that the Turing machine $\Compress$ is polynomial time follows from
  the fact that the Turing machines $\ComputeIntroVerifier$,\newline 
  $\ComputeAnsVerifier$, and $\ComputeParrepVerifier$ all run in polynomial time.

  The claimed number of levels and time complexity of the sampler
  $\sampler^\compr$ follows from the complexity bounds in Table~\ref{tab:params},
  which in turn only depend on the parameter $\lambda$.
  In particular, the sampler $\sampler^\compr$ is {independent} of the input
  verifier $\verifier$, as shown in
  Lemma~\ref{lem:compress-independent-samplers}.
  Finally, the claimed time complexity to compute the description of
  $\sampler^\compr$ also follows from Lemma~\ref{lem:compress-independent-samplers}.

  The claimed time complexity of $\decider^\compr$ follows from the preceding
paral\-lel-repetition analysis.  Indeed, $\decider^\compr\allowbreak =\allowbreak \decider^{(3)}$, and
$\decider^{(3)}$ is in standard timeout-counter form by
Item~\ref{enu:pr-complexity} of Theorem~\ref{thm:repetition}.  Its timeout
bound is $\poly(n,\allowbreak \lambda)$, because $k(n)\allowbreak =\allowbreak (\lambda n)^\tau$ and
$B_{\decider^{(2)}}(n)\allowbreak \leq\allowbreak (\lambda n)^{C_2}$ by \eqref{eq:c_2}.  On inputs for
which the tuple parsing succeeds, the repeated decider runs the $k(n)$ anchored
decider calls described in Section~\ref{sec:anchored-repetition}.  On arbitrary
inputs, including inputs with very long or malformed strings $x,y,a,b$, the
timeout counter gives the same worst-case $\poly(n,\allowbreak \lambda)$ bound, with
$\TIMEOUT$ treated as rejection.

  We now establish the completeness and soundness properties of
  $\verifier^\compr$.
  Assume that the input verifier $\verifier$ is $9$-level and $\lambda$-bounded. 
  The completeness property (Item~\ref{enu:compr-completeness} in
  Theorem~\ref{thm:compression}) follows from chaining together the completeness
  properties of $\verifier^{(3)}_n$, $\verifier^{(2)}_n$, $\verifier^{(1)}_n$
  and the assumption that $\verifier_N$ has a value-$1$ PCC strategy.

  We now establish the soundness property (Item~\ref{enu:compr-soundness} in
  Theorem~\ref{thm:compression}). Set 
    \begin{equation}
  \label{eq:c_0-recursive}
  	C_0 = \max \{ C_1, C_2, C_\ar,\tau \} \;.
  \end{equation}
  Since $C_1, C_2, C_\ar, \tau$ are all universal constants, $C_0$ is also a universal constant.
 By chaining together the soundness properties of $\verifier^{(3)}_n$, $\verifier^{(2)}_n$, $\verifier^{(1)}_n$, 
 we get that 
 \[
 	\Ent \big(
    \verifier_n^{(3)}, \frac{1}{2} \big) \geq \Ent(\verifier_n^{(2)}, 1 - \eps_2)
 \geq \frac12\Ent(\verifier_n^{(1)}, 1 - \eps_1) \geq \frac12\max \big
    \{\Ent(\verifier_N, \frac{1}{2}), 2^{N^\lambda - 1} \big \}
 \]
 for all $n \geq C_0$ as desired.

 Finally, the sampler dimension conclusion of Theorem~\ref{thm:compression} follows from \eqref{eq:sampler_3_dimension}. This concludes the proof of the theorem.
\end{proof}

\subsection{An $\mathsf{MIP}^*$ protocol for the halting problem}
\label{sec:halt}

For every Turing machine $\cal{M}$, we construct a nonlocal game $\game$ such
that $\val^*(\game) = 1$ if $\cal{M}$ halts and $\val^*(\game) \le \frac{1}{2}$
otherwise.  In what follows, to explicitly disambiguate between 
a Turing machine (which is a tuple consisting of a set of states, transition rules, etc) and its description 
(which is a binary string), we use calligraphic letters to denote the Turing machine $\cal{M}$ and
an overline $\overline{\cal{M}}$ to denote a \emph{description} of $\cal{M}$ (see 
Section~\ref{sec:tms} for a discussion of Turing machines and their descriptions).


\begin{figure}[!htp]
  \centering
  \begin{gamespec}
    Input: $(\desc{R}, \desc{M}, \lambda, n, x, y, a, b)$ where $\desc{R}$ is
    the description of an $8$-input Turing machine and $\desc{M}$ is a
    description of a Turing machine $\cal{M}$.

    \begin{enumerate}
      \item Compute the description $\overline{\sampler} =
      \ComputeSampler(\lambda)$.

    \item Run $\cal{M}$ on the blank input for $n$ steps.
      Accept if it halts and continue otherwise.
      
    \item Compute the description $\overline{\decider}$ of the $5$-input Turing
  machine $\decider$ defined as follows:
  \[
    \decider(n',x',y',a',b')
    =
    \cal{R}(\desc{R},\desc{M},\lambda,n',x',y',a',b')
  \]
  i.e., on input $(n',x',y',a',b')$ runs $\cal{R}$ with input
  $(\desc{R},\desc{M},\lambda,n',x',y',a',b')$.
    
    \item Compute the description
      $\overline{\verifier^\compr} = \Compress(\overline{\verifier}, \lambda)$
      where $\overline{\verifier} = (\overline{\sampler}, \overline{\decider})$.

    \item Let $\decider^\compr$ be the decider of $\verifier^\compr$, and compute its
  timeout bound $B_{\decider^\compr}(n)$ from its description. Simulate
  $\decider^\compr$ on $(n,x,y,a,b)$ if the input tuple $(n,x,y,a,b)$ has length at most $B_{\decider^\compr}(n)$ (if it has length exceeding this, then simulate $\decider^\compr$ on a truncation of the input tuple to at most $B_{\decider^\compr}(n)$ bits). Accept if and only if $\decider^\compr$ accepts.
    \end{enumerate}
  \end{gamespec}
  \caption{Description of Turing machine $\cal{F}$.
    The Turing machines $\ComputeSampler$ and $\Compress$ are given in
    Lemmas~\ref{lem:compress-independent-samplers} and ~\ref{thm:compression}. The Turing machine simulations in Step 2 and 5 are performed using the universal Turing machine $\cal{U}$ from Theorem~\ref{thm:universal-tm}.}
  \label{fig:halt_f}
\end{figure}

First, we define a Turing machine $\cal F$ as in
Figure~\ref{fig:halt_f}. For all Turing machines $\cal M$ and parameters
$\lambda\in\mathbb N$, define $\decider^\halt_{\cal M,\lambda}$ to be the
$5$-input Turing machine
\[
  \decider^\halt_{\cal M,\lambda}(n,x,y,a,b)
  =
  \cal F(\desc F,\desc M,\lambda,n,x,y,a,b).
\]
That is, on input $(n,x,y,a,b)$ it runs $\cal F$ with the first three inputs hardwired to $\desc F, \desc M, \lambda$, respectively, and the other inputs $(n,x,y,a,b)$ are read directly from the input tape of $\decider^\halt_{\cal M,\lambda}$. 

We generally omit the subscripts $\cal M,\lambda$ and denote this decider by
$\decider^\halt$. Define
\[
  \sampler^\halt=\ComputeSampler(\lambda),
  \qquad
  \verifier^\halt=(\sampler^\halt,\decider^\halt).
\]
By inspection of Figure~\ref{fig:halt_f}, the description
$\overline{\decider}$ computed when $\decider^\halt$ executes Step~3 is the
description of $\decider^\halt$ itself.\footnote{Alternatively, one may use
Kleene's recursion theorem~\cite{Kleene1954} to obtain the same fixed point.
This approach was taken in an earlier version of the paper.}
Therefore, the input to $\Compress$ in Step~4 is the description of
$\verifier^\halt$ together with $\lambda$.

For all Turing machines $\cal M$ and integers $\lambda$, the machine
$\decider^\halt$ halts on every input. Steps~1--4 of $\cal F$ terminate because
$\ComputeSampler$ and $\Compress$ are total procedures. In Step~5, the machine
simulates $\decider^\compr$ using the universal Turing machine $\cal{U}$ from
Theorem~\ref{thm:universal-tm}, truncating the input tuple $(n,x,y,a,b)$ to the timeout bound $B_{\decider^\compr}(n)$ if necessary. Since $\decider^\compr$ is in standard timeout-counter form by
Theorem~\ref{thm:compression}, Step~5 also
terminates. Since $\ComputeSampler(\lambda)$ always outputs a sampler,
$\verifier^\halt$ is a normal form verifier.

 We have the following lemma for $\verifier^\halt$.





\begin{lemma}\label{lem:dhalt-values}
  Let $\cal{M}$ be a Turing machine and $\lambda$ integer.
  Then the verifier $\verifier^\halt$ corresponding to $\cal{M}$ and $\lambda$ has the following properties. For all $n \in \N$,
  \begin{enumerate}
  \item If $\cal{M}$ halts in $n$ steps then
    $\val^*(\verifier^\halt_n)=1$.
   In this case, there is a value-$1$ PCC strategy for the game
    $\verifier^\halt_n$.
  \item If $\cal{M}$ does not halt in $n$ steps then
    $\verifier^\halt_n$ has a value-$1$ PCC strategy if and only if
    $\verifier^\compr_n$ does.
    Furthermore, under the same assumption (that $\cal{M}$ does not halt in $n$ steps) it holds that 
    \begin{equation*}
      \Ent\big(\verifier^\halt_n, \frac{1}{2}\big) \,=\,
      \Ent\big(\verifier^\compr_n, \frac{1}{2}\big)\;.
    \end{equation*}
  \end{enumerate}
\end{lemma}
		
\begin{proof}

  Suppose $\cal{M}$ halts in $n$ steps.
  Then from the definition of $\cal{F}$ and $\decider^\halt$, it follows that
  $\decider^{\halt}$ accepts on input $(n, x, y, a, b)$ for all valid sampled questions $x,y$ and for all $a,b$, and
  hence $\val^*(\verifier^\halt_n) = 1$. A value-$1$ PCC
  strategy for $\verifier^\halt_n$ is the following trivial strategy: for all questions, the players always
  return a fixed answer (e.g., $0$).

  If $\cal{M}$ does not halt in $n$ steps, then $\decider^{\halt}$ accepts on
  input $(n, x, y, a,b)$ if and only if $\decider^\compr$ accepts on (the truncation of) $(n, x, y,
  a,b)$.  Since $\verifier^\halt_n$ and $\verifier^\compr_n$ use the same sampler $\sampler^\halt = \ComputeSampler(\lambda)$
  by construction, any strategy for $\verifier^\halt_n$ is a strategy (with the same winning probability) for $\verifier^\compr_n$ and vice versa. This implies that there is a value-$1$ PCC strategy for $\verifier^\halt_n$ if and only if
  there is a value-$1$ PCC strategy for $\verifier^\compr_n$. The ``Furthermore'' statement also follows directly from this.
\end{proof}

We show in Lemma~\ref{lem:lambda} below that for all Turing machine $\cal{M}$ there is a choice of
$\lambda$ such that the corresponding $\verifier^\halt = (\sampler^\halt, \decider^\halt)$ is
$\lambda$-bounded. We first show a technical
lemma.

\begin{lemma}
  \label{lem:lambda-bound}
  \leavevmode\par Suppose $C,\allowbreak  C' \in \N$ are integers. 
  Then for all $\lambda \ge 4 \max \{ (4C)^{8C},\allowbreak  C \log C' \} $ and $n \ge 2$,
  \begin{equation*}
    C(C' \cdot \lambda n)^C \le n^\lambda.
  \end{equation*}
\end{lemma}

\begin{proof}
We aim to find a function $f(C,C')$ such that for all $\lambda \geq f(C,C')$, it holds that $C(C' \cdot \lambda n)^C \le n^\lambda$ for all $n \geq 2$. For this it suffices that 
\begin{equation}
\label{eq:lambda-bound-1}
	\lambda \geq 4 \max \{ \log C, C, C \log C', C \log \lambda \} \qquad \text{for all $\lambda \geq f(C,C')$}
\end{equation}
This is because~\eqref{eq:lambda-bound-1} implies that $\lambda \geq \log C + C + C\log C' + C\log \lambda$ which, by exponentiating both sides, gives
\[
	2^\lambda \geq 2^{\log C + C + C\log C' + C\log \lambda} = C \cdot 2^C \cdot (C' \lambda)^C\;.
\]
Since $n^{\lambda - C} \geq 2^{\lambda -C}$ for all $n \geq 2$, we get that $n^{\lambda - C} \geq C (C' \lambda)^C$ which is the desired inequality. It is easy to verify that for all $\lambda \geq (4C)^{8C}$, we have $\lambda \geq 4 C \log \lambda$. Thus, setting
\[
	f(C,C') = 4 \max \{ (4C)^{8C}, C \log C' \}~,
\]
we obtain~\eqref{eq:lambda-bound-1}, which concludes the lemma.
\end{proof}

\begin{lemma}
  \label{lem:lambda}
  \leavevmode\par There is an integer parameter $\lambda$, polynomial-time computable from the
  description of the Turing machine $\cal{M}$, and
  scaling as $\poly(|\overline{\machine}|)$, such that the verifier
  $\verifier^\halt$ corresponding to $\cal{M}$ and parameter $\lambda$ is
  $\lambda$-bounded and, moreover, $\TIME_{\decider^\halt}(n),\allowbreak \TIME_{\sampler^\halt}(n) = \poly(n,\allowbreak  \abs{\overline{\cal{M}}})$.
\end{lemma}

\begin{proof}

  We first do an accounting of the time complexity of $\decider^\halt$.
  By definition, for all $\cal M$ and $\lambda$, the decider $\decider^\halt$
runs $\cal F$ on input $(\desc F,\allowbreak \desc M,\allowbreak \lambda,\allowbreak n,\allowbreak x,\allowbreak y,\allowbreak a,\allowbreak b)$. Its running time is, therefore, polynomial in the time bounds of the following
steps.

  \begin{enumerate}
  \item Computing $\overline{\sampler}=\ComputeSampler(\lambda)$ takes
    $\polylog(\lambda)$ time by
    Lemma~\ref{lem:compress-independent-samplers}. 

  \item Simulating $\cal{M}$: this takes $\poly(\abs{\overline{\machine}}, n)$ time.
  \item \label{enu:d} Computing the description $\overline{\decider}$: this
    takes $\poly(\abs{\overline{\machine}}, \abs{\overline{\cal{F}}},\log \lambda)$ time.
  \item \label{enu:cdc} Computing the description $\overline{\decider^\compr}$: this takes
    $\poly(\abs{\overline{\verifier}}, \log\lambda)$ time (by Theorem~\ref{thm:compression}). 
  
    \item \label{enu:rdc} The parameters defining the timeout counter $B_{\decider^\compr}(n)$ can be recovered from the description $\overline{\decider^\compr}$ in $\poly(|\overline{\decider^\compr}|)$ time. Simulating $\decider^\compr$ on input $(n,x,y,a,b)$ (or a truncation of it to $B_{\decider^\compr}(n)$ bits if necessary), 
by Theorems~\ref{thm:compression} and~\ref{thm:universal-tm}, has time complexity
$\poly(n,\lambda)$. 
  \end{enumerate}

  As $\overline{\decider}$ is computed in time $\poly(\abs{\overline{\machine}},
  \abs{\overline{\cal{F}}},\log \lambda)$, we have that the description length of $\decider$ is bounded by $\abs{\overline{\decider}} \le
  \poly(\abs{\overline{\cal{M}}}, \abs{\overline{\cal{F}}},\log \lambda)$.
  Similarly, the description length of the sampler is bounded by $\abs{\overline{\sampler}} \le \polylog(\lambda)$ as
  $\ComputeSampler$ runs in polynomial time in the bit-length of $\lambda$.
  Therefore, the size of the verifier $\abs{\overline{\verifier}}$ is bounded by
  $\poly(\abs{\overline{\cal{M}}}, \abs{\overline{\cal{F}}}, \log \lambda)$.
  So the time complexity bound in Item~\ref{enu:cdc} is at most
  $\poly(\abs{\overline{\cal{M}}}, \abs{\overline{\cal{F}}}, \log \lambda)$. 
  We emphasize that we did not use the $\lambda$-boundedness of the verifier
  $\verifier$, as this is what we want to prove here!


Overall, the total time complexity of $\decider^\halt$ can be bounded by
\begin{equation}\label{eq:125-dec}
  \TIME_{\decider^\halt}(n)
  \leq
  C_{\labelstyle D}
  \bigl(
    \abs{\overline{\cal M}} \cdot 
    \abs{\overline{\cal F}} \cdot 
    \lambda \cdot n
  \bigr)^{C_{\labelstyle D}},
\end{equation}
for some universal integer constant $C_{\labelstyle D}>0$. This bound holds
on every input $(n,x,y,a,b)$, independently of the lengths of $x,y,a,b$. By Lemma~\ref{lem:lambda-bound}, we have for all $\lambda \ge \lambda_{0} := 4 \max \{ (4C_{\labelstyle{D}})^{8C_{\labelstyle{D}}} ,\allowbreak  C_\labelstyle{D} \log (\abs{\overline{\cal{M}}} \cdot \abs{\overline{\cal{F}}}) \}$ and $n \ge 2$, we have $\TIME_{\decider^\halt}(n) \le n^\lambda$.

  Similarly, by Theorem~\ref{thm:compression} the sampler $\sampler^\halt$ has running time $\poly(n, \lambda)$
  and, therefore, there is a universal constant $C_{\labelstyle{S}}>0$ such that
  \begin{equation}\label{eq:125-samp}
    \TIME_{\sampler^\halt}(n) \le C_{\labelstyle{S}}(\lambda \cdot n)^{C_{\labelstyle{S}}},
  \end{equation}
  which, again by Lemma~\ref{lem:lambda-bound}, is at most $n^\lambda$ for $\lambda
  \ge \lambda_1 := 4 \cdot (4C_{\labelstyle{S}})^{8C_{\labelstyle{S}}}$ and $n \ge 2$.

  As we have shown $\abs{\overline{\verifier}} = \poly(\abs{\overline{\cal{M}}}, \abs{\overline{\cal{F}}}, \log
  \lambda)$ and the description of the verifier $\verifier$ is the same as that
  of $\verifier^\halt$, there exists an integer $C_{\labelstyle{V}}>0$ such that
  \begin{equation*}
    \abs{\verifier^\halt} \le C_{\labelstyle{V}}(\abs{\overline{\cal{M}}} \cdot \abs{\overline{\cal{F}}} \cdot \log
    \lambda)^{C_{\labelstyle{V}}}.
  \end{equation*}

  Define $\lambda_2 :\allowbreak =\allowbreak  \max\left\{
      2^{4 (4C_{\labelstyle V})^{8C_{\labelstyle V}}},\allowbreak 
      (\abs{\overline{\cal{M}}} \cdot \abs{\overline{\cal{F}}})^{4 C_{\labelstyle V} }
    \right\}$. Then applying Lemma \ref{lem:lambda-bound} with $\lambda \allowbreak \mapsto\allowbreak  \log \lambda$, $n \allowbreak \mapsto\allowbreak  2$, $C \allowbreak \mapsto\allowbreak  C_{\labelstyle{V}}$ and $C' \allowbreak \mapsto\allowbreak  \abs{\overline{\cal{M}}} \cdot \abs{\overline{\cal{F}}}$, we get that
    for all $\lambda \allowbreak \geq\allowbreak  \lambda_2$, we have that $\abs{\verifier^\halt}  \allowbreak \leq\allowbreak  \lambda$.

  Finally, taking $\lambda = \big\lceil\max\{\lambda_0,\lambda_1,\lambda_2 \} \big \rceil$ completes the proof.
	The quantity $\abs{\overline{\cal{F}}}$ is a universal constant, because the Turing machine $\cal{F}$ does not depend on $\lambda$ or $\cal{M}$. The quantities $	C_{\labelstyle{D}}, C_{\labelstyle{S}}, C_{\labelstyle{V}}$ are universal constants; therefore, $\lambda_0, \lambda_1, \lambda_2$ and thus $\lambda$ can be computed from $|\overline{\machine}|$. It is clear that $\lambda = \poly(|\overline{\machine}|)$. Finally, for this value of $\lambda$ one immediately verifies from~\eqref{eq:125-dec} and~\eqref{eq:125-samp} that $\TIME_{\decider^\halt}(n),\allowbreak \TIME_{\sampler^\halt}(n) = \poly(n,\allowbreak  \abs{\overline{\cal{M}}})$, as claimed. 
	
\end{proof}




Putting things together we obtain the following result. 

\begin{theorem}
  \label{thm:halting}
  There exists a polynomial-time computable map from Turing machines $\machine$ to games $\game$ that satisfies the ``efficiency'' requirement from Definition~\ref{def:mipstar} and is
	such that
	\begin{enumerate}
  \item If $\machine$ halts on the empty input, then $\val^*(\game) = 1$. Moreover, in this case there is a (finite-dimensional) PCC strategy for the players in $\game$ that wins with certainty. 
  \item If $\machine$ does not halt on the empty input, then $\val^*(\game) \leq
    \frac{1}{2}$.
	\end{enumerate}
\end{theorem}

\begin{proof}
  The map from Turing machines to games is computed as follows. 
  Given a Turing machine $\machine$, first compute the parameter $\lambda$ as given by
  Lemma \ref{lem:lambda}. This takes time at most $\poly(|\overline{\machine}|)$, and, furthermore, $\lambda = \poly(|\overline{\machine}|)$.

  Then compute the description of the recursive verifier
\[
  \verifier^\halt=(\sampler^\halt,\decider^\halt)
\]
corresponding to $\machine$ and $\lambda$. First compute
$\sampler^\halt=\ComputeSampler(\lambda)$, and then compute the description of
the $5$-input Turing machine
\[
  \decider^\halt(n,x,y,a,b)
  =
  \cal F(\desc F,\desc M,\lambda,n,x,y,a,b).
\]
Finally, output the description of the game
\[
  \game=\verifier^{\halt}_{C_0},
\]
where $C_0$ is the universal constant from
Theorem~\ref{thm:compression}.

Because $\lambda=\poly(\abs{\overline{\machine}})$ and
$C_0,C_{\labelstyle D},\abs{\overline{\cal F}}$ are universal constants, this
construction takes time $\poly(\abs{\overline{\machine}})$. Furthermore, the
sampler and decider of the output game run in
$\poly(\abs{\overline{\machine}})$ time.
  
%
		
	Now fix a Turing machine $\machine$ 
  and let $\game_n$ denote the $n$-th game $\verifier^\halt_n$ of the verifier $\verifier^\halt$ (so, in particular, the output game $\game$ is $\game_{C_0}$). Suppose that $\machine$ halts on the empty input; let $T$ be the minimum number
  of time steps required for $\machine$ to halt on the empty input.
  Observe that for all $n \geq T$, by Lemma~\ref{lem:dhalt-values} it holds that
  $\game_n$ has a value-$1$ PCC strategy.
  We will use this to show inductively that $\game_n$ also has a value-$1$ PCC
  strategy, for all $ C_0 \leq n < T$.

  \begin{claim}
    Let $ C_0 \leq n < T$. Suppose that $\game_{m}$ has a value $1$ PCC
    strategy for all
    $m > n$. Then $\game_n$ also has a value-$1$ PCC strategy.
    \label{claim:induction-game}
  \end{claim}
  \begin{proof}
    Since by assumption $\cal{M}$ does not halt in $n$ steps, by
    Lemma~\ref{lem:dhalt-values} it holds that $\game_n$ has a value-$1$ PCC
    strategy if $\game^{\compr}_n$ does.
    Since $\verifier^\halt$ is $\lambda$-bounded (by Lemma~\ref{lem:lambda}) and $n \geq C_0$, by
    Theorem~\ref{thm:compression} it follows that $\game^{\compr}_n$ has a value-$1$ PCC
    strategy if $\game_{2^n}$ does.
    Since $2^n > n$, this is true by the hypothesis of the claim.
    Thus, $\game_n$ has a value-$1$ PCC strategy as claimed.
  \end{proof}
  By Claim~\ref{claim:induction-game} and downwards induction on $n$ (with the base
  case $n = T$), we have that $\game_n$ has a value-$1$ PCC strategy for all $n
  \geq C_0$.
  In particular, we have $\val^*(\game) = \val^*(\game_{C_0}) = 1$.
  This shows the first item in the theorem statement.

	Now suppose that $\machine$ does not halt on the empty input.
  We have that for all $n \geq C_0$:
  \begin{equation*}
		\Ent\big(\game_n,\frac{1}{2}\big) = \Ent\big(\verifier^\compr_n,\frac{1}{2}\big) \geq
\frac{1}{2}    \Ent\big(\game_{N},\frac{1}{2}\big)\;,
  \end{equation*}
	where the equality follows from Lemma~\ref{lem:dhalt-values} and the
  inequality follows from Theorem~\ref{thm:compression} (again using the
  $\lambda$-bounded property of $\verifier^\halt$).
  By induction, for every $k\in\N$ we obtain
\begin{equation*}
\begin{aligned}
&\Ent\big(\game,\frac{1}{2} \big)\\
 &=
  \Ent\big(\game_{C_0},\frac{1}{2} \big)\\
 &\geq
  2^{-k}\Ent\Bigl(\game_{g^{(k)}(C_0)},\frac{1}{2} \Bigr)\\
 &=
  2^{-k}\Ent\Bigl(
    \verifier^\compr_{g^{(k)}(C_0)},\frac{1}{2}
  \Bigr)\\
 &\geq
  2^{-k-2}\,
  2^{(g^{(k+1)}(C_0))^\lambda}.
\end{aligned}
\end{equation*}
Here $g^{(k)}$ denotes the $k$-fold composition of $g(n)=2^n$.  The first
inequality follows by iterating the preceding one-step inequality $k$ times,
the equality follows from Lemma~\ref{lem:dhalt-values}, and the final
inequality follows from Theorem~\ref{thm:compression}.  Since
\[
  2^{-k-2}\,2^{(g^{(k+1)}(C_0))^\lambda}
  \longrightarrow \infty
  \qquad\text{as }k\to\infty,
\]
there is no finite upper bound on $\Ent(\game,\frac{1}{2})$.  Therefore, no
finite-dimensional strategy for $\game$ can have success probability greater
than $\frac{1}{{2}}$.
\end{proof}

Recall the definition of the complexity class $\mathsf{RE}$, which stands for
the set of \emph{recursively enumerable} languages (also called
\emph{Turing-recognizable} languages).
Precisely, a language $L \subseteq \{0,1\}^*$ is in $\RE$ if and only if there
exists a Turing machine $\cal{M}$ such that if $x \in L$, then $\cal{M}(x)$
halts and outputs $1$, and if $x \notin L$, then either $\cal{M}(x)$ outputs~$0$
or it does not halt.
The Halting Problem is the language that contains descriptions of Turing
machines that halt on the empty input input.
The following well-known lemma shows that the Halting Problem is complete for
$\RE$, meaning that every language $L \in \RE$ can be polynomial-time reduced
to the Halting Problem. We include the simple proof for convenience. 

\begin{lemma}
  The Halting Problem is complete for $\RE$ via Karp reductions.\footnote{A \emph{Karp reduction} between languages $L$ and $K$ is a polynomial-time Turing machine $R$ such that $x \in L$ if and only if $R(x) \in K$ for all instances $x \in \{0,1\}^*$.}
\end{lemma}

\begin{proof}
  To see that the Halting Problem is in $\RE$, define $\cal{M}$ to take as input
  an $x$ that represents a Turing machine $\cal{N} = [x]$, and runs the
  universal Turing machine to simulate $\cal{N}$ on the empty input; if $\cal{N}$
  halts with a $1$ then so does $\cal{M}$.

  To show that the Halting problem is complete for $\RE$, let $L\in \RE$ and
  $\cal{M}$ a Turing machine such that if $x \in L$, then $\cal{M}(x)$ halts and
  outputs $1$.
  For an input $x$, let $\cal{N}_x$ be the following Turing machine.
  $\cal{N}_x$ first runs $\cal{M}$ on input $x$.
  If $\cal{M}$ accepts, then $\cal{N}_x$ accepts.
  On all other outcomes, $\cal{N}_x$ goes into an infinite loop.
  Thus $\cal{N}_x$ halts if and only if $x \in L$. 
\end{proof}

\begin{corollary}\label{cor:mip-re}
	$\mathsf{MIP}^* = \mathsf{RE}$.
\end{corollary}
\begin{proof}
	Since the Halting Problem is complete for $\mathsf{RE}$, and by
  Theorem~\ref{thm:halting} is contained in $\mathsf{MIP}^*_{1,1/2}\subseteq \MIP^*$ (see Definition~\ref{def:mipstar}), we have the
  inclusion $\mathsf{RE} \subseteq \mathsf{MIP}^*$.
  The reverse inclusion $\mathsf{MIP}^* \subseteq \mathsf{RE}$ follows from the
  following observation.
  Let $L \in \mathsf{MIP}^*$ .
  From the definition of $\MIP^*$ (see, e.g.,~\cite[Section
  6.1]{vidick2016quantum}, from which we borrow the terminology used here) it
  follows that there exists a polynomial-time Turing machine $\cal{R}$ such that
  for all $x \in \{0,1\}^*$, $\cal{R}(x)$ is the description of an $m$-turn
  verifier $V_x$ interacting with $k$ provers, where $m$ and $k$ are both
  polynomial functions of $|x|$ and such that
	\begin{gather*}
		\val^*(V_x) \geq 2/3 \qquad \text{ if $x \in L$ } \\
		\val^*(V_x)	\leq 1/3 \qquad \text{ if $x \notin L$ }
	\end{gather*}
	Consider the following Turing machine $\cal{A}$: on input $x$, it computes
  $V_x = \cal{R}(x)$, and then exhaustively searches over tensor-product
  strategies of increasing dimension and increasing accuracy to evaluate a lower
  bound on $\val^*(V_x)$.
  If $\val^*(V_x) \geq 2/3$, then for arbitrarily small $\delta$ there exists a
  finite-dimensional tensor-product strategy for the players that achieves value
  $2/3 - \delta > 1/3$.
  When the Turing machine $\cal{A}$ identifies such a strategy it terminates,
  outputting $1$.
  If there is no such strategy, then $\cal{A}$ never halts.
  This implies that $L \in \mathsf{RE}$.
\end{proof}
	
\subsection{An explicit separation}
\label{sec:separation}

As discussed in Section~\ref{sec:consequences}, Theorem~\ref{thm:halting}
implies that $C_{qa}$, the set of approximately finite-dimensional correlations,
is a strict subset of $C_{qc}$, the set of commuting-operator correlations.
This is because if $C_{qa} = C_{qc}$, then there exists an algorithm to
approximate the entangled value of a given nonlocal game $\game$ to arbitrary
accuracy.
On the other hand, Theorem~\ref{thm:halting} shows that deciding whether a game
has entangled value $1$ or at most $1/2$, promised that one is the case, is
undecidable.
Therefore, the correlation sets must be different. 

In fact, Theorem~\ref{thm:halting} implies that there is an infinite family
$\mathscr{M}$ of Turing machines that do not halt on an empty input such
that for all $\cal{M} \in \mathscr{M}$, the corresponding game $\game_\machine$
has $\val^*(\game_\machine) < \valco(\game_\machine)$, where recall that
$\valco(\game_\machine)$ denotes the \emph{commuting-operator value} of
$\game_\machine$, which is the supremum of success probabilities over all
commuting-operator strategies for $\game_\machine$.\footnote{To see why this
  holds, observe that if it were the case that $\val^*(\game_\machine) =
  \valco(\game_\machine)$ for all but finitely many nonhalting $\machine$
  then we could construct an algorithm $\cal{A}$ to decide the Halting problem
  as follows.
  On input $\cal{M}$, $\cal{A}$ first checks if $\cal{M}$ is one of the finitely
  many Turing machines for which $\val^*(\game_\machine) <
  \valco(\game_\machine)$; if so, then it outputs a hard-coded answer for
  whether $\cal{M}$ halts on the empty tape or not.
  Otherwise, $\cal{A}$ computes the nonlocal game $\game_\machine$ and executes
  the aforementioned algorithm for approximating the entangled value of games
  assuming that $C_{qa} = C_{qc}$.
}
However, it is not immediately clear, given a \emph{specific} nonhalting Turing
machine $\cal{M}$, whether the associated game $\game_\machine$ separates the
commuting-operator model from the tensor product model of strategies.
While Theorem~\ref{thm:halting} implies that $\val^*(\game_\machine) \leq
\frac{1}{2}$, it could also be the case that $\valco(\game_\machine) =
\val^*(\game_\machine)$ in that particular instance.
We conjecture that $\valco(\game_\machine) = 1$ for \emph{all} nonhalting
Turing machines $\cal{M}$, but it appears to be difficult to identify an
explicit value-$1$ commuting operator strategy that demonstrates this.

In this section we identify an explicit game $\game$ that separates the tensor
product model from the commuting-operator model; we show that $\val^*(\game)
\leq \frac{1}{2}$ but $\valco(\game) = 1$.
Interestingly, the proof does not exhibit an explicit value-$1$
commuting-operator strategy for $\game$.

We construct the separating game in a similar manner to the games constructed in
Section~\ref{sec:halt}.
Let $\cal{A}$ denote the following $1$-input Turing machine: it takes as input a
description of a nonlocal game $\game$ and runs the semidefinite programming
hierarchy of~\cite{navascues2008convergent,doherty2008quantum} to compute a
nonincreasing sequence $\alpha_1,\alpha_2,\ldots$ of upper bounds on
$\valco(\game)$ such that $\lim_{n \to \infty} \alpha_n = \valco(\game)$.
The Turing machine $\cal{A}$ halts if it obtains a bound $\alpha_n < 1$.
Thus this algorithm eventually halts whenever $\valco(\game) < 1$, and
otherwise it runs forever.

Consider the Turing machine $\cal{N}$ in Fig.~\ref{fig:separation}.
We follow the same steps as in Section~\ref{sec:halt}.
Let $\decider^\sep_\lambda$ be a decider that, on input $(n,x,y,a,b)$, runs $\cal N$ on input
$(\desc{\cal N},\lambda,n,x,y,a,b)$. We equip $\decider^\sep_\lambda$ with timeout bound
\[
  B_{\sep,\cal N,\lambda}(n)
  =
  C_{\sep}\bigl(|\overline{\cal N}|\cdot \lambda\cdot n\bigr)^{C_{\sep}}.
\]
for some universal constant $C_{\sep}$ chosen large enough so that this timeout bound dominates the running time of $\cal N$. We sometimes leave the parameter
$\lambda$ implicit.
Define $\sampler^\sep = \ComputeSampler(\lambda)$ and $\verifier^\sep =
(\sampler^\sep, \decider^\sep)$.  By the definition of the Turing machine
$\cal N$ in Fig.~\ref{fig:separation}, when $\decider^\sep$ runs $\cal N$ with
first input $\desc{\cal N}$, Step~2 computes the description of the decider that
runs $\cal N$ on input $(\desc{\cal N},\lambda,n',x',y',a',b')$, equipped with
timeout bound $B_{\sep,\cal N,\lambda}(n')$.  This is precisely
$\decider^\sep$.  Thus the verifier $\overline{\verifier}$ computed inside
$\decider^\sep$ is the description of $\verifier^\sep$ itself, including the
timeout counter of $\decider^\sep$, and the input to $\Compress$ is the
description of $\verifier^\sep$ and $\lambda$.

Define the game $\game^\sep = \verifier^\sep_{C_0}$ where $C_0$ is the constant
from Theorem~\ref{thm:compression}.
Then $\game^\sep$ is the game $\game$ computed in $\decider^\sep$.

A similar argument to that of Lemma~\ref{lem:lambda}, using the timeout counter
of $\decider^\sep$, shows that there exist choices of $\lambda$ computable from
the description of $\cal A$ such that $\verifier^\sep$ is $\lambda$-bounded.


\begin{figure}[!htb]
  \centering
  \begin{gamespec}
    Input: $(\desc{R}, \lambda, n, x, y, a, b)$ where $\desc{R}$ is
    the description of a $7$-input Turing machine.
    \begin{enumerate}
    \item Compute the description $\overline{\sampler} = \ComputeSampler(\lambda)$.
    \item Compute the description $\overline{\decider}$ of the $5$-input Turing
      machine $\decider$ that on input $(n', x', y', a', b')$ runs $\cal{R}$ with
      input $(\desc{R}, \lambda, n', x', y', a', b')$. It is equipped with a timeout counter with bound $B_{\sep,\cal R,\lambda}(n')$. 
      Let $\overline{\verifier} = (\overline{\sampler}, \overline{\decider})$.
    \item Compute the description of the game $\game = \verifier_{C_0}$ where
      $C_0$ is the integer from Theorem~\ref{thm:compression}.
    \item Simulate $\cal{A}$ on input $\game$ for $n$ steps and accept if
      $\cal{A}$ halts.
    \item Compute the description $\overline{\verifier^\compr} =
      \Compress(\overline{\verifier}, \lambda)$.
    \item Accept if the decider $\decider^\compr$ of verifier
      $\verifier^\compr$ accepts $(n, x, y, a, b)$.
    \end{enumerate}
  \end{gamespec}
  \caption{Description of Turing machine $\cal{N}$.
    The Turing machines $\ComputeSampler$ and $\Compress$ are given in
    Lemmas~\ref{lem:compress-independent-samplers} and~\ref{thm:compression}.}
  \label{fig:separation}
\end{figure}


\begin{theorem}
  \label{thm:separation}
  For the game $\game^\sep = \verifier^\sep_{C_0}$ it holds that
  $\val^*(\game^\sep) \leq \frac{1}{2}$ and $\valco(\game^\sep) = 1$.
\end{theorem}

\tnote{added:}This theorem implies a negative answer to Tsirelson's problem, as stated in the introduction: the sets $C_{qc}$ and $C_{qa}$ are different. As noted in the introduction, the same consequence already follows from the undecidability result shown in Theorem~\ref{thm:halting}. This was already known to experts before our work and is formally shown in e.g.,~\cite[Section 7]{goldbring2024universal}.

\begin{proof}
  Suppose that $\valco(\game^\sep) = 1$.
  The invocation of the Turing machine $\cal{A}$ on input $\game^\sep$ never
  halts, and, therefore, the decider $\decider^\sep$ never accepts in Step 4 of
  Fig.~\ref{fig:separation}.
  Applying Theorem~\ref{thm:compression}, we get that
  $\Ent(\verifier^\sep_{n},\frac{1}{2}) \geq
  \Ent(\verifier^\sep_{2^n},\frac{1}{2})$ and
  \begin{equation*}
    \Ent \Bigl( \verifier^\sep_n, \frac{1}{2} \Bigr) \geq 2^{2^{\lambda n}-1}
  \end{equation*}
  for all $n \geq C_0$.
  An inductive argument implies that there is no finite upper bound on
  $\Ent(\verifier^\sep_{n},\frac{1}{2})$, and thus $\val^*(\game^\sep) =
  \val^*(\verifier^\sep_{C_0}) \leq \frac{1}{2}$, which implies the theorem.

  On the other hand, suppose that $\valco(\game^\sep) < 1$.
  Then there exists some $m \geq C_0$ such that $\cal{A}$ halts on input
  $\game^\sep$ after $m$ steps, so $\verifier^\sep_n$ has a value-$1$ PCC
  strategy for all $n \geq m$ (i.e.,
  the players do not respond with any answers).
  Thus by the completeness statement of Theorem~\ref{thm:compression} and an
  induction argument analogous to the one in the proof of Theorem~\ref{thm:halting}, we have that $\verifier^\sep_n$ has a value-$1$ PCC
  strategy for all $n \geq C_0$, which implies that $\val^*(\game^\sep) = 1$, a
  contradiction because of $\val^*(\game^\sep) \leq \valco(\game^\sep)$.
  This completes the theorem.
\end{proof}


\appendix

\newcommand{\ideg}{\mathrm{ideg}}
\newcommand{\Obs}{\mathcal{O}}
\newcommand{\calH}{\mathcal{H}}
\newcommand{\combine}{\mathrm{combine}}
\newcommand{\nqubits}{M}

\section{Analysis of the Pauli basis test}
\label{sec:qld-analysis}
In this appendix we give a proof of Theorem~\ref{thm:pauli}. For
convenience, we repeat the statement of the theorem here.

\begin{theorem}
  \label{thm:pauli-appendix}
There exists a function 
\[\delta_{\qld}(\eps,m,d,q) = a(md)^a(\eps^b + q^{-b} + 2^{-bmd})\]
 for universal constants $a \geq 1$ and $0 < b < 1$ such that the following holds. For all admissible parameter tuples $\qldparams = (q,m,d)$ and for all strategies $\strategy = (\ket{\psi},A,B)$ for the game $\game^\pauli_\qldparams$ that succeed with probability at least $1 - \eps$, there exist local isometries $\phi_\alice: \mH_\alice \to \mH_{\alice'} \otimes \mH_{\alice''},\phi_\bob: \mH_\bob \to \mH_{\bob'} \otimes \mH_{\bob''}$ (where $\ket{\psi} \in \mH_\alice \otimes \mH_\bob$ and $\mH_{\alice''},\mH_{\bob''} \cong (\C^q)^{\otimes M}$ with $M = 2^m$) and a state $\ket{\aux} \in \mH_{\alice'} \otimes \mH_{\bob'}$ such that 
\begin{enumerate}
    \item $\left \| \phi_\alice \otimes \phi_B \ket{\psi} - \ket{\aux} \otimes \ket{\EPR_q}^{\otimes M}  \right \| \leq \delta_{\qld}(\eps,m,d,q)$,
    \item Letting $\tilde{A}^x_a = \phi_A\, A^x_a \, \phi_A^\dagger$ and $\tilde{B}^y_b = \phi_B B^y_b \phi_B^\dagger$, we have for $W \in \{X,Z\}$
    \begin{gather*}
        \tilde{A}^{(\pauli,W)}_u \otimes I_{\bob'\bob''} \approx_{\delta_{\qld}} (\tau^W_u)_{\alice''} \otimes I_{\alice' \bob' \bob''} \\
        I_{\alice'\alice''} \otimes \tilde{B}^{(\pauli,W)}_u    \approx_{\delta_{\qld}} I_{\alice' \alice'' \bob'} \otimes (\tau^W_u)_{\bob''}\;,
    \end{gather*}
    where the $\approx_{\delta_{\qld}}$ statement holds with respect to the state $\ket{\aux}_{\alice'\bob'} \otimes \ket{\EPR_q}^{\otimes M}_{\alice''\bob''}$ and the answer summation is over $u \in \F_q^M$.
\end{enumerate}
\end{theorem}
We also recall the meaning of the parameters: $q$ is the field size,
$m$ the number of variables, $d$ the degree of the low-degree code,
and $\nqubits=2^m$ is the number of $\ket{\EPR_q}$ states being tested for.

This appendix is structured as follows. First, we establish some preliminary facts and definitions 
in Section~\ref{sec:qld-prelim}. In the remainder of the appendix, we
show Theorem~\ref{thm:pauli-appendix} by showing how to construct a representation of
the Pauli group on the provers' Hilbert space, starting from a
strategy $\strategy$ that succeeds in $\game^\pauli_\qldparams$ with
high probability. This is done in several stages: in
Section~\ref{sec:commutation}, we use the strategy measurements to define a
set of observables that correspond to a subset of the Pauli $X$ and
$Z$ observables, and show that these
approximately satisfy the associated Pauli group relations. In Section~\ref{sec:expanding},
we adjoin ancillary registers containing EPR states to the provers'
state $\ket{\psi}$, and define a new set of approximately-commuting
observables acting on the expanded state. In Section~\ref{sec:combining}, we
combine the approximately-commuting observables into a strategy for
the \emph{classical} low-degree test, and in Section~\ref{sec:apply-ldt}, we
apply the soundness theorem of the classical low-degree test to
construct a single projective measurement that simultaneously measures all of the
approximately-commuting observables. Finally, in
Section~\ref{sec:separating}, we use this single simultaneous measurement to
construct an exact representation of the Pauli group, show that it
is close to the original strategy observables on the expanded state, and deduce that the game $\game^\pauli_\qldparams$ is a self-test for the strategy $\strategy^\pauli$.

\subsection{Preliminaries}
\label{sec:qld-prelim}

\begin{definition}
Given integer $d,m$ and a prime power $q$ we denote by $\ideg_{d,m}(\F_q)$ the set of polynomials in $m$ variables
  with \emph{individual} degree at most $d$ in each variable. When the field is clear from context we write $\ideg_{d,m}$. 	
	When
  the number of variables is clear from context we
  write $\ideg_d$. We denote by $\deg_{d}$ the set of polynomials with
  \emph{total} degree at most $d$.
\end{definition}

Recall the low-degree code defined in Section~\ref{sec:ld-encoding}, in which vectors $h \in \F_q^\nqubits$ get mapped to polynomials $g_h \in \ideg_{d,m}(\F_q)$ such that for every $u \in \{0,1\}^m$, $g_h(u) = h_u$ where the coordinates of $h$ are indexed by binary strings of length $m$. For $x\in \F_q^m$, recall the definition of the vector $\ind_m(x) \in \F_q^\nqubits$ from Section~\ref{sec:ld-encoding}. Recall also the finite field trace $\tr(\cdot)=\tr_{q\to 2}(\cdot):\F_q\to\F_2$, where $q=2^k$ for odd $k$, defined in Section~\ref{sec:subfields}. 


Call a tuple $\omega = (u_\xpt,u_\zpt,r_\xpt,r_\zpt) \in (\F_q^m)^2
\times (\F_q)^2$ \emph{anticommuting} if the quantity 
 \[\gamma\,=\,\tr ((r_\xpt \cdot
\ind_m(u_\xpt)) \cdot (r_\zpt \cdot \ind_m(u_\zpt)))\]
introduced in~\eqref{eq:gamma-value} is not $0$, and \emph{commuting} if $\gamma=0$.

\begin{fact}
\label{fact:omega-anticomm-prob}
	The probability that a tuple $\omega= (u_\xpt,u_\zpt,r_\xpt,r_\zpt)$ sampled uniformly at random from $(\F_q^m)^2 \times (\F_q)^2$
	is anti-commuting is at least 
	\[
		(1 - q^{-m}) \cdot (1 - q^{-1}) \cdot \Big (1 - \frac{md}{q} \Big )  \cdot \frac{1}{2} \geq \Big (1 - \frac{3md}{q} \Big ) \cdot \frac{1}{2}.
	\]
	The probability that $\omega$ is commuting is also at least $\Big(1 - \frac{3md}{q} \Big ) \cdot \frac{1}{2}$.
\end{fact}
\begin{proof}
	First, note that $\ind_m(u_\xpt) = 0$ if and only if $u_\xpt = 0$. If $u_\xpt=0$ then $\omega$ is commuting. Otherwise, if $u_\xpt \neq 0$, then observe that for all $r_\xpt, r_\zpt, u_\zpt$,
	\begin{align*}
		\tr( (r_\xpt \cdot \ind_m(u_\xpt)) \cdot (r_\zpt \cdot
                \ind_m(u_\zpt))) &= \tr( r_\xpt r_\zpt (\ind_m(u_\xpt) \cdot \ind_m(u_\zpt))) \\
								&= \tr( (r_\xpt r_\zpt) \cdot g_{\ind_m(u_\xpt)}(u_\zpt))\;,
	\end{align*}
	where $g_{\ind_m(u_\xpt)}$ is a nonzero polynomial in $\ideg_{d,m}(\F_q)$.
	
	If $r_\xpt = 0$, then $\omega$ is commuting. Condition on $r_\xpt \neq 0$.  Since $g_{\ind_m(u_\xpt)}\neq 0$ has total degree at most $md$, by the Schwartz-Zippel lemma the probability that $g_{\ind_m(u_\xpt)}(u_\zpt)$ is zero over a uniformly random choice of $u_\zpt$ is at most $md/q$. Conditioned on $g_{\ind_m(u_\xpt)}(u_\zpt) \neq 0$, the product $(r_\xpt r_\zpt) \cdot g_{\ind_m(u_\xpt)}(u_\zpt)$ is a uniformly random element of $\F_q$ when $r_\zpt$ is chosen uniformly at random. 
	
	The probability that $u_\xpt \neq 0, r_\xpt \neq 0, g_{\ind_m(u_\xpt)}(u_\zpt) \neq 0$ is at least $(1 - q^{-m}) \cdot (1 - q^{-1}) \cdot (1 - md/q)$ since $u_\xpt, r_\xpt, u_\zpt$ are chosen independently.
	
	Since $q = 2^t$ is an admissible field size, we have that for all $a \in \F_q$, the trace of $a$ is equal to $\sum_i a_i$ where $(a_1,\ldots,a_t)=\kappa(a) \in \F_2^t$ are defined in Section~\ref{sec:subfields}. Thus the probability that a uniformly random element of $\F_q$ has trace $0$ is exactly $1/2$. 
\end{proof}

In this appendix we often use the ``$\approx_\delta$'' notation to specify closeness of general operators (not necessarily POVMs) that are indexed by questions but not answers: given sets of operators $\{A^x \}$ and $\{B^x\}$ indexed by questions $x \in \cal{X}$, we write $A^x \approx_\delta B^x$ to denote 
\[
	\E_{x \sim \mu} \bra{\psi} (A^x - B^x)^\dagger (A^x - B^x)
        \ket{\psi}\,\leq\, O(\delta)
      \]
where $\mu$ is a question distribution and $\ket{\psi}$ is a state that has been fixed by context. This will be useful for expressing closeness of observables, for example. The following two lemmas relate the closeness between sets of operators and weighted sums of those operators. 
\begin{lemma}
\label{lem:avg-closeness}
Let $\{A^x \}$ and $\{B^x\}$ be operators indexed by some finite set $\mathcal{X}$, and let $\mu$ be a probability distribution over $\mathcal{X}$. Let $\{\alpha_x \}$ be a set of complex numbers such that $|\alpha_x| \leq 1$. Then if $A^x \approx_\delta B^x$ on average over $x$ drawn from $\mu$, then $A \approx_\delta B$, where $A = \E_x \alpha_x A^x$ and $B= \E_x \alpha_x B^x$.
\end{lemma}
\begin{proof}
Expand
\begin{align*}
	\Big\| \E_x \alpha_x \big ( A^x - B^x \big) \ket{\psi} \Big\|^2 &\leq \Big( \E_x \norm{ (A^x -B^x) \ket{\psi}} \Big)^2 \\
	&\leq \E_x \norm{(A^x - B^x) \ket{\psi}}^2 \\
	&\leq \delta\;.
\end{align*}
The first inequality is the triangle inequality and the second inequality is Jensen's inequality. 
\end{proof}

\begin{lemma}
\label{lem:povm-to-obs}
Let $\mathcal{A}$ be a finite set. Let $\{A^x_a \}_{a \in \mathcal{A}}$ and $\{B^x_a\}_{a \in \mathcal{A}}$ be POVMs and let $\{ \alpha_a \}_{a \in \mathcal{A}}$ be a collection of complex numbers on the unit circle. Let $A^x = \sum_a \alpha_a A^x_a$ and $B^x = \sum_a  \alpha_a B^x_a$. Then 
\[
	A^x_a \approx_\delta B^x_a \quad \text{implies} \quad A^x \approx_{\delta'} B^x
\]
for $\delta' = |\mathcal{A}| \cdot \delta$.
\end{lemma}
\begin{proof}
Expand
	\begin{align*}
		\E_x \norm{(A^x - B^x) \ket{\psi}}^2 &= \E_x \Big\|\sum_a \alpha_a (A^x_a - B^x_a) \ket{\psi}\Big\|^2 \\
		&\leq \E_x \Big( \sum_a \norm{ (A^x_a - B^x_a) \ket{\psi}} \Big)^2 \\
		&\leq \E_x |\mathcal{A}| \cdot \sum_a \norm{ (A^x_a - B^x_a) \ket{\psi}}^2 \\
		&\leq O\big(|\mathcal{A}| \cdot \delta\big)\;.
	\end{align*}
The first inequality follows from the triangle inequality, the second inequality follows from Cauchy-Schwarz, and the last inequality follows from the assumption that $A^x_a \approx_\delta B^x_a$. 
\end{proof}

The following lemma shows that POVMs that are approximately projective (in the sense of being self-consistent) are close to exact projective measurements. The lemma first appears in~\cite{KV11}; for a self-contained proof, see~\cite{ML20}.

\begin{lemma}[Orthonormalization lemma]\label{lem:ortho}
Let $0\leq \delta\leq 1$. There exists a function $\eta_{ortho}(\delta)\,=\, O\big(\delta^{1/4}\big)$ such that the following holds.  Let $\ket{\psi}$ be a state on $\mH_\alice \otimes \mH_\bob$ where $\mH_\alice,\mH_\bob$ are finite dimensional. Let $\{Q_a\}$ and $\{R_a\}$ be POVM on $\mH_\alice$ and $\mH_\bob$, respectively, such that 
\begin{equation}\label{eq:cons-cond} 
Q_a \abc R_a \,.
\end{equation}
Then there exists a projective measurement $\{P_a\}$ such that
\begin{equation}\label{eq:cons-ccl}
  P_a \otimes \Id_\bob \,\approx_{\eta_{ortho}} Q_a \otimes \Id_\bob \;.
\end{equation}
\end{lemma}

\newcommand{\acom}{\typestyle{Anticom}}

\subsection{Strategies}
\label{sec:commutation}

  Let $\qldparams = (q,m,d)$ be an admissible parameter tuple (Definition~\ref{def:admissible}) and let
  $\strategy = (\ket{\psi},M)$ be a strategy for
  $\game^\pauli_\qldparams$ that succeeds with probability at least $1 - \eps$, where $\ket{\psi}$ is a bipartite state on registers $\alice$ and $\bob$, with associated Hilbert spaces $\mH_\alice$ and $\mH_\bob$, respectively. For notational convenience we use $M$ to denote the operators for both players; it will be clear from context which player's operators we are referring to (for example, we write $M^{(\Point,W),u}_a \otimes \Id_\bob$ to indicate that $M^{(\Point,W),u}_a$ is viewed as an operator acting on register $\alice$).
  
  Using Naimark's
  theorem (as formulated in~\cite[Theorem 5.1]{ML20}), at the cost of
  extending the state $\ket{\psi}$ by adding sufficiently many qubits
  initialized in the $\ket{0}$ state we may assume
  that the measurements in $\strategy$ are projective. Let
  \[ \ket{\psi'} = (\ket{\psi} \otimes \ket{0 \cdots 0} \otimes \ket{0
      \cdots 0})_{\alice \bob} \]
  be the extended state. In the remainder of this section, we
  will work with the projective strategy on the extended state, which we continue labeling $\ket{\psi}$ (instead of $\ket{\psi'}$) for notational convenience.

We introduce several additional notational shorthands. By definition the strategy $\strategy$ contains measurement
  operators for each of the possible question types and content
  summarized in Fig.~\ref{fig:decider_pauli}, such as $\{M^{(\Point,
    W),y}_a\}_{a\in\F_q}$ for all $y\in\F_q^m$, etc. 
For ``line'' questions (of type $\ALine$ or $\DLine$) we will use
 $\Line$ as a formal placeholder for either type $\ALine$ or $\DLine$;
  which type is meant will be clear from the nature of the
  line. Moreover, we will use $\ell$ to denote the description of
  either an axis-parallel line (given by a pair $\ell=(u_0, s)$) or a
  diagonal line (given by a triple $\ell=(u_0, s, v)$). We also introduce the following collection of observables constructed from the
  measurement operators in the strategy. 	
	For every $u \in
  \F_q^m$, $r \in \F_q$, and $W \in \{X, Z\}$, define
  \begin{equation}
    W^r(u) =\sum_{a\in\F_q} (-1)^{\tr(ar)}  M^{(\Point, W), u}_a\;. \label{eq:qld-strat-obs}
  \end{equation}
	The observable \allowbreak{}$W^r(u)$\allowbreak{} acts on register \allowbreak{}$\alice$\allowbreak{} or \allowbreak{}$\bob$\allowbreak{} depending on whether \linebreak[4]{}$M^{(\Point, W), u}_a$\allowbreak{} is viewed as being an operator acting on register \allowbreak{}$\alice$\allowbreak{} or \allowbreak{}$\bob$\allowbreak{}; this will be made clear from context.


Recall the definitions of (anti-)commuting tuple given at the start of Section~\ref{sec:qld-prelim}. 

\begin{lemma}
\label{lem:qld-win-implications}
Assume that $\strategy$ succeeds in $\game^\pauli_\qldparams$ with probability at least $1-\eps$, for some $\eps\geq 0$. Then the following hold: 
	\begin{enumerate}
		\item \label{enu:qld-cons} (\textbf{Consistency check}) On average over $(\tvar,x)$ sampled from the question distribution of $\game^\pauli_\qldparams$, we have that $M^{\tvar,x}_a \otimes \Id_\bob \simeq_\eps \Id_\alice \otimes M^{\tvar,x}_a$.
		\item \label{enu:qld-ld} (\textbf{Low-degree check})
                  For all $W \in \{X,Z\}$, on average over $(\line,u)$
                  drawn from the line-point distribution (see
                  Definition~\ref{def:line-point-dist}), we
                  have $M^{(\Line,W),\line}_{[\eval_u(\cdot) = a]} \otimes \Id_\bob \simeq_\eps \Id_\alice \otimes M^{(\Point,W),u}_a$, where the answer summation is over $a \in \F_q$. 
		
	\item \label{enu:qld-pauli-basis-cons} (\textbf{Pauli basis consistency check}) For all $W \in \{X,Z\}$, on average over $u \in \F_q^m$, we have $M^{(\Point,W),u}_a \otimes \Id_\bob \simeq_\eps \Id_\alice \otimes M^{(\Pauli,W)}_{[g_h(u)=a]}$ where $M^{(\Pauli,W)}_{[g_h(u)=a]} = \sum_{h \in \F_q^{\nqubits} : g_h(u) = a} M^{(\Pauli,W)}_h$ and $g_h$ is the low-degree encoding of $h$. 
	
		\item \label{enu:qld-comm} (\textbf{Commutation check}) For all $W \in \{X,Z\}$, on average over commuting $\omega = (u_\xpt,u_\zpt,r_\xpt,r_\zpt)$, we have $M^{(\Pair,W),\omega}_a \otimes \Id_\bob \simeq_\eps \Id_\alice \otimes M^{\Pair,\omega}_{[\beta_W = a]}$, where the answer summation is over $a \in \F_2$.
		\item \label{enu:qld-comm-cons} (\textbf{Commutation consistency check}) For all $W \in \{X,Z\}$, on average over commuting $\omega = (u_\xpt,u_\zpt,r_\xpt,r_\zpt)$, we have $M^{(\Point,W),u_W}_{[\tr(\cdot r_W) = a]} \otimes \Id_\bob \simeq_\eps \Id_\alice \otimes M^{(\Pair,W),\omega}_{a}$, where the answer summation is over $a\in \F_2$.
		
		\item\label{enu:qld-ms} (\textbf{Magic square check}) For any anticommuting $\omega = (u_\xpt,u_\zpt,r_\xpt,r_\zpt)$ let $\Lambda_\omega$ be the value of the strategy $\Big (\ket{\psi},\allowbreak  \{M^{\Constraint_i,\omega}_{\alpha_1,\alpha_2,\alpha_3}\} \cup \{M^{\Variable_j,\omega}_a \} \Big )$ in the game $\game^\ms$. Then 
		\[
			\E_{\omega \, : \, \tr((r_\xpt \cdot
                          u_\xpt) \cdot (r_\zpt \cdot u_\zpt)) \neq 0} \Lambda_\omega = 1 - O(\eps)\;.
		\]

	\item \label{enu:qld-ms-cons} (\textbf{Magic square consistency check}) For all $W \in \{X,Z\}$, on average over anticommuting $\omega = (u_\xpt,u_\zpt,r_\xpt,r_\zpt)$,
	\begin{gather}
		M^{(\Point,X),u_\xpt}_{[\tr(\cdot r_\xpt) = a]} \otimes \Id_\bob \simeq_\eps \Id_\alice \otimes M^{\Variable_1,\omega}_{a} \;,\\
		M^{(\Point,Z),u_\zpt}_{[\tr(\cdot r_\zpt) = a]} \otimes \Id_\bob \simeq_\eps \Id_\alice \otimes M^{\Variable_5,\omega}_{a}\;,
	\end{gather}	
	where the answer summations are over $\F_2$.
	
	\end{enumerate}
	Furthermore, all of the approximations above hold with ``$\approx_\eps$'' instead of ``$\simeq_\eps$,'' and they also hold with the tensor factors interchanged.
\end{lemma}

\begin{proof}
	Since the question types are sampled according to the type graph $G^\Pauli$ of
  finite size in Fig.~\ref{fig:type-graph-pauli} in the game
  $\game^\pauli_\qldparams$, the probability of any of the subtests of
  $\decider^\pauli_\qldparams$ (described in Fig.~\ref{fig:decider_pauli})
  being executed is at least some universal constant.
  Thus the consistency conditions expressed in Items
  \ref{enu:qld-cons},~\ref{enu:qld-ld},~\ref{enu:qld-pauli-basis-cons} of the
  lemma follow directly from the corresponding tests in
  Fig.~\ref{fig:decider_pauli} (where Item~\ref{enu:qld-ld} follows from the
  consistency test that is a subtest of $\decider^\ld_\ldparams$) and the fact
  that the strategy $\strategy$ is assumed to succeed with probability at least
  $1 - \eps$ in $\game^\pauli_\qldparams$.
				
  For the remaining items, recall that by Fact~\ref{fact:omega-anticomm-prob}
  the probability that a tuple $\omega=(u_\xpt,u_\zpt,r_\xpt,r_\zpt)$ sampled as
  question content for a question of type $\Pair$, $(\Pair,W)$, $\Constraint_i$
  or $\Variable_j$ is commuting is at least $(1 - 3md/q) \cdot (1/2)$, which is
  at least $1/4$ assuming $6md \leq q$. We claim this assumption holds
  without loss of generality. Indeed, observe that the function
  $\delta$ in Theorem~\ref{thm:pauli-appendix} is monotonically increasing in $a$, so without loss of generality we may assume that $a
  \geq 6$.  Then it follows that we may assume that $6md \leq q$
  without loss of generality, since otherwise, $\delta \geq a md /q
  \geq 1$ and the conclusion of the theorem holds trivially.
  The same lower bound holds for the probability that $\omega$ is anticommuting.
  Therefore, the consistency conditions in
  Items~\ref{enu:qld-comm},~\ref{enu:qld-comm-cons} and~\ref{enu:qld-ms-cons} of
  the lemma also follow directly from the corresponding tests in
  Fig.~\ref{fig:decider_pauli}.
  Finally, Item~\ref{enu:qld-ms} in the lemma follows from the fact that the Magic
  Square check in Fig.~\ref{fig:decider_pauli} is executed with constant probability
  over the choice of a pair of questions.
				
That all of the approximations in the Lemma statement also hold with
$\simeq_\eps$ replaced by $\approx_\eps$ is immediate from Fact~\ref{fact:agreement}.
\end{proof}

\begin{lemma}
\label{len:qld-win-implications-obs}
	The following hold for the observables defined in~\eqref{eq:qld-strat-obs}. 
	\begin{itemize}
		\item For all $W \in \{X,Z\}$, $r \in \F_q$, on average over uniformly random $u \in \F_q^m$ we have
		\begin{equation}
		\label{eq:pts-obs-consistency}
			W^r(u) \otimes \Id_\bob \approx_\eps \Id_\alice \otimes W^r(u).
		\end{equation}
		\item On average over uniformly random $\omega = (u_\xpt,u_\zpt,r_\xpt,r_\zpt)$, 
		\begin{align}
&\begin{aligned}[t]
  &X^{r_\xpt}(u_\xpt) Z^{r_\zpt}(u_\zpt) \otimes \Id_\bob
\end{aligned} \notag \\
&\begin{aligned}[t]
  &\approx_{\sqrt{\eps}} (-1)^{\tr ((r_\xpt \cdot \ind_m(u_\xpt)) \cdot (r_\zpt \cdot \ind_m(u_\zpt)))} \, Z^{r_\zpt}(u_\zpt) X^{r_\xpt}(u_\xpt) \otimes \Id_\bob \;.
\end{aligned} \label{eq:pts-obs-commutation}
\end{align}
	\end{itemize}
	Furthermore, all approximations above hold with the tensor factors interchanged.
\end{lemma}
\begin{proof}
We first establish~\eqref{eq:pts-obs-consistency}. Fix a $W \in \{X,Z\}$ and $r \in \F_q$. From item~\ref{enu:qld-cons} of Lemma~\ref{lem:qld-win-implications} and the data processing inequality (Fact~\ref{fact:data-processing}), using that the question type $(\Point,W)$ has constant probability of being sampled by $\sampler^\pauli$, we get that on average over $u \in \F_q$,
\[
	M^{(\Point,W),u}_{[\tr(\cdot r) = a]} \otimes \Id_\bob \approx_\eps \Id_\alice \otimes M^{(\Point,W),u}_{[\tr(\cdot r) = a]}\;,
\]
where the answer summation is over $a \in \F_2$. Equation~\eqref{eq:pts-obs-consistency} follows directly from~Lemma~\ref{lem:povm-to-obs}. 

We now establish~\eqref{eq:pts-obs-commutation}. We have that on average over {commuting} $\omega = (u_\xpt,u_\zpt,r_\xpt,r_\zpt)$ each of the approximate consistency relations hold, for any $W\in\{X,Z\}$:
\begin{align*}
	&& M^{(\Point,W),u_W}_{[\tr(\cdot r_W)=a]} \otimes \Id_\bob &\simeq_\eps \Id_\alice \otimes M^{(\Pair,W),\omega}_a && \text{(Item~\ref{enu:qld-comm-cons} of Lemma~\ref{lem:qld-win-implications})}\\
	&& &\simeq_\eps M^{\Pair,\omega}_{[\beta_W = a]} \otimes \Id_\bob && \text{(Item~\ref{enu:qld-comm})} \\
	&& &\simeq_\eps \Id_\alice \otimes M^{\Pair,\omega}_{[\beta_W = a]}\;. && \text{(Item~\ref{enu:qld-cons} and Fact~\ref{fact:data-processing})}
\end{align*}
Fact~\ref{fact:agreement} then implies that on average over commuting $\omega$, 
\begin{equation}\label{eq:lc-11}
	M^{(\Point,W),u_W}_{[\tr(\cdot r_W)=a]} \otimes \Id_\bob \approx_\eps \Id_\alice \otimes M^{\Pair,\omega}_{[\beta_W = a]}\;.
\end{equation}
Since
$\{M^{\Pair,\omega}_{\beta_X,\beta_Z} \}$ is projective,
\eqref{eq:lc-11} puts us in
a position to apply Lemma~\ref{lem:commutation-analysis}, setting
$A^{x}_{a,b} = M^{(\Point, X), u_X}_{[\tr(\cdot r_X) = b]}$,
$C^{x}_{a,c} = M^{(\Point, Z), u_Z}_{[\tr(\cdot r_Z) = c]}$, and
$B^{x}_{a,b,c} = M^{\Pair, \omega}_{b,c}$. It follows that
\begin{equation}
\label{eq:qld-obs-comm}
	 M^{(\Point,X),u_\xpt}_{[\tr(\cdot r_\xpt)=b]} M^{(\Point,Z),u_\zpt}_{[\tr(\cdot r_\zpt)=c]} \otimes \Id_\bob \approx_\eps  M^{(\Point,Z),u_\zpt}_{[\tr(\cdot r_\zpt)=c]} M^{(\Point,X),u_\xpt}_{[\tr(\cdot r_\xpt)=b]}  \otimes \Id_\bob
\end{equation}
on average over commuting $\omega$. Then for every $\omega = (u_\xpt,u_\zpt,r_\xpt,r_\zpt)$, since the measurements $\{M^{\tvar,x}_a\}$ are projective we get that
\begin{align}
&\begin{aligned}[t]
  &X^{r_\xpt}(u_\xpt) Z^{r_\zpt}(u_\zpt) = \Big ( \Id - 2 M^{(\Point,X),u_\xpt}_{[\tr(\cdot r_\xpt)=1]} \Big ) \Big ( \Id - 2 M^{(\Point,Z),u_\zpt}_{[\tr(\cdot r_\zpt)=1]} \Big )
\end{aligned} \notag \\
&\begin{aligned}[t]
  &= I - 2 M^{(\Point,X),u_\xpt}_{[\tr(\cdot r_\xpt)=1]} - 2 M^{(\Point,Z),u_\zpt}_{[\tr(\cdot r_\zpt)=1]} + 4 M^{(\Point,X),u_\xpt}_{[\tr(\cdot r_\xpt)=1]} M^{(\Point,Z),u_\zpt}_{[\tr(\cdot r_\zpt)=1]}\;.
\end{aligned} \label{eq:lc-12}
\end{align}
Applying~\eqref{eq:qld-obs-comm} with $b = c = 1$ to commute the
measurement operators in the the fourth term
in~\eqref{eq:lc-12}, and performing the same steps in reverse, we deduce that on average over commuting $\omega$,
\[
	X^{r_\xpt}(u_\xpt) Z^{r_\zpt}(u_\zpt) \approx_\eps  Z^{r_\zpt}(u_\zpt) X^{r_\xpt}(u_\xpt).
\]
This shows the ``commuting'' part of~\eqref{eq:pts-obs-commutation}.

Now we consider the case that $\omega$ is anticommuting. Item~\ref{enu:qld-ms-cons} of Lemma~\ref{lem:qld-win-implications}, combined with Lemma~\ref{lem:povm-to-obs}, implies that the following two approximations hold on average over anticommuting $\omega$:
\begin{gather}
	X^{r_\xpt}(u_\xpt) \otimes \Id_\bob \approx_\eps \Id_\alice \otimes M^{\Variable_1,\omega} \;,\label{eq:lc-11a}\\
	Z^{r_\zpt}(u_\zpt) \otimes \Id_\bob \approx_\eps \Id_\alice \otimes M^{\Variable_5,\omega}\;,\label{eq:lc-11b}
\end{gather}
where for $j\in\{1,5\}$, $M^{\Variable_j,\omega} = M^{\Variable_j,\omega}_0-M^{\Variable_j,\omega}_1$.

For all anticommuting $\omega$, let $\eps_\omega = 1 -
\Lambda_\omega$, where $\Lambda_\omega$ is defined in
item~\ref{enu:qld-ms} of Lemma~\ref{lem:qld-win-implications}. From
item~\ref{enu:qld-ms} of Lemma~\ref{lem:qld-win-implications} we get
that $\E_{\omega \text{ anticomm.}} \eps_\omega \leq
O(\eps)$. For each anticommuting $\omega$,
Theorem~\ref{thm:ms-rigidity} implies that
\[ \Id_\alice \otimes M^{\Variable_1,\omega} M^{\Variable_5,\omega}
  \approx_{\sqrt{\eps_\omega}} -\Id_\alice \otimes M^{\Variable_5,\omega}
  M^{\Variable_1,\omega}\;. \]
Thus, averaging over anticommuting $\omega$, and applying Jensen's
inequality to the definition of $\approx_{\eps}$, we obtain the same
anticommutation relation with error $ \E_{\omega \text{anticomm.}}
\eps_{\omega}^{1/2} \leq (\E_{\omega \text{anticomm.}}
\eps_\omega)^{1/2} = O(\eps^{1/2})$. In summary, on average over anticommuting $\omega$, 
\begin{equation}
\label{eq:qld-implication-ms-anticomm}
\Id_\alice \otimes M^{\Variable_1,\omega} M^{\Variable_5,\omega} \approx_{\sqrt{\eps}} -\Id_\alice \otimes M^{\Variable_5,\omega} M^{\Variable_1,\omega}\;.
\end{equation}
Using \ref{fact:add-a-proj} twice we get from~\eqref{eq:lc-11a} and~\eqref{eq:lc-11b} that on average over anticommuting $\omega$,
\begin{gather}
	X^{r_\xpt}(u_\xpt) Z^{r_\zpt}(u_\zpt) \otimes \Id_\bob \approx_\eps \Id_\alice \otimes M^{\Variable_1,\omega}  M^{\Variable_5,\omega}\;, \\
	Z^{r_\xpt}(u_\xpt) X^{r_\zpt}(u_\zpt) \otimes \Id_\bob \approx_\eps \Id_\alice \otimes M^{\Variable_5,\omega}  M^{\Variable_1,\omega}\;.
\end{gather}
Using~\eqref{eq:qld-implication-ms-anticomm} we get that on average over anticommuting $\omega$, we have
\[
X^{r_\xpt}(u_\xpt) Z^{r_\zpt}(u_\zpt) \otimes \Id_\bob \approx_{\sqrt{\eps}} - Z^{r_\zpt}(u_\zpt)X^{r_\xpt}(u_\xpt)  \otimes \Id_\bob\;.
\]
This shows the ``anticommuting'' part of~\eqref{eq:pts-obs-commutation}.
\end{proof}

\subsection{Expanding the Hilbert space and defining commuting observables}
\label{sec:expanding}

\paragraph{Expanded state.} We introduce registers $\alice', \alice'', \bob', \bob''$ that are each isomorphic to $(\C^q)^{\otimes \nqubits}$. Define the state 
\begin{equation}\label{eq:def-psihat}
	\ket{\hat{\psi}}_{\alice \alice' \alice'' \bob \bob' \bob''} = (\ket{\psi})_{\alice \bob} \otimes \ket{\EPR_q}^{\otimes \nqubits}_{\alice' \alice''} \otimes \ket{\EPR_q}^{\otimes \nqubits}_{\bob' \bob''}\;,
\end{equation}
where we have added maximally entangled states in registers
$\alice' \alice''$ and $\bob' \bob''$; recall also that the state $\ket{\psi}$ is already implicitly  appended with sufficiently many ancilla qubits initialized in the $\ket{0}$ state. 
For the remainder of the Appendix all approximations will be evaluated with respect to this state.

\paragraph{Expanded observables.} For all $W \in \{X,Z\}$, $r \in \F_q$, and $u \in \F_q^m$ define the following observable $\hat{W}^r(u)$: 
\begin{equation}\label{eq:lc-23}
	\hat{W}^r(u) = W^r(u) \otimes \tau^W(r \cdot \ind_m(u))\;,
\end{equation}
where $W^r(u)$ is the observable defined in \eqref{eq:qld-strat-obs} and $\tau^W(r \cdot \ind_m(u))$ is the generalized Pauli observable (see Section~\ref{sec:generalized-pauli}) acting on $(\C^q)^{\otimes \nqubits}$. 

For $a\in \F_q$ define corresponding projections
\[
	\hat{M}^{(\Point,W),u}_a = \E_{r \in \F_q} (-1)^{\tr(ar)} \, \hat{W}^r(u)\;.
\]
Note that $\hat{M}^{(\Point,W),u}_a$ can be equivalently
written as a sum of projectors by applying Fourier
identities. Specifically, for $W \in \{X, Z\}$, $u \in \F_q^{m}$, and
$a \in \F_q$, define the projector
\begin{equation} \tau^{W, u}_{a} =
  \tau^{W}_{[ g_{\cdot}(u) = a]} = \sum_{h \in \F_q^{\nqubits}: h
    \cdot \ind_m(u) = a} \tau^{W}_{h}\;. \label{eq:qld-point-obs-def} \end{equation}
Then we can write $\hat{M}^{(\Point, W), u}_a$ as follows:
\begin{align*}
&\begin{aligned}[t]
  &\hat{M}^{(\Point,W),u}_a = \E_{r \in \F_q} (-1)^{\tr(ar)} \, W^r(u) \otimes \tau^W(r \cdot \ind_m(u))
\end{aligned}  \\
&\begin{aligned}[t]
  &= \E_{r \in \F_q} (-1)^{\tr(ar)} \, \Big(\sum_{a' \in \F_q}(-1)^{\tr (a' r)} M^{(\Point,W),u}_{a'} \Big) \\
  &\otimes \Big( \sum_{h \in \F_q^n} (-1)^{\tr( r( h \cdot \ind_m(u)))} \tau^{W}_{h} \Big)
\end{aligned}  \\
&\begin{aligned}[t]
  &= \E_{r \in \F_q} (-1)^{\tr(ar)} \, \Big(\sum_{a' \in \F_q}(-1)^{\tr (a' r)} M^{(\Point,W),u}_{a'} \Big) \otimes \Big( \sum_{a'' \in \F_q} (-1)^{\tr(a'' r)} \tau^{W,u}_{a''} \Big)
\end{aligned}  \\
&\begin{aligned}[t]
  &= \sum_{a',a'' \in \F_q} \E_{r \in \F_q} (-1)^{\tr((a + a' + a'')r)} \, M^{(\Point,W),u}_{a'} \otimes \tau^{W,u}_{a''}
\end{aligned}  \\
&\begin{aligned}[t]
  &= \sum_{\substack{a',a'' \in \F_q \\ a' + a'' = a}} \, M^{(\Point,W),u}_{a'} \otimes \tau^{W,u}_{a''}\;,
\end{aligned} 
\end{align*}
where in the second line we have used \eqref{eq:qld-strat-obs} on the
first tensor factor and \eqref{eq:pauli-obs-proj} on the second, and in the third line we
have used \eqref{eq:qld-point-obs-def}.
For a fixed $u\in\F_q^m$, $r \in \F_q$ and $b \in \F_2$, define a projection
\begin{equation}
	\hat{M}^{(\Point,W),u,r}_b = \sum_{a \in \F_q : \tr(ar) = b}
        \hat{M}^{(\Point,W),u}_a\;, \label{eq:qld-def-mptur}
\end{equation}
which can be equivalently written as
\begin{equation}\label{eq:lc-22}
\hat{M}^{(\Point,W),u,r}_b = \frac{1}{2} \Big ( \Id + (-1)^b \hat{W}^r(u) \Big)\;.
\end{equation}

The registers that the operators $\hat{W}^r(u)$,
$\hat{M}^{(\Point,W),u}_a$, and $\hat{M}^{(\Point,W),u,r}_b$ act on
depend on context. For example if $W^r(u)$ is viewed as an operator
acting on register $\alice$, then we will view $\hat{W}^r(u)$ as
acting on registers $\alice \alice'$ (with the $\tau^W$ observable
acting on register $\alice'$). We generally indicate via subscripts
which registers the operators act on (for example we will write
$(\hat{W}^r(u))_{\alice \alice'}$ or $(\hat{W}^r(u))_{\bob
  \alice''}$).

\paragraph{Partitioning the registers, and symmetries}
In the remainder of this section as well as
Sections~\ref{sec:combining} and~\ref{sec:apply-ldt}, we will partition the six registers
$\alice \alice' \alice'' \bob \bob' \bob''$ into
two ``parties'' in two different ways. The first is to consider
$\alice \alice'$ together to form  a single party, and $\bob \alice''$
to form the second party. The second is to consider $\bob \bob'$ to
form the first party, and $\alice \bob''$ to form the second
party. Thanks to the symmetry of the Pauli basis test under exchanging the
two players, as well as the symmetry of the ancilla states in
registers $\alice' \alice''$ and $\bob' \bob''$, every ``bipartite'' consistency relation we derive between two
operators $M, N$ will hold for both of the two bipartitioning schemes,
and with the first and second party interchanged. That is, whenever we
derive a bipartite relation of the form
\[ M_{\alice \alice'} \approx_\delta N_{\bob \alice''}, \]
the same steps, with the registers appropriately changed, will also
yield
\begin{align*}
  N_{\alice \alice'} &\approx_{\delta} M_{\bob \alice''}, \\
  M_{\bob \bob'} &\approx_{\delta} N_{\alice \bob''}, \\
  N_{\bob \bob'} &\approx_{\delta} M_{\alice \bob''}.
\end{align*}

Similarly, whenever we derive a single-party relation of the form
$M_{\alice \alice'} \approx N_{\alice \alice'}$, the same steps with
the registers appropriately changed will also yield the same relation
on $\bob\alice''$, $\bob \bob'$ and $\alice \bob''$. We refer to these
relations as \emph{symmetric equivalents}.

\begin{lemma}
\label{lem:qld-comm-cons}
	There exists a function $\delta_{\acom}(\eps) = \poly(\eps)$ such that the following holds. 
	
\begin{enumerate}
	\item (\textbf{Self-consistency}) For all $W \in \{X,Z\}$, on average over $r \in \F_q$ and $u\in \F_q^m$, we have 
	\[
		(\hat{M}^{(\Point,W),u}_a)_{\alice \alice'} \approx_\eps (\hat{M}^{(\Point,W),u}_a)_{\bob \alice''}\,.
		\]
	\item (\textbf{Approximate commutation}) On average over
          uniformly random $\omega = (u_\xpt,u_\zpt,r_\xpt,r_\zpt)$,
          we have 
	\begin{equation*}
\begin{aligned}[t]
  &(\hat{M}^{(\Point,\xpt),u_\xpt,r_\xpt}_{b} \hat{M}^{(\Point,\zpt),u_\zpt,r_\zpt}_{b'})_{\alice \alice'} \\
  &\approx_{\delta_{\acom}} (\hat{M}^{(\Point,\zpt),u_\zpt,r_\zpt}_{b'} \hat{M}^{(\Point,\xpt),u_\xpt,r_\xpt}_{b})_{\alice \alice'} \;.
\end{aligned} 
\end{equation*}
      \end{enumerate}
      Furthermore, all symmetric equivalents of these approximations
      also hold.
\end{lemma}

\begin{proof}
Item 1, the self-consistency of $\{\hat{M}^{(\Point,W),u}_a \}$, follows from the fact that the original points measurements $\{ M^{(\Point,W),u}_a \}$ are approximately self-consistent between registers $\alice$ and $\bob$ in $\ket{\hat{\psi}}$ (item~\ref{enu:qld-cons} in Lemma~\ref{lem:qld-win-implications}), the Pauli projections $\{\tau^W_a(\ind_m(u))\}$ are perfectly self-consistent between registers $\alice'$ and $\alice''$, and~Fact~\ref{fact:data-processing}. 
	
	We now establish the approximate commutation relations. On average over uniformly random $\omega = (u_\xpt,u_\zpt,r_\xpt,r_\zpt)$, we have
	\begin{align*}
	&\hat{M}^{(\Point,\xpt),u_\xpt,r_\xpt}_{b} \hat{M}^{(\Point,\zpt),u_\zpt,r_\zpt}_{b'}  \label{eq:qld-m-x-m-z} \\
	&= \frac{1}{4} \Big ( \Id + (-1)^b \hat{X}^{r_\xpt}(u_\xpt) \Big ) \Big ( \Id + (-1)^{b'} \hat{Z}^{r_\zpt}(u_\zpt) \Big ) \\
	&= \frac{1}{4} \Big ( \Id + (-1)^b \hat{X}^{r_\xpt}(u_\xpt) + (-1)^{b'} \hat{Z}^{r_\zpt}(u_\zpt) + (-1)^{b + b'} \hat{X}^{r_\xpt}(u_\xpt) \hat{Z}^{r_\zpt}(u_\zpt) \Big)\;,
	\end{align*}
	where the first equality follows from~\eqref{eq:lc-22}.	
	Note that
	\begin{align*}
	\hat{X}^{r_\xpt}(u_\xpt)
         & \hat{Z}^{r_\zpt}(u_\zpt) \\
				&=
                                             X^{r_\xpt}(u_\xpt)
                                             Z^{r_\zpt}(u_\zpt)
                                             \otimes \tau^X(r_\xpt
                                             \cdot \ind_m(u_\xpt))
                                             \tau^Z(r_\zpt \cdot \ind_m(u_\zpt)) \\
	&\approx_{\sqrt{\eps}} Z^{r_\zpt}(u_\zpt)
   X^{r_\xpt}(u_\xpt)  \otimes \tau^Z(r_\zpt\cdot \ind_m( u_\zpt))
   \tau^X(r_\xpt \cdot \ind_m(u_\xpt)) \\
	&= \hat{Z}^{r_\zpt}(u_\zpt) \hat{X}^{r_\xpt}(u_\xpt)\;, 
	\end{align*}
	where the first equality follows from~\eqref{eq:lc-23} and for the second line we used the approximate (anti-)commutation
        relation~\eqref{eq:pts-obs-commutation} for the
        $X^{r_\xpt}(u_\xpt)$ and  $Z^{r_\zpt}(u_\zpt)$
        observables, as well as the exact (anti-)commutation relations
        for the $\tau^Z(r_\zpt \cdot \ind_m(u_\zpt))$ and
        $\tau^X(r_\xpt \cdot \ind_m(u_\xpt))$ observables. This shows
        the desired approximate commutation relation, with
        $\delta_{\acom} = \sqrt{\eps}$.
\end{proof}

\begin{lemma}
\label{lem:qld-comm-line-cons}
There exists a function $\delta_{\Line}(\eps) = \poly(\eps)$ such that the following holds. 
	For all $W \in \{X,Z\}$ and lines $\line$ there exists a projective measurement $\{ \hat{M}^{(\Line,W),\line}_f \}_{f \in \deg_{md}(\line)}$, acting on registers $\alice\alice'$ or $\bob\alice''$, such that:
	\begin{enumerate}
		\item  (\textbf{Self-consistency})\label{enu:qld-comm-line-self-cons}  For all $W \in \{X,Z\}$ and $r \in \F_q$, on average over a uniformly random  pair $(\line, u)$
                drawn from the line-point distribution
                (Definition~\ref{def:line-point-dist}),
		\[
			(\hat{M}^{(\Line,W),\line}_f)_{\alice \alice'} \approx_{\delta_\Line} (\hat{M}^{(\Line,W),\line}_f)_{\bob \alice''}\;.
		\]
		
		\item (\textbf{Consistency with points measurements I}) \label{eq:qld-comm-line-pt-cons}
		On average over a uniformly random pair $(\line, u)$
                drawn from the line-point distribution, 
		\[
			(\hat{M}^{(\Line,W),\line}_f)_{\alice \alice'} \approx_{\delta_\Line} (\hat{M}^{(\Line,W),\line}_f)_{\alice \alice'} \otimes (\hat{M}^{(\Point,W),u}_{f(u)})_{\bob \alice''}\;.
		\]		
		\item (\textbf{Consistency with points measurements II}) \label{eq:qld-comm-line-pt-cons2}
		On average over a uniformly random pair $(\line, u)$
                drawn from the line-point distribution,
		\[
			(\hat{M}^{(\Line,W),\line}_{[\eval_u(\cdot) = a]})_{\alice \alice'} \approx_{\delta_\Line}  (\hat{M}^{(\Point,W),u}_{a})_{\bob \alice''}\;.
		\]
              \end{enumerate}
                    Furthermore, all symmetric equivalents of these approximations
      also hold.
\end{lemma}

\begin{proof}
	Define the operator
	\[
		\hat{M}^{(\Line,W),\line}_f = \sum_{\substack{f',f'' \in \deg_{md}(\line) : \\ f' + f'' = f}} M^{(\Line,W),\line}_{f'} \otimes \tau^{W,\line}_{f''}\;,
	\]
	where the Pauli operator $\tau^{W,\line}_{f''}$ acting on register $\alice'$ is obtained by measuring all $\nqubits$ qudits in the basis $W$ to obtain an outcome $h \in \F_q^\nqubits$ and then returning the restriction $f''$ of $g_h: \F_q^m \to \F_q$ to the line $\line$. Here, $g_h$ is the degree-$d$ low-degree extension of $h$ defined in~\eqref{eq:ld-encoding}. 
	
	Item~\ref{enu:qld-comm-line-self-cons},	the self-consistency of $\{\hat{M}^{(\Line,W),\line}_f \}$, follows from the fact that the original lines measurement $\{ M^{(\Line,W),\line}_f \}$ is self-consistent between registers $\alice$ and $\bob$ in $\ket{\hat{\psi}}$ (Item~\ref{enu:qld-cons} in Lemma~\ref{lem:qld-win-implications}), the Pauli projections $\{\tau^{W,\line}_{f''}\}$ are perfectly self-consistent on between registers $\alice'$ and $\alice''$, and~Fact~\ref{fact:data-processing}. 

Equation~\eqref{eq:qld-comm-line-pt-cons}, the consistency with the points measurements, follows because on average over $(\line,u)$ drawn from the line-point distribution we have
\begin{align}
&\begin{aligned}[t]
  &(\hat{M}^{(\Line,W),\line}_f)_{\alice \alice'}
\end{aligned}  \\
&\begin{aligned}[t]
  &= \sum_{\substack{f',f'' \in \deg_{md}(\line) : \\ f' + f'' = f}} (M^{(\Line,W),\line}_{f'} \otimes \tau^{W,\line}_{f''})_{\alice \alice'}
\end{aligned}  \\
&\begin{aligned}[t]
  &= \sum_{\substack{f', f'' \in \deg_{md}(\line): \\ f' + f'' = f}} ((M^{(\Line, W), \line}_{f'} \otimes \tau^{W, \line}_{f''}) \cdot (M^{(\Line, W), \line}_{[\eval_u(\cdot) = f'(u)]} )_{\alice \alice'}
\end{aligned} \label{eq:qld-mhat-line-1} \\
&\begin{aligned}[t]
  &\approx_\eps \sum_{\substack{f',f'' \in \deg_{md}(\line) : \\ f' + f'' = f}} (M^{(\Line,W),\line}_{f'} \otimes \tau^{W,\line}_{f''})_{\alice \alice'} \otimes (M^{(\Point,W),u}_{f'(u)})_{\bob \alice''}
\end{aligned} \label{eq:qld-mhat-line-2} \\
&\begin{aligned}[t]
  &= \sum_{\substack{f',f'' \in \deg_{md}(\line) : \\ f' + f'' = f}} (M^{(\Line,W),\line}_{f'} \otimes \tau^{W,\line}_{f''})_{\alice \alice'} \otimes (M^{(\Point,W),u}_{f'(u)} \otimes \tau^{W,u}_{f''(u)})_{\bob \alice''}
\end{aligned} \label{eq:qld-mhat-line-3} \\
&\begin{aligned}[t]
  &= \sum_{\substack{f',f'' \in \deg_{md}(\line) : \\ f' + f'' = f}} (M^{(\Line,W),\line}_{f'} \otimes \tau^{W,\line}_{f''})_{\alice \alice'} \\
  &\otimes \Big (\sum_{\substack{a', a'' : \\ a' + a'' = f'(u) + f''(u)}} M^{(\Point,W),u}_{a'} \otimes \tau^{W,u}_{a''} \Big )_{\bob \alice''}
\end{aligned} \label{eq:qld-mhat-line-4} \\
&\begin{aligned}[t]
  &= (\hat{M}^{(\Line,W),\line}_f)_{\alice \alice'} \otimes (\hat{M}^{(\Point,W),u}_{f(u)})_{\bob \alice''}\;.
\end{aligned} \label{eq:qld-comm-line-pt-cons-eps}
\end{align}
Equation~\eqref{eq:qld-mhat-line-1} follows from the projectivity of
$\{\hat{M}^{(\Line, W), \line}_{f}\}$. Equation~\eqref{eq:qld-mhat-line-2} follows from the consistency between the line
and point measurements (Item~\ref{enu:qld-ld} in
Lemma~\ref{lem:qld-win-implications}) together with
Fact~\ref{fact:add-a-proj2}. Equation~\eqref{eq:qld-mhat-line-3}
follows from the exact consistency between the $\{ \tau^{W,\line}_{f''} \}$ and $\{
\tau^{W,u}_{f''(u)} \}$ measurements on the $\alice'$ and $\alice''$ registers. Equation~\eqref{eq:qld-mhat-line-4} follows because
the consistency of the $\tau$ measurements force $a' = f'(u)$ and $a''
= f''(u)$ . This establishes \eqref{eq:qld-comm-line-pt-cons}, with
approximation error $\eps$.

To show \eqref{eq:qld-comm-line-pt-cons2} we would like to first
apply~Fact~\ref{fact:data-processing} to \eqref{eq:qld-comm-line-pt-cons-eps},
to sum over all functions $f$ that evaluate to the same value at a given
point $u$. Unfortunately, we cannot use the Fact as written, since the
bound in \eqref{eq:qld-comm-line-pt-cons} is in terms of $\approx$, not
$\simeq$, and, moreover, the right-hand side does not have the form $I
\ot B$. Instead, we show the desired bound by direct calculation from \eqref{eq:qld-comm-line-pt-cons-eps}
\begin{align}
  \eps &\geq \E_{\omega} \sum_{f} \bra{\hat{\psi}} (\hat{M}^{(\Line, W),
  \line}_f )^2 \ot (I - \hat{M}^{(\Point, W), u}_{f(u)})^2 \ket{\hat{\psi}}
  \\
                 &= \E_{\omega} \sum_{f} \bra{\hat{\psi}} \hat{M}^{(\Line, W),
                   \line}_f  \ot (I - \hat{M}^{(\Point, W), u}_{f(u)})
                   \ket{\hat{\psi}} \\
                 &= \E_{\omega} \sum_{a} \sum_{f: f(u) = a} \bra{\hat{\psi}} \hat{M}^{(\Line, W),
                   \line}_f  \ot (I - \hat{M}^{(\Point, W), u}_{a})
                   \ket{\hat{\psi}} \\
                 &=\E_{\omega} \sum_{a}  \bra{\hat{\psi}} \hat{M}^{(\Line, W),
                   \line}_{[\eval_u(\cdot) = a]}  \ot (I - \hat{M}^{(\Point, W), u}_{a})
                   \ket{\hat{\psi}},
\end{align}
where in going to the second line we have used the projectivity of the
measurements $\{\hat{M}^{(\Line, W), \line}_f\}$ and
$\{\hat{M}^{(\Point, W), u}_{a}\}$.
Thus, we have deduced that
\begin{equation}
\label{eq:qld-comm-line-1}
(\hat{M}^{(\Line,W),\line}_{[\eval_u(\cdot) = a]})_{\alice \alice'} \approx_\eps (\hat{M}^{(\Line,W),\line}_{[\eval_u(\cdot) = a]})_{\alice \alice'} \otimes (\hat{M}^{(\Point,W),u}_{a})_{\bob \alice''}\;.
\end{equation}
Since both the $\{ \hat{M}^{(\Line,W),\line}_f\}$ and $\{ \hat{M}^{(\Point,W),u}_a \}$ measurements are projective, we also get 
\begin{align}
\Id &= \sum_a (\hat{M}^{(\Line,W),\line}_{[\eval_u(\cdot) =
      a]})_{\alice \alice'} \otimes I_{\bob \alice''}\notag\\
&= \sum_a \Big((\hat{M}^{(\Line,W),\line}_{[\eval_u(\cdot) =
                                                               a]})_{\alice
                                                               \alice'}
                                                               \ot
                                                               I_{\bob
                                                               \alice''}-(\hat{M}^{(\Line,W),\line}_{[\eval_u(\cdot) = a]})_{\alice \alice'} \otimes (\hat{M}^{(\Point,W),u}_{a})_{\bob \alice''}\Big) \notag\\
&\qquad\qquad + \sum_a (\hat{M}^{(\Line,W),\line}_{[\eval_u(\cdot) = a]})_{\alice \alice'} \otimes (\hat{M}^{(\Point,W),u}_{a})_{\bob \alice''}\;.\label{eq:com-1a}
\end{align}
For projective sub-measurements $\{A_i\}$ and $\{B_i\}$ and a state $\ket{\varphi}$,
\begin{align*}
\sum_i \bra{\varphi}\big(A_i-B_i\big)\ket{\varphi} &= \sum_i \bra{\varphi}A_i(A_i-B_i)\ket{\varphi} + \sum_i \bra{\varphi}(A_i-B_i)B_i\ket{\varphi}\\
&\leq \Big(\sum_i \bra{\varphi}A_i^2\ket{\varphi} \Big)^{1/2}\Big(\sum_i \bra{\varphi}(A_i-B_i)^2\ket{\varphi}\Big)^{1/2} \\
&\qquad\qquad+ \Big(\sum_i \bra{\varphi}(A_i-B_i)^2 \ket{\varphi}\Big)^{1/2}\Big(\sum_i \bra{\varphi}B_i^2\ket{\varphi}\Big)^{1/2}\\
&\leq 2\Big(\sum_i \bra{\varphi}(A_i-B_i)^2 \ket{\varphi}\Big)^{1/2}\;.
\end{align*}
Applying this fact to the projective sub-measurements $\{(\hat{M}^{(\Line,W),\line}_{[\eval_u(\cdot) = a]})_{\alice \alice'}\}_a$ and $\{(\hat{M}^{(\Line,W),\line}_{[\eval_u(\cdot) = a]})_{\alice \alice'} \otimes (\hat{M}^{(\Point,W),u}_{a})_{\bob \alice''}\}$ in~\eqref{eq:com-1a} and using~\eqref{eq:qld-comm-line-1} gives
\begin{equation}
\label{eq:qld-comm-line-2}
	\Id \approx_{\sqrt{\eps}} \sum_a (\hat{M}^{(\Line,W),\line}_{[\eval_u(\cdot) = a]})_{\alice \alice'} \otimes (\hat{M}^{(\Point,W),u}_{a})_{\bob \alice''}\;.
      \end{equation}
      To go from~\eqref{eq:qld-comm-line-2} to \eqref{eq:qld-comm-line-pt-cons2} of the lemma we use another fact about
      projective measurements. For families of projective measurements $\{A^x_i\}$
      and $\{B^x_i\}$ acting on separate registers of a bipartite state
      $\ket{\phi}$, suppose $I \approx_{\delta} \sum_i A^x_i \ot B^x_i$. Then
      \begin{align*}
        \delta &\geq \E_{x} \Big\| \Big(\Id - \sum_i A^x_i \ot B^x_i\Big) \ket{\phi}\Big\|^2  \\
               &= \E_{x} \Big(1 + \bra{\phi} \Big(\sum_i A^x_i \ot
                 B^x_i\Big)^2 \ket{\phi} - 2 \sum_i \bra{\phi} A^x_i \ot B^x_i
                 \ket{\phi} \Big) \\
               &= 1 - \E_{x} \sum_i \bra{\phi} A^x_i \ot B^x_i \ket{\phi}\;,
      \end{align*}
      where we used the projectivity in going from the second to
      the third line, to remove the square. Thus
      \[ \E_{x} \sum_{i \neq j} \bra{\phi} A^x_i \ot B^x_j \ket{\phi} \leq
        \delta\;, \]
      and $A^x_i \ot I \simeq_{\delta} I \ot B^x_i$. By
      Item~1 of Fact~\ref{fact:agreement}, this in turn implies that $A^x_i \ot I
      \approx_\delta I \ot B^x_i$. Applying this fact to the projective
      measurements $\{ (\hat{M}^{(\Line, W), \line}_{[\eval_u(\cdot) =
        a]})_{\alice \alice'} \}_a$ and $\{(\hat{M}^{(\Point,W),
        u}_{a})_{\bob \alice''}\}$ in~\eqref{eq:qld-comm-line-2} gives
      \[
	(\hat{M}^{(\Line,W),\line}_{[\eval_u(\cdot) = a]})_{\alice \alice'} \approx_{\sqrt{\eps}} (\hat{M}^{(\Point,W),u}_{a})_{\bob \alice''}.
      \]
      Taking the function $\delta_{\Line}$ in the lemma to be $\sqrt{\eps}$,
      this yields the conclusion of \eqref{eq:qld-comm-line-pt-cons2}
      of the lemma.
\end{proof}


\subsection{Combining the $X$ and $Z$ measurements}
\label{sec:combining}

In this section our goal is to create a set of combined measurements
that simultaneously measure the approximately-commuting measurements
constructed in the previous section. We do this first for the points.
Recall that $\mH_\alice$ and $\mH_\bob$ denote the Hilbert spaces corresponding to
registers $\alice$ and $\bob$, respectively.

\begin{lemma}
\label{lem:qld-4-10}
There exists a function $\delta_Q(\eps) = \poly(\eps)$ such that the following holds. 
	For every $x,z \in \F_q^m$ and $\mH \in \{ \mH_\alice, \mH_\bob\}$ there are projective measurements $\{ \hat{Q}^{x,z}_{a,b} \}_{a,b \in \F_q}$ acting on $\mathcal{H} \otimes (\C^q)^{\otimes \nqubits}$ such that:
	\begin{enumerate}
		\item (\textbf{Self-consistency}) On average over uniformly random $x,z \in \F_q^m$,
		\begin{equation}
			\label{eq:qld-q-self-cons}
			(\hat{Q}^{x,z}_{a,b})_{\alice \alice'}  \approx_{\delta_Q} (\hat{Q}^{x,z}_{a,b})_{\bob \alice''}\,.
		\end{equation}
		
		\item (\textbf{Consistency with $\hat{M}$}) On average over uniformly random $x,z \in \F_q^m$,
		\begin{gather}
			\label{eq:qld-q-cons-m-hat-xz}
				(\hat{Q}^{x,z}_{a,b})_{\alice \alice'} \approx_{\delta_Q} (\hat{M}^{(\Point,X),x}_{a} \hat{M}^{(\Point,Z),z}_b)_{\bob \alice''}\;, \\
							\label{eq:qld-q-cons-m-hat-zx}
				(\hat{Q}^{x,z}_{a,b})_{\alice \alice'} \approx_{\delta_Q} (\hat{M}^{(\Point,Z),z}_b  \hat{M}^{(\Point,X),x}_{a} )_{\bob \alice''}\;.
		\end{gather}
              \end{enumerate}
                    Furthermore, all symmetric equivalents of these approximations
      also hold.
\end{lemma}

The construction of the measurements $\{ \hat{Q}^{x,z}_{a,b} \}$ whose existence is guaranteed in the lemma proceeds in three steps. In the first step, for every $(r,s)\in\F_q^2$ we combine the binary-outcome measurements $\hat{M}^{(\Point,X),x,r}_a$ and $\hat{M}^{(\Point,X),z,s}_b$; this is made possible by their approximate commutation as shown in Lemma~\ref{lem:qld-comm-cons}. In a second step, we show that the measurements thus constructed have a particular ``approximate linearity'' property, when seen as a function of the pair $(r,s)$. This allows us, in the third step, to combine them into a single measurement $\{ \hat{Q}^{x,z}_{a,b} \}$ with outcomes in $\F_q^2\cong \F_2^{2t}$ by applying the following  result from~\cite{NV17}:
	
	\begin{theorem}[Quantum linearity test]
	\label{thm:linearity}
		Let $0 \leq \delta \leq 1$, let $t$ be a positive integer, and let $\ket{\psi}$ be a state in a Hilbert space $\mathcal{H}_\alice \otimes \mathcal{H}_\bob$. Let $\{\mathcal{O}^u \}$ be a set of observables indexed by $u \in \F_2^t$ that act on $\mathcal{H}_\alice$. Suppose that on average over $u, u'$ chosen uniformly from $\F_2^t$, we have the following ``approximate linearity'' holds:
		\[
			(\mathcal{O}^u \mathcal{O}^{u'})_{\alice} \approx_\delta (\mathcal{O}^{u+u'})_{\alice}
		\]
		where the $\approx$ statement is taken with respect to the state $\ket{\psi}$. 
		Then there exists an extended state $\ket{\psi'}_{\alice \alice' \bob} = \ket{\psi}_{\alice \bob} \otimes \ket{0}_{\alice'}$ in an extended space $\mathcal{H}_{\alice} \otimes \mathcal{H}_\bob \otimes \mathcal{H}_{\alice'}$ and observables $\{ \mathcal{L}^u \}$ acting on $\mathcal{H}_\alice \otimes \mathcal{H}_{\alice'}$ such that for all $u, u' \in \F_2^t$, 
		\[
			\mathcal{L}^u \mathcal{L}^{u'} = \mathcal{L}^{u + u'} \qquad \text{and} \qquad (\mathcal{L}^u)_{\alice \alice'} \approx_\delta (\mathcal{O}^u)_\alice \,.
		\]
		Furthermore, the observable $\mathcal{L}^u = \Id$ when $u = 0$.
	\end{theorem}
		
		We now give the proof of Lemma~\ref{lem:qld-4-10}.
		
\begin{proof}[Proof of Lemma~\ref{lem:qld-4-10}]
For convenience in the proof we slightly abuse the notation $\approx_\delta$ to mean $\approx_{\poly(\eps)}$ for some polynomial function $\delta=\poly(\eps)$ that may differ each time the notation is used. 

\textbf{First step: construction of projective binary measurements \allowbreak{}$\{ \hat{R}^\omega_{a,b} \}$\allowbreak{}.}
	For every \allowbreak{}$\omega \allowbreak = (x,\allowbreak z,\allowbreak r,\allowbreak s)$\allowbreak{} and \allowbreak{}$a,\allowbreak b \allowbreak \in \F_2$\allowbreak{}, define a measurement operator
	\[
		\hat{R}^\omega_{a,b} = \hat{M}^{(\Point,\zpt),z,s}_{b} \cdot \hat{M}^{(\Point,\xpt),x,r}_{a} \cdot \hat{M}^{(\Point,\zpt),z,s}_{b} \,.
	\]
	It is clear that $\{ \hat{R}^\omega_{a,b} \}_{a,b\in\F_2}$ is a POVM. Observe that on average over uniformly random $\omega$, 
	\begin{align}
	\label{eq:qld-r-2}
	\hat{R}^\omega_{a,b} \approx_{\sqrt{\delta_{\acom}}} \hat{M}^{(\Point,\zpt),z,s}_{b} \cdot \hat{M}^{(\Point,\xpt),x,r}_{a}\;,
	\end{align}
	which follows from the approximate commutation relation of Lemma~\ref{lem:qld-comm-cons} and the projectivity of $\{ \hat{M}^{(\Point,\zpt),z,s}_{b} \}$, together with Fact~\ref{fact:add-a-proj}. Next we show that $\{ \hat{R}^\omega_{a,b} \}$ is approximately self-consistent:  we aim to show that on average over $\omega$, it holds that $(\hat{R}^\omega_{a,b})_{\alice \alice'} \simeq_\delta (\hat{R}^\omega_{a,b})_{\bob \alice''}$ for some polynomial function $\delta = \poly(\eps)$. To show this, we perform the following calculation: (recall the state $\ket{\hat{\psi}}$ defined in~\eqref{eq:def-psihat})
	\begin{align}
&\begin{aligned}[t]
  &\E_\omega \sum_{a,b} \bra{\hat{\psi}} (\hat{R}^\omega_{a,b})_{\alice \alice'} \otimes (\hat{R}^\omega_{a,b})_{\bob \alice''} \ket{\hat{\psi}}
\end{aligned} \notag \\
&\begin{aligned}[t]
  &\approx_{\delta_{\acom}^{1/4}} \E_\omega \sum_{a,b} \bra{\hat{\psi}} (\hat{M}^{(\Point,\zpt),z,s}_{b} \cdot \hat{M}^{(\Point,\xpt),x,r}_{a})_{\alice \alice'} \otimes (\hat{R}^\omega_{a,b})_{\bob \alice''} \ket{\hat{\psi}}
\end{aligned} \label{eq:qld-rw-self-cons-1} \\
&\begin{aligned}[t]
  &\approx_{\delta_{\acom}^{1/4}} \E_\omega \sum_{a,b} \bra{\hat{\psi}} (\hat{M}^{(\Point,\zpt),z,s}_{b} \cdot \hat{M}^{(\Point,\xpt),x,r}_{a})_{\alice \alice'} \\
  &\otimes (\hat{M}^{(\Point,\zpt),z,s}_{b} \cdot \hat{M}^{(\Point,\xpt),x,r}_{a})_{\bob \alice''} \ket{\hat{\psi}}
\end{aligned} \label{eq:qld-rw-self-cons-2} \\
&\begin{aligned}[t]
  &\approx_{\sqrt{\eps}} \E_\omega \sum_{a,b} \bra{\hat{\psi}} (\hat{M}^{(\Point,\zpt),z,s}_{b})_{\alice \alice'} \otimes (\hat{M}^{(\Point,\zpt),z,s}_{b} \cdot \hat{M}^{(\Point,\xpt),x,r}_{a})_{\bob \alice''} \ket{\hat{\psi}}
\end{aligned} \label{eq:qld-rw-self-cons-3} \\
&\begin{aligned}[t]
  &= \E_\omega \sum_{b} \bra{\hat{\psi}} (\hat{M}^{(\Point,\zpt),z,s}_{b})_{\alice \alice'} \otimes (\hat{M}^{(\Point,\zpt),z,s}_{b})_{\bob \alice''} \ket{\hat{\psi}}
\end{aligned} \notag \\
&\begin{aligned}[t]
  &\approx_{\eps} \E_\omega \sum_{b} \bra{\hat{\psi}} (\hat{M}^{(\Point,\zpt),z,s}_{b})_{\alice \alice'} \ket{\hat{\psi}}
\end{aligned} \label{eq:qld-rw-self-cons-4} \\
&\begin{aligned}[t]
  &= 1\;.
\end{aligned} \notag
\end{align}
   	The approximation in~\eqref{eq:qld-rw-self-cons-1} is derived by bounding the magnitude of the difference:
	\begin{align*}
&\begin{aligned}[t]
  &\left | \E_\omega \sum_{a,b} \bra{\hat{\psi}} (\hat{R}^\omega_{a,b} - \hat{M}^{(\Point,\zpt),z,s}_{b} \cdot \hat{M}^{(\Point,\xpt),x,r}_{a})_{\alice \alice'} \otimes (\hat{R}^\omega_{a,b})_{\bob \alice''} \ket{\hat{\psi}} \right |
\end{aligned}  \\
&\begin{aligned}[t]
  &\leq \sqrt{\E_\omega \sum_{a,b} \bra{\hat{\psi}} (\hat{R}^\omega_{a,b} - \hat{M}^{(\Point,\zpt),z,s}_{b} \cdot \hat{M}^{(\Point,\xpt),x,r}_{a})_{\alice \alice'}^2 \ket{\hat{\psi}} } \\
  &\cdot \sqrt{\E_\omega \sum_{a,b} \bra{\psi} (\hat{R}^\omega_{a,b})^2_{\bob \alice''} \ket{\psi} }
\end{aligned}  \\
&\begin{aligned}[t]
  &\leq \sqrt{\delta_{\acom}^{1/2}} \cdot \sqrt{1}\;,
\end{aligned} 
\end{align*}
	where the second line follows from Cauchy-Schwarz, and the third line follows from~\eqref{eq:qld-r-2} and the fact that $\sum_{a,b} (\hat{R}^\omega_{a,b})^2 \leq \Id$. The approximation in~\eqref{eq:qld-rw-self-cons-2} follows from a similar calculation. 
	The approximation in~\eqref{eq:qld-rw-self-cons-3} is derived by bounding the magnitude of the difference:
	\begin{align}
&\begin{aligned}[t]
  &\left |\begin{aligned}[t]
  &\E_\omega \sum_{a,b} \bra{\hat{\psi}} (\hat{M}^{(\Point,\zpt),z,s}_{b} \cdot (\Id - \hat{M}^{(\Point,\xpt),x,r}_{a}))_{\alice \alice'} \\
  &\otimes (\hat{M}^{(\Point,\zpt),z,s}_{b} \cdot \hat{M}^{(\Point,\xpt),x,r}_{a})_{\bob \alice''} \ket{\hat{\psi}}
\end{aligned}\right |
\end{aligned} \notag \\
&\begin{aligned}[t]
  &= \left |\begin{aligned}[t]
  &\E_\omega \bra{\hat{\psi}} \Big(\sum_b \hat{M}^{(\Point,\zpt),z,s}_{b} \otimes \hat{M}^{(\Point,\zpt),z,s}_{b}\Big) \\
  &\cdot \Big(\sum_{a} (\Id - \hat{M}^{(\Point,\xpt),x,r}_{a}) \otimes \hat{M}^{(\Point,\xpt),x,r}_{a} \Big) \ket{\hat{\psi}}
\end{aligned}\right |
\end{aligned}  \\
&\begin{aligned}[t]
  &\leq \sqrt{\E_\omega \sum_{a} \bra{\hat{\psi}} (\Id - \hat{M}^{(\Point,\xpt),x,r}_{a})_{\alice \alice'}^2 \otimes (\hat{M}^{(\Point,\xpt),x,r}_{a})_{\bob \alice''}^2 \ket{\hat{\psi}}}
\end{aligned} \label{eq:qld-rw-self-cons-3-1} \\
&\begin{aligned}[t]
  &= \sqrt{\E_\omega \sum_{a} \bra{\hat{\psi}} (\Id - \hat{M}^{(\Point,\xpt),x,r}_{a})_{\alice \alice'} \otimes (\hat{M}^{(\Point,\xpt),x,r}_{a})_{\bob \alice''} \ket{\hat{\psi}}}
\end{aligned} \label{eq:qld-rw-self-cons-3-2} \\
&\begin{aligned}[t]
  &= \sqrt{ 1 - \E_{\omega} \sum_a \bra{\hat{\psi}} (\hat{M}^{(\Point,\xpt),x,r}_{a})_{\alice \alice'} \ot (\hat{M}^{(\Point,\xpt),x,r}_{a})_{\bob \alice''} \ket{\hat{\psi}}}
\end{aligned} \label{eq:qld-rw-self-cons-3-3} \\
&\begin{aligned}[t]
  &=\sqrt{\begin{aligned}[t]
  & \frac{1}{2} \E_{\omega} \sum_a \bra{\hat{\psi}} ( (\hat{M}^{(\Point,\xpt),x,r}_{a})_{\alice \alice'} \ot I_{\bob \alice''} \\
  &- I_{\alice \alice'} \ot (\hat{M}^{(\Point,\xpt),x,r}_{a})_{\bob \alice''})^2 \ket{\hat{\psi}}
\end{aligned}}
\end{aligned} \label{eq:qld-rw-self-cons-3-4} \\
&\begin{aligned}[t]
  &\leq \sqrt{\eps}\;.
\end{aligned} \label{eq:qld-rw-self-cons-3-5}
\end{align}
	Here \eqref{eq:qld-rw-self-cons-3-1} follows from the Cauchy-Schwarz inequality, the projectivity of $\hat{M}^{(\Point,\xpt),x,r}_{a}$, and the fact that $\sum_b \hat{M}^{(\Point,\zpt),z,s}_{b} \otimes \hat{M}^{(\Point,\zpt),z,s}_{b} \leq \Id$. Equations~\eqref{eq:qld-rw-self-cons-3-2}, \eqref{eq:qld-rw-self-cons-3-3}, and~\eqref{eq:qld-rw-self-cons-3-4} follow from projectivity and completeness of $\hat{M}^{(\Point, \xpt), x, r}_{a}$. \eqref{eq:qld-rw-self-cons-3-5} follows from the self-consistency of $\hat{M}^{(\Point,\xpt),x,r}_{a}$, established in Lemma~\ref{lem:qld-comm-cons}. 

        The approximation in~\eqref{eq:qld-rw-self-cons-4} follows from the self-consistency of \linebreak[4]{}$\hat{M}^{(\Point, \xpt), x, r}_a$\allowbreak{}, again from Lemma~\ref{lem:qld-comm-cons}, together with projectivity and Fact~\ref{fact:agreement}.

	This shows that $(\hat{R}^\omega_{a,b})_{\alice \alice'} \simeq_\delta (\hat{R}^\omega_{a,b})_{\bob \alice''}$, as desired. Combined with Fact~\ref{fact:agreement}, we have
	\begin{equation}
		\label{eq:rw-sc}
		(\hat{R}^\omega_{a,b})_{\alice \alice'} \approx_\delta (\hat{R}^\omega_{a,b})_{\bob \alice''}\;.
	\end{equation}
%
	
		Finally we apply the orthonormalization Lemma (Lemma~\ref{lem:ortho}) to $\{ \hat{R}^\omega_{a,b} \}$ in order to obtain projective measurements $\{ \hat{L}^\omega_{a,b} \}$ such that on average over $\omega$,
	\begin{equation}\label{eq:rl-close}
		\hat{L}^\omega_{a,b} \approx_{\delta} \hat{R}^\omega_{a,b}\;.
	\end{equation}
	
	\textbf{Second step: approximate linearity relations for $\{ \hat{R}^\omega_{a,b} \}$.}
	We proceed in three steps. First we establish linearity relations for $\{ \hat{M}^{(\Point,W),u,r}_b \}$. For all $W \in \{X,Z\}$, $u \in \F_q^m$, $r, r' \in \F_q$ and $c \in \F_2$, 
	\begin{align}
		\sum_{\substack{b,b' \in \F_2 : \\ b + b' = c }} \hat{M}^{(\Point,W),u,r}_b \, \hat{M}^{(\Point,W),u,r'}_{b'} &= \sum_{\substack{a,a' \in \F_q: \\ \tr(ar + a'r') = c}} \hat{M}^{(\Point,W),u}_a \cdot \hat{M}^{(\Point,W),u}_{a'} \notag\\
		&= \sum_{a : \tr(a (r +r')) = c } \hat{M}^{(\Point,W),u}_a \notag\\
		&= \hat{M}^{(\Point,W),u,r+r'}_c\;,\label{eq:m-lin}
	\end{align}
	where the second line follows from projectivity of $\{\hat{M}^{(\Point,W),u}_a\}$. 
	
	Second we show approximate linearity of the $\{ \hat{R}^\omega_{a,b} \}$. On average over $x,z \in \F_q^m$ and $r,r',s,s' \in \F_q$, 
	\begin{align}
&\begin{aligned}[t]
  &\sum_{\substack{a,b,a',b' : \\ a + a' = c \\ b + b' = d}} \Big (\hat{R}^{x,z,r,s}_{a,b} \cdot \hat{R}^{x,z,r',s'}_{a',b'} \Big)_{\alice \alice'}
\end{aligned} \notag \\
&\begin{aligned}[t]
  &\approx_\delta \sum_{\substack{a,b,a',b' : \\ a + a' = c \\ b + b' = d}} \Big( \hat{R}^{x,z,r,s}_{a,b} \Big)_{\alice \alice'} \otimes \Big( \hat{R}^{x,z,r',s'}_{a',b'} \Big)_{\bob \alice''}
\end{aligned} \label{eq:qld-r-52} \\
&\begin{aligned}[t]
  &\approx_\delta \sum_{\substack{a,b,a',b': \\ a + a' = c \\ b + b' = d}} \Big( \hat{M}^{(\Point,\zpt),z,s}_{b} \\
  &\cdot \hat{M}^{(\Point,\xpt),x,r}_{a} \Big)_{\alice \alice'} \otimes \Big(\hat{M}^{(\Point,\zpt),z,s'}_{b'} \cdot \hat{M}^{(\Point,\xpt),x,r'}_{a'} \Big)_{\bob \alice''}
\end{aligned} \label{eq:qld-r-53} \\
&\begin{aligned}[t]
  &\approx_\delta \sum_{\substack{b,b': \\ b + b' = d}} \Big( \hat{M}^{(\Point,\zpt),z,s}_{b} \Big)_{\alice \alice'} \otimes \Big(\hat{M}^{(\Point,\zpt),z,s'}_{b'} \cdot \hat{M}^{(\Point,\xpt),x,r+r'}_{c} \Big)_{\bob \alice''}
\end{aligned} \label{eq:qld-r-54} \\
&\begin{aligned}[t]
  &\approx_\delta \sum_{\substack{b,b' : \\ b + b' = d}} \Big( \hat{M}^{(\Point,\zpt),z,s}_{b} \Big)_{\alice \alice'} \otimes \Big(\hat{M}^{(\Point,\xpt),x,r+r'}_{c} \cdot \hat{M}^{(\Point,\zpt),z,s'}_{b'} \Big)_{\bob \alice''} \;.
\end{aligned} \label{eq:qld-r-55}
\end{align}
	In each approximation, the answer summation is over $c,d \in \F_2$. Each approximation follows by applying a previously derived approximation to part of the expression, together with Fact~\ref{fact:add-a-proj}.\anote{Eventual TODO: split up all these steps} Specifically, line~\eqref{eq:qld-r-52} uses the self-consistency of $\{ \hat{R}^\omega_{a,b} \}$ shown in~\eqref{eq:rw-sc}. Line~\eqref{eq:qld-r-53} follows from using~\eqref{eq:qld-r-2} twice. Line~\eqref{eq:qld-r-54} follows from the self-consistency and linearity properties of $\{ \hat{M}^{(\Point,X),x,r}_a \}$ shown in Lemma~\ref{lem:qld-comm-cons} and~\eqref{eq:m-lin}, respectively. Line~\eqref{eq:qld-r-55} follows from the approximate commutativity relation shown in Lemma~\ref{lem:qld-comm-cons}.  Continuing,
 \begin{align}
\text{\eqref{eq:qld-r-55}} &\approx_\delta \Big(\hat{M}^{(\Point,\xpt),x,r+r'}_{c} \cdot \hat{M}^{(\Point,\zpt),z,s+s'}_{d}  \Big)_{\bob \alice''} \label{eq:qld-r-56} \\
 &\approx_\delta \Big ( \hat{R}^{x,z,r+r',s+s'}_{c,d} \Big)_{\bob \alice''} \label{eq:qld-r-57} \\
 &\approx_\delta \Big ( \hat{R}^{x,z,r+r',s+s'}_{c,d} \Big)_{\alice \alice'} \label{eq:qld-r-58}.
	\end{align}	
Line~\eqref{eq:qld-r-56} follows from self-consistency and linearity properties of \linebreak[4]{}$\{ \hat{M}^{(\Point,Z),z,s}_b \}$\allowbreak{}. Line~\eqref{eq:qld-r-57} follows from~\eqref{eq:qld-r-2} and Lemma~\ref{lem:qld-comm-cons}, and line~\eqref{eq:qld-r-58} follows from the self-consistency of \allowbreak{}$\{ \hat{R}^\omega_{a,b} \}$\allowbreak{} from Eq.~\eqref{eq:rw-sc}. 
	
Third we deduce approximate linearity of the $\{ \hat{L}^\omega_{a,b} \}$. On average over $x,z \in \F_q^m$, and $r,r',s,s' \in \F_q$, we have
	\begin{align}
	\sum_{\substack{a,b,a',b': \\ a + a' = c \\ b + b' = d}}  \Big ( \hat{L}^{x,z,r,s}_{a,b} \, \hat{L}^{x,z,r',s'}_{a',b'} \Big)_{\alice \alice'} &\approx_\delta  \sum_{\substack{a,b,a',b': \\ a + a' = c \\ b + b' = d}} \Big( \hat{L}^{x,z,r,s}_{a,b} \Big)_{\alice \alice'} \otimes \Big (\hat{L}^{x,z,r',s'}_{a',b'} \Big)_{\bob \alice''}\notag \\
		&\approx_\delta  \sum_{\substack{a,b,a',b' : \\ a + a' = c \\ b + b' = d}}  \Big( \hat{R}^{x,z,r,s}_{a,b} \Big)_{\alice \alice'} \otimes \Big( \hat{R}^{x,z,r',s'}_{a',b'} \Big)_{\bob \alice''} \notag\\
		&\approx_\delta \Big ( \hat{R}^{x,z,r+r',s+s'}_{c,d} \Big)_{\alice \alice'} \notag\\
		&\approx_\delta \Big ( \hat{L}^{x,z,r+r',s+s'}_{c,d} \Big)_{\alice \alice'}\;.\label{eq:l-al}
	\end{align}
	The first two approximations follow from the closeness of the $\{ \hat{R}^\omega_{a,b} \}$ and $\{ \hat{L}^\omega_{a,b} \}$  guaranteed by~\eqref{eq:rl-close} as well as the self-consistency of $\{ \hat{R}^\omega_{a,b} \}$ shown in~\eqref{eq:rw-sc}. The third approximation follows from the linearity properties of $\{ \hat{R}^\omega_{a,b} \}$ in~\eqref{eq:qld-r-55}. 
	
	\textbf{Third step: construction of $\{ \hat{Q}^{x,z}_{a,b} \}$.}
	For each $\omega$ define an observable $\mathcal{O}^\omega = \sum_{a,b \in \F_2} (-1)^{a+ b} \hat{L}^\omega_{a,b}$.  The self-consistency of $\mathcal{O}^\omega$ between registers $\alice \alice'$ and $\bob \alice''$ follows from the self-consistency of the $\{ \hat{L}^\omega_{a,b} \}$, which is obtained by combining~\eqref{eq:rw-sc} and~\eqref{eq:rl-close}. We now verify approximate linearity. On average over $x,z \in \F_q^m$, and $r,r',s,s' \in \F_q$, 
	\begin{align*}
		\mathcal{O}^{x,z,r,s} \, \mathcal{O}^{x,z,r',s'} &= \sum_{a,b,a',b'} (-1)^{a + b + a' + b'} \hat{L}^{x,z,r,s}_{a,b} \, \hat{L}^{x,z,r',s'}_{a',b'} \\
		&\approx_\delta \sum_{c,c'} (-1)^{c+c'} \hat{L}^{x,z,r+r',s+s'}_{c,c'} \\
		&= \mathcal{O}^{x,z,r+r',s+s'}\,
	\end{align*}
	where the approximation follows from the approximate linearity properties of $\{ \hat{L}^\omega_{a,b} \}$ shown in~\eqref{eq:l-al}.
	Identify the field $\F_q$ with the vector space $\F_2^t$, and thus we treat pairs $(r,s)$ as vectors in $\F_2^{2t}$. For every $x,z \in \F_q^m$, we can apply Theorem~\ref{thm:linearity} to the collection of observables $\{ \mathcal{O}^{x,z,r,s} \}_{r,s \in \F_q}$ to obtain the existence of ``{exactly linear}'' observables $\{ \mathcal{L}^{x,z,r,s} \}$ that are $\delta$-close to $\{ \mathcal{O}^{x,z,r,s} \}$, on average over the choice of $x,z,r,s$. By assuming that $\ket{\hat{\psi}}$ has sufficiently many ancilla zero qubits, we do not need to extend the state further. 
	
	We are ready to define the measurements $\{\hat{Q}^{x,z}_{a,b} \}$ whose existence is claimed in the statement of the lemma. For all $x,z\in \F_q^m$ and $a,b \in \F_q$, define
	\[
		\hat{Q}^{x,z}_{a,b} = \E_{r,s} (-1)^{\tr(ra + sb)} \mathcal{L}^{x,z,r,s}\;.
	\]
	This operator is a projection (and thus positive semidefinite):
	\begin{align*}
	(\hat{Q}^{x,z}_{a,b})^2 &= \E_{r,s,r',s'} (-1)^{\tr((r+r')a + (s+s')b)} \mathcal{L}^{x,z,r,s} \, \mathcal{L}^{x,z,r',s'}\\ 
	&= \E_{r,s,r',s'} (-1)^{\tr((r+r')a + (s+s')b)} \mathcal{L}^{x,z,r+r',s+s'} \\
	&= \E_{r,s} (-1)^{\tr(ra + sb)}  \mathcal{L}^{x,z,r,s}\;.
	\end{align*}
	The second equality follows from the exact linearity of the $\{ \mathcal{L}^{x,z,r,s} \}$. Next, the operators sum to identity:
	\begin{align*}
		\sum_{a,b} \hat{Q}^{x,z}_{a,b} &= \E_{r,s} \sum_{a,b} (-1)^{\tr(ra + sb)} \mathcal{L}^{x,z,r,s} \\
		&= \mathcal{L}^{x,z,0,0} \\
		&= \Id\;.
	\end{align*}
	Thus $\{\hat{Q}^{x,z}_{a,b}\}$ forms a projective measurement.
	
	We establish self-consistency:
	\begin{align*}
		\Big( \hat{Q}^{x,z}_{a,b} \Big)_{\alice \alice'} &= \Big( \E_{r,s} (-1)^{\tr(ra + sb)} \mathcal{L}^{x,z,r,s} \Big)_{\alice \alice'} \\
		&\approx_\delta \Big( \E_{r,s} (-1)^{\tr(ra + sb)} \mathcal{O}^{x,z,r,s} \Big)_{\alice \alice'} \\
		&\approx_\delta \Big( \E_{r,s} (-1)^{\tr(ra + sb)} \mathcal{O}^{x,z,r,s} \Big)_{\bob \alice''} \\
		&\approx_\delta \Big( \hat{Q}^{x,z}_{a,b} \Big)_{\bob \alice''}\,.
	\end{align*}
	The first approximation is due to~Lemma~\ref{lem:avg-closeness}. The second approximation is due to~Lemma~\ref{lem:avg-closeness} and the self-consistency of the $\{\mathcal{O}^{x,z,r,s}\}$ observables. The third approximation follows from~Lemma~\ref{lem:avg-closeness} again. 
	
	Finally we show consistency with the $\{ \mathcal{M}^{(\Point,W),u}_a\}$. 
	\begin{align*}
&\begin{aligned}[t]
  &\Big( \hat{Q}^{x,z}_{a,b} \Big)_{\alice \alice'}\approx_\delta \Big( \E_{r,s} (-1)^{\tr(ra + sb)} \mathcal{O}^{x,z,r,s} \Big)_{\bob \alice''}
\end{aligned}  \\
&\begin{aligned}[t]
  &= \Big( \E_{r,s} \sum_{c,d} (-1)^{\tr(ra + sb) + c + d} \, \hat{L}^{x,z,r,s}_{c,d} \Big)_{\bob \alice''}
\end{aligned}  \\
&\begin{aligned}[t]
  &\approx_\delta \Big( \E_{r,s} \sum_{c,d} (-1)^{\tr(ra + sb) + c + d} \, \hat{R}^{x,z,r,s}_{c,d} \Big)_{\bob \alice''}
\end{aligned}  \\
&\begin{aligned}[t]
  &\approx_\delta \Big( \E_{r,s} \sum_{c,d} (-1)^{\tr(ra + sb) + c + d} \, \hat{M}^{(\Point,\zpt),z,s}_d \, \hat{M}^{(\Point,\xpt),x,r}_c \Big)_{\bob \alice''}
\end{aligned}  \\
&\begin{aligned}[t]
  &= \Big ( \E_s \sum_d (-1)^{\tr(sb) + d} \hat{M}^{(\Point,\zpt),z,s}_d \Big) \cdot \Big (\E_r \sum_c (-1)^{\tr(ra) + c} \hat{M}^{(\Point,\xpt),x,r}_c \Big)_{\bob \alice''}
\end{aligned}  \\
&\begin{aligned}[t]
  &= \Big( \E_s \sum_d \sum_{b', a'} (-1)^{\tr(sb) + \tr(sb')} \hat{M}^{(\Point, \zpt), z}_{b'} \Big)_{\bob \alice''} \\
  &\cdot \Big( \E_r \sum_c (-1)^{\tr(ra) + \tr(r a')} \hat{M}^{(\Point, \xpt), x}_{b'} \Big)_{\bob \alice''}
\end{aligned}  \\
&\begin{aligned}[t]
  &= \Big ( \hat{M}^{(\Point,\zpt),z}_b \cdot \hat{M}^{(\Point,\xpt),x}_a \Big)_{\bob \alice''}\;.
\end{aligned} 
\end{align*}
	The second line follows from the definition of $\{ \mathcal{O}^{x,z,r,s} \}$. The third line follows from the closeness of $\{\hat{L}^\omega_{c,d}\}$ and $\{\hat{R}^\omega_{c,d} \}$. The fourth line follows from~\eqref{eq:qld-r-2}.  The fifth line follows from~\eqref{eq:qld-def-mptur}, and the last line from Lemma~\ref{lem:cancellation}.

        The above calculation establishes \eqref{eq:qld-q-cons-m-hat-zx} of the Lemma. To obtain \eqref{eq:qld-q-cons-m-hat-xz}, which is the same  consistency relation with $X$ and $Z$ swapped, we may use Lemma~\ref{lem:qld-comm-cons} immediately after the fourth line to swap the two operators in the product, and then proceed as above.
	
	An analogous derivation shows the same properties of $\{ \hat{Q}^{x,z}_{a,b} \}$ with the tensor factors switched to $\bob \bob'$ and $\alice \bob''$, respectively. This concludes the proof of the lemma.
	
	
\end{proof}

\begin{lemma}\label{lem:qld-xz-lines}
  There exists a function $\delta_P(\eps,m,d,q) = \poly(\eps,
  md/q)$ such that the following holds. For $\mH \in \{\mH_\alice,\mH_\bob\}$ and for every pair of lines $\line_X$ and $\line_Z$ there exists a POVM $\{T^{\line_X, \line_Z}_{f_X, f_Z}\}$ acting on
  $\mathcal{H} \otimes (\C^q)^{\otimes \nqubits}$ with answers
  consisting of pairs of polynomials $f_X, f_Z$ in $\deg_{md}(\line_X)
  \times \deg_{md}(\line_Z)$
  such that on average over pairs $(\line_X, x)$ and $(\line_Z, z)$
  independently sampled from the line-point distribution,
  \begin{align*}
    (T^{\line_X, \line_Z}_{[ \eval_x(\cdot) =a, \eval_z(\cdot) = b]})_{\alice \alice'} &\simeq_{\delta_P} (\hat{Q}^{x,z}_{a, b})_{\bob \alice''}, \\
  \end{align*}
  as well as all symmetric equivalents of this approximation. Moreover, if $\line_W$ is axis-parallel, then $f_W \in \deg_{d}(\line_W)$.
\end{lemma}

\begin{proof}
  We construct the required measurements by combining the measurements
  $\hat{M}^{(\Line,W),\line_W}_{f_W}$ for $W \in \{X, Z\}$. For convenience in the proof we slightly abuse the notation $\approx_\delta$ to mean $\approx_{\poly(\eps)}$ for some polynomial function $\delta=\poly(\eps)$ that may differ each time the notation is used.

  	We first argue that the measurements $\{ \hat{Q}^{x,z}_{a,b} \}$ obtained in Lemma~\ref{lem:qld-4-10} are consistent with the $\{\hat{M}^{(\Line,W),\line_W}_{f_W} \}$. 

	Applying Fact~\ref{fact:add-a-proj} to~\eqref{eq:qld-q-cons-m-hat-zx}, we get that on average over uniformly random $x,z$,
	\begin{align}
		\Big( \hat{Q}^{x,z}_{a,b} \Big)_{\alice \alice'} \cdot \Big( \hat{M}^{(\Point,\zpt),z}_{b} \Big)_{\bob \alice''}
&=	\Big( \hat{M}^{(\Point,\zpt),z}_{b} \Big)_{\bob \alice''}\cdot	\Big( \hat{Q}^{x,z}_{a,b} \Big)_{\alice \alice'} \notag\\
& \approx_\delta \Big (\hat{M}^{(\Point,\zpt),z}_{b} \hat{M}^{(\Point,\xpt),x}_{a} \Big)_{\bob \alice''} \notag\\
&\approx_\delta \Big( \hat{Q}^{x,z}_{a,b} \Big)_{\alice \alice'}\;.\notag
	\end{align}
	Since $\{\hat{Q}^{x,z}_{a,b}\}$ is a projective measurement, applying Lemma~\ref{lem:cool-closeness-fact} with $(\hat{Q}^{x,z}_{a,b})_{\alice, \alice'}$ playing the role of $A^{x}_a$ and $(\hat{Q}^{x,z}_{a,b} )_{\alice, \alice'} \cdot (\hat{M}^{(\Point, \zpt), z}_{b})_{\bob \alice''}$ playing the role of $B^{x}_a$, we obtain that
	\begin{equation}
	\Big( \hat{Q}^{x,z}_{a} \Big)_{\alice \alice'} \approx_\delta \sum_b \Big( \hat{Q}^{x,z}_{a,b} \Big)_{\alice \alice'} \cdot \Big( \hat{M}^{(\Point,\zpt),z}_{b} \Big)_{\bob \alice''}\;.\label{eq:qqm}
	\end{equation}
	We now show that on average over uniformly random $x,z$, 
	\begin{equation}
	\label{eq:qld-qxz-close-to-point}
	\Big ( \hat{Q}^{x,z}_{a} \Big)_{\alice \alice'} \approx_{\delta_Q} \Big (\hat{M}^{(\Point,X),x}_{a} \Big)_{\bob \alice''} \;.
	\end{equation}
	This follows from~\eqref{eq:qqm} and the following calculation:
	\begin{align*}
&\begin{aligned}[t]
  &\E_{x,z} \sum_a \norm{ \sum_b \Big( \hat{Q}^{x,z}_{a,b} \Big)_{\alice \alice'} \cdot \Big ( \hat{M}^{(\Point,\zpt),z}_{b} \Big)_{\bob \alice''} - \Big ( \hat{M}^{(\Point,\xpt),x}_{a} \Big)_{\bob \alice''} \ket{\hat{\psi}} }^2
\end{aligned}  \\
&\begin{aligned}[t]
  &=\E_{x,z} \sum_a \norm{ \sum_b \Big ( \hat{M}^{(\Point,\zpt),z}_{b} \Big)_{\bob \alice''} \cdot \Big( \Big( \hat{Q}^{x,z}_{a,b} \Big)_{\alice \alice'} - \Big ( \hat{M}^{(\Point,\xpt),x}_{a} \Big)_{\bob \alice''} \Big) \ket{\hat{\psi}} }^2
\end{aligned}  \\
&\begin{aligned}[t]
  &= \E_{x,z} \sum_{a,b} \norm{ \Big ( \hat{M}^{(\Point,\zpt),z}_{b} \Big)_{\bob \alice''} \cdot \Big( \Big( \hat{Q}^{x,z}_{a,b} \Big)_{\alice \alice'} - \Big ( \hat{M}^{(\Point,\xpt),x}_{a} \Big)_{\bob \alice''} \Big) \ket{\hat{\psi}} }^2
\end{aligned}  \\
&\begin{aligned}[t]
  &= \E_{x,z} \sum_{a,b} \norm{\begin{aligned}[t]
  & \Big( \hat{M}^{(\Point,\zpt),z}_{b} \Big)_{\bob \alice''} \cdot \Big( \Big( \hat{Q}^{x,z}_{a,b} \Big)_{\alice \alice'} \\
  &- \Big( \hat{M}^{(\Point, \zpt),z}_{b} \cdot \hat{M}^{(\Point,\xpt),x}_{a} \Big)_{\bob \alice''} \Big) \ket{\hat{\psi}} 
\end{aligned}}^2
\end{aligned}  \\
&\begin{aligned}[t]
  &\leq \E_{x,z} \sum_{a,b} \norm{ \Big( \Big( \hat{Q}^{x,z}_{a,b} \Big)_{\alice \alice'} - \Big( \hat{M}^{(\Point, \zpt),z}_{b} \cdot \hat{M}^{(\Point,\xpt),x}_{a} \Big)_{\bob \alice''} \Big) \ket{\hat{\psi}} }^2
\end{aligned}  \\
&\begin{aligned}[t]
  &= O(\delta)\;,
\end{aligned} 
\end{align*} 
	where the second line uses the property \allowbreak{}$\sum_b \hat{M}^{(\Point,\zpt),z}_b \allowbreak = I$\allowbreak{}, the third and fourth lines uses the projectivity of \allowbreak{}$\{ \hat{M}^{(\Point,\zpt),z}_{b} \}$\allowbreak{}, the fifth line uses \allowbreak{}$\hat{M}^{(\Point, \zpt), z}_{b} \allowbreak \leq I$\allowbreak{}, and the last line uses \eqref{eq:qld-q-cons-m-hat-zx}. This shows~\eqref{eq:qld-qxz-close-to-point}. A similar argument shows that
	\begin{equation}\label{eq:qld-qxz-close-to-point-2}
	\Big ( \hat{Q}^{x,z}_{b} \Big)_{\alice \alice'} \approx_{\delta_Q} \Big (\hat{M}^{(\Point,Z),z}_{b} \Big)_{\bob \alice''}\;.
	\end{equation}
	Next, from Lemma~\ref{lem:qld-comm-line-cons} we get that on average over $(\line_X,x)$ and $(\line_Z,z)$ sampled independently from the line-point distribution (Definition~\ref{def:line-point-dist}),
	\begin{gather}
	(\hat{M}^{(\Line,X),\line_\xpt}_{[\eval_x(\cdot) = a]})_{\bob \alice''} \approx_\delta  (\hat{M}^{(\Point,\xpt),x}_{a})_{\bob \alice''} \;,\\
	( \hat{M}^{(\Line,Z),\line_\zpt}_{[\eval_z(\cdot) = b]})_{\bob \alice''} \approx_\delta  (\hat{M}^{(\Point,\zpt),z}_{b})_{\bob \alice''}\;.
	\end{gather}
	where we used both the self-consistency of the line measurements (item 1 in Lemma~\ref{lem:qld-comm-line-cons}) as well as their consistency with the points measurements (item 3). To deduce both of these relations, we used Fact~\ref{fact:agreement} to convert between $\approx$ and $\simeq$ distance, and Fact~\ref{fact:data-processing} to sum over outcomes to the lines measurement. Since the $\{\hat{M}^{(\Line,W),\line}_f \}$ and $\{ \hat{Q}^{x,z}_{a,b} \}$ are projective, together with~\eqref{eq:qld-qxz-close-to-point} and~\eqref{eq:qld-qxz-close-to-point-2}
	by Fact~\ref{fact:agreement} we get
	\begin{align}
&\begin{aligned}[t]
  &\Big ( \hat{Q}^{x,z}_{a} \Big)_{\alice \alice'} \simeq_{\delta_Q} \Big (\hat{M}^{(\Line,\xpt),\line_\xpt}_{[\eval_x(\cdot) = a]} \Big)_{\bob \alice''} \qquad \text{and} \\
  &\qquad \Big ( \hat{Q}^{x,z}_{b} \Big)_{\alice \alice'} \simeq_{\delta_Q} \Big (\hat{M}^{(\Line,\zpt),\line_\zpt}_{[\eval_z(\cdot) = b]} \Big )_{\bob \alice''} \;.
\end{aligned} \label{eq:pasting-q1}
\end{align}

	For all lines $\line_\xpt,\line_\zpt$ and $(f_\xpt,f_\zpt) \in \deg_{md}(\line_\xpt) \times \deg_{md}(\line_\zpt)$ define the operator
	\[
		T^{\line_\xpt,\line_\zpt}_{f_\xpt,f_\zpt} = \hat{M}^{(\Line,\xpt),\line_\xpt}_{f_\xpt} \cdot \hat{M}^{(\Line,\zpt),\line_\zpt}_{f_\zpt}  \cdot \hat{M}^{(\Line,\xpt),\line_\xpt}_{f_\xpt} \;.
	\]
	Then the collection $\{ T^{\line_\xpt,\line_\zpt}_{f_\xpt,f_\zpt} \}$ forms a valid POVM. Moreover, by the fact that $\{\hat{M}^{(\Line, W), \line_W}_{f_W}\}$ comes from a valid strategy for the low-degree test, it holds that $f_W \in \deg_{d}(\line_W)$ whenver $\line_W$ is axis-parallel, and thus property holds for $T$ as well. 

        It only remains to bound the consistency of $\{T^{\line_\xpt, \line_\zpt}_{f_\xpt, f_\zpt}\}$ with $\{\hat{Q}^{x,z}_{a,b}\}$. Using the following choices of measurements,
	\begin{equation*}
\begin{aligned}[t]
  &``A^{x,y_1,y_2}_{a_1,a_2}" : \hat{Q}^{x,z}_{a,b} \;, \quad ``(G_1)^x_g" : \hat{M}^{(\Line,\xpt),\line_\xpt}_{f_\xpt} \;, \\
  &\quad ``(G_2)^x_g" : \hat{M}^{(\Line,\zpt),\line_\zpt}_{f_\zpt} \;, \quad ``J^x_{g_1,g_2}": T^{\line_\xpt,\line_\zpt}_{f_\xpt,f_\zpt} \;,
\end{aligned} 
\end{equation*}
	Then the hypotheses of Lemma~\ref{lem:pasting-updated} hold, with $\delta$ in the lemma set to $\delta_Q$, and $\eta$ in the lemma set to $md/q$ (this follows from the Schwartz-Zippel lemma applied to the polynomials $f_X, f_Z$, which are of total degree at most $md$). Applying the lemma, we conclude that
	\begin{equation}
	\label{eq:qld-4-13-1}
		\Big( \hat{Q}^{x,z}_{a,b} \Big)_{\alice \alice'} \simeq_{\delta_P} \Big( T^{\line_\xpt,\line_\zpt}_{[\eval_x(\cdot) = a, \eval_z(\cdot) = b]} \Big)_{\bob \alice''}\;,
	\end{equation}
	where $\delta_P = \poly(\eps,md/q)$ is the $\delta_{pasting}$ as given by Lemma~\ref{lem:pasting-updated}. In applying the lemma, the condition~\eqref{eq:gonna-compare-like-a-pear} and its symmetric equivalent \eqref{eq:gonna-compare-like-a-pear-backwards} are given by~\eqref{eq:pasting-q1} and its symmetric equivalent, and the condition~\eqref{eq:spend-some-time-reflecting} is given by~\eqref{eq:qld-q-self-cons}, respectively.
The distribution over $(\ell_\xpt,\ell_\zpt)$ is given by two independently and uniformly chosen lines. This is a symmetric equivalent of the consistency relation in the conclusion of the current Lemma. Thus, by applying an identical argument with the appropriate modifications to the registers, all symmetric equivalents of this relation hold, establishing the conclusion of the Lemma.

      \end{proof}

      \subsection{Applying the classical low-degree test}
      \label{sec:apply-ldt}
      In the previous section we showed how to combine the
      approximately commuting $X$ and $Z$ basis point and line
      measurements into a joint point measurement and a joint line
      measurement. In this section we show how these measurements can be combined
			in a single measurement
      that returns a pair of global polynomials encoding the $X$-basis and
      $Z$-basis information, respectively, by applying the lemma that states quantum soundness of the classical low-degree test, Theorem~\ref{lem:ld-soundness}.
			
      Introduce a ``combining'' map that
      takes two polynomials \allowbreak{}$f,\allowbreak  g \allowbreak \in \ideg_{d,m}(\F_q)$\allowbreak{} and returns a new polynomial \allowbreak{}$\combine_{f,g} \allowbreak \in \ideg_{d,
        2m+2}(\F_q)$\allowbreak{} defined by
      \[ \combine_{f,g}(\underbrace{x}_{\in \F_q^{m}},
        \underbrace{z}_{\in \F_q^m}, \underbrace{\alpha}_{\in \F_q},
        \underbrace{\beta}_{\in \F_q}) = \alpha f(x) + \beta g(z)\;. \]
      This combining map can also be defined for polynomials restricted to lines. Let \allowbreak{}$\line$\allowbreak{} be a line in \allowbreak{}$\F_q^{2m+2}$\allowbreak{} with intercept \allowbreak{}$u \allowbreak = (u_X,\allowbreak  u_Z,\allowbreak  u_\alpha,\allowbreak  u_\beta)$\allowbreak{} and slope \allowbreak{}$v \allowbreak = (v_X,\allowbreak  v_Z,\allowbreak  v_\alpha,\allowbreak  v_\beta)$\allowbreak{}. Let \allowbreak{}$\line_X$\allowbreak{} and \allowbreak{}$\line_Z$\allowbreak{} be lines in \allowbreak{}$\F_q^{m}$\allowbreak{} such that for any point \allowbreak{}$(x,\allowbreak z,\allowbreak \alpha,\allowbreak \beta) \allowbreak \in \line$\allowbreak{}, it holds that \allowbreak{}$x \allowbreak \in \line_X$\allowbreak{} and \allowbreak{}$z \allowbreak \in \line_Z$\allowbreak{}. Then for any two polynomials \allowbreak{}$f \allowbreak \in \deg_{dm}(\line_X)$\allowbreak{} and \allowbreak{}$g \allowbreak \in \deg_{dm}(\line_Z)$\allowbreak{}, we define \allowbreak{}$\combine_{f,g} \allowbreak \in \deg_{dm + 1}(\line)$\allowbreak{} by
      \begin{equation}
        \combine_{f,g}(u + t \cdot v) = (u_\alpha + t\cdot v_\alpha) f(u_X + t \cdot v_X) + (u_\beta + t \cdot v_\beta) g(u_Z + t \cdot v_Z)\;. \label{eq:combine-lines}
      \end{equation}

      In the following two lemma we define combined point and line
      measurements according to this definition of $\combine_{f,g}$.

\begin{lemma}
\label{lem:qld-4-12}
	For all $x,z \in \F_q^m$ and $\alpha,\beta \in \F_q$ and $\mH \in \{ \mH_\alice, \mH_\bob\}$ there exists a projective measurement $\{ \hat{Q}^{x,z,\alpha,\beta}_c \}_{c \in \F_q}$ acting on $\mathcal{H} \otimes (\C^q)^{\otimes \nqubits}$ such that the following holds:
	\begin{enumerate}
		\item (\textbf{Self-consistency})\newline On average over uniformly random \allowbreak{}$(x,\allowbreak z,\allowbreak \alpha,\allowbreak \beta) \allowbreak \in \F_q^{2m + 2}$\allowbreak{}:
		\begin{equation}
			\label{eq:qld-4-12-self-cons}
			(\hat{Q}^{x,z,\alpha,\beta}_{c})_{\alice \alice'} \approx_{\delta_Q}  (\hat{Q}^{x,z,\alpha,\beta}_{c})_{\bob \alice''}\,.
		\end{equation}
		\item (\textbf{Consistency with $\hat{M}$}) On average over uniformly random $(x,z,\alpha,\beta) \in \F_q^{2m + 2}$:
		\begin{gather}
			\label{eq:qld-4-12-cons-m-hat}
			(\hat{Q}^{x,z,\alpha,\beta}_{c})_{\alice \alice'} \approx_{\delta_Q} \Big ( \sum_{\substack{a,b: \\ \alpha a + \beta b = c}} \hat{M}^{(\Point,X),x}_{a} \hat{M}^{(\Point,Z),z}_{b} \Big )_{\bob \alice''} \;, \\
			\label{eq:qld-4-12-cons-m-hat2}
			(\hat{Q}^{x,z,\alpha,\beta}_{c})_{\alice \alice'} \approx_{\delta_Q}  \Big ( \sum_{\substack{a,b: \\ \alpha a + \beta b = c}} \hat{M}^{(\Point,Z),z}_{b} \, \hat{M}^{(\Point,X),x}_{a}  \Big )_{\bob \alice''}  \,.			
		\end{gather}
              \end{enumerate}
                    Furthermore, all symmetric equivalents of these approximations
      also hold.
\end{lemma}
\begin{proof}
	Define 
	\[
		\hat{Q}^{x,z,\alpha,\beta}_{c} = \sum_{\substack{a,b:\\ \alpha a + \beta b = c}} \hat{Q}^{x,z}_{a,b}.
	\]
	The self-consistency and consistency properties of $\{ \hat{Q}^{x,z,\alpha,\beta}_{c} \}$ follow from the consistency properties of $\{ \hat{Q}^{x,z}_{a,b} \}$ established by Lemma~\ref{lem:qld-4-10}, and Fact~\ref{fact:data-processing}. 
\end{proof}

\begin{lemma}
\label{lem:qld-4-13}
There exists a function \allowbreak{}$\delta_{\combine}(\eps,\allowbreak m,\allowbreak d,\allowbreak q) \allowbreak = \poly(m^2\eps,\allowbreak md/q)$\allowbreak{} such that the following holds. 
For every line \allowbreak{}$\line$\allowbreak{} in
\allowbreak{}$\F_q^{2m + 2}$\allowbreak{} and \allowbreak{}$\mH \allowbreak \in \{ \mH_\alice,\allowbreak \mH_\bob\}$\allowbreak{} there exists a POVM \allowbreak{}$\{ \hat{Q}^\line_f \}$\allowbreak{} acting on
\allowbreak{}$\mathcal{H} \allowbreak \otimes (\C^q)^{\otimes \nqubits}$\allowbreak{} with outcomes \allowbreak{}$f\allowbreak \in
\deg_{md+1}(\line)$\allowbreak{} such that on average over a uniformly random
pair \allowbreak{}$(\line,\allowbreak  (x,\allowbreak z,\allowbreak \alpha,\allowbreak  \beta))$\allowbreak{} drawn from the line-point
distribution over \allowbreak{}$\F_q^{2m+2}$\allowbreak{},
\begin{gather}
\label{eq:qld-4-13}
(\hat{Q}^\line_{[\eval_{(x,z,\alpha,\beta)}(\cdot) = a]})_{\alice \alice'} \simeq_{\delta_{\combine}} (\hat{Q}^{x,z,\alpha,\beta}_{a} )_{\bob \alice''}\;, \\
(\hat{Q}^\line_{[\eval_{(x,z,\alpha,\beta)}(\cdot) = a]})_{\bob \bob'} \simeq_{\delta_{\combine}} ( \hat{Q}^{x,z,\alpha,\beta}_{a} )_{\alice \bob''}\;,
\end{gather}
where the answer summation is over $a \in \F_q$. Moreover, if $\line$ is axis-parallel, then the outcome $f \in \deg_{d}(\line)$.
\end{lemma}

Before proving the lemma we give definitions regarding distributions over lines and points.

\begin{definition}\label{def:ith-restricted-line}
  For any $i \in \{1, \dots,
  m\}$ the \emph{$i$-th restricted axis parallel line distribution $D_{\ALine,
    i}$} is the restriction of $D_{\ALine}$ to pairs  $(\line, u)$
  where the line $\line$ is along the direction $e_i$ (i.e., is of the
  form $\line = (u_0, e_i)$. Similarly, the \emph{$i$-th restricted
  diagonal line distribution $D_{\DLine, i}$} is the restriction of
  $D_{\DLine}$ to pairs $(\line, u)$, where the line $\line = (u_0,
  s, v)$ is such that $\chi(s) = i$, and coordinates $1, \dots, i-1$ of
  $v$ are $0$. (Recall that $\chi$ is defined in~\eqref{eq:chi-func}.)
\end{definition}
  Observe that $D_{\ALine}$ is a uniform mixture over the
  distributions $D_{\ALine, i}$ for $i \in \{1, \dots, m\}$, and
  likewise for $D_{\DLine}$.
  As a consequence, any approximation that
  holds on average over the line-point distribution $D_{\Line}$ with
  error $\delta$ holds over each of the $i$-th restricted distributions
  $D_{\ALine, i}$ and $D_{\DLine, i}$ with error $2m\delta$. Similarly,
  any approximation that holds over $D_{\Line} \times D_{\Line}$ holds
  over any product of the $i$-th restricted distributions with error
  $4m^2 \delta$. For example, by Lemma~\ref{lem:qld-xz-lines} for any $\tvar_1, \tvar_2 \in \{\ALine, \DLine\}$ and $i, j \in \{1, \dots, m\}$, we have
  \begin{equation}
\begin{aligned}[t]
  &\E_{(\line_X, x) \sim D_{\tvar_1, i}} \E_{(\line_Z, z) \sim D_{\tvar_2, j}} \sum_{f_X, f_Z} \bra{\hat{\psi}} (T^{\line_X, \line_Z}_{f_X, f_Z})_{\alice \alice'} \\
  &\ot (I - (\hat{Q}^{x,z}_{f_X(x), f_Z(z)}))_{\bob \alice''} \ket{\hat{\psi}} = O(m^2 \delta_P)\;.
\end{aligned} \label{eq:qld-xz-lines-restricted}
\end{equation}

    \begin{lemma}\label{lem:qld-sublines}
      There exists a distribution $D$ over tuples $(\line, \line_X, \line_Z)$ with the following properties:
      \begin{enumerate}
      \item The marginal distribution over $\line$ is the same as the marginal of the line-point distribution over $\F_q^{2m+2}$.
      \item For any point of the form $u = (u_X, u_Z, \alpha, \beta)$ lying on $\line$, it holds that $u_X \in \line_X$ and $u_Z \in \line_Z$. Moreover, for $u$ chosen uniformly at random in $\line$ the marginal distribution over $(\line_X,\line_Z, u_X)$ (resp.\ over $(\line_X,\line_Z, u_Z)$) is such that the marginal over $(\ell_X,\ell_Z)$ is a mixture of $D_{\ALine, i}\times D_{\ALine, j}$ and $D_{\DLine, i}\times D_{\DLine, j}$ for $i,j\in\{1,\ldots,m\}$ and, moreover, conditioned on $(\ell_X,\ell_Z)$, $u_X$ is uniformly random on $\line_X$ (resp. $u_Z$ uniformly random on $\line_Z$). 
      \end{enumerate}
    \end{lemma}
		
    \begin{proof}
      We describe $D$ by giving a procedure to sample from it. We start by sampling $(\line, u)$ from the line-point distribution over $\F_q^{2m+2}$. Write $u = (u_X, u_Z, \alpha, \beta)$. 
	There are two cases to consider: the case where $\line$ is
        axis-parallel and the case where $\line$ is diagonal. In each case, we sample lines $\line_\xpt$ and $\line_\zpt$ such that the tuples  $(\line_\xpt,
        \line_\zpt,u_\xpt)$ and $(\line_\xpt, \line_\zpt,u_\zpt)$ satisfy Property~2 of the lemma.
        \begin{enumerate}
          \item Suppose $\line$ is an  axis-parallel line $(v,e_j)$ where
            $v = (v_\xpt,v_\zpt,\alpha_0,\beta_0)$. Then we sample $(v'_\xpt,\allowbreak  i_\xpt),\allowbreak  (v'_\zpt,\allowbreak  i_\zpt) \in \F_q^m \times \{1,\allowbreak 2,\allowbreak  \ldots,\allowbreak  m\}$ as described below, and set $\line_\xpt =
            (v'_\xpt,e_{i_\xpt}),\line_\zpt = (v'_\zpt, e_{i_\zpt})$.
            \begin{enumerate}
              \item If $1 \leq j \leq m$, then set $i_\xpt = j$ and
                choose $i_\zpt$ to be uniformly random in $\{1, \dots,
                m\}$.
              \item If  $m+1 \leq j \leq 2m$, then set $i_\xpt$ to be
                uniformly random in $\{1, \dots, m\}$ and $i_\zpt = j - m$.
              \item If $2m + 1 \leq j \leq 2m+2$, then set $i_\xpt$
                and $i_\zpt$ to both be uniformly random in $\{1, \dots, m\}$.
              \item Let $v'_\xpt$ be the canonical representative
                $L_{e_{i_\xpt}}^\lnf(v_\xpt)$, and
                likewise let $v'_\zpt$ be the canonical
                representative $L_{e_{i_\zpt}}^\lnf(v_{\zpt})$ (see
                Definition~\ref{def:line-representative} for the definition of
                the canonical representative of a line). Note that
                $v_\xpt$ lies on the line $\line_\xpt = (v'_\xpt,
                e_{i_\xpt})$, and similarly for $v_\zpt$ and
                $\line_{\zpt} = (v'_\zpt, e_{i_\zpt})$.
              \end{enumerate}
              We claim that this construction satisfies Property~2 of the lemma. We show this first $(\line_X,\line_Z)$ and $u_X$, the other case being symmetric. If we are in case (a), then $u_X$ is a uniformly random point on the line $\line_X=(v_X, e_j)$ and $\line_Z=(v_Z,e_{i_Z})$, which is independent from $\line_X$. Thus, the marginal distribution over $(\line_X,u_X)$ conditioned on $j$ in this case is equal to $D_{\ALine, j}$ and, furthermore, even conditioned on $(\line_X,u_X)$, $\line_Z$ is distributed according to $D_{\ALine,i_Z}$ for a uniformly random $i_Z$. Cases (b) and (c) are analogous except that in this case even conditioned on $j$, $i_X$ is uniform. 
      \item Suppose $\ell$ is a random diagonal line $(v, s, w)$,
        where $v = (v_\xpt, v_\zpt, \alpha, \beta)$, and
        where $w = (w_\xpt, w_\zpt, \gamma, \delta) $ is a direction vector whose first nonzero coordinate is
         $j = \chi(s)$ (where $\chi$ is defined in \eqref{eq:chi-func}). We sample $\line_\xpt =  (v'_\xpt, s_\xpt,
        w'_\xpt)$ and $\line_\zpt = (v'_\zpt, s_\zpt, w'_\zpt)$ in the following
        manner:
        \begin{enumerate}
          \item If $1 \leq j \leq m$, then $w'_\xpt = w_\xpt$ and $w'_\zpt = w_\zpt$. We choose $s_\xpt$ and $s_\zpt$ such that $\chi(s_\xpt) = j$ and $\chi(s_\zpt) = 1$.
          \item If $m+1 \leq j \leq 2m$, then $w'_\xpt$ is a random
            diagonal line direction (and $s_{\xpt}$ is chosen accordingly), and $w'_\zpt = w_\zpt$, with $s_\zpt$ such that $\chi(s_\zpt) = j - m$.
          \item If $2m+1 \leq j \leq 2m+2$ then both $w_\xpt$, and
            $w_\zpt$ are uniformly random, with $s_\xpt$ and $s_\zpt$ chosen accordingly.
          \item Set $v'_\xpt = L_{w'_{\xpt}}^\lnf(v_\xpt)$ and $v'_\zpt = L_{w'_{\zpt}}^\lnf(v_\zpt)$.
          \end{enumerate}
          We claim that this construction satisfies Property~2 of the Lemma. We show this first $(\line_X,\line_Z)$ and $u_X$, the other case being symmetric.  If we are in case (a), then $u_X$ is a uniformly random point on the diagonal line $(v_X, s_X, w_X)$ which is equal to $\line_X$. Thus, the marginal distribution over $(\line_X, u_X)$ conditioned on $j$ in this case is equal to $D_{\DLine, j}$. Furthermore, conditioned on $(\line_X,u_X)$, $\line_Z$ is distributed according to $D_{\DLine,1}$. The other cases can be analyed analogously, with the only difference being how the indices $i,j$ of $\line_X$ and $\line_Z$ distributions are chosen. 
					    \end{enumerate}
	Let $D$ be the distribution over tuples $(\line, \line_X, \line_Z)$ induced by this procedure. This distribution $D$ satisfies Property 1 in the lemma by  construction, and Property 2 as argued above.
    \end{proof}

\begin{proof}[Proof of Lemma~\ref{lem:qld-4-13}]
We construct the measurements $\hat{Q}^\line_f$
out of the lines measurements $T^{\line_X, \line_Z}_{f_X, f_Z}$
guaranteed by Lemma~\ref{lem:qld-xz-lines}. To do so we first show how
to map any line-point pair $(\line, u)$ in the expanded space
$\F_q^{2m+2}$ to two pairs  $(\line_X, u_X)$ and  $(\line_Z, u_Z)$ in $\F_q^m$. We will 
define $\hat{Q}^\line$ in terms of $T^{\line_X, \line_Z}_{f_X, f_Z}$ and
show that the resulting measurement is consistent with the combined
points measurement $\hat{Q}^{x,z,\alpha, \beta}$ from Lemma~\ref{lem:qld-4-12}.

To define $\hat{Q}^\line$ we make use of the distribution $D$ from Lemma~\ref{lem:qld-sublines}. Let $D_{|\line}$ denote the distribution derived from $D$ by conditioning on the first element being $\line$. Then for every line $\line$ in $\F_q^{2m+2}$ and function $f \in \deg_{(md+1)}(\line)$, define
\[ \hat{Q}^\line_f = \E_{(\line_X, \line_Z) \sim D_{|\line}} \sum_{\substack{f_X \in \deg_{md}(\line_X), f_Z \in \deg_{md}(\line_Z):\\ f = \combine_{f_X, f_Z}}} T^{\line_X, \line_Z}_{f_X, f_Z}\;, \]
where $\combine_{f_X, f_Z}$ is defined in~\eqref{eq:combine-lines}. Note that $\{\hat{Q}^\line_f\}$ forms a valid POVM for every choice of line $\line$, and, moreover, that $\combine_{f_X, f_Z}$ is well-defined for the lines $\line, \line_X, \line_Z$ because of Property 2 of Lemma~\ref{lem:qld-sublines}. 

Observe that by Property~3 of Lemma~\ref{lem:qld-sublines}, whenever $\line$ is axis-parallel, $\line_X$ and $\line_Z$ are axis-parallel also. Thus, in this case, by Lemma~\ref{lem:qld-xz-lines} it holds that the outcomes $f_X$ and $f_Z$ both have degree $d$. By definition of $\combine_{f_X, f_Z}$ it holds that the outcome $f$ of $\{\hat{Q}^\line_f\}$ has degree $d$ as well.

We argue the following bound:
\begin{align}
&\begin{aligned}[t]
  &\E_{(\line, \line_X, \line_Z) \sim D} \E_{(x,z,\alpha, \beta) \in \line} \sum_{f_X, f_Z} \bra{\hat{\psi}} T^{\line_X, \line_Z}_{f_X, f_Z} \ot (I \\
  &- \hat{Q}^{x, z }_{f_X(x),f_Z(z)}) \ket{\hat{\psi}} \,=\, O\big( m \cdot (\eps^{1/4} + \delta_P^{1/4} + \delta_Q^{1/4} + \delta_{\Line}^{1/2})\big)\;.
\end{aligned} \label{eq:qld-combined-lines-consistency}
\end{align}
This bound will follow from chaining the following three claims. 

\begin{claim}\label{claim:17-1}
\begin{align}
&\begin{aligned}[t]
  &\E_{(\line, \line_X, \line_Z) \sim D} \E_{(x,z,\alpha, \beta) \in \line} \sum_{f_X, f_Z} \bra{\hat{\psi}} T^{\line_X, \line_Z}_{f_X, f_Z} \ot \hat{Q}^{x, z }_{f_X(x),f_Z(z)} \ket{\hat{\psi}}
\end{aligned} \notag \\
&\begin{aligned}[t]
  &\approx_{\delta_Q^{1/2}} \E_{(\line, \line_X, \line_Z) \sim D} \E_{(x,z,\alpha, \beta) \in \line} \sum_{f_X, f_Z} \bra{\hat{\psi}} T^{\line_X, \line_Z}_{f_X, f_Z} \ot \hat{M}^{(\Point,Z),z}_{f_Z(z)} \hat{M}^{(\Point,X),x}_{f_X(x)} \ket{\hat{\psi}}\;.
\end{aligned} 
\end{align}
\end{claim}

\begin{proof}
Form the difference and apply the Cauchy Schwarz inequality to obtain
\begin{align*}
&\begin{aligned}[t]
  &\Big| \E_{(\line, \line_X, \line_Z) \sim D} \E_{(x,z,\alpha, \beta) \in \line} \sum_{f_X, f_Z}\bra{\hat{\psi}} T^{\line_X, \line_Z}_{f_X, f_Z} \\
  &\ot \big(\hat{Q}^{x, z }_{f_X(x),f_Z(z)}-\hat{M}^{(\Point,Z),z}_{f_Z(z)}\hat{M}^{(\Point,X),x}_{f_X(x)} \big) \ket{\hat{\psi}}\Big|
\end{aligned}  \\
&\begin{aligned}[t]
  &\leq \Big( \E_{(\line, \line_X, \line_Z) \sim D} \E_{(x,z,\alpha, \beta) \in \line} \sum_{f_X, f_Z} \big\|\sqrt{T^{\line_X, \line_Z}_{f_X, f_Z} }\ot \Id \ket{\hat{\psi}} \big\|^2 \Big)^{1/2}
\end{aligned}  \\
&\begin{aligned}[t]
  &\qquad\cdot \Big( \E_{(\line, \line_X, \line_Z) \sim D} \E_{(x,z,\alpha, \beta) \in \line} \sum_{f_X, f_Z}\big\| \sqrt{T^{\line_X, \line_Z}_{f_X, f_Z}} \\
  &\ot \big(\hat{Q}^{x, z }_{f_X(x),f_Z(z)}-\hat{M}^{(\Point,Z),z}_{f_Z(z)}\hat{M}^{(\Point,X),x}_{f_X(x)} \big) \ket{\hat{\psi}}\big\|^2 \Big)^{1/2}\;.
\end{aligned} 
\end{align*}
The first term above is at most $1$. For the second term, writing 
\[W^{x,z}_{a,b} = \big(\hat{Q}^{x, z }_{a,b}- \hat{M}^{(\Point,Z),z}_{b}\hat{M}^{(\Point,X),x}_{a}\big)\]
for short
we bound it as 
\begin{align*}
&\begin{aligned}[t]
  &\E_{(\line, \line_X, \line_Z) \sim D} \E_{(x,z,\alpha, \beta) \in \line} \sum_{f_X, f_Z}\big\| \sqrt{T^{\line_X, \line_Z}_{f_X, f_Z}} \ot W^{x,z}_{f_X(x),f_Z(z)} \ket{\hat{\psi}}\big\|^2 \\
  &\leq \E_{(\line, \line_X, \line_Z) \sim D} \E_{(x,z,\alpha, \beta) \in \line} \sum_{a,b}\bra{\hat{\psi}} \Id \ot (W^{x,z}_{a,b})^2 \ket{\hat{\psi}}
\end{aligned}  \\
&\begin{aligned}[t]
  &= \E_{(x,z)} \sum_{a,b}\bra{\hat{\psi}} \Id \ot (W^{x,z}_{a,b})^2 \ket{\hat{\psi}}
\end{aligned}  \\
&\begin{aligned}[t]
  &\leq \delta_Q\;.
\end{aligned} 
\end{align*}
Here the first inequality uses that for any \allowbreak{}$x,\allowbreak z,\allowbreak a,\allowbreak b$\allowbreak{}, \allowbreak{}$\sum_{f_X,f_X: f_X(x)=a,f_Z(z)=b} T^{\line_X, \line_Z}_{f_X, f_Z}\allowbreak \leq \Id$\allowbreak{}. 
In the equality in the before-last line, the expectation is over independent uniform \allowbreak{}$x,\allowbreak z\allowbreak \in \F_q^m$\allowbreak{}. The equality holds because for a line \allowbreak{}$\ell$\allowbreak{} sampled from the line-point distribution \allowbreak{}$D$\allowbreak{}, a uniformly random point \allowbreak{}$(x,\allowbreak z,\allowbreak \alpha,\allowbreak \beta)$\allowbreak{} on \allowbreak{}$\ell$\allowbreak{} is distributed as a uniformly random point in \allowbreak{}$\F_q^{2m+2}$\allowbreak{}. The last inequality is by~\eqref{eq:qld-q-cons-m-hat-zx}.
\end{proof}

\begin{claim}\label{claim:17-2}
\begin{align}
&\begin{aligned}[t]
  &\E_{(\line, \line_X, \line_Z) \sim D} \E_{(x,z,\alpha, \beta) \in \line} \sum_{f_X, f_Z} \bra{\hat{\psi}} T^{\line_X, \line_Z}_{f_X, f_Z} \ot \hat{M}^{(\Point,Z),z}_{f_Z(z)} \hat{M}^{(\Point,X),x}_{f_X(x)} \ket{\hat{\psi}}
\end{aligned} \notag \\
&\begin{aligned}[t]
  &\approx_{m\delta_\Line^{1/2}} \E_{(\line, \line_X, \line_Z) \sim D} \E_{(x,z,\alpha, \beta) \in \line} \sum_{f_X,f_Z} \bra{\hat{\psi}} T^{\line_X, \line_Z}_{f_X, f_Z} \ot \hat{M}^{(\Point,Z),Z}_{f_Z(z)} \ket{\hat{\psi}}\;.
\end{aligned} 
\end{align}
\end{claim}

\begin{proof}
We form the difference and apply the Cauchy-Schwarz inequality
\begin{align*}
&\begin{aligned}[t]
  &\Big|\E_{(\line, \line_X, \line_Z) \sim D} \E_{(x,z,\alpha, \beta) \in \line} \sum_{f_X, f_Z} \bra{\hat{\psi}} T^{\line_X, \line_Z}_{f_X, f_Z} \ot \hat{M}^{(\Point,Z),z}_{f_Z(z)} (\Id-\hat{M}^{(\Point,X),x}_{f_X(x)} ) \ket{\hat{\psi}}\Big|
\end{aligned}  \\
&\begin{aligned}[t]
  &\leq \Big(\E_{(\line, \line_X, \line_Z) \sim D} \E_{(x,z,\alpha, \beta) \in \line} \sum_{f_X, f_Z} \big\| \sqrt{ T^{\line_X, \line_Z}_{f_X, f_Z}} \ot \hat{M}^{(\Point,Z),z}_{f_Z(z)} \ket{\hat{\psi}}\big\|^2 \Big)^{1/2}
\end{aligned}  \\
&\begin{aligned}[t]
  &\qquad\cdot \Big(\E_{(\line, \line_X, \line_Z) \sim D} \E_{(x,z,\alpha, \beta) \in \line} \sum_{f_X, f_Z} \big\| \sqrt{ T^{\line_X, \line_Z}_{f_X, f_Z}} \ot (\Id-\hat{M}^{(\Point,X),x}_{f_X(x)}) \ket{\hat{\psi}}\big\|^2 \Big)^{1/2}\;.
\end{aligned} 
\end{align*}
The first term above is at most $1$. The second term can be bounded as
\begin{align}
&\begin{aligned}[t]
  &\E_{(\line, \line_X, \line_Z) \sim D} \E_{(x,z,\alpha, \beta) \in \line} \sum_{f_X, f_Z} \big\| \sqrt{ T^{\line_X, \line_Z}_{f_X, f_Z}} \ot (\Id-\hat{M}^{(\Point,X),x}_{f_X(x)}) \ket{\hat{\psi}}\big\|^2
\end{aligned} \notag \\
&\begin{aligned}[t]
  &= \E_{(\line, \line_X, \line_Z) \sim D} \E_{(x,z,\alpha, \beta) \in \line} \sum_{a} \bra{\hat{\psi}} T^{\line_X, \line_Z,x}_{a} \ot (\Id-\hat{M}^{(\Point,X),x}_{a})^2 \ket{\hat{\psi}}
\end{aligned} \notag \\
&\begin{aligned}[t]
  &= 1-\E_{(\line, \line_X, \line_Z) \sim D} \E_{(x,z,\alpha, \beta) \in \line} \sum_{a} \bra{\hat{\psi}} T^{\line_X, \line_Z,x}_{a} \ot \hat{M}^{(\Point,X),x}_{a} \ket{\hat{\psi}}\;,
\end{aligned} \label{eq:17-4}
\end{align}
where we used the shorthand $T^{\line_X, \line_Z,x}_{a} = \sum_{f_X,f_Z: f_X(x)=a} T^{\line_X, \line_Z}_{f_X, f_Z}$ and the second equality uses that $\{\hat{M}^{(\Point,X),x}_{a}\}$ is a projective measurement. Using the definition 
\[	T^{\line_\xpt,\line_\zpt}_{f_\xpt,f_\zpt} = \hat{M}^{(\Line,\xpt),\line_\xpt}_{f_\xpt} \cdot \hat{M}^{(\Line,\zpt),\line_\zpt}_{f_\zpt}  \cdot \hat{M}^{(\Line,\xpt),\line_\xpt}_{f_\xpt} \]
and the fact that $\{\hat{M}^{(\Line,\xpt),\line_\xpt}_{f_\xpt} \}$ is projective we get
\begin{align}
1-\eqref{eq:17-4}&=\E_{(\line, \line_X, \line_Z) \sim D} \E_{(x,z,\alpha, \beta) \in \line} \sum_{a} \bra{\hat{\psi}}  \hat{M}^{(\Line,\xpt),\line_\xpt}_{[\eval_x(\cdot)=a]} \ot \hat{M}^{(\Point,X),x}_{a} \ket{\hat{\psi}}\;.\label{eq:a17-8}
\end{align}
By Property~2 of Lemma~\ref{lem:qld-sublines} the right-hand side of~\eqref{eq:a17-8}
 is a mixture of terms of the form
\begin{equation}\label{eq:a17-9}
 \E_{(\line_X, x) \sim D_{\tvar, i}}  \sum_{a} \bra{\hat{\psi}}  \hat{M}^{(\Line,\xpt),\line_\xpt}_{[\eval_x(\cdot)=a]} \ot \hat{M}^{(\Point,X),x}_{a} \ket{\hat{\psi}}\;, 
\end{equation}
for $\tvar \in \{\ALine, \DLine\}$ and $i \in \{1, \dots, m\}$. 
Hence by the discussion following Definition~\ref{def:ith-restricted-line} and  item 3 of Lemma~\ref{lem:qld-comm-line-cons} it follows that
\[ \eqref{eq:a17-9} \,=\, O(m^2 \delta_\Line)\;. \]
\end{proof}

\begin{claim}\label{claim:17-3}
\begin{align}
  \E_{(\line, \line_X, \line_Z) \sim D}\E_{(x,z,\alpha, \beta) \in \line}\sum_{f_X,f_Z} \bra{\hat{\psi}} T^{\line_X, \line_Z}_{f_X, f_Z} \ot  \hat{M}^{(\Point,Z),z}_{f_Z(z)} \ket{\hat{\psi}} 
       &\approx_{\sqrt{m}\cdot(\delta_P^{1/4}+\delta_Q^{1/4}+\eps^{1/4})}1 \;. 
\end{align}
\end{claim}

\begin{proof}
At first we proceed similarly to the proof of Claim~\ref{claim:17-2}, bounding the difference using Cauchy-Schwarz as
\begin{align*}
&\Big| \E_{(\line, \line_X, \line_Z) \sim D} \E_{(x,z,\alpha, \beta) \in \line} \sum_{f_X,f_Z} \bra{\hat{\psi}} T^{\line_X, \line_Z}_{f_X, f_Z} \ot  (\Id-\hat{M}^{(\Point,Z),z}_{f_Z(z)}) \ket{\hat{\psi}}\Big|\\
&\leq\Big(\E_{(\line, \line_X, \line_Z) \sim D} \E_{(x,z,\alpha, \beta) \in \line} \sum_{f_X, f_Z} \big\| \sqrt{ T^{\line_X, \line_Z}_{f_X, f_Z}} \ot  \Id \ket{\hat{\psi}}\big\|^2 \Big)^{1/2}\\
&\qquad\cdot \Big(\E_{(\line, \line_X, \line_Z) \sim D} \E_{(x,z,\alpha, \beta) \in \line} \sum_{f_X, f_Z} \big\| \sqrt{ T^{\line_X, \line_Z}_{f_X, f_Z}} \ot  (\Id-\hat{M}^{(\Point,Z),z}_{f_Z(z)}) \ket{\hat{\psi}}\big\|^2 \Big)^{1/2}\;.
\end{align*}
The first term is at most $1$. The second, as in the proof of Claim~\ref{claim:17-2}, equals
\begin{align}
&\begin{aligned}[t]
  &1-\E_{(\line, \line_X, \line_Z) \sim D} \E_{(x,z,\alpha, \beta) \in \line} \sum_{b} \bra{\hat{\psi}} T^{\line_X, \line_Z, z}_{b} \ot \hat{M}^{(\Point,Z),z}_{b} \ket{\hat{\psi}}
\end{aligned} \notag \\
&\begin{aligned}[t]
  &=1-\E_{(\line_X,\line_Z)\sim D_{\tvar,i}\times D_{\tvar,j}} \E_{z\in \line_Z } \sum_{b} \bra{\hat{\psi}} T^{\line_X, \line_Z, z}_{b} \ot \hat{M}^{(\Point,Z),z}_{b} \ket{\hat{\psi}}
\end{aligned} \label{eq:a17-5} \\
&\begin{aligned}[t]
  &=1-\E_{(\line_X,\line_Z)\sim D_{\tvar,i}\times D_{\tvar,j}} \E_{z\in \line_Z }\E_{x\in \line_X} \sum_{a,b} \\
  &\bra{\hat{\psi}} T^{\line_X, \line_Z}_{[\eval_x(\cdot)=a,\eval_z(\cdot)=b]} \ot \hat{M}^{(\Point,Z),z}_{b} \ket{\hat{\psi}}
\end{aligned} \label{eq:a17-6}
\end{align}
where we use the shorthand $T^{\line_X, \line_Z, z}_{b} = \sum_{f_X} T^{\line_X, \line_Z}_{f_X,[\eval_z(\cdot)=b]}$ and $D_{\tvar,i}\times D_{\tvar,j}$ to denote a mixture over all such distributions where $\tvar\in\{\ALine,\DLine\}$ and $i,j\in\{1,\ldots,m\}$. Equality holds in~\eqref{eq:a17-5} because by item 2 of Lemma~\ref{lem:qld-sublines} for $(\ell,\ell_X,\ell_Z)\sim D$ the marginal distribution on $(\ell_X,\ell_Z)$ is a mixture of the product distributions  $D_{\ALine, i}\times D_{\ALine, j}$ and $D_{\DLine, i}\times D_{\DLine,j}$, for $i,j\in \{1,\ldots,m\}$. Moreover, conditioned on $(\ell_X,\ell_Z)$, $z$ is a uniformly random point on $\ell_X$. The last equality in~\eqref{eq:a17-6} holds by definition of 
\begin{align*}
T^{\line_X, \line_Z, z}_{b}&= \sum_{f_X} T^{\line_X, \line_Z}_{f_X,[\eval_z(\cdot)=b]}\\
&= \Es{x \in \line_X} \sum_a \sum_{f_X: f_X(x)=a} T^{\line_X, \line_Z}_{f_X,[\eval_z(\cdot)=b]}\\
&= \Es{x \in \line_X} \sum_a  T^{\line_X, \line_Z}_{[\eval_x(\cdot)=a,\eval_z(\cdot)=b]}\;.
\end{align*}
We next write
\begin{align*}
&\begin{aligned}[t]
  &1-\eqref{eq:a17-6}\approx_{m(\sqrt{\delta_P}+\sqrt{\delta_Q})}\E_{(\line_X,\line_Z)\sim D_{\tvar,i}\times D_{\tvar,j}}\E_{x\in \line_X} \\
  &\E_{z\in \line_Z } \sum_{a,b} \bra{\hat{\psi}} \hat{M}^{(\Point,X),x}_{a}\hat{M}^{(\Point,Z),z}_{b} \ot \hat{M}^{(\Point,Z),z}_{b} \ket{\hat{\psi}}
\end{aligned}  \\
&\begin{aligned}[t]
  &= \sum_{b} \bra{\hat{\psi}} \hat{M}^{(\Point,Z),z}_{b} \ot \hat{M}^{(\Point,Z),z}_{b} \ket{\hat{\psi}}
\end{aligned}  \\
&\begin{aligned}[t]
  &\geq 1-\sqrt{\eps}\;,
\end{aligned} 
\end{align*}
where the first approximation follows from the discussion following Definition~\ref{def:ith-restricted-line} applied to the bounds from Lemma~\ref{lem:qld-sublines} and Lemma~\ref{lem:qld-4-10}, the second line uses that $\{\hat{M}^{(\Point,X),x}_{a}\}$ is a measurement, and the last is by self-consistency of the points measurements (Lemma~\ref{lem:qld-comm-cons}).
\end{proof}

Together, Claim~\ref{claim:17-1}, Claim~\ref{claim:17-2} and Claim~\ref{claim:17-3} show~\eqref{eq:qld-combined-lines-consistency}. To complete the proof of~\eqref{eq:qld-4-13}, observe that by Property~2 of Lemma~\ref{lem:qld-sublines} the right-hand side of \eqref{eq:qld-combined-lines-consistency} is a mixture of terms of the form
Together, Claim~\ref{claim:17-1}, Claim~\ref{claim:17-2} and Claim~\ref{claim:17-3} show {eq:qld-combined-lines-consistency}. To complete the proof of~\eqref{eq:qld-4-13}, observe that by Property~2 of Lemma~\ref{lem:qld-sublines} the right-hand side of \eqref{eq:qld-combined-lines-consistency} is a mixture of terms of the form
\[ \E_{(\line_X, x) \sim D_{\tvar_1, i}} \E_{(\line_Z, z) \sim D_{\tvar_2, j}} \sum_{f_X, f_Z} \bra{\hat{\psi}} (T^{\line_X, \line_Z}_{f_X, f_Z})_{\alice \alice'} \ot (I - (\hat{Q}^{x,z}_{f_X(x), f_Z(z)})) \ket{\hat{\psi}} \]
for $\tvar_1, \tvar_2 \in \{\ALine, \DLine\}$ and $i, j \in \{1, \dots, m\}$. Hence using~\eqref{eq:qld-xz-lines-restricted} it follows that
\[ \eqref{eq:qld-combined-lines-consistency} \,=\, O(m^2 \delta_P)\;. \]
This establishes the first relation~\eqref{eq:qld-4-13}. An analogous argument shows the second relation.

\end{proof}

\begin{lemma}
\label{lem:qld-4-7}
There exists a function $\delta_S(\eps,m,d,q) = a(md)^a (\eps^b + q^{-b} + 2^{-bmd})$ for some universal constants $a > 1, 0 < b < 1$ such that the following holds. For $\mH \in \{\mH_\alice,\mH_\bob\}$ there exists a projective measurement $\{ \hat{S}_{g_\xpt,g_\zpt} \}$ acting on $\mathcal{H} \otimes (\C^q)^{\otimes \nqubits}$ with outcomes consisting of pairs $(g_\xpt,g_\zpt)$ of polynomials each in $\ideg_{d,m}(\F_q)$ such that for $W \in \{X,Z\}$, on average over uniformly random $u \in \F_q^m$,
\begin{align}
  (\hat{S}_{[\eval_{u}(\cdot_{W}) = a]})_{\alice \alice'} &\simeq_{\delta_S} (\hat{M}^{(\Point, W), u}_{a})_{\bob \alice''} \label{eq:qld-s-point-con-alice} \\ 
    (\hat{S}_{[\eval_{u}(\cdot_{W}) = a]})_{\bob \bob'} &\simeq_{\delta_S} (\hat{M}^{(\Point, W), u}_{a})_{\alice \bob''} \label{eq:qld-s-point-con-bob},
\end{align}
where the notation $\eval_{u}(\cdot_W)$ means the evaluation of the first outcome $g_X$ of $\hat{S}$ if $W = X$, or the second outcome $g_Z$ if $W = Z$, at the point $u$.
\end{lemma}

\begin{proof}
	Lemma~\ref{lem:qld-4-13} establishes the existence of a strategy \allowbreak{}$\strategy_\ld \allowbreak = (\hat{\psi},\allowbreak  \{
        \hat{Q}^\line_f \} \allowbreak \cup \{ \hat{Q}^{x,z,\alpha,\beta}_a \} )$\allowbreak{}
        for the classical low-degree test \allowbreak{}$\game^\ld_\ldparams$\allowbreak{} for
        \allowbreak{}$\ldparams \allowbreak = (q,\allowbreak 2m \allowbreak + 2,\allowbreak d,\allowbreak  1)$\allowbreak{}, with value \allowbreak{}$1 \allowbreak -
        \delta_{\combine}$\allowbreak{}. This strategy is not necessarily projective because the measurement \allowbreak{}$\{\hat{Q}^{\line}_f\}$\allowbreak{} is not guaranteed to be; however,\allowbreak  we may projectivize it by applying Naimark's theorem to this measurement. This preserves the consistency (\allowbreak{}$\allowbreak \simeq$\allowbreak{}) relation obtained as the conclusion of Lemma~\ref{lem:qld-4-13}. Theorem~\ref{lem:ld-soundness} applied to this projectivized strategy implies the
        existence of a POVM \allowbreak{}$\{ \hat{S}_g \}$\allowbreak{} with
        outcomes \allowbreak{}$g \allowbreak \in \ideg_{d,2m+2}(\F_q)$\allowbreak{} that is
        \allowbreak{}$\delta_\ld$\allowbreak{}-self-consistent and consistent with the
        ``points'' measurements \allowbreak{}$\{ \hat{Q}^{x,z,\alpha,\beta}_a\}$\allowbreak{},
        where \allowbreak{}$\delta_\ld \allowbreak = a(md)^a(\delta_\combine^b \allowbreak + q^{-b} \allowbreak + 2^{-bmd})$\allowbreak{} for some universal constants \allowbreak{}$a > 1,\allowbreak  0 < b < 1$\allowbreak{} is given
        by Theorem~\ref{lem:ld-soundness}. We may apply Naimark's theorem once again to \allowbreak{}$\{\hat{S}_g\}$\allowbreak{} to obtain a projective measurement with the same consistency guarantees, and henceforth, we will use \allowbreak{}$\{\hat{S}_g\}$\allowbreak{} to refer to this projective measurement.
	
	We now argue that with high probability, the polynomial $g$ returned by the measurement $\{\hat{S}_g \}$ has the form $g(x,z,\alpha,\beta) = \alpha g_\xpt(x) + \beta g_\zpt (z)$ for polynomials $g_\xpt, g_\zpt \in \ideg_{d,m}(\F_q)$. Let $\mathcal{G}$ denote the set of such polynomials. On average over uniformly random $(x,z,\alpha,\beta)$, 
	\begin{align}
          (\hat{S}_g)_{\alice \alice'} &= (\hat{S}_g)^2_{\alice \alice'} \notag \\
                                       &\approx_{\delta_\ld} (\hat{S}_g)_{\alice \alice'} \otimes (\hat{Q}^{x,z,\alpha,\beta}_{g(x,z,\alpha,\beta)})_{\bob \alice''}\notag\\
		&\approx_{\delta_Q} (\hat{S}_g)_{\alice \alice'} (\hat{Q}^{x,z,\alpha,\beta}_{g(x,z,\alpha,\beta)})_{\alice \alice'}\notag \\
		&\approx_{\delta_Q} (\hat{S}_g)_{\alice \alice'} \otimes \Big ( \sum_{b,b'} 1_{\alpha b + \beta b' = g(x,z,\alpha,\beta)} \, \hat{M}^{(\Point,X),x}_{b} \hat{M}^{(\Point,Z),z}_{b'} \Big)_{\bob \alice''} \label{eq:qld-g-42} \\
		&\approx_{\delta_Q} (\hat{S}_g)_{\alice \alice'} \otimes \Big ( \sum_{b,b'} 1_{\alpha b + \beta b' = g(x,z,\alpha,\beta)} \, \hat{M}^{(\Point,Z),z}_{b'} \hat{M}^{(\Point,X),x}_{b} \Big)_{\bob \alice''} \;.\label{eq:qld-g-43} 		
	\end{align}
	 Here the first approximation uses the projectivity of $\{\hat{S}_g\}$, and the first approximation uses the consistency of $\{\hat{S}_g\}$ with  $\{\hat{Q}^{x,z,\alpha,\beta}_a\}$ from Theorem~\ref{lem:ld-soundness}, together with Fact~\ref{fact:agreement} to switch from $\simeq$ to $\approx$, and Fact~\ref{fact:add-a-proj2}. The second, third and fourth approximations follow from the consistency properties of $\{\hat{Q}^{x,z,\alpha,\beta}_a\}$ from Lemma~\ref{lem:qld-4-12} combined with Fact~\ref{fact:add-a-proj2}. Note that we will need both \eqref{eq:qld-g-42} and \eqref{eq:qld-g-43} for the subsequent calculations.
	
	For all $g$ define $ \ket{\hat{\psi}_g} =(\hat{S}_g)_{\alice \alice'} \ket{\hat{\psi}}$. 
	Then on average over uniformly random $(x,z,\alpha,\beta)$, 
	  \begin{align}
&\begin{aligned}[t]
  &\E_{x,z,\alpha,\beta} \norm{ \Big ( \sum_{b,b'} 1_{\alpha b + \beta b' = g(x,z,\alpha,\beta)} \, \hat{M}^{(\Point,X),x}_{b} \hat{M}^{(\Point,Z),z}_{b'} \Big)_{\bob \alice''} \ket{\hat{\psi}_g} }^2
\end{aligned} \notag \\
&\begin{aligned}[t]
  &=_{\phantom{1/q}} \E_{x,z,\alpha,\beta} \sum_{b,b',b''} 1_{\alpha b + \beta b' = g(x,z,\alpha,\beta)} 1_{\alpha b + \beta b'' = g(x,z,\alpha,\beta)} \\
  &\bra{\hat{\psi}_g} \hat{M}^{(\Point,Z),z}_{b''} \hat{M}^{(\Point,X),x}_{b} \hat{M}^{(\Point,Z),z}_{b'} \ket{\hat{\psi}_g}
\end{aligned} \notag \\
&\begin{aligned}[t]
  &\approx_{1/q} \E_{x,z,\alpha,\beta} \sum_{b,b'} 1_{\alpha b + \beta b' = g(x,z,\alpha,\beta)} \bra{\hat{\psi}_g} \\
  &\hat{M}^{(\Point,Z),z}_{b'} \hat{M}^{(\Point,X),x}_{b} \hat{M}^{(\Point,Z),z}_{b'} \ket{\hat{\psi}_g}
\end{aligned} \label{eq:qld-g-prime}
\end{align}
	  where in the first equality, we used the fact that $ \{  \hat{M}^{(\Point,X),x}_{b} \}$ is projective, and in the approximation, we used that for fixed $x,z,\alpha,\beta$ with $\beta \neq 0$, if $\alpha b + \beta b' = g(x,z,\alpha,\beta)$ and $\alpha b + \beta b'' = g(x,z,\alpha,\beta)$, then $b' = b''$. The probability that $\beta = 0$ is $1/q$, and the term under the approximation has an absolute value of at most $1$, so the error incurred is at most $1/q$.

	 We will now show that the right-hand side of \eqref{eq:qld-g-prime} is small for polynomials $g$ that are not of the desired form. 
	Fix $b,b' \in \F_q$, and let $h_{b,b'}(x,z,\alpha,\beta) = g(x,z,\alpha,\beta) - (\alpha b + \beta b')$. Observe that since $g$ has individual degree $d$, so does $h_{b,b'}$. Write 
	\[
	h_{b,b'}(x,z,\alpha,\beta) = \sum_{i,j=0}^{d}\, h_{b,b',i,j}(x,z) \,  \alpha^i \beta^j 
	\]
	for some polynomials \allowbreak{}$h_{b,b',i,j}(x,\allowbreak z)$\allowbreak{} of individual degree at most \allowbreak{}$d$\allowbreak{}. 
	Let \allowbreak{}$\mathcal{G}' \subset \ideg_{d,2m+2}(\F_q)$\allowbreak{} denote the polynomials that are linear in \allowbreak{}$\alpha,\allowbreak  \beta$\allowbreak{}, i.e., \allowbreak{}$g \allowbreak = \alpha g_1(x,\allowbreak z) \allowbreak + \beta g_2(x,\allowbreak z)$\allowbreak{}. Fix a \allowbreak{}$g \notin \mathcal{G}'$\allowbreak{}. Then \allowbreak{}$h_{b,b'}(x,\allowbreak z,\allowbreak \alpha,\allowbreak \beta)$\allowbreak{} is a nonzero polynomial of total degree at most \allowbreak{}$(2m\allowbreak +2)d$\allowbreak{}, and, in particular, there must exist an \allowbreak{}$(i,\allowbreak j)$\allowbreak{} such that the polynomial \allowbreak{}$h_{b,b',i,j}(x,\allowbreak z)$\allowbreak{} is not the identically zero polynomial. Call a pair \allowbreak{}$(x,\allowbreak z) \allowbreak \in \F_q^2$\allowbreak{} \emph{good} if \allowbreak{}$h_{b,b'}(x,\allowbreak z,\allowbreak \alpha,\allowbreak \beta)$\allowbreak{} is a nonzero polynomial function of \allowbreak{}$\alpha,\allowbreak \beta$\allowbreak{}, and \emph{bad} otherwise. The probability that \allowbreak{}$(x,\allowbreak z)$\allowbreak{} is bad is at most \allowbreak{}$(2m\allowbreak +2)d/q$\allowbreak{} by the Schwartz-Zippel lemma. Otherwise, conditioned on a good \allowbreak{}$(x,\allowbreak z)$\allowbreak{} pair, the probability that \allowbreak{}$h_{b,b'}(x,\allowbreak z,\allowbreak \alpha,\allowbreak \beta)\allowbreak =0$\allowbreak{} over the choice of \allowbreak{}$\alpha,\allowbreak \beta$\allowbreak{} is at most \allowbreak{}$(2m\allowbreak +2)d/q$\allowbreak{}, again by the Schwartz-Zippel lemma. Thus for \allowbreak{}$g \notin \mathcal{G}'$\allowbreak{} we can upper-bound~\eqref{eq:qld-g-prime} by \allowbreak{}$\frac{2(2m+2)d}{q} \norm{ \ket{\hat{\psi}_g}}^2$\allowbreak{}. Combined with~\eqref{eq:qld-g-42} we get that
	\begin{equation}
	\label{eq:qld-g-prime-bound}
		\sum_{g \notin \mathcal{G}'} \norm{ (\hat{S}_g)_{\alice \alice'} \ket{\hat{\psi}}}^2 \leq O(\delta_\ld + \delta_Q + md/q).
	\end{equation}
	
	
	
	Next, we argue that not only is $g$ linear in $\alpha, \beta$
        with high probability, but also that $g = \alpha g_\xpt +
        \beta g_\zpt$ where $g_\xpt$ depends on $x$ only and $g_\zpt$
        depends on $z$ only. Fix a polynomial $g \in \mathcal{G}'$,
        i.e., $g(x,z,\alpha,\beta) = \alpha g_1(x,z) + \beta g_2(x,z)$
        for polynomials $g_1, g_2 \in \ideg_{d,2m}(\F_q)$. For all $x,z \in \F_q^m$, $\alpha, \beta, b,b' \in \F_q$,
	\begin{align*}
&\begin{aligned}[t]
  &1_{\alpha b + \beta b' = \alpha g_1(x,z) + \beta g_2(x,z)} = 1_{b = g_1(x,z)} 1_{b' = g_2(x,z)} \\
  &+ 1_{b \neq g_1(x,z) \, \vee \, b' \neq g_2(x,z)} \cdot 1_{\alpha b + \beta b' = \alpha g_1(x,z) + \beta g_2(x,z)}\;.
\end{aligned} 
\end{align*}
	By the Schwartz-Zippel lemma,  
	\begin{equation}
	\label{eq:qld-g-separable}
	\E_{\alpha, \beta} 1_{b \neq g_1(x,z) \, \vee \, b' \neq g_2(x,z)} \cdot 1_{\alpha b + \beta b' = \alpha g_1(x,z) + \beta g_2(x,z)} \leq 1/q\;.
	\end{equation}
	Let $\mathcal{G}' = \mathcal{G}'_\xpt \cup \mathcal{G}'_\zpt \cup \mathcal{G}$, where $\mathcal{G}'_\xpt$ (resp. $\mathcal{G'}_\zpt$) is the set of polynomials $g \in \mathcal{G}'$ where $g_1(x,z)$ depends nontrivially on $z$ (resp. $g_2(x,z)$ depends nontrivially on $x$). Fix $g \in \mathcal{G}'_\zpt$. We can upper-bound~\eqref{eq:qld-g-prime} by
\begin{align}
&\begin{aligned}[t]
  &\frac{1}{q} \norm{ \ket{\hat{\psi}_g}}^2 + \E_{x,z} \sum_{b,b'} 1_{b = g_1(x,z)} 1_{b' = g_2(x,z)} \bra{\hat{\psi}_g} \\
  &\hat{M}^{(\Point,Z),z}_{b'} \hat{M}^{(\Point,X),x}_{b} \hat{M}^{(\Point,Z),z}_{b'} \ket{\hat{\psi}_g}
\end{aligned} \label{eq:qld-g-48} \\
&\begin{aligned}[t]
  &\leq \frac{1}{q} \norm{ \ket{\hat{\psi}_g}}^2 + \E_{z} \sum_{b'} \Big(\E_x 1_{b' = g_2(x,z)} \Big) \, \bra{\hat{\psi}_g} \hat{M}^{(\Point,Z),z}_{b'} \ket{\hat{\psi}_g}
\end{aligned} \label{eq:qld-g-2}
\end{align}
	where the indicator $1_{b = g_1(x,z)}$ is removed using
        positivity in the inequality. To bound the second term
      in~\eqref{eq:qld-g-2}, we use an argument similar to the one we used to
      derive~\eqref{eq:qld-g-prime-bound}. That is, say that a value
      of $z$ is \emph{good} if $g_2(\cdot ,z)$ is not a constant
      function. The probability that $z$ is good is at least $1 -
      (2m+2)d/q$ by the Schwartz-Zippel lemma and the assumption that $g \in \mathcal{G}'_{\zpt}$ (and so $g_2$ depends nontrivially on $x$). Moreover, for each good $z$ and
      for each $b'$ the expectation $\E_x 1_{b' = g_2(x,z)}$ is at
      most $(2m+2)d/q$. Hence, we conclude that if $g_2(x,z)$ depends on $x$, then~\eqref{eq:qld-g-2} is at most $ \frac{2(2m+2)d + 1}{q} \norm {\ket{\hat{\psi}_g}}^2$, and thus 
	\begin{equation}
	\label{eq:qld-g-prime-xpt-bound}
		\sum_{g \in \mathcal{G}'_\zpt} \norm{ (\hat{S}_g)_{\alice \alice'} \ket{\hat{\psi}}}^2 \leq O(\delta_\ld + \delta_Q + md/q).
	\end{equation}	
	
	By starting with~\eqref{eq:qld-g-43} instead (i.e., the order of \linebreak[4]{}$M^{(\Point,Z),z}_{b'}$\allowbreak{} and \allowbreak{}$M^{(\Point,X),x}_b$\allowbreak{} are switched), we can perform similar reasoning to deduce that
	\begin{equation}
	\label{eq:qld-g-prime-zpt-bound}
		\sum_{g \in \mathcal{G}'_\xpt} \norm{ (\hat{S}_g)_{\alice \alice'} \ket{\hat{\psi}}}^2 \leq O(\delta_\ld + \delta_Q + md/q).
	\end{equation}	
	We thus obtain
	\begin{align}
&\begin{aligned}[t]
  &\sum_{g \notin \mathcal{G}} \norm{ (\hat{S}_g)_{\alice \alice'} \ket{\hat{\psi}}}^2 \leq \sum_{g \notin \mathcal{G}'}\norm{ (\hat{S}_g)_{\alice \alice'} \ket{\hat{\psi}}}^2 \\
  &+ \sum_{g \in \mathcal{G}'_\xpt}\norm{ (\hat{S}_g)_{\alice \alice'} \ket{\hat{\psi}}}^2 + \sum_{g \in \mathcal{G}'_\zpt}\norm{ (\hat{S}_g)_{\alice \alice'} \ket{\hat{\psi}}}^2
\end{aligned}  \\
&\begin{aligned}[t]
  &\leq O(\delta_\ld + \delta_Q + md/q).
\end{aligned} \label{eq:qld-g-non-separable}
\end{align}
        
	Let $\delta_{\mathcal{G}}$ be the error in the right hand side of \eqref{eq:qld-g-non-separable}. Define a projective sub-measurement $\{\hat{S}_{g_X, g_Z}\}$ with outcomes $g_X, g_Z \in \ideg_{d,m}(\F_q)$ by
        \begin{equation}
          \hat{S}_{g_X, g_Z} = \hat{S}_{\combine_{g_X, g_Z}}.
        \end{equation}
        A bound on the incompleteness of $\hat{S}_{g_X, g_Z}$ follows from~\eqref{eq:qld-g-non-separable} and the projectivity of $\hat{S}_g$:
        \begin{equation} 1 - \sum_{g_X, g_Z} \bra{\hat{\psi}} (\hat{S}_{g_X, g_Z})_{\alice \alice'} \ket{\hat{\psi}} = \sum_{g \notin \mathcal{G}} \bra{\hat{\psi}} (\hat{S}_g)_{\alice \alice'} \ket{\hat{\psi}} = \sum_{g \notin \mathcal{G}} \norm{ (\hat{S}_g)_{\alice \alice'} \ket{\hat{\psi}} }^2 \leq\delta _{\mG}. \label{eq:qld-sgg-completeness} \end{equation}
        We now bound the consistency of $\{\hat{S}_{g_X, g_Z}\}$ with the points measurements $\{\hat{M}^{(\Point, W), u}_{a}\}$.
	\begin{align}
&\begin{aligned}[t]
  &1 = \sum_g\bra{\hat{\psi}} (\hat{S}_g)_{\alice \alice'} \ket{\hat{\psi}}
\end{aligned} \label{eq:qld-s-good-and-bad} \\
&\begin{aligned}[t]
  &\approx_{\delta_{\mG}} \sum_{g \in \mathcal{G}} \bra{\hat{\psi}} (\hat{S}_g)_{\alice \alice'} \ket{\hat{\psi}}
\end{aligned} \label{eq:qld-s-good-g} \\
&\begin{aligned}[t]
  &\approx_{(\delta_Q)^{1/2}} \E_{x,z,\alpha,\beta} \sum_{g_\xpt, g_\zpt}\bra{\hat{\psi}} (\hat{S}_{g_\xpt,g_\zpt})_{\alice \alice'} \\
  &\otimes \Big ( \sum_{b,b'} 1_{\alpha b + \beta b' = \alpha g_\xpt (x) + \beta g_\zpt(z)} \, \hat{M}^{(\Point,Z),z}_{b'} \hat{M}^{(\Point,X),x}_{b} \Big)_{\bob \alice''} \ket{\hat{\psi}}
\end{aligned} \label{eq:qld-s-mm} \\
&\begin{aligned}[t]
  &\approx _{(\delta_Q)^{1/2}} \E_{x,z,\alpha,\beta} \sum_{g_\xpt, g_\zpt}\bra{\hat{\psi}} (\hat{S}_{g_\xpt,g_\zpt})_{\alice \alice'} \otimes \Big ( \sum_{b,b', b''} 1_{\alpha b + \beta b' = \alpha g_\xpt (x) + \beta g_\zpt(z) = \alpha b'' + \beta b'} \cdot\,
\end{aligned} \notag \\
&\begin{aligned}[t]
  &\hat{M}^{(\Point, X), x}_{b''} \hat{M}^{(\Point,Z),z}_{b'} \hat{M}^{(\Point,X),x}_{b} \Big)_{\bob \alice''} \ket{\hat{\psi}}
\end{aligned} \label{eq:qld-s-mmm} \\
&\begin{aligned}[t]
  &\approx _{1/q} \E_{x,z,\alpha,\beta} \sum_{g_\xpt, g_\zpt}\bra{\hat{\psi}} (\hat{S}_{g_\xpt,g_\zpt})_{\alice \alice'} \otimes \Big ( \sum_{b,b'} 1_{\alpha b + \beta b' = \alpha g_\xpt (x) + \beta g_\zpt(z) } \cdot \,
\end{aligned} \notag \\
&\begin{aligned}[t]
  &\hat{M}^{(\Point, X), x}_{b} \hat{M}^{(\Point,Z),z}_{b'} \hat{M}^{(\Point,X),x}_{b} \Big)_{\bob \alice''} \ket{\hat{\psi}}
\end{aligned} \label{eq:qld-s-mmm-positive} \\
&\begin{aligned}[t]
  &\approx_{ 1/q}\E_{x,z} \sum_{g_\xpt, g_\zpt} \bra{\hat{\psi}} (\hat{S}_{g_\xpt,g_\zpt})_{\alice \alice'} \\
  &\otimes \Big ( \hat{M}^{(\Point, X), x}_{g_\xpt(x)} \hat{M}^{(\Point,Z),z}_{g_\zpt(z)} \hat{M}^{(\Point,X),x}_{g_\xpt(x)} \Big)_{\bob \alice''} \ket{\hat{\psi}}.
\end{aligned} \label{eq:qld-xzx-disagree}
\end{align}
	Here \eqref{eq:qld-s-good-g} follows from~\eqref{eq:qld-sgg-completeness}. \eqref{eq:qld-s-mm} follows by approximation~\eqref{eq:qld-g-43} together with the Cauchy-Schwarz inequality and the projectivity of $\hat{S}_g$:
        \begin{align*}
&\begin{aligned}[t]
  &\Big| \E_{x,z,\alpha,\beta} \sum_{g_\xpt, g_\zpt} \bra{\hat{\psi}} \Big( (\hat{S}_{g_\xpt, g_\zpt})_{\alice \alice'} \cdot (I \\
  &- \Big( \sum_{b,b'} 1_{\alpha b + \beta b' = \alpha g_\xpt (x) + \beta g_\zpt(z)} \, \hat{M}^{(\Point,Z),z}_{b'} \hat{M}^{(\Point,X),x}_{b} \Big)_{\bob \alice''}\Big) \ket{\hat{\psi}} \Big|
\end{aligned}  \\
&\begin{aligned}[t]
  &\leq \sqrt{\E_{x,z,\alpha,\beta} \sum_{g_\xpt, g_\zpt} \bra{\hat{\psi}} (\hat{S}_{g_\xpt, g_\zpt})_{\alice \alice'}^2 \ket{\hat{\psi}}}
\end{aligned}  \\
&\begin{aligned}[t]
  &\quad \cdot \sqrt{\begin{aligned}[t]
  &\E_{x,z,\alpha,\beta} \sum_{g_\xpt, g_\zpt} \\
  &\norm{\begin{aligned}[t]
  & (\hat{S}_{g_\xpt, g_\zpt})_{\alice \alice'} \ot \Big(I - \Big( \sum_{b,b'} 1_{\alpha b + \beta b' = \alpha g_\xpt(x) + \beta g_\zpt(z)} \, \\
  &\hat{M}^{(\Point, Z),z}_{b'} \hat{M}^{(\Point, X), x}_{b} \Big)_{\bob \alice''} \Big)\ket{\hat{\psi}} 
\end{aligned}}^2 
\end{aligned}}
\end{aligned}  \\
&\begin{aligned}[t]
  &\leq 1 \cdot \sqrt{\delta_Q}.
\end{aligned} 
\end{align*}
        \eqref{eq:qld-s-mmm} follows by a similar calculation\anote{TODO}. To pass from here to \eqref{eq:qld-s-mmm-positive}, we argue that for $\alpha \neq 0$, the indicator forces $b = b''$, and the terms where $\alpha = 0$ are bounded in absolute value by $1/q$. Finally, to reach \eqref{eq:qld-xzx-disagree}, we apply \eqref{eq:qld-g-separable} to the expectation of the indicator over $\alpha, \beta$. \eqref{eq:qld-xzx-disagree} tells us that
        \begin{equation}
\begin{aligned}[t]
  &\big(\hat{S}_{[\eval_{x}(\cdot) = a, \eval_{z}(\cdot) = b]}\big)_{\alice \alice'} \\
  &\simeq_{O(\delta_{\mG} + (\delta_Q)^{1/2} + 1/q)} \big( \hat{M}^{(\Point, X), x}_{a} \hat{M}^{(\Point, Z), z}_{b} \hat{M}^{(\Point, X), x}_{a}\big)_{\bob \alice''}.
\end{aligned} \label{eq:qld-sgg-mhat-sandwich}
\end{equation}

        This is almost the desired conclusion of the lemma. The only issue is that $\{\hat{S}_{g_X, g_Z}\}$ is a sub-measurement. To address this, we complete $\hat{S}_{g_X, g_Z}$ by adding $1 - \sum_{g_X, g_Z} \hat{S}_{g_X, g_Z}$ to an arbitrary measurement element. By \eqref{eq:qld-sgg-completeness}, it follows that the consistency relation in \eqref{eq:qld-sgg-mhat-sandwich} holds for the completed measurement with an error increased by $\delta_{\mG}$. Let $\delta_S$ be this new, increased, error, and observe that $\delta_S = O(\delta_{\mG} + (\delta_Q)^{1/2} + d/q)$. Substituting in the bound on $\delta_{\mG}$ from \eqref{eq:qld-g-non-separable}, we obtain that $\delta_S = O(\delta_\ld + (\delta_Q)^{1/2} + md/q)$ has the form $a' (md)^{a'} (\eps^{b'} + q^{-b'} + 2^{-b'md})$ for some universal constants $a' > 1, 0 < b' < 1$. We thus conclude by an application of Fact~\ref{fact:data-processing} that \eqref{eq:qld-s-point-con-alice} holds for $W = X$.

        To obtain the same result for $W = Z$, we perform exactly the same steps starting from \eqref{eq:qld-s-good-and-bad} but with $X$ and $Z$ interchanged (so we use \eqref{eq:qld-g-42} instead of \eqref{eq:qld-g-43}). Finally, entirely analogous calculations with the registers swapped yield \eqref{eq:qld-s-point-con-bob}.

\end{proof}


\subsection{Pulling the $X$ and $Z$ measurements apart}
\label{sec:separating}

Recall the decoding map of the low-degree code defined in Section~\ref{sec:ld-encoding}. This map takes in a polynomial $g \in \ideg_{d,m}(\F_q)$ and returns a string $\coded(g) \in \F_q^{\nqubits}$ consisting of the evaluations of $g$ on the points in the hypercube $\{0,1\}^m$.
We now use this decoding map to construct a new set of measurement operators out of $\hat{S}$.
First, we introduce some notation. For all $W \in \{X,Z\}$ and all degree $d$ polynomials $g: \F_q^m \to \F_q$ define projectors
\[ \hat{S}^X_g = \sum_{g_\zpt} \hat{S}_{g,g_\zpt} \quad  \text{and}  \quad \hat{S}^Z_g = \sum_{g_\xpt} \hat{S}_{g_\xpt,g}\;,\]
where the measurement $\{\hat{S}_{g_\xpt,g_\zpt} \}$  is given by Lemma~\ref{lem:qld-4-7}.

For all $W \in \{X, Z\}$, $\tilde{u} \in \F_q^{\nqubits}$, and $a \in \F_q$, define the projector
\begin{equation}\label{eq:def-tauwu}
 \tau^{W}_{a}(\tilde{u}) = \sum_{h \in \F_q^{\nqubits} \,:\, h \cdot \tilde{u} = a} \tau^W_{h}
\end{equation}
where recall that $\tau^W_h$ is the projector corresponding to measuring $(\C^q)^{\otimes \nqubits}$ in the $W$ basis and obtaining $h \in \F_q^\nqubits$ as the outcome. For all $\tilde{u} \in \F_q^\nqubits$ define the projective measurement $\{ \tilde{M}^{W,\tilde{u}}_a \}_{a \in \F_q}$ by 
	\begin{equation}
	\label{eq:tilde_M}
	\tilde{M}^{W,\tilde{u}}_a =  \sum_{g} \hat{S}^W_g \otimes \tau_{(\coded(g) \cdot \tilde{u}) - a}^W(\tilde{u})\;.
	\end{equation}
	If $\hat{S}^W_g$ is viewed as an operator acting on registers $\alice \alice'$ (resp. $\bob \bob'$) then we view $\tilde{M}^{W,\tilde{u}}_a$ as an operator acting on registers $\alice \alice' \alice''$ (resp. $\bob \bob' \bob''$). (When necessary we will indicate via subscripts which registers the operators are acting on.)
The $\tilde{M}^{W,\tilde{u}}_a$ are projective because the $\hat{S}^W_g$ and $\tau^W$ operators are projective.
        
	For all $j \in \{1,\ldots,t\}$ define the observable
	\begin{equation}\label{eq:def-tildewj}
		\tilde{W}^j(\tilde{u}) = \sum_a (-1)^{\tr(e_j a)} \, \tilde{M}_a^{W,\tilde{u}} \;,
	\end{equation}
	where $\{e_j\}_{j \in \{1,\ldots,t\}}$ denotes the self-dual basis for $\F_q$ over $\F_2$ specified in Section~\ref{sec:subfields}. Let $\tau^W$ be the generalized Pauli observable defined in Section~\ref{sec:generalized-pauli}. Observe that
	\begin{align*}
&\begin{aligned}[t]
  &\Big ( \sum_{g_\xpt,g_\zpt} (-1)^{\tr(e_j (\coded(g_W) \cdot \tilde{u}))} \hat{S}_{g_\xpt,g_\zpt} \Big ) \otimes \tau^W(e_j \tilde{u}) \\
  &= \sum_{g_\xpt,g_\zpt} \sum_h (-1)^{\tr(e_j (\coded(g_W) \cdot \tilde{u}))} (-1)^{\tr(e_j(\tilde{u} \cdot h))} \hat{S}_{g_\xpt,g_\zpt} \otimes \tau^W_h
\end{aligned}  \\
&\begin{aligned}[t]
  &= \sum_{g_W} \sum_a (-1)^{\tr(e_j (\tilde{u} \cdot\coded(g_W) + a))} \hat{S}_{g_W}^W \otimes \Big( \sum_{h:\, h\cdot \tilde{u} = a}\tau^W_h\Big)
\end{aligned}  \\
&\begin{aligned}[t]
  &= \sum_{g_W} \sum_a (-1)^{\tr(e_j (\tilde{u} \cdot\coded(g_W) + a))} \hat{S}_{g_W}^W \otimes \tau^W_a(\tilde{u})
\end{aligned}  \\
&\begin{aligned}[t]
  &= \tilde{W}^j(\tilde{u}) \;,
\end{aligned} 
\end{align*}
	where the first equality uses the definition of $\tau^W(e_j \tilde{u})$, the second uses the definition of $\hat{S}^W_{g_W}$ and regroups terms, the third uses the definition~\eqref{eq:def-tauwu} of $\tau^W_a(\tilde{u})$, and the last is by a change of variables. These equations, in particular, show that the observable $\tilde{W}^j(\tilde{u})$ is Hermitian and squares to the identity, because the measurement $\{ \hat{S}_{g_\xpt,g_\zpt} \}$ is projective. 
	Furthermore, it can be verified by direct calculation that the observables $\tilde{W}^j(\tilde{u})$ satisfy the same relations as the generalized Pauli group. Specifically, for all $j,j' \in \{1,\ldots,t\}$, $\tilde{u}, \tilde{v} \in \F_q^\nqubits$,
	\[
		\tilde{X}^j(\tilde{u}) \tilde{Z}^{j'}(\tilde{v}) = \begin{cases}
		\tilde{Z}^{j'}(\tilde{v}) \tilde{X}^j(\tilde{u})  & \qquad j \neq j' \\
		(-1)^{\tr((e_j \tilde{u}) \cdot (e_j \tilde{v}))} \tilde{Z}^{j}(\tilde{v}) \tilde{X}^j(\tilde{u}) & \qquad j = j'
		\end{cases}~.
	\]
	The next lemma shows that the measurements $\{ \tilde{M}^{W,\tilde{u}}_a \}$  and observables $\tilde{W}^j(\tilde{u})$  are consistent with the original points measurements, and are self-consistent. 
	

\begin{lemma}
  \label{lem:qld-construct-the-paulis}
Let $\delta_S$ as in Lemma~\ref{lem:qld-4-7}.
For all $W \in \{X,Z\}$, all $\tilde{u} \in \F_q^\nqubits$ and $j \in \{1,\ldots,t\}$, the measurements $\{ \tilde{M}^{W,\tilde{u}}_a \}_{a \in \F_q}$ and observables $\tilde{W}^j(\tilde{u})$ satisfy the following properties:
\begin{enumerate}
	\item (\textbf{Consistency with points measurements}) For all $W \in \{X,Z\}$, for all $j \in \{1,\ldots,t\}$, and on average over uniformly random $u \in \F_q^m$,
	\[
		\Id_{\alice \alice' \alice''} \otimes \tilde{M}^{W,\ind_m(u)}_a \simeq_{\delta_S} M^{(\Point,W),u}_a \otimes \Id_{\bob \bob' \bob''}
	\]
	and
	\[
		\tilde{M}^{W,\ind_m(u)}_a \otimes \Id_{\bob \bob' \bob''} \simeq_{\delta_S} \Id_{\alice \alice' \alice''} \otimes M^{(\Point,W),u}_a \;.
	\]	
	\item (\textbf{Self-consistency}) On average over uniformly random $\tilde{u} \in \F_q^\nqubits$,
	\[
		\tilde{W}^j(\tilde{u}) \otimes \Id_{\bob \bob' \bob''} \approx_{\delta_S} \Id_{\alice \alice' \alice''} \otimes \tilde{W}^j(\tilde{u})\,.
	\]
\end{enumerate}
\end{lemma}

\begin{proof}
We show the first item of the lemma for the case when $M^{(\Point,W),u}_a$ acts on register $\alice$ and $\tilde{M}^{W,\ind_m(u)}_a$ acts on registers $\bob \bob' \bob''$ (the argument for when the $\alice$ and $\bob$ registers are interchanged proceeds analogously):
	\begin{align*}
		&\E_{u \in \F_q^m} \sum_a \bra{\hat{\psi}} M^{(\Point,W),u}_a \otimes \tilde{M}^{W,\ind_m(u)}_a \ket{\hat{\psi}} \\
		&= \E_{u \in \F_q^m} \sum_{g,a} \bra{\hat{\psi}} (M^{(\Point,W),u}_a)_{\alice} \otimes (\hat{S}^W_g)_{\bob \bob'} \otimes (\tau^W_{\coded(g) \cdot \ind_m(u) -a}(\ind_m(u)))_{\bob''} \ket{\hat{\psi}} \\
		&= \E_{u \in \F_q^m} \sum_{g} \bra{\hat{\psi}} (\hat{M}^{(\Point,W),u}_{\coded(g) \cdot \ind_m(u)})_{\alice \bob''} \otimes (\hat{S}^W_g)_{\bob \bob'} \ket{\hat{\psi}} \\
		&= \E_{u \in \F_q^m} \sum_{g} \bra{\hat{\psi}} (\hat{M}^{(\Point,W),u}_{g(u)})_{\alice \bob''} \otimes (\hat{S}^W_g)_{\bob \bob'} \ket{\hat{\psi}} \\
		&\geq 1 - \delta_S\;.
	\end{align*}
	The second line follows from expanding the definition of $\tilde{M}^{W,\ind_m(u)}_a$ using~\eqref{eq:tilde_M}, the third line follows from the definition of $\hat{M}^{(\Point,W),u}_a$ given in Section~\ref{sec:expanding}, the fourth line follows from the fact that $\coded(g) \cdot \ind_m(u) = g(u)$, and the fifth line follows from Lemma~\ref{lem:qld-4-7}. This implies the first item of the Lemma.
	To show the second item we first argue that, on average over $\tilde{u}$, 
	\begin{equation}
		\label{eq:qld-pulling-cons}
		\Big ( \tilde{M}^{W,\tilde{u}}_a \Big)_{\alice \alice' \alice''} \approx_{\delta_S} \Big ( \tilde{M}^{W,\tilde{u}}_a \Big)_{\bob \bob' \bob''}.
	\end{equation}
	This follows because on average over $u \in \F_q^m$ and $\tilde{u} \in \F_q^\nqubits$ we have
	\begin{align}
&\begin{aligned}[t]
  &\Big ( \tilde{M}^{W,\tilde{u}}_a \Big)_{\alice \alice' \alice''}
\end{aligned} \label{eq:qld-pulling-0} \\
&\begin{aligned}[t]
  &\approx_{\delta_S} \Big ( \tilde{M}^{W,\tilde{u}}_a \Big)_{\alice \alice'\alice''} \cdot \Big ( \sum_g (\hat{S}^W_g)_{\alice \alice'} \otimes (\hat{M}^{(\Point,W),u}_{g(u)})_{\bob \alice''} \Big)
\end{aligned} \label{eq:qld-pulling-1} \\
&\begin{aligned}[t]
  &=_{\phantom{\delta_S}} \sum_g \Big ( (\hat{S}^W_g)_{\alice \alice'} \otimes (\tau^W_{(\coded(g) \cdot \tilde{u}) - a}(\tilde{u}))_{\alice''} \Big) \\
  &\cdot \Big(\sum_{r,s: r + s = g(u)} (M^{(\Point,W),u}_r)_{\bob} \otimes (\tau_s^W(\ind_m(u)))_{\alice''} \Big)
\end{aligned} \label{eq:qld-pulling-2} \\
&\begin{aligned}[t]
  &=_{\phantom{\delta_S}} \sum_{\substack{g,h: \\ (\coded(g)-h)\cdot \tilde{u} = a}} (\hat{S}^W_g)_{\alice \alice'} \otimes (\tau^W_h)_{\alice''} \otimes (M^{(\Point,W),u}_{(g-g_h)(u)})_{\bob}~.
\end{aligned} \label{eq:qld-pulling-2b}
\end{align}
	The first inequality follows from right-multiplying~\eqref{eq:qld-pulling-0} by the approximation~\eqref{eq:qld-sg-cons} in Lemma~\ref{lem:qld-constructing-the-paulis-helper} (to be proved below) and using Fact~\ref{fact:add-a-proj}. Line~\eqref{eq:qld-pulling-2} follows from expanding the definition of $\tilde{M}^{W,\tilde{u}}_a$.  To obtain Line~\eqref{eq:qld-pulling-2b}, we expanded the definitions of $\tau^W_{(\coded(g) \cdot \tilde{u}) - a}(\tilde{u})$ and $\tau^W_s(\ind_m(u))$ using~\eqref{eq:def-tauwu}, and the fact that $h \cdot \ind_m(u) = g_h(u)$ (due to \eqref{eq:low-degree-encoding-definition}). Thus for all $g \in \ideg_{d,m}(\F_q)$ and $s \in \F_q$ we have
	\[
		\tau^W_{(\coded(g) \cdot \tilde{u}) - a}(\tilde{u}) \cdot \tau^W_s(\ind_m(u)) = \sum_{\substack{h: \\ (\coded(g) - h) \cdot \tilde{u} = a \\ g_h(u) = s}} \tau^W_h~.
	\]
	
	Using the self-consistency of $\{M^{(\Point,W),u}_a\}$ shown in Lemma~\ref{lem:qld-win-implications}, we then obtain
	\begin{align}
&\begin{aligned}[t]
  &\text{\eqref{eq:qld-pulling-2b}} \approx_{\eps} \sum_{\substack{g,h: \\ (\coded(g)-h)\cdot \tilde{u} = a}} (\hat{S}^W_g)_{\alice \alice'} \\
  &\cdot \Big((M^{(\Point,W),u}_{(g-g_h)(u)})_{\alice} \otimes (\tau_h^W)_{\alice''} \Big) \cdot \Big ( M^{(\Point,W),u}_{(g-g_h)(u)} \Big)_{\bob}
\end{aligned} \label{eq:qld-pulling-3} \\
&\begin{aligned}[t]
  &\approx_{0}\sum_{\substack{g,h: \\ (\coded(g)-h)\cdot \tilde{u} = a}} (\hat{S}^W_g)_{\alice \alice'} \cdot \Big((M^{(\Point,W),u}_{(g-g_h)(u)})_{\alice} \\
  &\otimes (\tau_h^W)_{\alice'} \otimes (\tau_h^W)_{\alice''} \Big) \cdot \Big ( M^{(\Point,W),u}_{(g-g_h)(u)} \Big)_{\bob}
\end{aligned} \label{eq:qld-pulling-3a} \\
&\begin{aligned}[t]
  &= \sum_{\substack{g,h: \\ (\coded(g)-h)\cdot \tilde{u} = a}} (\hat{S}^W_g)_{\alice \alice'} \cdot \Big((\hat{M}^{(\Point,W),u}_{g(u)})_{\alice \alice'} \\
  &\otimes (\tau_h^W)_{\alice''} \Big) \cdot \Big ( M^{(\Point,W),u}_{(g-g_h)(u)} \Big)_{\bob} \;.
\end{aligned} \label{eq:qld-pulling-3b}
\end{align}
	The ``$\approx_0$'' in Line~\eqref{eq:qld-pulling-3a} indicates that the left hand side is equal to the right hand side when applied to the state $\ket{\hat{\psi}}$. Line~\eqref{eq:qld-pulling-3a} follows from the self-consistency of the operators $\{\tau^W_h\}$, and Line~\eqref{eq:qld-pulling-3b} follows from the definition of $\{ \hat{M}^{(\Point,W),u}_a\}$. Using the fact that $\ket{\hat{\psi}} = \sum_{h'} (\tau^W_{h'})_{\bob'} \otimes (\tau^W_{h'})_{\bob''} \ket{\hat{\psi}}$, we get that 
\begin{align}
&\begin{aligned}[t]
  &\text{\eqref{eq:qld-pulling-3b}} \approx_0 \sum_{\substack{g,h: \\ (\coded(g)-h)\cdot \tilde{u} = a}} \Big((\hat{S}^W_g \cdot \hat{M}^{(\Point,W),u}_{g(u)})_{\alice \alice'} \\
  &\otimes (\tau_h^W)_{\alice''} \Big) \cdot \Big( M^{(\Point,W),u}_{(g-g_h)(u)} \otimes \sum_{h'} \tau_{h'}^W \otimes \tau_{h'}^W\Big)_{\bob \bob' \bob''}
\end{aligned} \label{eq:qld-pulling-4} \\
&\begin{aligned}[t]
  &= \sum_{\substack{g,h,h': \\ (\coded(g)-h)\cdot \tilde{u} = a}} \Big((\hat{S}^W_g \cdot \hat{M}^{(\Point,W),u}_{g(u)})_{\alice \alice'} \\
  &\otimes (\tau_h^W)_{\alice''} \Big) \cdot \Big( (\hat{M}^{(\Point,W),u}_{(g-g_h+g_{h'})(u)})_{\bob \bob'} \otimes (\tau_{h'}^W)_{\bob''} \Big) \;.
\end{aligned} \label{eq:qld-pulling-5}
\end{align}
	Line~\eqref{eq:qld-pulling-5} follows from~\eqref{eq:qld-pulling-4} by expanding the definition of $\{ \hat{M}^{(\Point,W),u}_a\}$. 
		Next, we argue that, on average over $u \in \F_q^m$,
%
%
		\begin{align}
&\begin{aligned}[t]
  &\text{\eqref{eq:qld-pulling-5}} \approx_{\delta_S} \sum_{\substack{g,h,h': \\ (\coded(g)-h)\cdot \tilde{u} = a}} \Big((\hat{S}^W_g)_{\alice \alice'} \\
  &\otimes (\tau_h^W)_{\alice''} \Big) \otimes \Big( (\hat{M}^{(\Point,W),u}_{(g-g_h+g_{h'})(u)})_{\bob \bob'} \otimes (\tau_{h'}^W)_{\bob''} \Big)~.
\end{aligned} \label{eq:qld-pulling-7}
\end{align}
		We show this by bounding the magnitude of the difference. Using the fact that $\hat{S}^W_g$ and $\tau^W_h$ are projective, we have:
		\begin{align}
&\begin{aligned}[t]
  &\E_u \sum_a \Big \|\sum_{\substack{g,h,h': \\ (\coded(g)-h)\cdot \tilde{u} = a}} (\hat{S}^W_g \cdot (\Id - \hat{M}^{(\Point,W),u}_{g(u)}))_{\alice \alice'} \\
  &\otimes (\tau_h^W)_{\alice''} \otimes (\hat{M}^{(\Point,W),u}_{(g-g_h+g_{h'})(u)})_{\bob \bob'} \otimes (\tau_{h'}^W)_{\bob''} \, \ket{\hat{\psi}} \Big \|^2
\end{aligned} \notag \\
&\begin{aligned}[t]
  &= \E_u \sum_{\substack{g,h,h'}} \bra{\hat{\psi}} \Big(((\Id - \hat{M}^{(\Point,W),u}_{g(u)}) \cdot \hat{S}^W_g \cdot (\Id - \hat{M}^{(\Point,W),u}_{g(u)}))_{\alice \alice'} \otimes (\tau_h^W)_{\alice''} \Big)
\end{aligned} \notag \\
&\begin{aligned}[t]
  &\cdot \Big( (\hat{M}^{(\Point,W),u}_{(g-g_h+g_{h'})(u)})^2_{\bob \bob'} \otimes (\tau_{h'}^W)_{\bob''} \Big) \ket{\hat{\psi}}
\end{aligned} \notag \\
&\begin{aligned}[t]
  &\leq \E_u \sum_{g} \bra{\hat{\psi}} ((\Id - \hat{M}^{(\Point,W),u}_{g(u)}) \cdot \hat{S}^W_g \cdot (\Id - \hat{M}^{(\Point,W),u}_{g(u)}))_{\alice \alice'} \ket{\hat{\psi}}
\end{aligned} \label{eq:qld-pulling-8} \\
&\begin{aligned}[t]
  &\leq \delta_S\;.
\end{aligned} \label{eq:qld-pulling-9}
\end{align}
		The inequality in Line~\eqref{eq:qld-pulling-8} follows from the fact that for all $g,h$, the sum $\sum_{h'} (\hat{M}^{(\Point,W),u}_{(g-g_h+g_{h'})(u)})^2_{\bob \bob'} \otimes (\tau_{h'}^W)_{\bob''}$ is at most $\Id$. The inequality in Line~\eqref{eq:qld-pulling-9} follows from the second item of Lemma~\ref{lem:qld-constructing-the-paulis-helper} (to be proved below). 
		Next, we show that on average over $u \in \F_q^m$,
		\begin{align}
&\begin{aligned}[t]
  &\text{\eqref{eq:qld-pulling-7}} =\sum_{\substack{g,h,g',h': \\ (\coded(g)-h)\cdot \tilde{u} = a}} \Big((\hat{S}^W_g)_{\alice \alice'} \\
  &\otimes (\tau_h^W)_{\alice''} \Big) \cdot \Big( (\hat{S}^W_{g'} \cdot \hat{M}^{(\Point,W),u}_{(g-g_h+g_{h'})(u)})_{\bob \bob'} \otimes (\tau_{h'}^W)_{\bob''} \Big)
\end{aligned} \label{eq:qld-pulling-9a} \\
&\begin{aligned}[t]
  &\approx_{O(\delta_S + \sqrt{\eps})} \sum_{\substack{g,h,g',h': \\ (g' - g_{h'})(u) = (g -g_h)(u) \\ (\coded(g)-h)\cdot \tilde{u} = a}} \Big((\hat{S}^W_g)_{\alice \alice'} \\
  &\otimes (\tau_h^W)_{\alice''} \Big) \cdot \Big( (\hat{S}^W_{g'} \cdot \hat{M}^{(\Point,W),u}_{g'(u)})_{\bob \bob'} \otimes (\tau_{h'}^W)_{\bob''} \Big)
\end{aligned} \label{eq:qld-pulling-10} \\
&\begin{aligned}[t]
  &\approx_{\delta_S} \sum_{\substack{g,h,g',h': \\ (g' - g_{h'})(u) = (g -g_h)(u) \\ (\coded(g)-h)\cdot \tilde{u} = a}} \Big((\hat{S}^W_g)_{\alice \alice'} \otimes (\tau_h^W)_{\alice''} \Big) \cdot \Big( (\hat{S}^W_{g'})_{\bob \bob'} \otimes (\tau_{h'}^W)_{\bob''} \Big)
\end{aligned} \label{eq:qld-pulling-11} \\
&\begin{aligned}[t]
  &\approx_{md/q} \sum_{\substack{g,h,g',h': \\ g' - g_{h'} = g -g_h \\ (\coded(g)-h)\cdot \tilde{u} = a}} \Big((\hat{S}^W_g)_{\alice \alice'} \otimes (\tau_h^W)_{\alice''} \Big) \cdot \Big( (\hat{S}^W_{g'})_{\bob \bob'} \otimes (\tau_{h'}^W)_{\bob''} \Big)~.
\end{aligned} \label{eq:qld-pulling-12}
\end{align}
		Before we justify the sequence of steps to go from Line~\eqref{eq:qld-pulling-7} to Line~\eqref{eq:qld-pulling-12}, we first argue how this establishes Line~\eqref{eq:qld-pulling-cons}, and hence the second item of the Lemma statement.
		
		Absorbing the error terms $O(md/q)$ and $O(\sqrt{\eps})$ into $\delta_S$, we have shown that 
		\[
		\Big ( \tilde{M}^{W,\tilde{u}}_a \Big)_{\alice \alice' \alice''} \approx_{\delta_S} \sum_{\substack{g,h,g',h': \\ g' - g_{h'} = g -g_h \\ (\coded(g)-h)\cdot \tilde{u} = a}} \Big((\hat{S}^W_g)_{\alice \alice'} \otimes (\tau_h^W)_{\alice''} \Big) \cdot \Big( (\hat{S}^W_{g'})_{\bob \bob'} \otimes (\tau_{h'}^W)_{\bob''} \Big).
		\]
		Note that the conditions $(\coded(g)-h)\cdot \tilde{u} = a$ and $g - g_h = g' - g_{h'}$ are equivalent to the conditions $(\coded(g') - h') \cdot \tilde{u} = a$ and $g - g_h = g' - g_{h'}$. Thus the expression in~\eqref{eq:qld-pulling-12} is symmetric between $g,h$ and $g',h'$, and so an analogous derivation shows that $\Big ( \tilde{M}^{W,\tilde{u}}_a \Big)_{\bob \bob' \bob''}$ is $\delta_S$-close to~\eqref{eq:qld-pulling-12}, and thus $\delta_S$-close to $\Big ( \tilde{M}^{W,\tilde{u}}_a \Big)_{\alice \alice' \alice''}$. This shows~\eqref{eq:qld-pulling-cons}. Since $\{ \tilde{M}^{W,\tilde{u}}_a \}$ is projective, by using Item 2 of Fact~\ref{fact:agreement}, followed by Fact~\ref{fact:data-processing}, and then followed by Item 1 of Fact~\ref{fact:agreement}, we get for all $j \in \{1,\ldots,t\}$, on average over $\tilde{u} \in \F_q^\nqubits$,
\[
	(\tilde{M}^{W,\tilde{u}}_{[\tr(e_j \cdot) = b]})_{\alice \alice' \alice''} \approx_{\delta_S} ( \tilde{M}^{W,\tilde{u}}_{[\tr(e_j \cdot) = b]})_{\bob \bob' \bob''}
\]
where the answer summation is over $b \in \F_2$. Finally, the second item of the Lemma follows from applying~Lemma~\ref{lem:povm-to-obs}.

		We now return to justifying the sequence of steps between Lines~\eqref{eq:qld-pulling-7} and~\eqref{eq:qld-pulling-12}. The equality in Line~\eqref{eq:qld-pulling-9a} follows by left-multiplying~\eqref{eq:qld-pulling-7} by $I_{\bob \bob'} = \sum_{g'} (\hat{S}^W_{g'})_{\bob \bob'}$. 
		
		\paragraph{The approximation in Line~\eqref{eq:qld-pulling-10}.} The approximation in Line~\eqref{eq:qld-pulling-10} follows by bounding the magnitude of the difference:	
\begin{align}
&\begin{aligned}[t]
  &\E_u \sum_a \Big \| \sum_{\substack{g,h,g',h': \\ (g' - g_{h'})(u) \neq (g -g_h)(u) \\ (\coded(g)-h)\cdot \tilde{u} = a}} \Big((\hat{S}^W_g)_{\alice \alice'} \\
  &\otimes (\tau_h^W)_{\alice''} \Big) \cdot \Big( (\hat{S}^W_{g'} \cdot \hat{M}^{(\Point,W),u}_{(g-g_h+g_{h'})(u)})_{\bob \bob'} \otimes (\tau_{h'}^W)_{\bob''} \Big) \ket{\hat{\psi}} \Big \|^2
\end{aligned} \notag \\
&\begin{aligned}[t]
  &= \E_u \sum_{\substack{g,h,g',h': \\ (g' - g_{h'})(u) \neq (g -g_h)(u)}} \bra{\hat{\psi}} \Big(\hat{S}^W_g)_{\alice \alice'} \\
  &\otimes (\tau_h^W)_{\alice''} \Big) \cdot \Big( (\hat{M}^{(\Point,W),u}_{(g-g_h+g_{h'})(u)} \cdot \hat{S}^W_{g'} \cdot
\end{aligned} \notag \\
&\begin{aligned}[t]
  &\hat{M}^{(\Point,W),u}_{(g-g_h+g_{h'})(u)})_{\bob \bob'} \otimes (\tau_{h'}^W)_{\bob''} \Big) \ket{\hat{\psi}}
\end{aligned} \notag \\
&\begin{aligned}[t]
  &\leq \E_u \sum_{\substack{g,h,g',h',a': \\ a' \neq g'(u)}} \bra{\hat{\psi}} \Big(\hat{S}^W_g)_{\alice \alice'} \otimes (\tau_h^W)_{\alice''} \Big) \\
  &\cdot \Big( (\hat{M}^{(\Point,W),u}_{a'} \cdot \hat{S}^W_{g'} \cdot \hat{M}^{(\Point,W),u}_{a'})_{\bob \bob'} \otimes (\tau_{h'}^W)_{\bob''} \Big) \ket{\hat{\psi}}
\end{aligned} \notag \\
&\begin{aligned}[t]
  &= \E_u \sum_{g,a: a \neq g(u)} \bra{\hat{\psi}} (\hat{M}^{(\Point,W),u}_{a} \\
  &\cdot \hat{S}^W_{g} \cdot \hat{M}^{(\Point,W),u}_{a})_{\bob \bob'} \ket{\hat{\psi}}
\end{aligned} \label{eq:qld-pulling-13a} \\
&\begin{aligned}[t]
  &\approx_{O(\sqrt{\eps})} \E_u \sum_{g} \bra{\hat{\psi}} (\Id - \hat{M}^{(\Point,W),u}_{g(u)})_{\alice \bob''} \otimes ( \hat{S}^W_{g})_{\bob \bob'} \ket{\hat{\psi}}
\end{aligned} \label{eq:qld-pulling-13} \\
&\begin{aligned}[t]
  &\leq \delta_S \;.
\end{aligned} \label{eq:qld-pulling-14}
\end{align}
The approximation in Line~\eqref{eq:qld-pulling-13} follows from the following calculation:
\begin{align}
&\begin{aligned}[t]
  &\Big | \text{\eqref{eq:qld-pulling-13}} - \text{\eqref{eq:qld-pulling-13a}} \Big |
\end{aligned} \notag \\
&\begin{aligned}[t]
  &\leq \Big | \E_u \sum_{g,a: a \neq g(u)} \bra{\hat{\psi}} (\hat{M}^{(\Point,W),u}_{a} \cdot \hat{S}^W_{g})_{\bob \bob'} \\
  &\cdot ((\hat{M}^{(\Point,W),u}_{a})_{\alice \bob''} - (\hat{M}^{(\Point,W),u}_{a})_{\bob \bob'}) \ket{\hat{\psi}} \Big |
\end{aligned} \label{eq:qld-pulling-15} \\
&\begin{aligned}[t]
  &\qquad + \Big | \E_u \sum_{g,a: a \neq g(u)} \bra{\hat{\psi}} ((\hat{M}^{(\Point,W),u}_{a})_{\alice \bob''} - (\hat{M}^{(\Point,W),u}_{a})_{\bob \bob'}) \\
  &\cdot (\hat{S}^W_{g})_{\bob \bob'} \cdot (\hat{M}^{(\Point,W),u}_{a})_{\alice \bob''} \ket{\hat{\psi}} \Big |
\end{aligned} \label{eq:qld-pulling-15a}
\end{align}
Using Cauchy-Schwarz, we can bound the first term on the right hand side of the inequality by
\begin{align}
	&\Biggl( \E_u \sum_{g,a: a \neq g(u)} \bra{\hat{\psi}} (\hat{M}^{(\Point,W),u}_{a} \cdot \hat{S}^W_{g} \cdot \hat{M}^{(\Point,W),u}_{a})_{\bob \bob'} \ket{\hat{\psi}}  \Biggr)^{1/2} \cdot \notag \\
	&\Biggl( \E_u \sum_{g,a: a \neq g(u)} \bra{\hat{\psi}} ((\hat{M}^{(\Point,W),u}_{a})_{\alice \bob''} - (\hat{M}^{(\Point,W),u}_{a})_{\bob \bob'}) \cdot (\hat{S}^W_{g})_{\bob \bob'} \cdot \notag\\
  & ((\hat{M}^{(\Point,W),u}_{a})_{\alice \bob''} - (\hat{M}^{(\Point,W),u}_{a})_{\bob \bob'})  \ket{\hat{\psi}}  \Biggr)^{1/2} \notag \\
	\leq & \sqrt{1} \cdot \sqrt{\eps} \notag
\end{align}
The inequality follows from the self-consistency of $\hat{M}^{(\Point,W),u}_{a}$, as established in Lemma~\ref{lem:qld-comm-cons}. Similarly, we can bound the second term in Line~\eqref{eq:qld-pulling-15a} by $\sqrt{\eps}$.
This establishes the approximation in Line~\eqref{eq:qld-pulling-13}. 

The inequality in Line~\eqref{eq:qld-pulling-14} follows from Line~\eqref{eq:qld-s-point-con-bob} of Lemma~\ref{lem:qld-4-7}. This concludes the derivation of the approximation in Line~\eqref{eq:qld-pulling-10}.

\paragraph{The approximation in Line~\eqref{eq:qld-pulling-11}.} We establish this approximation by bounding the magnitude of the difference:
\begin{align*}
&\begin{aligned}[t]
  &\E_u \sum_a \Big \| \sum_{\substack{g,h,g',h': \\ (g' - g_{h'})(u) = (g -g_h)(u) \\ (\coded(g)-h)\cdot \tilde{u} = a}} \Big((\hat{S}^W_g)_{\alice \alice'} \\
  &\otimes (\tau_h^W)_{\alice''} \Big) \cdot \Big( (\hat{S}^W_{g'} \cdot (\Id - \hat{M}^{(\Point,W),u}_{g'(u)}))_{\bob \bob'} \otimes (\tau_{h'}^W)_{\bob''} \Big) \ket{\hat{\psi}} \Big \|^2
\end{aligned}  \\
&\begin{aligned}[t]
  &= \E_u \sum_{\substack{g,h,g',h': \\ (g' - g_{h'})(u) \\ \quad = (g -g_h)(u)}} \bra{\hat{\psi}} (\hat{S}^W_g)_{\alice \alice'} \otimes (\tau_h^W)_{\alice''} \otimes ((\Id - \hat{M}^{(\Point,W),u}_{g'(u)}) \cdot \hat{S}^W_{g'} \cdot
\end{aligned} \notag \\
&\begin{aligned}[t]
  &(\Id - \hat{M}^{(\Point,W),u}_{g'(u)}))_{\bob \bob'} \otimes (\tau_{h'}^W)_{\bob''} \ket{\hat{\psi}}
\end{aligned}  \\
&\begin{aligned}[t]
  &\leq \E_u \sum_{g'} \bra{\hat{\psi}} ((\Id - \hat{M}^{(\Point,W),u}_{g'(u)}) \cdot \hat{S}^W_{g'} \cdot (\Id - \hat{M}^{(\Point,W),u}_{g'(u)}))_{\bob \bob'} \ket{\hat{\psi}}
\end{aligned}  \\
&\begin{aligned}[t]
  &\leq \delta_S \;.
\end{aligned} 
\end{align*}
The last inequality follows from the second item of Lemma~\ref{lem:qld-constructing-the-paulis-helper} (to be proved below). 

\paragraph{The approximation in Line~\eqref{eq:qld-pulling-12}.} We establish this approximation by bounding the magnitude of the difference:
\begin{align*}
&\begin{aligned}[t]
  &\E_u \sum_a \Big \| \sum_{\substack{g,h,g',h': \\ g' - g_{h'} \neq g -g_h \\ (g' - g_{h'})(u) = (g -g_h)(u) \\ (\coded(g)-h)\cdot \tilde{u} = a}} \Big((\hat{S}^W_g)_{\alice \alice'} \\
  &\otimes (\tau_h^W)_{\alice''} \Big) \cdot \Big( (\hat{S}^W_{g'})_{\bob \bob'} \otimes (\tau_{h'}^W)_{\bob''} \Big) \ket{\hat{\psi}} \Big \|^2
\end{aligned}  \\
&\begin{aligned}[t]
  &= \sum_{\substack{g,h,g',h': \\ g' - g_{h'} \neq g -g_h}} \bra{\hat{\psi}} \Big((\hat{S}^W_g)_{\alice \alice'} \otimes (\tau_h^W)_{\alice''} \Big) \\
  &\cdot \Big( (\hat{S}^W_{g'})_{\bob \bob'} \otimes (\tau_{h'}^W)_{\bob''} \Big) \ket{\hat{\psi}} \cdot \E_u \mathbf{1}[(g' - g_{h'})(u) = (g -g_h)(u)]
\end{aligned}  \\
&\begin{aligned}[t]
  &\leq md/q.
\end{aligned} 
\end{align*}
The last inequality follows from the Schwartz-Zippel lemma: two distinct polynomials of total degree at most $md$ cannot agree on more than $md/q$ fraction of points.

\end{proof}

We now give a proof of the following Lemma, which was used in several derivations in Lemma~\ref{lem:qld-construct-the-paulis}. 
\begin{lemma}
\label{lem:qld-constructing-the-paulis-helper}
	For $W \in \{X,Z\}$, on average over $u \in \F_q^m$, we have that
	\begin{gather}
		\label{eq:qld-sg-cons}
		\sum_g (\hat{S}^W_g)_{\alice \alice'} \otimes (\hat{M}^{(\Point,W),u}_{g(u)})_{\bob \alice''} \approx_{\delta_S} \Id
	\end{gather}
	and
	\begin{gather}
		\label{eq:qld-sg-cons2}
		(\hat{S}^W_g \cdot (\Id - \hat{M}^{(\Point, W), u}_{g(u)}))_{\alice \alice'} \approx_{\delta_S} 0
	\end{gather}
	where the answer summation is over $g \in \ideg_{d,m}(\F_q)$.
      Furthermore, all symmetric equivalents of these approximations
      also hold.
\end{lemma}
\begin{proof}
We first prove \eqref{eq:qld-sg-cons}. From Line~\eqref{eq:qld-s-point-con-alice} of Lemma~\ref{lem:qld-4-7} we have that on average over $u \in \F_q^m$,
\begin{equation}
\label{eq:qld-constructing-the-paulis-helper0}
	  \bra{\hat{\psi}} \sum_g (\hat{S}^W_g)_{\alice \alice'} \otimes (\hat{M}^{(\Point, W), u}_{g(u)})_{\bob \alice''} \ket{\hat{\psi}} \geq 1 - \delta_S
\end{equation}
which is equivalent to $\sum_g (\hat{S}^W_g)_{\alice \alice'} \otimes (\hat{M}^{(\Point, W), u}_{g(u)})_{\bob \alice''}  \simeq_{\delta_S} \Id$. By Fact~\ref{fact:agreement}, we have that
\[
\sum_g (\hat{S}^W_g)_{\alice \alice'} \otimes (\hat{M}^{(\Point, W), u}_{g(u)})_{\bob \alice''} \approx_{\delta_S} \Id.
\]

We now prove \eqref{eq:qld-sg-cons2}. Line~\eqref{eq:qld-constructing-the-paulis-helper0} and Fact~\ref{fact:agreement} show that
\[ (\hat{S}^W_{[g(u)=a]})_{\alice \alice'} \approx_{\delta_S} (\hat{M}^{(\Point, W), u}_{a})_{\bob \alice''}\]
where the answer summation is over $a \in \F_q$. Left-multiplying this expression by $(\hat{S}^W_{[g(u)=a]})_{\alice \alice'}$ and using Fact~\ref{fact:add-a-proj} we get
\begin{align*}
(\hat{S}^W_{[g(u)=a]})_{\alice \alice'} &\approx_{\delta_S} (\hat{S}^W_{[g(u)=a]})_{\alice \alice'} \otimes (\hat{M}^{(\Point, W), u}_{a})_{\bob \alice''} \\
&\approx_\eps (\hat{S}^W_{[g(u)=a]} \cdot \hat{M}^{(\Point, W), u}_{a})_{\alice \alice'}
\end{align*}
where the second approximation follows from the following calculation:
\begin{align}
	& \E_u \sum_a \left \| (\hat{S}^W_{[g(u)=a]})_{\alice \alice'} \cdot \Big ( (\hat{M}^{(\Point, W), u}_{a})_{\alice \alice'} - (\hat{M}^{(\Point, W), u}_{a})_{\bob \alice''} \Big ) \ket{\hat{\psi}} \right \|^2  \\
	\leq & \E_u \sum_a \bra{\hat{\psi}} \Big ( (\hat{M}^{(\Point, W), u}_{a})_{\alice \alice'} - (\hat{M}^{(\Point, W), u}_{a})_{\bob \alice''} \Big )^2 \ket{\hat{\psi}} \\
	= & \E_u \sum_a \left \| (\hat{M}^{(\Point, W), u}_{a})_{\alice \alice'} - (\hat{M}^{(\Point, W), u}_{a})_{\bob \alice''} \ket{\hat{\psi}} \right \|^2 \\
	\leq & \eps
\end{align}
The first inequality follows from the fact that $\hat{S}^W_{[g(u)=a]} \leq \Id$ for all $a$, and the second inequality follows from the self-consistency of $\hat{M}^{(\Point, W), u}_{a}$ (established in Lemma~\ref{lem:qld-comm-cons}). Thus
\[
	(\hat{S}^W_{[g(u)=a]})_{\alice \alice'} \approx_{O(\delta_S + \eps)} (\hat{S}^W_{[g(u)=a]} \cdot \hat{M}^{(\Point, W), u}_{a})_{\alice \alice'}~.
\]
On the other hand, notice that
\begin{align*}
	&\E_u \sum_g \left \| (\hat{S}^W_{g} \cdot (\Id - \hat{M}^{(\Point, W), u}_{g(u)}))_{\alice \alice'} \ket{\hat{\psi}} \right \|^2 \\
	=& \E_u \sum_g \bra{\hat{\psi}} ((\Id - \hat{M}^{(\Point, W), u}_{g(u)}) \cdot \hat{S}^W_{g} \cdot (\Id - \hat{M}^{(\Point, W), u}_{g(u)}))_{\alice \alice'} \ket{\hat{\psi}} \\
	=& \E_u \sum_a \bra{\hat{\psi}} ((\Id - \hat{M}^{(\Point, W), u}_{a}) \cdot \hat{S}^W_{[g(u)=a]} \cdot (\Id - \hat{M}^{(\Point, W), u}_{a}))_{\alice \alice'} \ket{\hat{\psi}} \\
	\leq& O(\delta_S + \eps).
\end{align*}
Line~\eqref{eq:qld-sg-cons2} is thus obtained by absorbing $\eps$ into the function $\delta_S$.  
\end{proof}


The next lemma shows that we can construct local unitaries $V_A,V_B$ from the measurements $\hat{S}$  such that the shared state $\ket{\hat{\psi}}$ is close to $\ket{\aux} \otimes \ket{\EPR_q}^{\otimes \nqubits}$ for some state $\ket{\aux}$, and the observables $\tilde{W}^j(\tilde{u})$ are mapped to Pauli observables $\tau^W(e_j \tilde{u})$ acting on $\ket{\EPR_q}^{\otimes \nqubits}$. Using the consistency between the $\{\tilde{M}^{W,\tilde{u}}_a\}_{a \in \F_q}$ measurements and the points measurements $\{M^{(\Point,W),u}_a\}_{a \in \F_q}$ established in Lemma~\ref{lem:qld-construct-the-paulis}, and the consistency between the points measurements and the ``total Pauli measurements'' $\{M^{(\Pauli,W)}_h\}_{h \in \F_q^\nqubits}$ established in Lemma~\ref{lem:qld-win-implications}, we deduce that the total Pauli measurements must be close (up to the local unitaries $V_A,V_B$) to the Pauli projectors $\{\tau^W_h\}_{h \in \F_q^\nqubits}$ acting on $\ket{\EPR_q}^{\otimes \nqubits}$.

\begin{lemma}
\label{lem:qld-unitary}
There exists a function $\delta_\qld(\eps,m,d,q) = O(\delta_S^{1/4} + md/q)$ (where $\delta_S$ is defined in Lemma~\ref{lem:qld-4-7}) and unitaries $V_\alice$ acting on registers $\alice \alice' \alice''$ and $V_\bob$ acting on $\bob \bob' \bob''$ such that 
\begin{enumerate}
	\item There exists a state $\ket{\aux}$ on registers $\alice \alice' \bob \bob'$ such that
	\[
		\left \| V_\alice \otimes V_\bob \ket{\hat{\psi}}_{\alice \alice' \alice'' \bob \bob' \bob''} - \ket{\aux}_{\alice \alice' \bob \bob'} \otimes \ket{\EPR_q}^{\otimes \nqubits}_{\alice'' \bob''} \right \|^2 \leq \delta_\qld.
	\]
	
	\item For all $W \in \{X,Z\}, h \in \F_q^\nqubits$
	\begin{align*}
		V_\alice \, \Big(M^{(\Pauli,W)}_h \Big)_{\alice} \, V_\alice^\dagger &\approx_{\delta_\qld} \Id_{\alice \alice'} \otimes (\tau^W_h)_{\alice''} \\
		 V_\bob \, \Big(M^{(\Pauli,W)}_h  \Big)_{\bob}\, V_\bob^\dagger &\approx_{\delta_\qld} \Id_{\bob \bob'} \otimes (\tau^W_h)_{\bob''}
	\end{align*}
	where the \allowbreak{}$\allowbreak \approx$\allowbreak{} statements hold with respect to the state \[
\ket{\aux}_{\alice \alice' \bob \bob'} \allowbreak \otimes \ket{\EPR_q}^{\otimes \nqubits}_{\alice'' \bob''}.
\] 
\end{enumerate}
\end{lemma}

\begin{proof}

Define the linear map
\[
	V_\alice = \sum_{g_X,g_Z} (\hat{S}_{g_X,g_Z})_{\alice \alice'} \otimes (\tau^X(\coded(g_Z)) \tau^Z(\coded(g_X)))_{\alice ''}\;,
\]
where the $\tau^X(\cdot),\tau^Z(\cdot)$ act on $(\C^q)^{\otimes \nqubits}$. (Also note that the argument of $\tau^X$ is $\coded(g_Z)$ and the argument of $\tau^Z$ is $\coded(g_X)$.) We verify the unitarity of $V_\alice$:
\begin{align*}
&\begin{aligned}[t]
  &V_\alice V_\alice^\dagger = \sum_{g_X,g_Z} (\hat{S}_{g_X,g_Z})_{\alice \alice'} \\
  &\otimes (\tau^X(\coded(g_Z)) \tau^Z(\coded(g_X)) \tau^Z(\coded(g_X)) \tau^X(\coded(g_Z)) )_{\alice ''}
\end{aligned}  \\
&\begin{aligned}[t]
  &= \sum_{g_X,g_Z} (\hat{S}_{g_X,g_Z})_{\alice \alice'} \otimes \Id_{\alice ''}
\end{aligned}  \\
&\begin{aligned}[t]
  &= \Id_{\alice \alice' \alice''}
\end{aligned} 
\end{align*}
where in the first line we used that the measurement $\{ \hat{S}_{g_\xpt,g_\zpt} \}$  is projective and in the second line we used that the $\tau^Z$ and $\tau^X$ observables are self-inverse. The unitary $V_\bob$ is defined analogously.

Let $\{e_j\}_{j \in \{1,\ldots,t\}}$ denote the self-dual basis for $\F_q$ over $\F_2$ specified in Section~\ref{sec:subfields}, and let $\tilde{W}^j$ be the observables introduced in~\eqref{eq:def-tildewj}. Conjugating the observable $\tilde{W}^j$ acting on the registers $\alice \alice' \alice''$ by the unitary $V_\alice$ we get
\begin{align*}
	&V_\alice \, \tilde{W}^j(\tilde{u}) \, V_\alice^\dagger \\
  =& \sum_{g_X, g_Z} (-1)^{\tr( e_j(\coded(g_W) \cdot \tilde{u}))} \Big( \hat{S}_{g_X,g_Z}  \Big)_{\alice \alice'} \otimes \Big ( \tau^X(\coded(g_Z)) \tau^Z(\coded(g_X)) \tau^W(e_j \tilde{u}) \\
  &\tau^Z(\coded(g_X))^\dagger \tau^X(\coded(g_Z))^\dagger \Big)_{\alice ''} \\
	 =& \sum_{g_X, g_Z} \Big( \hat{S}_{g_X,g_Z}  \Big)_{\alice \alice'} \otimes \Big ( \tau^W(e_j \tilde{u}) \Big)_{\alice ''} \\
	 =& \Id_{\alice \alice'} \otimes \Big( \tau^W(e_j \tilde{u}) \Big)_{\alice ''}\;,
\end{align*}
where in the third line we used the (anti-)commutation relation
\begin{equation*}
\begin{aligned}[t]
  &\tau^X(\coded(g_Z)) \tau^Z(\coded(g_X)) \tau^W(e_j \tilde{u}) = \\
  &(-1)^{\tr( e_j(\coded(g_W) \cdot \tilde{u}))} \tau^W(e_j \tilde{u}) \tau^X(\coded(g_Z)) \tau^Z(\coded(g_X)) \;.
\end{aligned} 
\end{equation*}
An entirely analogous calculation shows that \[
V_\bob \, \tilde{W}^j(\tilde{u}) \, V_\bob^\dagger \allowbreak = \Id_{\bob \bob'} \allowbreak \otimes \Big( \tau^W(e_j \tilde{u}) \Big)_{\bob ''}.
\]

Next, from the self-consistency property of the observables $\tilde{W}^j$ specified in Lemma~\ref{lem:qld-construct-the-paulis} we get that for $W \in \{X,Z\}$,
\begin{equation*}
\begin{aligned}[t]
  &\E_{j \in \{1,\ldots,t\}} \E_{\tilde{u} \in \F_q^\nqubits} \\
  &\bra{\hat{\psi}} \left ( \tilde{W}^j(\tilde{u}) \otimes \Id - \Id \otimes \tilde{W}^j(\tilde{u}) \right)^\dagger \left ( \tilde{W}^j(\tilde{u}) \otimes \Id - \Id \otimes \tilde{W}^j(\tilde{u}) \right) \ket{\hat{\psi}} \leq \delta_S\;,
\end{aligned} 
\end{equation*}
which, since $\tilde{W}^j$ is Hermitian, is equivalent to
\begin{equation}
\label{eq:qld-unitary-1}
\E_{j \in \{1,\ldots,t\}} \E_{\tilde{u} \in \F_q^\nqubits} \bra{\hat{\psi}} \tilde{W}^j(\tilde{u}) \otimes \tilde{W}^j(\tilde{u}) \ket{\hat{\psi}} \geq 1 - \delta_S/2\;.
\end{equation}
Let $\ket{\vartheta} = V_\alice \otimes V_\bob \ket{\hat{\psi}}$, and for $W \in \{X,Z\}$ define the operator
\begin{align*}
	H_W &= \E_{j \in \{1,\ldots,t\}} \E_{\tilde{u} \in \F_q^\nqubits} \tau^W(e_j \tilde{u}) \otimes \tau^W(e_j \tilde{u}) \\
		&= \E_{j \in \{1,\ldots,t\}} \Big ( \E_{s \in \F_q} \tau^W(e_j s) \otimes \tau^W(e_j s) \Big)^{\otimes \nqubits} \\
		&= \Big ( \E_{s \in \F_q} \tau^W(s) \otimes \tau^W(s) \Big)^{\otimes \nqubits}
\end{align*}
where in the second line we used that for all $\tilde{u} \in \F_q^\nqubits$, $\tau^W(\tilde{u}) = \bigotimes_{i = 1}^\nqubits \tau^W(\tilde{u}_i)$ with $\tau^W(\tilde{u}_i)$ being the generalized Pauli $W$ observable acting on the $i$-th qudit. In the third line, we used the fact that for every fixed $j$, we have $\E_{s \in \F_q} \tau^W(e_j s) \otimes \tau^W(e_j s) = \E_{s \in \F_q} \tau^W(s) \otimes \tau^W(s)$. \eqref{eq:qld-unitary-1} is then equivalent to 
\[
	\bra{\vartheta} H_W \ket{\vartheta} \geq 1 - \delta_S/2\;,
\]
for $W \in \{X,Z\}$. 

Observe that
\begin{align*}
&\begin{aligned}[t]
  &\E_{s,s'} \tau^X(s) \tau^Z(s') \otimes \tau^X(s) \tau^Z(s') = \E_{s,s'} \sum_{a,b \in \F_q} \\
  &(-1)^{\tr(as')} \ketbra{a+s}{a} \otimes (-1)^{\tr(bs')} \ketbra{b+s}{b}
\end{aligned}  \\
&\begin{aligned}[t]
  &= \E_{s} \sum_{a\in \F_q} \ketbra{a+s}{a} \otimes \ketbra{a+s}{a}
\end{aligned}  \\
&\begin{aligned}[t]
  &= \proj{\EPR_q}
\end{aligned} 
\end{align*}
where to go from the first to the second line we used Lemma~\ref{lem:cancellation}. This implies that $H_X H_Z = \Big ( \E_{s,s' \in \F_q} \tau^X(s)\tau^Z(s') \otimes \tau^X(s)\tau^Z(s') \Big)^{\otimes \nqubits} = \ketbra{\EPR_q}{\EPR_q}^{\otimes \nqubits}$.
%
%
By the triangle inequality, we have
\[
	\left \| H_X \ket{\vartheta} - H_Z \ket{\vartheta} \right\| \leq \left \| \ket{\vartheta} - H_X \ket{\vartheta} \right\| + \left \| \ket{\vartheta} - H_Z \ket{\vartheta} \right\|\;.
\]
Note that for $W \in \{X,Z\}$,
\[
	\left \| \ket{\vartheta} - H_W \ket{\vartheta} \right\|^2 = 1 - 2\bra{\vartheta} H_W \ket{\vartheta} + \bra{\vartheta} H_W^2 \ket{\vartheta} \leq 2 - 2\bra{\vartheta} H_W \ket{\vartheta} \leq \delta_S\;,
\]
where in the inequality we used that $H_W \leq \Id$. Furthermore,
\begin{align*}
	\bra{\vartheta} H_W^2 \ket{\vartheta} &=\| H_W \ket{\vartheta} \|^2\\
	 &= \| \ket{\vartheta} + (H_W - \Id) \ket{\vartheta} \|^2 \\
	 &\geq (\|\ket{\vartheta}\| - \| (H_W - \Id) \ket{\vartheta} \|)^2 \\
	 &\geq (1 - \delta_S)^2 \\
	 &\geq 1 - 2\delta_S\;. 
\end{align*}
Therefore, 
\begin{align*}
2\sqrt{\delta_S} &\geq \left \| H_X \ket{\vartheta} - H_Z \ket{\vartheta} \right\|^2  \\
&= \bra{\vartheta} H_X^2 \ket{\vartheta} + \bra{\vartheta} H_Z^2 \ket{\vartheta} - 2\bra{\vartheta} H_X H_Z \ket{\vartheta} \\
&\geq 2 - 4\delta_S - 2\bra{\vartheta} H_X H_Z \ket{\vartheta} \;,
\end{align*}
which implies that 
%
\[
	\bra{\vartheta} (\Id_{\alice \alice' \bob \bob'} \otimes (\proj{\EPR_q}^{\otimes \nqubits})_{\alice'' \bob''}) \ket{\vartheta} \geq 1 - 2\delta_S - \sqrt{\delta_S} \;.
\]
Define the unnormalized state $\ket{\aux_0}$ on the registers $\alice \alice' \bob \bob'$ as
\[
	\ket{\aux_0}_{\alice \alice' \bob \bob'} = (\bra{\EPR_q}^{\otimes \nqubits})_{\alice'' \bob''} \cdot \ket{\vartheta}_{\alice \alice' \alice'' \bob \bob' \bob''}
\]
and define the normalized state $\ket{\aux} = \frac{1}{\| \, \ket{\aux_0} \, \|} \ket{\aux_0}$. 
The inner product between $\ket{\vartheta}$ and $\ket{\aux} \otimes \ket{\EPR_q}^{\otimes \nqubits}$ can thus be evaluated as
\begin{align*}
&\begin{aligned}[t]
  &\bra{\vartheta} (\ket{\aux} \otimes \ket{\EPR_q}^{\otimes \nqubits}) = \frac{1}{\| \, \ket{\aux_0} \, \|} \cdot \bra{\vartheta} (\ket{\aux_0} \otimes \ket{\EPR_q}^{\otimes \nqubits})
\end{aligned}  \\
&\begin{aligned}[t]
  &= \frac{1}{\| \, \ket{\aux_0} \, \|} \cdot \bra{\vartheta} (\Id_{\alice \alice' \bob \bob'} \otimes (\proj{\EPR_q}^{\otimes \nqubits})_{\alice'' \bob''}) \ket{\vartheta}
\end{aligned}  \\
&\begin{aligned}[t]
  &= \frac{1}{\| \, \ket{\aux_0} \, \|} \cdot \| \, \ket{\aux_0} \, \|^2
\end{aligned}  \\
&\begin{aligned}[t]
  &= \| \, \ket{\aux_0} \, \|
\end{aligned}  \\
&\begin{aligned}[t]
  &\geq \sqrt{1 - 2\delta_S - \sqrt{\delta_S}}
\end{aligned}  \\
&\begin{aligned}[t]
  &\geq 1 - O(\sqrt{\delta_S})\;.
\end{aligned} 
\end{align*}
This implies $\| \ket{\vartheta} - \ket{\aux} \otimes \ket{\EPR_q}^{\otimes \nqubits} \|^2 \leq O(\sqrt{\delta_S})$, which establishes the first item of the lemma.

We now establish the second item of the lemma. In what follows, all $\simeq$ and $\approx$ statements hold with respect to the state $\ket{\hat{\psi}}$. From Lemma~\ref{lem:qld-win-implications} we have the following implications: for all $W \in \{X,Z\}$ and on average over $u \in \F_q^m$,
	\begin{align}
	\label{eq:qld-unitary-2}
	M^{(\Pauli,W)}_{[g_h(u)=a]} \otimes \Id_{\bob \bob' \bob''} &\simeq_\eps \Id_{\alice \alice' \alice''} \otimes M^{(\Point,W),u}_a\;, \\
	\label{eq:qld-unitary-3}
	\Id_{\alice \alice' \alice''} \otimes M^{(\Point,W),u}_a &\simeq_\eps  M^{(\Point,W),u}_a \otimes \Id_{\bob \bob' \bob''}\;,
	\end{align}
	where $M^{(\Pauli,W)}_{[g_h(u)=a]} = \sum_{h : g_h(u) = a} M^{(\Pauli,W)}_h$. 
From Lemma~\ref{lem:qld-construct-the-paulis} we get that
	\begin{equation}
	\label{eq:qld-unitary-4}
		M^{(\Point,W),u}_a \otimes \Id_{\bob \bob' \bob''} \simeq_{\delta_S} \Id_{\alice \alice' \alice''}  \otimes \tilde{M}^{W,\ind_m(u)}_a\;.
	\end{equation}
	Using Fact~\ref{fact:triangle-for-simeq} with Equations \eqref{eq:qld-unitary-2}, \eqref{eq:qld-unitary-3}, and \eqref{eq:qld-unitary-4}, we get 
	\begin{equation}
		\label{eq:qld-unitary-5}
		M^{(\Pauli,W)}_{[g_h(u)=a]} \otimes \Id_{\bob \bob' \bob''} \simeq_{\sqrt{\delta_S}} \Id_{\alice \alice' \alice''}  \otimes \tilde{M}^{W,\ind_m(u)}_a\;,
	\end{equation}
	where we used that $\delta_S \geq \eps$. 
	
	Next, we have for all $u \in \F_q^m$ 
\begin{align}
&\begin{aligned}[t]
  &V_\bob \, \tilde{M}^{W,\ind_m(u)}_a \, V_\bob^\dagger
\end{aligned} \notag \\
&\begin{aligned}[t]
  &= \sum_{g_X, g_Z} \Big( \hat{S}_{g_X,g_Z} \Big)_{\bob \bob'} \otimes \Big ( \tau^X(\coded(g_Z)) \tau^Z(\coded(g_X)) \\
  &\tau^W_{g_W(u) - a}(\ind_m(u)) \tau^Z(\coded(g_X))^\dagger \tau^X(\coded(g_Z))^\dagger \Big)_{\bob ''}
\end{aligned} \notag \\
&\begin{aligned}[t]
  &= \sum_{g_X, g_Z} \Big( \hat{S}_{g_X,g_Z} \Big)_{\bob \bob'} \otimes \sum_{h : g_h(u) = g_W(u) - a} \\
  &\Big ( \tau^X(\coded(g_Z)) \tau^Z(\coded(g_X)) \tau^W_h \tau^Z(\coded(g_X))^\dagger \tau^X(\coded(g_Z))^\dagger \Big)_{\bob ''}
\end{aligned} \notag \\
&\begin{aligned}[t]
  &= \sum_{g_X, g_Z} \Big( \hat{S}_{g_X,g_Z} \Big)_{\bob \bob'} \otimes \sum_{h : g_h(u) = g_W(u) - a} \Big ( \tau^W_{h + \coded(g_W)} \Big)_{\bob ''}
\end{aligned} \notag \\
&\begin{aligned}[t]
  &= \sum_{g_X, g_Z} \Big( \hat{S}_{g_X,g_Z} \Big)_{\bob \bob'} \otimes \sum_{h : g_h(u) = a} \Big ( \tau^W_{h} \Big)_{\bob ''}
\end{aligned} \notag \\
&\begin{aligned}[t]
  &= \Id_{\bob \bob'} \otimes (\tau^W_{[g_h(u)=a]})_{\bob''}\;,
\end{aligned} \label{eq:qld-unitary-6}
\end{align}
	where in the second line we expanded out the definitions of \allowbreak{}$V_\bob$\allowbreak{} and \allowbreak{}$\tilde{M}^{W,\ind_m(u)}_a$\allowbreak{} (defined in \eqref{eq:tilde_M}); in the third line we expanded out the definition of \linebreak[4]{}$\tau^W_{g_W(u) - a}(\ind_m(u))$\allowbreak{} (defined in \eqref{eq:def-tauwu}) and used the fact that for all \allowbreak{}$h \allowbreak \in \F_q^M$\allowbreak{}, the inner product \allowbreak{}$h \cdot \ind_m(u)$\allowbreak{} is equal to \allowbreak{}$g_h(u)$\allowbreak{}; in the fourth line we used the identities
	\[
		\tau^X(s) \tau^Z_h \tau^X(s)^\dagger = \tau^Z_{h + s} \qquad \text{and} \qquad \tau^Z(s) \tau^X_h \tau^Z(s)^\dagger = \tau^X_{h + s}~;
	\]
	and in the fifth line is due to the following short calculation: for all fixed $g_W$ we have
	\begin{align*}
&\begin{aligned}[t]
  &\sum_{h : g_h(u) = g_W(u) - a} \tau^W_{h + \coded(g_W)} = \sum_{h: g_W(u) - g_h(u) = a} \tau^W_{h + \coded(g_W)} \\
  &= \sum_{h: g_W(u) + g_h(u) = a} \tau^W_{h + \coded(g_W)} = \sum_{h': g_{h'}(u) = a} \tau^W_{h'}~.
\end{aligned} 
\end{align*}
	The second equality from the fact that $\F_q$ is a field of characteristic $2$ (so subtraction is the same as addition), and the third equality follows from the fact that letting $h' = h + \coded(g_W)$, we have $g_{h'} = g_h + g_W$ (in other words, the low-degree encoding $h \mapsto g_h$ is linear in $h$).

	
	Let $\ket{\Delta} = \ket{\aux} \otimes \ket{\EPR_q}^{\otimes \nqubits}$. We show that $V_\alice \, \Big(M^{(\Pauli,W)}_h \Big)_{\alice} \, V_\alice^\dagger \approx_{\delta_\qld} \Id_{\alice \alice'} \otimes (\tau^W_h)_{\alice''}$ (the argument for the operators acting on registers $\bob \bob' \bob''$ proceeds identically). 
	\begin{align}
		&\sum_h \bra{\Delta} \left ( V_\alice \, M^{(\Pauli,W)}_h \, V_\alice^\dagger - (\tau^W_h)_{\alice''}  \right)^\dagger \left ( V_\alice \, M^{(\Pauli,W)}_h \, V_\alice^\dagger - (\tau^W_h)_{\alice''} \right) \ket{\Delta} \notag \\
		&= 2 - 2 \Re \, \sum_h \bra{\Delta} V_\alice\, M^{(\Pauli,W)}_h \, V_\alice^\dagger  \, (\tau^W_h)_{\alice''} \ket{\Delta} \notag \\
		&= 2 - 2 \, \sum_h \bra{\Delta} V_\alice \, M^{(\Pauli,W)}_h \, V_\alice^\dagger  \, \otimes (\tau^W_h)_{\bob''} \ket{\Delta} \label{eq:qld-unitary-7}
	\end{align}
	where in the last line we used that $(\tau^W_h)_{\alice''} \ket{\EPR_q}^{\otimes \nqubits} = (\tau^W_h)_{\bob''} \ket{\EPR_q}^{\otimes \nqubits}$.
	We now claim that
	\begin{align}
	&  \, \sum_h \bra{\Delta} V_\alice \, M^{(\Pauli,W)}_h \, V_\alice^\dagger  \, \otimes (\tau^W_h)_{\bob''} \ket{\Delta}  \notag \\
	&\approx_{md/q}  \E_u \sum_{\substack{h, h' : \\ g_h(u) = g_{h'}(u)}} \bra{\Delta} V_\alice \, M^{(\Pauli,W)}_h \, V_\alice^\dagger  \, \otimes (\tau^W_{h'})_{\bob''} \ket{\Delta} \label{eq:qld-unitary-8} \\
	&=  \E_u \sum_{a \in \F_q} \bra{\Delta} V_\alice \, M^{(\Pauli,W)}_{[g_h(u)=a]} \, V_\alice^\dagger \, \otimes (\tau^W_{[g_h(u)=a]})_{\bob''} \ket{\Delta} \notag \\
	&=  \E_u \sum_{a \in \F_q} \bra{\Delta} V_\alice \, M^{(\Pauli,W)}_{[g_h(u)=a]} \, V_\alice^\dagger \, \otimes V_\bob  \, \tilde{M}^{W,\ind_m(u)}_a \, V_\bob^\dagger \ket{\Delta} \label{eq:qld-unitary-9}
	\end{align}
	where the equality in \eqref{eq:qld-unitary-9} follows from the identity in \eqref{eq:qld-unitary-6}, and the approximation in \eqref{eq:qld-unitary-8} follows from the following calculation:
	\begin{align*}
		& \E_u \sum_{\substack{h \neq h' : \\ g_h(u) = g_{h'}(u)}} \bra{\Delta} V_\alice \, M^{(\Pauli,W)}_h \, V_\alice^\dagger  \, \otimes (\tau^W_{h'})_{\bob''} \ket{\Delta}  \\
		&= \sum_{h \neq h'} \left( \E_u \mathbf{1}[g_h(u) = g_{h'}(u)] \right) \bra{\Delta} V_\alice \, M^{(\Pauli,W)}_h \, V_\alice^\dagger  \, \otimes (\tau^W_{h'})_{\bob''} \ket{\Delta} \\
		&\leq \frac{md}{q}\;.
	\end{align*}
	In the last line we used that for for distinct $h \neq h'$, the polynomials $g_h \neq g_{h'}$ and thus by the Schwartz-Zippel lemma can only agree on at most $md/q$ fraction of points $u \in \F_q^m$. 
	
	Continuing from \eqref{eq:qld-unitary-9}, we use the first item of the Lemma which shows that $\| \ket{\Delta} - \ket{\vartheta} \|^2 \leq O(\sqrt{\delta_S})$ and equivalently $\| V_\alice^\dagger \otimes V_\bob^\dagger \ket{\Delta} - \ket{\hat{\psi}} \|^2 \leq O(\sqrt{\delta_S})$, so 
	\begin{align*}
		\text{\eqref{eq:qld-unitary-9}} &\approx_{O(\delta_S^{1/4})} \E_u \sum_{a \in \F_q} \bra{\hat{\psi}} M^{(\Pauli,W)}_{[g_h(u)=a]} \otimes \tilde{M}^{W,\ind_m(u)}_a \ket{\hat{\psi}} \\
		&\geq 1 - O(\delta_S^{1/2})\;,
	\end{align*}
        where the last line comes from \eqref{eq:qld-unitary-5}.
	Together with \eqref{eq:qld-unitary-7}, 
	\[
	\text{\eqref{eq:qld-unitary-7}} \leq O(\delta_S^{1/4} + md/q)\;.	
	\]
	This concludes the proof of the second item of the lemma.
\end{proof}

We now prove Theorem~\ref{thm:pauli-appendix}. 
\begin{proof}[Proof of Theorem~\ref{thm:pauli-appendix}]
	Let \allowbreak{}$\strategy \allowbreak = (\psi,\allowbreak M)$\allowbreak{} be a projective strategy for \allowbreak{}$\game^\pauli_\qldparams$\allowbreak{} that succeeds with probability at least \allowbreak{}$1 \allowbreak - \eps$\allowbreak{}, where \allowbreak{}$\ket{\psi}$\allowbreak{} is a bipartite state on registers \allowbreak{}$\alice \bob$\allowbreak{}. As mentioned at the beginning of Section~\ref{sec:commutation}, we assume that the state \allowbreak{}$\ket{\psi}$\allowbreak{} is padded with sufficiently many ancilla \allowbreak{}$\ket{0}$\allowbreak{} qubits. Let \allowbreak{}$\phi_\alice$\allowbreak{} be the following isometry mapping the register \allowbreak{}$\alice$\allowbreak{} to the registers \allowbreak{}$\alice \alice' \alice''$\allowbreak{} where \allowbreak{}$\alice'$\allowbreak{} and \allowbreak{}$\alice''$\allowbreak{} are isomorphic to \allowbreak{}$(\C^q)^{\otimes \nqubits}$\allowbreak{}:
	\[
		\phi_\alice: \ket{\theta}_\alice \mapsto V_\alice (\ket{\theta}_\alice \otimes (\ket{\EPR_q}^{\otimes \nqubits})_{\alice' \alice''})
	\]
	where $V_\alice$ is the unitary acting on $\alice \alice' \alice''$ from Lemma~\ref{lem:qld-unitary}. Define the isometry $\phi_\bob$ analogously. Lemma~\ref{lem:qld-unitary} then implies that there exists a state $\ket{\aux}$ on registers $\alice \alice' \bob \bob'$ such that
	\[
		\left \| \phi_\alice \otimes \phi_\bob \ket{\psi} - \ket{\aux} \otimes \ket{\EPR_q}^{\otimes \nqubits} \right \|^2 \leq \delta_\qld
	\]
	and for $W \in \{X,Z\}$,
	\begin{equation*}
\begin{aligned}[t]
  &\phi_\alice \, (M^{(\Pauli,W)}_h)_\alice \, \phi_\alice^\dagger \approx_{\delta_\qld} (\tau^W_h)_{\alice''} \\
  &\qquad \text{and} \qquad \phi_\bob \, (M^{(\Pauli,W)}_h)_\bob \, \phi_\bob^\dagger \approx_{\delta_\qld} (\tau^W_h)_{\bob''}
\end{aligned} 
\end{equation*}
	where the $\approx$ statement holds with respect to the state $\ket{\aux} \otimes \ket{\EPR_q}^{\otimes \nqubits}$. This shows that $\game^\pauli_\qldparams$ is a self-test for the strategy $\strategy^\pauli$ with robustness $\delta_\qld(\eps,m,d,q)$. We conclude the proof by making the identification of registers $\alice \alice'$ with $\alice'$ and $\bob \bob'$ with $\bob'$ to be consistent with the statement of Theorem~\ref{thm:pauli-appendix}. We also note that, when unpacking the dependence on parameters $\eps, m, d, q$, the function $\delta_\qld(\eps,m,d,q) = O(\delta_S^{1/4} + md/q)$ has the form $a(md)^a(\eps^b + q^{-b} + 2^{-bmd})$ for some universal constants $a > 1, 0 < b < 1$, as desired.
\end{proof}









\bibliographystyle{arxiv-alpha}
\bibliography{compression}

\end{document}